\documentclass[a4paper,fleqn]{cas-sc}

\usepackage[numbers]{natbib}
\usepackage[capitalize]{cleveref}
\usepackage{bm}
\usepackage{mathtools}
\usepackage{suffix}

\def\tsc#1{\csdef{#1}{\textsc{\lowercase{#1}}\xspace}}
\tsc{WGM}
\tsc{QE}
\tsc{EP}
\tsc{PMS}
\tsc{BEC}
\tsc{DE}

\RequirePackage[framemethod=default]{mdframed}
\newmdenv[skipabove=7pt,
skipbelow=7pt,
rightline=false,
leftline=true,
topline=false,
bottomline=false,
linecolor=gray,
backgroundcolor=black!5,
innerleftmargin=5pt,
innerrightmargin=5pt,
innertopmargin=5pt,
leftmargin=0.1cm,
rightmargin=0.1cm,
linewidth=4pt,
innerbottommargin=5pt]{cBox}

\newcounter{dummy} 
\numberwithin{dummy}{section}

\newtheorem{corollaryT}[dummy]{Note\linebreak}

\newenvironment{summary}{\begin{cBox} \textbf{Summary:}}{\end{cBox}}	

\DeclarePairedDelimiterX\MeijerM[3]{\lparen}{\rparen}%
{\begin{smallmatrix}#1 \\ #2\end{smallmatrix}\delimsize\vert\,#3}

\newcommand\MeijerG[8][]{%
  G^{\,#2,#3}_{#4,#5}\MeijerM[#1]{#6}{#7}{#8}}

\WithSuffix\newcommand\MeijerG*[7]{%
  G^{\,#1,#2}_{#3,#4}\MeijerM*{#5}{#6}{#7}}

\begin{document}
\let\WriteBookmarks\relax
\def\floatpagepagefraction{1}
\def\textpagefraction{.001}
\shorttitle{Heavy Quarkonium in Extreme Conditions}
\shortauthors{Alexander Rothkopf}
%\begin{frontmatter}

\title [mode = title]{Heavy Quarkonium in Extreme Conditions}                      
%\tnotemark[1,2]

%\tnotetext[1]{This document is the results of the research
%   project funded by the National Science Foundation.}

%\tnotetext[2]{The second title footnote which is a longer text matter
%   to fill through the whole text width and overflow into
%   another line in the footnotes area of the first page.}

\author[1]{Alexander Rothkopf}[type=author,
                        auid=000,bioid=1,
                        prefix=Dr.,
                        %role=Associate Professor,
                        orcid=0000-0002-5526-0809]
%\cormark[1]
\ead{alexander.rothkopf@uis.no}
\ead[url]{http://www.alexrothkopf.de}
\ead[url]{http://deeprtp.uis.no}
\ead[url]{http://www.ux.uis.no/~rothkopf}

\credit{Conceptualization and implementation of this review}

\address[1]{Faculty of Science and Technology, University of Stavanger, 4021 Stavanger, Norway}

\begin{abstract}
In this report we review recent progress achieved in the understanding of heavy quarkonium under extreme conditions from a theory perspective. Its focus lies both on quarkonium properties in thermal equilibrium, as well as recent developments towards a genuine real-time description, valid also out-of-equilibrium. We will give an overview of the theory tools developed and deployed over the last decade, including effective field theories, lattice field theory simulations, modern methods for spectral reconstructions and the the open-quantum systems paradigm. The report will discuss in detail the concept of quarkonium melting, providing the reader with a contemporary perspective. In order to judge where future progress is needed we will also discuss recent results from experiments and phenomenological modeling of quarkonium in relativistic heavy-ion collisions.
\end{abstract}

%\includegraphics{figs/grabs.pdf}

%\item First principles lattice QCD computations confirm that quarkonium in-medium modification is hierarchically ordered with their vacuum binding energy.
%\item Vital properties of in-medium quarkonium can be obtained from a QCD derived and non-perturbatively evaluated complex in-medium potential.
%\item The open quantum systems paradigm allows the derivation of a real-time description of heavy quarkonium in and out of equilibrium from first principles QCD using well defined approximations.
%\item Quarkonium melting is a genuinely dynamical process arising from the combination of screening and scattering processes, leading to environment induced decoherence.
%\item Quarkonium suppression in high energy heavy-ion collisions results from the subtle interplay of  quarkonium melting and recombination of quark antiquark pairs produced in the early stages .


\maketitle

\section{Introduction and intuition building}
\label{sec:intro}

It is not often to find a single species of particles whose study has propelled our understanding of the strong interactions as thoroughly as heavy quarkonium did and does. The bound states of a heavy quark and its anti-quark ($c$ and $\bar c$ is called charmonium, $b$ and $\bar{b}$ bottomonium), more than 40 years after their first observation, still elicit fascination from both experimentalists and theorists due to their versatile role played in high energy nuclear physics. In this review we will consider one of the central foci in contemporary theoretical quarkonium studies, i.e. their behavior under extreme environmental conditions, which typically refers to energy and baryon densities of the order of mega electron volts and above. It is timely to take account of the status of the field, as over the past decade it has seen a boost in activity, taking on the challenge to go beyond the purely static notion of in-medium quarkonium it has refined over the years and to embark on a quest towards a microscopic understanding of quarkonium real-time dynamics. 

Already in vacuum, heavy quarkonium constitutes a versatile laboratory of strongly interacting matter and of the underlying field theory of quantum chromo dynamics (QCD) (for a comprehensive review see \cite{Brambilla:2010cs}). As these color singlet states due to the OZI rule decay only into three gluons (single gluon decays are prohibited due to color, two gluon decays due to symmetries of the wavefunction) they represent exceptionally stable bound states with inverse life times or equivalently spectral widths of $\tau^{-1} = \Gamma\sim {\cal O}({\rm keV})$. This in turn leads to a significant fraction of their decays into dileptons $e^+e^-$ and $\mu^+ \mu^-$, which act as well accessible channels for their experimental detection. High precision data on their $T=0$ properties, including masses and widths have been accumulated, as compiled in the PDG in Ref.~\cite{Tanabashi:2018oca}, by a wealth of experiments, such as BELLE (KEK), BARBAR (SLAC), CLEO (FLAB), BESII+III (BEPC) and LHCb (CERN) to name only recent ones. Below the open heavy flavor threshold these states and their production in elementary collisions are well captured by the naive quark model and its quantum numbers for spin and color $[ \frac{1}{2}\otimes \frac{1}{2} = 0 \oplus 1, \; 3\otimes \bar 3 = 1\oplus8]$. Quarkonium like states above threshold, which exhibit quantum numbers beyond those available in the quark model, so called exotic XYZ states, are a current focus of quarkonium research at T=0. 

The large charm and bottom quark mass compared to the intrinsic scale of quantum fluctuations in QCD, $(m_Q/\Lambda_{\rm QCD}\ll 1)$, means that quarkonium in vacuum is amenable to a {\it non-relativistic} treatment (this fact will be further exploited in \cref{sec:EFTs}). In turn quarkonium states can be classified according to the well known scheme from atomic physics using spin, angular momentum and total spin as labels $^{2S+1}L_{J}$. Examples of so called S-wave states are the $^3S_1$ $\Upsilon$ and $J/\psi$, the bottomonium and charmonium vector channel ground states. The $\chi_{c1}$ and $\chi_{b1}$ states on the other hand are classified as P-wave states with $^3P_1$. Thanks to the different values of the charm and bottom mass a variety of  states exist, which exhibit a wide range of binding energies and related spatial extents, ranging from the deeply bound $\Upsilon$ with $E_{\rm bind}=1.1$GeV over $\chi_{b1}$ and $J/\psi$ with nearly degenerate $E_{\rm bind}\approx0.64$GeV to more weakly bound $\chi_{c1}$ with $E_{\rm bind}\approx0.2$GeV. The non-relativistic nature of $T=0$ quarkonium makes it possible to capture the properties of these states with good accuracy in a simple potential model, consisting of a Coulombic part at small separation distances and a linearly rising part at large distances, the so called {\it Cornell potential}. This model exhibits the two hallmarks of QCD, {\it asymptotic freedom} and {\it confinement}.  Consequently the many different quarkonium states, depending on their depth of binding, allow to explore in detail the physics associated with both phenomena. Heavy quarkonium at $T=0$ has also been explored using lattice QCD simulations where high precision post- and predictions of bound state masses have been achieved (see e.g. Fig.20 in Ref.\cite{Dowdall:2011wh}), providing a direct link between the microscopic theory of QCD and experiment.

The motivation to study heavy quarkonium under extreme conditions is intimately related to exploring the physics of strongly interacting matter in the early universe (for an introduction to the topic see e.g. \cite{Yagi:2005yb}). At around $10^{-20}$s after the Big Bang and at correspondingly high temperatures of $100$s of MeV the universe is expected to have been filled with nuclear matter in the form of its microscopic building blocks the quarks and gluons, forming a strongly correlated {\it quark-gluon plasma} (QGP). In order to understand how the strong interactions behave under such extreme conditions quarkonium can again play an important role. Due to the deep binding of e.g. $\Upsilon$, this bound state is expected to exist deep into the QGP phase, meaning that it can still act as well defined experimental observable there. At the same time the separation of scales between binding energy and temperature underlying this stability means that even under such extreme conditions theoretical tools developed originally in vacuum have a chance to be extended to provide meaningful insight. I.e. quarkonium promises to retains its role as well controlled QCD laboratory even in the context of strongly interacting matter in the early universe. 

Experimental efforts to recreate the conditions in the early universe have lead to the construction of a series of collider facilities and accompanying experiments specialized in relativistic heavy-ion collisions. Starting with SPS at CERN ($\sqrt{s_{NN}}\approx0.15$TeV), followed by RHIC at BNL ($\sqrt{s_{NN}}=0.2$TeV) and most recently the LHC again located at CERN ($\sqrt{s_{NN}}=2.76,5.02$TeV), higher and higher energy ranges for the collision of heavy nuclei, mainly gold and lead, have been explored. In the near future new accelerator facilities devoted to heavy-ion collisions are going online, among them NICA at JINR and FAIR at GSI, possible also a machine at JPARC. All of them feature experiments that are devoted toward the measurement of quarkonium properties under extreme conditions.

\begin{figure}
\centering
\includegraphics[scale=0.5,clip=true,trim=3cm 3cm 5cm 1cm]{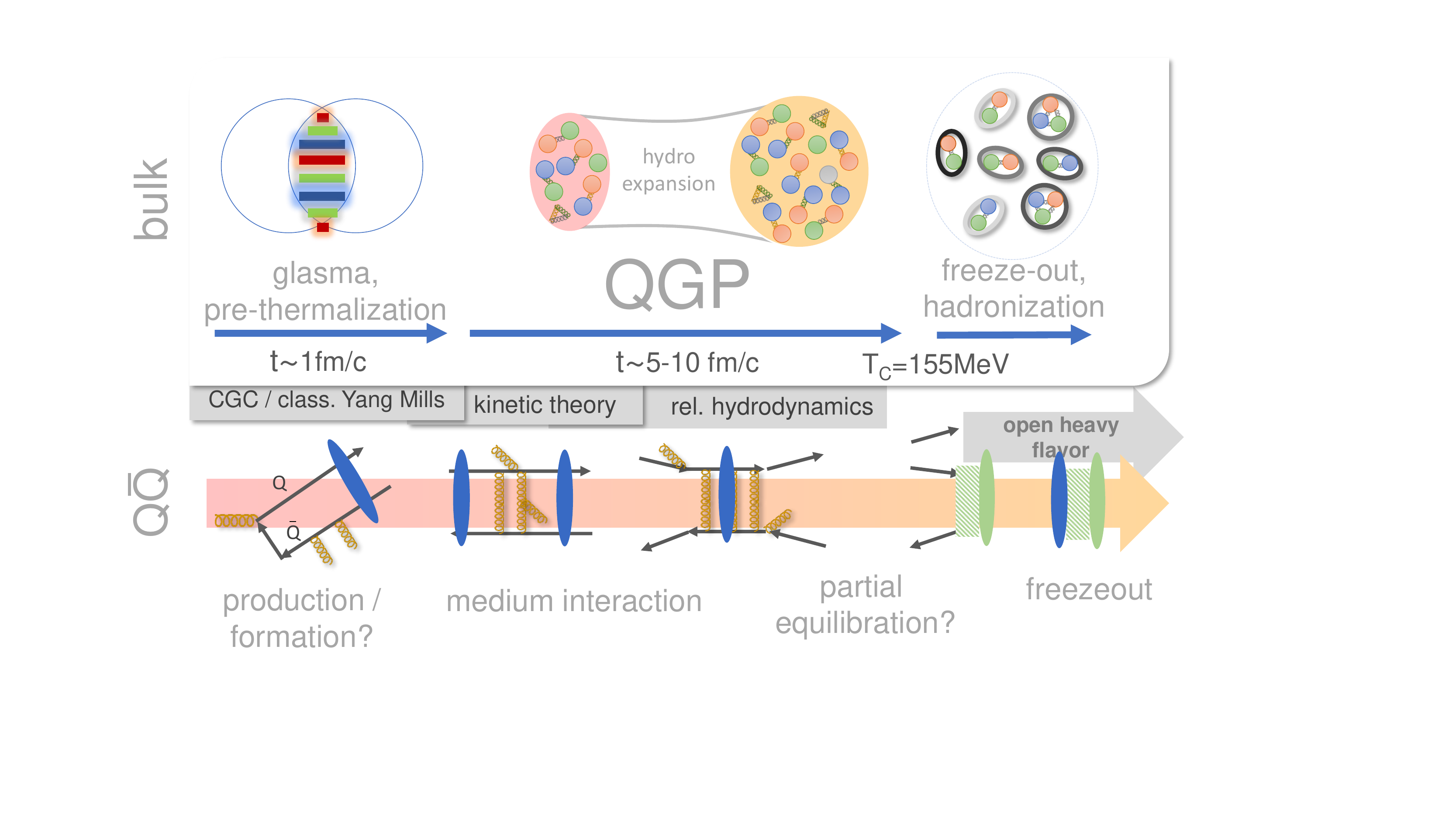}
\caption{(top) Sketch of the different stages of a relativistic heavy-ion collision, featuring the initial stages dominated by coherent color electric and magnetic fields of the glasma and the intermediate QGP phase, which terminates at the crossover transition to the hadronic phase at $T_c=155$MeV. In grey a chain of effective field theories is shown, which provide a dynamical description of the bulk medium physics. (bottom) The tentative life cycle of heavy quarkonium in the context of a relativistic heavy-ion collision.}\label{fig:HICSketch}
\end{figure}

Over the past decade the interplay of experiment, phenomenology and theory has lead to an improved understanding of the different stages of a heavy-ion collision as sketched in \cref{fig:HICSketch}. In the instant of the collision, the partons within the highly Lorentz contracted nuclear projectiles, which have formed a {\it color glass condensate} (CGC), are able to interact, leading to the generation of strong coherent color electric and color magnetic fields called the {\it glasma}. These subsequently fragment into light quarks and gluons, which efficiently exchange energy and momentum, so that after a a short pre-thermalization phase of around $1$fm a locally equilibrated collection of deconfined quarks and gluons emerges. This strongly correlated quark-gluon-plasma expands and cools over a period of $5-10$fm before it reaches the crossover transition at $T_C=155$MeV, where colored partons have to combine into color neutral hadrons. While the chemical abundances of these hadrons are established at around this temperature in what is called chemical freezeout, the resulting gas of hadrons may still interact and exchange energy and momentum until kinetic freezeout is reached.

How did we end up with this dynamical picture of a heavy-ion collisions. While QCD contains all the physics necessary to describe the evolution of the light bulk matter it has not yet been established how to compute the dynamics of a heavy-ion collision from first principles QCD based on e.g. lattice QCD simulations, due to the notorious sign problem. The fact that the QGP in a contemporary heavy-ion collision is created at temperatures of up to just $T\lesssim600$MeV, as deduced from hydrodynamic modeling of the bulk also prevents the use of weak-coupling methods, which are the bedrock of theory computations in elementary particle collisions.

Instead theory has developed {\it effective field theory} descriptions, capturing only the physics that is relevant at a particular stage of the collision. By focusing on a subset of degrees of freedom the problem at hand can be simplified enough so that a dynamical description may be directly derived from QCD, often in a non-perturbative manner. If such effective descriptions furthermore share a common range of validity they can be chained together to provide a consistent description of the dynamics. In the context of bulk matter a combination of classical statistical simulations of Yang-Mills fields in the earliest stage, followed by kinetic theory, which in turn smoothly matches onto a relativistic hydrodynamic description of the QGP has proven this strategy successful. In this report we will explore several avenues of how effective field theories can also support a comprehensive understanding of quarkonium in a heavy-ion collisions.

The life cycle of heavy quarkonium is aligned with the different stages of a contemporary heavy-ion collision. Due to the required energies and virtualities, only in the earliest moments of the collision can $c\bar{c}$ and $b\bar{b}$ pairs be created. At RHIC ${\cal O}(10)$ $c\bar{c}$ and ${\cal O}(1)$ $b\bar{b}$ pairs are expected to arise, while at LHC their number already increases to ${\cal O}(100)$ and ${\cal O}(10)$ respectively. A simple estimate based on the binding energy of individual states and the uncertainty principle suggests that the deeply bound quarkonium can form early on in the pre-thermal stage of the collision. Whether such early formation indeed takes place however remains an active field of research. 

A pre-formed state that finds itself immersed in the medium consisting of light bulk matter then interacts with this hot medium during the QGP phase, which will in general weaken its binding. Depending on the energy and time scales present, quarkonium may either survive the medium or its constituents may become decorrelated, i.e. the state melts. The inverse process is also fathomable if enough quark antiquark pairs are present, i.e. quarkonium states may be regenerated. If the lifetime of the medium is long enough the heavy quarks may even become kinetically equilibrated with the surrounding medium. They however will remain out of chemical equilibrium as their numbers do not change in a medium at temperatures $T\ll m_Q$. How exactly quarkonium states interact with a deconfined medium and in turn approach equilibrium with their surrounding is one of the central questions theory sets out to answer. 

At hadronization most of the decorrelated heavy quarks combine with a light quark to form open heavy flavor particles. On the other hand if a large enough number of pairs has been created early on, they may also recombine now into quarkonium. If an in-medium bound state exists (primordial or regenerated) it will transition into a vacuum state. Note that the dynamics of hadronization are among the least well understood aspects of quarkonium in heavy-collisions. After leaving the QGP a collection of quarkonium states will have formed, some of which may still decay into a lower lying state on the way to the detector. I.e. the final abundances as measured by experiment will come from the dilepton signals of vacuum states long after the QGP has ceased to exist. For theory the challenge consists of translating the knowledge gained about heavy quarkonium in a medium, relevant for the QGP phase, into such vacuum abundances in the end, if a connection to the measured yields is to be made.

As we will argue in this report, only a comprehensive microscopic understanding of all the different stages of the collision will reveal the nature of heavy quarkonium production. I.e. in order to improve our understanding of heavy quarkonium we are incentivised to also gain a better understanding of many other aspects of strongly interacting matter, be it the composition of the incoming nuclei projectiles or the general dynamics of hadronization.

One of the most influential early theory studies regarding heavy quarkonium in heavy-ion collisions is Ref.~\cite{Matsui:1986dk} by Matsui and Satz, which shapes intuition about quarkonium production to this day. The paper contains two ideas. The first is based on an analogy with electromagnetic plasma in thermal equilibrium. There, the presence of freely moving light charges screens static electric fields (Debye screening), i.e. fields generated by heavy test charges inserted into the medium. In turn it was argued that the presence of a quark gluon plasma will also weaken the binding between heavy quarks and thus destabilize existing quarkonium, as well as prevent the formation from thermal quark antiquark pairs. As we will discuss in this report, by now the presence of screening in a hot QCD medium has been thoroughly established and its strength explored using different non-perturbative means. 

The picture that Ref.~\cite{Matsui:1986dk} paints and which was made more quantitative in Ref.~\cite{Karsch:2005nk} is one of sequential quarkonium melting. Using a completely static notion of kinetically equilibrated in-medium quarkonium as time independent eigenstates of a hermitean in-medium Hamiltonian, the authors argued that there exist well defined melting temperatures at which quarkonium states instantaneously turn from bound to scattering state. This in turn invited the intuitive picture of quarkonium as a thermometer, indicating the temperature of its environment by what states remain bound and which do not. In this report we will discuss how the theory developments of the past decade have led to an overhaul of this static thermometer picture. Among others it turns out that the concept of melting temperature is not uniquely defined and thus plays a less decisive role as originally thought. While the intuition that those states, which are more weakly bound in vacuum, are more easily destroyed in a medium of course remains true, we will see that the physics of quarkonium melting cannot be cast into a static language and instead required a genuine dynamical treatment. This dynamical treatment will also unify how we think about the melting of charmonium and bottomonium. 

Even in everyday life there are two ways how to measure temperature. One may either bring into contact with the object of study, say a cup of hot earl gray tea, a second smaller system, a thermometer. This thermometer, after mutual equilibration is removed and its internal state interrogated on the, by that time, common temperature. This form of temperature measurement requires the two systems to be in contact for long enough that full thermalization is achieved. It is this static picture that is promoted by Ref.~\cite{Matsui:1986dk} for the case of quarkonium as thermometer. Note that while equilibration of charm quarks turns out to be realized to some degree at the highest available collider energies today, bottomonium so far does not show significant signs of equilibration.

\begin{figure}
\centering
\includegraphics[scale=0.5,clip=true,trim=0cm 9cm 9cm 0cm]{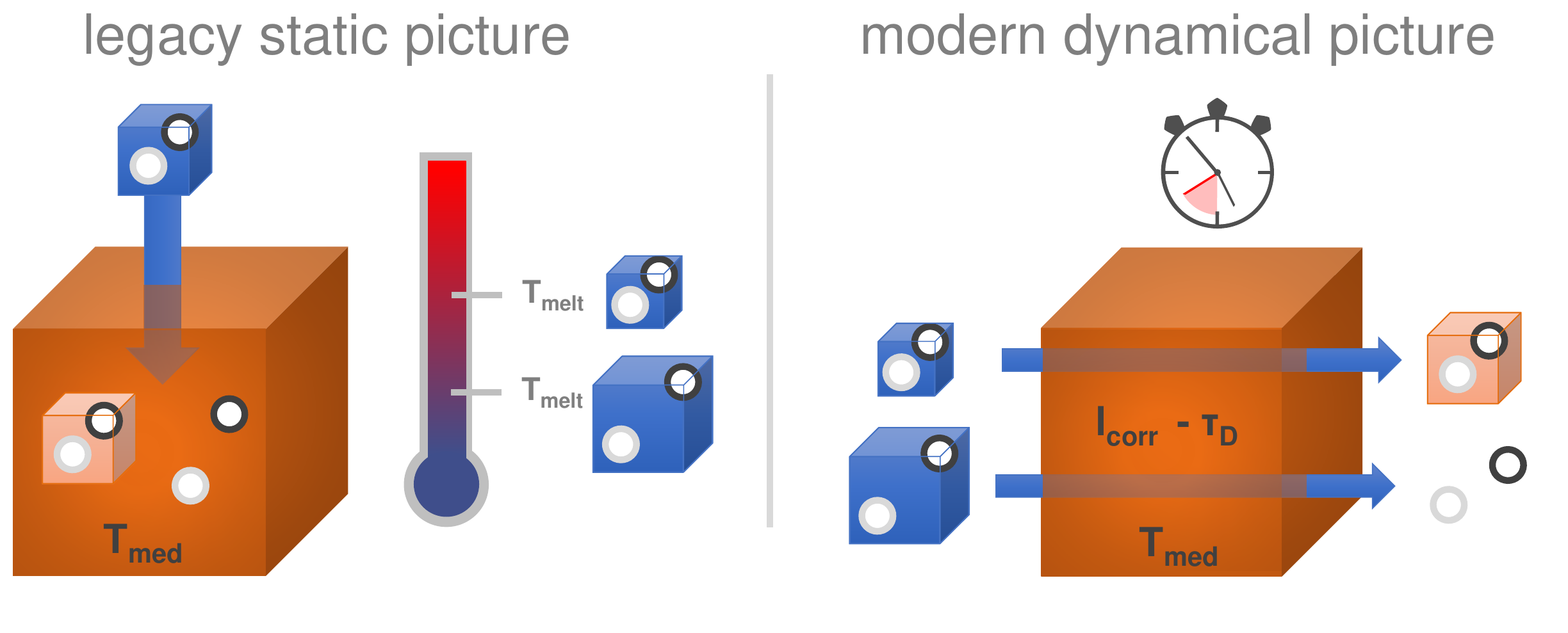}
\caption{(left) The legacy picture of heavy quarkonium as static thermometer. After fully kinetically equilibrating the heavy quark state one askes about the probability of its survival in the hot environment. (right) The modern picture of quarkonium melting as a dynamical process, in which the medium acts as a sieve that filters out quarkonium states over time, depending on the strength of their binding.}\label{fig:MeltingOldNew}
\end{figure}

On the other hand we may use a dynamical process to measure temperature. For our cup of tea it amounts to inserting e.g. a sugar cube and to observe how over time the chemical bonds are dissolved one by one. The speed by which this process happens informs us about the temperature present. The important difference to the static approach is that the measurement here can be performed even without reaching mutual equilibrium between sugar cube and tea. Besides the temperature of the environment it is now the timescales for which the process is allowed to proceed, which determine the fate of the sugar cube. Removing it quickly from a very hot surrounding will leave the bonds relatively intact, while even in a cold environment the sugar cube may melt as long as we wait long enough.

The change in intuition for quarkonium melting we invite the reader to explore in this report, and which is sketched in \cref{fig:MeltingOldNew} is analogous. When a quarkonium vacuum state is immersed in a hot medium, its binding will over time become weakened. It is then the interplay of how strongly the medium interferes with the binding at any instant, as well as the time spent in the medium that determines the survival of the quarkonium state. The hot medium in turn becomes a sieve that dynamically filters out more weakly bound states first while allows more strongly bound state to survive for longer time. One exciting recent development in quarkonium theory is that by using the {\it open-quantum-systems} framework, originally developed in the context of condensed matter theory, a dynamical description of such a small quarkonium system immersed in a hot QCD environment can be systematically derived from microscopic QCD. It turns out that quarkonium melting is actually intimately connected with the phenomenon of decoherence.

Let us now come to the second idea  of Ref.~\cite{Matsui:1986dk}, which took the melting picture and applied it straight forwardly to quarkonium production in heavy-ion collision. In turn it suggested an equally sequential suppression of quarkonium yields in the presence of a hot QGP medium in the collision center. This suppression arises both from the actual destabilization of a bound state, as well as diminished feed-down from higher lying states that melt more easily. In turn quarkonium suppression was positioned as a gold plated signal for the creation of a deconfined quark gluon plasma in heavy-ion collisions. For bottomonium only a small number of quark antiquark pairs are produced in the initial stages of current heavy-ion. Its production is thus dominated by primordial quarkonium traveling through a QGP in the collision center and indeed clear signs for such sequential suppression are observed.

On the other hand, the system originally discussed by Ref.~\cite{Matsui:1986dk}, i.e. charmonium, turns out to be an instructive tale how in general in a heavy-ion collision the physics of all stages need to be considered carefully to arrive at a comprehensive picture of quarkonium production. I.e. the production of $c\bar{c}$ pairs in the initial stages can significantly alter the production mechanism. Indeed as we understand today, quarkonium yields may be replenished by the recombination of quark antiquark pairs both during the QGP phase as well as at the moment of hadronization. This mechanism was noted early on by Matsui in Ref.~\cite{Matsui:1987im} but at that time not followed up further. With the prospect of the RHIC collider on the horizon, regeneration and recombination were considered in much more detail and have over time become vital ingredients in the explanation of measured charmonium yields at RHIC and most prominently at LHC. 

In other words, while the concept of sequential quarkonium melting in its modern dynamical fashion remains a valid guiding principle for the understanding of heavy quarkonium in extreme conditions, quarkonium suppression in heavy ion-collision arises from the subtle interplay of several different mechanisms, requiring insight into all different stages of the collision from heavy quark pair production over the QGP phase to hadronization.
 
What remains then of the role of quarkonium as probe of the quark-gluon plasma? While not as simple as a static thermometer, charmonium and bottomonium provide vital insight into the hot matter created in a heavy-ion collision. The fact that bottomonium at LHC appears to behave still as a genuine test-particle not equilibrated with its surroundings allows it to sample the full time evolution of the QGP. On the other hand charmonium at LHC shows clear signs of kinetic equilibration with its surroundings. This entails a loss of memory about the initial conditions of its evolution, positioning it as a probe of the late stages of the collision. Again the availability of bound states of different sizes and binding energies allows quarkonium to be a versatile laboratory of the strong interactions even under extreme conditions.

In this report we focus on the theory developments over the last decade that have significantly improved our understanding of the dynamical nature of heavy quarkonium in a hot medium and in turn in heavy-ion collisions. We start out in \cref{sec:theorytools} with a review of theory tools that are vital ingredients in contemporary studies of quarkonium in medium. These include a brief recap of (non-)equilibrium quantum field theory and the concept of spectral functions in \cref{sec:qftandspec}, effective field theories for heavy quarkonium in \cref{sec:EFTs}, lattice QCD in \cref{sec:latQCD}, modern methods for spectral function reconstruction in \cref{sec:specrec}, as well as the open quantum systems framework in \cref{sec:OQS}.

At first we will consider the idealized setting of fully kinetically equilibrated heavy-quarkonium immersed into a static medium in \cref{sec:qqbarequil}. Here it is possible to investigate the questions of screening in QCD (\cref{sec:screeningQCD}), the concept of the complex valued in-medium potential (\cref{sec:inmedpot}) and most importantly the in-medium spectral properties of quarkonium (\cref{sec:QQbarequilprop}) directly from first principles by utilizing modern concepts of effective field theories and lattice QCD. The concept of quarkonium melting in thermal equilibrium is discussed in \cref{sec:QQbarmelting} and its dynamical nature is emphasized.

In order to do justice to the dynamical nature of quarkonium even in thermal equilibrium \cref{sec:qqbarrealtime} sets out to review recent exciting progress made in the theory community in deriving genuine real-time descriptions for heavy quarkonium from first principles QCD. The open quantum systems framework is instrumental in this task and leads to so called {\it Lindblad equations}. From the recent activities in the community, we will highlight two derivations of quarkonium evolution equations, one that can be expressed in the language of wavefunctions (in \cref{sec:FVIFOQS}) and another in the language of distribution functions (in \cref{sec:BoltzmannEqOQS}). The former, for the first time, allows to construct a stochastic non-linear Schr\"odinger equation for quarkonium from first principles, which previously had been proposed on purely phenomenological grounds. The latter provides a direct connection between QCD and the Boltzmann equation deployed in many transport models of in-medium quarkonium. Not only has it become possible to better understand the dynamical role played by the imaginary part of the in-medium potential in the evolution of the quarkonium wavefunction but also to connect quarkonium melting to the phenomenon of wavefunction decoherence, as discussed in \cref{sec:DecDynMelt}.

In \cref{sec:qqbarhic} we return to the question of quarkonium production on heavy-ion collisions. In order to showcase the complexities of the task at hand we first discuss the physics of quarkonium production in $pp$ collisions in \cref{sec:qqbarpp} and briefly touch on cold nuclear matter effects in \cref{sec:qqbarcnm}. The main discussion of quarkonium production in nucleus-nucleus collisions follows in \cref{sec:qqbarprodhic} considering separately charmonium in \cref{sec:ccbarprodhic} and bottomonium in \cref{sec:bbbarprodhic}. Here we will scout for opportunities how progress in the real-time description of heavy quarkonium can help to discriminate between current models and eventually allows us to arrive at a genuine first principles based description of the measured production yields. 

This report aims at highlighting the recent progress achieved in the theory community and builds on the cumulative efforts of many research groups (for previous reviews on in-medium quarkonium see Refs.~\cite{Bazavov:2009us,Mocsy:2013syh}). Therefore the author attempts to provide the appropriate references for each discussed topic, and strongly encourages the citation of that original research.

\section{Theory tools}
\label{sec:theorytools}

In the following five sections we will give an overview to central theory tools used today in the study of quarkonium in extreme conditions. Their purpose is to provide a reference on concepts, techniques and quantities of interest prevalent in the literature. The included references also provide a starting point for newcomers to the field to embark on a more in-depth exploration.

In general, theoretical studies of quarkonium aim at an understanding of its physics from first principles. I.e. their starting point is a microscopic description of the strong interactions, the quantum field theory quantum-chromo-dynamics, defined with the following gauge invariant classical action
\begin{align}
S_{\rm QCD}[A,\psi,\bar{\psi},Q,\bar{Q}]=\int d^3x \Big( - \frac{1}{4}\big([D_\mu,D_\nu]\big)^2 +  \bar{\psi}(x)\big( i\gamma^\mu D_\mu - m_q\big)\psi(x) + \bar{Q}(x)\big( i\gamma^\mu D_\mu - m_Q\big)Q(x)  \Big)
\end{align}
formulated in terms of matrix valued gluon fields $A_\mu=A_\mu^a T^a$ where $T^a$ refers the generators of $SU(3)$. Their covariant derivative reads $D_\mu=\partial_\mu+igA_\mu$ with $g$ the strong coupling. We denote light quarks with $\psi$ and heavy quarks with $Q$. This action remains invariant under local rotations $\Lambda(x)$ in $SU(3)$ color space, acting as $\psi_a^\prime(x)=\Lambda_{ab}(x)\psi_b(x)$ and $(A^\prime_\mu)^{ab}(x)=\Lambda^{ac}A_\mu^{cd}(x)(\Lambda^\dagger)^{db}(x) +i (\partial_\mu\Lambda^{ac}(x))(\Lambda^{cb})^\dagger(x)$. The distinction between light and heavy quarks will be exploited in the section on effective field theories and open quantum systems. It is from fully quantized QCD that one wishes to deduce quarkonium properties in thermal equilibrium, as well as its real-time evolution in an evolving medium. 

\subsection{(Non-)equilibrium QFT and quarkonium spectral functions}
\label{sec:qftandspec}

In this section we will summarize how quarkonium can be described in the language of (non-)thermal field theory and introduce the concept of spectral functions. The quark anti-quark pair shall be immersed in a medium of quarks and gluons, which are not necessarily in thermal equilibrium. Excellent introductions to quantum fields in and out-of thermal equilibrium can be found in reviews \cite{Chou:1984es,Landsman:1986uw,Berges:2004yj} and textbooks \cite{Calzetta:2008iqa,Bellac:2011kqa,Laine:2016hma}.

In quantum field theory particles are understood as excitations of quantum fields, which propagate with a well defined {\it dispersion relation}. I.e. in contrast to quantum mechanics, particles do not appear as individual degrees of freedom and instead emerge from fluctuations of the fields. Thus to describe quarkonium particles, one needs a quantity, which encodes the real-time evolution of fluctuations of a heavy quark $\hat Q(x)$ and antiquark field $\hat{\bar{Q}}(x)$ (here $\hat Q$ refers to a Dirac spinor field describing either charm or bottom). The simplest possible candidate is the meson current $\hat M(x)=\hat{ \bar{Q}}(x)\hat Q(x)$. Two-point correlation functions of such a meson operator, similar to the variance of random variables, provide vital insight into the strength and form of quantum and statistical fluctuations that encode the properties of quarkonium particles. As an example let us consider the time ordered correlator $D$
\begin{align}
&D(\mathbf{x},\mathbf{x}_0,t,t_0)=\langle {\cal T}\hat M(\mathbf{x},t)\hat M^\dagger (\mathbf{x}_0,t_0) \rangle = {\rm Tr}\big[\hat \sigma {\cal T}\hat M(\mathbf{x},t)\hat M^\dagger (\mathbf{x}_0,t_0)\big]= \sum_n\langle n| \hat \sigma {\cal T}\hat M(\mathbf{x},t)\hat M^\dagger (\mathbf{x}_0,t_0) |n\rangle \label{eq:corrSK}\\
\nonumber &= \underbracket{\int  d[A^+,\ldots]\int d[A^-,\ldots] \langle A^+,\ldots|\hat \sigma|A^-,\ldots \rangle}_{{\rm initial\, conditions}}  \underbracket{\int_{A^+,\ldots}^{A^-,\ldots} {\cal D}[A,\psi,\bar{\psi},Q,\bar{Q}] {\cal T} M(\mathbf{x},t)M^\dagger (\mathbf{x}_0,t_0) e^{iS_{\rm QCD}[A,\psi,\bar{\psi},Q,\bar{Q}]}}_{\rm quantum \, dynamics}
\end{align}
Here $\hat \sigma$ denotes the initial density matrix of the system and the trace has been formally rewritten in the eigenstates $|n\rangle$ of the system Hamiltonian $\hat H_{\rm QCD}$.

\begin{figure}
\centering
\includegraphics[scale=0.5,clip=true,trim=0 14cm 12cm 0]{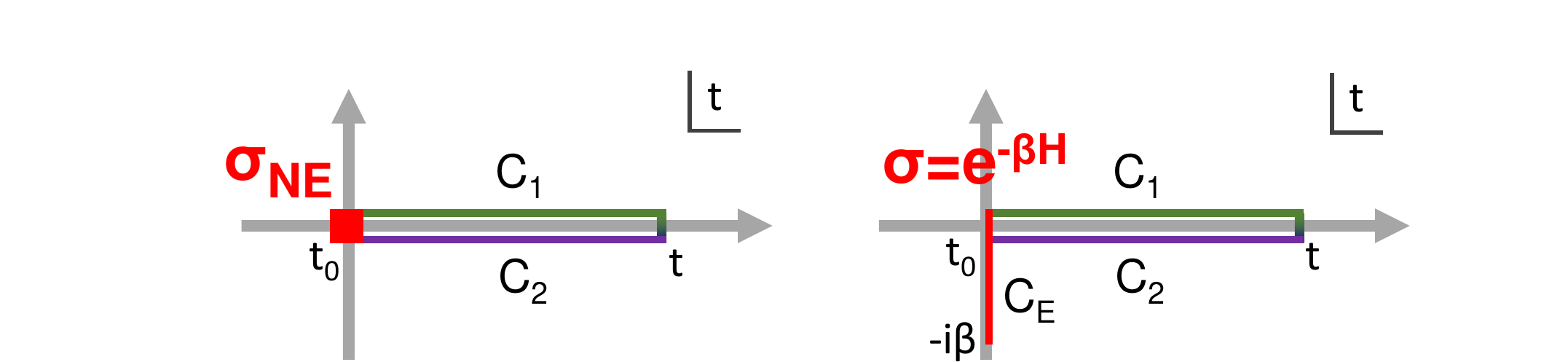}
\caption{(left) The general formulation of the quantum initial-value problem on the Schwinger Keldysh (SK) contour for an arbitrary density matrix $\hat \sigma$. Fields populate both the forward ${\cal C}_1$ and backward ${\cal C}_2$ branch. (right) Particular SK contour for a system in thermal equilibrium. The initial conditions $\hat \sigma=e^{-\beta \hat H}$ may be sampled efficiently as a path integral along a compact imaginary time branch ${\cal C}_E$.}\label{fig:SchwingerKeldysh}
\end{figure}

The path integral formulation in the second line makes explicit that one deals with an initial value problem. It is appropriately formulated along the Schwinger-Keldysh contour with a forward branch ${\cal C}_1$ and backward branch ${\cal C}_2$, as sketched in \cref{fig:SchwingerKeldysh}. The time ordering operator is denoted by ${\cal T}$. The fields $A,\psi,\bar{\psi},Q,\bar{Q}$ live on both branches of the contour, whose initial conditions ($^+$ fields for the forward and $^-$ fields for the backward contour) are sampled over statistically, weighted by the matrix elements of the density matrix. 

There are several different combinations of the meson operators, which we can consider on the Keldysh contour, all of which provide access to different facets of the field fluctuations. Depending on which part of the real-time contour the meson operator resides on we have
\begin{align}
&D_{11}= D(\mathbf{x},\mathbf{x}_0,t,t_0) = \langle {\cal T}\hat M(\mathbf{x},t)\hat M^\dagger (\mathbf{x}_0,t_0) \rangle, \label{eq:TOCorr}\\
&D_{12}= D^{>}(\mathbf{x},\mathbf{x}_0,t,t_0)= \langle \hat M(\mathbf{x},t) \hat M^\dagger (\mathbf{x}_0,t_0) \rangle, \label{eq:FWCorr}\\
&D_{21}= D^{<}(\mathbf{x},\mathbf{x}_0,t,t_0)= \langle \hat M^\dagger (\mathbf{x}_0,t_0) \hat M(\mathbf{x},t) \rangle, \label{eq:BWCorr}\\
&D_{22}= \tilde D(\mathbf{x},\mathbf{x}_0,t,t_0) = \langle \tilde{\cal T} \hat M(\mathbf{x},t)\hat M^\dagger (\mathbf{x}_0,t_0) \rangle. \label{eq:ATOCorr}
\end{align}
$D^>$ and $D^<$ are referred to as Wightman functions. In addition it is useful to define the retarded and advanced correlator, which involve the commutator of meson operators $[\hat A, \hat B]=\hat A \hat B- \hat B \hat A$, as
\begin{align}
D_{\rm R}&=  \langle[\hat M(\mathbf{x},t), \hat M^\dagger (\mathbf{x}_0,t_0)] \rangle \theta(t-t_0), \label{eq:RCorr}\\
D_{\rm A}&= -\langle [ \hat M^\dagger (\mathbf{x},t), \hat M (\mathbf{x}_0,t_0) ]\rangle \theta(t_0-t). \label{eq:ACorr}
\end{align}
The forward correlator $D^>$ e.g. describes the transition amplitude of finding a state created at the point $(\mathbf{x}_0,t_0)$ by $\hat M^\dagger$ at a later point in time $t$ at the position $\mathbf{x}$. The symmetric correlator on the other hand can be defined from the anticommutator $\{\hat A,\hat B\}=\hat A\hat B + \hat B \hat A$ as
\begin{align}
D_{\rm S}&=  \langle\{\hat M(\mathbf{x},t), \hat M^\dagger (\mathbf{x}_0,t_0)\} \rangle. \label{eq:SCorr}
\end{align}

The explicit form of $\hat M$ encodes, what kind of meson particles we are dealing with. In the study of heavy quarkonium two kinds of meson operators play a central role, local meson currents, as well as point split operators
\begin{align}
&\hat M_\Gamma(\mathbf{x},t)= \hat{ \bar{ Q}}(\mathbf{x},t)\Gamma \hat Q(\mathbf{x},t),\\
&\hat M_\Gamma(\mathbf{x},\mathbf{y},t)= \hat{ \bar{ Q}}(\mathbf{x},t)\Gamma W(\mathbf{x},\mathbf{y},t) \hat{ Q}(\mathbf{y},t).
\end{align}
In order for the correlator of point split operators to remain gauge invariant one conventionally inserts a straight {\it Wilson line} between the quark and antiquark fields, trading gauge dependence for an easier parametrizable path dependence
\begin{align}
W(\mathbf{x},\mathbf{y},t)={\rm exp}\Big[ - i g  \int_{\mathbf{x}}^{ \mathbf{y}} dz^\mu A_\mu(z) \Big]. 
\end{align}
The correlator of local currents encodes directly the properties of heavy quarkonium mesons while we will encounter the point split operators in the context of defining a potential acting between the heavy quark and antiquark.

The quantities $\Gamma$ residing in the meson currents are referred to as vertex operators. They allow us to select the spin and angular momentum properties of the fluctuations contributing to a correlator. Since \cref{eq:corrSK} involves a path integral, the propagation of the system over time includes all possible field configurations accessible to the system, weighted with the Feynman phase ${\rm exp}\big[iS_{\rm QCD}\big]$. It follows that all mesons states with quantum numbers compatible to $\Gamma$ contribute to this quantity and in turn computing $D$ will allow us to access the information of what states of a particular quantum number are present in the system.

To be specific, we may select under which representation of the Lorentz group the meson transforms. One example is the vector current with $\Gamma=\gamma_\mu$, which eponymously transforms as a Lorentz four-vector. Since we wish to select angular momentum and spin quantum numbers, the focus in the following lies on transformations in the rotation subgroup of the Lorentz group. Hence using for the meson operator the spatial components $\Gamma=\gamma_i$, its correlator receives contributions from all $Q\bar Q$ mesons with quantum numbers $J^{PC}=1^{--}$. In the case of bottomonium in vacuum this would correspond to at least three stable meson states $\Upsilon$, $\Upsilon^\prime$ and $\Upsilon^{\prime\prime}$. Access to channels with different quantum numbers requires an appropriate choice for $\Gamma$, some of which are listed in \cref{tab:mesonqn}. Since local $\Gamma$'s give access to only a limited set of quantum numbers, the computation of e.g. $\chi_{c2}$ with $J^{PC}=2^{++}$ requires the use of additional covariant derivative operators $ \stackrel{\leftrightarrow}{\mbox{ $D$}} = \stackrel{\leftarrow}{\mbox{ $D$}} -\stackrel{\rightarrow}{\mbox{ $D$}}$, as discussed in Refs.~\cite{Dudek:2010wm,Thomas:2011rh}. 
\begin{table}
\begin{tabular}[c]{|c|c|c|c|c|c|}\hline
$\Gamma$ & $J^{PC}$ & $^{2s+1}S_J$ & channel &$c\bar{c}$ & $b\bar{b}$\\\hline
$\gamma_5$ & $0^{-+}$ & $^1S_0(n)$& Pseudoscalar& $\eta_c$,$\eta_c^\prime$& $\eta_b$,$\eta_b^\prime$\\
$\gamma_j$ & $1^{--}$ & $^3S_1(n)$ & Vector ${\rm V}_{\rm ii}$& $J/\psi$, $\psi^\prime$& $\Upsilon$, $\Upsilon^\prime$, $\Upsilon^{\prime\prime}$\\
$\gamma_j\gamma_k$  & $1^{+-}$ & $^1P_0(n)$ & Tensor  & $h_c$ & $h_b$\\
1 & $0^{++}$ & $^3P_0(n)$ & Scalar &$\chi_{c0}$& $\chi_{b0}$, $\chi_{b0}^\prime$\\
$\gamma_5\gamma_j$ & $1^{++}$ & $^3P_1(n)$ & Axial-vector ${\rm AV}_{\rm ii}$ & $\chi_{c1}$& $\chi_{b1}$,$\chi_{b1}^\prime$\\
$\gamma_k i \stackrel{\leftrightarrow}{\mbox{ $D$}}_j + \gamma_j i \stackrel{\leftrightarrow}{\mbox{ $D$}}_k$& $2^{++}$& $^3P_2(n)$& &$\chi_{c2}$& $\chi_{b2}$,$\chi_{b2}^\prime$ \\\hline
\end{tabular}
\caption{A selection of common vertex operators and their quantum numbers. In addition we list the lowest lying quarkonium states contributing to the corresponding channel. (adapted from \cite{Bazavov:2009us,Dudek:2010wm})}\label{tab:mesonqn}
\end{table}

In order to access the information about which states contribute to the evolution of the meson operators, let us decompose the time ordered correlator into its commutator and anticommutator
\begin{align}
D(\mathbf{x},\mathbf{x}_0,t,t_0) &= \frac{1}{2} \langle \{\hat M(\mathbf{x},t),\hat M^\dagger (\mathbf{x}_0,t_0)\} \rangle + \frac{1}{2} \langle [\hat M(\mathbf{x},t),\hat M^\dagger (\mathbf{x}_0,t_0)]\rangle {\rm sign}_{\cal C}(t-t_0) \\
&= F(\mathbf{x},t,\mathbf{x}_0,t_0) - \frac{i}{2} \rho(\mathbf{x},t,\mathbf{x}_0,t_0) {\rm sign}_{\cal C}(t-t_0).
\end{align}
Here ${\rm sign}_{\cal C}(t-t_0)$ tracks the location of $t$ and $t_0$ on the SK contour.
We will see below that the spectral function $\rho$ encodes what states are accessible in the system, whereas the statistical function $F$ encodes how strongly those states are populated. 

The most intuitive representation of that information is found when introducing a Wigner transform, i.e. expressing the correlator in terms of relative and center of mass coordinates $r=x-x_0$ and $s=(x+x_0)/2$ and Fourier transforming in the former.
\begin{align}
D(\mathbf{s},t_s;\mathbf{p},\omega)=\int d^3r e^{i\mathbf{p} \mathbf{r}} \, \int_{-2t_s}^{2t_s} dt_r  e^{i\omega t_r} D(\mathbf{s},t_s;\mathbf{r},t_r).
\end{align}
The Wigner space spectral function $\rho(\mathbf{s},t_s;\mathbf{p},\omega)$ when viewed as a function of $\mathbf{p}$ and $\omega$ will represent a particle-like excitation as a narrow shell, located along a strip of $\omega(\mathbf{p})$. This is how the dispersion relation of the particle is read-off. As we are dealing with an initial value problem out of equilibrium, one has to keep in mind that the frequency resolution that can be achieved for the spectral function is limited by how much time $t_s$ has passed, as intuitively expected from the uncertainty principle. As it will come in handy later on, we note that the spectral function may be computed from different combinations of correlators introduced above. I.e. we may either express it in terms of the forward and backward correlator or as the imaginary part of the retarded correlator
\begin{align}
\rho(\mathbf{s},t_s;\mathbf{p},\omega)=\frac{1}{2} \big(D^>(\mathbf{s},t_s;\mathbf{p},\omega)-D^<(\mathbf{s},t_s;\mathbf{p},\omega)\big), \quad \rho(\mathbf{s},t_s;\mathbf{p},\omega)=-{\rm Im}\big[ D_{\rm R}(\mathbf{s},t_s;\mathbf{p},\omega)\big].
\end{align}

The Wigner transformed spectral function also allows us to connect directly with its counterpart in thermal equilibrium. Starting out of equilibrium one will observe that the function $\rho$ changes with the center of mass coordinate. As one approaches equilibrium this dependence weakens and eventually the fully thermal system will become independent of it, leaving us with $\rho(\mathbf{p},\omega)$. Let us also define a generalized occupation number $n$
\begin{align}
F(\mathbf{s},t_s;\mathbf{p},\omega)= \big( n(\mathbf{s},t_s;\mathbf{p},\omega) +\frac{1}{2} \big) \rho(\mathbf{s},t_s;\mathbf{p},\omega),
\end{align} 
which reduces to the standard occupation number in thermal equilibrium. This tells us that an enormous simplification of the system takes place in that $n$ not only becomes independent of the center-of-mass coordinates but furthermore becomes a function of $\omega$ only, leading to the Bose-Einstein distribution for a bosonic correlation function. I.e. in equilibrium, knowledge of the statistical function already implies knowledge of $\rho$. 

In thermal equilibrium the density matrix is given by $\hat \sigma={\rm exp}[-\beta \hat H]$, with $\beta=1/T$ the inverse temperature and $\hat H$ the Hamiltonian of the system. In that case the initial conditions part of \cref{eq:corrSK} may be rewritten in the form of a second path integral along a compact imaginary time axis spanning from $t=0$ to $t=-i\beta$. The start and end points of that imaginary time axis provide the initial conditions for the forward and backward branch of the Schwinger-Keldysh real-time contour. For operators residing on the imaginary time axis we may consider the Euclidean correlator
\begin{align}
D_{\rm E}(\mathbf{x}, \tau) =\langle M(\mathbf{x},\tau)M^\dagger (\mathbf{0},0) \rangle = D^{>}(\mathbf{x},-i\tau).
\end{align}
This correlator is related to the real-time forward correlator via analytic continuation. Due to the compactness of the imaginary time axis, a theory formulated in Euclidean time, after Fourier transform, only has access to the correlator on discrete imaginary frequencies, the so called Matsubara frequencies $\omega_n=2\pi n T$ 
\begin{align}
D_{\rm M}(\mathbf{x}, \omega_n) = \frac{1}{2\pi} \int_0^\beta d\tau  e^{i\omega_n\tau} D_{\rm E}(\mathbf{x}, \tau).
\end{align}
In addition the KMS relation tells us that the real-time correlators themselves become related to each other via 
\begin{align}
D^{>}(t,t_0)=D^{<}(t+i\beta,t_0).
\end{align}
Out of equilibrium we need to determine the off-diagonal correlators $D^>=D_{12}$ and $D^<=D_{21}$ separately to compute $\rho$. Via KMS we learn that $D^<(\omega)=e^{-\beta \omega}D^>(\omega)$ and already $D^>$ suffices.

We can pin down some of the properties of the spectral function in thermal equilibrium by explicitly computing $D^>$ and $D^<$, suppressing in the following the spatial dependence. Writing the trace as a sum over a complete set of eigenstates of the Hamiltonian $\langle n|$ and inserting a representation of unity we get
\begin{align}
D^>(t)=\frac{1}{{\rm Tr}[e^{-\beta \hat H}]} {\rm Tr}[e^{-\beta \hat H} \hat M(t) \hat M^\dagger(t_0)] = \frac{1}{Z[\beta]}\sum_{n,m} e^{-\beta E_n}e^{iE_n(t-t_0)}e^{-iE_m(t-t_0)}|\langle n| \hat M(t_0) |m\rangle |^2,
\end{align}
which subsequently leads to 
\begin{align}
\rho(\omega)=\frac{1}{Z[\beta]}\sum_{n,m} e^{-\beta E_n}\big[ \delta(\omega+E_n-E_m)-\delta(\omega+E_m-E_n)\big] |\langle n| \hat M(t_0) |m\rangle |^2. \label{eq:specfunc}
\end{align}
This expression first tells us that the spectral function is anti symmetric around the frequency origin $\rho(-\omega)=-\rho(\omega)$. As long as the product/contraction of $\Gamma\Gamma^\dagger$ remains positive (e.g. for $\Gamma_i$) and we utilize the same meson operator for creation and annihilation, $\rho$ is positive semi-definite for $\omega>0$. 

The vector channel spectral function on the other hand may contain both positive and negative contributions. To see this let us decompose it into its transverse and longitudinal components for general $\Gamma^\mu\Gamma^{\nu\dagger}$
\begin{align}
\rho^{\mu\nu}(\mathbf{p},\omega)=P^{\mu\nu}_{\rm T}\rho_T(\mathbf{p},\omega)+P^{\mu\nu}_{\rm L}\rho_L(\mathbf{p},\omega),
\end{align}
where the following projection operators are used
\begin{align}
P^{00}_{\rm T}=P^{0i}_{\rm T}=P^{i0}_{\rm T}=0, \quad P^{ij}_{\rm T}=\delta^{ij}-\frac{p^ip^j}{p^2}, \quad P^{\mu\nu}_{\rm L}= \frac{p^\mu p^\nu}{p^2}-g^{\mu\nu}-P^{\mu\nu}_{\rm T}.
\end{align}
For a particular choice of e.g. $\mathbf{p}=(p,0,0)$ the following relations are obtained
\begin{align}
&\rho_{\rm T}(\mathbf{p},\omega)=\frac{1}{2}\big(\rho^{22}(\mathbf{p},\omega)+ \rho^{33}(\mathbf{p},\omega)\big), \quad \rho_{\rm L}(\mathbf{p},\omega)=\frac{\omega^2-p^2}{\omega^2} \rho^{11}(\mathbf{p},\omega).\label{eq:translongdecomp}
\end{align} 
This tells us that while $\rho_{\rm T}\geq 0$ for positive frequencies, $\rho_{\rm L}$ can become negative below the light cone.  

Based on the canonical dimension of the naively defined composite meson operators ${\rm dim}[M]$ we can deduce the dimension of $\rho$. This provides a first guess of how the spectral function behaves at high frequencies, where $\omega$ is the dominant scale. I.e.
\begin{align}
\rho(\mathbf{p},\omega) \overset{\omega\gg \mathbf{p}}{\sim} \omega^\gamma,
\end{align}
with $\gamma={\rm dim}[M]+{\rm dim}[M^\dagger]-4$, which for quarkonium mesons turns out to be $\gamma=2$. For very heavy quarks the spectral expression simplifies, since neither quantum fluctuations nor statistical fluctuations are able to spontaneously produce a $Q\bar{Q}$ pair. More concretely one can argue \cite{Rothkopf:2009pk}, that for $m_Q \gg T$ the large mass of the heavy quarks leads to a suppression of Boltzmann factors with $E_n$ or $E_{m}$, whenever the intermediate states $|n\rangle$ or $|m\rangle$ in the spectral decomposition contain heavy quarks
\begin{align}
\rho(\omega)=(1- e^{-\beta \omega } ) \tilde{D}^>(\omega)
 =\frac{1}{Z}\sum_{m,n} |\langle m|\hat M|n\rangle|^2 \Big(e^{-\beta E_m}
 -e^{-\beta E_{n}}\Big)
 \delta(\omega-(E_{n}-E_m)).
\end{align}

In \cref{eq:specfunc} we have assumed the whole spectrum of the Hamiltonian to be discrete, which is the case when the theory is e.g. regularized on a finite space-time lattice. Then we understand that $\rho$ is simply composed of a sum of delta peaks. A peak exists at a certain frequency if the system Hamiltonian admits an energy level at that value and the matrix element $|\langle M \rangle|^2$ is non-vanishing. At zero temperature therefore one expects there to be a well defined ground state peak, clearly separated from higher lying excited states and eventually followed by densely spaced peaks above the continuum threshold, corresponding to unbound states. In thermal equilibrium, due to the sum over the medium states $|m\rangle$, the delta peaks below the threshold can cluster (i.e. the difference between $E_n$ and $E_m$ can be small) which leads to peak structures whose envelope exhibits a finite thermal width.

Well defined peak structures can be related to (quasi-)particle properties. 
Depending on the relative momentum $\mathbf{p}$ the position of a sharp spectral peak traces out the dispersion relation of that particle. At 
$\mathbf{p}=0$ it is simply the rest mass of that particle. Its binding energy can be read-off from the distance of the bound state peak from the onset of the continuum structure. On the other hand the width of a peak is related to its inverse lifetime, as it translates into a dampening in e.g. the forward correlator $D^>$. At finite temperature a finite thermal width does not necessarily imply that the state decays via annihilation of the constituent quark antiquark pair. Instead due to energy and momentum exchange with the medium degrees of freedom the particle may be excited into another state within the same color channel (singlet or octet) or even outside of that channel, signaling decoherence over time. The information of which state the particle transitions into however is encoded in higher correlation functions of meson operators and thus cannot be disentangled from an inspection of the two-point correlator spectral function.

If we consider $\Gamma_V=\gamma_\mu$, the area under the spectral peaks in the corresponding $\rho_V$, can be straight forwardly related to experimentally relevant decay processes involving dileptons \cite{Shifman:1998rb}. It is exactly the vector current $j_\mu^{\rm em}=q\bar Q\gamma_\mu Q$ with electric charge $q$ that couples to the photon field. At $T=0$ the prefactor to a spectral delta peak, to first order in QED perturbation theory, encodes the probability of the bound state to decay to a dilepton pair, via annihilation into a virtual photon. I.e. the R-ratio of decay into $e^+ e^-$ pairs is given by 
\begin{align}
R(s)= -\frac{4\pi^2}{s}\rho_V(s), \quad s=k_0^2-\mathbf{k}^2.
\end{align}  
In Ref.~\cite{Bodwin:1994jh} the decay probability of quarkonium at $T=0$ has been connected to a simple picture of a non-relativistic wavefunction determined by a potential, using the concepts of effective field theories discussed in the next section. The strength of spectral features for an individual state is related to the properties of the corresponding radial wavefunction at the origin.

At finite temperature it is the area under an in-medium spectral feature, weighted by the Bose-Einstein distribution, which can be related to the dilepton emission rate from fully thermalized heavy quarkonium \cite{McLerran:1984ay,Braaten:1990wp,Weldon:1990iw}. In the center of momentum frame of the emitted dileptons $\mathbf{p}_{\ell^-}=-\mathbf{p}_{\ell^+}$ and assuming that the energy of the emitted particles $\omega=E_{\ell^-}+E_{\ell^+}$ is sizably larger than twice their masses one obtains
\begin{align}
 \frac{{\rm d} N_{\ell^-\ell^+}}{{\rm d}^4x\,{\rm d}^4 P} = 
 \frac{2 q^2 e^4}{3 (2\pi)^5 \omega^2} % \theta(Q^2 - 4 m_\ell^2) 
 \big( 1 + \frac{2 m_\ell^2}{\omega^2}
 \big)
 \big(
 1 - \frac{4 m_\ell^2}{\omega^2} 
 \big)^\frac{1}{2} n_B(\omega) \Big[- \rho_V (\mathbf{p}=0,\omega) \Big]
 \;, 
\end{align}
where depending on whether charm or bottom quarks are involved the different electric charges $q_c=\frac{2}{3}$ and $q_b=-\frac{1}{3}$ need to be taken into account.

Another interesting property is encoded in the spatial component of the vector channel spectral function (i.e. using $\Gamma=\gamma^i$). The spectral structures in the low frequency regime may be related via linear-response theory \cite{Fetter,Kapusta:2006pm,Petreczky:2005nh} to a so called Kubo-formula for the heavy quark diffusion coefficient
\begin{align}
D_{Q}=\frac{1}{6\chi^{00}}\lim_{\omega\to 0}\frac{1}{\omega}\rho_{V_{ii}}(\omega,\mathbf{p}=0),
\end{align}
with $\chi^{00}$ the quark number susceptibility. Note that the order of limits is important here, first we need to take the spatial momentum to zero and then inspect the low frequency regime of the spectral function.

Besides encoding the particle content of a theory, spectral functions serve another important technical role. They allow us to relate the many different correlators introduced above via appropriate integral transformations. In particular it turns out that the correlator formulated in imaginary time is governed by the same spectral function as the real-time correlator. This fact will become essential when trying to extract real-time information from numerical lattice QCD simulations later on.

The first relation we consider is that between the spectral function and the Matsubara correlator
\begin{align}
D_M(\mathbf{p},\omega_n)=\int_{-\infty}^{\infty} \frac{d\omega}{\pi} \frac{\rho(\mathbf{p},\omega)}{\omega - i \omega_n}=\int_0^\infty \frac{d\omega}{\pi} \frac{2\omega \rho(\mathbf{p},\omega)}{\omega_n^2+\omega^2}.\label{eq:KLrep}
\end{align}
In the second step the antisymmetry of the spectral function has been used. Since the integral kernel here only decays with $1/\omega$, one has to make sure that the correlation function is still well defined, given the canonical dimension of the spectral function. That means that in practice UV divergent contributions to the spectral functions need to be subtracted for \cref{eq:KLrep} to make sense. Note that this spectral decomposition also tells us that while the Euclidean formulation of thermal field theory does not have access to the values of $D_M$ in between the Matsubara frequencies, they are well defined. Using the analytic continuation of the spectral decomposition we may obtain expressions for the retarded and advanced correlators
\begin{align}
D_{R/A}(\mathbf{p},p_0)=\pm i\int_{-\infty}^{\infty} \frac{d\omega}{\pi} \frac{\rho(\mathbf{p},\omega)}{p_0-\omega\pm i\epsilon} = \mp i D_M(\mathbf{p},i(p_0\pm i\epsilon)).
\end{align}

To arrive at the correlator in Euclidean time, the Fourier series over Matsubara frequencies needs to be carried out. Using the relation 
\begin{align}
\frac{1}{\beta}\sum_n\frac{e^{i\omega_n\tau}}{\omega-i\omega_n}=\frac{e^{-\omega\tau}}{1-e^{-\beta\omega}},
\end{align}
one obtains the following spectral decomposition
\begin{align}
D_E(\mathbf{p},\tau)&= \frac{1}{\pi} \int_{-\infty}^{\infty} \frac{e^{-\tau\omega}}{1-e^{-\beta\omega}}\rho(\mathbf{p},\omega)=\frac{1}{\pi} \int_0^\infty \frac{{\rm cosh}(\omega(\tau-\beta/2))}{{\rm sinh}(\omega\beta/2)}\rho(\mathbf{p},\omega),\label{eq:Euclrep}
\end{align}
where in the second line we have used the antisymmetry of the of the meson spectral function. The finite temperature kernel diverges as $\omega\to 0$, which is a manifestation of the fact that the spectral function is antisymmetric and thus has to vanish at the origin. It also reduces to a simple exponential falloff at vanishing temperature. Note that in case of the Matsubara correlator, the integral kernel itself is not temperature dependent, while the kernel of the Euclidean correlator is. 

Now that we have summarized the formal relations between different correlators and the spectral function, let us consider the non-interacting limit as an annalytically accessible example \cite{Karsch:2003wy,Aarts:2005hg}. Due to asymptotic freedom in QCD, this limit agrees with the infinite temperature limit.
Since at high temperatures binding of quarks into bound states will be impossible the expectation is that only the {\it continuum} persists above $\omega>2m_Q$. Thus contrary to a single-particle spectral functions that exhibit a simple delta peak at the mass of the quantum field, the free meson spectral function shows a non-trivial broad structure at high frequencies. In addition it can also feature a remnant structure at low frequencies related to the {\it transport peak} in an interacting theory. At vanishing momentum the explicit form reads
\begin{align}
\rho_\Gamma(\omega,T) = \frac{N_c}{8 \pi^2} \,
\Theta(\omega^2- 4 m ^2)\, \omega^2 \, \tanh\left(\frac{\omega}{4T}\right)
\sqrt{1-\left(\frac{2m}{\omega}\right)^2}\cdot\;\left[a_\Gamma + \left(\frac{2m}{\omega}\right)^2 b_\Gamma \right] 
+ \frac{N_c}{3} \frac{T^2}{2} f_\Gamma\;
\omega \; \delta (\omega)
\quad ,\label{eq:freespecfunccont}
\end{align}
where the coefficients $a_\Gamma$, $b_\Gamma$ and $f_\Gamma$ take on specific values for different channels. At asymptotically high energies most channels, including the vector and axial vector one show the scaling $\rho(\omega)\sim \omega^2$ expected from dimensional grounds. Only for $\Gamma=\gamma_0$ and $\gamma_0\gamma_5$ a cancellation occurs and the dependence on $\omega$ drops out. 

At vanishing frequencies a remnant of the physics below the light cone persists in the form of a delta peak in the vector and axial vector channel, which would correspond to an infinite diffusion constant. At finite coupling it is expected that both channels contain a washed out counterpart of this delta peak close to zero frequencies. This is the transport peak related to heavy quark diffusion. Modeling its shape based on Brownian motion and a Langevin equation \cite{Petreczky:2005nh} predicts a Breit-Wigner form, where the width of the peak is inversely proportional to the diffusion constant $D$. Note that an explicit delta peak at the origin leads to a constant contribution in the Euclidean correlator, which, as we will discuss later may lead to complications in the extraction of spectral functions from lattice QCD simulations. In contrast the constant $f_\Gamma$ vanishes for the scalar and pseudoscalar channel. In turn it is expected that also in the interacting theory these channels do not contain a transport peak.

\begin{summary} Quarkonium particles in quantum field theory are described by local meson current correlators, different combinations of which can be formally constructed. All particles with quantum numbers compatible with those selected by vertex operators $\Gamma$ contribute to such a meson correlator. We can relate different correlators (retarded, Matsubara, etc.) to a common spectral function $\rho$ via integral transforms. Expressed in relative momentum and relative frequency $\rho(\mathbf{p},\omega)$ encodes bound state particles as peaked features and in turn allows us to read-off their masses, binding energies and lifetimes. At large frequencies the spectral function exhibits a continuous structure related to unbound pairs, which in most channels diverges with $\omega^2$ in accordance with dimensionality. Some channels, such as the vector channel, at vanishing spatial momentum also contain an additional structure close to $\omega=0$ related to the diffusion of heavy quarks, the so called transport peak.
\end{summary}

\subsection{Effective field theories of heavy quarkonium}
\label{sec:EFTs}

In this section we review how the separation between energy scales in the quarkonium system allows us to simplify its description with non-relativistic language. To this end we will consider the two effective field theories Non-relativistic QCD (NRQCD) \cite{Caswell:1985ui} and potential NRQCD (pNRQCD) \cite{Brambilla:1999xf}, which have found various applications in the study of in-medium heavy quarkonium. For a pedagogical introduction to the general concept of EFT's see e.g. \cite{Manohar:2018aog}, to NRQCD at $T=0$ in particular e.g.  Ref.~\cite{Grinstein:1998xb}. A comprehensive review of both NRQCD and pNRQCD can be found in Ref.~\cite{Brambilla:2004jw}.  

Heavy quarkonium is exceptional among mesons, as the masses of charm and bottom quarks arrange ideally so that a hierarchy of well separated scales emerges
\begin{align}
m_Q \gg m_Q v \gg m_Q v^2, \quad m_Q\gg \Lambda_{\rm QCD}\sim 0.2-0.5{\rm GeV}, \quad  m_Q\gg\epsilon_{\rm medium} \;.
\end{align}
Let us consider the characteristic relative velocity of the heavy quark within a bound state $v=|\mathbf{p}|/m_Q$. Attributing the mass splitting of e.g. the lowest lying bottomonium and charmonium S-wave states of roughly $\Delta\sim 500$MeV to the average kinetic energy available $\langle m_Q v^2\rangle $ it follows \cite{Thacker:1990bm} that $\langle v_b^2\rangle\approx0.1$ and $\langle v_c^2\rangle\approx 0.3$. In turn we find that the so called {\it hard scale} $m_Q$ of the rest mass lies well above the {\it soft scale} $m_Q v$, which is related to the momentum exchanged between the quark antiquark pair. For systems that permit a perturbative description, it can be shown that the soft scale is related to the inverse Bohr radius of the bound state.  At even lower energies one finds the {\it ultrasoft scale} $m_Q v^2$ related to the binding energy of the two-body system. In addition, the rest mass is much larger than the intrinsic scale of quantum fluctuations in QCD, $\Lambda_{\rm QCD}$. We will later on consider quarkonium in a heavy-ion collision, where at current collider facilities temperatures up to around $T\approx 0.6$GeV have been achieved. We may thus for the time being assume another separation of scales to hold, i.e. between the heavy quark mass and the characteristic energy density of the environment $\epsilon_{\rm medium}$.

At asymptotically high temperature another scale hierarchy emerges (for a more detailed discussion see e.g. \cite{Arnold:2007pg}), involving the temperature $T$ and the QCD coupling $g$
\begin{align}
T \gg gT \gg g^2 T \gg g^4 T.
\end{align}
In this weak-coupling context at the scale $gT$ color electric fields begin to be screened within the thermal medium. Thus $gT\sim m_D$ is related to the concept of the electric or Debye screening mass of gluons. The next lower scale is $g^2T$, where also color magnetic fields become screened. The physics of magnetic screening even at high temperatures is genuinely non-perturbative. The lowest of the scales $g^4 T$ is the scale of the inverse mean free path at which a hydrodynamic long-wavelength description of the thermal medium becomes viable. 

When considering effective field theory descriptions for quarkonium in a thermal medium we will have to deal with the confluence of many of the scales listed above. While at high temperatures their separation is often apparent, in the non-perturbative context of quarkonium in heavy-ion collisions they may become entangled and care must be taken to ensure that scale separation arguments hold.

In general, effective field theories exploit hierarchies of scales in a systematic fashion in order to simplify the description of physical processes relevant to the user. In the context of heavy quarkonium we are e.g. interested in learning about whether a quark antiquark pair immersed into a hot medium can form a bound state or at least temporarily coalesce into a resonance. To understand the binding properties of such a two-body system, we are not concerned with how the heavy quark pair came into being in the first place. This is where EFT's play their strength: instead of having to deal with the whole intricacies of relativistic Dirac spinor fields we will see that the physics of bound state formation, involving energies at the order of the binding energy of such a state can be described instead by non-relativistic Pauli spinors. The physics of heavy quark creation and annihilation at a much higher energy scale in this sense is not relevant and thus not treated explicitly, it is said to be {\it integrated out}.

In order to set up an EFT description of heavy quarkonium four steps are required:
\begin{enumerate}
\item[A.] Identify the energy scale of interest
\item[B.] Identify the degrees of freedom relevant at that energy scale
\item[C.] Construct the most general Lagrangian from these d.o.f. compatible with the symmetries of underlying QCD. Assign each term an in general complex prefactor, a {\it Wilson coefficient}.
\item[D.] Determine the values of the Wilson coefficients by {\it matching}
\end{enumerate}

\subsubsection{Non-relativistic QCD}

To understand the basic ingredients of the construction of the effective field theory NRQCD, let us start out by considering processes that occur below some energy $\Lambda_{\rm EFT}$ scale which itself lies firmly below the hard scale $m_Q$ (A). I.e. the energies of the quark and gluon fields involved are smaller than what is necessary to create a heavy quark anti-quark pair. Since pair creation is a hallmark of relativistic field theory, its absence intuitively tells us that eventually a simpler non-relativistic description should emerge. To proceed one needs to determine what degrees of freedom are relevant in such as scenario (B). The Foldy-Tani transform \cite{Foldy:1949wa,Tani:1951ab}, known from the derivation of the relativistic corrections of the hydrogen atom, proves helpful in this context. Starting out from the relativistic Dirac Lagrangian (where for the time being explicit factors of c have been reinstated)
\begin{align}
\bar{Q}(x)\big( i\gamma^\mu D_\mu - mc\big)Q(x)
\end{align} 
with $D_\mu=\partial_\mu+\frac{ig}{c}A_\mu$, one introduces a unitary  field redefinition 
\begin{align}
&Q={\rm exp}\big[-\frac{i}{2m_Q c}\gamma^i D_i\big] Q_1(x),
\end{align}
in the form of an exponential, which contains a small dimensionless expansion parameter due to $m_Q$ in the denominator. A subsequent second field redefinition of the form
\begin{align}
&Q_1= {\rm exp}\big[ \frac{g\gamma_0\gamma_i}{4m_Q^2c^3}E_i\big] Q_2,
\end{align}
with $E_i=F_{0i}$, the electric field, defined via the temporal components of the field strength tensor $F_{\mu\nu}=\partial_\mu A_\nu-\partial_\nu A_\mu + \frac{ig}{c} \big[A_\mu,A_\nu\big]$, then leads to the following \cite{Barchielli:1986zs} approximate Dirac Lagrangian 
\begin{align}
{\cal L}_P=\bar{Q}_2(x)\Big[ \Big( \left( \begin{array}{cc} 1 & 0 \\ 0& -1 \end{array} \right)D_0 -mc \Big) + \frac{1}{2m_Qc^2} \left( \begin{array}{cc} \sigma_i & 0 \\ 0& \sigma_i \end{array} \right) B_i + \frac{1}{2 m_Q c} D_i^2 \Big]Q_2(x) + {\cal O}(1/m_Q^2).
\end{align}
This Lagrangian is the Pauli Lagrangian familiar from non-relativistic quantum mechanics, where to the order in the expansion considered here, the upper and lower components of the original Dirac four spinor $Q=(\psi,\chi)$ are completely decoupled. Since the rest mass only enters as a constant it too can be eliminated by a field redefinition. No pair creation processes are possible at this stage. I.e. we conclude that as long as the rest mass of the quarks is much larger than the characteristic canonical momentum $D_i/m_Q c \ll 1$ Pauli spinors should provide us with an appropriate set of degrees of freedom. 

In order to construct the most general Lagrangian of Pauli spinors (C) a consistent power counting scheme needs to be developed. Two strategies have been followed in the literature. On the one hand an expansion formulated in powers of "$\Lambda_{\rm EFT}/m_Q$" has lead to {\it heavy-quark effective theory} (HQET) (for a review see \cite{Grozin:2004yc}), which has been successfully deployed in the study of heavy-light particles, such as B and D mesons. On the other hand NRQCD has been developed based on organizing the expansion in the dimensionless small parameter $v$, the relative heavy quark velocity. At the lowest orders it agrees with the Foldy Tani result but allows to systematically extend the series to higher orders, eventually reproducing the QCD Dirac Lagrangian. Its lowest order terms read explicitly for the $\psi$ component
\begin{align}
{\cal L}_{\psi}&=
\psi^{\dagger} \Biggl\{ i D_0 + {c_k^{(1)}\over 2 m_1} {\bf D}^2 + {c_4^{(1)} \over 8 m_1^3} {\bf D}^4 
+ {c_F^{(1)} \over 2 m_1} { { \bm \sigma} \cdot g{\bf B}} 
+ { c_D^{(1)} \over 8 m_1^2} \left({\bf D} \cdot g{\bf E} - g{\bf E} \cdot {\bf D} \right) 
+ i \, { c_S^{(1)} \over 8 m_1^2} 
{\bm \sigma \cdot \left({\bf D} \times g{\bf E} -g{\bf E} \times {\bf D}\right) }
\Biggr\} \psi. \label{eq:NRQCDLagCont}
\end{align}
The NRQCD action for the anti-quark field ${\cal L}_\chi$ is obtained from the charge conjugation of ${\cal L}_\psi$ with $\psi^c=-i\sigma^2\chi^{*}$ and $A^c_\mu=-A^t_\mu$, as antiquarks transform under the $\bar{\mathbf{3}}$ representation of $SU(3)$. 

Note that each term has been assigned a complex valued prefactor $c_i$, a so called Wilson coefficients. These play an important conceptual and practical role in the construction of an EFT. Since we include in the EFT only d.o.f. with energies at or below $\Lambda_{\rm EFT}$ the remnants of the physics of those d.o.f. at higher energies must be able to manifest itself. This is where Wilson coefficients come into play (for the underlying theory of the renormalization group see \cite{Wilson:1974mb}). For the EFT to reproduce the physics of QCD faithfully below $\Lambda_{\rm EFT}$ the Wilson coefficients need to be tuned in a procedure called matching (D). I.e. one computes correlation functions of heavy quark fields both in QCD and the EFT and then require that they agree at energies below $\Lambda_{\rm EFT}$. In addition the symmetries of the microscopic theory provide apriori constraints on some of them. E.g. Lorentz invariance subtly reappears in NRQCD as the constraints $c_k=c_4=1$ and $2c_F-c_S-1=0$. If  $\Lambda_{\rm EFT}\gg \Lambda_{\rm QCD}$ the matching procedure can be carried out perturbatively, otherwise it requires fully non-perturbative methods, such as lattice QCD simulations, discussed in the following section. 

\Cref{eq:NRQCDLagCont} however is not yet all that contributes to the heavy quark dynamics at order $v^2$. Indeed somehow the the pair creation processes eventually need to find their way back in, if NRQCD is a systematic approximation of QCD. This is taken care of by the contributions of additional color singlet and color octet four-fermion interaction terms,
\begin{align}
{\cal L}_{\psi\chi}=\frac{f_1(^1S_0)}{m_Q^2}\psi^\dagger \chi \chi^\dagger \psi + \frac{f_1(^3S_1)}{m_Q^2}\psi^\dagger \mathbf{\sigma} \chi \chi^\dagger \mathbf{\sigma} \psi+ \frac{f_8(^1S_0)}{m_Q^2}\psi^\dagger T^a \chi \chi^\dagger  T^a \psi + \frac{f_8(^3S_1)}{m_Q^2}\psi^\dagger T^a \mathbf{\sigma} \chi \chi^\dagger T^a \mathbf{\sigma} \psi, \label{eq:fourfermicont}
\end{align}
whose prefactors encode the physics of gluons with energies of the order of the hard scale. Similarly the Fermi constant encodes the explicit physics of the weak gauge bosons. One should keep in mind that to consistently formulate NRQCD also the light degrees of freedom in QCD need to be restricted in their energy below $\Lambda_{\rm EFT}$. In turn additional interaction terms appear also for the light quarks and gluons, each with their own Wilson coefficient, which however are suppressed by inverse powers of the EFT cutoff scale. 

At least for bottom quarks, a perturbative determination of the Wilson coefficients is often possible, which tells us that the $c_i$'s within \cref{eq:NRQCDLagCont} start at unity and the first correction goes linearly in the strong coupling $c=1+{\cal O}(\alpha_S)$ including, as expected, logarithmic dependencies on the EFT cutoff. Most of the $f$'s in \cref{eq:fourfermicont} start out at ${\cal O}(\alpha_S^2)$ except for the octet $f_8(^3S_1)$, which goes as ${\cal O}(\alpha_S)$.

When speaking about NRQCD one always refers to a specific cutoff $\Lambda_{\rm QCD}$. Once we change its value, in principle, all Wilson coefficients need to be reevaluated to produce a consistent effective description. 

In QCD, quarkonium particles are described by correlation functions of meson operators as discussed in \cref{sec:qftandspec}. Using a similar stratgey as the Foldy-Tani transform in the Lagrangian, the form of the NRQCD counterparts of these operators can be derived. In the spirit of EFTs each term in the expansion is assigned a Wilson coefficient. For the vector and axial vector channel the explicit expressions read
\begin{align}
M^{\rm V}_k = b^{\rm V}_1\big( \psi^\dagger \sigma_k \chi\big) -\frac{b_2^{\rm V}}{6m_Q^2}\Big[ \psi^\dagger \sigma_k\big(-\frac{i}{2}\stackrel{\leftrightarrow}{\mbox{\boldmath $D$}}\big)^2\chi\Big] + {\cal O}\big(\frac{D^4}{m_Q^4}\big), \quad M^{\rm AV}_k =\frac{b^{\rm AV}_1}{m_Q} \Big[ \psi^\dagger \big(-\frac{i}{2}\stackrel{\leftrightarrow}{\mbox{\boldmath $D$}}\times \bf \sigma\big)_k\chi\Big] + {\cal O}\big(\frac{D^3}{m_Q^3}\big),
\end{align}
where the symmetric covariant derivative reads $\psi^{\dagger} \stackrel{\leftrightarrow}{\mbox{\boldmath $D$}} \chi 
\equiv \psi^{\dagger} ({\bf D} \chi)-({\bf D} \psi)^{\dagger}\chi$. 

How does the behavior of non-relativistic correlators differ from that of their QCD counterparts and in consequence what are the corresponding differences in the underlying spectral functions? Intuitively what we have done in setting up NRQCD is to separate the forward and backward propagating contributions to the spectral function in \cref{eq:specfunc}. I.e. similar to what happens when one introduces a large chemical potential for the heavy quark fields, only spectral structures at positive frequencies contribute. In addition we introduced a field redefinition with a phase ${\rm exp}[-im_Q t]$ to get rid of the rest mass term in the NRQCD Lagrangian. This in turn leads to a frequency shift of $2m_Q$ in the NRQCD quarkonium spectral function. I.e. the frequency origin of the NRQCD spectral function lies at the former threshold $2m_Q$ and a spectrum of bound states can now in principle extend to negative frequencies in this new coordinate system. The transport peak itself is not encoded anymore in these spectra. This leads to a simplification in the spectral representation, in particular when considering the in-medium Euclidean correlator, which is now connected to the equilibrium spectral function by a temperature independent integral kernel
\begin{align}
D_E^{\rm NRQCD}(\tau)=\int_{0>\omega_{\rm min}\gg -2m_Q}^{\Lambda_{\rm EFT}} \frac{d\omega}{2\pi} e^{-\omega\tau}\rho(\omega).\label{eq:EuclrepNRQCD}
\end{align}
Note that even though this correlator does not feature the periodicity usually associated with thermal relativistic correlators, it encodes the physics of a quarkonium particle fully kinetically equilibrated with its surroundings. 

Let us have a look at the explicit form of the S-wave and P-wave quarkonium correlators in the non-interacting theory \cite{Beraudo:2007ky}. They differ in that the vertex operator of the latter introduces additional factors of the momentum operator
\begin{align} 
&D^{\rm free\, NRQCD}_{\rm S-wave}(\tau,\mathbf{p})=2N_c\int \frac{d^3q}{(2\pi)^3} e^{-(E_{\mathbf{q}}+E_{\mathbf{p}+\mathbf{q}})\tau} = \frac{N_c}{4\pi^{3/2}}e^{-\tau p^2/4M}\Big(\frac{M}{\tau}\big)^{3/2},\\
&D^{\rm free\, NRQCD}_{\rm P-wave}(\tau,\mathbf{p})=2N_c\int \frac{d^3q}{(2\pi)^3} (\mathbf{q}+\mathbf{p}/2)^2 e^{-(E_{\mathbf{q}}+E_{\mathbf{p}+\mathbf{q}})\tau} = \frac{3N_c}{8\pi^{3/2}}e^{-\tau p^2/4M}\Big(\frac{M}{\tau}\big)^{5/2}.
\end{align}
This translates \cite{Aarts:2011sm} into the non-interacting spectral functions 
\begin{align}
&\rho^{\rm free\, NRQCD}_{\rm S-wave}(\omega,\mathbf{p}=0)=4\pi N_c\int \frac{d^3q}{(2\pi)^3} \delta(\omega^\prime-2 E_{\mathbf{q}}) = \frac{N_c}{\pi}M^{3/2}(\omega^\prime)^{1/2}\Theta(\omega^\prime),\\
&\rho^{\rm free\, NRQCD}_{\rm P-wave}(\omega,\mathbf{p}=0)=4\pi N_c\int \frac{d^3q}{(2\pi)^3} \mathbf{p}^2 \delta(\omega^\prime-2 E_{\mathbf{q}}) = \frac{N_c}{\pi}M^{5/2}(\omega^\prime)^{3/2}\Theta(\omega^\prime),
\end{align}
where the finite momentum simply induces a shift of the threshold $\omega^\prime=\omega-{\bf p}^2/4M$. In contrast to the relativistic spectral functions the high frequency behavior now differs between the S-wave and the P-wave, the latter containing a much stronger contribution at high frequencies. Note also that the free spectra start off from $\omega=0$, which corresponds to the threshold $2m_Q$ in the original unshifted frequency axis of QCD.  

The specific properties of the medium so far did not occur in the discussion of the setup of NRQCD. As long as $\Lambda_{\rm EFT}\geq T$, which is the case in current studies of heavy quarkonium, this is fine, since only the separation of scale $m_Q \gg \Lambda_{\rm EFT}$ has been exploited in setting up the EFT. 

\subsubsection{Potential Non-relativistic QCD}
\label{sec:pNRQCD}

If quarkonium binding properties determined by the physics at the ultrasoft scale are our only concern, then NRQCD still contains more explicit d.o.f. than necessary.  I.e. we can set the energy cutoff below the characteristic momentum $m_Q v \sim |\mathbf{p}| > \Lambda_{\rm EFT}$ and wish to integrate out those d.o.f. that cannot be excited spontaneously above $E\sim m_Q v^2$. The resulting EFT is called potential NRQCD, where the term potential formally refers to non-local Wilson coefficients, which are a hallmark of how pNRQCD describes heavy quarkonium physics.

The characteristic scale of quantum fluctuations in QCD $\Lambda_{\rm QCD}$ lies in between the various values of binding energies of different vacuum quarkonium states. At the same time the temperatures created in relativistic heavy-ion collisions can also easily reach the same magnitude. Therefore both selecting the relevant degrees of freedom and the matching of the corresponding Wilson coefficients differs, depending on the exact hierarchy of scales present. In case that $m_Q v \gg \Lambda_{\rm QCD}$ where a perturbative approach to integrating out the soft scale is applicable, the setup of pNRQCD has been thoroughly established. In the non-perturbative regime an equally robust understanding is still outstanding and remains an active field of research. 

Let us briefly summarize the most important ingredients to weakly coupled pNRQCD. In contrast to NRQCD, one now consider two cutoffs, one for the relative spatial momentum $m_Q > \Lambda_{\rm EFT}^{|{\bf p}|} > m_Q v $, which is the same as in NRQCD and one for the energy of the heavy quarks which is now restricted to $ m_Q v>  \Lambda_{\rm EFT}^{E}$. One further assumes that the relevant d.o.f. are again quarks and gluons, i.e. the actual particle content remains the same between NRQCD and weakly coupled pNRQCD. 

While one could stay with the field $\psi$ and $\chi$, it turns out that in order to write the pNRQCD Lagrangian in an intuitive fashion and to establish a systematic power counting, it is helpful to go over to what in this context is called a quarkonium wavefunction
\begin{align}
\Psi(\mathbf{x}_1,\mathbf{x}_2,t)_{ab} \sim \psi(\mathbf{x}_1,t)_a\chi^\dagger (\mathbf{x}_2,t)_b,
\end{align}
given by a point split product of NRQCD fields. Both the color structure and the spatial dependencies of this object may now be decomposed leading to an expression in terms of a color singlet $S(\mathbf{r},\mathbf{s},t)$ and color octet wavefunction $O(\mathbf{r},\mathbf{s},t)$
\begin{align}
&\Psi(\mathbf{x}_1,\mathbf{x}_2,t)= \Psi(\mathbf{r},\mathbf{s},t)= \\
&{\cal P}\Big[ {\rm exp}\Big\{ig\int_{\mathbf{x}_1}^{\mathbf{x}_2}\mathbf{A} d\mathbf{z}\Big\}\Big]]S(\mathbf{r},\mathbf{s},t) + {\cal P}\Big[ {\rm exp}\Big\{ig\int_{\mathbf{R}}^{\mathbf{x}_1}\mathbf{A} d\mathbf{z}\Big\}\Big] O(\mathbf{r},\mathbf{s},t) {\cal P}\Big[ {\rm exp}\Big\{ig\int^{\mathbf{R}}_{\mathbf{x}_2}\mathbf{A} d\mathbf{z}\Big\}\Big].
\end{align}
The appropriate choice of quark mass to use in these expressions is an active research topic, having lead to the definition of the renormalon subtracted mass \cite{Pineda:2001zq}. This form  of $\Psi(\mathbf{r},\mathbf{s},t)$ is advantageous, since the different transformation properties of $S$ and $O$ under color rotations induced by ultrasoft gluons are explicit. At the same time $\Lambda_{\rm EFT}^{p} > \Lambda_{\rm EFT}^{E}$ translates into the fact that relative distances $\mathbf{r}$ are always smaller than the the typical length scales of the light degrees of freedom. In turn the gauge fields that remain active as explicit degrees of freedom in pNRQCD enter the Langrangian via a multipole expansion.

The most general combination of color singlet and color octet heavy quark wavefunctions in the presence of ultrasoft light degrees of freedom can thus be written as
\begin{align}
\nonumber L_{\rm pNRQCD}=\int d^3\mathbf{r} {\rm Tr}\Big[& S^\dagger \big[ i\partial_0 - \big( \Big\{ c_1^S(r) ,\frac{\mathbf{p}^2}{2 \mu}\Big\} + c_2^S(r) \frac{\mathbf{P}^2}{2 M} + V_S^{(0)} + \frac{V_S^{(0)}}{m_Q} + \frac{V_S^{(1)}}{m_Q^2}\big) \big]S \\
\nonumber +& O^\dagger \big[ iD_0 - \big( \Big\{ c_1^O(r) ,\frac{\mathbf{p}^2}{2 \mu}\Big\} + c_2^O(r) \frac{\mathbf{P}^2}{2 M} + V_O^{(0)} + \frac{V_S^{(0)}}{m_Q} + \frac{V_S^{(1)}}{m_Q^2}\big) \big]O\Big]\\
+&V_A(r){\rm Tr}\Big[  O^\dagger \mathbf{r} g \mathbf{E} S + S^\dagger \mathbf{r} g \mathbf{E} O \Big]+V_B(r){\rm Tr}\Big[  O^\dagger \mathbf{r} g \mathbf{E} O + O^\dagger O\mathbf{r} g \mathbf{E}  \Big] +{\cal O}\big(r^2,\frac{1}{M^3}\big),\label{eq:pNRQCDcont}
\end{align}
with $\mu=m_Q/2$ the reduced mass and $M=2m_Q$ the total mass of the two heavy quarks. The nonlocal Wilson coefficients $V$ can carry a dependence on both the relative distance $\mathbf{r}$ the corresponding relative momentum $\mathbf{p}$, the center of mass momentum $\mathbf{P}$, as well as the spin operators of both quark and antiquark $\mathbf{S}_1$ and $\mathbf{S}_2$.

Let us have a look at the form of this Lagrangian. On the one hand \cref{eq:pNRQCDcont} exhibits the simple form of a Schr\"odinger Lagrangian in the first two lines, telling us that the evolution of the singlet and octet wavefunctions are governed by a potential. The terms $V_{S,O}^{(0)}$ refer to static potentials, which act even in the case of static quarks. The other terms are momentum and spin dependent corrections to these static potentials. On the other hand we are dealing with a genuine field theory here in which ultrasoft gluons are still contributing. Their influence is seen in the third line, inducing dipole-like transitions between the color singlet and color octet states and also within the color octet states. I.e. in pNRQCD even for static quarks, singlet and octet quarkonium states do not automatically evolve separately with a simple Schr\"odinger equation. As we will see later, we may however find situations where the effects of the dipole exchange can be summarized by a time independent contribution to the in-medium potential.

We must now answer the question how to determine the values of the potential terms. For the implementation of direct perturbative matching of pNRQCD using resummed hard-thermal loop perturbation theory see Ref.~\cite{Brambilla:2008cx}. 

One versatile strategy for matching is to relate the potential terms to expressions involving the real-time QCD {\it Wilson loop}
\begin{align}
W_\square(r,t)={\cal P} {\rm exp}\big[ig\oint_{r\times t} A^\mu dz_\mu\big],
\end{align}
where the gauge field is integrated over a rectangular path with spatial extent $r$ and temporal extent $t$. The starting point is to consider the correlator of point split meson operators, the NRQCD counterpart to the pNRQCD correlator of singlet wavefunctions.
\begin{align}
&\left\langle \int {\cal D}[S,O] S(x_1,x_2) S^\dagger(y_1,y_2) e^{iS_{\rm pNRQCD}} \right\rangle_{\rm medium}\\
 &= \left\langle \int {\cal D} [\psi,\chi]{\rm Tr}_{\rm color}\big[ \psi^\dagger (x_1) U(x_1,x_2) \chi(x_2) \chi^\dagger(y_2) U(y_2,y_1) \psi(y_1) \big] e^{iS_{\rm NRQCD}} \right\rangle_{\rm medium}\\
&= \left\langle G_\psi(y_1,x_1)U(x_1,x_2)G_\chi(x_2,y_2)U(y_2,y_1)   \right\rangle_{\rm medium}. \label{eq:singcorr}
\end{align}
Since the NRQCD Lagrangian is quadratic in the heavy quark fields, the path integral has been performed in the second line and one ends up with expressions in terms of heavy quark and antiquark propagators. 

The propagator $G_\psi(x,y)=\langle \psi^\dagger(x) \psi(y) \rangle$ is defined using the NRQCD Lagrangian in the standard way. If ${\cal L}_\psi=\int dx \int dy\psi^\dagger(x) K(x,y) \psi(y)$ then
\begin{align}
\int dz K(x,y)G_\psi(y,z) =\delta^4(x-y)
\end{align} 
and vice versa for $G_\chi$.
The form of G can be determined explicitly in the static case
\begin{align}
\big(i\partial_0-gA^0(x)\big)G_\psi(x,y)=\delta^{(4)}(x-y), \quad \Rightarrow \quad G(x,y)= \delta^{(3)}(\mathbf{x}-\mathbf{y})\theta(t_x-t_y){\rm exp}\Big\{ig\int_{t_x}^{t_y} A^0(s) ds\Big\},\label{eq:staticpropq}
\end{align}
where it reduces to a temporal {\it Wilson line} and in turn \cref{eq:singcorr} reduces to the Wilson loop. At the same time in pNRQCD at zeroth order in the multipole expansion one obtains a simple exponential in the static limit. Singling out the ultrasoft energy regime by considering late times, we may formally write 
\begin{align}
&\left\langle \int {\cal D}[S,O] S(x_1,x_2) S^\dagger(y_1,y_2) e^{iS_{\rm pNRQCD}} \right\rangle_{\rm medium} =  Z_s^0(r)\delta^{(3)}(\mathbf{x}_1-\mathbf{y}_1)\delta^{(3)}(\mathbf{x}_2-\mathbf{y}_2){\rm exp}\big[-it V_S^{(0)}(r)\big]\\
&\overset{t\gg 1/\Lambda_{\rm EFT}^E}{=} \left\langle \delta^{(3)}(\mathbf{x}_1-\mathbf{y}_1)\delta^{(3)}(\mathbf{x}_2-\mathbf{y}_2){\rm Tr}[W_\square] \right\rangle_{\rm medium}.
\end{align}
In turn we may connect the late time behavior of the real-time Wilson loop with the values of the static heavy quark potential
\begin{align}
V_S^{(0)}(r)=\lim_{t\to\infty} \frac{i\partial_t W_\square(r,t)}{W_\square(r,t)}.\label{eq:realtimepotedef}
\end{align}
Before we turn our attention to the evaluation of this expression a few remarks are in order. First of all \cref{eq:realtimepotedef} required us to stay within the lowest order of the multipole expansion. It is not apriori clear whether this approximation is justified and its validity has to be ascertained depending on e.g. the energy density of the medium surrounding the heavy quark fields. In contrast to the definition of the heavy quark potential often encountered in the lattice QCD literature \cref{eq:realtimepotedef} is formulated in Minkowski time and not in imaginary time. I.e. the Wilson loops are oscillatory functions with a real and an imaginary part and thus the value of the potential $V_S$ can be in general complex. 

Up to this point we have only considered the static potential for the singlet. Note that matching of the octet potential, as well as momentum dependent and spin dependent corrections is in principle possible. Following e.g. Ref.~\cite{Barchielli:1986zs} one can derive expressions for these potentials by rewriting the propagator $G$ in terms of a non-relativistic quantum mechanical path integral. In vacuum several finite mass correction terms to the singlet potential have already been determined in lattice QCD in Refs.~\cite{Koma:2006si,Koma:2006fw} and a non-perturbative definition of a color adjoint potential has been discussed in Ref.~\cite{Philipsen:2013ysa}. Similar results at finite temperature are however still outstanding.

In the following section we will turn our attention to lattice QCD simulations, which will provide us with the means to compute the potential and meson correlation functions in general in a genuinely non-perturbative fashion.

\begin{figure}
\centering
\includegraphics[scale=0.35, clip=true, trim= 0 5cm 8cm 0 ]{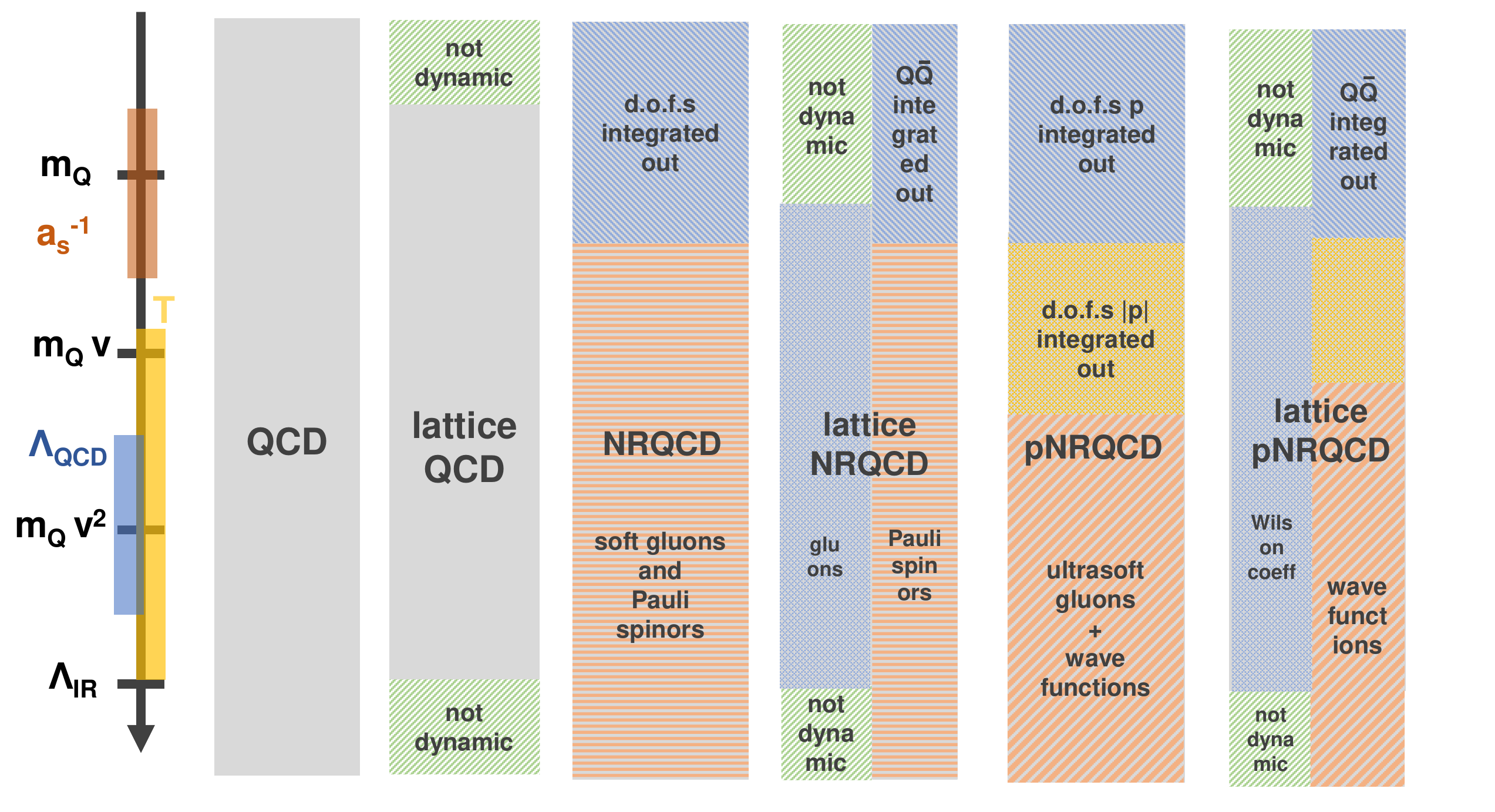}
\caption{Overview of the different scales and corresponding EFT constructions used in the study of heavy quarkonium. The underlying theory of QCD is valid at any energy scale, its regularization on a lattice introduces an infrared and UV cutoff given by the physical lattice volume and the lattice spacing respectively. In NRQCD both medium d.o.f. and heavy quarks are integrated out down to momenta below $m_Q$. The active d.o.f. freedom are thus soft gluons and Pauli spinors. The lattice implementation is formulated also in Pauli spinors with appropriate Wilson coefficients capturing the physics of the medium d.o.f. above $a_s^{-1}$. In pNRQCD the second energy cutoff now reaches down below $m_Q v$, below which wavefunctions and ultrasoft medium d.o.f. are active. In the current lattice implementation no active medium d.o.f. remain above $\Lambda_{\rm IR}$ and  the gluons and light quarks on the lattice  are used to compute the non-local Wilson coefficients.}\label{fig:EFTsketch}
\end{figure}

\begin{summary} The natural hierarchy of scales within quarkonium $m_Q\gg m_Q v\gg m_Q v^2$ as well as the fact that $m_Q \gg \Lambda_{\rm QCD}$ and in practice $m_Q \gg \epsilon_{\rm medium}$ allow us to simplify the description of quarkonium using non-relativistic language. In the EFT NRQCD the hard scale is integrated out and the relevant d.o.f. are Pauli spinors. In pNRQCD also the soft scale is integrated out. As long as the same relevant d.o.f. can be identified as in NRQCD, one may straightforwardly go over to a description in terms of color singlet and octet wavefunctions whose Lagrangian contains non-local Wilson coefficients called potentials. The propagation of these wavefunctions is determined by both potential and non-potential contributions. If the former dominate we can match the values of the static potential to the late time evolution of the real-time Wilson loop in QCD. The different energy scales, as well as the corresponding EFT setups are sketched in \cref{fig:EFTsketch}. 
\end{summary}

\subsection{Lattice QCD}
\label{sec:latQCD}

In this section we summarize relevant ingredients to non-perturbative numerical simulations of quarkonium physics, based on lattice regularized QCD. Several excellent textbooks provide a comprehensive introduction to this field \cite{Montvay:1994cy,Rothe:1992nt,Gattringer:2010zz}. The need for genuine non-perturbative methods in the study of quarkonium in extreme conditions is twofold. On the one hand, already at $T=0$ the physics of most quarkonium states, in particular charmonium, cannot be reliably captured using perturbation theory. On the other hand when we wish to understand heavy quark binding in a heavy-ion collision, the temperatures encountered at today's colliders are so close to the QCD crossover transition that a non-perturbative approach to the quark and gluon d.o.f. in the quarkonium environment is warranted. One indication is the large value of the trace anomaly, also called the interaction measure, in that temperature region \cite{Borsanyi:2013bia,Bazavov:2014pvz}. 

The starting point for lattice QCD is the discovery of Wilson \cite{Wilson:1974sk} that QCD can be regularized in a gauge invariant manner by placing its d.o.f. on a compact four dimensional spacetime grid $N_x \times N_y \times N_z \times N_t$ with lattice spacing $a_\mu$. Most often isotropic $a_\mu=a$ lattices are considered, but also anisotropies in temporal direction are used in practice with $a_0=a_t$ and $a_i=a_s$. Discretized fermion fields $\psi^{a,\alpha}(\mathbf{x})$ reside on the nodes of the grid. The role of gauge fields as parallel transporters for the quarks d.o.f. is made explicit and they are placed on the links of the lattice in the form of so called link variables $U^{\mu}(\mathbf{x})={\rm exp}[ig A_c^\mu(\mathbf{x}) T_c a_\mu]$ where no summation over $\mu$ is implied. These take on values in the group of $SU(3)$, while the gauge fields $A^\mu_c(\mathbf{x})T_c$ are elements of the generator algebra spanned by the Gell-Mann matrices $\lambda_c/2=T_c$. The finite lattice spacing introduces a UV cutoff, the finite box size an IR cutoff in the available momenta. The eigenvalues of the momentum operator corresponding to the central finite difference hence become
\begin{align}
\tilde p_i=\frac{2}{a_s}{\rm sin}\big(\frac{\pi n_j}{N_s}\big), \quad -N_s/2 < n_i \leq N_s/2.
\end{align}
As the number of degrees of freedom is finite, the corresponding Feynman path integral is well defined. In turn correlation functions that exhibit divergences in continuum computations also come out finite. In particular, the quarkonium spectral functions defined in \cref{sec:qftandspec} consist of only a finite number of delta peaks. The challenge of course lies in eventually having to take the continuum limit $a\to0$ and the thermodynamics limit $V\to \infty$ to recover the continuum theory of QCD. It is at latest at this point where a careful consideration of the renormalization of lattice regularized operators becomes essential.

In the coordinate space regularization it is possible to set up a numerical simulation prescription, which approximates Feynman's path integral in a non-perturbative fashion. There however exists an important restriction. While we wish to compute real-time correlation functions and the associated spectral functions on the real-time branches ${\cal C}_1$ and ${\cal C}_2$ of the SK contour, conventional lattice QCD simulations have access only to the compact imaginary time branch ${\cal C}_E$. On it, bosonic fields, according to the KMS relation, have to obey periodic boundary conditions in $\tau$, fermionic fields anti-periodic ones. I.e. while the finite extent of the box in spatial direction is a discretization artifact, the finite extent in imaginary time direction encodes vital physics, i.e. the inverse temperature of the system under consideration. Since the box extent in a numerical simulation is always finite, lattice QCD simulations are always performed at a finite temperature. In what is called a $T=0$ simulation the Euclidean time extent is made large enough that the induced temperature is negligible.

Analytically continuing real-time $t$ to Euclidean time $\tau$ we may express the partition function of the theory as
\begin{align}
Z=\int {\cal D}[U] \int{\cal D}[\psi,\bar\psi,Q,\bar{Q}] e^{-S_E},\label{eq:latPI}
\end{align}
where $S_E$ denotes the Euclidean QCD action and $DU$ represents the Haar measure integrating over the link variables in $SU(3)$ group space. The reason for restricting to ${\cal C}_E$ lies in the fact that only on the imaginary time axis the Feynman weight ${\rm exp}[iS]$ becomes purely real and bounded. In turn ${\rm exp}[-S_E]$ may be interpreted as an unnormalized probability distribution, opening up the toolbox of stochastic Monte-Carlo simulations.
This provides a practical path to evaluate the quantum statistical expectation values of operators
\begin{align}
\langle O(\tau_1,\tau_2,\ldots) \rangle = \frac{1}{Z} {\rm Tr}[e^{-\beta H}\hat O(\tau_1,\tau_2,\ldots) ]=\frac{1}{Z} \int {\cal D}[U] \int{\cal D}[\psi,\bar\psi,Q,\bar{Q}] O(U,\psi,\bar\psi,Q,\bar{Q};\tau_1,\tau_2,\ldots) e^{-S_E}. \label{eq:latOpPI}
\end{align}
The efforts related to simulating directly in real-time, as well as at finite Baryon density, both of which leads to a complex Feynman weight, are summarized under the label {\it sign problem} (for a review see e.g. Ref.~\cite{Gattringer:2016kco}).

The Euclidean action contains a term for the gauge fields and for the fermion degrees of freedom $S_E=S_E^g+S_E^f$. Many different implementations are possible in the discretized theory, all of which lead to the same continuum limit. The choice of discretization however determines how efficiently this limit is approached as the lattice spacing is reduced. Better convergence usually requires adding further terms to so called improved actions, increasing the numerical cost for their evaluation. The most naive choice for the gluons is the Wilson plaquette action, which for anisotropic lattices \cite{Klassen:1998ua} reads
\begin{align}
\quad S_E^g= \frac{\beta}{N_c}\sum_{\mathbf{n}}\sum_{i=1}^4\sum_{j<i}\Big( \frac{1}{\xi_0}{\rm ReTr}[1-U_{ij}(\mathbf{n})] + \xi_0 {\rm ReTr}[1-U_{0i}(\mathbf{n})]\big)
\end{align}
The vectors $\mathbf{n}=(x/a_s,y/a_s,z/a_s,\tau/a_\tau)$ and $\mathbf{m}$ denote the position on the four dimensional grid, the vectors $\mathbf{h}=(x/a_s,y/a_s,z/a_s)$ and $\mathbf{l}$ the spatial part. The central quantities $U_{{\mu\nu}}(\mathbf{n})=U_{\mu}(\mathbf{n})U_{\nu}(\mathbf{x}+\hat\mu)U_{\mu}^\dagger(\mathbf{n}+\hat{\nu}) U_{\nu}^\dagger(\mathbf{n})$ are called plaquettes, the closed products of links around the unit loop. The unit vector in $\mu$ direction is denoted with $\hat\mu$. Here $\beta=2N_c/g^2$ does not refer to the inverse temperature but stands for the inverse bare coupling, as is convention in the lattice community. The bare anisotropy parameter reads $\xi_0$. The Wilson action is invariant under local gauge transformations $G(\mathbf{n})\in SU(3)$, which act on the link variables as $U^\prime_\mu(\mathbf{n})=G(\mathbf{n})U_\mu(\mathbf{n})G^\dagger(\mathbf{n}+\hat\mu)$. Different improved actions for the gauge sector, such as the Iwasaki action \cite{Iwasaki:1985we}, have been developed, following the Szymanzik improvement program introduced in Refs.~\cite{Symanzik:1983dc,Symanzik:1983gh,Luscher:1984xn}

%\footnote{ For completeness let us note that perferct actions (see e.g. Ref.~\cite{Hasenfratz:1993sp}) on the classical level as well as improved actions without classical continuum limit (for a recent study see e.g. Ref.~\cite{Bietenholz:2010xg}) have been designed based on universality arguments.}.

In general the gauge invariant fermionic part of the Euclidean action can be expressed as a bilinear in terms of Grassmann valued quark fields. Since explicit matrix representations of Grassmann numbers in terms of complex numbers are numerically too costly, one instead carries out the Gaussian integral apriori and ends up with a fermion determinant. Most efficient simulation prescriptions exploit further that such a determinant can be expressed as a path integral over auxiliary bosonic fields.
\begin{align}
Z=&\int {\cal D}[U] \int {\cal D}[\bar \psi,\psi] {\rm exp}\Big\{- \int dx \bar\psi K \psi\Big\} e^{-S_E^g} = \int {\cal D}[U] {\rm det }K[U] e^{-S_E^g} = \int {\cal D}[U] \frac{1}{{\rm det }K^{-1}[U]} e^{-S_E^g} \\
&= \int {\cal D}[U]\int {\cal D}[\phi] e^{-S_E^g - \phi^* K^{-1}[U] \phi}.
\end{align}
The so called pseudo fermion field $\phi$ can be straight forwardly accommodated in numerical simulations. Note that while the matrix $K$ usually has a sparse banded structure its inverse is generally dense.

The treatment of light fermionic d.o.f. on the lattice, i.e. quarks of the thermal QCD medium, is complicated by the so called doubler problem. The deformed dispersion relation from the discretized Dirac equation leads to artificially light modes within the first Brilloin zone. This issue is intimately related to the question of how to implement a discretized form of chiral symmetry on the lattice. In this context the discretization of the light fermions also leads to an artificially large mass $m_\pi$ to the pionic degrees of freedom. In order to keep the numerical cost of simulations under control the largest lattice QCD collabroations working at finite temperature have chosen the staggered quark discretization, e.g. the highly-improved staggered quarks (HISQ) \cite{Follana:2006rc} or the so called stout action \cite{Aoki:2005vt}. As staggered quarks preserve a remnant of chiral symmetry one also has to deal with the issue of doublers. Other collaboration have opted for the more costly but formally advantageous Wilson fermions (see e.g. \cite{AliKhan:2001ek}) or the Wilson fermion derived twisted mass formulation \cite{Shindler:2007vp}.

For the treatment of the heavy quark degrees of freedom chiral symmetry does not play an equally important role. Instead it is the fact that since discretization artifacts in the most naive formulation scale with $m_Qa$ that a very fine lattice spacing is required. Improved actions that allow for a more advantageous scaling are therefore often deployed \cite{Luscher:1996sc}. Based on the staggered formulation, the ${\cal O}(a^2)$ Szymanzik improved HISQ fermions have been introduced in Ref.~\cite{Follana:2006rc} with heavy quarks at $T=0$ in mind. In finite temperature studies a popular relativistic action for quarkonium is the clover improved Wilson action (also known as the Shekholeslami-Wohlert action and starting point for the Fermilab action \cite{ElKhadra:1996mp}), which for anisotropic lattices \cite{Chen:2000ej} reads
\begin{align}
\nonumber S_F^{\xi}
  &= a_\tau a_s^3  
      \sum_{\mathbf{n}} \bar{Q}(\mathbf{n})  \left[ 
      m_0 + 
      \nu_\tau [
      \gamma_\tau \nabla_\tau - 
      \frac{a_\tau}{2} \nabla^2_\tau ]  +
      \nu_s \sum_s [
      \gamma_s \nabla_s -\frac{a_s}{2} \nabla^2_s ]   
       \right.\\
      & \left. - \frac{a_s}{2} [ 
      C_{\rm sw}^\tau  \sum_{s} \sigma_{\tau s} F_{\tau s} + 
      C_{\rm sw}^s  \sum_{s<s\prime} \sigma_{ss\prime} F_{ss\prime} ]
      \right] Q(\mathbf{n})\\                  
  &= a_\tau a_s^3 
      \sum_{\mathbf{n}} \bar{Q}(\mathbf{n})  \left[ 
      m_0 + 
      \nu_\tau  \!\not\! D^{\rm Wilson}_\tau +
      \sum_s \nu_s  \!\not\! D^{\rm Wilson}_s 
      - \frac{a_s}{2} [ 
      C_{\rm sw}^\tau  \sum_{s} \sigma_{\tau s} F_{\tau s} + 
      C_{\rm sw}^s  \sum_{s<s\prime} \sigma_{ss\prime} F_{ss\prime} ]
      \right] Q(\mathbf{n}) , 
\end{align}
where $D^{\rm Wilson}$ denotes the original Wilson Dirac operator. The covariant lattice derivatives are
\begin{align}
 \nabla_\mu Q(\mathbf{n})  & =  
 {1\over 2a_\mu}\, \biggl[ U_\mu(\mathbf{n}) Q(\mathbf{n}+\hat\mu) - U_{-\mu}(\mathbf{n}) 
           Q(\mathbf{n}-\hat \mu)\biggr], \\ 
 \Delta_\mu Q(\mathbf{n})  & =  
 {1\over a_\mu^2} \, \biggl[ U_\mu(\mathbf{n}) Q(\mathbf{n}+\hat\mu) + U_{-\mu}(\mathbf{n}) Q(\mathbf{n}-\hat\mu)-2 Q(\mathbf{n}) \biggr]   \, ,
\end{align}
writing concisely $U_{-\mu}(\mathbf{n}) \equiv U_\mu(\mathbf{n}-\hat \mu)^\dagger$. The additional contribution that implements the improvement is called the clover term which contains the field strength tensor, discretized by four plaquette terms
\begin{align}
4 S_{\mu\nu}(\mathbf{n})  & =  
                   U_\mu(\mathbf{n}) U_\nu(\mathbf{n}+\hat\mu) U^\dagger_\mu(\mathbf{n}+\hat\nu) 
                   U^\dagger_\nu(\mathbf{n}) + U_\nu(\mathbf{n}) U^\dagger_\mu(\mathbf{n}-\hat\mu+\hat\nu) 
	           U^\dagger_\nu(\mathbf{n}-\hat\mu) U(\mathbf{n}-\hat\mu) +
                                                           \\
	       &  U^\dagger_\mu(\mathbf{n}-\hat\mu) U^\dagger_\nu(\mathbf{n}-\hat\mu-\hat\nu) 
                   U_\mu(\mathbf{n}-\hat\mu-\hat\nu) U_\nu(\mathbf{n}-\hat\nu)+
             U^\dagger_\nu(\mathbf{n}-\hat\nu) U_\mu(\mathbf{n}-\hat\nu) 
		   U_\nu(\mathbf{n}+\hat\mu-\hat\nu) U^\dagger_\mu(\mathbf{n}) ,        
                                                           \\
F_{\mu\nu}(\mathbf{n})   & =  
                   \frac{-i}{2a^2}[S_{\mu\nu}(\mathbf{n}) - S_{\mu\nu}^\dagger(\mathbf{n})] .
\end{align}
By appropriate tuning of the parameters $\nu_i$ and $C_{\rm SW}$ the ${\cal O}(a)$ lattice artifacts can be made to vanish on the level of the classical action, in turn suppressing the discretization artifacts in the full quantum theory. An empirical but well established non-perturbative procedure to tuning the action parameters is so called tadpole improvement, in which the mean value of link variables is used to improve the convergence of the simulation results to the continuum limit.

In practice the path integral in \cref{eq:latOpPI} is approximated stochastically (see e.g. \cite{Kennedy:2006ax}). I.e. one designs a stochastic process in computer time $t_{\rm MC}$ also called Monte Carlo time, which generates successive sets of 4-dimensional field configurations with a probability distribution according to the Euclidean Feynman weight. In order to obtain an ensemble of gauge configurations that accurately represents the quantum probability distribution, the space of configurations must be efficiently traversed. To this end hybrid Monte Carlo algorithms are currently deployed, where a stochastic update is combined with a classical evolution of Hamilton's equation of motion for gluon and pseudofermion fields. The necessity to solve a large dense system of linear equation at each computer time step constitutes the main numerical cost. 

To keep costs manageable, often it is only the light quarks $u,d$ and $s$ that are treated fully dynamically, with some collaborations starting to include $c$ quarks. The dynamical fermion content is indicated in lattice simulations conventionally by a code such as $N_f=2+1$ indicating in that case two mass degenerate $u$ and $d$ quarks and a more massive $s$ quark to be present. At low enough temperatures top and bottom quarks do not significantly contribute to virtual processes and their determinant can be approximated to be unity, they are said to be {\it quenched}.

The quantum statistical expectation value of an observable $\langle O\rangle$, i.e. of a gauge invariant (composite) operator, is approximated by computing the value of $O$ on each realization within one ensemble. Since one can often use volume averaging in determining the value of $O$ on each lattice configuration, the outcome constitutes a subaverage. If the variance of the these subaverages $O^k$, what the lattice community often calls a measurement, is finite, then thanks to the central limit theorem their distribution will become approximately Gaussian if enough configurations are available. In that case the mean is taken as simple estimator for the expectation value
\begin{align}
\langle O \rangle = \frac{1}{N_{\rm conf}}\sum_l O^k + (\epsilon_{\rm MC}/\sqrt{N_{\rm conf}}).
\end{align}
If there are no residual {\it autocorreations} between the generated configurations the statistical error in the end result decreases with $1/\sqrt{N}$. 

Note that at no step above a particular gauge had to be chosen in the process of simulating $\langle O \rangle$. There can however arise situations in the study of quarkonium where an evaluation of gauge dependent quantities is of interest. By now there exist standard iterative methods (see e.g. Ref.\cite{davies:1988}) to generate appropriate sets of gauge transformation matrices $G(\mathbf{x})$ in order to approximate the (local) extremum of the gauge fixing functional $F[A^\mu]=0$ that encodes a discretized variant of a gauge condition, such as Landau $\partial_\mu A_\mu=0$ or Coulomb gauge $\bm \nabla \cdot \mathbf{A}=0$.

At this point some remarks are in order on how to judge the reliability of a lattice QCD computation result. The outcome of a simulation is an estimate of an imaginary time corelation function with a certain statistical uncertainty, which depending on the computation resources available can be made arbitrary small. At the same time the discretization related artifacts need to be kept in mind according to the following checklist inspired by the FLAG criteria \cite{Aoki:2019cca}:
\begin{itemize}
\item For the temperature range considered, are all relevant quarks d.o.f. dynamically included?
\item Is the pion mass at the physical point $m_\pi=140$MeV and related, is the crossover temperature at $T_c=155$MeV?
\item Has the continuum limit $a\to0$ been taken?
\item Has the thermodynamic limit $V\to\infty$ been taken?
\end{itemize}
The first item in current simulations is usually satisfied. On the other hand the second to fourth item still require considerable resources often preventing these conditions from being fulfilled. If e.g. the limits are not taken one has to make sure that the relevant energy scales are located reasonably far away from the corresponding IR and UV cutoffs. In essence this also tells us that only if all criteria are met, two lattice simulations can be expected to agree within their statistical uncertainties. 

As discussed in \cref{sec:qftandspec} we wish to compute quarkonium current correlators according to fixed quantum numbers. Inspecting their spectral functions allows us to identify possible bound states and their in-medium properties. The evaluation of these correlators requires additional care on the lattice. One reason is that the introduction of the hypercubic grid breaks the rotational symmetry of the continuum theory and reduces it to the octahedral symmetry group $O_h$ \cite{Johnson:1982yq}. Instead of the good quantum number spin $J^{PC}$ labeling one of the infinite numbers of irreducible representations of rotations, we have to deal with $\Lambda^{PC}$ which refers to a finite number of lattice irreducible representations of $O_h$. For integer spin there exist exactly five irreducible representations $A_1,T_1,T_2,E,A_2$. The identification of particles with physical $J>1$, such as e.g. $\chi_2$, which appear as admixtures in different channels set by $\Lambda^{PC}$ requires particular care as described e.g. in Ref.~\cite{Dudek:2009qf}.

To evaluate a quarkonium current correlator on the lattice we have to consider the lattice counterpart of the continuum expression
\begin{align}
M^{\rm cont}_\Gamma=Z_\Gamma(a,m_Q,\mu=1/a) a_s^{-3}M^{\rm lat}_\Gamma= Z_\Gamma a_s^{-3}\bar\psi \Gamma W \psi.
\end{align}
The renormalization factor $Z_\Gamma$ can be computed in lattice perturbation theory \cite{Gockeler:1996hg}, where $\mu$ denotes the scale at which the strong coupling is evaluated in such a computation.

The Euclidean correlator for a general meson operator is obtained after one carries out the heavy quark Grassmann integrals 
\begin{align}
&\langle M_\Gamma(\tau) M^\dagger_\Gamma(0) \rangle\\
 &= \int {\cal D}[U,\bar\psi,\psi] \int {\cal D}[\bar Q,Q]\big\{ \bar{Q}(\mathbf{h}_1,0)\Gamma W(\mathbf{h}_1,\mathbf{l}_1,0)Q(\mathbf{l}_1,0)\bar{Q}(\mathbf{l}_2,\tau)\Gamma^\dagger W^\dagger(\mathbf{l}_2,\mathbf{h}_2,\tau)Q(\mathbf{h}_2,\tau)\big\} e^{-S_E}\\
 &=\int {\cal D}[U,\bar\psi,\psi]\big\{ {\rm Tr}\big[ K^{-1}_Q(\mathbf{h}_2,\tau,\mathbf{h}_1,0)\Gamma W(\mathbf{h}_1,\mathbf{l}_1,0)K^{-1}_Q(\mathbf{l}_1,0,\mathbf{h}_2,\tau)\Gamma^\dagger W^\dagger(\mathbf{h}_2,\mathbf{l}_2,\tau) \big] \\
& - {\rm Tr}\big[ K^{-1}_Q(\mathbf{l}_1,0,\mathbf{h}_1,0)\Gamma W(\mathbf{h}_1,\mathbf{l}_1,0)\big] {\rm Tr}\big[ K^{-1}_Q(\mathbf{h}_2,\tau,\mathbf{l}_2,\tau)\Gamma^\dagger W^\dagger(\mathbf{h}_2,\mathbf{l}_2,\tau) \big]
\big\} {\rm det}[K_Q] e^{-S_E[U,\bar\psi,\psi]}.
\end{align}
Since the mass of the heavy fermion suppresses its contributions to virtual processes we neglect ${\rm det}[K_Q]\approx1$. The contributions with an overall trace on the outside are referred to as connected diagrams, the ones including two individual traces as disconnected ones. The latter are often neglected for heavy quarks since they are suppressed due to the OZI rule and as their contribution has been found to be small in practice (see e.g. \cite{deForcrand:2004ia}). The computation of the quark propagators $K^{-1}$  constitutes a boundary value problem and involves solving the defining equation 
\begin{align}
\sum _{z,\gamma,c} K_{xz,\alpha\gamma,ac}K^{-1}_{zy,\gamma\beta,cb} = \delta_{xy} \delta_{\alpha\beta} \delta_{ab},\label{eq:proprel}
\end{align}
which constitutes a large system of linear equations.

\begin{figure}
\centering
\includegraphics[width=7.5cm]{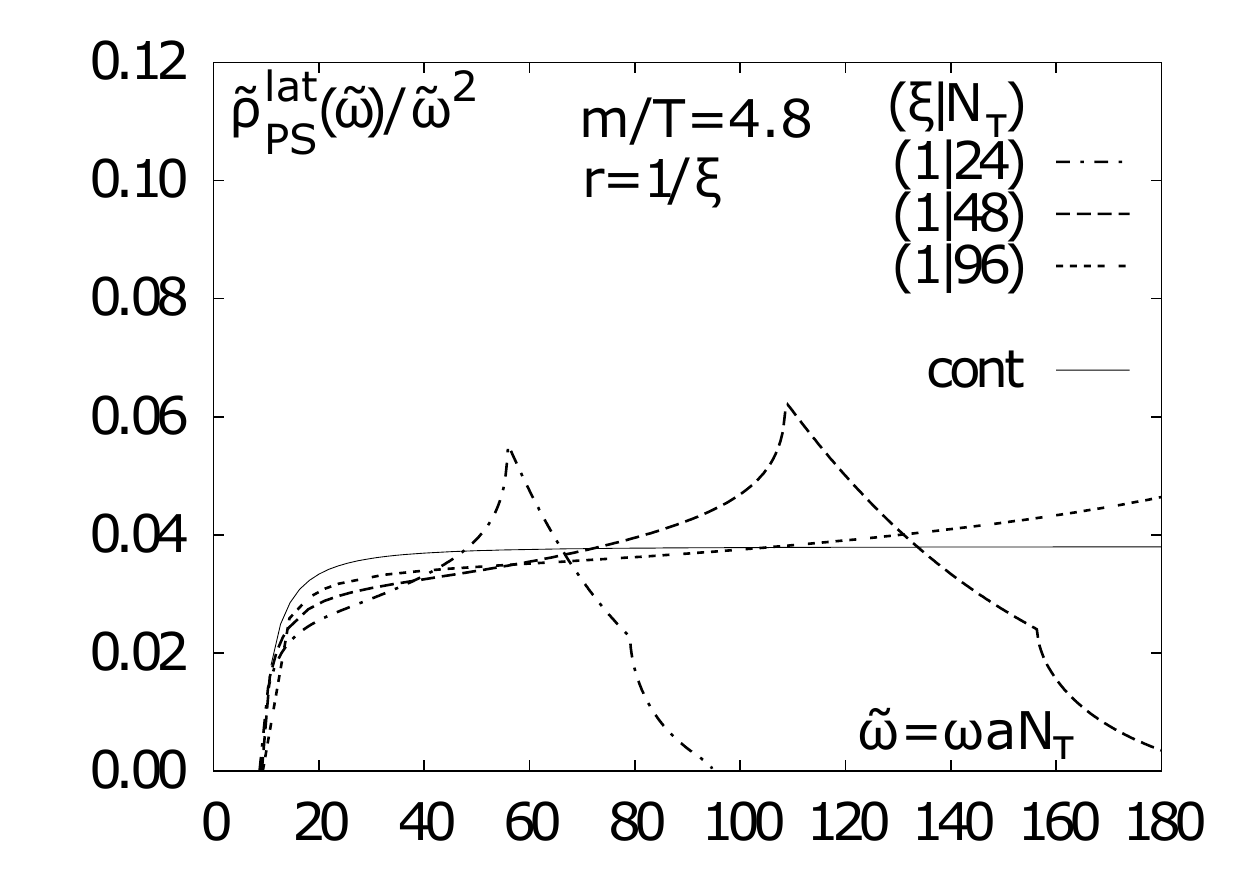}
\caption{The non-interacting spectral function on isotropic lattices, normalized to the continuum asymptotic behavior in the pseudoscalar channel for finite mass $m/T=4.8$, adapted from \cite{Karsch:2003wy}. Note both the approach to the continuum limit for increasing resolution with $N_\tau$ as well as the presence of discretization artifacts at high momentum.}\label{fig:freeWilsonspec}
\end{figure}

We can learn about some of the effects of the lattice discretization by considering the meson correlator in terms of Wilson fermions in the non-interacting limit. In that case explicit expressions for the Euclidean correlator and the spectral function have been computed in Ref.~\cite{Karsch:2003wy,Aarts:2005hg}. The differences to the continuum theory are cleanly illustrated for the $\mathbf{p}=0$ correlator in \cref{fig:freeWilsonspec}.

For small frequencies the free lattice spectral function lies close to the functional form of its continuum counterpart. On the other hand the UV part is severely distorted and the spectral function vanishes above a finite frequency $\omega_{\rm max}$. On isotropic lattices it is given by $\omega^{\rm iso}_{\rm max} a= 2{\rm ln} 7$, corresponding to the largest energy available to a meson with lattice momentum $\mathbf{p}=(\pi/a,\pi/a,\pi/a)$. Whenever the lattice momentum reaches a corner of the Brilloin zone, i.e. at $\mathbf{p}=(\pi/a,0,0)$ and $\mathbf{p}=(\pi/a,\pi/a,0)$, corresponding to $\omega_1a=2{\rm ln}3$ and $\omega_2a=2{\rm ln}5$ kinks appear in the spectral function, the lowest one often exhibiting an amplitude above the continuum result.

It is useful at this point to consider briefly the analysis of interacting quarkonium correlation functions in $T=0$ lattice QCD and discuss differences to the $T>0$ case. Summing over all spatial lattice points of $\langle \hat M_\Gamma(\mathbf{n}, \tau) \hat M^\dagger_\Gamma(\mathbf{0},0) \rangle$, the $\mathbf{p}=0$ correlator is obtained. Since for quarkonium the vacuum ground states are expected to be well separated from excited states and the underlying spectral function is simply a sum of delta peaks, the correlator will exhibit a well pronounced single exponential falloff at the region $\tau=\beta/2$ (for periodic boundary condition the correlator is still symmetric). The exponent is simply the rest mass of the heavy quarkonium particle. It is either determined via an exponential fit to the $\tau<\beta/2$ correlator or by considering the so called {\it effective mass}
\begin{align}
m_{\rm eff}(\tau)={\rm log}\Big[\frac{D(\tau+a_\tau)}{D(\tau)}\Big],\label{eq:effmass}
\end{align}  
which will approach a plateau with the value of the ground state rest mass $M_0$ at large enough $\tau$. For positive definite spectra $m_{\rm eff}$ will be convex and monotonous. Using correlators at different momenta $m_{\rm eff}(\tau\to\infty)$ traces out the dispersion relation $E(\mathbf{p})$. At $T>0$ the peak structures are not as well separated and the resulting correlator will not exhibit clear exponential falloffs. This induces curvature in the effective mass and a similarly straight forward interpretation in terms of spectral features is unavailable.

Already at $T=0$ spectral contributions from excited states induce curvature in $m_{\rm eff}(\tau)$ at early $\tau$'s. The stronger the the effect, the later $m_{\rm eff}(\tau)$ reaches a plateau. Since the signal to noise ratio decreases with $\tau$ this complicates the determination of $M_0$. At $T=0$ we are only interested in the mass and the ground state is a single delta function. Thus we may improve the signal to noise by increasing the amplitude of that peak structure. This requires modifying the {\it overlap}, i.e. the magnitude of the ground state matrix element of the meson operator in the spectral decomposition. There are many different strategies to do so on the market, among them so called {\it smearing}. It either refers to replacing the delta function on the r.h.s of \cref{eq:proprel} by an extended object and/or a smoothing prescription of the gauge fields in spatial direction. In the latter the UV modes of the gauge fields are successively damped away. In turn the excited state and continuum contributions to the spectral function diminish and the ground state peak is relatively more pronounced. Since a bound state at $T=0$ is represented by a single delta peak, this procedure often achieves its goal. However at finite temperature where clusters of peaks populate the spectral function and their envelope encodes vital properties, such as thermal widths, the application of smearing requires additional care. It is not apriori clear how the amplitude of individual peaks is affected by smearing and in turn the envelope may be distorted, leading to uncontrolled changes in its position and width. To avoid these additional systematic uncertainties smearing is largely avoided in studies of spectral structures at $T>0$.

The last item we have not touched so far is the subject of how to select the bare values of the parameters in the lattice action (strong coupling, quark masses, anisotropies). The quantum fluctuations present in the discretized path integral lead to a renormalization of the bare parameters and the values of the physical parameters need to be determined by comparison with a physical measurement a posteriori. One important part is the setting of the absolute energy scale in a lattice simulation \cite{Sommer:2014mea}, most often quoted as a lattice spacing in units of fm. Since a lattice QCD simulation is formulated in dimensionless fields indeed external input is needed to assign such a physical scale. 

The most direct approach would be to measure the mass of a stable Baryon but this often is prohibitively expensive, numerically. Instead until recently it has been customary to use the so called Sommer scales $r_0$ and $r_1$ to set the scale. These quantities are derived from a phenomenological Cornell type model of heavy quarkonium, fixing in essence the slope of the string like part of the potential extracted from a $T=0$ lattice simulations to a model dependent physical value. While easy to implement, the disadvantage of this approach is that it prevents an independent determination of the heavy quark potential on the same lattice. More recently a different high precision scale setting approach \cite{Borsanyi:2012zs} has been developed based on the Gradient flow technique \cite{Luscher:2010iy}. It provides a scale $\omega_0$ derived from smoothed correlators of the field strength tensor, which, if once determined to high precision on a lattice using e.g. a Baryon mass, can function as straight forwardly evaluable standard in other lattice simulations. 

In a simulation without dynamical quarks only the strong coupling $\beta$ needs to be adjusted to select a particular lattice spacing. In dynamical simulations also the light quark parameters have to be tuned. This requires as additional inputs e.g. the pion decay constant \cite{Fukugita:1992np} and the physical mass ratio between s and u/d quarks. Instead of trying to reproduce the pion mass, one enforces that the so called partially conserved axial current (PCAC) mass vanishes. At that point the residual Ward identity for the remnant chiral symmetry on the lattice is fulfilled. In turn the pion mass emerges as a dynamical property of the simulation and not as external input. 

For the heavy quarks the mass and anisotropy parameters in the action can be tuned independently from those of the dynamical quarks and gluons. On the one hand, we have to make sure that the ground state mass $M_0$ of one chosen meson channel agrees with its PDG value and at the same time that the relativistic dispersion relation is fulfilled
\begin{align}
E^2(\mathbf{\tilde p})=M^2_0 + \frac{M_1}{M_0} \mathbf{ \tilde p}^2 + {\cal O}(\mathbf{ \tilde p}^4), \quad \frac{1}{M_1}=\left. \frac{\partial^2 E}{\partial \tilde p_i^2}\right|_{ \mathbf{ \tilde p}=0},
\end{align}
i.e. the kinetic mass $M_1$ needs to equal the rest mass $M_0$. Here the PDG quarkonium mass is the actual external input.

The temperature in a simulation depends on the extent of the imaginary time axis. It can be varied either by changing the lattice coupling $\beta$, i.e. the physical lattice spacing ({\it fixed box} approach) or by changing the number of grid points ({\it fixed scale} approach). Since varying the coupling modifies the renormalization scale of the simulation, accompanying $T=0$ simulations need to be carried out for scale setting and the subtraction of temperature independent UV divergent terms. The fixed scale approach requires just one $T=0$ simulation but temperatures can only be varied in integer steps of $N_\tau$ and the number of Euclidean time steps at which the correlator can be computed diminishes as one increases temperature. 

\subsubsection{Lattice NRQCD}
\label{sec:latNRQCD}

The effective field theory NRQCD provides an alternative discretization prescription for heavy quarks on the lattice \cite{Thacker:1990bm,Lepage:1992tx}. The NRQCD expansion on the lattice is formulated in terms of increasing powers of $v \sim {\mathbf p}/(m_Qa)$, in contrast to an expansion directly in $(m_Qa)^{-1}$, as in the EFT called heavy quark effective theory \cite{HQET1,HQET2}. Lattice NRQCD has been successfully used for precision spectroscopy at $T=0$ (see e.g. \cite{Dowdall:2011wh}) and found application to in-medium quarkonium, as will be discussed in detail in later sections. Instead of populating the spacetime grid with both heavy and light degrees of freedom, which requires very fine lattice spacings, the quarks and gluons of the QCD medium will be treated separately in a simulation without heavy d.o.f. at a coarser spacing. The evolution of the heavy quarks is implemented in terms of Pauli spinors propagating in the background of the light quarks and gluons.

On the one hand this strategy provides three clear practical advantages. First, the NRQCD correlator is not periodic in Euclidean time and thus in contrast to a relativistic formulation provides additional independent information at $\tau>1/2T$. I.e. access to the ground state properties dominant at large $\tau$ is improved. Secondly, the absence of a transport peak contribution simplifies the extraction of spectra from Euclidean correlators. Third, the computational cost of the Euclidean heavy quark propagator is significantly reduced. Instead of having to solve a 4d boundary value problem as in \cref{eq:proprel}, one instead faces an initial value problem in Euclidean time and evolves the propagator step by step. 

On the other hand there are some caveats that require additional care. The process of integrating out the hard scale by shifting the frequency origin on the lattice becomes lattice spacing dependent. I.e. while scale setting fixes the relative frequency scale, an additional comparison with experimental data is needed to fix its absolute scale. Due a steeper rise of the free spectral functions at intermediate frequencies in NRQCD compared to the relativistic computation, the contribution of the continuum may manifest itself more strongly also in the interacting correlator, thus requiring a high signal to noise ratio for e.g. the P-wave channels.

A conceptually distinct aspect of a lattice EFT is that as a non-renormalizable theory no naive continuum limit exists. Naively, as the continuum limit $a\to 0$ is approached, more and more terms in the expansion need to be taken into account and for each a Wilson coefficients has to be determined. Iz.e. eventually predictability is lost. What allows EFTs to nevertheless be an important tool of precision physics is that we are usually interested in results only up to a finite accuracy. This can be achieved by working to a certain finite order $n$ in the expansion $v^{n}$. The set of Wilson coefficients for a given order and given lattice spacing need to be computed via (lattice) perturbation theory and eventually compared to experimentally measured quantities. Such radiative corrections have been studied in detail for several discretization schemes of accompanying light quarks and gluons (see e.g. \cite{Davies:1993ec,Hammant:2013sca}).

Following \cite{Lepage:1992tx} the lattice Hamiltonian that governs the heavy quark fields to ${\cal  O}(v^6)$ is given by
\begin{align}
H= H_0 + H^{(4)} + H^{(6)}, \quad H_0 = -\frac{\Delta^{(2)}}{2m_Q},
\end{align}
with the naive kinetic term given by $H_0$ being of order ${\cal O}(v^2)$ and
\begin{align}
\Delta^{(2)} &= \sum_i \Delta_i^{(2)}, \quad a^2 \Delta_i^{(2)} \psi(\mathbf{n}) = U_i(\mathbf{n}) \psi(\mathbf{n} + \hat i) + U_i^{\dagger}(\mathbf{n} - \hat i) \psi(\mathbf{n} - \hat i) - 2 \psi(\mathbf{n}).
\end{align}
The contributions to ${\cal  O}(v^4)$ read 
\begin{align}
\delta H^{(4)} & = -c_1\frac{\left(\Delta^{(2)}\right)^2}{8M_b^3} +
  c_2\frac{ig}{8M_b^2}\left(\boldmath\Delta\cdot E -
  E\cdot\boldmath\Delta\right) - c_3 \frac{g}{8M_b^2}\bm \sigma\cdot\left( \tilde\Delta\times\tilde {\bf E} -
  \tilde{\bf E}\times\tilde\Delta\right) - c_4\frac{g}{2M_b}\bm \sigma\cdot
  \tilde{\bf B} \\
  & + c_5 \frac{a^2\Delta^{(4)}}{24M_b} - c_6
  \frac{a\left(\Delta^{(2)}\right)^2}{16nM_b^2}.
\end{align}
Some studies include the effects of spin dependent corrections up to order ${\cal  O}(v^6)$
\begin{align}
\delta H^{(6)} & =  - c_7 \frac{g}{8M_b^3}\lbrace
 \Delta^{(2)},\bm\sigma\cdot {\bf B}\rbrace
  - c_8 \frac{3g}{64M_b^4} \lbrace
  \Delta^{(2)},\bm\sigma\cdot\left(\Delta\times{\bf E} - {\bf
  E}\times\Delta\right)\rbrace 
  - c_9 \frac{ig^2}{8M_b^3}\bm\sigma\cdot{\bf E}\times {\bf E},
\end{align}
which play an important role in the precision study of the hyperfine splitting of quarkonium states at $T=0$. To define the color $E$ and $B$ fields the clover discretized field strength tensor is used
\begin{align}
F_{\mu\nu}(x) &= - \frac{1}{4} \sum_{\Box}\left( \frac{U_{\mu
     \nu}(\mathbf{n}) - U_{\mu\nu}^{\dagger}(\mathbf{n})}{2i} - 
\frac{1}{3} {\rm Tr}({\rm Im} U_{\mu\nu}(\mathbf{n})) \right), \quad 
 E^{i}  =  F^{i0}, \quad B^{i} =  -\frac{1}{2}\epsilon_{ijk}F^{jk}.
\end{align}
In the above definition of the Hamiltonian several corrections terms have already been included that remove ${\cal O}\left(a^2 v^4\right)$ discretization errors. This has been achieved by using a higher order implementation of the field strength tensor in the $c_3$ and $c_4$ terms
\begin{align}
\tilde{F}_{\mu \nu}(\mathbf{n}) &= \frac{5}{3} F_{\mu \nu}(\mathbf{n}) - \frac{1}{6}
\left(U_\mu(\mathbf{n}) F_{\mu \nu} (\mathbf{n} + \hat{\mu}) U_{\mu}^{\dagger}(\mathbf{n}) + U_{\mu}^{\dagger}(\mathbf{n} - \hat{\mu})F_{\mu \nu}(\mathbf{n} - \hat{\mu}) U_\mu(\mathbf{n} -
  \hat{\mu}) - (\mu \leftrightarrow \nu) \right),
\end{align}
as well as a higher order derivative in $c_3$
\begin{align}
\tilde{\Delta}_i &= \Delta_i - \frac{a^2}{6} \Delta_i^{(+)} \Delta_i
\Delta_i^{(-)}, \quad a \Delta_i^{(+)}G(\mathbf{n}) = U_i(\mathbf{n}) G(\mathbf{n} + \hat i) - G(\mathbf{n}) \nonumber \\
a \Delta_i^{(-)}G(\mathbf{n}) &= G(\mathbf{n}) - U_i^{\dagger}(\mathbf{n} -  \hat i)G(\mathbf{n} - 
\hat i), \quad a \Delta_i G(\mathbf{n}) = \frac{1}{2} \left(U_i(x) G(\mathbf{n} +  \hat i) -
  U_i^{\dagger}(\mathbf{n} - \hat i)G(\mathbf{n} -  \hat i) \right).
\end{align}
Additional discretization corrections are included as the last two terms in $\delta H^{(4)}$, where $\Delta^{(4)} = \sum_i \left( \Delta_i^{(2)} \right)^2$.
In order to improve agreement with the continuum theory, tadpole improvement is often implemented on the level of the Hamiltonian. In this empirical approach to matching, all links entering the Hamiltonian are divided by the fourth root of the single plaquette $u_0=\langle{\frac{1}{3}\rm Tr}U_{\mu\nu}\rangle^{\frac{1}{4}}$ and at the same time the Wilson coefficients $c_i$ are set to unity.

Since to this order in NRQCD the propagation of quark and antiquark fields is decoupled $G_\psi$ is computed by solving an initial value problem. It is common to deploy the naive Euler scheme for its discretization which in Euclidean time leads to 
\begin{align}
  G(\tau+1) & = \left( 1 - \frac{aH_0}{2n}\right)^n U_4^{\dag}\left( 1 -
  \frac{aH_0}{2n}\right)^n\left( 1 - a\delta H \right) G(\tau), \quad 
  G(0) = S(\mathbf{h},0) .\label{eq:EulerNRQCD}
\end{align}
As the simple forward discretization does not possess favorable stability properties it is customary to stabilize it by artificially reducing the temporal step size via the so called Lepage parameter $n$. It was shown \cite{Davies:1991py} that on isotropic lattices $(\xi=1)$ a value of $n=1$ leads to a well defined UV behavior of the evolution if $a_s m_Q>3$ or $a_s m_Q>1.5$ if $n=2$. There are efforts underway to implement higher order stable solvers for the evolution of $G$ based on e.g. the Crank-Nicholson scheme \cite{Lehmann:2019}. 

The initial source can be a point source $S=\delta_{{\bf x},0}$, or for an improved signal to noise ratio a complex valued stochastic source 
\begin{align}
S^{\Upsilon}(\mathbf{h},0)=\eta(\mathbf{h},0), \quad  \langle \eta^\dagger(\mathbf{h},0) \eta(\mathbf{l},0) \rangle=\delta_{\mathbf{h}\mathbf{l}}, \label{Eq:sourcesNRQCD}
\end{align}
diagonal in spin and color. Note that the explicit choice of the source determines the normalization of $D_E$ on the lattice and thus the normalization of the underlying spectral function.

The quarkonium correlation function is composed of heavy quark propagators and quantum numbers are chosen by NRQCD vertex operators
\begin{align}
D_E(\mathbf{p},\tau)=\sum_{\mathbf h} e^{i\mathbf{p}\mathbf{h}} {\rm Tr}\big[ G^\dagger_\psi (\mathbf{h},\tau) \Gamma_{\rm NRQCD} G_\psi(\mathbf{h},\tau)\Gamma_{\rm NRQCD}\big].
\end{align}
We have listed common vertex operators for S-wave and P-wave states in \cref{tab:mesonqnNRQCD}, where the following definition of the symmetric derivative has been used $\chi^\dagger\overset{\leftrightarrow_s}{\Delta}_i\psi=-\Big[\frac{1}{4}\big(\Delta^+_i+\Delta^-_i\big)\chi\Big]^\dagger\psi +\chi^\dagger\Big[\frac{1}{4}\big(\Delta^+_i+\Delta^-_i\big)\psi\Big]$. Note that Euclidean NRQCD is simpler than its Minkowski time counterpart in that only the propagator of one field needs to be computed to construct $D_E$.

\begin{table}
\begin{tabular}[c]{|c|c|c|c|}\hline
$^{2s+1}S_J$ & lattice irrep. & $\sqrt{2N_c}M$ &  $\approx$ cont. $A_0$ \\\hline\hline
$^1S_0$ & $A_1^{-+}$ & $\chi^\dagger \psi$ & $R(0)/\sqrt{4\pi}$\\
$^3S_1$ & $T_1^{--}$ & $\chi^\dagger \sigma_i \psi$ & $R(0)/\sqrt{4\pi}$\\
$^1P_1$ & $T_1^{+-}$ & $\chi^\dagger \stackrel{\leftrightarrow}{\mbox{$\Delta_i$}}  \psi$ & $aR'(0)/\sqrt{4\pi/3}$\\
$^3P_0$ & $A_1^{++}$ & $\chi^\dagger \sum_i\sigma_i\stackrel{\leftrightarrow}{\mbox{$\Delta_i$}} \psi$ & $aR'(0)/\sqrt{4\pi/9}$\\
$^3P_1$ & $T_1^{++}$ & $\chi^\dagger ( \stackrel{\leftrightarrow}{\mbox{$\Delta_j$}}\sigma_i - \stackrel{\leftrightarrow}{\mbox{$\Delta_i$}}\sigma_j)\psi$ & $aR'(0)/\sqrt{2\pi/3}$\\
$^3P_2$ & $E^{++}$ & $\chi^\dagger  (\stackrel{\leftrightarrow}{\mbox{$\Delta_i$}}\sigma_i -  \stackrel{\leftrightarrow}{\mbox{$\Delta_j$}}\sigma_j)  \psi$ & $aR'(0)/\sqrt{2\pi/3}$\\
	    & $T^{++}$ & $\chi^\dagger ( \stackrel{\leftrightarrow}{\mbox{$\Delta_j$}}\sigma_i + \stackrel{\leftrightarrow}{\mbox{$\Delta_i$}}\sigma_j ) \psi$ & $aR'(0)/\sqrt{2\pi/3}$\\\hline
\end{tabular}
\caption{A selection of common meson operators in NRQCD, their quantum numbers and corresponding lattice irreducible representations adapted from \cite{Thacker:1990bm}. In addition we list the approximate relation of the ground state overlap $A^2_0=|\langle 0 |M M^\dagger | 0 \rangle_{\mathbf{p}=0}|^2$ with the continuum radial wavefunction.}\label{tab:mesonqnNRQCD}
\end{table}

The calibration of NRQCD simulations first requires selecting an appropriate mass parameter $m_Q a$. The state-of-the art presription \cite{Dowdall:2011wh} involves fixing the spin averaged kinetic mass $M_2(\overline{1S})=(M_2(\eta_b)+3M_2(\Upsilon))/4$ from the dispersion relation
\begin{align}
a_\tau E(\tilde p^2)=a_\tau \Delta E(0) + \frac{a_s^2\tilde p^2}{2\xi^2 a_\tau M_2}
\end{align}
The scale setting in the simulation of the light d.o.f. allows us to assign to the quantity $\Delta E(0)$ physical units. However, as indicated by writing $\Delta E(0)$ instead of $E(0)$, the mass of the ground state in NRQCD does not coincide with the physical rest mass of the quarkonium state. Removing the $2m_Q$ term in constructing NRQCD introduces a scale dependent energy shift, which needs to be accounted for by using one of the experimentally known quarkonium masses as additional input. 

\begin{figure}
\centering
\includegraphics[scale=0.35]{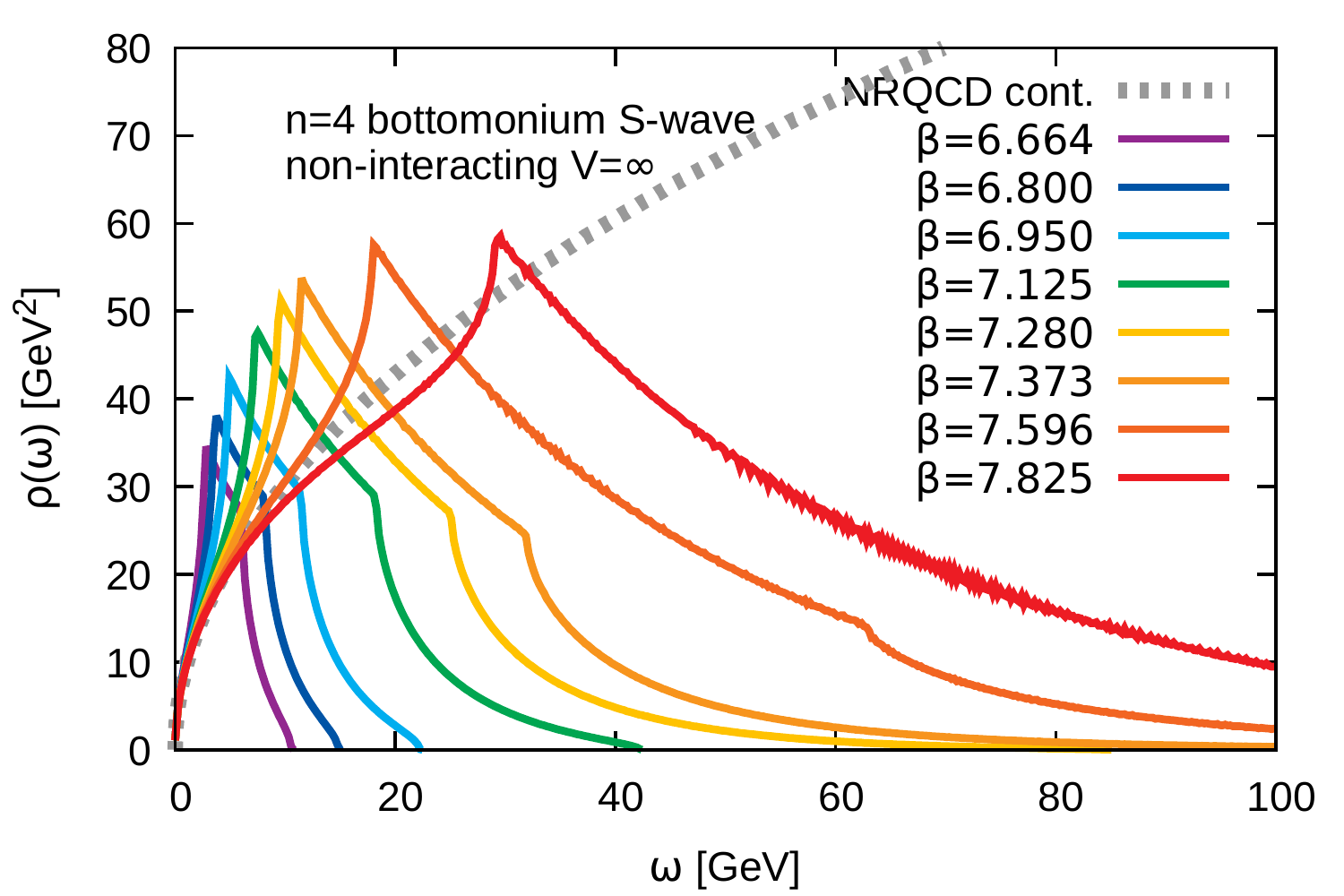}
\includegraphics[scale=0.35]{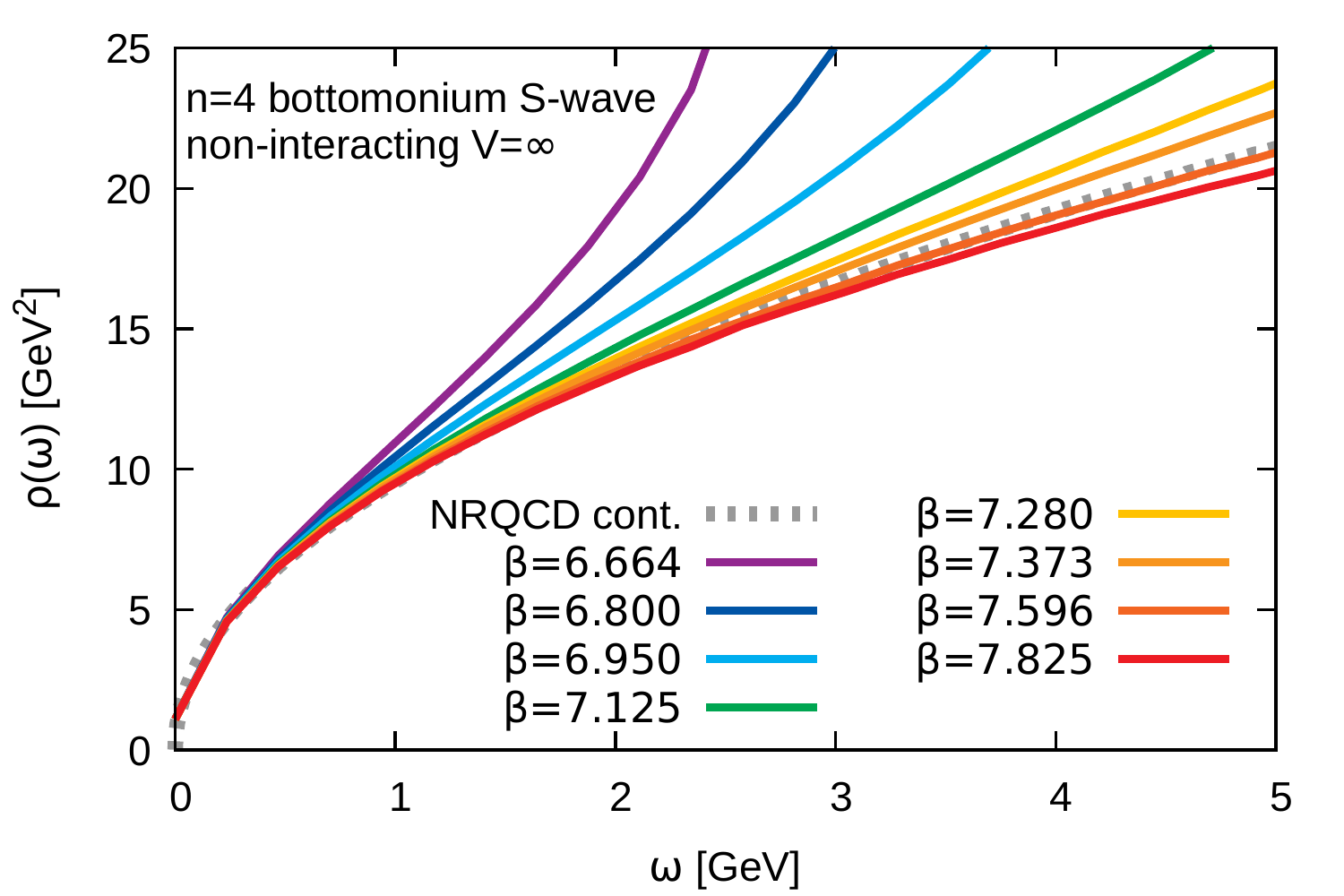}
\caption{A selection of noninteracting NRQCD spectral functions in the thermodynamic limit, zoomed in region around the origin on the right. The smaller the lattice spacing, the better the continuum spectral function is reproduced at small frequencies but at the same time the lattice artifacts extent to higher and higher frequencies.}\label{fig:FreeSpecFunc}
\end{figure}

As for the continuum theory let us also inspect the free spectral functions on the lattice. Using the non-interacting NRQCD Hamiltonian one obtains the following dispersion relation
\begin{align}
a_{\tau}&E_{\tilde{\mathbf{p}}} = 2n {\rm Log}\big[1-\frac{1}{2}\frac{\tilde{\mathbf{p}}^2}{ 2 n\xi a_s M_b }\big]+ {\rm Log}\big[1 + \frac{ (\tilde{\mathbf{p}}^2)^2 }{ 16 n\xi (a_s M_b)^2 } +  \frac{ (\tilde{\mathbf{p}}^2)^2 }{ 8 \xi (a_s M_b)^3 }  - \frac{ \tilde{\mathbf{p}}^4 }{ 24 \xi (a_s M_b) } \big]\label{Eq:NRQCDDispRel}
 \end{align}
and the spectral functions
\begin{align}
 \rho_S(\omega)=\frac{4\pi N_c}{N_s^3}\sum_{\tilde{\mathbf{p}}}\delta(\omega-2E_{\tilde{\mathbf{p}}} ),  \rho_P(\omega)=\frac{4\pi N_c}{N_s^3}\sum_{\tilde{\mathbf{p}}} \tilde{\mathbf{p}}^2\delta(\omega-2E_{\tilde{\mathbf{p}}} )\label{Eq:AnalytFreeSpec}
\end{align}
can be evaluated numerically (see e.g.\cite{Aarts:2014cda}). Note that the summation runs over all accessible lattice momenta $\tilde{\mathbf{p}}$ As the spectral functions are composed out of a large number of infinitely thin delta peaks, binning is required for their numerical evaluation. $\rho_{S,P}$ are independent of temperature and are fully specified by the mass parameter $\hat{M}=a_s m_Q$, the physical anisotropy $\xi$, the Lepage parameter $n$ and the lattice volume. $a_s$ denotes the physical spatial lattice spacing, $\xi$ the renormalized anisotropy and $n$ the Lepage parameter. Often the free spectral function in the thermodynamic limit is quoted in the literature,
examples of which we plot in the left panel of \cref{fig:FreeSpecFunc}. The normalization of this spectral function differs from that naively computed
on a lattice with unit links, due to differences in the choice of initial sources in the propagator evolution.

\begin{figure}
\centering
\includegraphics[scale=0.35]{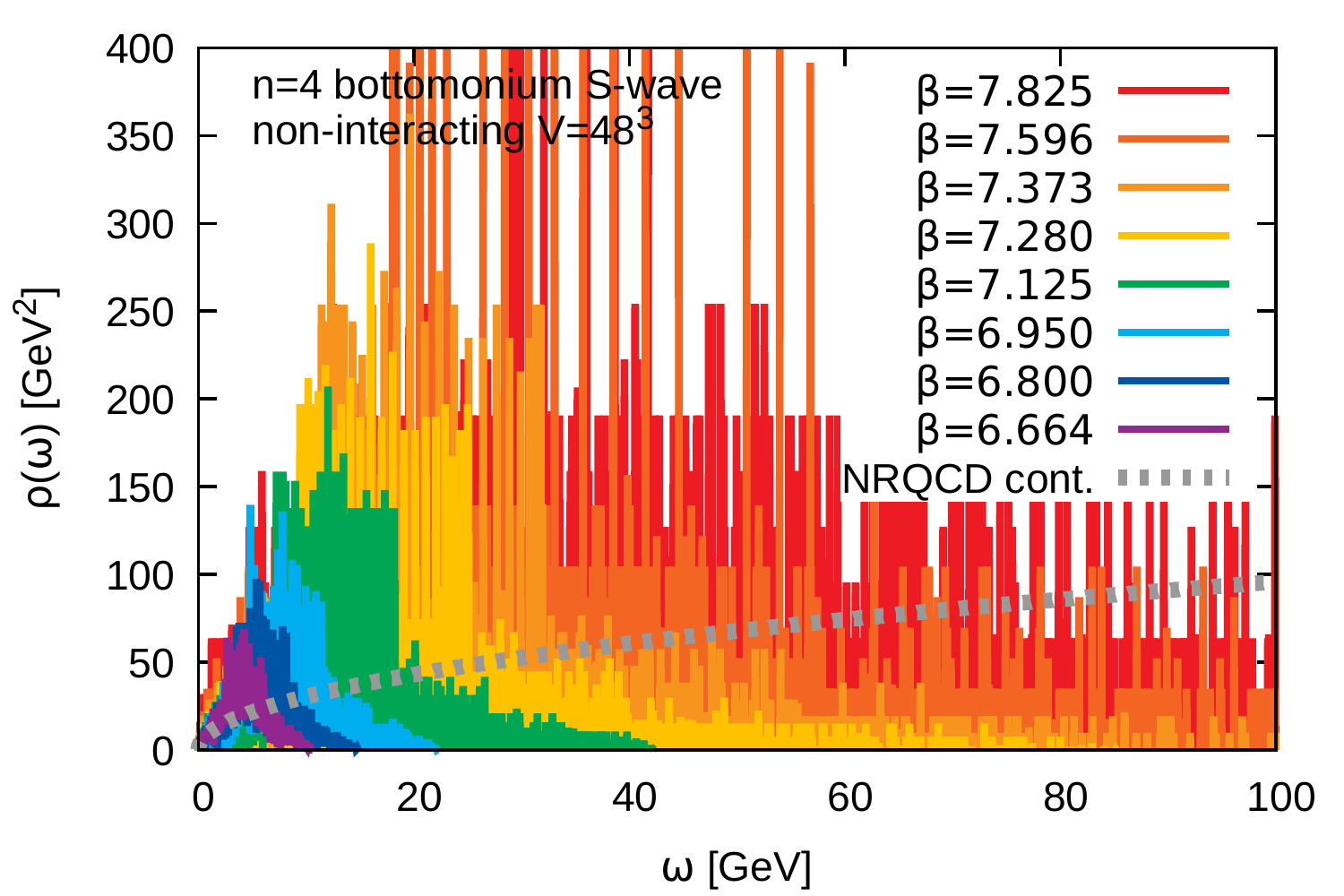}
\includegraphics[scale=0.35]{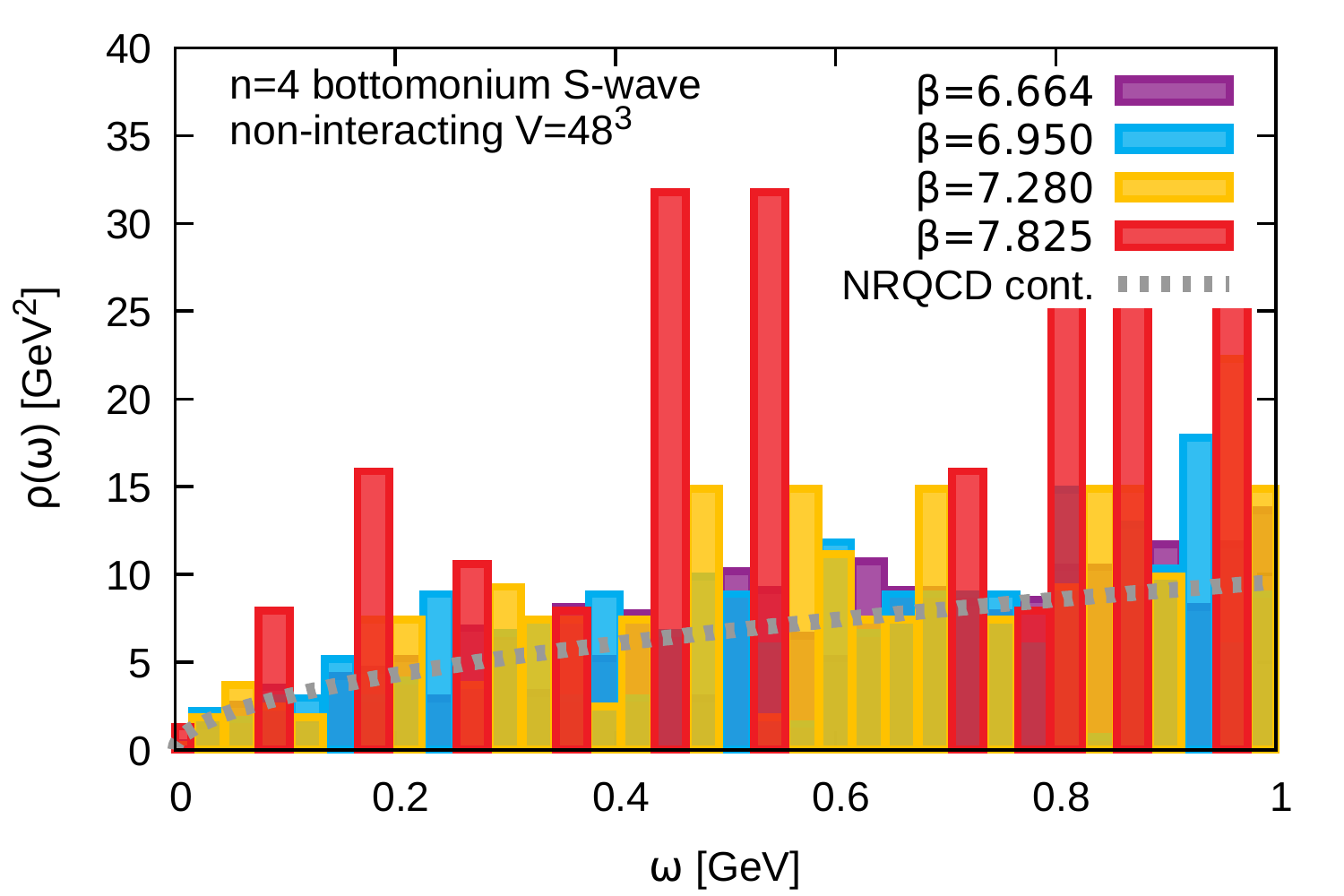}
\caption{A selection of noninteracting NRQCD spectral functions in the at a realistic finite volume $48^3$, zoomed in region around the origin on the right. The binning resolution is chosen as $\Delta \omega=30$MeV to illustrate the individual delta peak character of $\rho$. Note the sparseness in the available peaks at small momenta, indicating that also in the interacting theory broad peak structures will be coarsely resolved.}\label{fig:FreeSpecFunc2}
\end{figure}

We find that in contrast to the relativistic formulations the free spectral function does not have a UV cutoff that is fixed by the inverse lattice spacing. The smaller the lattice spacing, the stronger NRQCD approximation artifacts contribute at high frequencies. On the other hand the right panel shows that decreasing the lattice spacing allows lattice NRQCD to better reproduce the the continuum spectral function at small frequencies. Due to the Euler discretization there exists a smallest value of $\hat M$ beyond which the time evolution becomes unstable and the spectrum is artificially populated up to infinity.

The effects of finite volume are displayed in Fig.\ref{fig:FreeSpecFunc2} for a realistic scenario on a $48^3$ grid using a binning resolution of $\Delta \omega=30$MeV. This visualization clearly reflects the peaked character of the spectral function. In particular the zoom-in on the right emphasizes that the sparsity of available peaks at small momenta needs to be kept in mind when attempting to interpret how coarsely the broad peak structures of the thermodynamic limit are approximated in a finite box simulation. This closes our discussion of the lattice NRQCD approach.

\begin{summary} Monte-Carlo simulations of QCD on a hypercubic lattice with an imaginary time direction offer non-perturbative access to Euclidean meson correlation functions in QCD. There exist different discretization prescription for heavy quarkonium correlators either based on relativistic or non-relativistic formalisms. While the former offer direct access to the continuum limit they are numerically costly and due to the KMS relation provide only $N_\tau/2$ independent correlator points along $\tau$ from which to extract spectral information. On the other hand the NRQCD prescription offer reduced numerical cost and $N_\tau$ independent correlator points. To connect to continuum QCD is more involved as it requires precision matching of the different Wilson coefficients of the theory. In general when comparing results from different lattice computations it is important to establish, whether the continuum and thermodynamic limits have been reached, as otherwise differences in the results may reflect differences in discretization and not differences in the physics.
\end{summary}

\subsection{Reconstruction of spectral functions}
\label{sec:specrec}

In the previous section we have seen how lattice QCD simulations provide non-perturbative access to correlation functions in Euclidean time $D_E(\tau)$. In this section we will explore how to access the physical information encoded in these quantities by extracting the underlying spectral function.

In a zero temperature lattice simulation the low lying structures of the spectral function, as a sum of well separated delta function peaks, are relatively simple. In turn it is possible to extract the position and amplitude of such peaks straight forwardly by exponential fits to the Euclidean correlator. I.e. we use domain knowledge of QCD, which strongly restricts the possible structures present in the spectrum to reduce the difficulty in determining the relevant spectral properties.

At finite temperature the presence of light degrees of freedom in the heat bath makes it possible for the delta peaks in the spectral function to cluster. This leads to structures, which in the infinite volume limit, approach a continuous broad peak with a finite width. While still required to be positive definite, their actual functional form may be quite involved. 

At $T=0$, peak position and height are the only relevant spectral peak properties. At $T>0$ we have to distinguish between central position, width, enclosed spectral area and possibly skewness. QCD domain knowledge in general does not allow us to reduce the determination of these properties to a similarly simple fit ansatz as the exponentials in vacuum. Thus we require more flexible schemes to extract the spectral features from in-medium Euclidean correlators. This however leads us towards an inherently ill-posed inverse problem, which needs to be solved.

Indeed, the difficulty is directly visible if the general linear relation between correlator and spectral function found in QCD (\cref{eq:KLrep}, \cref{eq:Euclrep} and \cref{eq:EuclrepNRQCD}) is written in its discretized form
\begin{align}
D_j=\sum_{l=1}^{N_\omega} \,\Delta \omega_l\, K_{jl}\,\rho_l, \quad j\in[1,N_\tau],\, dD_j/D_j > 0. \label{eq:discrspecdec}
\end{align}
A lattice QCD simulation provides at most $N_\tau$ discrete averaged values for the correlator $D(\tau_j)=D_j$ along a compact imaginary times axis $\tau_j=a_\tau\cdot j$. As the result of the stochastic Monte-Carlo based simulation algorithm, each $D_j$ carries a finite statistical error $\Delta D_j$. On the other hand in anticipation of the intricate features present in the spectral function the function $\rho(\omega)$ is discretized along $N_\omega$ frequency bins $\Delta\omega$ with $N_\omega\gg N_\tau$. To discretize divergent kernels, such as ${\rm cosh}/{\rm sinh}$ at $T>0$, one can e.g. go over to consider instead $\rho^\prime=\rho/\omega$ and $K^\prime=K\omega$, both of which in turn are well behaved.

The confluence of two issues make the inversion of the above relation ill-posed. On the one hand the number of available correlator input points $D_i$ in today's simulations lies between 10-200, while it is not uncommon that a frequency discretization with $N_\omega\sim{\cal O}(1000)$ is required. I.e. there are more parameters to fix than input data points. Discretizing the kernel $K_{ij}$ over a finite range of Euclidean times and frequencies usually leads to a strong decay of the singular values of the corresponding matrix. In addition, in QCD already the form of the continuum kernel contains an exponential decay, leading also to an intrinsic decay of the singular values. This makes the inversion task ill-conditioned. On the other hand the finite uncertainty in the input data means that many different combinations of the parameters $\rho_l$ are able to reproduce the input data within one standard deviation. Therefore at first sight an infinite number of degenerate solutions exist for the inverse problem, making it genuinely ill-posed. I.e. a naive $\chi^2$ fit of the $\rho_l$'s is meaningless.

In order to give meaning to this inverse problem, i.e. to select a unique answer, requires to incorporate additional domain knowledge, often called prior information. This prior information may be provided by QCD, at least it should be motivated by general physical arguments. This also means that the answer will in part be determined by the simulation input in part by the prior information. It is therefore paramount to quantify, how robust the extracted spectral information is with respect to uncertainty both in the data and the choice of prior information that has been included.

In the following we will discuss in detail two different approaches, which have found application to the extraction of spectral properties in the quarkonium community.

\subsubsection{Bayesian spectral reconstruction}
\label{sec:BayesRec}

The Bayesian approach to spectral function extraction utilizes methods of Bayesian inference to regularize the ill-posed nature of the task at hand. This is achieved by providing a systematic prescription on how to incorporate prior information, i.e. domain knowledge, into the inversion task. By formulating statements of confidence in the language of probabilities, Bayesian statistics provides a versatile language to formulate and solve the ill-posed problem (see e.g. Ref.~\cite{Rothkopf:2019dzu}). For an excellent introduction to Bayesian statistics see e.g. Refs.~\cite{gelmanbda04,mcelreath2016statistical}, for statistical inference see e.g. Refs.~\cite{MacKay:2003,Bishop:2006}. We will discuss below both the general ingredients to the Bayesian approach, as well as the particular strengths and weaknesses of three implementations that have found application in the literature on in-medium quarkonium. Particular attention will be placed on an understanding of reconstruction artifacts, which is mandatory for Bayesian methods to fulfil their potential as quantitative precision tools for the investigation of spectral functions.

The central quantity of Bayesian statistics relevant for spectral reconstruction is the posterior probability $P[\rho|D,I]$, which denotes the probability for a test function $\rho$ to be the correct spectral function, given the simulation data $D$ and any further prior information $I$. Starting from the joint probability distribution $P[\rho,D,I]$ and using the rules of conditional probabilities, the posterior can be expressed in terms of three quantities, a relation known as {\it Bayes theorem}
\begin{align}
P[\rho|D,I]=\frac{P[D|\rho,I]P[\rho|I]}{P[D|I]}.
\end{align}
The Bayesian strategy answers the inversion problem by interrogating the posterior for which function $\rho$ is most probable, given simulation data and domain knowledge.

The first term on the right $P[D|\rho,I]$ is called the \textit{likelihood} probability and encodes all information about how the simulation data has been obtained. This distribution describes how the values of the individual (often subaveraged) $D_j$'s vary among the different lattice configurations. Thanks to the central limit theorem, one often finds that a Gaussian distribution emerges. In that case $P[D|\rho,I]$, the probability for the data given a known value of the mean and variance, can be written as
\begin{align}
P[D|\rho,I]={\cal N}_L{\rm exp}[-L],\quad L=\frac{1}{2}\sum_{jk} (D_j-D^\rho_j)C^{-1}_{jk}(D_k-D^\rho_k), \quad {\cal N}_L=(2\pi)^{-N_{\rm data}/2}({\rm det}[C])^{-1/2}.
\end{align}
Here $D^\rho_j$ denotes the correlator values obtained from inserting the test function $\rho$ into \cref{eq:discrspecdec} and the indices $i$ and $j$ refer to the imaginary times that are included in the reconstruction $\tau_{\rm min}/a \leq i,j \leq \tau_{\rm max}/a$, the total number of supplied input points is denoted by $N_{\rm data}$. $C$, the covariance matrix of the data average with respect to the true value, in the absence of autocorrelations among lattice configurations reads
\begin{align}
C_{jk}=\frac{1}{N_{\rm conf}(N_{\rm conf}-1)}\sum_{m=1}^{N_{\rm conf}} (D_j^m-D_j)(D_k^m-D_m),
\end{align}
where $D^m_k$ refers to the m-th realization of the k-th correlator point and $D_k=\sum_m D^m_k/N_{\rm conf}$ their average. Note that in case of finite autocorrelations in Monte-Carlo time, $C$ needs to be multiplied with the product of autocorrelation times for $D_j$ and $D_k$. 

Besides possible autocorrelations in Monte-Carlo time, there exist correlations between the $D_j$'s at different imaginary times. These correlations manifest themselves as off-diagonal entries in C. One has to keep in mind that in order for these off-diagonal elements to be robustly estimated, one requires significantly more realizations $N_{\rm conf}\gg N_{\rm data}$ than the number of points along imaginary time considered. Is the number of realizations smaller than $N_{\rm data}$, then $C^{-1}$ even becomes singular due to exactly vanishing eigenvalues.

In order to accelerate the evaluation of the likelihood probability, one transforms the data and the kernel into a basis, in which the correlation matrix is diagonal $C=R\,{\rm diag}[c_k]\,R^{-1}$. This reduced the likelihood to
\begin{align}
L=\frac{1}{2}\sum_k (\tilde D_k - \tilde D^\rho_k)/c_k, \quad \tilde D_k=\sum_j R^{-1}_{kj}D_j, \quad \tilde K_{jl} = \sum_k R^{-1}_{jk}K_{kl},
\end{align}
and correspondingly the kernel $K$ in \cref{eq:discrspecdec} is replaced by $\tilde K$. Note that in case of correlators with highly constrained values, such as the trace of the Wilson loop, the Gaussian distribution may not be an adequate description of the likelihood and the histogram of the input data needs to be consulted.

There exists prior information already about the data generation process that has been used in the literature to modify the likelihood. E.g. some studies use that if Euclidean data from a known spectral function is sampled along $N_\tau$ points with Gaussian noise, then on average the likelihood, evaluated for the true underlying spectral function will take on the value $L=N_\tau$.

The second and decisive term in Bayes theorem is the \textit{prior} probability 
\begin{align}
P[\rho|D,I]={\cal N}_S{\rm exp}[\alpha S], \label{eq:BayesTheorem}
\end{align}
which encodes how compatible the test function $\rho$ is to the available prior information. The choice of this distribution is the central ingredient in setting up a Bayesian analysis and differs among the multiple implementations found in the literature. Commonly the parameters of the prior are encoded in terms of the so called default model $m(\omega)$, which denotes the extremum $\delta S/\delta \rho|_{\rho=m} =0$.  For a convex prior functional this is the only extremum. The hyperparameter $\alpha$ on the other hand encodes the overall uncertainty present in the values of the default model. This historic way of parametrizing the prior arose from choosing originally as prior a Gaussian distribution for each frequency bin $\rho_l$ with individual mean and one overall variance (Tikhonov regularization). In a modern Bayesian analysis, domain knowledge with appropriately quantified uncertainties will provide the mean and variance (and possibly higher nontrivial moments) for the prior distribution individually for each of the discretized $\rho_l$.

In the study of heavy quarkonium one most often encounters spectral functions that are positive definite. Positivity is a form of prior information obtained from the QCD spectral decomposition of hadronic correlators based on identical source and sink operators. Thus in the following we will focus on implementations of the prior probability that explicitly incorporate this fact (for Bayesian methods applicable to non-positive spectra see e.g. \cite{Rothkopf:2016luz} and references therein. For the Backus-Gilbert method see e.g. \cite{Hansen:2019idp,Tripolt:2017pzb}).

The most well known implementation of the Bayesian approach is the \textit{Maximum Entropy Method} (MEM) \cite{ Bryan:1990, Jarrell:1996rrw,skilling1991}, which was developed originally for inverse problems in the context of two-dimensional image restoration in astronomy. It has been introduced in the context of lattice QCD simulations for the first time in Ref.\cite{Asakawa:2000tr} and subsequently been used in a variety of studies of quarkonium properties. Its application to continuum QCD was first considered in \cite{Nickel:2006mm}. The MEM is based on four theorems: locality, coordinate invariance, system independence and the Bayesian requirement that the default model describes the extremum of the prior. The combination of the second and third theorem constrains the function $\rho$ to be positive definite. Taken together the Shannon-Jaynes entropy emerges 
\begin{align}
S_{\rm SJ}=\int_{\omega_{\rm min}}^{\omega_{\rm max}}\, d\omega\, \Big( \rho(\omega) - m(\omega) - \rho(\omega){\rm log}\Big[\frac{\rho(\omega)}{m(\omega)}\Big]\Big) \, <0,\quad {\cal N}_S\approx \prod_{l=1}^{N_\omega} (\alpha\Delta\omega_l)^{1/2}/(2\pi)^{-1/2}.
\end{align}

In a more recently developed implementation of the Bayesian strategy, simply christened \textit{Bayesian Reconstruction} (BR) \cite{Burnier:2013nla}, the prior is constructed specifically for the one-dimensional inversion problem of \cref{eq:discrspecdec}. While it shares with the MEM the first and last axiom it replaces the two others by a smoothness and a scale invariance axiom. The former requires that the output is a smooth function, where data has not introduced sharp peaked structures. The latter, in contrast to the axioms of the MEM, guarantees that the units assigned to the correlator do not affect the end result. This is achieved as only ratios of $\rho$ and $m$ enter, both of which have to be assigned the same units. Another consequence is that the spectral function must be positive definite. The corresponding regulator functional reads
\begin{align}
S_{\rm BR}= \int_{\omega_{\rm min}}^{\omega_{\rm max}}\, d\omega\, \Big( 1- \frac{\rho(\omega)}{m(\omega)} + {\rm log}\Big[\frac{\rho(\omega)}{m(\omega)}\Big]\Big) \, <0,\quad {\cal N}_S= \prod_{l=1}^{N_\omega} {\rm exp}[\alpha \Delta\omega_l](\alpha\Delta\omega_l)^{-\alpha \Delta\omega_l}m_l\Gamma(\alpha \Delta\omega_l),
\end{align} 
which corresponds to $P[\rho|I]$ taking on the form of a gamma distribution. 

Even though both the MEM and the BR method have been designed in a way that intuitively suggests that they favor smooth functions over wiggly functions, it turns out (see e.g. Ref.~\cite{Fischer:2017kbq}) that this is not the case in general.
\begin{figure}
\centering
\includegraphics[scale=0.5]{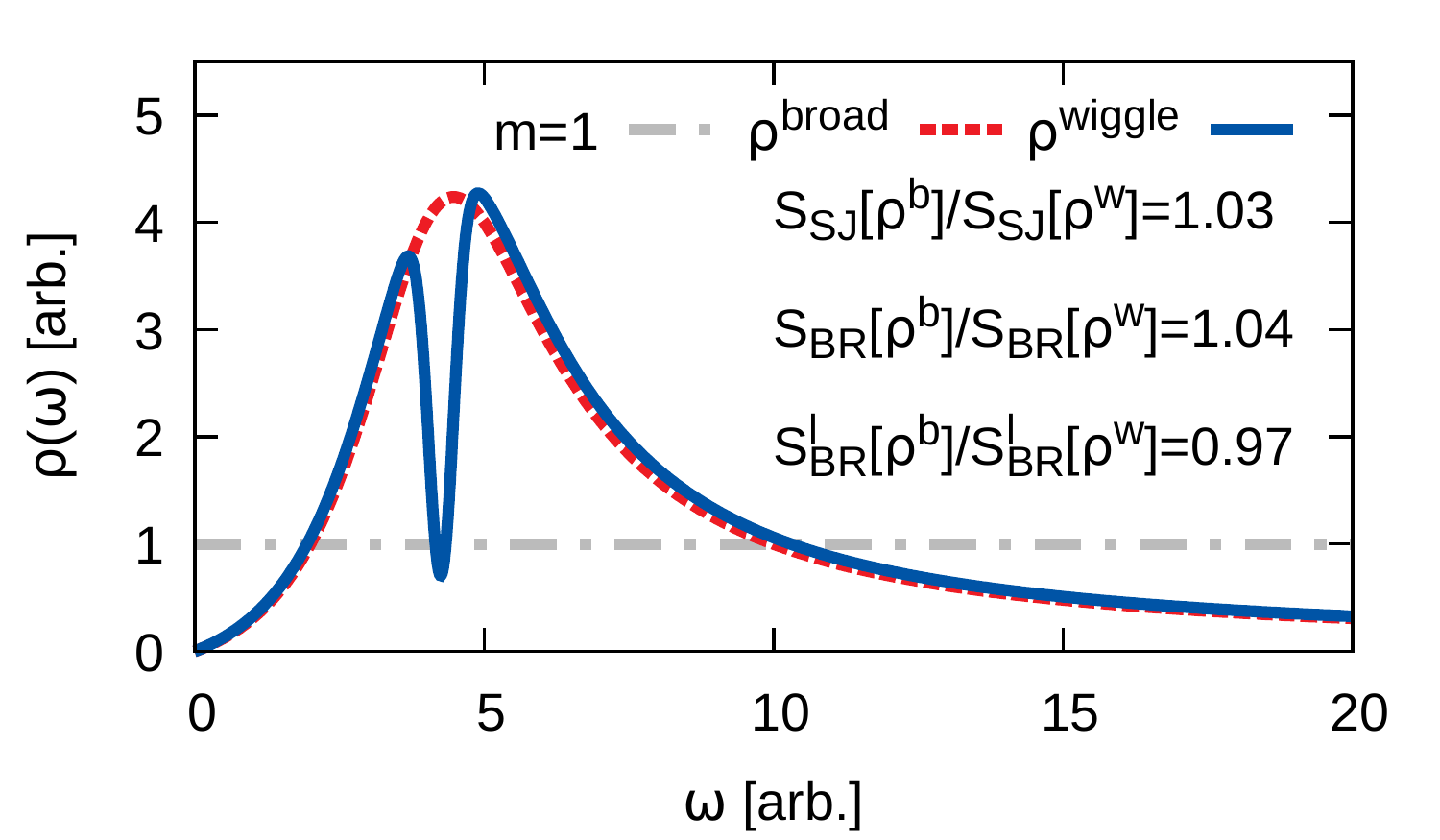}
\caption{Comparison of different regulators and their response to wiggly structures.}\label{fig:ringingcheck}
\end{figure}
In \cref{fig:ringingcheck} we provide an explicit example of the penalties assigned to a function with (blue solid) and without wiggles (red dashed) in case of a constant default model. The two curves enclose the same area. Both the Shannon Jaynes entropy and the BR regulator lead to a higher probability for the wiggly structure (since they are negative, a ratio larger than one means that the wiggly line is favored). This is an issue affecting purely local regulator functionals and may lead to the appearance of artificial ringing in reconstructed spectral functions. Remember that already in well conditioned inverse problems, such as the inverse Fourier series, so called Gibbs ringing occurs. Reconstructing a sharp feature with compact support from a finite number of exactly known Fourier coefficients will lead to a result, which oscillates even where the original input is exactly zero, only converging to the correct result for an increasing number of datapoints. In order to accurately determine the physics content of a reconstructed spectral functions, ringing needs to be reliably controlled.

In order to overcome this shortcoming, the state-of-the art implementation of the MEM by Bryan \cite{Bryan:1990} proposes to limit the functional space from which to choose the function $\rho$ to a low dimensional subspace of smooth functions around the default model. In that case the strength of the smoothing is directly related to the number of available input datapoints. For the BR method a different strategy has been proposed that implements smoothing explicitly in the prior functional $S$, keeping in line with the Bayesian philosophy. I.e. one adds to the standard BR prior an additional term that penalizes the arc length of the function $\rho$, one possible criterion to quantify the wiggliness of the spectrum. A naive way of constructing the corresponding regulator has been proposed in Ref.~\cite{Fischer:2017kbq} leading to 
\begin{align}
S_{\rm BR}^{\rm smooth}=\int_{\omega_{\rm min}}^{\omega_{\rm max}}\, d\omega\, \Big( -\kappa\Big[\frac{\partial \rho}{\partial \omega} - \frac{\partial m}{\partial \omega}\Big]^2 + 1 - \frac{\rho(\omega)}{m(\omega)} + \frac{\rho(\omega)}{m(\omega)}\Big) \, <0.
\end{align}

\begin{figure}
\centering
\includegraphics[scale=0.45]{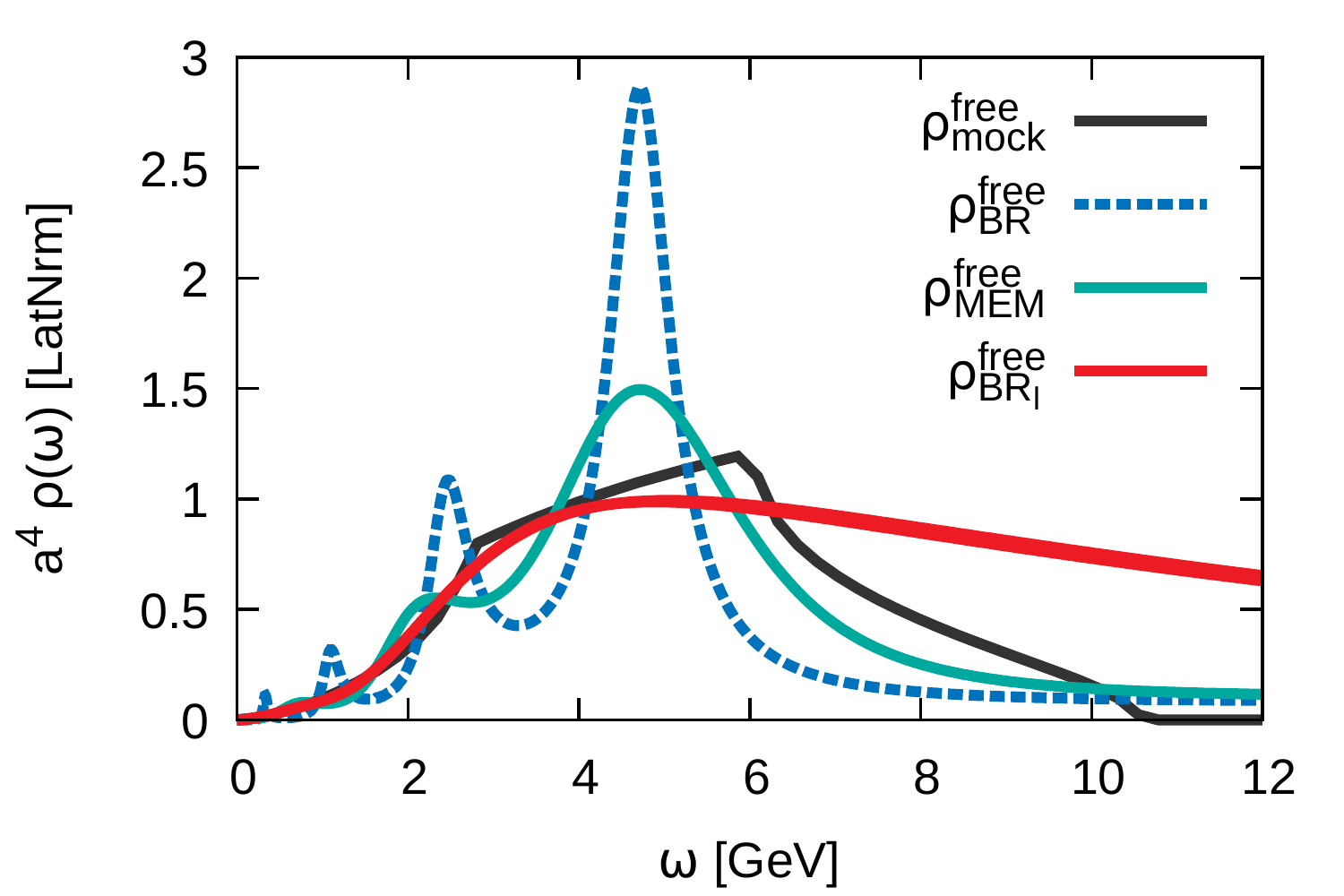}
\caption{Example of tuning of the smoothing hyperparameter $\kappa$ in the smooth BR method based on prior information in the form of the analytically known free P-wave spectral function in lattice NRQCD (black). The reconstruction of this structure from $N_\tau=12$ datapoints along an Euclidean extent of $\beta=1.4$fm via the standard BR method with $\kappa=0$ (blue dashed), the MEM (green solid) and the smoothed BR method with $\kappa=1$ are plotted.}\label{fig:kappatuning}
\end{figure}

The strength of smoothing in the end result is now made explicit by the additional hyperparameter $\kappa$. The value it takes on depends on the problem at hand and needs to be set in a self consistent manner. In practice prior information is used to do so, with one pertinent example being the use of analytically known free spectral functions. In \cref{fig:kappatuning} the procedure for the example of quarkonium in lattice NRQCD is sketched. The black curve denotes the free P-wave spectral function in the thermodynamic limit, which contains no peak structures and extends over a relatively large frequency range. After computing the Euclidean correlator of this spectrum, discretized with a small number of $N_\tau=12$ points along a realistic Euclidean range of $\beta=1.4$fm the reconstruction outcomes of different methods can be compared. Using $\kappa=0$, i.e. the standard BR method one finds significant ringing to be present (dashed blue), due to the small number of input points. The MEM (solid green) already shows much less ringing. Increasing the value of $\kappa$ in the smooth BR method one will arrive at intermediate values at a reconstruction result, which is quite similar to the MEM. Eventually when (in this case) one reaches $\kappa=1$ (red solid) no more remnants of ringing are present and the range of frequencies up to $3$GeV is faithfully reproduced. In all reconstructions a default model with unit value is used, which all reconstruction eventually approach at high frequencies. 

Returning to the general discussion, we see that prior information enters Bayes theorem in multiple ways. Not only does one need to choose the functional form of the prior probability depending on which sets of axioms appear more appropriate. One also has to set the parameters governing that distribution, i.e. the default model using QCD domain knowledge. Note that already selecting the frequency range along which to discretize $\rho(\omega)$ constitutes prior information.

The third term in Bayes theorem $P[D|I]$ historically christened the {\it evidence} corresponds to the $\rho$ independent normalization of the numerator of \cref{eq:BayesTheorem}.  While often neglected in practice it can play a role in the self consistent determination of the hyperparameter $\alpha$ in the MEM.

With all the ingredients on the r.h.s. of Bayes theorem defined, we are ready to solve the inverse problem in a Bayesian fashion, i.e. by interrogating the resulting posterior. The modern form to do so is to deploy Monte-Carlo methods (e.g. the MC-STAN \cite{JSSv076i01} library \url{https://mc-stan.org/} ) to sample the full distribution $P[\rho|D,I]$ through which we may define the most probable spectral function via the expectation value
\begin{align}
\rho_l^{\rm Bayes}=\int d\rho \, \rho\, P[\rho|D,I]
\end{align}
An important feature of the Bayesian approach is that it is very easy to incorporate new input data once a higher quality in the lattice simulations has been achieved. We may then use the posterior distribution of a previous analysis as the prior for the next analysis, incrementally improving the accuracy of the inversion. 

In the past, due to the computational cost involved in sampling the posterior it has been common to simply compute the value of $\rho$ at the extremum of the posterior. This \textit{maximum aposteriori} solution $\rho^{\rm MAP}$ satisfies
\begin{align}
\left.\frac{\delta P[\rho|D,I]}{\delta \rho}\right|_{\rho=\rho^{\rm MAP}}=0 \quad \Leftrightarrow \quad \left.\frac{\delta}{\delta \rho}\big( -L+\alpha S\big)\right|_{\rho=\rho^{\rm MAP}}=0
\end{align}
and is found by carrying out a numerical optimization procedure. Since most optimization algorithms are designed to find minima, one often considers $L-\alpha S$ instead. ( For the stochastic analytic inference approach see e.g. \cite{Ding:2017std}.)

Since the MAP is still the most common approach to Bayesian reconstruction in the study of heavy quarkonium, let us briefly discuss some of its aspects. The extremum $\rho^{\rm MAP}$ results from a competition between a likelihood with many degenerate local minima and the prior. As long as the prior is convex it can be shown that if an extremum of the posterior exists it is unique (for a proof see \cite{Asakawa:2000tr}). This is the case for both the MEM and the standard BR method. Note that the derivative term in the smooth BR method can lead to the appearance of additional local extrema and thus more sophisticated optimization, such as simulated annealing is in general called for. Note also that for the same choice of $m$, the curvature of $S_{\rm BR}$ is weaker than that of $S_{\rm SJ}$, meaning that in a MAP procedure the BR method will imprint the prior information more weakly on the end result than the MEM.

Let us remark on the fact that it is possible to obtain a unique result for the $N_\omega \gg N_\tau$ parameters $\rho_l$ here. The reason is that we have provided not just the $N_\tau$ pieces of information via the simulated correlator but in addition $N_\omega$ pieces of information via the default model, as well as information due to the shape of the regulator functional. Contrary to statements previously made in the literature, i.e. there is now more information present than free parameters in the inverse problem. I.e. determining the Bayesian answer is well posed in terms of the $N_\omega$ d.o.f. $\rho_l$.

The hyperparameter $\alpha$ is handled differently in the MEM and BR method. In the {\it historic MEM} prescription the extremum of the posterior is obtained with alpha chosen such that for $\rho^{\rm MAP}_{\rm hist}$ the likelihood takes on the value $L=N_\tau$. This choice reflects the fact that on average the correct spectral function sampled with Gaussian noise would lead to exactly this value. On the other hand in the so called {\it modern MEM}, several $\rho^{\rm MAP}$'s are computed for a range of different $\alpha$ value and subsequently averaged, weighted according to which $\alpha$ leads to an extremal evidence. Since the evidence cannot be computed analytically this is conventionally implemented by using a simple Gaussian approximation of $P[D|I]$. In the BR method on the other hand one assumes complete ignorance of the value of $\alpha$ and marginalizes $\alpha$ apriori. In practice this step is implemented semin-analytically or fully numerically. In addition the BR method also enforces $L=N_\tau$ as part of the prior information entering the likelihood.

Note that removing the prior information by setting $P[\rho|I]=1$ above, one returns to an underdetermined maximum likelihood problem, equivalent to a naive $\chi^2$ fit. On the other hand, since the prior is defined to have a global extremum at $\rho=m$, the function $m(\omega)$ by definition is the Bayesian answer to the inverse problem in the absence of simulation data.

As long as the input data is finite in number and carries a non-vanishing uncertainty $\rho^{\rm Bayes}$ the solution of the Bayesian extraction will in general depend on the prior information included in $P[\rho|I]$. On the other hand Bayes theorem assures us that in the {\it Bayesian continuum limit} of concurrently increasing the number of input data $N_\tau\to\infty$ and reducing the uncertainty $dD/D\to0$ one converges to a unique result in which the influence of the prior information becomes negligible.

As we will see in the investigation of heavy quarkonium spectra, for a given quality of input data, some features in the reconstruction may already be stably reproduced (e.g. the position of the ground state peak), while others are not (e.g. excited state peaks and ground state width). The choice of $P[\rho|I]$ determines in what way we will approach the true result, as the input data quality is improved. I.e. it is imperative to understand what artifacts are present and how they are related to the regularization. The use of mock input data, i.e. correlators computed from known input spectra play an important role in doing so.

Both MEM and BR feature a convex prior, leading in principle to a unique extremum of the posterior. For the MEM, when studying spectra with sharp peaks and otherwise vanishing spectral weight, the convergence to the global extremum is impeded by the fact that the Shannon Jaynes entropy takes on a finite value $-m$ as $\rho\to0$, leading effectively to a flat direction, where numerical optimization algorithms are not efficiently pulled towards the global extremum. In the BR method the prior diverges as $\rho\to0$, avoiding this issue. 

Since in quarkonium studies from lattice QCD we are faced with the problem that only a relatively small number of datapoints will be available in the near future $N_\tau\sim{\cal O}(10-40)$ the most relevant artifact in the BR method and the MEM is the presence of ringing. The appearance of unphysical wiggly structures interferes with the identification of the relevant physical structures, which also come in the form of peaks. Different strategies exist on how to mitigate the effects of ringing.

Let us first discuss the BR method. Due to its weak curvature the BR prior is more susceptible to ringing than the Shannon Jaynes entropy. In order to obtain meaningful results in the presence of only a small number of datapoints, the smooth BR prior has been proposed. The strategy will be to tune the smoothing hyperparameter $\kappa$ based on a mock-data analysis involving realistic and analytically known input spectral functions. In practice these will often be the non-interacting spectral functions. Reconstructing such spectra featuring broad structures from correlator data that is discretized in the same fashion as the actual simulation data, one finds a minimum value for $\kappa$ at which it effectively suppresses ringing, while still allowing us to pick up peaked structures that are actually encoded in the input data. 

Let us now discuss the MEM. Dealing with ringing in the MEM has traditionally been avoided by incorporating an additional smoothing in the MAP procedure. There are two independent proposals in the literature on how to do so. Both propose that the global extremum of the posterior is located in a subspace spanned by a collection of smooth functions related to the transpose Kernel $K^t$. 

In the state-of-the-art implementation by Bryan \cite{Bryan:1990}, the spectral function is parametrized in terms of deviations from the default model $\rho_l=m_l{\rm exp}[a_l]$. It is then argued that the parameters $a_l$ are  restricted to a functional space corresponding to the first $N_\tau$ columns of the matrix $U$ arising from the singular value decomposition of the transpose kernel $K^t=U\Sigma V^t$. Here $U$ refers to an $N_\omega\times N_\omega$ matrix containing the orthonormal basis of eigenvectors of $K^t K$, $\Sigma$ is a $N_\omega\times N_\tau$ diagonal matrix containing the $N_\tau$ finite singular values. $V^t$ denotes an orthonormal $N_\tau\times N_\tau$ matrix. The more recent proposal by Jakovac \cite{Jakovac:2006sf} implicitly solves for the MAP solution leading to a restricted search space for the $a_l$'s spanned directly by the columns of $K^t$. If the reconstruction problem were purely linear (the forward problem obviously is) then the two formulations are equivalent. I.e. the columns of $K^t$ span the same image space as the first $N_\tau$ columns of the matrix $U$ obtained by the SVD. In both cases the smoothness of the solution is directly related to the number of input datapoints provided. 

Jakovac already observed that the two prescriptions lead to different results in practice (see Fig.28 in \cite{Jakovac:2006sf}), indicating that they are actually not equivalent. In turn the question arises whether the global extremum in general is contained within the image space of $K^t$. This question is further emphasized by a counterexample to the SVD restriction, which has been put forward in Ref.~\cite{Rothkopf:2011ef}.

Since it will help us understand the systematics of the MEM more clearly, let us discuss the counterexample in more detail by considering the historic MEM. The SVD search space is fully specified once the range and discretization of $\tau$ and $\omega$ have been selected. For $N_\tau$ input data the search space is parametrized by the first $N_\tau$ columns of $U$. This statement is made independent of the input data and independently from the recipe of how to choose the value $\alpha$. 

Now let us choose instead the $N_\tau+1$st column of $U$ as mock spectrum $\rho$ and compute from it the corresponding Euclidean data. Then, by construction, this data cannot be reproduced within one sigma from within the SVD search space, while it is still possible to reproduce it in the full search space. Taking the errors on that mock input data to zero, the minimal value of $L$ in the SVD subspace can be made arbitrarily large. And since the Shannon Jaynes entropy is negative definite it cannot compensate the large values of L. In turn the posterior probability of that $\rho$ in the SVD subspace can be made smaller than in the full search space, disproving the general claim that the extremum of the posterior always lies in the SVD subspace.

Thus the claim of the counterexample is that the smoothing introduced by restricting the search space to the image of $K^t$ is ad-hoc and may lead to artificially smoothed results. In practice it turns out that the SVD prescription works well for problems where accurate prior information is already available. This corresponds to cases where the true global extremum indeed lies within the SVD subspace, as the default model is located close enough to it in parameter space apriori. However in situations where both only a relatively small number of input data is available (e.g. ${\cal O}(10)$) and the prior information is limited, the SVD search space may not provide the Bayesian answer to the inversion task. Of course, since the global extremum of the MEM posterior is unique if it exists, we can still solve for it in the full search space. Systematically extending the search space has been considered in \cite{Rothkopf:2011ef,Rothkopf:2012vv}.

The last point to consider is how to estimate the uncertainties in the spectral reconstruction. Bayes theorem makes explicit that the posterior depends on two ingredients, data and prior information. If one samples the posterior using Monte-Carlo methods, the spread in that distribution encodes the combined uncertainty arising from spread in both the likelihood and prior. If on the other hand one only computes a point estimate, such as the MAP, the uncertainty must be estimated in addition. One proposal in the literature \cite{Asakawa:2000tr} is to evaluate how pronounced the maximum of the posterior is, taking its curvature as a measure for the robustness of the solution. While intuitive, it has been found in practice that this criterion may underestimate the full uncertainty budget.

In general to estimate the statistical uncertainties of the result, one of the many different bootstrapping methods can be deployed. E.g. the blocked {\it Jackkknife} procedure remains a simple and computationally cheap option \cite{Montvay:1994cy}. Instead of using all available lattice realizations of the correlator $D$ for the reconstruction, one forms $N_J$ so called Jackknife averages $D^j$, where for each $j$ a consecutive subset of $N_{\rm conf}/N_J$ realizations of the correlator have been excluded from the average. This leads to $N_J$ reconstructions of the spectral function $\rho^j$. The variance of the spectral function computed from the full statistics average is then estimated as 
\begin{align}
\sigma_\rho^2 = \frac{N_J-1}{N_J}\sum_{j=1}^{N_J} (\rho^j-\rho)^2.
\end{align}

Quantifying the systematic uncertainties arising from prior information is less formalized. E.g. one often does not possess a reliable estimate of the uncertainty of the values of the default model.  However since Bayes theorem makes the influence of the prior information explicit, it may also be varied in a straight forward fashion. I.e. one should repeat the reconstruction using default models, which posses different functional forms in those regions, where accurate information on $\rho$ is absent. In addition deploying different prior distributions that are applicable to the problem at hand provides additional insight on how the regularization affects the end result. The variation among the reconstructions from different default models and priors is then summarized into a systematic error bar. The prescription on how to do so differs between studies in the literature, some taking the maximum deviation, others the quadratic mean.
 
\subsubsection{Pade approximation based reconstruction}
\label{sec:Pade}

The second class of reconstruction prescriptions that find application in quarkonium studies falls into the category of projection based methods. I.e. in these methods the simulation data in Euclidean time $\tau$ is first Fourier transformed into imaginary frequencies $q^0$ before being projected onto a set of analytically known functions. In essence one is expanding the data in a particular basis set. The basis functions can then be analytically continued in an appropriate fashion to yield the retarded propagator $D^{R}(\omega)$ in real-time frequencies, whose imaginary part is directly related to the spectral function ${\rm Im}[D^{R}(\omega)]\propto \rho(\omega)$.

Carrying out a projection does not require prior information. On the other hand, the fact that one is projecting not only the true Euclidean data but also noise means that these methods suffer strongly from statistical uncertainties in the input. The ill-conditioned-ness of the analytic continuation can be understood intuitively. While the basis functions in Euclidean time are mostly monotonously damped functions, in real-time they behave oscillatory, often with strongly growing amplitudes. Thus to linearly combine these basis functions into a finite and damped real-time correlator the coefficients obtained from the projection need to be computed with extremely high precision.

In the case of relativistic Euclidean correlators, which exhibit periodic behavior along Euclidean time, a projection method based on the so called Pollaczek polynomials has been proposed in Ref.~\cite{Cuniberti:2001hm}. The prescription is straight forward, i.e. the expansion parameters can be explicitly computed from the input data, as can be the analytically continued approximant in terms of Laguerre polynomials. Unfortunately it turned out that in practice the fact that only a finite number of input datapoints are available and that statistical errors are present, leads to a significant deterioration of the reconstruction results \cite{Burnier:2011jq}. Hence this method has so far not been applied at a large scale.

A more general approach, also applicable to Euclidean correlators without specific symmetry properties, is the Pade approximation. It expresses the correlator of interest in terms of a rational function. Depending on the choice of the highest monomial contributing to the numerator $n$ or denominator $m$ one refers to a $(n,m)$ Pade approximation $R_{(n,m)}$. The $(n,0)$ approximation corresponds to the simple Taylor series to order $n$. Interest in this method has recently been rekindled by studies of spectral reconstructions from approximate non-perturbative analytic computations \cite{Tripolt:2017pzb,Tripolt:2018xeo} and the arrival of very high statistics lattice QCD ensembles.

In the context of spectral function reconstruction the close relation between Pade approximants and continued fractions is exploited in the following way. In order for the analytic continuation of the approximant to be stable one wishes to construct expressions with either the same highest monomial power in the numerator and denominator $(n,n)$ or even better, with one power less in the numerator $(n-1,n)$. For $N_{\rm data}$ input data points $D_j$ along imaginary frequencies $q^0_j$, one can construct the approximant 
\begin{align}
\nonumber \tilde D_{N_{\rm data}}(q^0)=\frac{D_M(q^0_0)}{1+} &\frac{a_0(q^0-q^0_0)}{1+}\frac{a_1(q^0-q^0_1)}{1+}\ldots\frac{a_{N_{\rm data}-1}(q^0-q^0_{N_{\rm data}a-1})}{1+},\label{Eq:ContFrac}
\end{align}
which corresponds to the $((N_{\rm data}-1)/2,(N_{\rm data}-1)/2)$ approximation in case that $N_{\rm data}$ is odd and to $(N_{\rm data}/2-1,N_{\rm data}/2)$ if $N_{\rm data}$ is even. The expansion coefficients $a_i$ can be determined recursively \cite{Schlessinger:1968} by the Schlessinger method
\begin{align}
a_l(&q^0_{l+1}-q^0_l)=-\Big\{ 1+\frac{a_{l-1}(q^0_{l+1}-q^0_{l-1})}{1+}+\frac{a_{l-2}(q^0_{l+1}-q^0_{l-2})}{1+}\cdots\frac{a_{0}(q^0_{l+1}-q^0_{0})}{1-[D_M(q^0_0)-D_M(q^0_{l+1})]}\Big\}.
\end{align}
The resulting interpolation $\tilde D_{N_{\rm data}}(i\omega)$ exactly reproduces the input data $D_M$ at the imaginary frequencies $q^0_l$ provided. The retarded correlator needed to compute the spectral function may now be obtained by Wick rotating the imaginary frequency argument of the approximant $D^R(\omega)\approx-\tilde D_{N_{\rm data}}(q^0\to \omega - i\epsilon)$ approaching $\epsilon\to0^+$ from above.

As in any direct projection method, cancellations occur in the evaluation of the continued fraction, in particularly if there are symmetries in the input data. This requires the evaluation of the intermediate steps in the computation to be carried out with high precision arithmetic. As we will show in the explicit application of the Pade approximation to the determination of the in-medium heavy quark potential, currently available lattice QCD data is of high enough quality for this direct projection method to be of use.

While the Pade reconstruction does not presuppose explicit prior information, analyticity of the approximant is required for meaningful results. I.e. if the correlator contains divergences or otherwise non-analytic features the accuracy of the Pade based reconstruction may suffer. In the context of studying the spectral functions of gluons it has been found \cite{Cyrol:2018xeq} that the Pade reconstruction often violates the spectral decomposition of the correlator. I.e. plugging the reconstructed spectrum back into \cref{eq:discrspecdec} will produce a Euclidean correlator deviating from the input data by significantly more than one standard deviation. In addition, since positive definiteness never entered the analysis, the Pade reconstruction may lead to artificial excursions into negative values. The stability of reconstructed spectral features needs to be ascertained with a Jackknife analysis, as well as by testing how the removal of input datapoints affects the end result.

These artifact however do not spell doom for the method, if our goal is to study only subsets of the spectral features. As will be discussed, the position of e.g. the ground state peak may be well captured by the Pade reconstructed spectral function, while its width and the higher lying structures are not accurately reproduced. It is therefore paramount to test in each individual reconstruction scenario whether the Pade method is able to capture the spectral features of interest, given a certain quality of input data.

\begin{summary} Bayesian inference of spectral functions provides a systematic prescription of how to incorporate prior domain knowledge into the regularization of the otherwise ill-posed inversion task. By interrogating the posterior $P[\rho|D,I]$ expressed in terms of the likelihood $P[D|\rho,I]$ and prior $P[\rho|I]$, the most probable spectral function, given simulation data and prior information, may be found. Prior information on the data generation may enter the likelihood, most prior information however resides in the prior probability itself. The MEM and BR method provide two different implementations of $P[\rho|I]$, based on different underlying axioms. They have in common that positivity of $\rho$ is enforced. While for a finite number of datapoints with finite uncertainty the most probable spectral function may differ between the methods, they will converge to a unique answer in the Bayesian continuum limit.  For reconstructions based on a small number  of ${\cal O}(10-40)$ datapoints, similar to what is often encountered in $T>0$ quarkonium studies, ringing constitutes an important numerical artifact. In the MEM it is suppressed by restricting the solution ad-hoc to a smooth functional subspace while in the smooth BR method it is treated by self-consistently tuning an additional hyperparameter $\kappa$. Since prior information is made explicit, its role in the total uncertainty budget can be straight forwardly be estimated by comparing results based on different default models and prior probabilities. In case of high precision input data also direct projection based methods become viable. The Pade approximation e.g. exploits the analyticity of the correlator and does not require further prior information. Its sensitivity to statistical fluctuations however also requires careful error estimation. 
\end{summary}

\subsection{Open Quantum Systems}
\label{sec:OQS}

In the preceding sections we have discussed how the presence of a separation of energy scales allows us to simplify the description of heavy quarkonium using a non-relativistic language. So far we restricted ourselves to a fully equilibrated scenario, since lattice QCD simulations are limited to Euclidean time. Extracting spectral functions and e.g. thermal widths from these simulations gave first insight into the real-time dynamics of heavy quarkonium in equilibrium with its environment. In the context of a heavy-ion collision however we need to go beyond equilibrium to understand the physics of heavy quarkonium. Indeed insight is required on how a heavy quarkonium state reacts to a QCD environment with which it initially is not equilibrated at all.

One of the exciting developments of the past decade is the realization that heavy quarkonium is an ideal example of an open quantum system (OQS) and we may understand many aspects of its real-time properties in and out-of equilibrium from this viewpoint. One example is the concept of decoherence that provides a new perspective on the dynamics of quarkonium melting. The framework of OQS has a long history in condensed matter physics and we thus benefit from many established results in that field. For an excellent introduction to OQS see e.g. Ref.~\cite{BRE02}. On the other hand, the strongly coupled nature of QCD and the particular physics of the color quantum number introduce additional complexities, leading to rich phenomenology not present in other OQS. While considerations of a separation in energy scales forms the basis for deriving effective descriptions in thermal equilibrium it is the separation of timescales that will allow us to come up with effective descriptions of the quarkonium real-time dynamics. Bridging the language of the effective field theories NRQCD and pNRQCD and the OQS framework is a central focus of current research on in-medium quarkonium dynamics. 

A system is well suited to the open-quantum-systems approach if its d.o.f. can be separated into an environment $E$ and a small subsystem $S$ coupled via common interactions. In case that $E$ consists of an infinite number of degrees of freedom it is usually referred to as a reservoir, if in addition it is in thermal equilibrium we may call it a heat bath. It is the presence of a continuous distribution of modes in a reservoir and the fact that its dynamics do not show recurrence that will lead to the emergence of  dissipative effects when considering the dynamics of the small subsystem. In our case the environment is represented by a thermal QCD medium and the small subsytem is formed by the quarkonium two-body system. 

In the following we will give an introduction to the aspects of OQS relevant for the treatment of quarkonium, following closely the exposition in Ref.~\cite{BRE02}. As the concept of the open system is not commonly treated in high energy physics contexts, we will discuss in detail the derivation of two equations of motion arising from two different time scale hierarchies: the quantum optical and quantum Brownian motion limit.

\begin{figure}
\centering
\includegraphics[scale=0.35,clip=true,trim=0cm 8cm 1cm 0]{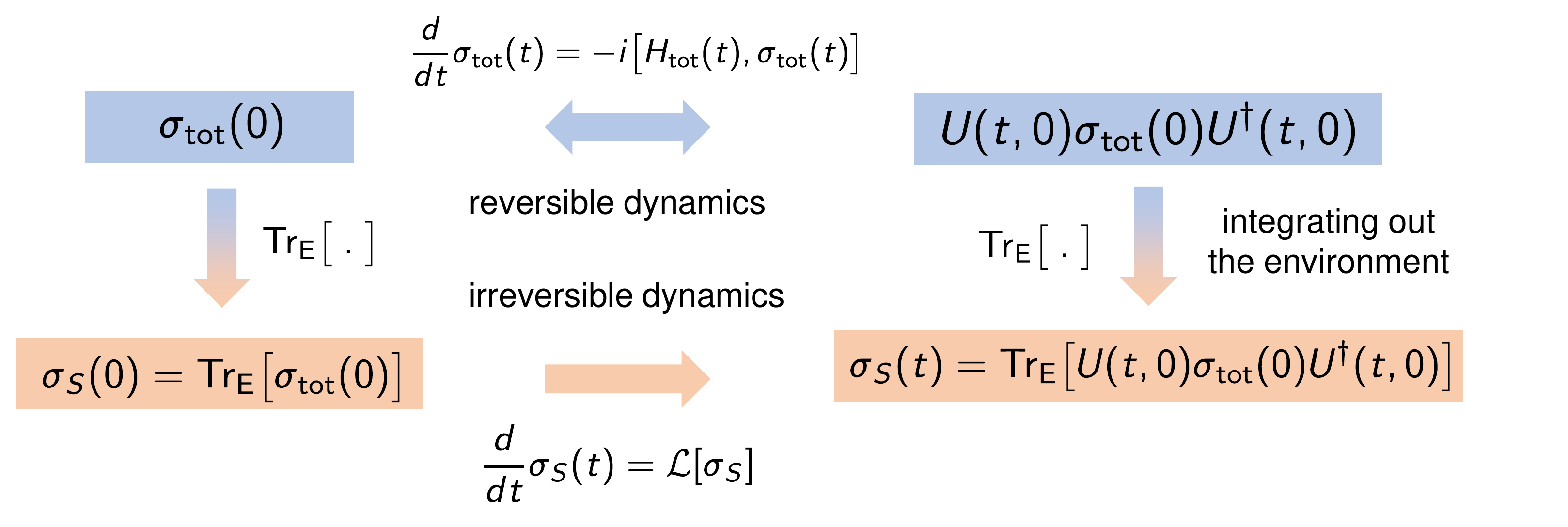}
\caption{Sketch of the two formulations of the time evolution of the quarkonium systems. The top row operates on the level of the total system including both medium and subsystem d.o.f., which leads to reversible unitary time evolution. The bottom row follows from integrating out the environment and leads to irreversible dynamics, formulated in terms of the reduced density matrix.}\label{fig:OQSSketch}
\end{figure}

The overall system consisting of medium and heavy quark d.o.f. is of course closed and possesses a hermitean Hamiltonian. It can be decomposed in terms of a Hamiltonian acting only on the environment d.o.f. $H_{\rm E}$, one only in the subsystem $H_{\rm S}$ and an interaction term $H_{\rm int}$ that connects both
\begin{align}
H_{\rm tot}=H_{\rm S}\otimes I_{\rm E} + I_{\rm S} \otimes H_{\rm E} + H_{\rm int}. \label{eq:hamdec}
\end{align}
The density matrix of the full system $\sigma_{\rm tot}$ initialized using state vectors $|\psi_k(t_0)\rangle$ of the full system
\begin{align}
\sigma_{\rm tot}(t_0)=\sum_k p_k |\psi_k(t_0)\rangle \langle \psi_k(t_0)|, \quad \frac{d}{dt}\sigma_{\rm tot}(t)=-i[H_{\rm tot}(t),\sigma_{\rm tot}(t)], \quad \sigma_{\rm tot}(t)=U(t,0)\sigma_{\rm tot}(t_0)U^\dagger(t,0),
\end{align}
evolves according to the well known von-Neumann equation, captured in a unitary time evolution operator $U(t,0)={\cal T}{\rm exp}[-i\int dt H_{\rm int}(t)]$. While solving the dynamics of the whole system may be too demanding it is also not our goal. Instead we wish to focus on the evolution of the quarkonium system only, which we may formally achieve by tracing out all medium degrees of freedom. This leads to the reduced density matrix 
\begin{align}
\sigma_{\rm S}(t)={\rm Tr}_{\rm E}[\sigma_{\rm tot}(t)], \quad \sigma_{\rm S}(t)=V(t)\sigma_{\rm S}(0).
\end{align}
The quantity $\sigma_{\rm S}$ is the central focus in the open quantum systems approach and we will set out to formulate its explicit equation of motion, a so called master equation. Time evolution of $\sigma_{\rm S}$ is formally implemented via a so called dynamical map $V$. While $V(t)$ can in general be a very complicated operator there is a class of systems, where its form simplifies to that of a semi-group, i.e. $V(t_1)V(t_2)=V(t_1+t_2)$. This is the case when the dynamics of the system is Markovian, i.e. if the next infinitesimal step in the system evolution only depends on the current state of the system. Neglecting memory effects is admissible if a separation of timescales exists between the fast damping of correlations in the environment $\tau_{\rm E}$ and the relaxation scale of the subsystem $\tau_{\rm rel}$. In that case the linear map may be written in terms of a generator of the dynamical semi-group ${\cal L}$ and a linear evolution equation for $\sigma_S$ emerges
\begin{align}
V(t)={\rm exp}\big[{\cal L}t\big],\quad \frac{d}{dt}\sigma_{\rm S}(t)={\cal L}\sigma_{\rm S}.
\end{align} 
The most general form of such a Markovian master equation has been derived independently by two groups \cite{Gorini:1975nb,Lindblad:1975ef} and is known as the GKS or Lindblad equation
\begin{align}
\frac{d}{dt}\sigma_{\rm S}= -i[\tilde H_{\rm S},\sigma_{\rm S}]+\sum_k \gamma_k\Big( L_k \sigma_{\rm S} L^\dagger_k -\frac{1}{2} L^\dagger_kL_k\sigma_{\rm S} - \frac{1}{2}\sigma_{\rm S}L^\dagger_k L_k\Big).\label{eq:Lindblad}
\end{align}
The quantities $L$ are called Lindblad operators and encode the coupling of the system to its environment. They can always be chosen to be traceless. The Hamiltonian $\tilde H_{\rm S}$ not necessarily agrees with the Hamiltonian $H_{\rm S}$ in \cref{eq:hamdec}. If the $L$'s are made dimensionless, the quantities $\gamma_k >0$ take on the dimensions of $1/{\rm time}$ and represent relaxation rates for modes that decay over time in $S$. In practice the $\gamma$'s are given by correlation functions of the environment degrees of freedom. Special care needs to be taken if strong external fields are present (for oscillatory fields e.g. Floquet theory is invoked). Note that the time evolution implemented by \cref{eq:Lindblad} is irreversible, as sketched in \cref{fig:OQSSketch} and we will show explicitly that entropy increases with time inside the small subsystem.

Formulating the dynamics of an OQS in terms of a Lindblad equation is of particular interest as it can be proven that the resulting time evolution preserves the fundamental properties of the reduced density matrix, i.e. positivity, hermiticity and unitarity
\begin{align}
\langle n | \sigma_{\rm S} |n \rangle >0, \forall n, \quad \sigma_{\rm S}^\dagger = \sigma_{\rm S}, \quad {\rm Tr}[\sigma_{\rm S}]=1.
\end{align}
If the Lindblad operators $L$ do not explicitly depend on time, the time evolution of a Heisenberg operator $O(t)$ can be formulated by simply replacing $\sigma_{\rm S}$ by $O(t)$ in \cref{eq:Lindblad}.

To understand the different types of approximations often invoked in the derivations of master equations in practice, it is instructive to consider a simple weakly coupled example. We will encounter three different timescales, whose hierarchy determines what kind of dynamics emerges. Considering $H_{\rm int}=\sum_m \Sigma_m \otimes  \Xi_m$ in terms of operators in the environment $\Xi_m$ and the subsystem $\Sigma_m$ we have:
\begin{itemize}
\item $\tau_{\rm E}$: the timescale along which correlations in the medium $\langle \Xi_m(t)\Xi_n(0)\rangle$ decay.
\item $\tau_{\rm S}$: intrinsic timescale of the evolution in the subsystem, defined by a typical value of neighboring frequencies contributing in the spectral decomposition $\tau_{\rm S}=|\omega-\omega^\prime|^{-1}$. 
\item $\tau_{\rm rel}$: the subsystem relaxation time, which e.g. for a single heavy particle subsystem is defined from the randomization timescale of its momentum $\langle p\rangle (t)\propto e^{-t/\tau_{\rm rel}}\langle p\rangle (0)$
\end{itemize}

In the interaction picture, where Heisenberg operators evolve via the interaction Hamiltonian, let us consider the full density matrix written in integral form 
\begin{align}
\sigma_{\rm tot}(t)=\sigma_{\rm tot}(0)-i\int_0^t ds [H_{\rm int}(s),\sigma_{\rm tot}(s)],
\end{align}
from which, assuming ${\rm Tr}_{\rm E}[[H_{\rm int}(t),\sigma_{\rm tot}(0)]]=0$, the evolution of the reduced density matrix follows as
\begin{align}
\frac{d}{dt}\sigma_{\rm S}(t)=-\int_0^t ds {\rm Tr}_{\rm E}\big[[H_{\rm int}(t),[H_{\rm int}(s),\sigma_{\rm tot}(s)]]\big].
\end{align}
The first approximation made to simplify the equation of motion is the {\it Born} approximation, which amounts to a weak-coupling assumption between the system and environment leading to $\sigma_{\rm tot}(t)\approx \sigma_{\rm S}(t)\otimes \sigma_{\rm E}$
\begin{align}
\frac{d}{dt}\sigma_{\rm S}(t)=-\int_0^t ds {\rm Tr}_{\rm E}\big[[H_I(t),[H_I(s),\sigma_{\rm S}(s)\otimes \sigma_{\rm E}]]\big].
\end{align}
Note that this approximation is less strict as may seem at first sight, since it does not automatically require that the d.o.f. of the environment or the medium themselves are weakly interacting.
The next simplification, the {\it Markovian} approximation invokes a separation of timescales. If excitations of the environment decay much faster than what the dynamics of the subsystem resolves, we may replace $\sigma_{\rm S}(s)$ by $\sigma_{\rm S}(t)$. In addition, if the integrand decays appreciably for times $s\gg\tau_{\rm E}$ we may also extend the integral boundary to infinity after taking $s\to t-s$. This leads to a master equation coarse grained in time of the form
\begin{align}
\frac{d}{dt}\sigma_{\rm S}(t)=-\int_0^\infty ds {\rm Tr}_{\rm E}\big[[H_I(t),[H_I(t-s),\sigma_{\rm S}(t)\otimes \sigma_{\rm E}]]\big].
\end{align}
In order to arrive at a Lindblad equation one often invokes another approximation, which involves the intrinsic time scale $\tau_{\rm S}$. In order to see its role, one can introduce a spectral decomposition in terms of the system Hamiltonian $H_S$, so that $H_{\rm int}(t)=\sum_{m,\omega}e^{-i\omega t}\Sigma_m(\omega)\otimes \Xi_m(\omega)$. The resulting e.o.m. reads
\begin{align}
\frac{d}{dt}\sigma_{\rm S}(t)=\sum_{\omega,\omega^\prime}\sum_{m,n} e^{i(\omega^\prime-\omega)t}\Gamma_{mn}(\omega) \Big( \Sigma_n(\omega)\sigma_{\rm S}(t)\Sigma_m(\omega^\prime)-\Sigma_m^\dagger(\omega^\prime)\Sigma_n(\omega)\sigma_{\rm s}(t)\Big),
\end{align}
where 
\begin{align}
\Gamma_{mn}(\omega)=\int_0^\infty ds e^{is\omega}\langle \Xi_m^\dagger(t)\Xi_n(t-s)\rangle \quad {\rm and} \quad \Gamma_{mn}(\omega)=\frac{1}{2}\gamma_{mn} + i S_{mn}(\omega). \label{eq:DefGamma}
\end{align}
If $\tau_{\rm rel}\gg \tau_{\rm E}$ only the diagonal terms with $\omega=\omega^\prime$ contribute. This is called the {\it rotating wave approximation}. Further, decomposing $\Gamma_{mn}$ into a hermitean part $S_{mn}=(\Gamma_{mn}-\Gamma_{mn}^*)/2i$ and the positive matrix $\gamma_{mn}=(\Gamma_{mn}+\Gamma_{mn}^*)$, the following master equation is obtained
\begin{align}
\frac{d}{dt}\sigma_{\rm S}(t)= -i[\tilde H_{\rm S},\sigma_{\rm S}(t)]+\sum_\omega\sum_{m,n} \gamma_{mn}(\omega) \Big( \Sigma_n(\omega) \sigma_{\rm S} (t)\Sigma^\dagger_m(\omega) -\frac{1}{2}\big\{ \Sigma^\dagger_m(\omega) \Sigma_n(\omega),\sigma_{\rm S}(t)\big\}\Big),\label{eq:opticalme}
\end{align}
where $\tilde H_{\rm S}=\sum_{\omega}\sum_{mn}S_{mn}(\omega)\Sigma^\dagger_m(\omega)\Sigma_n(\omega)$. \Cref{eq:opticalme} goes over into Lindblad form, if the matrix $\gamma_{mn}$ is diagonalized. This master equation is based on the separation of scales
\begin{align}
\tau_{\rm S} \ll \tau_{\rm rel}, \quad \tau_{\rm E} \ll \tau_{\rm rel}\quad {\rm the\, quantum\, optical\, limit.}
\end{align}
As a consistency check one can convince oneself that the late time limit of the the above master equation is indeed the thermal distribution $\sigma_{\rm S}^{\rm therm}={\rm exp}[-\beta \tilde H_S]/{\rm Tr}_{\rm S}[ e^{-\beta \tilde H_S} ]$. 

While the above discussion provided an easily accessible example of the derivation of a microscopic master equation the underlying separation of timescales did not apply to quarkonium. Indeed, due to the large rest mass of the constituent quarks we expect that the intrinsic dynamics of the subsystem are actually slow compared to that of the medium. In this case the Markovian approximation may hold but the following separation of scales is considered
\begin{align}
\tau_{\rm E} \ll \tau_{\rm S}, \quad \tau_{\rm E} \ll \tau_{\rm rel}\quad {\rm the\, quantum\, brownian\, motion \,  limit,}
\end{align}
i.e. the rotating wave approximation may not apply.

A well known model of Markovian quantum Brownian motion is the Caldeira-Leggett model at high temperature. In this model a (heavy) point particle is coupled to a bath consisting of a large number (eventually infinitely many) of light harmonic oscillators. It thus represents a very crude but intuitive analogy for a heavy quark. The model Hamiltonian reads
\begin{align}
H_{\rm S}=\frac{p^2}{2m}+V_c(x),\quad H_{\rm E}= \sum_n\hbar \nu_n\big(b_m^\dagger b_n +\frac{1}{2}\big),\quad H_{\rm int}=-x\sum_n\kappa_n x_n= -x\sum_n \kappa_n\sqrt{\frac{\hbar}{2m_n\nu_n}}(b_n+b_n^\dagger)=-x\Xi.
\end{align}
Here the creation and annihilation operators of the harmonic oscillators with mass $m_n$ and frequency $\nu_n$ are denoted by $b_n$ and $b^\dagger_n$. A linear interaction term between the point particle and the environment is used. If continuous reservoir modes are considered, the interactions leads to a renormalization of the potential. This effect is compensated by adding to the heavy particle potential $V(x)$ a correction $V_c(x)=V(x)+x^2\sum_n \kappa_n/(2m_n\nu_n^2)=V(x)+H_c(x)$.

The derivation of the master equation at first proceeds very similarly, invoking the Born and Markov approximations. In this model we assume that the particle is immersed in a heat bath of temperature $\beta=1/k_BT$ so that $\sigma_{\rm E}={\rm exp}[-\beta H_{\rm E}]/{\rm Tr}_{\rm E}{\rm exp}[-\beta H_{\rm E}]$. We had seen that correlation functions of the medium d.o.f. play an important role which motivates introducing the following thermally averaged commutator and anti-commutator 
\begin{align}
&D(s)=i\langle [\Xi,\Xi(-s)]\rangle=2\hbar \int _0^\infty d\omega J(\omega) {\rm sin}(\omega s),\\
&D_1(s)=\langle \{ \Xi,\Xi(-s)\}\rangle = 2\hbar \int _0^\infty d\omega J(\omega) {\rm coth}[\hbar\omega/2k_BT]{\rm cos}(\omega s),
\end{align}
which are known as the dissipation and noise kernel respectively. All the information about the medium in this model is encoded in the two kernel functions, which may be written in terms of the bath spectral density
\begin{align}
J(\omega)=\sum_n\frac{\kappa_n}{2m_n\nu_n}\delta(\omega-\nu_n).
\end{align}
The equation of motion so far takes the form
\begin{align}
\frac{d}{dt}\sigma_{\rm S}(t)=-\frac{i}{\hbar}[H_S+H_c,\sigma_{\rm S}(t)]+\frac{1}{\hbar^2}\int_0^\infty ds \Big(\frac{i}{2}D(s)[x,\{x(-s),\sigma_{\rm S}(t)\}]-\frac{1}{2}D_1(s)[x,[x(-s),\sigma_{\rm S}(t)]]\Big).\label{eq:clmasterpre}
\end{align}

We are interested in a reservoir with infinitely many d.o.f. in order to observe genuine damping of modes. Thus we will consider the continuous counterpart of $J(\omega)$, whose $\omega$ dependence is determined from the particular form of the interactions $\kappa_n$. An analytically solvable case is obtained with the following model ansatz
\begin{align}
J(\omega)=\frac{2m\gamma}{\pi}\omega \frac{\Omega}{\Omega^2+\omega^2}.
\end{align}
Here the small frequency regime behaves linear, modeling an Ohmic damping of modes in the subsystem with damping constant $\gamma$. The high frequency regime on the other hand is regularized by a cutoff function. Let push the analogy with QCD a bit further at this point. Since the coupling of the heavy quark to the bath occurs via gluons the spectral density considered above will be linked to a gluon spectral function, while the kernels $D$ and $D_1$ are the corresponding gluon correlators. In case of quarkonium it will turn out that since the gluon correlators at high temperature are also the building blocks to describe the real- and imaginary part of the in-medium heavy quark potential, $D$ and $D_1$ are intimately related to ${\rm Im}[V]$.

The master equation \cref{eq:clmasterpre} can be further simplified by invoking the time scale separation of Brownian motion. In case of the model spectral density $J(\omega)$ the decay of the kernel functions contains contributions from all finite Matsubara frequencies $\omega_n=2\pi n T$ with $n>0$ up to the UV cutoff $\Lambda_{\rm UV}$. I.e. the medium relaxation scale is given by $\tau_{\rm E}={\rm max}[\Lambda_{\rm UV}^{-1}, \hbar/\omega_1]$. Instead of the rotating wave approximation, we here consider the case that the typical timescale of the system is much larger than that of the medium.

Carefully evaluating all expressions in \cref{eq:clmasterpre} under $\tau_{\rm E}\ll\tau_{\rm S}$ one arrives at the Caldeira-Leggett master equation 
\begin{align}
\frac{d}{dt}\sigma_{\rm S}(t)=-\frac{i}{\hbar}[H_S,\sigma_{\rm S}(t)]-\frac{i\gamma}{\hbar}[x,\{ p,\sigma_{\rm S}(t)\}]-\frac{2m\gamma k_B T}{\hbar^2} [x,[x,\sigma_{\rm S}(t)]]. \label{eq:clmasterpre}
\end{align}
The von-Neumann like term describes the coherent dynamics of the subsystem similar to when no environment is present. The second term arises from the contributions associated with the function $D$ and encodes dissipation of energy from the system to the medium. Its strength is governed by the damping rate $\gamma$ that enters $J$. The effects of thermal fluctuations on the other hand are encoded in the third term related to $D_1$. At this stage the master equation is not in Lindblad form but it can be made such by adding a term that is parametrically small at high temperatures $-\frac{\gamma}{8mk_BT}[p,[p,\sigma_{\rm S}]]$. In order for fluctuations and dissipation to work in tandem it makes sense that $\gamma$ will appear also in the noise related terms. This yields a Lindblad master equation with a single relaxation rate $\gamma$ and the Lindblad operator
\begin{align}
L=\sqrt{\frac{4mk_BT}{\hbar^2}}x+i\sqrt{\frac{1}{4mk_BT}}p.
\end{align}
We will find Lindblad operators of similar form when considering heavy quarks and heavy quarkonium in QCD. Let us briefly list the Ehrenfest equations of motion for position and momentum of the heavy particle in this model
\begin{align}
\frac{d}{dt}\langle x\rangle  =\frac{1}{m} \langle p\rangle, \quad \frac{d}{dt}\langle p\rangle  = -\langle V^\prime (x)\rangle- 2\gamma \langle p\rangle.
\end{align}
For a free particle we find 
\begin{align}
\langle x(t)\rangle= \langle x(0)\rangle +\frac{1}{2m\gamma}(1-e^{-2\gamma t}) \langle p(0)\rangle, \quad \langle p(t)\rangle = e^{-2\gamma t}\langle p(0)\rangle,
\end{align}
so that the average momentum relaxes to zero exponentially with a rate of $2\gamma$ and the average position at late times is displaced by the value $\langle \dot{x}(0)\rangle/2\gamma$.

While already much closer to the case of quarkonium, the Caldeira-Leggett model also does not in general accommodate its physics, since the above derivation assumes that the extend of the heavy particle is always smaller than the correlation length of the medium. Since quarkonium possesses an internal structure, its radius can however easily be of the same order or larger than the medium correlation length in practice.
  
In practice we will often encounter the situation where the medium d.o.f. are formulated in terms of a path integral. Feynman and Vernon \cite{Feynman:1963fq} developed a systematic treatment of how to derive the system-medium interactions in such a case. Consider the expression for the density matrix evolution on the Schwinger-Keldysh contour, where we denote the d.o.f. in the environment with capital letters, those in the subsystem with lowercase letters
\begin{align}
\sigma_{\rm tot}(x,y,X,Y,t)=\int d[x_0,y_0]\int d[X_0,Y_0] \langle x_0,X_0|  \hat\sigma_{\rm tot}(0) | y_0,Y_0\rangle \int_{x_0}^{y_0} {\cal D}[x,y] \int_{X_0}^{Y_0} {\cal D}[X,Y] e^{iS[x,X] - iS[y,Y]}.
\end{align}
The forward path houses the $x,X$ d.o.f., while the backward path contains $y,Y$. The trace operation over the environment in the language of functional integrals is implemented by integrating over $X$ and $Y$ and inserting a delta function, assuming $\sigma_{\rm tot}(0)=\sigma_{\rm S}\otimes\sigma_{\rm E}$ we find
\begin{align}
\sigma_{\rm S}(x,y)=\int {\cal D} [X,Y] \delta(X-Y) \sigma_{\rm tot}(x,y,X,Y,t)= \int d[x_0,y_0] \langle x_0 | \hat \sigma_{\rm S}(0)|y_0\rangle \int_{x_0}^{y_0} {\cal D}[x,y]  e^{iS_{\rm S}[x] - iS_{\rm S}[y]} F[x,y].\label{eq:FVfunctional}
\end{align}
The Feynman-Vernon influence functional $F[x,y]$ together with the subsystem action  $S_{\rm S}$ encodes all the information on the subsystem and its interaction with the medium, required to construct a master equation. It can equally well be written as an effective action contribution in the path integral weight $S_{\rm FV}[x,y]={\rm log}[ F[x,y] ]$. When expressing the time evolution of the reduced density matrix in terms of its Greens function $K_\sigma$ 
\begin{align}
\sigma_{\rm S}(x_f,y_f,t)=\int dx_i \int dx_i^\prime  K_\sigma(x_f,y_f,t_f,x_i,y_i,t_i)\sigma_{\rm S}(x_i,y_i,t_i),
\end{align}
we can read off its path integral representation. The challenge in practice then lies in reverse engineering the operator expressions for the corresponding equation of motion for $\sigma_{\rm S}$ starting from the c-number valued path integral. It is the inverse operation to the original construction of the path integral starting from the Greens function for an individual wavefunction. A similar task exists in lattice QCD studies, where the Hamilton operator for a lattice discretized path integral is constructed via the so called transfer matrix. For the Caldeira-Leggett model the Feynman-Vernon influence functional has been considered in \cite{Caldeira:1982iu}.

After having gained some insight into the description of open quantum systems via Lindblad type master equations we turn our attention to one of the characteristic phenomena occurring in open quantum systems: {\it Decoherence}. 
Whenever our subsystem evolves in contact with its environment, correlations among the d.o.f. arise. Very often this leads to a particular behavior in the subsystem, once the environment is traced out. I.e. certain sets of states in $S$ are singled out by the interaction, a so called preferred basis, which in turn remain stable under time evolution. Any superpositions of these states on the other hand are damped away. The timescale of this damping $\tau_D$ is often much shorter than the intrinsic timescales of the subsystem. Decoherence in the context of heavy quarkonium will be often be found to be accompanied by a decay of populations of states but in general such decay is not necessarily a byproduct of decoherence.

The dynamics of decoherence is often characterized by a so-called decoherence function $\Gamma$, which describes the decay of the off-diagonal entries of the reduced density matrix, expressed in the preferred basis. Let us consider the particularly chosen interaction $H_{\rm int}=\sum_m \Sigma_m\otimes\Xi_m= \sum_m |m\rangle\langle m|\otimes \xi_n$, where the states $|m\rangle$ are eigenstates of the subsystem Hamiltonian $H_{\rm S}$. Starting out from an initial state given by $|\psi(0)\rangle =\sum_m c_m |m\rangle\otimes|\phi_m\rangle$ the system will evolve into
\begin{align}
|\psi(t)\rangle =\sum_m c_m |m\rangle\otimes|\phi_m(t)\rangle, \quad \sigma_{\rm S}(t)={\rm Tr}[ |\psi(t)\rangle \langle\psi(t)|] = \sum_{m,n}c_mc_n^*|m\rangle\langle n| \langle \phi_n(t)|\phi_m(t)\rangle.
\end{align}
Due to unitarity, the norm of $\langle \phi_m(t)|\phi_m(t)\rangle=1$ is preserved and the diagonal entries of $\sigma_{\rm S}$ remain constant. I.e. only the off-diagonal terms are affected by the evolution and we may define the decoherence function $\Gamma_{mn}$ via
\begin{align}
|\langle \phi_n(t)|\phi_m(t)\rangle|={\rm exp}[\Gamma_{mn}(t)], \quad n\neq m, \quad  \Gamma_{mn}(t)\leq 0,
\end{align} 
whose values are of course highly dependent on the system parameters and also the choice of initial state. The typical timescale over which the overlap $\langle \phi_n(t)|\phi_m(t)\rangle$ decreases is denoted by $\tau_D$ the {\it decoherence time}. If the overlap approaches zero the reduced density matrix takes on the form of an incoherent mixture of the states $|m\rangle$. We can consider the linear entropy $S_l[\sigma]=1-{\rm Tr}[\sigma^2]$ in the subsystem, which starts out as $S_l[ \sigma_{S}(0)]=0$ but evolves to
\begin{align}
S_l[\sigma_{S}(t)]=1-\sum_{mn}|c_m|^2|c_n|^2{\rm exp}[2\Gamma_{mn}(t)],
\end{align}
approaching at late times $S_l[\rho_{S}(t\to\infty)]=1-\sum_m|c_m|^4$.

One interesting effect of decoherence is the localization of the states in position space, which occurs e.g. in quantum Brownian motion at high temperature. Neglecting the effects of dissipation, i.e. taking the {\it recoilless limit}, let us consider the master equation
\begin{align}
\frac{d}{dt}\sigma_{\rm S}(t)=-i[H_{\rm S},\sigma_{\rm S}(t)]-\frac{2m\gamma k_BT}{\hbar^2}[\mathbf{x},[\mathbf{x},\sigma_{\rm S}(t)]]
\end{align}
with a simple $H_{\rm S}=\mathbf{p}^2/2m$. It turns out that the decay of the off-diagonals occurs much faster than the evolution of the diagonals and the free part, hence we may neglect the $H_{\rm S}$ contribution and solve approximately
\begin{align}
\sigma_{\rm S}(\mathbf{x},\mathbf{x}^\prime,t)\approx {\rm exp}\big[ -\frac{2m\gamma k_BT}{\hbar^2} (\mathbf{x}-\mathbf{x}^\prime)^2t\big]\sigma_{\rm S}(\mathbf{x},\mathbf{x}^\prime,0).
\end{align}
The resulting decoherence function, which damps away off-diagonal terms in position space, is linear in time and we can read off the corresponding decoherence time
\begin{align}
\Gamma(t)=-\frac{2m\gamma k_BT}{\hbar^2}\Delta x^2 t,\quad \tau_D=\frac{\hbar^2}{2m\gamma k_BT\Delta x^2}.
\end{align}
It remains to get an impression of how the decoherence time relates to the relaxation time of the subsystem. The latter is defined from the relaxation of the square of the momentum, which from the Ehrenfest e.o.m. is $\tau_{\rm rel}=1/4\gamma$. Expressed in the {\it thermal wavelength} of the medium $\lambda_{\rm th}=\hbar/\sqrt{2m k_B T}$ we then find
\begin{align}
\frac{\tau_D}{\tau_{\rm rel}}=4\Big(\frac{\lambda_{\rm th}}{\Delta x}\Big)^2,
\end{align}
which especially for macroscopic objects can be an extremely small ratio.

In order to observe the phenomenon of decoherence and its consequences in practice, i.e. in realistic systems such as quarkonium, a numerical implementation of the Lindblad equation is required. On the one hand one could attack the problem head on and setup a simulation of $\langle x| \sigma_{\rm S} |y\rangle$ in coordinate space and discretize the resulting partial differential equation. In three dimensions this leads to a six dimensional matrix equation which is computationally very demanding. On the other hand one can {\it unravel} \cref{eq:Lindblad} into the stochastic evolution of an ensemble of wavefunctions $|\psi(t)\rangle$ whose fluctuations allow us to reconstruct the density matrix
\begin{align}
\sigma_{\rm{S}} = \left\langle |\psi(t)\rangle\langle\psi(t)| \right\rangle_{\rm ensemble}.
\end{align}
In that case one has to deal with the evolution of three-dimensional wavefunctions only. It is important to note that the time evolution described by \cref{eq:Lindblad} in general cannot be unravelled in terms of a deterministic Schr\"odinger equation. Introducing a gradient expansion it may be possible to unravel it into unitary and linear time evolution of a wavefunction with stochastic noise contributions. For the full dynamics, the Quantum State Diffusion (QSD) approach has proven to be an important tool \cite{gisin1992quantum}. It tells us that in general to unravel any Lindblad master equation the following non-linear stochastic Schr\"odinger equation can be considered
\begin{align}
|d\psi\rangle = -i \tilde H_{\rm S} + \sum_m \Bigr(2\langle L^{\dagger}_m\rangle_{\psi} L_m-L^{\dagger}_m L_m-\langle L^{\dagger}_m \rangle_{\psi}\langle L_m \rangle_{\psi}
\Bigr)|\psi(t)\rangle dt +\sum_m \Bigr(L_m- \langle L_m \rangle_{\psi}\Bigr)|\psi(t)\rangle d\xi_i.
\end{align}
The stochastic nature of the evolution is implemented via complex valued Gaussian noise of the form
\begin{align}
&\left\langle d\xi_i \right\rangle_{\rm ensemble} = \left\langle {\rm Re}(d\xi_i)\right\rangle_{\rm ensemble}=\left\langle {\rm Im}(d\xi_j)\right\rangle_{\rm ensemble}=0, \\
&\left\langle {\rm Re}(d\xi_i){\rm Re}(d\xi_j) \right\rangle_{\rm ensemble} =\left\langle {\rm Im}(d\xi_i){\rm Im}(d\xi_j)\right\rangle_{\rm ensemble}=\delta_{ij} dt.
\end{align}
The non-linearity on the other hand arises from the fact that in terms such as $\langle L_m \rangle_{\psi}$ the expectation value with respect to the wavefunction is computed. Once the Lindblad operators are specified the above prescription offers a straight forward recipe for implementation of numerical simulations.

\begin{summary} The open quantum systems approach considers systems, which can be decomposed in a small subsystem $S$ and a large environment $E$ and thus naturally accommodates in-medium quarkonium. By tracing out the medium degrees of freedom a simplified description of the irreversible dynamics of the subsystem is developed, formulated as a master equation for the reduced density matrix $\sigma_{\rm S}$. Interactions between $S$ and $E$ induce correlations in their d.o.f. which leads to the phenomenon of decoherence in the subsystem. A set of preferred basis states remains stable but superpositions are damped away. Exploiting a separation of time scales between the subsystem and the medium $\tau_{\rm E} \ll \tau_{\rm rel}$ Markovian master equations in the Lindblad form can be derived, which preserve the positivity and hermiticity of the reduced density matrix.  Simulating the time evolution often involves unravelling of the master equation in terms of stochastic evolution of an ensemble of wavefunctions. The quantum state diffusion method provides a straight forward implementation.
\end{summary}

\section{Quarkonium in thermal equilibrium}
\label{sec:qqbarequil}

Our exploration of heavy quarkonium in extreme conditions begins with its physics in thermal equilibrium. Here the quark--anti-quark pair is immersed into an infinitely extended QCD medium at a constant temperature $T$ (and in this report we restrict to vanishing Baryo-chemical potential $\mu_B=0$). The advantage of investigating such an idealized scenario is that quantum field theoretical methods, such as lattice QCD, are capable of providing direct non-perturbative insight from first principles. I.e. no modeling assumptions need to be made. We may learn about various aspects of in-medium heavy quarkonium, which will support the analysis of its physics in more complicated and experimentally relevant scenarios, such as in relativistic heavy ion collisions. 

In the following we will assume that enough time has passed so that the momentum distribution of the individual quarks has fully equilibrated, becoming Fermi-Dirac distributed. At very high energies $T\gg 2m_Q$ heavy quark pairs may be spontaneously created from thermal fluctuations, i.e. their occupation number also follows the thermal distribution. They system is said to be chemically equilibrated. Since $E_{\rm bind}<m_Q$ for quarkonium, one does not expect stable bound states to exist in this regime. On the other hand at lower temperatures $E_{\rm bind}>T$ where quarkonium may survive, the thermal occupancy for heavy quarks is very low. If we consider a $Q\bar{Q}$ in this regime, some other process must have produced it and placed it in the medium. I.e. the system is kinetically but not chemically equilibrated. Such settings are commonly considered today, as they are the idealized counterpart to a heavy-ion collision, where it is the hard partonic processes in the early stages that create a quark antiquark pair, which subsequently finds itself surrounded by the hot bulk matter.

The central goal will be to understand how the binding properties of quarkonium vacuum states are altered in the presence of a thermal medium. We will explore this question from several angles: screening of color fields in thermal QCD, the modification of binding in terms of an in-medium potential and the manifestation of binding in the spectral properties of quarkonium states. On the way we will learn to connect the weakening of interquark binding to the concept of a Debye mass and eventually consider what the insight of the past decade tells us about the dynamical process of quarkonium melting.

\subsection{Screening of color fields in thermal QCD}
\label{sec:screeningQCD}

Heavy quarkonium bound states in vacuum are sustained by an exchange of gluons among the constituent charm or bottom quarks. Since the heavy quark velocities are small it is expected that color electric interactions dominate over color magnetic ones. If immersed into a thermal QCD medium one expects that the exchanged gluons will interact with the medium d.o.f. and that their properties become modified. In analogy with electromagnetically interacting media (see e.g. Debye-H\"uckel theory \cite{huckel1923theory}) it is furthermore expected that this leads to a weakening of the strength of binding. Our goal in this section is thus to gain an understanding of how the medium influences the propagation of QCD color fields. Note that at this point we are asking about a property of the medium and not yet one of the quarkonium system. 

Consideration of limiting cases, whether physical or not, is a central tool of the physicist toolbox to explore the logical consequences of hypotheses. In classical electrodynamics the idealized concept of a test charge plays an important role to shed light on field configurations and their modification in medium. In the study of QCD a similar role is taken up by considering infinitely heavy quarks. We have already seen in \cref{eq:staticpropq} that in this {\it static limit} the propagation of a quark in time reduces simply to a change in its $SU(3)$ color phase, given by a temporal Wilson line. The spatial position of the quark is obviously fixed.

In QED, screening properties of static electric and magnetic fields are well understood in a gauge invariant manner. The gluons responsible for sustaining the electric field acquire a mass from interactions with the medium. For a medium of fermions with charge $e$ this leads to an electric photon self energy
\begin{align}
\lim_{{\bf p} \to 0} \Pi^{\rm QED}_{00}(\omega=0,{\bf p}) = m_D, \quad m_D=\frac{1}{\sqrt{3}}eT+{\cal O}(e^2T^2)
\end{align}
and in turn the electric field becomes short ranged. Correlation functions between static electric fields are damped exponentially $\langle {\bf E}({\bf x}) {\bf E}({\bf y})\rangle \sim {\rm exp}[-m_D |{\bf x} - {\bf y}|]$ at large distances with the in-medium mass, christened the {\it Debye mass}. It is possible to compute the QED Debye mass perturbatively, with the lowest order being proportional to $eT$. Magnetic fields on the other hand remain unscreened $\lim_{{\bf p} \to 0} \Pi^{\rm QED}_{ij}(\omega=0,{\bf p}) =0$ and remain effective on large distance scales.

In QCD on the other hand screening is more involved. The first stumbling block is that the concept of color electric and magnetic field is not gauge invariant anymore. So a simple inspection of gluon self energies does not lead to an unambiguous concept of electric or magnetic screening, as one component can be rotated into another. Instead one has to define adequate gauge invariant quantities, whose symmetry properties agree with the symmetries associated with an electric and magnetic field. 

It turns out that at asymptotically high temperatures, where QCD is weakly coupled, color electric fields are screened at a scale $gT$, where $g$ is the strong coupling. I.e. there exists a QCD counterpart to the Debye mass, which at leading order scales as $m_D\sim gT$. In contrast to QED magnetic fields also become screened at a lower energy scale $m_M\sim g^2T$ but the physics of this magnetic screening is fully non-perturbative (even at high temperatures). Not only is $m_M$ not accessible via perturbative means but also $m_D$ receives non-perturbative contributions beyond the first logarithmic correction that itself can be evaluated in perturbation theory. The general result for $SU(N_c)$ gauge theory with $N_f$ active fermion flavors in the medium reads
\begin{align}
m_D= \sqrt{\frac{N_c}{3} +\frac{N_f}{6}}gT + \frac{1}{4\pi}N_cg^2T{\rm log}\Big[\sqrt{\frac{N_c}{3} +\frac{N_f}{6}}\frac{1}{g}\Big] + \kappa g^2T +{\cal O}(g^3 T), \label{eq:DebMassNLO}
\end{align}
where $\kappa$ already escapes perturbative evaluation.

There are currently two main approaches found in the literature to define and determine screening masses in QCD non-perturbatively. Both rely on considering static quarks as test color charges and deploy specifically constructed correlation functions between such a static quark and antiquark pair to reveal how their interactions are damped when the spatial distance between them is increased. Interestingly the intimate relation between Euclidean QFT at finite temperature and statistical physics allows us to connect the evolution of static quarks in imaginary time to a thermodynamic property of the system, its free energy. One may ask how a genuinely gauge dependent concept, such as a color phase rotation can lead to a gauge invariant thermodynamic quantity. As it turns out the compactness of the imaginary time axis plays a key role.

Let us recall how the free energy is connected to static quark propagation, as first discussed in Ref.~\cite{McLerran:1981pb}. The free energies $F(\mathbf{x}_1,\ldots,\mathbf{x}_n,\mathbf{x}_{n+1},\ldots,\mathbf{x}_{n+m})$ of a system with $n$ static quarks and $m$ antiquarks present, located at positions $\mathbf{x}_i$ may be related to the partition function via
\begin{align}
{\rm exp}[-\beta F ] = \frac{1}{{N_c}^{n+m}}\sum_s\langle s|e^{-\beta H} |s \rangle=\frac{1}{{N_c}^{n,m}}\sum_{q}\langle q|Q(\mathbf{x}_1,0)\ldots  Q(\mathbf{x}_n,0)\,e^{-\beta H}\,\bar{Q}(\mathbf{x}_{n+1},0)\ldots \bar Q(\mathbf{x}_{n_m},0) |q \rangle.
\end{align}
Here the states $|s\rangle$ are those that contain both the medium degrees of freedom an the static quarks and $|q\rangle$ only refer to the medium. Using the fact that $e^{-\beta H}$ is the generator of imaginary time translations and that these time translations only change the affected fields by a color rotation, we may write
\begin{align}
{\rm exp}[-\beta F_{n,m} ] = \frac{1}{N_c^{n+m}}{\rm Tr}_{\rm medium}\Big[ {\rm Tr}_c\big[ L(\mathbf{x}_1)\big]\ldots{\rm Tr}_c\big[L(\mathbf{x}_n)\big] {\rm Tr}_c\big[L^\dagger(\mathbf{x}_{n+1})\big]\ldots {\rm Tr}_c\big[L^\dagger(\mathbf{x}_{n_m})\big]\Big].\label{eq:freenrg}
\end{align}
The above expression is written in terms of the color trace of the {\it Polyakov loop}
\begin{align}
L(\mathbf{x})= {\cal T} {\rm exp}\Big[ i \int_0^\beta d\tau A_4(\mathbf{x},\tau)\Big] =  \prod_{i=1}^{N_\tau} U_4(\mathbf{l},i),\label{eq:polloop}
\end{align}
which on the lattice is given by a product of links in imaginary time direction closing around the compact Euclidean domain. Note that if the color trace is taken ${\rm Tr}_c L$ is indeed is gauge invariant. The Polyakov loop is hence related to the free energy of a single quark in the medium. It is furthermore the order parameter of the $SU(3)$ center symmetry and in the absence of light quark degrees of freedom $\langle L(0)\rangle$ in turn takes the role of an order parameter for the, in that case well defined, confinement-deconfinement phase transition. In QCD with dynamical fermions the change of the Polyakov loop with temperature still provides a clear signal for the crossover transition from hadrons to the quark-gluon-plasma.

In order to get access to the free energies on the lattice, we need to rewrite \cref{eq:freenrg} in terms of an actual expectation value, i.e. we need to normalize by the partition function of the medium without quarks present. Hence we compute the difference $\Delta F_{n,m}=F_{n,m}-F_{0,0}$ in the free energy between a system with and without the static quarks present
\begin{align}
{\rm exp}[-\beta \Delta F_{n,m} ]=\frac{1}{N_c^{n+m}}\langle {\rm Tr}_c\big[ L(\mathbf{x}_1)\big],\ldots,{\rm Tr}_c\big[L(\mathbf{x}_n)\big] {\rm Tr}_c\big[L^\dagger(\mathbf{x}_{n+1})\big]\ldots {\rm Tr}_c\big[L^\dagger(\mathbf{x}_{n_m})\big]\rangle.
\end{align}

The main interest in this section lies in the interactions between one static quark and one antiquark, i.e. we consider the spatial correlations of a Polyakov loop and its complex conjugated counterpart 
\begin{align}
P_2(r=|\mathbf{x}-\mathbf{y}|)={\rm exp}[-\beta \Delta F_{1,1} ]=\langle {\rm Tr}_c\big[L(\mathbf{x})\big]{\rm Tr}_c\big[L^\dagger(\mathbf{y})\big]\rangle, \qquad \lim_{r\to\infty} P_2(r) = |\langle {\rm Tr}_c\big[L(0)\big]\rangle|^2.
\end{align}
We may understand the corresponding free energy as encoding (after proper renormalization) the amount of work that is required to pull apart the two quark anti-quark pair to a distance $r$. At very large distances the correlator $P_2$ will go over into a constant, given by the expectation value of a single Polyakov loop squared. In pure gauge theory at low temperatures this quantity diverges as color confinement prevents the separation of the two static quarks. In QCD with dynamical fermions the phenomenon of string breaking occurs even in vacuum, where the energy stored in the color field between the static quarks is transformed into a light quark--antiquark pair, which in turn forms a color neutral state with the static color sources.

\subsubsection*{Screening masses from electric and magnetic correlators}

Let us consider the first strategy for defining and extracting the QCD screening masses. It relies on the work of Ref.~\cite{Arnold:1995bh}, where it has been shown that transformation properties under Euclidean time reversal ${\cal R}$ and charge conjugation ${\cal C}$ allow one to make the distinction between color electric and magnetic operators in a non-perturbative fashion. In short, an operator even (odd) under ${\cal R}$ is considered to be magnetic (electric). The large distance spatial decay of correlators of such operators are expected to reveal the corresponding magnetic (electric) {\it screening masses}. This concept has already been used to study QCD at very high temperature (remember that even there magnetic screening is non-perturbative) using the dimensionally reduced effective field theory EQCD in Ref.~\cite{Hart:2000ha}. We will now discuss how it can be used in the study of screening in fully dynamical lattice QCD.

Ref.~\cite{Arnold:1995bh} first pointed out that since the Polyakov loop as written in \cref{eq:polloop} does not have well defined transformation properties under Euclidean time reversal  ${\cal R}L = L^\dagger$ and charge conjugation ${\cal C}L = L^*$ it mixes both color electric and color magnetic contributions. If its correlator, i.e. the free energy difference $\Delta F_{1,1}(r)$ shows exponential decay at large distances it is most probably dominated by the magnetic screening mass $m_M < m_E$, as electric contributions have already died out before. Note that if we wish to connect the electric screening mass obtained in that fashion to the perturbatively defined concept of Debye mass we will have to divide with a factor two. The reason is that at small distances perturbation theory tells us \cite{Nadkarni:1986cz} that $\Delta F_{1,1}(r)$ is dominated by two-gluon exchange. In turn a $-\alpha_S^2{\rm exp}(-2m_D r)/r^2$ behavior arises, from which we take it that the screening masses found from such traced out Polyakov loop correlators will represent twice the value of the corresponding gluon mass.

As proposed first in the lattice study of Ref.~\cite{Maezawa:2010vj} (WHOT collaboration) we can instead consider 
\begin{align}
L_M=\frac{1}{2}(L+L^\dagger), \quad L_E=\frac{1}{2}(L-L^\dagger),\qquad L_{M\pm}=\frac{1}{2}(L_M\pm L_M^*), \quad L_{E\pm}=\frac{1}{2}(L_E\pm E_M^*),
\end{align}
where the only two non-vanishing contributions ${\rm Tr}[L_{M+}]$ and ${\rm Tr}[L_{E-}]$ represent the real and imaginary part of the Polyakov loop respectively. The corresponding magnetic and electric correlation functions are written as
\begin{align}
&C_{M+}(r)=\left\langle \sum_{\bf x} {\rm Tr}_c\big[ L_{M+}(\mathbf{x})\big]{\rm Tr}_c\big[ L_{M+}(\mathbf{x}+\mathbf{r})\big]\right\rangle - \left| \sum_{\bf x} {\rm Tr}_c\big[ L(\mathbf{x})\big]\big]\right|^2,\\
&C_{E-}(r)=\left\langle \sum_{\bf x} {\rm Tr}_c\big[ L_{E-}(\mathbf{x})\big]{\rm Tr}_c\big[ L_{E-}(\mathbf{x}+\mathbf{r})\big]\right\rangle.
\end{align}
If the above defined correlators show a large distance exponential decay we may use them to extract the magnetic and electric screening masses in a non-perturbative fashion. The high temperature effective field theory of dimensionally reduced QCD, called EQCD, suggests \cite{Braaten:1994qx} that the correlator takes on the asymptotic form
\begin{align}
C_{M+}(r\to\infty)\sim \gamma_M \frac{e^{-m_M r}}{rT}, \quad C_{E-}(r\to\infty)\sim \gamma_E \frac{e^{-m_D r}}{rT},
\end{align}
which has been found to describe lattice QCD simulations well even at temperatures around the crossover transition.

While several studies have considered correlators of Polyakov loops in lattice QCD over the past two decades, it is only recently that fully continuum extrapolated results with dynamical quarks close to the physical pion mass have been obtained in the temperature range up to $T=350$MeV in Ref.~\cite{Borsanyi:2015yka} and up to $T=1451$MeV in Ref.~\cite{Bazavov:2018wmo}. The electric and magnetic correlators have been investigated previously only in Ref.~\cite{Maezawa:2010vj}.

Let us consider the electric and magnetic screening masses obtained in Ref.~\cite{Borsanyi:2015yka} (Budapest-Wuppertal collaboration). The simulations in that study are based on a Symanzik improved gauge action and a stout smearing improved staggered quark action tuned to reproduce physical quark masses. Scale setting along the line of constant physics has been implemented using the gradient flow scale $\omega_0$. The continuum limit is taken on lattices with different imaginary time extent from $N_\tau=6\ldots 16$, where the physical volume is kept approximately constant with a ratio of $N_s/N_\tau\approx 5\ldots 6$. As a fixed box approach, temperature is varied by changing the lattice spacing.

Since the naive Polyakov loop diverges in the continuum limit, so does its correlator at large separation distances and hence the free energies. I.e. to take the continuum limit in a meaningful way this divergence needs to be subtracted. One possibility (for a detailed discussion see \cite{Borsanyi:2015yka}) is to define the renormalized quantity
\begin{align}
\Delta F_{1,1}^{\rm ren} = \Delta F_{1,1}(r,T)-\Delta F_{1,1}(r\to\infty,T_0)
\end{align}
at a fixed reference temperature $T_0$. Since the divergence is UV dominated it is the same at any temperature and by choosing one fixed $T_0$ the renormalization procedure remains temperature independent, as required.

\begin{figure}
\centering
\includegraphics[scale=0.25]{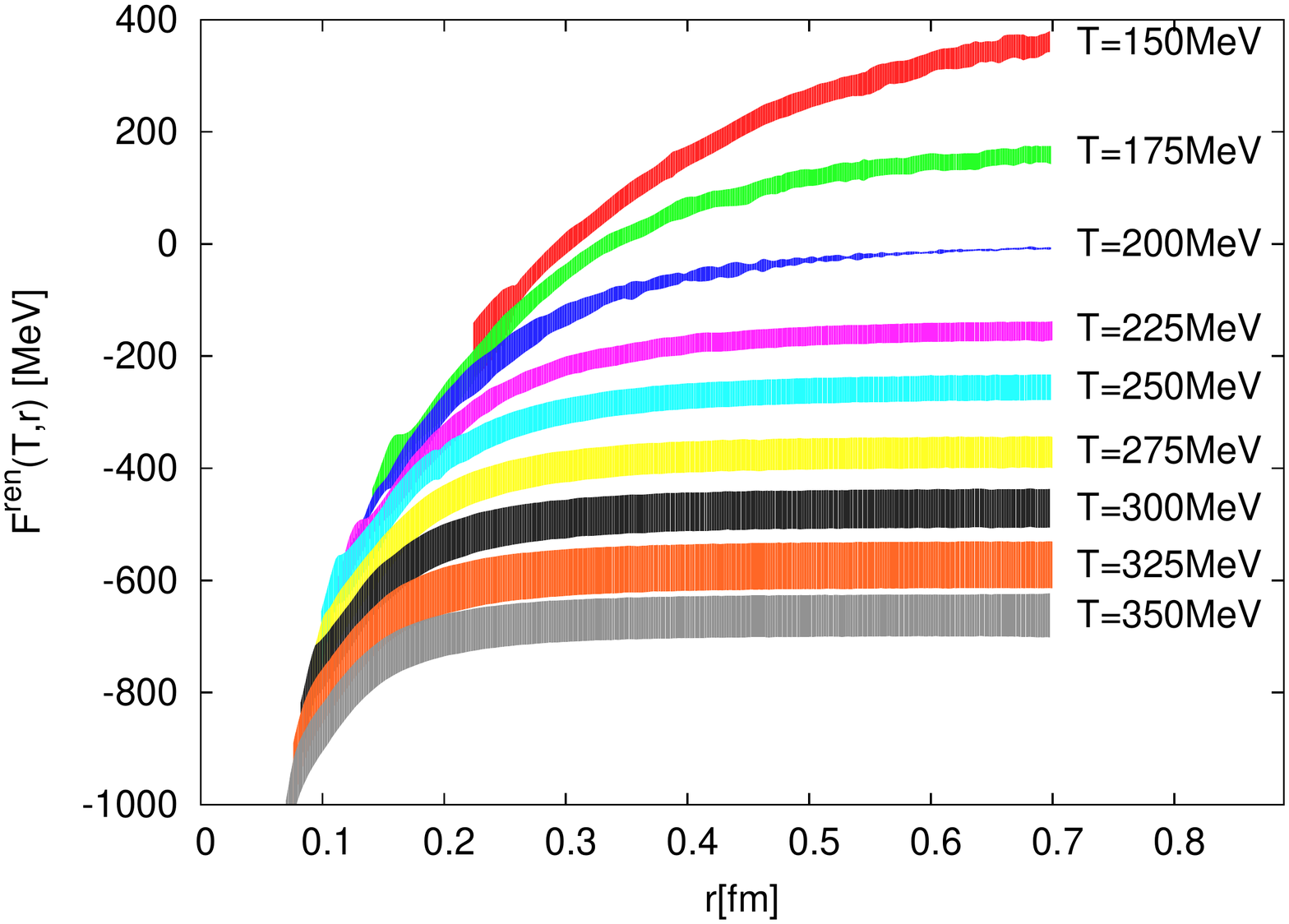}
\includegraphics[scale=0.25]{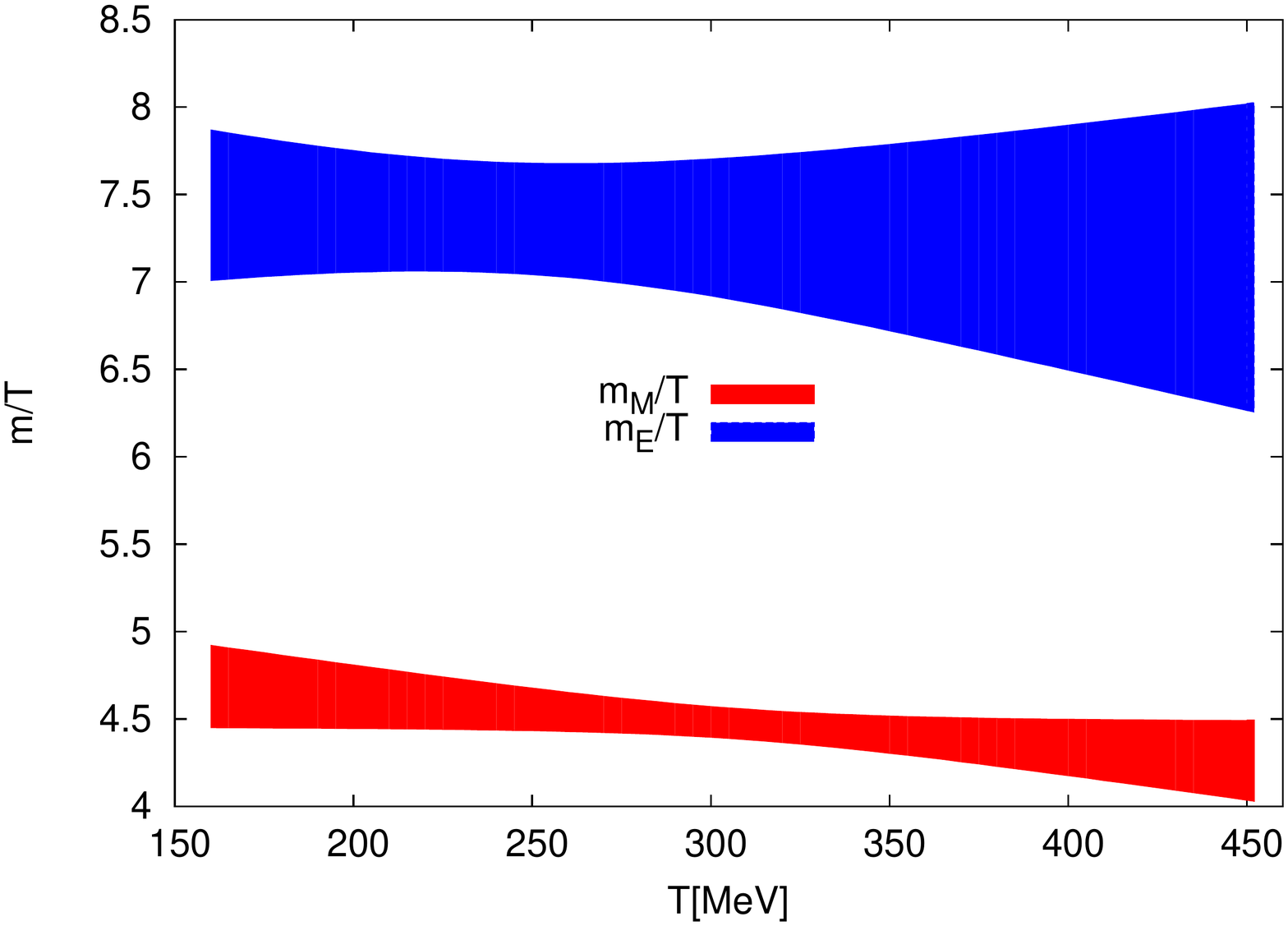}
\caption{(left) The continuum extrapolated and renormalized free energy difference for a system with and without a static $Q\bar Q$ pair inserted. (right) The continuum extrapolated values for the electric (blue, top) and magnetic (red, bottom) screening masses from Yukawa fits to the large distance decay of the electric and magnetic correlation functions. Figure adapted from Ref.~\cite{Borsanyi:2015yka}.}\label{fig:FrenAndScreenMass}
\end{figure}
 
In the left plot of \cref{fig:FrenAndScreenMass} the continuum extrapolated results for $\Delta F_{1,1}^{\rm ren}$ from Ref.~\cite{Borsanyi:2015yka} are presented.  At every temperature considered, the values approach a common low temperature form for small enough distances, where the influence of the thermal medium becomes insignificant. At large distances and high temperatures on the other hand we find that the free energy difference asymptotes to a constant value, which indicates that indeed screening is present. It manifests itself also as an exponential decay in the underlying correlation function of Polyakov loops. It is clear that screening weakens as the temperature is reduced. At lower temperatures around and below the crossover transition $T_c=155$MeV the effects of screening will eventually become mixed with those arising from string breaking. A recent analysis of string breaking effects in $T=0$ lattice QCD (yet without continuum extrapolation) in Ref.~\cite{Bulava:2019iut} indicates a string breaking distance of $r_c=1.224(15)$fm, so that the flattening off in the $T=150$ MeV result shown here up to $r=0.7$fm is most likely still an indication of the onset of screening.

In order to extract the electric and magnetic screening masses, the Yukawa like decay of the corresponding correlators $C_{E-}(r)$ and $C_{M+}(r)$ is fitted and the results are given in the right plot of \cref{fig:FrenAndScreenMass} as ratios with the temperature. The values of the magnetic mass $m_M/T$ together with the full error budget including statistical and continuum extrapolation uncertainty are plotted as the red band, the electric mass $m_D/T$ as blue band. One finds that the separation of $m_E>m_M$ holds with $m_E/m_M(T=300{\rm MeV})=1.63(8)$, even in the regime just above the crossover temperature. While not shown in \cref{fig:FrenAndScreenMass}, the value of the magnetic screening mass obtained from $C_{M+}(r)$ is expected to agree with that from $\Delta F_{1,1}$, as had been observed previously in Ref.~\cite{Maezawa:2010vj}. 

Within the errorbars the results do not show a significant dependence on temperature, which suggests that the masses indeed scale linearly with temperature. The results obtained in this analysis are consistent with previous studies using EQCD, keeping in mind that applying EQCD at such low temperatures pushes the envelope of its range of validity. EQCD with both $N_f=2$ and $N_f=3$ massless quarks in the medium predicts $m_E/m_M(T=300{\rm MeV})=1.8(2)$.

As we argued that the electric screening mass extracted here non-perturbatively relates to the Debye mass as $m_E=2 m_D$, we can compare its values to the predictions of NLO perturbation theory in \cref{eq:DebMassNLO}. One finds that the lattice values are systematically larger than the weak-coupling results by a factor between $1.8$ and $2.0$.

\subsubsection*{Screening masses from color singlet and averaged free energies}

The second approach followed in the literature considers the transformation properties of the Polyakov loops under color rotations instead of Euclidean time reversal. In the non-interacting theory a product of two color matrices indeed can be unambiguously decomposed according to $3\otimes \bar 3 = 1\oplus 8$ into a color singlet and a color octet (adjoint) contribution
\begin{align}
{\rm exp}[-\beta \Delta F_1]&= \frac{1}{N_c}\langle{\rm Tr}_c\big[ L(\mathbf{x})L^\dagger(\mathbf{y})\big] \rangle \label{eq:singletFELat}\\
{\rm exp}[-\beta \Delta F_a]&= \frac{1}{N_c^2-1}\langle{\rm Tr}_c\big[ L(\mathbf{x})T^a L^\dagger(\mathbf{y})T^a\big] \rangle\\
 &= \frac{1}{N_c^2-1} \langle{\rm Tr}_c\big[ L(\mathbf{x})\big] {\rm Tr}\big[L^\dagger(\mathbf{y})\big] \rangle- \frac{1}{N_c(N_c^2-1)}\langle{\rm Tr}_c\big[ L(\mathbf{x})L^\dagger(\mathbf{y})\big] \rangle
\end{align}
The quantities, where the trace is on the outside of the product of loops are by themselves not gauge invariant. One conventionally chooses Coulomb gauge when determining their values on the lattice. The physical free energy thus contains both singlet and adjoint contributions
\begin{align}
&{\rm exp}[-\beta \Delta F_{1,1}]= \frac{1}{N_c^2}{\rm exp}[-\beta F_1]+ \frac{N_c^2-1}{N_c^2}{\rm exp}[-\beta F_a]
\end{align}

In the interacting theory the above decomposition may become modified. Following the discussion of Ref.~\cite{Bazavov:2018wmo} (and references therein) there exist two linear combinations of the color singlet and adjoint contributions that renormalize multiplicatively. Since each comes with their own renormalization constant, distilling the renormalized singlet contribution from the differently traced Polyakov loop correlators in principle requires knowledge of both of them. I.e. depending on the scale, the naively defined singlet correlator of \cref{eq:singletFELat} may in principle receive admixtures from the adjoint sector, even if these are in practice small over the phenomenologically relevant distance scales.

The temperature dependence of these quantities has been studied various lattice QCD scenarios in the literature. Starting with $SU(2)$ gauge theory in Refs.~\cite{Irback:1991eh,LaCock:1991hh,Datta:1999yu,Fiore:2003yw, Digal:2003jc,Bazavov:2008rw} in $SU(3)$ pure gauge in \cite{Karkkainen:1992jh, Kaczmarek:1999mm,Petreczky:2001pd,Kaczmarek:2002mc,nakamura:2004,Umeda:2008kz,Akerlund:2013cga} and dynamical QCD with $N_f=2$ in \cite{Detar:1998qa, Karsch:2000kv,Kaczmarek:2005ui, Bornyakov:2004ii,Doring:2005ih,Maezawa:2007fc, Ejiri:2009hq} and $N_f=3$ in \cite{Petreczky:2004pz}. Continuum extrapolated results with $N_f=2+1$ dynamical flavors have been presented in \cite{Borsanyi:2015yka,Bazavov:2018wmo}.

The underlying idea to extract electric screening masses from $F_1$ is to realize that at $T=0$ the color singlet contribution agrees with the real-valued interquark potential $\Delta F_1(r,T=0)=V^{(0)}_S(r,T=0)$, which represents color electric field interactions among the constituent quarks. Hence by investigating the finite temperature counterpart $\Delta F_1(r,T>0)$ and how it asymptotes to a constant at large distances (manifest as an exponential decay in the corresponding correlation function), one expects to extract the electric screening mass. Considering $\Delta F_{1,1}$ itself, as argued before will reveal the magnetic screening mass.

Let us consider the latest results from Ref.~\cite{Bazavov:2018wmo} (TUMQCD and HotQCD collaboration) where the color singlet free energies $F_1$ and the gauge invariant $\Delta F_{1,1}$ have been investigated using a Szymanzik improved gauge action and the highly improved staggered (HISQ) action to describe $N_f=2+1$ dynamical light quark flavors. The scale along the line of constant physics has been set using the Sommer scale $r_1$ and light quark parameters are tuned to lead to a physical strange quark mass $m_s=20 m_l$ and a pion mass that is close to its physical value $m_\pi\approx 160$MeV. Using a fixed box approach, i.e. changing temperature with the lattice spacing, the study covers a temperature range between $116<T<5814$MeV. Some ensembles at temperatures above $T>400$MeV use larger pion mass $m_\pi\approx 320$MeV to speed up the simulations, the effects of which has been checked to be insignificant. The continuum limit is taken on lattices with temporal extent between $N_\tau=4\ldots 16$ keeping an aspect ratio of $N_s/N_\tau=4$ and $6$. 

The free energies are renormalized by first subtracting from their bare value twice the bare value of free energy of a single quark and then adding twice the value of the renormalized single quark free energy obtained from the renormalized Polyakov loop. This procedure leads to a well defined continuum limit for the color singlet free energy $F_1$ if performed in Coulomb gauge, while other gauge choices leave remnant divergencies in place \cite{Burnier:2009bk}. 

\begin{figure}
\centering
\includegraphics[scale=0.5]{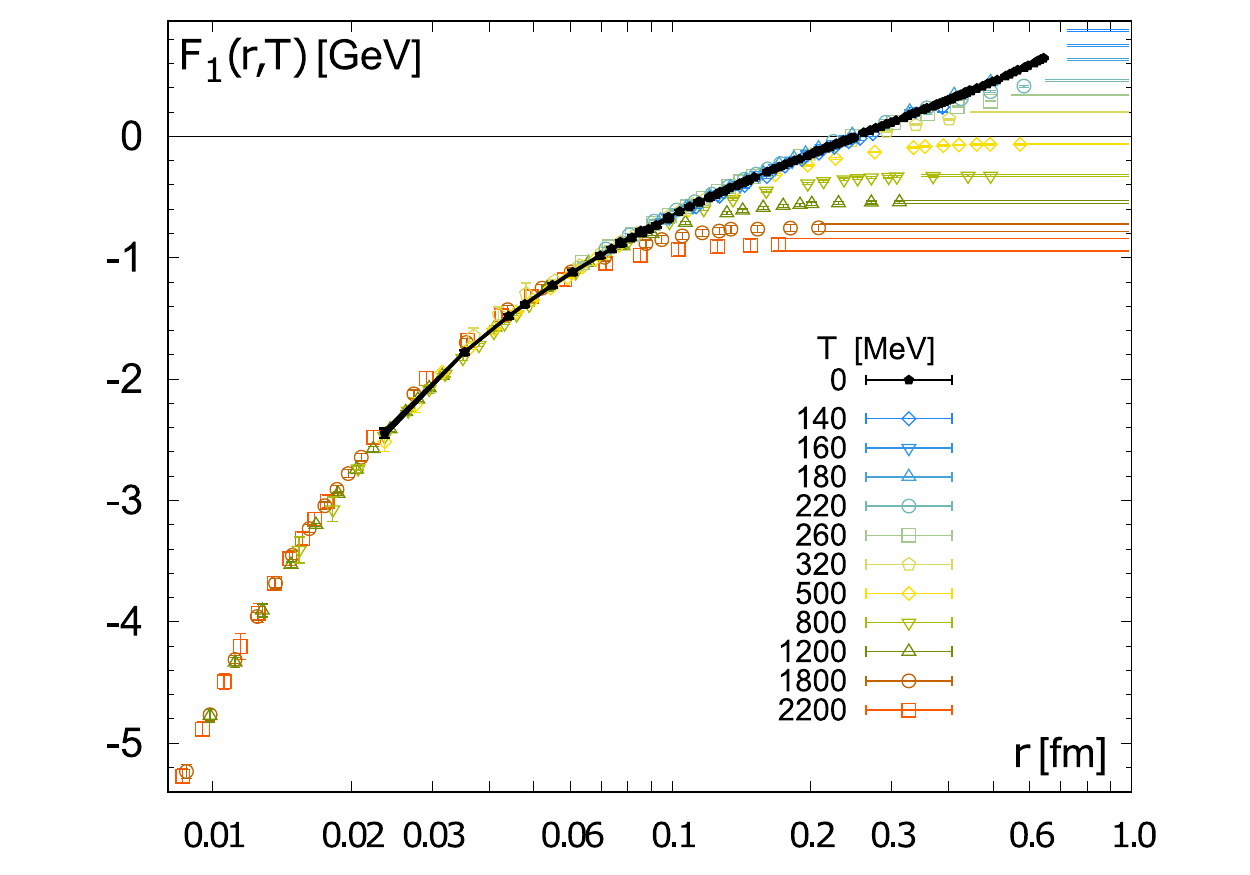}
\includegraphics[scale=0.5]{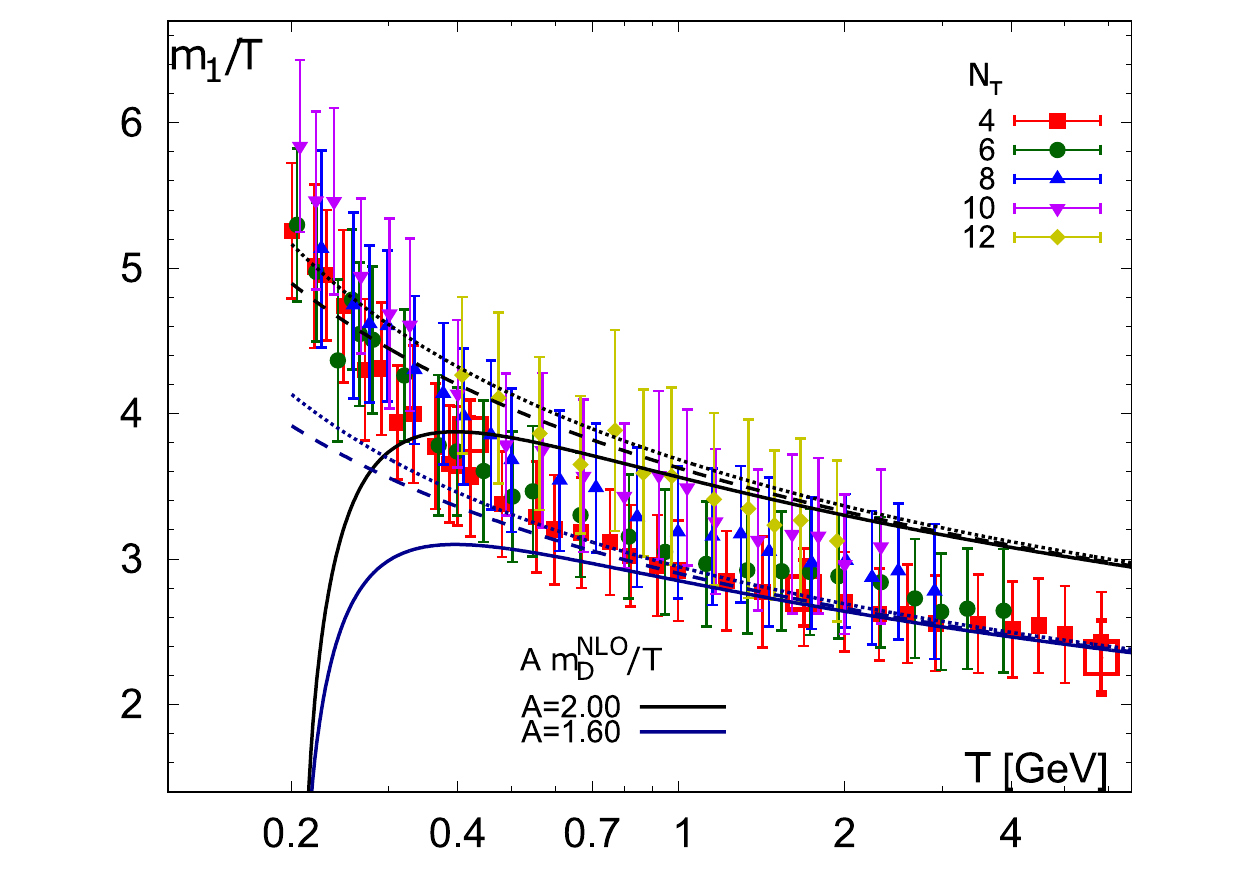}
\caption{(left) The continuum extrapolated and renormalized color singlet free energies $F_1$. (right) The continuum extrapolated values for the electric (blue, top) and magnetic (red, bottom) screening masses from Yukawa fits to the large distance decay of the electric and magnetic correlation functions. The solid, dotted and dashed lines correspond to the predictions of the NLO Debye mass using the scales $\mu=\pi T,2\pi T, 4\pi T$ respectively. Figure adapted from Ref.~\cite{Bazavov:2018wmo}.}\label{fig:SinglNrgAndScreenMass}
\end{figure}

In the right plot of \cref{fig:SinglNrgAndScreenMass} the continuum extrapolated values of the renormalized color singlet free energies $F_1$ are shown at different temperatures. At small distances, this quantity exhibits a $-\alpha_S{\rm exp}[-m_D r]/r$ behavior, as perturbation theory \cite{Nadkarni:1986as} tells us that it is dominated by single gluon exchange. Note the difference to $\Delta F_{1,1}$. We also clearly see that at small distances all simulation points fall on top of the $T=0$ curve. If considered in terms of rescaled distances $rT$ in-medium effects only become significant for $rT>0.3$. At large distances also $F_1$ shows screening, as it asymptotes to a constant value, indicated by the horizontal solid lines.

One of the goals of Ref.~\cite{Bazavov:2018wmo} is to elucidate in which region of distances and temperatures the lattice results on the different free energies become amenable to a perturbative description. It was shown that for $rT\ll 1$ the perturbatively matched EFT pNRQCD provides a good description, while for intermediate distances $rT \lesssim 1$ the effective theory EQCD agrees with the lattice data within its scale uncertainty. At the asymptotic large distances, where the screening masses are defined, however neither EQCD not perturbation theory can be deployed and the previously discussed Yukawa fits were used.

The screening mass $m_1$ obtained from the color singlet free energies divided by temperature is plotted in the right panel of \cref{fig:SinglNrgAndScreenMass}. The values at several different lattice spacings are shown and it is found that for $N_\tau>8$ no more cutoff effects remain, providing the result essentially in the continuum limit. Here a much larger temperature range is considered as in the study of $C_{E-}$. It appears that the ratio shows a tendency to decrease as temperature is increased, while the relatively large uncertainty at $N_\tau=12$ does not yet allow to make a definite statement.

Since the color singlet free energies perturbatively are dominated by single gluon exchange, the screening mass $m_1$ is expected to agree with the Debye screening mass itself at high temperatures. Hence its values may be directly compared to the NLO prediction of \cref{eq:DebMassNLO}. Interestingly, as shown by the dotted and dashed curve, the temperature dependence of the lattice quantities appears to resemble qualitatively the NLO prediction at the scale $\mu>\pi T$ but is systematically larger by a factor of $1.6$-$2.0$. The screening masses  obtained here from the renormalized $\Delta F_{1}$ are compatible with the values of Ref.~\cite{Borsanyi:2015yka} for temperatures above $T=300$MeV.

The study of asymptotic screening masses at temperatures below $T=200$MeV is challenging due to the low singal to noise ratio in the Polyakov loop correlators in that regime. Thus the errorbars of the extracted screening masses also increases as temperature is lowered. On the other hand the region around the phase transition and below is of particular interest in order to understand the fate of screening in the hadronic phase. In pure gauge theory, where there exists a genuine phase transition, one would expect that below $T_C$ screening is strongly suppressed, as it is only glueballs that may interfere with the binding of the (by now) confined static color charges. I.e. the expectation is that $m_D$ quite abruptly takes on very small values, probably even vanishes, as e.g. indicated by the results of Ref.~\cite{Kaczmarek:1999mm}. In full QCD without a genuine transition this change in $m_D$ will proceed more gently but eventually at $T=0$ it also has to vanish. Therefore it will be interesting to see how the lattice results, which down to $T=160$MeV show either a flat or even an increasing ratio of $m_D/T$ will behave at temperatures below $T_c$. A first indication of a downward trend in the ratio has been observed in $m_M$ in Ref.~\cite{Bazavov:2018wmo} however only at the lowest temperatures where a continuum limit is not yet available.

We have seen that two independent approaches to electric screening masses $m_E\equiv 2m_D$ and $m_1\equiv m_D$ provide consistent results on the values of the corresponding Debye mass in the QGP phase. In addition different lattice QCD setups agree within uncertainty on the values of the magnetic screening mass extracted from $\Delta F_{1,1}$ in the continuum limit. Quantitatively the Debye masses obtained from the lattice in the QGP phase, characterizing electric screening, turn out to be up to a factor 2 larger than what is predicted by NLO perturbation theory. This is pertinent information when it comes to understanding heavy quarkonium binding, as it tells us that the (color electric) interactions that sustain the bound states will receive significant modification in the deconfined medium. 

There are efforts ongoing in the lattice community to extend the computation of screening masses to a medium with finite Baryo-chemical potential $\mu_B$, via the Taylor expansion method \cite{Doring:2005ih} or analytic continuation from imaginary chemical potential \cite{Takahashi:2013mja,Andreoli:2017zie}. In the latter case charge conjugation symmetry is broken explicitly and the standard prescription discussed here to separate color electric and magnetic contributions is inapplicable. With a refined strategy it is found that at finite lattice spacing the presence of a small but finite $\mu_B$ increases both electric and magnetic screening masses. The change is of the same order of magnitude as expected for the Debye mass from lowest order perturbation theory.

\subsubsection*{Screening masses from parton properties}

In contrast to using gauge invariant correlators of static test charges, the screening properties of QCD may also be elucidated by investigating the in-medium behavior of gluons themselves (for studies of in-medium quark properties see e.g. Refs.~\cite{Karsch:2009tp,Fischer:2017kbq,Oliveira:2019erx}). This line of research is pursued using both non-perturbative lattice QCD, as well as analytic methods, such as the functional renormalization group and Dyson-Schwinger equations. 

\begin{figure}
\centering
\includegraphics[scale=0.4]{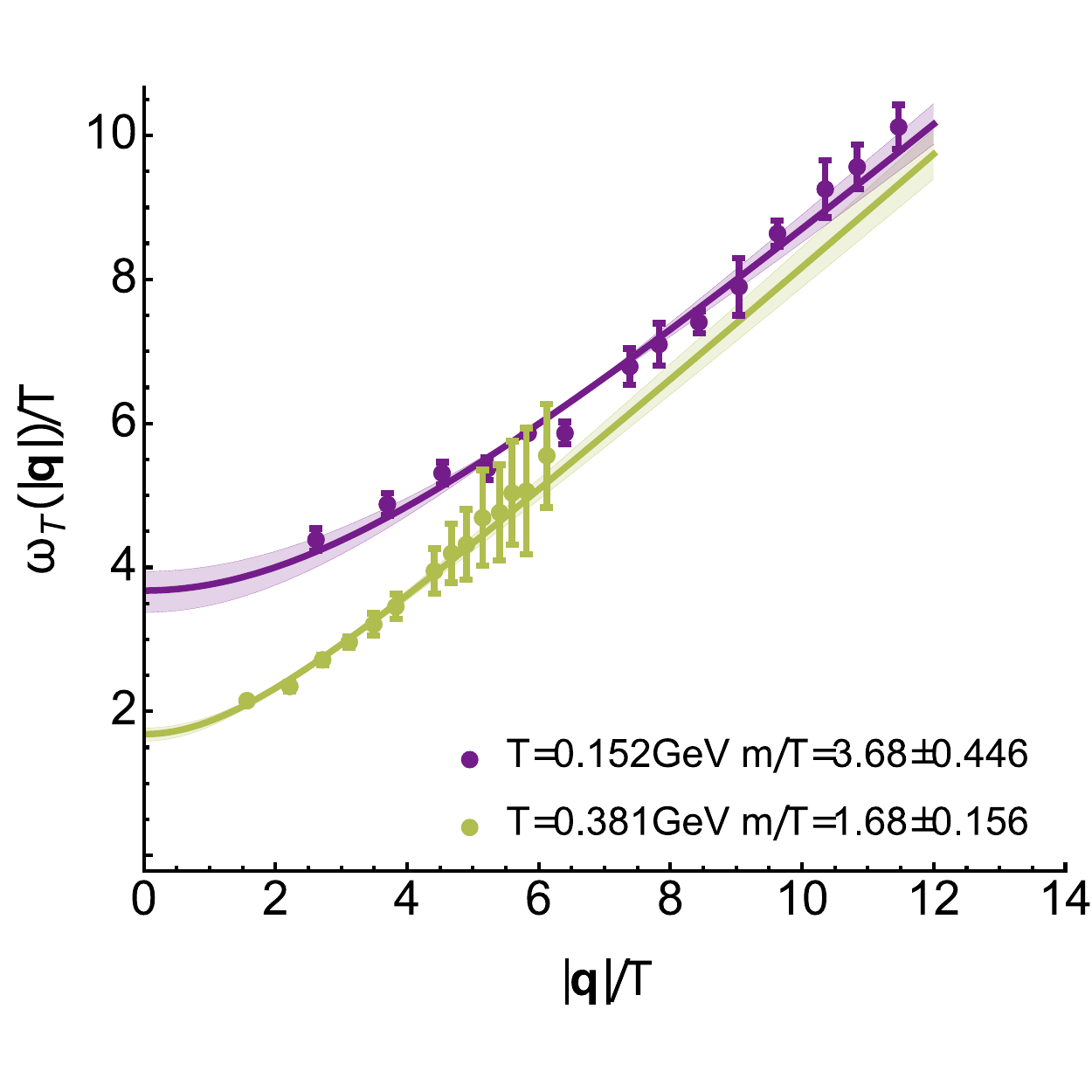}\hspace{1cm}
\includegraphics[scale=0.4]{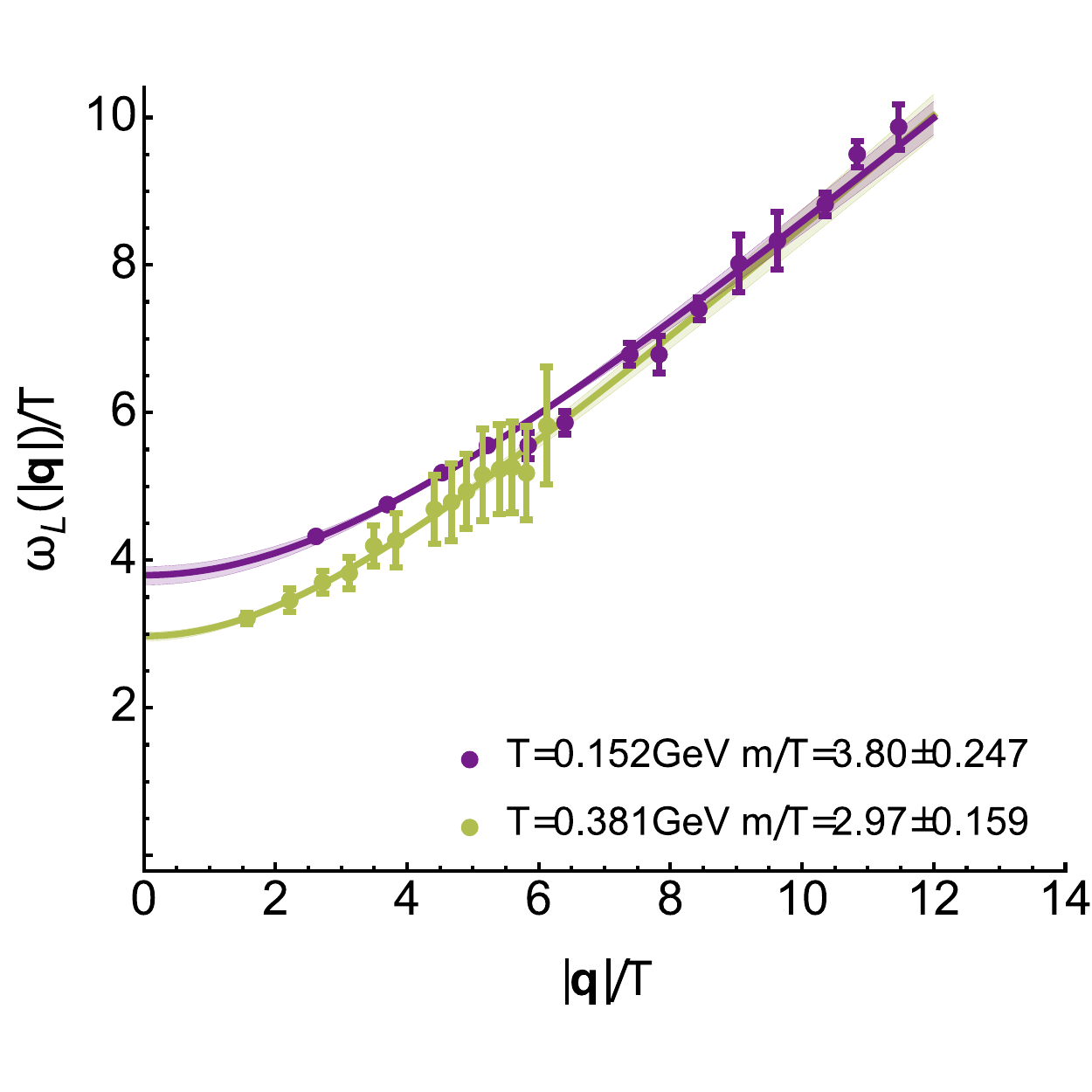}
\caption{(left) Dispersion relation according to the position of the magnetic gluon quasiparticle peak (colored symbols) identified in the transversal gluon spectral function in Landau gauge. Shown are the lowest and highest temperature results at $T=151$MeV (violet) and $T=381$MeV (green). Solid lines denote best fit with $\omega(|\mathbf{q}|)=A\sqrt{B^2+|\mathbf{q}|^2}$. The dashed red line shows the value of  (right) Same quantities shown for the electric gluon from the longitudinal gluon spectral function. Figures adapted from Ref.~\cite{Ilgenfritz:2017kkp}.}\label{fig:GluonDispRel}
\end{figure}

A study of gluon properties necessarily involves gauge fixing (usually to Landau gauge), where transverse (magnetic) and longitudinal (electric) contributions to the gauge fields can be separated by adequate projection. In high temperature QCD, gluons are understood as actual propagating quasiparticle d.o.f. and their interaction with the environment endows them with a dynamically generated mass. This picture underlies the perturbative definition of the Debye mass. Now if a quasiparticle-like structure can be identified in gluon spectral functions, obtained non-perturbatively e.g. from lattice QCD, changes in its position with temperature may substantiate the presence of such a thermal mass. 

Extracting gluon spectral functions is extremely challenging due to the additional technical complication that it is possibile for them to contain non-positive regions. This is in particular relevant in both the small frequency regime, as has been recently pointed out in Ref.~\cite{Cyrol:2018xeq} and as is known from perturbative arguments at very large frequencies. Hence so far only a few studies of in-medium gluon spectral functions on the lattice have been carried out (for two recent ones see e.g. Refs.~\cite{Silva:2017mds,Ilgenfritz:2017kkp}). 

Using simulations of the QCD medium by the tmft collaboration, which implements $N_f=2+1+1$ dynamical quark flavors in the twisted mass formalism,  gluon correlators and spectral functions in Landau gauge have been investigated in Ref.~\cite{Ilgenfritz:2017kkp}. The study focused on a set of lattices with relatively small lattice spacing $a=0.0646$fm but with a still unphysically high pion mass of $m_\pi=369$MeV. A fixed box approach allowed investigating a temperature range between $T=152-381$ MeV changing the number of Euclidean points between $N_\tau=20-8$. The deconfinement crossover temperature on these lattices is $T_C=193$Mev. Spectral reconstructions were performed using the generalized BR method \cite{Rothkopf:2016luz}. One needs to be aware that due to the KMS relation only half of the input points provide independent information so that the reconstructions are based on a relatively small number of input data. This drawback was partially compensated for by the very good signal to noise ratio in the underlying gluon correlators. A well defined lowest lying spectral feature was observed at all temperatures in both the longitudinal and transversal spectral functions, which in the QGP phase has been interpreted as representing a gluon quasi-particle. 

In \cref{fig:GluonDispRel} the dispersion relation for the magnetic (left) and electric (right) gluon quasiparticle is plotted normalized by temperature. The lattice values (colored symbols) are obtained from fitting the position of the dominant low lying spectral peak in the corresponding gluon spectral function. The lowest temperature $T=151$MeV (violet) and the highest temperature results $T=381$MeV (green) are shown together with fits (solid lines) based on the simple ansatz $\omega(|\mathbf{q}|)=A\sqrt{B^2+|\mathbf{q}|^2}$. The good agreement with the lattice values shows that while a linear dispersion relation is recovered at large momenta close to the origin the behavior is cut-off, leading to a finite intercept with the y-axis. Evaluating the fit at $T>T_C$ for $|\mathbf{q}|\to0$ provides an estimate for the corresponding screening masses in the QGP. One finds that even in this still non-perturbative regime the magnetic screening mass $m_T$ takes on significantly lower values than the electric screening mass $m_L$
\begin{align}
m_T/T(T=381{\rm MeV})=1.68\pm 0.156, \quad m_L/T(T=381{\rm MeV})=2.97\pm 0.159,
\end{align}
which itself is larger at that temperature than the prediction from NLO perturbation theory. A result that is qualitatively compatible with those obtained from the gauge invariant correlation functions discussed previously. 

\begin{summary}
The phenomenon of screening in a QCD medium can be investigated by considering the correlations among test color charges in the form of infinitely heavy quarks evolving in Euclidean time. In contrast to QED not only electric fields become screened but also magnetic ones, i.e. at large enough distances all gluons acquire a thermal mass. Expressed as Polyakov loops, the static quark correlation functions are intimately related to the free energy of static pairs in the medium. Electric and magnetic screening masses can be extracted from appropriately constructed Polyakov loop correlators. In the literature three independent approached are considered. The first two construct gauge invariant quantities with definite transformation properties w.r.t. Euclidean time reversal or w.r.t. color rotations. Corresponding screening masses have been computed in lattice QCD in the continuum limit and it has been found that the perturbative prediction of $m_E>m_M$, as well as $m_E/T \approx {\rm  const.}$ and $m_M/T \approx {\rm const.}$ holds approximately even in the non-perturbative regime close to the crossover transition. Quantitatively it has been established on the lattice that electric screening, most relevant for the binding of quarkonium states is more efficient than predicted by NLO perturbation theory by a factor of $1.6$ to $2.0$ in terms of the Debye mass. Alternatively screening may be investigated directly from the in-medium modification of gluon spectral functions, which yields qualitatively consistent behavior of electric and magnetic screening masses.
\end{summary}

\subsection{The in-medium static inter quark potential}
\label{sec:inmedpot}

In the previous section we have discussed the phenomenon of screening of static color fields in the QGP and found that in the relevant temperature range for current heavy ion collisions electric screening is not only present but it is stronger than predicted by NLO perturbation theory. This finding begs the question, how the medium affects the binding of actual heavy quarkonium. Due to the separation of scales between the heavy quark rest mass, as well as temperature $T/m_Q\ll 1$ and $\Lambda_{\rm QCD}/m_Q \ll 1$ we have seen in \cref{sec:pNRQCD} that significant aspects of the physics of quarkonium can be captured in a real-time potential, which corresponds to a non-local Wilson coefficient in the effective field theory pNRQCD. The leading order contribution to the color singlet potential in the heavy quark velocity expansion is given by the static interquark potential $V_S^{(0)}(r)$, which is independent of the heavy quark velocity $v$.

In this section we will consider evaluating this proper real-time  potential $V_S^{(0)}$ in QCD at finite temperature according to the defining relation based on the real-time Wilson loop
\begin{align}
V^{(0)}_S(r)=\lim_{t\to\infty}\frac{i\partial_t W_\square}{W_\square}.\label{eq:realtimeVdef2}
\end{align}

\subsubsection*{Perturbative determination of the in-medium potential}

\begin{figure}
\centering
\includegraphics[scale=0.35]{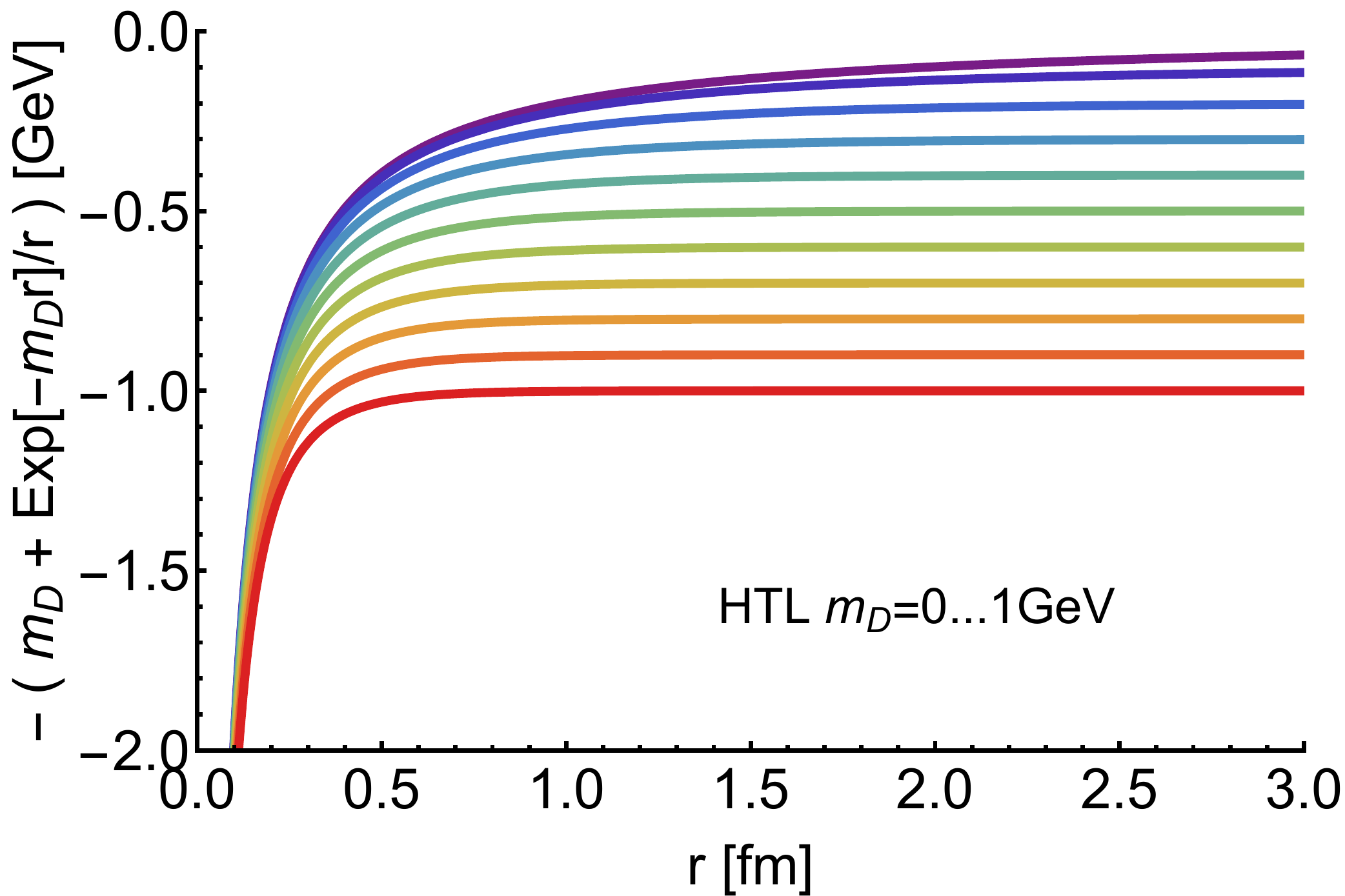}
\includegraphics[scale=0.35]{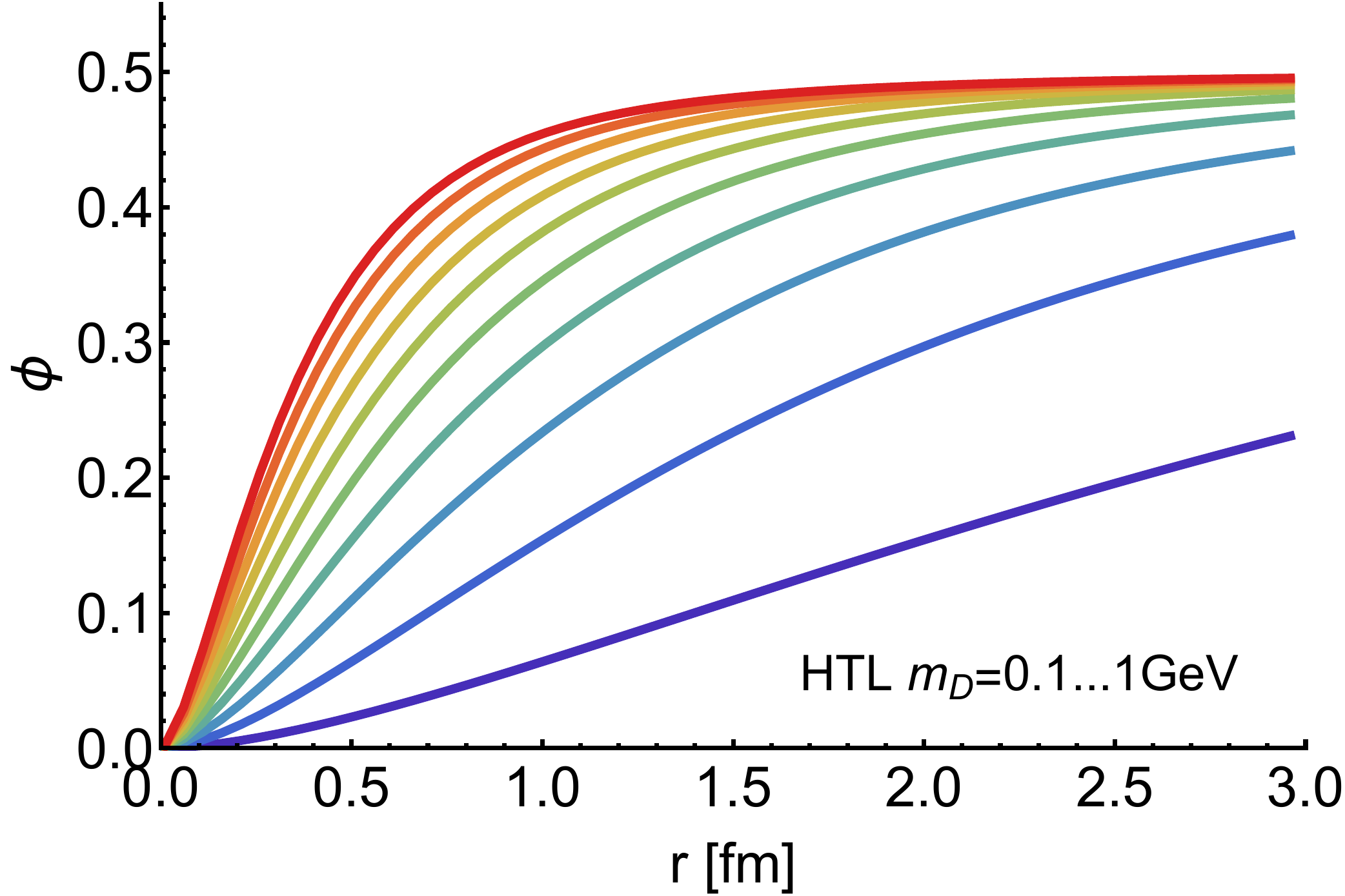}
\caption{(left) The real part of the real-time static interquark potential from HTL perturbation theory. We plot its values for a selection of Debye masses $m_D=0\ldots1$GeV, lowest values are denoted by the dark blue line, the largest value by the red line. (right) The function $\phi(m_D r)$ that governs the non-trivial separation dependence of the imaginary part of the static potential.}\label{fig:LainePot}
\end{figure}

The real-time definition of the static potential has been evaluated in thermal equilibrium for the first time in Ref.\cite{Laine:2006ns} using hard thermal loop perturbation theory \cite{Pisarski:1988vd}. For a pedagogical treatment see \cite{Beraudo:2007ky}.  The setting chosen in Ref.\cite{Laine:2006ns} corresponds to the particular scale hierarchy where $mv\sim gT$ \cite{Laine:2008cf}. Unexpected at that time, it turns out that the real-time potential besides its real-part also features a finite imaginary part, the explicit expressions of which are given by
\begin{align}
&{\rm Re}[V_{S,HTL}^{(0)}](r)=-\frac{g^2 C_F}{4\pi}\Big[ m_D+\frac{{\rm exp}(-m_D r)}{r}\Big]=-\alpha_S\Big[ m_D+\frac{{\rm exp}(-m_D r)}{r}\Big] ,\label{eq:RePotHTL}\\ 
&{\rm Im}[V_{S,HTL}^{(0)}](r)= -\alpha_S T \phi(m_D r), \quad \phi(x)= 2\int_0^\infty \frac{dz z}{(z^2+1)^2}\Big[1-\frac{{\rm sin}(zx)}{zx}\Big]. \label{eq:ImPotHTL}
\end{align}
where $\alpha_S= g^2 C_F/4\pi$ is used as definition of the strong coupling constant, similar to the convention in the quarkonium phenomenology literature. At intermediate distances, where the running of the strong coupling is not pronounced, the temperature dependence of the real part is solely determined by the value of $m_D$, while the imaginary part carries an explicit dependence on the temperature $T$. We plot the values for ${\rm Re}[V_{S,HTL}^{(0)}]$ in the left panel of \cref{fig:LainePot} and the values of the function $\phi(m_D r)$ on the right, spanning a range of Debye mass values from $m_D=0\ldots1$GeV. The smallest value of $m_D$ corresponds to the dark blue lines, the largest value to the red line. 

As expected from the discussion of screening in QCD, one finds that the real-part of the potential shows a Debye screened behavior. It asymptotes at large distances to a constant that represents twice the heavy quark in-medium mass shift \cite{Gava:1981qd}. Interestingly to this lowest order in the HTL approximation the expression for ${\rm Re}[V_{S, \rm HTL}^{(0)}]$ coincides with the color singlet free energies $F_{1, \rm HTL}(r)$ in Coulomb gauge. This agreement is however only coincidental, as from the next higher order in the perturbative approximation the two quantities are not equal anymore. On the other hand, as we will see, current lattice QCD results suggest that the difference between the two remains relatively small.

The imaginary part of \cref{eq:ImPotHTL} starts out quadratically and eventually asymptotes to a finite value, twice that of the heavy quark thermal width \cite{Beraudo:2007ky}. It had been already understood in \cite{Laine:2006ns} that in this scale hierarchy ${\rm Im}[V_{S,HTL}^{(0)}]$ arises from the phenomenon of Landau damping, which corresponds to the low-frequency gluon mediating the binding between the heavy quarks loosing energy to a medium parton with a higher energy $\sim T$. As in a perturbative calculation the origin of ${\rm Im}[V_{S,HTL}^{(0)}]$ can be traced back to particular diagrams, which appear also in other contexts, it is possible to relate it e.g. to the energy loss of single heavy quarks and their diffusion in a hot medium.

In other scale hierarchies, the values for ${\rm Re}[V]$ and ${\rm Im}[V]$ can take on significantly different values \cite{Brambilla:2008cx,Brambilla:2010vq,Hong:2018vgp}. The imaginary part may receive additional contributions from the phenomenon of gluo-dissociation, where a singlet state absorbs a colored gluon from the medium, changing into an octet and subsequently dissolves. Note that this is a particular case where the interaction with ultrasoft gluons can still be recast in the form of a time independent potential. At small distances, where the heavy quarks effectively do not see the thermal medium it turns out that ${\rm Im}[V]$ becomes exponentially suppressed.

It is important to stress at this point that the potential we have discussed so far governs the time evolution of the medium averaged color singlet correlator of point split meson operators. In particular it does not describe the evolution of the microscopic wavefunction of a heavy quarkonium state. This distinction is important in order to avoid misinterpretations of the imaginary part of the potential. 

${\rm Im}[V]$ has been derived in the static limit, where the heavy quarks are fixed to their spatial positions and never meet to annihilate. I.e. the probability to find the two heavy quarks in the system remains constant and ${\rm Im}[V]$ cannot tell us that the amplitude of the wavefunction of the two-body system decays exponentially. Instead if understood as implementing the damping of the unequal time correlator of singlet fields, the role of the imaginary part becomes clear. It encodes the loss of coherence between the state of the heavy quarkonium particle at initial time (when it was just inserted into the medium) and its state at later time after interacting with the medium. Intuitively, one expects the medium fluctuations to perturb the wavefunction of the quarkonium stochastically until it has equilibrated with its surroundings. Decoherence represented via ${\rm Im}[V]$ is an aspect of the thermalization of the quarkonium system. In the context of open-quantum-systems we will return to the question of how to implement the microscopic dynamics for the wavefunction of heavy quarkonium based on the knowledge of the complex static potential from pNRQCD.

While a lot has been learned about the potential acting between static quarks from perturbative studies, the question remains how quarkonium states behave in the non-perturbative regime just above the crossover transition, where perturbative methods are inapplicable. I.e. to approach an understanding of quarkonium in heavy-ion collisions we need to evaluate \cref{eq:realtimeVdef2} using lattice QCD.

\subsubsection*{Lattice QCD determination of the in-medium potential}

The real-time definition of the static in-medium potential in \cref{eq:realtimeVdef2} is formulated in Minkowski time and thus not directly amenable to an evaluation in lattice QCD. Bridging the real- and imaginary-time domain is however possible by resorting to the technical concept of spectral functions. We saw that all different types of meson correlation functions are ultimately governed by a single positive definite function, i.e. the correlator can be expressed in terms of a convolution of the spectral function with an appropriate kernel function.

Since the Wilson loop corresponds to the infinite mass limit of the forward correlator of point split meson operators, its spectral decomposition in Euclidean time, as shown in Ref.~\cite{Rothkopf:2009pk}, leads to the simple relation 
\begin{align}
W_\square(r,\tau)=\int_{-2m_Q}^{\Lambda_{\rm UV}} d\omega e^{-\omega \tau} \rho_\square(r,\omega), \quad \rho_\square(r,\omega)=\sum_{n,m}P_n(T)|M_{n,m}(r)|^2\delta(\omega-E_m^{Q\bar{Q}}(r)+E_n).\label{eq:WilsonSpecDec}
\end{align}
The spectral function (in a finite volume) is composed of a finite number of delta peaks. Here $E_n$ denotes the energy of an eigenstate $|n\rangle$ of the system without a $Q\bar{Q}$ pair and $E_m^{Q\bar{Q}}(r)$ the energy of a state with a static $Q\bar{Q}$ pair present at a distance $r$. The matrix element $M_{n,m}=\langle n|M(\mathbf{x}+\mathbf{r},\mathbf{x},t=0)|m\rangle$ also carries an $r$ dependence, while the Boltzmann factor $P_n(T)=e^{-E_n/T}/[\sum_n e^{-E_n/T}]$ only depends on temperature T. Note that the positions of the peaks in $\rho_\square$ are $T$ independent but their amplitude surely is, due to $P_n(T)$. In the infinite volume limit the delta peaks bunch and their envelope will form a continuous function along real-time frequencies $\omega$.

Analytic continuation of the Wilson loop to Minkowski time changes the Laplace-type kernel of \cref{eq:WilsonSpecDec} into nothing but a Fourier transform over the spectral function
\begin{align}
W_\square(r,t)=\int_{-2m_Q}^{\Lambda_{\rm UV}} d\omega e^{-i\omega t} \rho_\square(r,\omega). \label{eq:Wilsonrealspec}
\end{align}
In case that we have access to the spectral function we may then straight forwardly relate the value of the potential to $\rho_\square(\omega)$ by plugging \cref{eq:Wilsonrealspec} back into \cref{eq:realtimeVdef2}
\begin{align}
V^{(0)}_S(r)=\lim_{t\to\infty} \frac{\int d\omega \omega e^{-i\omega t} \rho_\square(r,\omega)}{\int d\omega e^{-i\omega t}\rho_\square (r,\omega)}.\label{eq:PotFromSpec}
\end{align}
This relation forms the basis for lattice studies of the proper static interquark potential. 

The fact that a late time limit is taken, indicates that only spectral features at small frequencies are relevant for the potential. I.e. at $T=0$ where there exists a single well separated lowest lying delta peak in $\rho_\square$, it alone determines the value of $V^{(0)}_S(r)$. If we plug a single delta peak at position $\omega_0$ into \cref{eq:PotFromSpec} we obtain a purely real ${\rm Re}[V]=\omega_0$ and ${\rm Im}[V]=0$ as expected. The single delta peak present in $\rho_\square$ at $T=0$ makes it possible to rewrite the real-time definition of the potential in terms of Euclidean quantities, recovering the conventional definition of the vacuum heavy quark potential $V^{(0)}_S(r,T=0)=\lim_{\tau\to\infty}\frac{1}{\tau}{\rm log}[W_\square(r,\tau)]$. I.e. the single delta peak in $\rho_\square$ leads to single exponential decay of the Euclidean Wilson loop at late imaginary times and its decay constant may be extracted by considering the logarithm of the Wilson loop. It is often customary to this end to compute the {\it effective potential}, which is just the effective mass (see \cref{eq:effmass}) of the Euclidean Wilson loop. Note that at finite temperature, due to the compact Euclidean domain we cannot take the $\tau\to\infty$ limit anymore. Furthermore it turns out that the Wilson loop at the latest available time $\tau=\beta$ is not related directly to the proper in-medium potential.

At finite but not too high temperature the single delta peak starts to broaden, while still being well separated from any higher lying spectral structures. In the simplest case we may expect a Breit-Wigner like shape to emerge in $\rho_\square$ at low frequencies. Inserting such a shape into \cref{eq:PotFromSpec} it tells us that the position of the peak provides the real- and the width of the peak the imaginary part of the potential. Note that finding in $\rho_\square(\omega)$  only a single Breit-Wigner would indicate that the evolution of the Wilson loop at any time is governed by a time independent potential.

Having derived the relation between the Wilson loop and the static potential at timescales, much longer than the characteristic scales of the medium, based on the lowest order of the multipole expansion, it has to be checked whether also non-potential effects can affect the time evolution of $W_\square$ in practice. Since early and late times both contribute to the Fourier transform it is important to disentangle them in order to identify the relevant spectral structures encoding the potential. This has been achieved in Ref.~\cite{Burnier:2012az}. Consider the general time evolution equation of the Wilson loop, which follows from its spectral decomposition
\begin{align}
 i\partial_tW_\square(r,t)=\Phi(r,t)W_\square(r,t).\label{Eq:WilsonLoopTimeEvol}
\end{align}
It contains a complex valued and time dependent function $\Phi(r,t)$. If the potential picture is valid $\Phi(r,t)$ will asymptote to a time independent quantity, we identify with the potential
\begin{align}
 \lim_{t\to\infty}\Phi(r,t)=V(r)\label{Eq:PotAsympt}.
\end{align}
We wish to understand how the time dependence in $\Phi(r,t)$ affects the shape of the lowest lying peak, which encodes the potential. To this end let us decompose $\Phi(r,t)=V(r)+\phi(r,t)$ into a time dependent part $\phi(r,t)$ which vanishes after a characteristic time $t_{Q\bar{Q}}$ has passed. Only after $t_{Q\bar{Q}}$ have enough gluon exchanges taken place that the retarded field theoretical interaction may be viewed through the coarse grained lens of an instantaneous exchange potential.

We can formally solve \cref{Eq:WilsonLoopTimeEvol} at positive times $t>0$. For the solution to be convergent at late times the imaginary part of the potential needs to be negative ${\rm Im}[V](r)<0$. One obtains
\begin{align}
 \nonumber W_\square(r,t)={\rm exp}\Bigg[ -i\Big(& {\rm Re}[V](t)t + {\rm Re}[\sigma](r,t)\Big) -|{\rm Im}[V](r)|t+{\rm Im}[\sigma](r,t) \Bigg]\label{Eq:WLPotMod}.
\end{align}
where the function $\sigma(r,t)=\int_0^t\phi(r,t)dt$ is defined from integrating over the time dependent contribution to $\Phi$, and $\sigma_\infty(r)=\sigma(r,|t|>t_{Q\bar{Q}})=\int_0^\infty\phi(r,t)dt$ denotes its asymptotic value. The solution of the Wilson loop at negative times can be directly obtained from the symmetry condition  $W_\square(r,-t)=W^*_\square(r,t)$. Now we are ready to solve for the corresponding spectral function, which reads
\begin{align}
\nonumber \rho_\square(r,\omega)=\frac{1}{2\pi}\int_{-\infty}^\infty & dt\; {\rm exp}\Bigg[ i \Big(\omega-{\rm Re}[V](r)\Big) t -i{\rm Re}[\sigma](r,|t|)\mathrm{sign}(t) -|{\rm Im}[V](r)||t|+ {\rm Im}[\sigma](r,|t|)\Bigg]\notag.
\end{align}
We can distinguish two contributions here, one from within the time range $-t_{Q\bar{Q}}<t<t_{Q\bar{Q}}$ and those beyond
 \begin{align}
 &\rho_\square(r,\omega)=\frac{1}{2\pi}e^{{\rm Im}[\sigma_\infty](r)}\int_{-\infty}^\infty  dt\; {\rm exp}\Bigg[ i \Big(\omega-{\rm Re}[V](r)\Big)t -|{\rm Im}[V](r)||t|  -i{\rm Re}[\sigma_\infty](r)\mathrm{sign}(t) \Bigg]\\
\nonumber &+\frac{1}{2\pi}\int_{-t_{Q\bar{Q}}}^{t_{Q\bar{Q}}}  dt\; {\rm exp}\Bigg[ i \Big(\omega-{\rm Re}[V](r)\Big) t  -|{\rm Im}[V](r)||t|\Bigg] \Bigg(e^{ -i{\rm Re}[\sigma](r,|t|)\mathrm{sign}(t) + {\rm Im}[\sigma](r,|t|)}-e^{-i{\rm Re}[\sigma_\infty](r)\mathrm{sign}(t) + {\rm Im}[\sigma_\infty](r)}\Bigg).
\end{align}
We can solve the integral in the first line analytically, as it extends over the whole time axis. It corresponds to a well defined peak structure, which encodes the values of the potential. It is this peak structure, which we wish to identify in actual lattice QCD spectral functions and fit the shape around its maximum in the region $(\omega-{\rm Re} V(r))t_{Q\bar{Q}}\ll1$. The integral in the second line will produce a spectral structure, which acts as background to the potential peak. In order to capture its effect we expand the first exponential in the second integral $\exp\left[i (\omega-{\rm Re}[V](r)) t\right]$  around the peak frequency. This leads to the following final expression
 \begin{eqnarray}
 \nonumber \rho_\square(r,\omega)=&&\frac{1}{\pi}e^{{\rm Im}[\sigma_\infty](r)} \frac{|{\rm Im}[V](r)|{\rm cos}[{\rm Re}[\sigma_\infty](r)]-({\rm Re}[V](r)-\omega){\rm sin}[{\rm Re}[\sigma_\infty](r)]}{ {\rm Im}[V](r)^2+ ({\rm Re}[V](r)-\omega)^2}\\&&+c_0(r)+c_1(r)t_{Q\bar Q}({\rm Re}[V](r)-\omega)+c_2(r)t_{Q\bar Q}^2({\rm Re}[V](r)-\omega)^2+\cdots\label{Eq:FitShapeFull}
\end{eqnarray}
The first term, related to the potential, takes the form of a skewed Breit Wigner and reduces to a naive Breit-Wigner peak only in case that non-potential effects are absent $\partial_t\Phi(r,t)=0$. In addition to the phase $\sigma_\infty$, when fitting the potential peak structure one needs to be aware of the background terms $c_i(r)$, also arising from the early time dependence of $\Phi(r,t)$. I.e. even in the region close to the tip of the peak, where all $c_i(r)$ with $i>0$ can be ignored, the influence of the early time physics influences the spectral shape through $c_0$ and ${\rm Re}[\sigma_\infty](r)$, which implies that these two coefficients have to be included no matter what fitting range is chosen. In early attempts of extracting the potential e.g. in Ref.~\cite{Rothkopf:2011db} the authors were not aware of this fact, which lead to results overestimating the values of the real part of the potential.

While at first sight it is not clear what kind of potential results from inserting \cref{Eq:FitShapeFull} into \cref{eq:PotFromSpec} it turns out that thanks to the late time limit, none of the coefficients $c_i(r)$, nor the terms with $\sigma_\infty$ contribute and a time independent potential emerges. In essence, only the the principal part from the skewed Lorentzian, i.e. the term with one negative power of $\omega$ survives.

\begin{figure}
\centering
\includegraphics[scale=0.85]{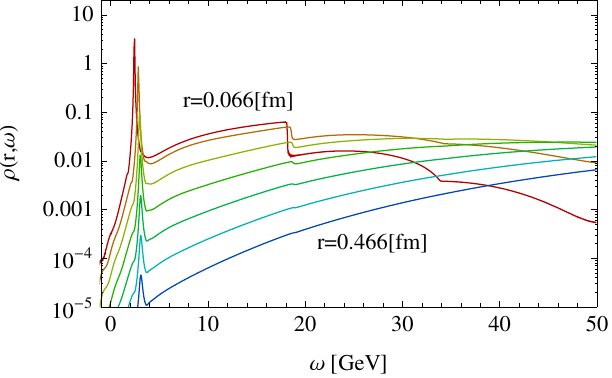}
\includegraphics[scale=0.7]{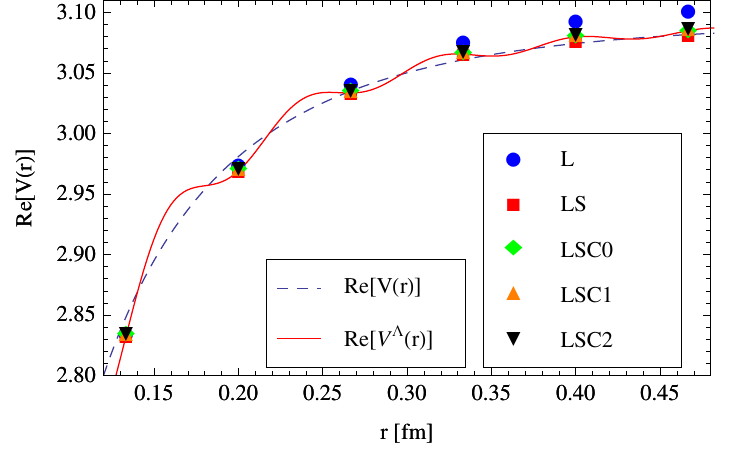}
\includegraphics[scale=0.7]{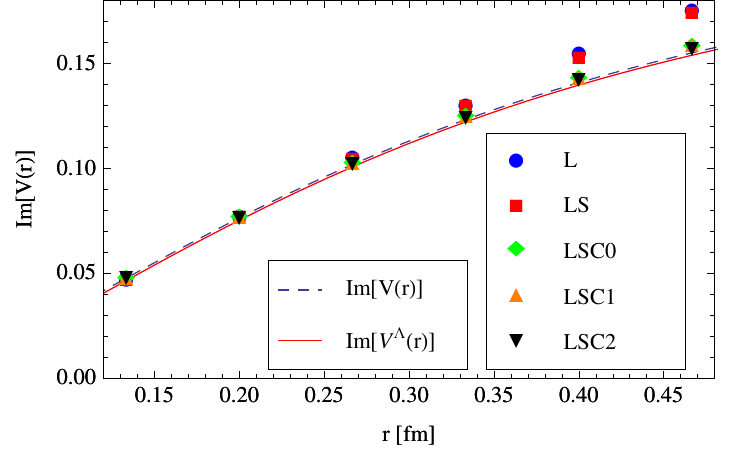}
\caption{(left) The Wilson loop spectral functions evaluated in Hard thermal loop perturbation theory at $T=2.33T_C$. One can clearly identify a well defined lowest lying peak, which is embedded in a significant background structure. (center) The real-part of the potential obtained from fitting the peak structure in the Wilson loop spectrum using different sets of fitting functions (filled symbols). The solid line corresponds to ${\rm Re}[V]$ for the same IR and UV regulator used in the perturbative computation. (left) The imaginary part of the potential fitted from the spectral peak (solid symbols). Figure reproduced from Ref.~\cite{Burnier:2013fca}}\label{fig:PotLoopHTL}
\end{figure}

Let us benchmark the extraction of the in-medium potential from Wilson loop spectral functions within HTL perturbation theory, where both the values of the potential and the spectral function are known analytically. To this end in Ref.~\cite{Burnier:2013fca} the authors computed $\rho_\square$ at $T=2.33T_C=630$MeV using a UV cutoff of $\Lambda_{\rm UV}=5\pi$GeV to mimic the effects of a lattice regularization, the outcome of which is plotted for different spatial separation distances $r=0.066-0.466$fm in the left panel of \cref{fig:PotLoopHTL}. At this temperature deep in the QGP phase one finds a clear signal for a lowest lying spectral peak, which however is immersed in a large background structure. If one now tries to extract the peak position and width from the spectrum to read out the values of ${\rm Re}[V]$ and ${\rm Im}[V]$ one finds that a naive Breit Wigner ansatz will systematically overestimate the real part, as seen in the blue circles in the center panel of \cref{fig:PotLoopHTL}. Once the skewing and constant background term are included the actual values are reproduced excellently. Here one should compare to the solid line, which represents the HTL result for ${\rm Re}[V]$ in the presence of the UV cutoff. The same holds for the extraction of the imaginary part, which, as shown in the left panel, clearly requires the inclusion of the constant background term to yields an accurate result.

\begin{figure}
\centering
\includegraphics[scale=0.9]{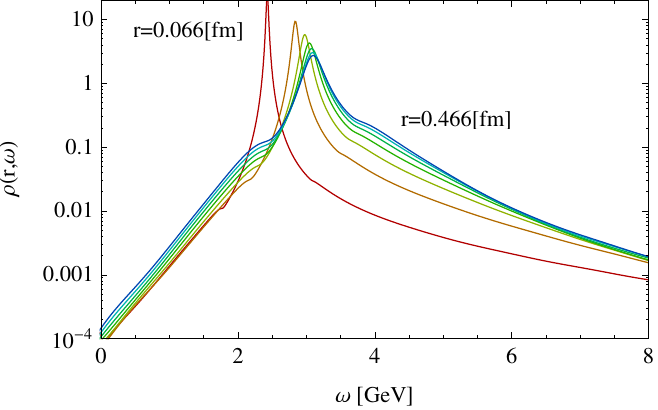}
\caption{Spectral function of the Wilson line correlator in Coulomb gauge computed in HTL perturbation theory at $T=2.33T_C$. Figure reproduced from Ref.~\cite{Burnier:2013fca}}\label{fig:WlineSpecHTL}
\end{figure}

On the one hand this analysis in principle bodes well for performing a potential reconstruction in lattice QCD. On the other hand the behavior of the Wilson loop spectrum tells us that the large UV spectral weight will lead to a strong suppression of the Euclidean Wilson loop and in turn it will be very difficult to obtain a good signal to noise ratio at intermediate $\tau$ values where the potential peak signal dominates. The reason for the large (unphysical) background structures in $\rho_\square$ is a class of UV divergences, called cusp divergencies \cite{Polyakov:1980ca,Berwein:2012mw}, which arise from the 90 degree corners of the Wilson loop. The authors of Ref.~\cite{Burnier:2013fca} therefore explored another quantity, the {\it Wilson line correlator} in Coulomb gauge $W_{||}$, which corresponds to the Wilson loop with its spatial Wilson lines removed. This in general gauge dependent quantity goes over into the Wilson loop when axial gauge $A_0=0$ is chosen. It was found that this quantity at leading order in HTL perturbation theory encodes exactly the same potential peak as the Wilson loop but exhibits a significantly reduced background, as shown in \cref{fig:WlineSpecHTL}. It is this quantity which currently forms the basis for the extraction of the proper in-medium potential on the lattice. (At $T=0$ it has been established  \cite{Luscher:2001up} that the Wilson line correlator and the Wilson loop encode the same potential also non-perturbatively.)

We have established that if a potential picture is valid its values can be extracted from spectral functions using appropriate fits to the lowest lying spectral structure. In practice we will turn this argument around and take the presence of a well defined lowest lying peak structure as indication that the late time behavior of the Wilson loop is indeed governed by a simple potential.

\begin{figure}
\centering
\includegraphics[scale=0.6, clip=true, trim= 1cm 11cm 8cm 1cm]{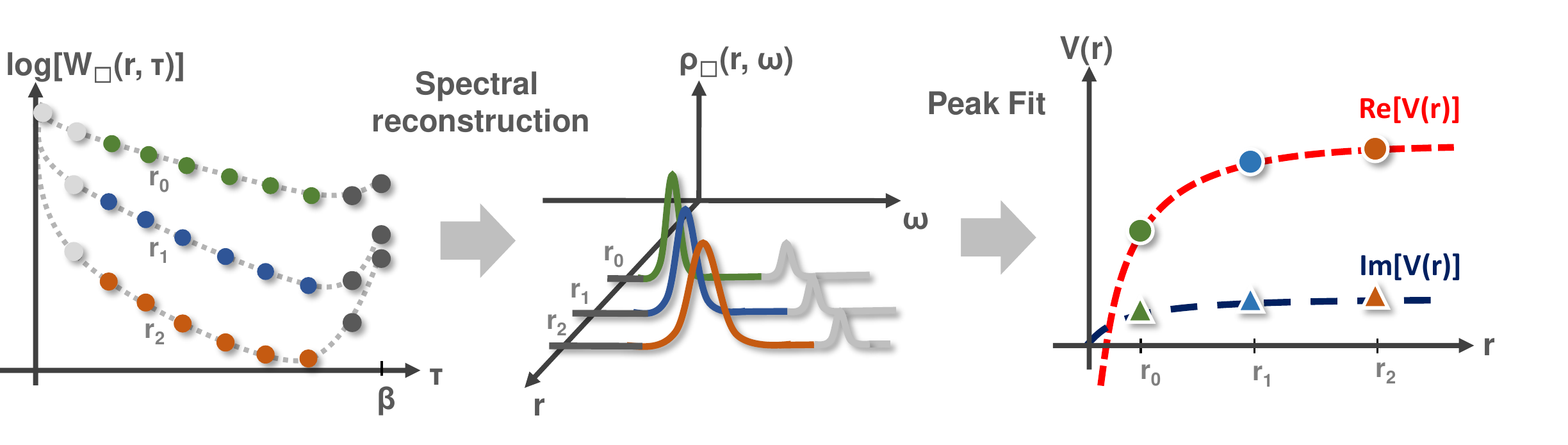}
\caption{Sketch of the extraction strategy for the proper real-time heavy quark potential from lattice QCD Wilson loop correlators. Figure adapted from Ref.~\cite{Rothkopf:2013ria}}\label{fig:ExtrStrat}
\end{figure}

The central challenge in lattice QCD based studies of the in-medium potential lies in the reconstruction of the spectral function of the Wilson loop or Wilson lines from the Euclidean time observables. As a simple spectral decomposition of the Wilson loop exists (see \cref{eq:WilsonSpecDec}) this task is but one example of the inverse problem discussed in \cref{sec:specrec} amenable to both Bayesian spectral reconstruction and the Pade method. The full recipe is sketched in \cref{fig:ExtrStrat}.

The first attempts of extracting spectral functions for the determination of the potential were based on the standard implementation of the MEM by Bryan. Since generally only a small number of simulation datapoints ${\cal O}(10-40)$ are available, it was found that the restricted SVD search space of the MEM was unable to accurately recover the narrow and strongly peaked spectral feature encoding the potential. Its position was systematically overestimated. The situation improved after an MEM implementation with an extended search space had been developed in Refs.~\cite{Rothkopf:2011ef,Rothkopf:2012vv}. As has been benchmarked with HTL correlator data in Ref.~\cite{Burnier:2013fca} this leads to more accurate values for ${\rm Re}[V]$. On the other hand even with an extended search space the MEM was found to produce spectral features, which resemble Gaussian peaks \cite{Rothkopf:2011db}, even if the underlying spectral function contains a Breit-Wigner like structure. This is a serious drawback, since a Gaussian peak inserted into \cref{eq:PotFromSpec} will lead to the incorrect conclusion that the imaginary part of the potential grows with time. It took the development of a completely redesigned reconstruction approach, the BR method, to achieve reconstructions that faithfully reproduce not only the overall functional form, but also quantitatively the position and the width of the encoded peaks (see Ref.~\cite{Burnier:2013nla}). Note that, as is common for all spectral reconstruction approaches, the robust determination of the peak width requires input data of a much higher quality than the determination of the peak position.

The first extraction of the potential from Wilson line correlators using both the BR method and the appropriate fitting strategy was presented in Ref.~\cite{Burnier:2014ssa}. This study on the one hand investigated lattice Wilson correlators from pure gauge $SU(3)$ fixed scale simulations on $N_s=32$ lattices with the naive anisotropic Wilson action. A relatively fine lattice spacing $a=0.039$fm at $\beta=7$ and renormalized anisotropy $\xi_r=4$ was chosen as originally introduced in Ref.~\cite{Asakawa:2003re}. On the other hand the study used fixed box simulations on isotropic $N_s=48$ grids with $N_f=2+1$ flavors of light quarks ($m_\pi\approx 300$Mev) described by the asqtad action originally deployed in Ref.~\cite{Bazavov:2011nk} (HotQCD collaboration). Updated results for quenched QCD on lattices with significantly larger physical volume using the parameter set $\beta=6.1$, $a=0.097$ and $\xi_b=4$ originally tuned in Ref.~\cite{Matsufuru:2001cp} were subsequently discussed in \cite{Burnier:2016mxc}, clarifying the effect of finite volume artifacts in the previous studies.

\begin{figure}
\centering
\includegraphics[scale=0.25]{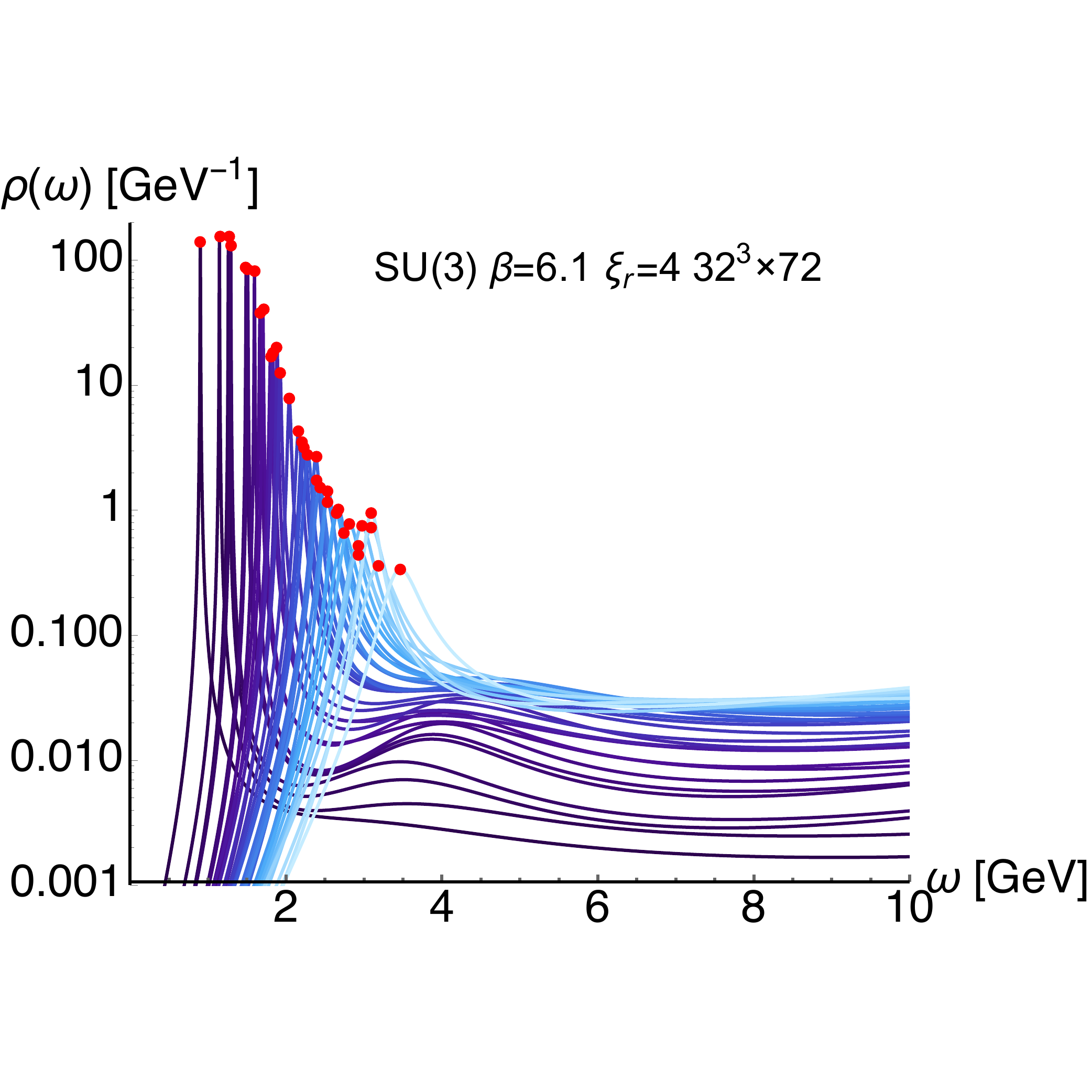}
\hspace{0.3cm}\includegraphics[scale=0.25]{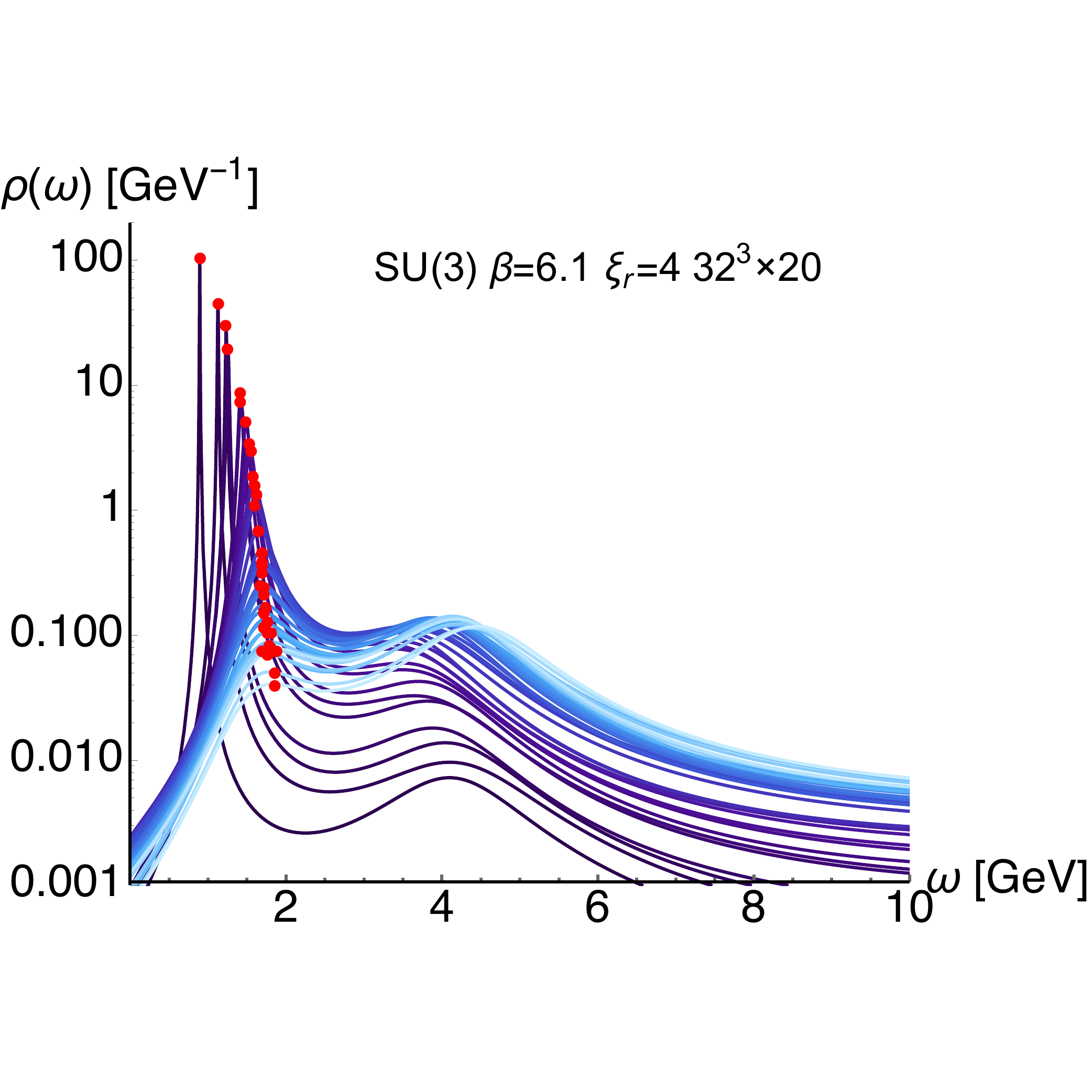}\\\includegraphics[scale=0.25]{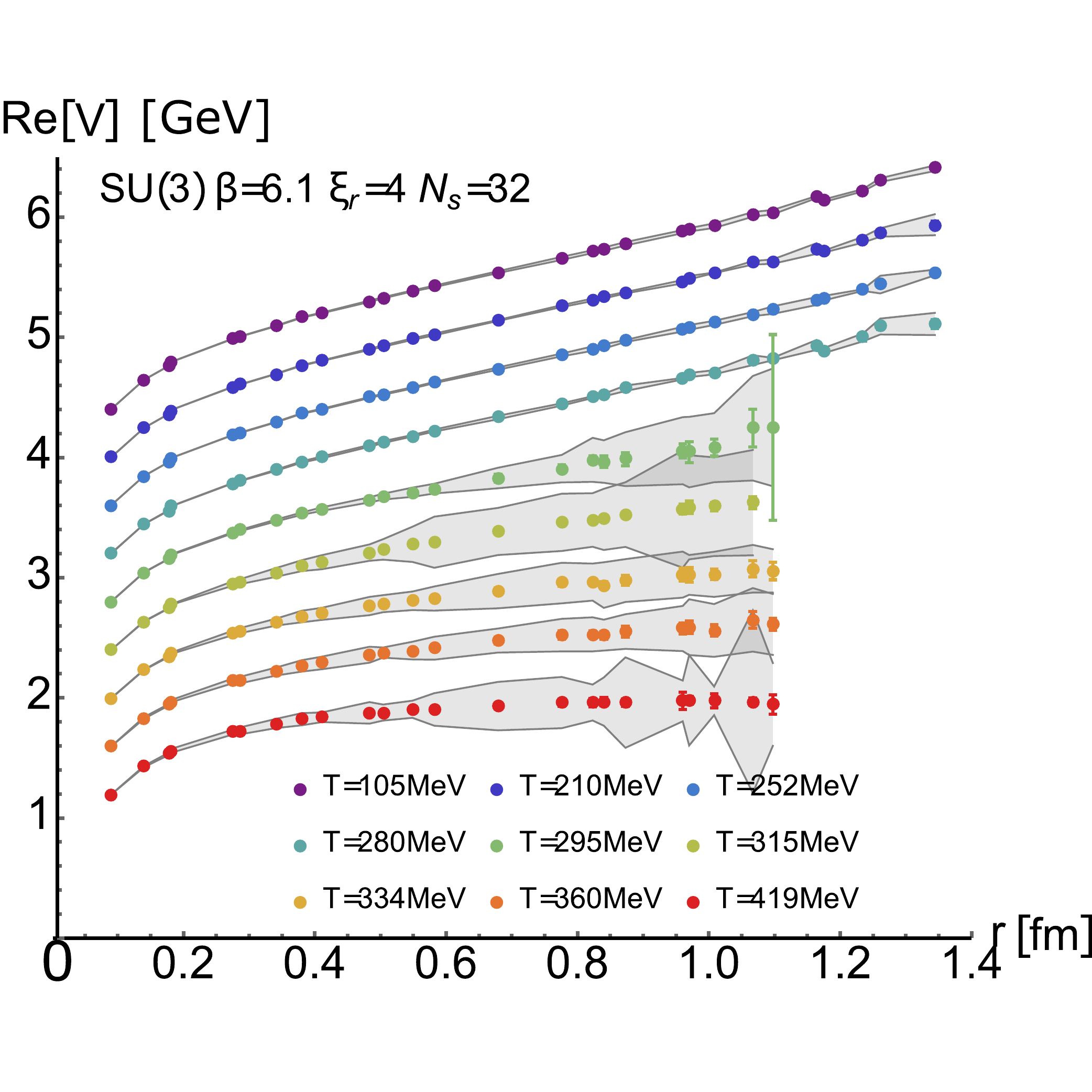}\hspace{0.5cm}\includegraphics[scale=0.25]{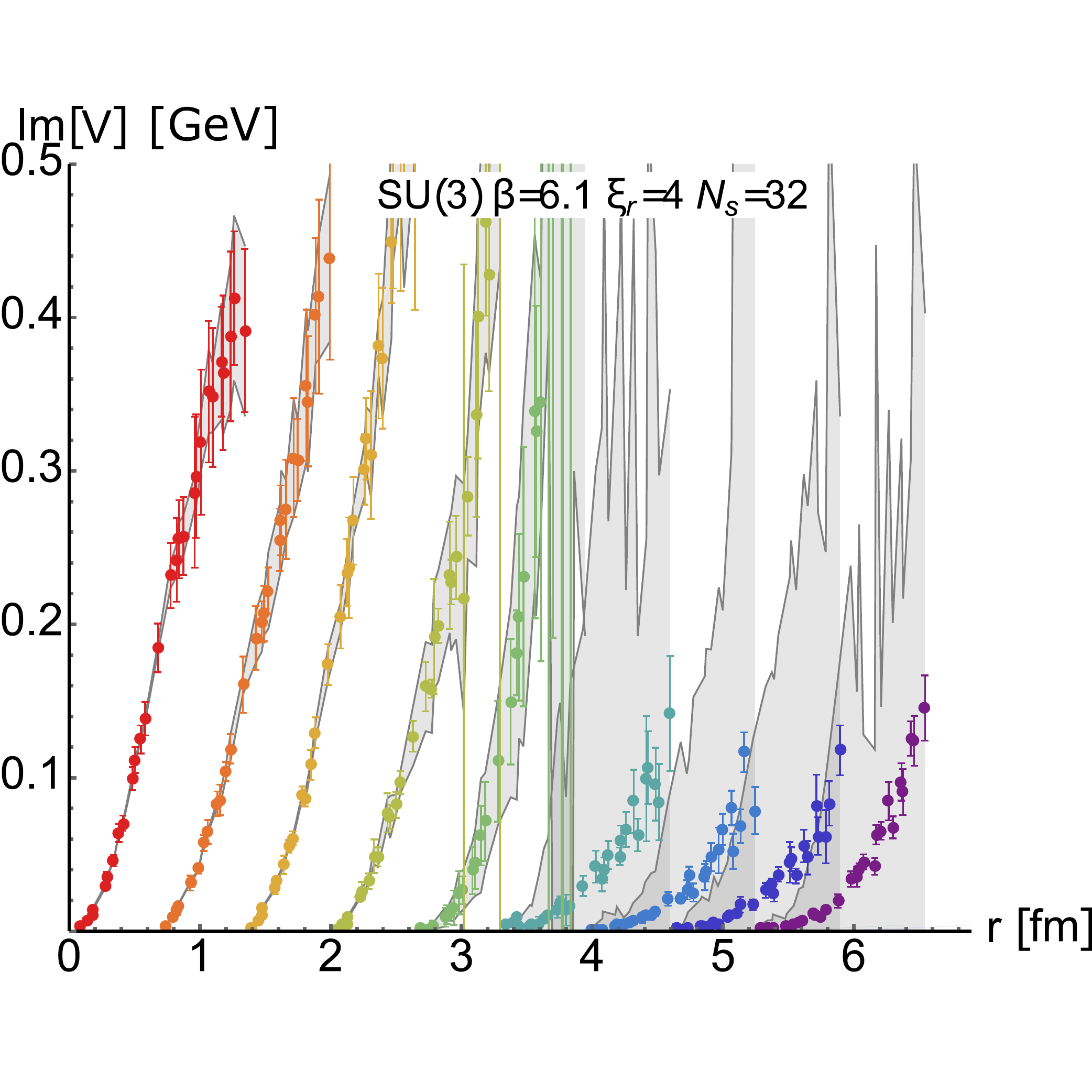}
\caption{(top row) Reconstructed spectral functions from anisotropic $SU(3)$ simulations of Wilson lines in Coulomb gauge at (left) low temperatures $T_c>T=113$MeV in the confined phase and (right) at high temperatures $T_c<T=406$MeV in the deconfined phase. In both cases a well defined lowest lying peak structure is visible at all separation distances $r/a\in[1,17]$. (bottom row) The corresponding values of the (left) real part and (right) imaginary part of the potential. For better readability the former has been shifted by hand in y- the latter in x-direction for different temperatures. Errorbars denote statistical uncertainty, while gray errorbands represent systematic uncertainties such as default model dependence. Figures adapted from Ref.~\cite{Burnier:2016mxc}}\label{fig:PotLoopQuenched}
\end{figure}

In \cref{fig:PotLoopQuenched} we show in the top row concrete examples of reconstructed spectral functions from quenched lattice simulations of Wilson line correlators in Coulomb gauge from Ref.~\cite{Burnier:2016mxc}. In order to avoid possible divergent terms, as well as to reduce the influence of lattice artifacts, the reconstruction is carried out by excluding the first $\tau=0$ and last correlator point $\tau=1/T$ along imaginary time. The effects of regularization are checked by carrying out the reconstruction not just with a constant default model but also deploying polynomials $m(\omega)=m_0(\omega_{\rm min}-1)^\gamma$ in frequency with positive and negative powers $\gamma$, as well as different amplitudes $m_0$. 

The left panel corresponds to a temperature of $T=113$MeV deep in the confined phase, while the right panel contains the results at $T=406$MeV, where the gluonic medium is already deconfined ($T_c^{\beta=6.1}=290$MeV). The curves of different shades correspond to different spatial separation distances between $r/a=1\ldots 17$, from darkest to lightest. One clearly sees that in all cases a well defined lowest lying peak is recovered, which has the form of a skewed Breit Wigner. In turn the spectral reconstruction indicates that indeed a potential description emerges at sufficiently late Minkowski times.  Note that due to the fixed scale approach there are much less datapoints $N_\tau=20$ vs. $N_\tau=72$ available at high temperatures but the BR method is still able to identify a well defined lowest lying peak, as well as indications of the shoulder structures at higher frequencies akin to what was found in the HTL computations in \cref{fig:WlineSpecHTL}. Using the fit ansatz of \cref{Eq:FitShapeFull} the real and imaginary part can be estimated, as shown in the left and right panel of the bottom row respectively. The datapoints of ${\rm Re}[V]$ are shifted by hand in y-direction for better readability. Correctly renormalized they take on the same values at small distances, recovering the $T=0$ behavior there. The ${\rm Im}[V]$ values on the other hand are shifted by hand in x-direction, they originally all vanish at distance $r=0$. The errorbars denote statistical uncertainties based on a Jackknife resampling, while the gray errorbands encode systematic uncertainty, among others the dependence on the choice of default model.

At this point let us focus on the qualitative behavior seen in the real- and imaginary parts. In the confined phase the potential exhibits a Cornell like behavior with a Coulombic part at small distances $r<0.3$fm and a linearly rising part at larger distances. In the absence of dynamical quarks no string breaking occurs. Interestingly the lattice results indicate that below $T_c$ in the absence of deconfined color charges no significant screening occurs. The heavy mass of the glueball excitations suppresses their interference with the binding of static quarks. Once the deconfined phase is reached at $T=295$MeV a very different behavior emerges, the real part starts to flatten off and approaches a constant at large distances, indicative of screening. Such a rather abrupt change in behavior is interpreted as a manifestation of the phase transition that emerges in the infinite volume limit.

Another interesting finding is that ${\rm Re}[V]$ lies quite close to the color singlet free energies extracted on the same lattices (not shown). This is a nontrivial finding, as the data point $W_{||}(r,\tau=\beta)$, which encodes the free energies is not even included in the reconstruction of the spectral function. There exists however a clear difference in behavior between $F_1$ and ${\rm Re}[V]$ in the confined phase of quenched QCD. As illustrated in Ref.~\cite{Kaczmarek:2003dp} the former shows a clear overshoot of its $T=0$ value. Since it is expected that for a microscopic interaction potential thermal effects lead at most to a weakening, it is reassuring that ${\rm Re}[V](T>0)$ indeed does not show a similar overshoot. 

The extraction of the imaginary part is much less robust. Below $T_c$ an artificial width is present in the reconstructed spectra, which changes with the choice of default model. I.e. no significant signal for a finite imaginary part is found below $T<290$MeV. On the other hand in the deconfined phase, the spectral width grows significantly and Ref.~\cite{Burnier:2016mxc} concludes that indeed a finite ${\rm Im}[V]$ has been observed. Note that these results are obtained at a finite lattice spacing and volume and that so far no continuum limit extrapolation for the proper potential in quenched QCD exists.

\begin{figure}
\centering
\includegraphics[scale=0.4]{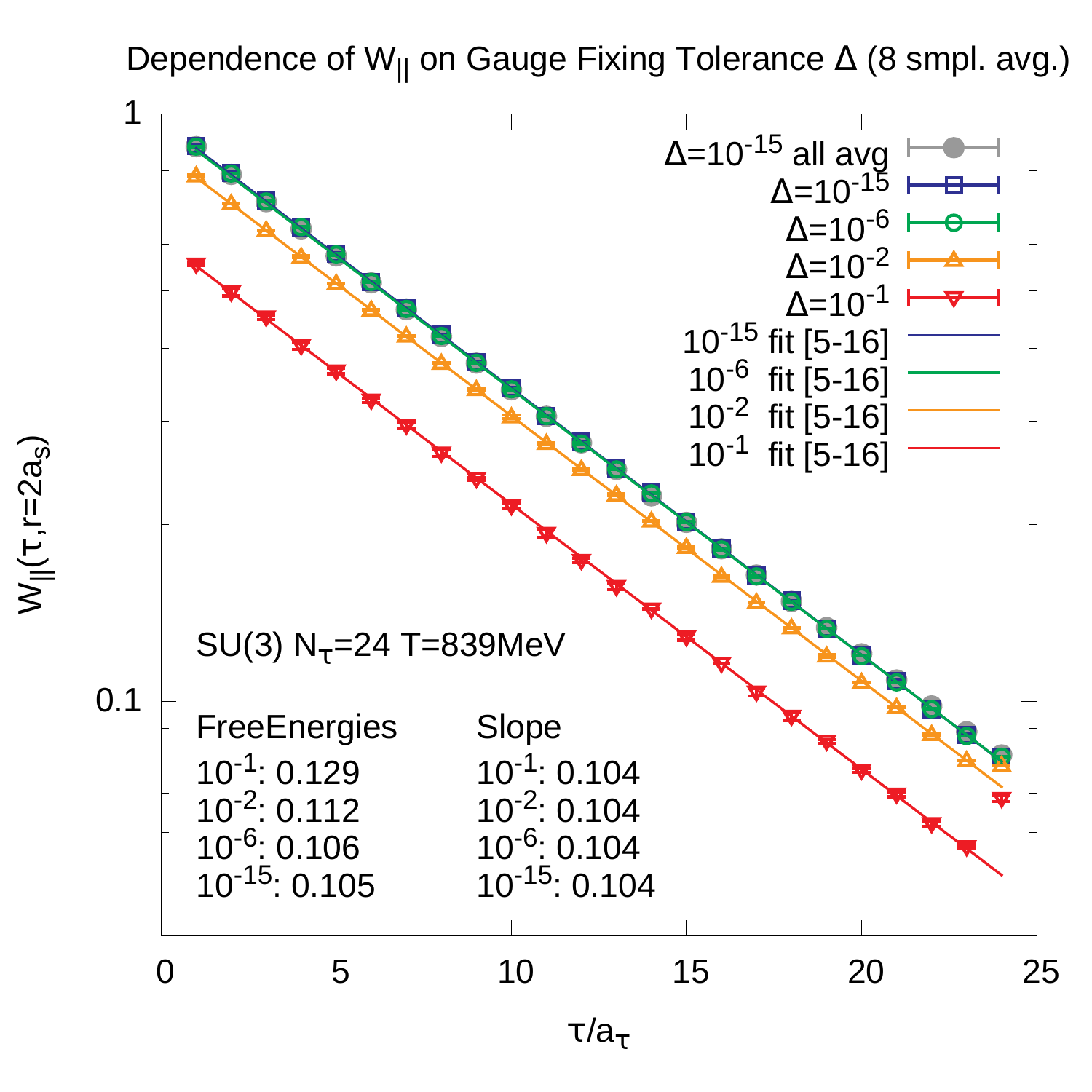}
\caption{ The Wilson line correlator at spatial separation distance $r/a_s=2$ in quenched QCD ($\beta=7, \chi=3.5, N_s=32, N_\tau=24$) at $T=839$MeV. The different datasets correspond to different deviations $\Delta$ from the Coulomb gauge fixing condition (colored points). The slope of the correlator, corresponding to the real-part of the potential is fitted by a single exponential in the range given in the legend and shown as solid line. One finds that while the color singlet free energies, defined from the correlator at $\tau=\beta$ depend on the position in gauge space beyond statistical errors, the slope remains unchanged. }\label{fig:GaugeDepCheck}
\end{figure}

\begin{figure}
\centering
\includegraphics[scale=0.25]{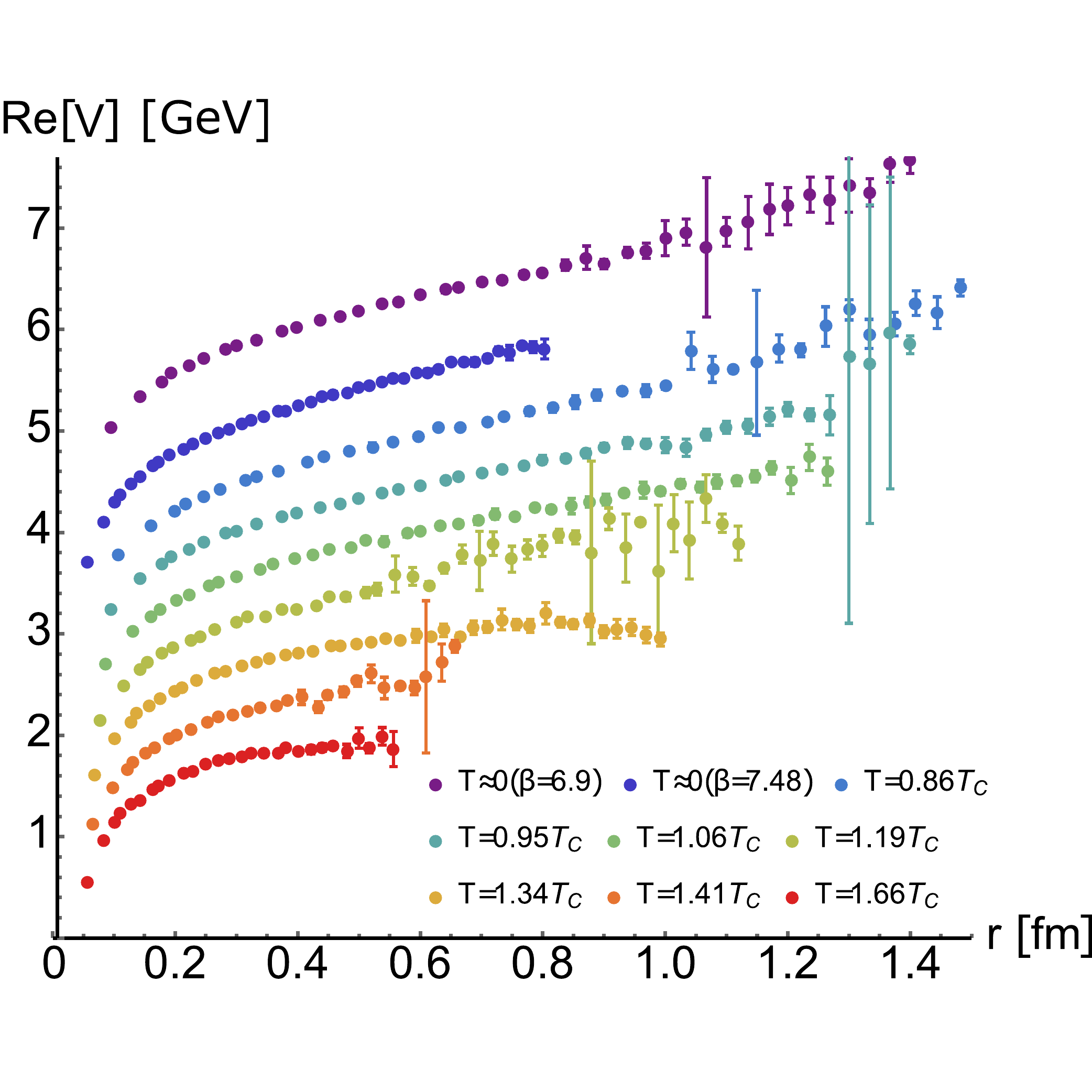}
\hspace{0.3cm}\includegraphics[scale=0.25]{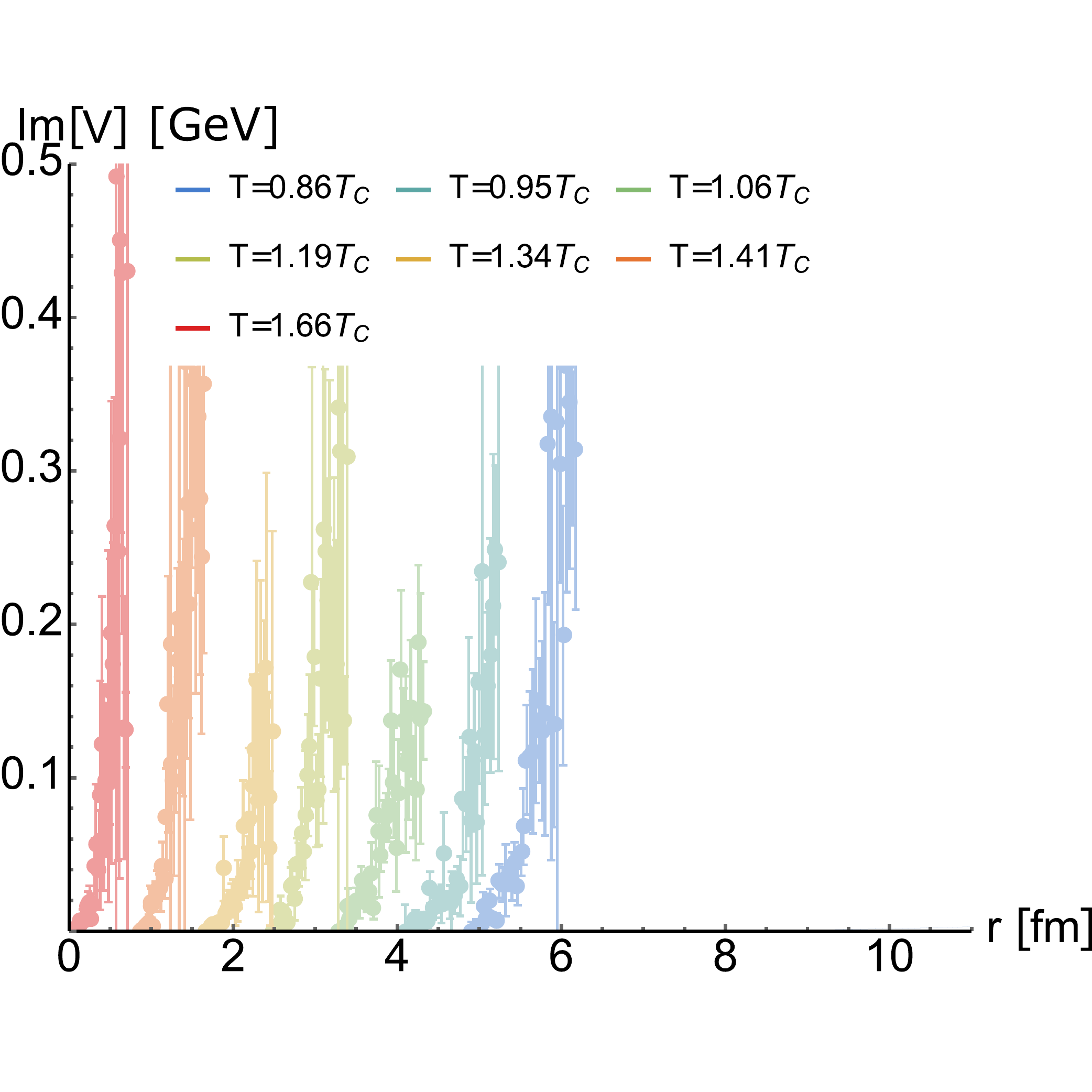}
\caption{Estimates for the real- (left) and imaginary part (right) of the real-time potential from dynamical QCD. The values of ${\rm Re}[V]$ are shifted manually in y-direction, those of ${\rm Im}[V]$ in x-direction for better readability. The top two datasets in the left panel correspond to two $T=0$ results at different lattice spacing. Since only $N_\tau=12$ datapoints underly the necessary spectral reconstruction the values of ${\rm Im}[V]$ are considered to be tentative only. Errorbars encode both the statistical, as well as estimates of the systematic uncertainty. Figures adapted from Ref.~\cite{Burnier:2015tda}}\label{fig:PotLoopAsqtad}
\end{figure}

The use of the Wilson line correlator to extract the potential begs the question of how the choice of gauge influences the analysis outcome. To shed light on this issue we present in \cref{fig:GaugeDepCheck} the Wilson line correlator in quenched QCD from an anisotropic $32^3\times24$ lattice at $\beta=7$ and $\xi_r=4$. The colored points correspond to a spatial separation of $r/a_s=2$. After iteratively fixing to Coulomb gauge down to a tolerance of $\Delta=10^{-15}$ (blue squares) we act with a random gauge transformation and again step towards Coulomb gauge computing the Wilson line correlator on the way for different tolerances $\Delta=10^{-1},10^{-2},10^{-6}$. One clearly sees that the absolute value of the correlator changes as the gauge is changed. This e.g. affects the value of the color singlet free energies, which are defined from the logarithm of the correlator at the latest available Euclidean time. Their values change beyond the statistical errorbars. At the same time, an inspection of the behavior of the correlator in the intermediate $\tau$ region, which is where the low lying spectral peak dominates, reveals that its slope remains unaffected by the deviation from Coulomb gauge. In turn the real-part of the potential in this example is equally unaffected. The picture that emerges is that the Wilson line correlator contains both gauge dependent and gauge independent information. The position of its lowest lying spectral peak appears to be insensitive to a change of gauge, while its tails at low frequencies (dominating the $\tau=\beta$ regime) as well as the spectral structures at high frequencies (dominating at small $\tau$) are modified significantly. This crosscheck, while far from a proof, is at least an indication that the extraction of the real part of the potential from Wilson line correlators in Coulomb gauge is fathomable.

First estimates for the real and imaginary part of the real-time potential in full QCD have been presented in Ref.~\cite{Burnier:2014ssa,Burnier:2015tda}. The results for the real part are shown in the left panel of \cref{fig:PotLoopAsqtad} with the values at different temperatures shifted by hand in y-direction for better readability. The two topmost datasets correspond to $T=0$ simulations at different lattice spacings. As only $N_\tau=12$ Euclidean time points were available in that study the extracted values for ${\rm Im}[V]$ are taken to be tentative only and are plotted in the right panel. Those values are shifted by hand in x-direction for different temperatures, as they all start out as zero at the origin.

At very low temperatures a Cornell-like behavior is found but no indications of string breaking up to $r=1.2$fm. The relatively heavy pion mass pushes the string breaking radius to higher values in this case. The in-medium behavior of ${\rm Re}[V]$ in the presence of dynamical fermions shows differences compared to quenched QCD, as expected. I.e. instead of an abrupt onset of screening one finds a smooth decrease of the string like part from the hadronic to the QGP regime. At the highest temperature $T=1.66T_c=286$MeV the values show a clear trend towards a constant at large distances, indicative of Debye screening. Again it is found that the values of ${\rm Re}[V]$ lie close to those of the color singlet free energies, without however matching exactly. In the presence of dynamical fermions $F_1$ also does not show an overshoot of its $T=0$ values anymore, behaving more in line with ${\rm Re}[V]$. There are signs of a finite imaginary part above $T_c$,  however the uncertainties, as indicated by the sizable errorbars, are still large.

The analysis of the real-time potential in full QCD is currently ongoing on realistic high statistic lattices from the HotQCD and TUMQCD collaboration, a status update has been presented at the 2018 Quark Matter conference in Ref.~\cite{Petreczky:2018xuh}. Using the HISQ action to implement $N_f=2+1$ light quarks flavors with almost physical pion mass $m_\pi=160$MeV on $N_s=48$ and $N_\tau=12$  (see \cite{Bazavov:2014pvz}), as well as $N_\tau=16$ grids (see \cite{Bazavov:2017dsy}), a large number of configurations ($N_{\rm conf}=4-9\times 10^3$) has been generated. Deploying a fixed box approach, the ensembles span a temperature range between $T=150-1451$MeV, much further than previous studies. In the regime $T>198$MeV spectral structures become extended enough that the BR method starts to exhibit ringing artifacts, as it tries to reconstruct these features from a very small number of input datapoints. Since Bayesian methods only improve as both the number of datapoints is increased and the errors on the input is reduced, the available high statistics alone cannot remedy this situation. However with sub percent errors in the correlators one may expect the Pade reconstruction approach of \cref{sec:Pade} to be a viable alternative. 

\begin{figure}
\centering
\includegraphics[scale=0.3]{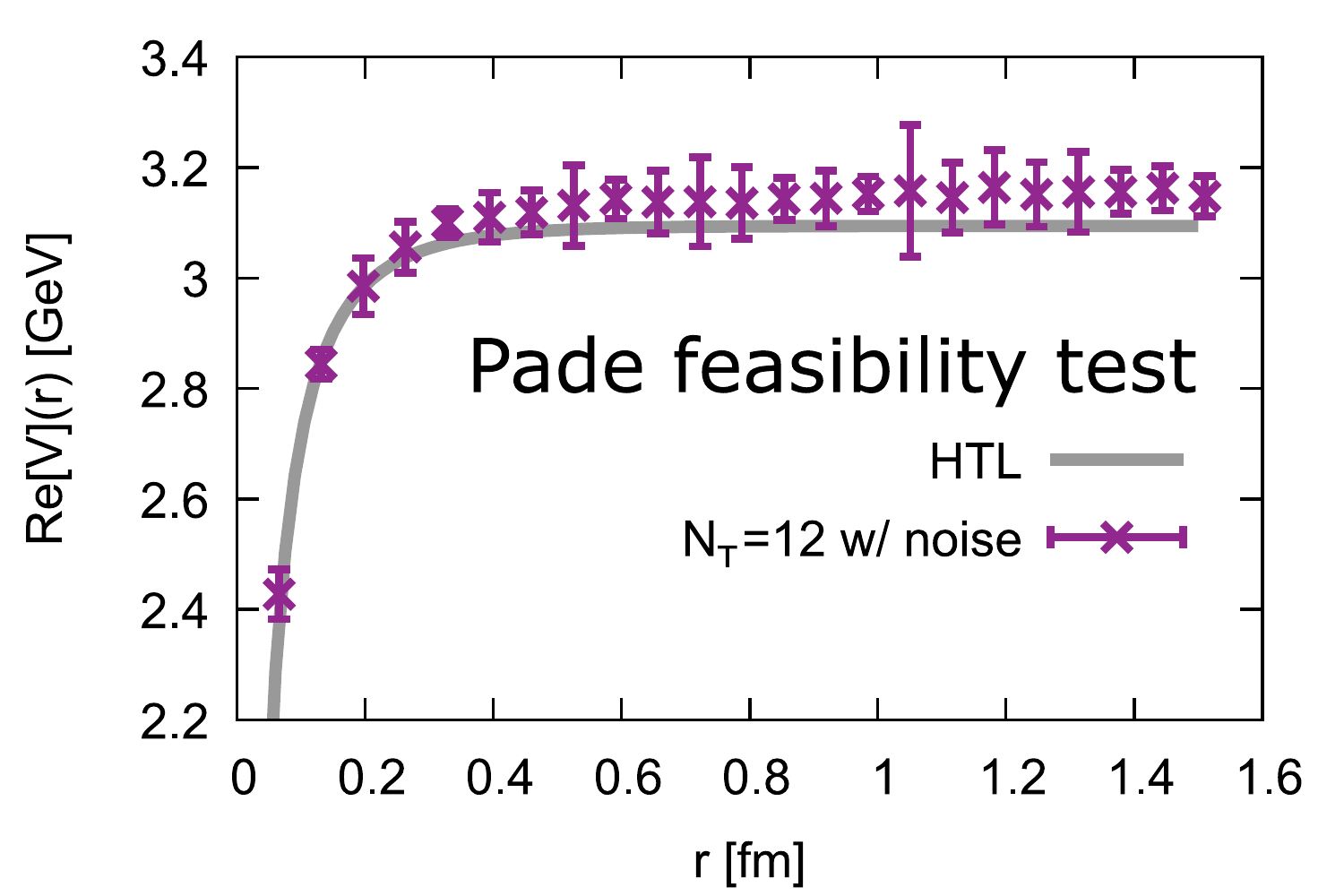}
\hspace{0.3cm}\includegraphics[scale=0.3]{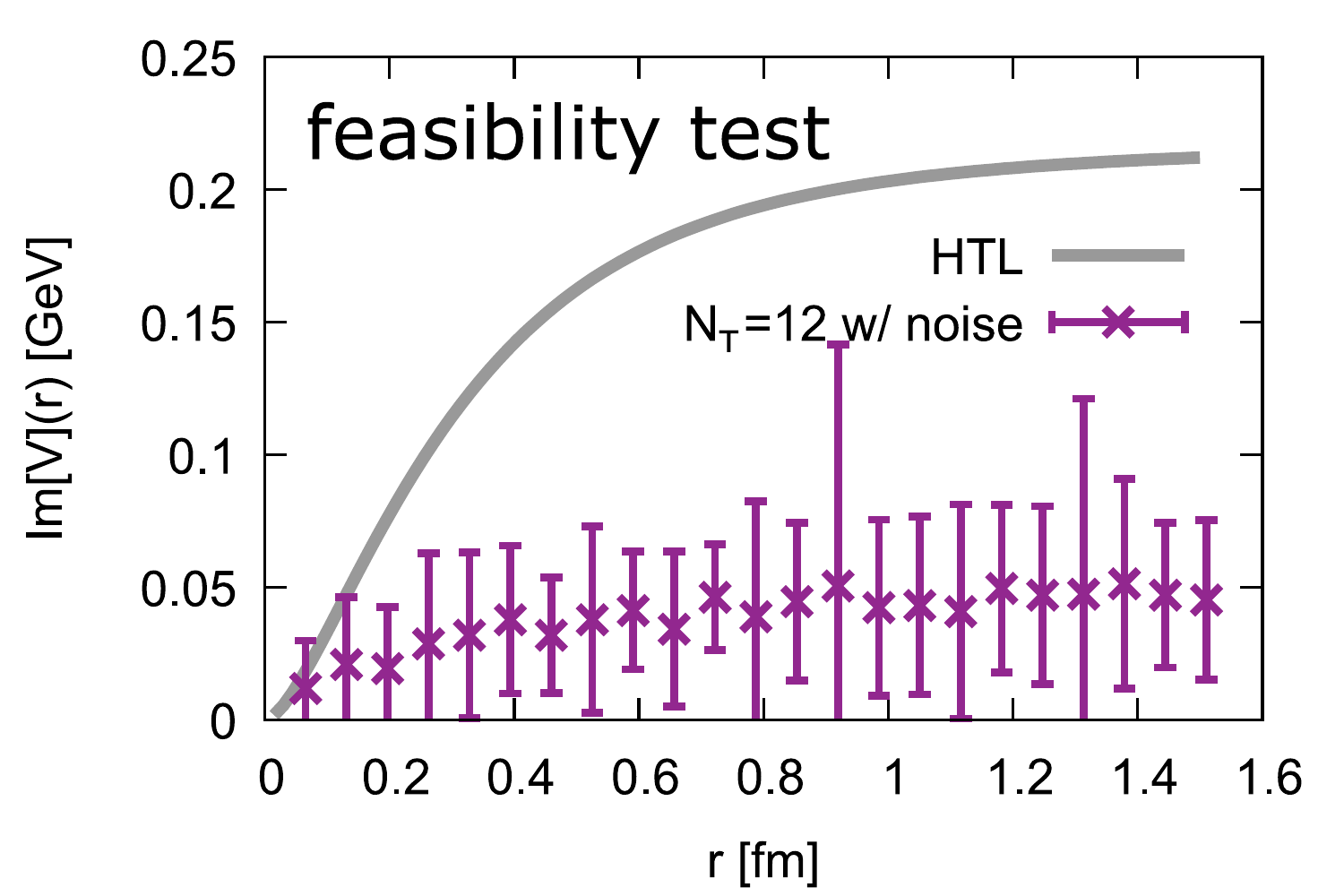}
\caption{Feasibility tests of the Pade reconstruction method based on Euclidean Wilson line correlators computed from HTL perturbation theory. The right panel shows the successfully reconstructed real-part from input data with $\Delta W/W=10^{-2}$ relative error (an error much worse than present in the actual lattice data). The right panel on the other hand shows that the Pade method is unable to capture the imaginary part from $N_\tau=12$ datapoints. Figures adapted from Ref.~\cite{Petreczky:2018xuh}}\label{fig:TstPadeExtr}
\end{figure}

The feasibility of the Pade approach is tested by using the HTL Wilson line correlators corresponding to the spectral functions at $T=2.33T_C$ given in \cref{fig:WlineSpecHTL}. Discretized with the same $N_\tau=12$ points available on the lattice and distorted with Gaussian noise that leads to constant relative errors $\Delta W/W=10^{-2}$, the real- and imaginary part of the HTL potential have been estimated in the left and right panel of \cref{fig:TstPadeExtr} respectively. Surprisingly even with uncertainties much larger than in the actual lattice data, the Pade method manages to capture ${\rm Re}[V]$ within its errorbars. The average value slightly overestimates the potential though. On the other hand the spectral width, i.e. the imaginary part of the potential is not captured well at all by the Pade with $N_\tau=12$. Further tests have revealed that if the number of input points is increased to $N_\tau=48$ an acceptable reconstruction of ${\rm Im}[V]$ can be achieved up to $r\approx0.4$fm.   

\begin{figure}
\centering
\includegraphics[scale=0.35]{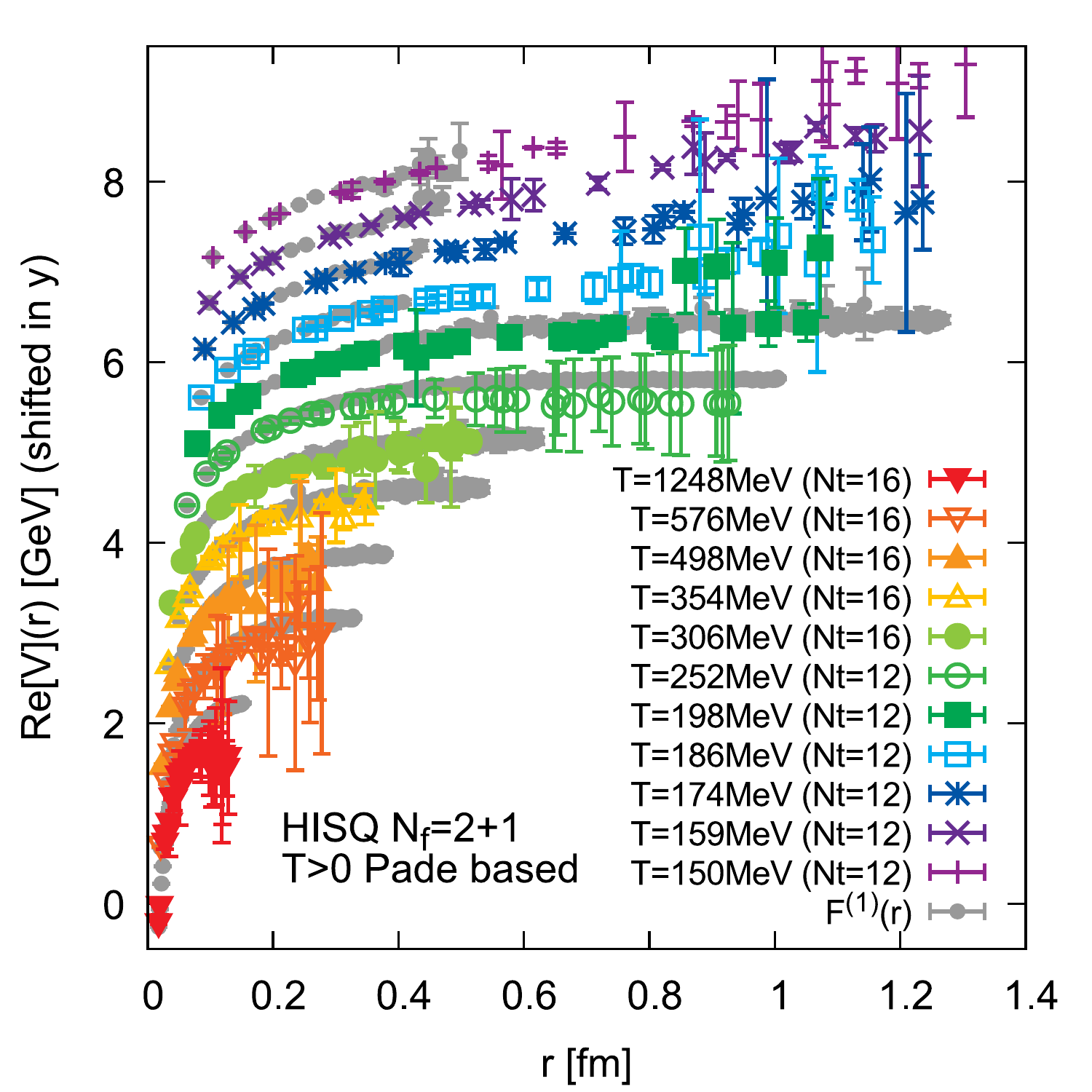}\hspace{0.5cm}\includegraphics[scale=0.35]{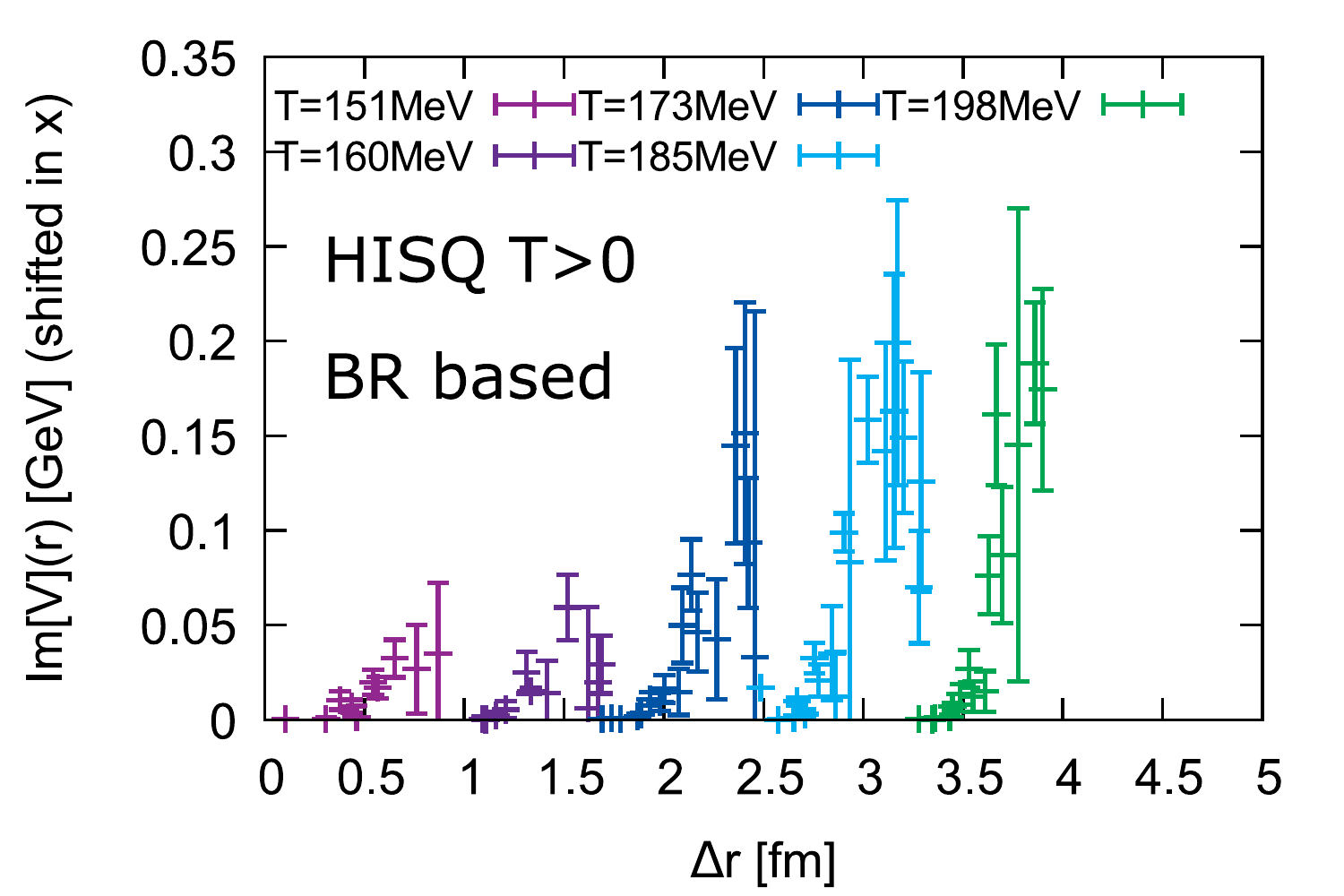}
\caption{The current best estimates of the real-time potential from realistic HISQ lattices. The real part (left) as obtained from the Pade reconstruction (colored symbols) together with the color singlet free energies (gray cricles). The tentative estimates of the imaginary part (right) are obtained with the BR method, at $T\leq 198$MeV where it is free from ringing artifacts. Figures adapted from Ref.~\cite{Petreczky:2018xuh}}\label{fig:PotLoopHISQ}
\end{figure}

Deploying the Pade reconstruction to the actual lattice data one obtains the values for ${\rm Re}[V]$ shown as colored symbols in the left panel of \cref{fig:PotLoopHISQ}. Again the values are shifted manually in y-direction for better readability. The smooth transition, observed already in simulations with higher pion masses, from a Cornell behavior at small temperatures (upper datasets) to a Debye screened behavior at high temperatures (lower datasets) persists. In the absence of a default model dependence, the errorbars here correspond to the combined statistical uncertainty from a Jackknife resampling, as well as from removing one or two datapoints from the large $\omega$ region of the Matsubara correlator used in the Pade method. Within those still sizable errorbars the values of ${\rm Re}[V]$ are compatible with the color singlet free energies shown as gray filled circles. Since the Pade approach at $N_\tau=12,16$ does not allow us to capture the imaginary part in a meaningful fashion, we deploy the BR method at those temperatures, where is is free of ringing artifacts, i.e. up to $T=198$MeV. The tentative values obtained after subtracting the artificial spectral width, induced by the spectral reconstruction method at $T=0$, are shown in the right panel of \cref{fig:PotLoopHISQ}, indicating the presence of a finite ${\rm Im}[V]$ in the QGP phase.


The values of the real-time potential on HISQ lattices so far have not been published, as the full uncertainty budget of the computation has not been established. In particular the question remains whether the spectral reconstruction from a discrete set of delta peaks leads to artifacts as of yet unaccounted for. This analysis is currently work in progress.

The spectral reconstruction presents an ill-posed inverse problem and both the Bayesian as well as the Pade reconstruction can introduce methods artefacts into the extracted potential values that may not have been spotted by mock data analyses so far. Thus there are ongoing efforts to devise extraction strategies that operate directly on the Euclidean correlator or its moments, denoted here as $\mu_n$. The first moment is nothing but the effective potential, the second moment is defined as $\mu_2=-\partial_\tau \mu_1$. A first study has been presented at the Hard Probes conference 2013 as Ref.~\cite{Bazavov:2014kva}, where the Wilson line correlator in the QGP phase was fitted using the HTL functional form of the spectral function with modified frequency arguments $\rho^{\rm HTL}_{||}(\lambda(\omega-E))$. $\lambda$ and $E$ are taken as fit parameters. At high temperatures, where QCD becomes weakly coupled and on very fine lattices, where the cutoff is far from the relevant spectral structures, this approach is promising. On the other hand around the crossover transition the question remains whether such an ansatz is well justified. In a follow up study presented at Quark Matter 2017 as Ref.~\cite{Petreczky:2017aiz} a Breit-Wigner fit without skewing has been performed in order to reproduce the first and second moments. In both studies the values for ${\rm Re}[V]$ that have been obtained are significantly larger than those from the direct spectral reconstruction and lie close to the $T=0$ results. On the one hand the use of a Breit-Wigner without skewing may partially contribute such a larger value. On the other hand these result pose the question whether in general fits of a discrete spectral function composed of delta peaks using a continuous ansatz suffer from sparseness in frequency. At the Lattice 2019 conference an interesting alternative proposal has been put forward \cite{bala2019}, which by combining Euclidean Wilson correlators in the forward $W(\tau)$ and backward $W(\beta-\tau)$ direction proposes to single out the real- and imaginary part from a plateau fit of the corresponding Euclidean data itself. Progress in direct methods would indeed be highly appreciated to reduce the systematic uncertainties arising from the spectral reconstructions.

\subsubsection{Analytic parametrization of the in-medium potential}
\label{sec:GaussLawModel}

At first sight the advent of a first principles definition of a proper real-time in-medium interquark potential and its extraction from lattice QCD appears to render {\it potential models} irrelevant. On closer inspection however it turns out that only their role has shifted. Lattice QCD results come in the form of discrete datapoints for ${\rm Re}[V]$ and ${\rm Im}[V]$ which cannot be directly used for solving e.g. a Schr\"odinger equation for the forward quarkonium correlator $D^>$. I.e. the first role of potential models today is to provide an {\it analytic parametrization} of lattice QCD results, best described by as small as possible a number of parameters. The main parameter responsible for the in-medium modification is often called the Debye mass, in analogy with the original works of Debye and H\"uckel \cite{huckel1923theory}. If secondly the construction of the potential model provides an intuitive physical picture to explain the temperature dependence of the lattice potential, it may help us interpret the otherwise "black box" results arising from the numerical simulation. The third role for model potentials can lie in opening up avenues to explore the values of the potential in parameter ranges, where lattice QCD simulations are currently unavailable, be it at finite Baryo-chemical potential or finite center of mass velocity.

Over the past two decades several studies have put forward proposals on how to construct appropriate analytic parameterizations of the interquark potential at finite temperature. Before the 2007 discovery that the potential is complex valued, the focus lay on capturing its real part. The starting point of all these models is the realization that the $T=0$ potential is very well described by the Cornell form 
\begin{equation}
\label{eq:cornell_pot}
V^{\mathrm{vac}}(r)=-\frac{\alpha_S}{r}+\sigma r + c,
\end{equation}
with \(\alpha_S=C_F\,g^2/4\pi\) the (running) strong coupling constant defined in accord with the phenomenology literature. \(\sigma\) denotes the string-tension and \(c\) is an additive constant. These three parameters will have to be determined using $T=0$ lattice QCD simulations.

The first work in this context is the study by Karsch, Mehr and Satz \cite{Karsch1988}. Using arguments based on the two-dimensional Schwinger model, an exponential screening of both the Coulombic and string like part was proposed
\begin{equation}
\label{eq:first_screening}
V^{\rm KMS}(r)=\frac{\sigma}{m_D}\left(1-e^{-m_D r}\right)-\frac{\alpha_S}{r}e^{-m_D r},
\end{equation}
which depends on a single temperature dependent parameter $m_D$ and reduces to the Cornell form as $m_D\to 0$. The in-medium string part of the KMS potential has also been argued to arise from a non vanishing gluon condensate as proposed in \cite{Megias:2005ve}.

The KMS expression for the string real part may be obtained by modelling the quark antiquark interactions as being effectively one-dimensional and string like as proposed in version 1 of Ref.~\cite{Guo:2018vwy}. In addition Ref.~\cite{Satz:2015jsa} suggests that an entropic force may influence the interquark binding. Using an ad-hoc identification of the real part of the potential with thermodynamic quantities such an entropic term may be added and as shown in Ref.~\cite{Guo:2018vwy} leads to 
\begin{equation}
\label{eq:Vs_ad_hoc}
\mathrm{Re}V^{\rm SB}(r)=\frac{2\sigma}{m_D}\left(1-e^{-m_Dr}\right)- \sigma r e^{-m_D r}.
\end{equation}
Compared to the KMS result this expression shows a weaker dependence on $m_D$. I.e. using the same value for $m_D$ KMS shows a stronger deviation from the $m_D=0$ behavior. The same form of the potential had been proposed by Bazow and Strickland in Ref.~\cite{Strickland:2011aa} using a different line of reasoning. They start from the KMS potential and subtract an entropy related term, arriving at the expression in \cref{eq:Vs_ad_hoc}.

Another set of studies have exploited the fact that the concept of Gauss-law, well known from the study of classical electrodynamics, can be extended (for a derivation see Ref.~\cite{Dixit1990}) to potentials with arbitrary monomial powers in the separation distance. In Ref.~\cite{Digal:2005ht} for the first time the Gauss-law was combined with Debye H{\"u}ckel theory to describe the in-medium screening of the Cornell potential. The resulting functional form did not yet allow to capture the behavior of the color singlet free energy in lattice QCD, which at that time was taken as a proxy for ${\rm Re}[V]$. Introducing an additional free fit parameter $\kappa$ improved the agreement with numerical data, but its relation to $m_D$ remained unclear.
 
The first step toward consistently parametrizing a complex valued potential was taken in Ref.~\cite{Thakur:2013nia}. It is based on linear response theory where the in-medium electric field is computed by dividing the vacuum potential by the static dielectric constant of the medium in momentum space (see e.g. Ref.~\cite{Kapusta:2006pm})
\begin{equation}
\label{eq:linrep_p}
V(\mathbf{p})=\frac{V^{\mathrm{vac}}(\mathbf{p})}{{\varepsilon(\mathbf{p},m_D)}}.
\end{equation}
The authors propose to use the permittivity of a weakly coupled QGP obtained from HTL perturbation theory
\begin{equation}
\label{eq:perm_med}
\varepsilon^{-1}\!\left(p,m_{D}\right)=\frac{p^2}{p^2+m_D^2}-i\pi T \frac{p m_D^2}{\left(p^2+m_D^2\right)^2}.
\end{equation}
In this expression all medium effects are governed by a single temperature dependent parameter, the Debye mass $m_D$. Since $\varepsilon$ is complex valued, so will be the in-medium potential. Computing the inverse Fourier transform separately for the Coulombic and string like part and subsequently adding the two contributions leads to 
\begin{align}
\label{eq:IITR_ReV_final}
\mathrm{Re}V^{\rm TKP}(r)&=-\alpha_S m_{D}\left(\frac{e^{-m_D r}}{r}+1\right)+\frac{2\sigma}{m_D}\left(\frac{e^{-m_D r}-1}{r}+1\right),
\end{align}
\begin{align}
\label{eq:IITR_ImV_final}
\mathrm{Im}V^{\rm TKP}(r)&=-\alpha_S T \phi(m_D r)+\frac{2\sigma T}{m_D^2}\chi_0(m_D r).
\end{align}
\(\phi\) is the same function that appears in the HTL result in \cref{eq:ImPotHTL} and \(\chi_0\) is given by 
\begin{equation}
\label{eq:ImV_div}
\chi_0(x)=2\int_{0}^{\infty}\frac{\mathrm{d}z}{z\left(z^2+1\right)^2}\left(1-\frac{\sin(xz)}{xz}\right).
\end{equation}
This potential model suffers from two drawbacks. The first and most critical is the fact that the in-medium real part arising from the string-like contribution of the Cornell potential exhibits an unscreened $1/r$ term, which is both counter intuitive and also does not reflect the actual behavior found in the lattice QCD potential. It appears that the linear response relation alone does not yet self consistently implement screening. The second issue is related to the fact that the imaginary part arising from the vacuum string diverges logarithmically. As we will discuss below this can be remedied by implementing string breaking. 

Another proposal using direct Fourier transforms was put forward in version 2 of \cite{Guo:2018vwy}. The authors add a term to the perturbative gluon propagator, which at $T=0$ behaves as $\propto p^{-4}$. At $T>0$ it is washed out in momentum space by appropriately placed factors of $m_D$. In vacuum this implements the Cornell potential ansatz. At finite temperature it represents an independent way of incorporating medium effects into the string part of the potential. In practice this setup recovers the KMS real-part and leads in addition to a modified imaginary part, which remains finite in the large distance limit.

In order to implement screening based on the HTL permittivity in a self consistent fashion Ref.~\cite{Burnier:2015nsa} proposed to return to the generalized Gauss law, combining the ideas of  the Debye H{\"u}ckel theory from \cite{Digal:2005ht} with the use of the HTL permittivity from \cite{Thakur:2013nia}. At that time the authors however were not able to solve the resulting differential equations without introducing an ad-hoc assumption about the form of the string part at finite temperature. Only recently Ref.~\cite{Lafferty:2019jpr} succeeded in that respect, starting from the generalized Gauss law
\begin{equation}
\label{eq:gauss_gen}
\nabla\cdot\left(\frac{\mathbf{E}^{\mathrm{vac}}}{r^{a+1}}\right)=4\pi q\delta\!\left(\mathbf{r}\right),
\end{equation}
applicable to electric fields \(\mathbf{E}^{\mathrm{vac}}\left(r\right)=-\nabla V^{\mathrm{vac}}\!\left(r\right)=qr^{a-1}\hat{r}\). For \(a=-1, q=\alpha_S\) it reduces to the standard Gauss law for the Coulomb potential, but also accommodates a linearly rinsing potential via \(a=1, q=\sigma\). The expression for general powers of the separation distance \(a\) in terms of the Gauss-law operator \(\mathcal{G}_a\) reads
\begin{equation}
\label{eq:gaus_gen_a}
\mathcal{G}^{\rm vac}_a[V] = -\frac{1}{r^{a+1}}\nabla^2V^{\mathrm{vac}}(r)+\frac{1+a}{r^{a+2}}\nabla V^{\mathrm{vac}}(r)=4\pi q\delta\!\left(\mathbf{r}\right),
\end{equation} 
The decisive step is to apply the Gauss law operator to the linear response relation of \cref{eq:linrep_p} 
\begin{align}
\label{eq:gauss_apply}
\mathcal{G}_a\left[V(r)\right]=\mathcal{G}_a \int d^3y \left(V^\mathrm{vac}(r-y)\varepsilon^{-1}(y)\right) =4\pi q\left(\delta*\varepsilon^{-1}\right)(r)=4\pi q\;\varepsilon^{-1}(r,m_D).
\end{align}
The second equal sign follow from the fact that the convolution commutes with \(\mathcal{G}_a\). Considering the Coulombic and string part separately one thus obtains two sets of differential equations for the in-medium potential
\begin{align}
-\nabla^2 V_C(r) &= 4\pi\alpha_S\;\varepsilon^{-1}(r,m_D),\label{eq:Gauss_coulomb_DL} \\[1.5ex]
-\frac{1}{r^2}\frac{\mathrm{d}^2V_S(r)}{\mathrm{d}r^2}&=4\pi\sigma\;\varepsilon^{-1}(r,m_D). \label{eq:Gauss_string_DL}
\end{align}
At this point the coordinate space expression for the HTL permittivity is needed, which may be computed from the explicit expression \cref{eq:perm_med}
\begin{align}
\label{eq:perm_med_re}
&\mathrm{Re}\;\varepsilon^{-1}\!\left(r,m_{D}\right)=-\frac{m_D^2e^{-m_D r}}{4\pi r},\\
\label{eq:perm_med_im}&\mathrm{Im}\;\varepsilon^{-1}\!\left(r,m_{D}\right)=-\frac{m_D T}{4r\sqrt{\pi}}\MeijerG[\Bigg]{2}{1}{1}{3}{-\frac{1}{2}}{-\frac{1}{2},-\frac{1}{2},0}{\frac{1}{4}m_D^2r^2}.
\end{align}
Using \cref{eq:perm_med_re,eq:perm_med_im} in \cref{eq:Gauss_coulomb_DL} reproduces the original HTL result \cite{Laine:2006ns}
\begin{align}
&\mathrm{Re}V_C\!\left(r\right)=-\alpha_S\left[m_D+\frac{e^{-m_Dr}}{r} \right], \label{eq:DL_ReVc} \\
&\mathrm{Im}V_C\!\left(r\right)=-\alpha_S\left[iT\phi(m_D r) \right], \quad \phi(x)=2\int_{0}^{\infty}\mathrm{d}z\;\frac{z}{\left(z^2+1\right)^2}\left(1-\frac{\sin(xz)}{xz}\right), \label{eq:DL_ImVc}
\end{align}
while for the string part one obtains
\begin{align}
\mathrm{Re}V_S\!\left(r\right)&=\frac{2\sigma}{m_D}-\frac{e^{-m_D r}\left(2+m_Dr\right)\sigma}{m_D}, \label{eq:DL_ReVs} \\
\mathrm{Im}V_S\!\left(r\right)&=\frac{\sqrt{\pi}}{4}m_D T\sigma\;r^3\;\MeijerG[\Bigg]{2}{2}{2}{4}{-\frac{1}{2},-\frac{1}{2}}{\frac{1}{2},\frac{1}{2},-\frac{3}{2},-1}{\frac{1}{4}m_D^2r^2}. \label{eq:DL_ImVs_unreg}
\end{align}
Note that the string in-medium real part obtained here, takes on the same form as \cref{eq:Vs_ad_hoc}. Interestingly this means that a contribution similar to an entropic force terms arises naturally in this model setup.

Added together we arrive at the Gauss-law expression for the complex in-medium potential.
\begin{align}
\mathrm{Re}V = \mathrm{Re}V_C + \mathrm{Re}V_S + c, \quad \mathrm{Im}V= \mathrm{Im}V_C + \mathrm{Im}V_S,
\end{align}
in which the values of both ${\rm Re}[V]$ and ${\rm Im}[V]$ are governed by a single temperature dependent parameter $m_D$. 

A first inspection of the real-part of the Gauss-law parameterization shows that it smoothly recovers the Cornell potential as \(m_D\to 0\). At short distances, where temperature effects are small it also recovers the $T=0$ form. Screening at large distances is reflected in an exponential flattening of ${\rm Re}[V]\sim e^{-m_Dr}$ governed by $m_D$. Consistent with expectations, the string contribution to the real part becomes increasingly suppressed at high temperatures and large values of $m_D$, as it carries an additional factor of $1/m_D$. I.e. eventually the pure HTL result will dominate at high $T$. For the Coulombic ${\rm Im}[V]$, consistent with Landau damping, we find that it asymptotes to a constant at large distances.

Similar to what we had seen in previous attempts of constructing parametrizations of the in-medium potential, also here the naively evaluated string contribution to ${\rm Im}[V]$ diverges at large distances. Ref.~\cite{Lafferty:2019jpr} proposed that this behavior is connected with the unphysical and unregulated rise in the Cornell potential at $T=0$ and can be remedied by modeling the phenomenon of string breaking. After identifying the IR divergence which underlies the unphysical behavior, it is regularized by introducing the term \(\Delta_D=\Delta/m_D\) that effectively cuts off very small momenta, where the Debye mass not already does so
\begin{equation}
\label{eq:ImVs_reg}
\mathrm{Im}V_S(r,\Delta_D)=\frac{\sigma T}{m_D^2} \chi(m_D r,\Delta_D), \quad \chi(x)=2\int_0^{\infty}\mathrm{d}p\;\frac{2-2\cos(px)-px\sin(px)}{\sqrt{p^2+\Delta_D^2}\left(p^2+1\right)^2}.
\end{equation}
The expression for ${\rm Im}[V]$ above has been rewritten in a very similar form to the Coulombic contribution, i.e. as a temperature dependent prefactor with dimensions of energy, which is multiplied by a dimensionless momentum integral. The regularization condition is simply that this dimensionless integral shall asymptote to unity for \(r\to\infty\), as does the Coulombic one. This leads to the temperature independent value of
\begin{equation}
\label{eq:reg_param}
\Delta_D=\Delta/m_D\simeq 3.0369. 
\end{equation}
Consistent with expectation, based on this regularization, the imaginary part goes over to the pure HTL result, as temperature is increased.

Up to this point the running of the strong coupling in the $T=0$ Cornell potential has not yet been considered. Upcoming lattice QCD results on grids with very fine lattice spacing however will require to take this effect into account, as already indicated by the study of color singlet free energies in \cite{Petreczky:2019ozv}. I.e. the parameter $\alpha_S$ needs to be elevated to a function of distance $\alpha_S(r)$, which for the purposes of the Gauss-law model is written as a power series
\begin{align}
\label{eq:cornell_running}V^{\mathrm{vac}}(r)&=-\frac{\alpha_S(r)}{r}+c+\sigma r =...-\frac{\alpha_S^{(-1)}}{r^2}-\frac{\alpha_S^{(0)}}{r}+\left(c-\alpha_S^{(1)}\right)+\left(\sigma-\alpha_S^{(2)}\right)r+...\;,
\end{align}
where contributions from \(\alpha_S^{(1)}\) and \(\alpha_S^{(2)}\) can be absorbed into the already present vacuum parameters. To accommodate the terms besides \(a=-1,1\) the corresponding Gauss law operator \(\mathcal{G}_a\) needs to be considered, which leads to the following closed expression for the in-medium ${\rm Re}[V]$ and ${\rm Im}[V]$
\begin{equation}
\label{eq:gauss_gen_sol_re}
\mathrm{Re}V_a(r)=c_0+c_a\frac{r^a}{a}-\frac{q}{\left(m_D\right)^a}\left[\Gamma(a,m_D r)+\Gamma(1+a,m_D r)\right], \qquad \Gamma(s,x)=\int_x^\infty\mathrm{d}t\;t^{s-1}e^{-t},
\end{equation}
\begin{equation}
\label{eq:gauss_gen_sol_im}
\mathrm{Im}V_a(r)=c_0+\frac{1}{m_D}\left[c_a\frac{m_Dr^a}{a}-\sqrt{\pi}qr^aT\MeijerG[\Bigg]{2}{2}{2}{4}{\frac{1}{2},1-\frac{a}{2}}{\frac{3}{2},\frac{3}{2},0,-\frac{a}{2}}{\frac{1}{4}m_D^2r^2}\right].
\end{equation}
The divergence of the in-medium imaginary part for \(a\geq 1\) requires a similar regularization strategy as discussed above for the string like part.

\begin{figure}
\centering
\begin{tabular}{*{2}{m{0.35\textwidth}}} \includegraphics[scale=0.3]{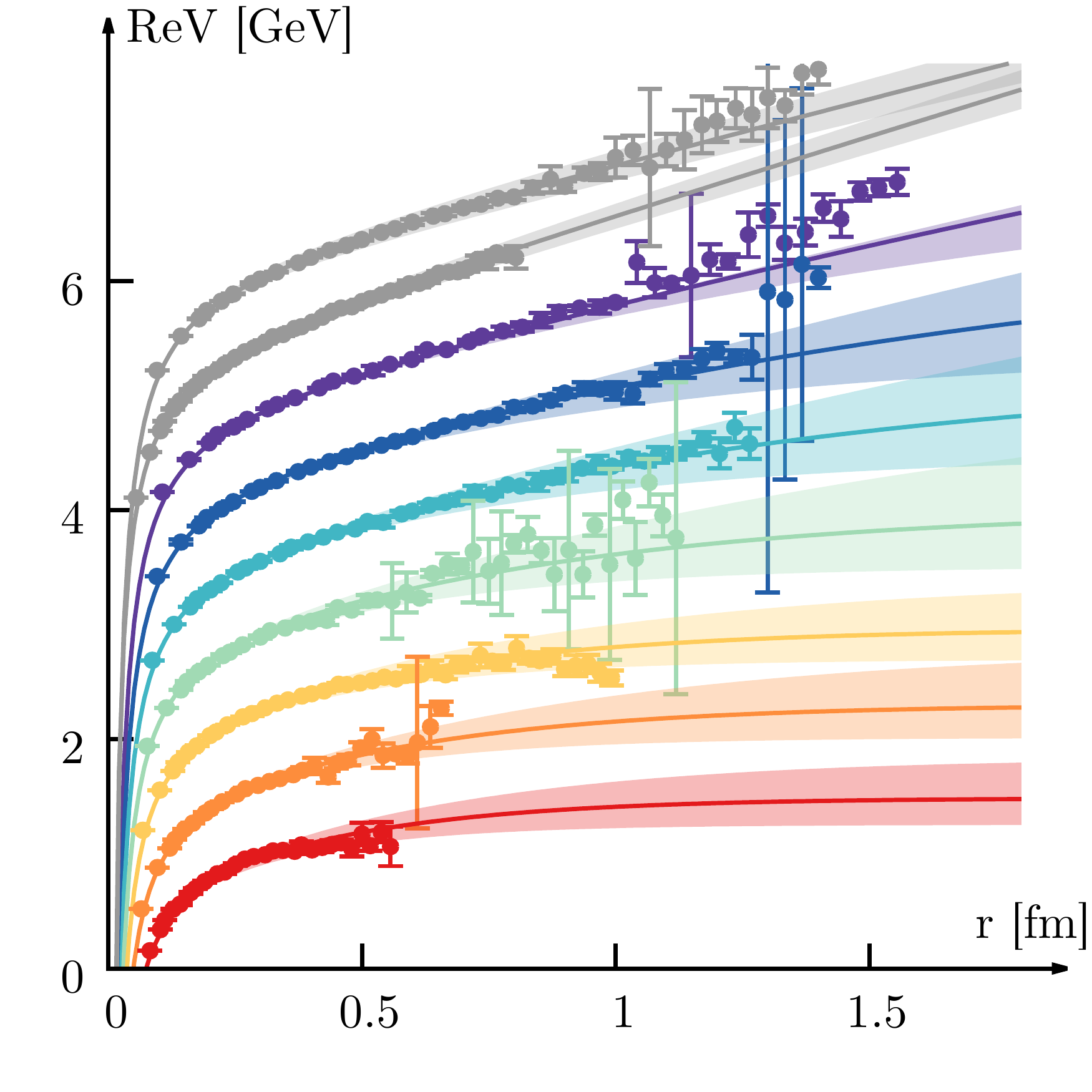}&
\includegraphics[scale=0.4]{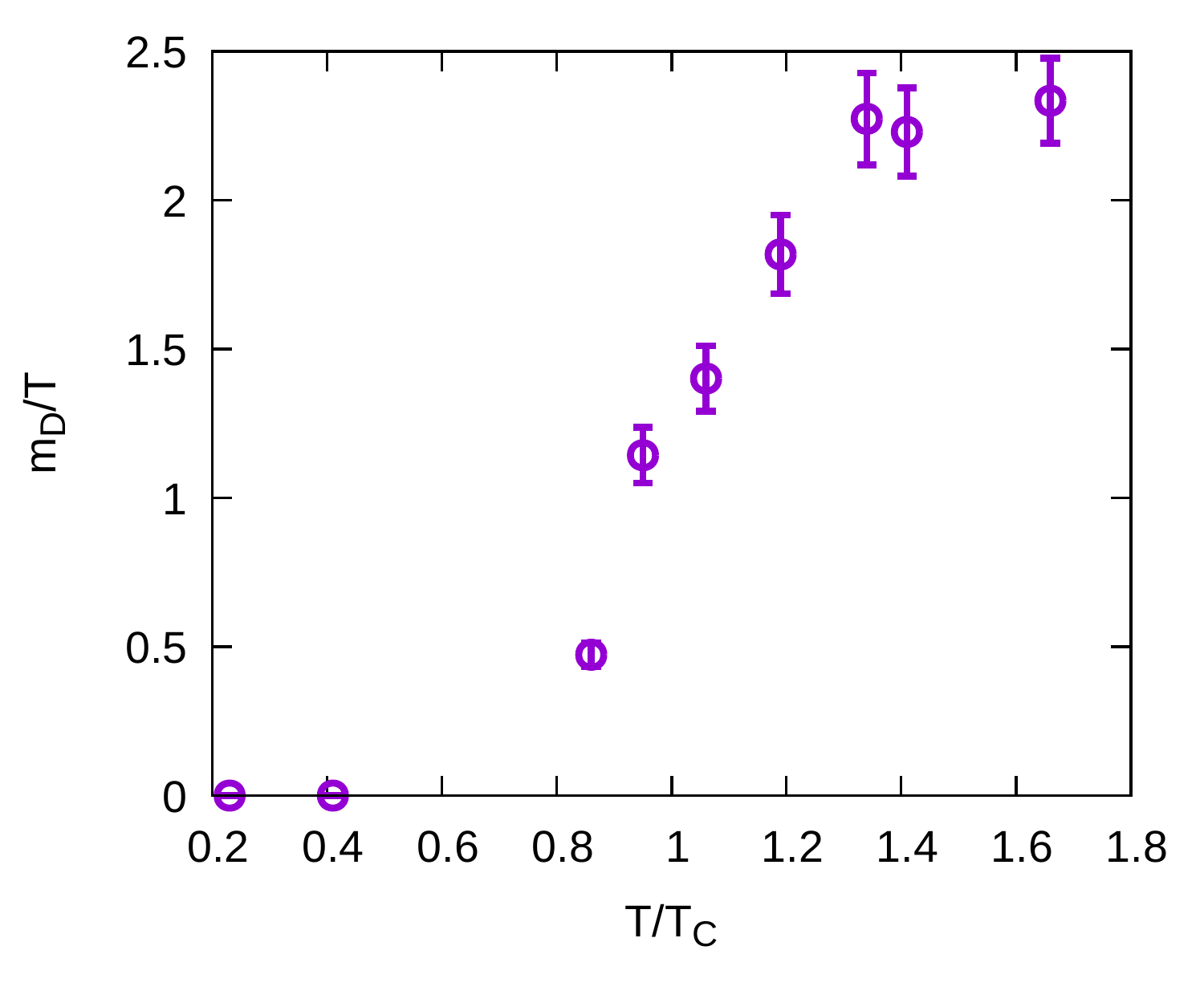} \end{tabular}
\caption{(left) The Gauss law fit of the real-part of the proper real-time interquark potential in the temperature range of $T=0.85\ldots 1.64T_C=148\ldots286$MeV (colored solid lines). Combined uncertainty from $T=0$ and $T>0$ fits shown as errorbands. The $T=0$ calibration (gray solid lines) is based on the two dataset depicted by gray symbols. (right) The corresponding best fit values of the Debye mass parameter $m_D$. Note that the crossover transition $T_C\approx174$MeV is occurs at higher than physical values on these lattices. Figures adapted from Ref.~\cite{Lafferty:2019jpr}.}\label{fig:GaussLawVet}
\end{figure}

The main benchmark for any potential parametrization is whether it is able to faithfully reproduce the non-perturbative lattice QCD data. Since the model presented here operates with a single temperature dependent parameter $m_D$ this is non-trivial. Based on the published data from \cite{Burnier:2015tda,Burnier:2014ssa}, plotted in \cref{fig:PotLoopAsqtad}, the Gauss-law parametrization presented here has shown to succeed better in this task than the proposals of Ref.~\cite{Thakur:2013nia} and Ref.~\cite{Burnier:2015nsa}. For the determination of the vacuum parameters \(\alpha_S\), \(\sigma\) and \(c\) of the Cornell potential, two low temperature ensembles were utilized. It was found that at the distance scales $0.1<r<1.2$fm available on these lattices and within the uncertainties of the reconstruction the running of the coupling is negligible. I.e. as shown in the left panel of \cref{fig:GaussLawVet} the naive Cornell ansatz reproduces the two $T=0$ datasets (gray) very well. Since the results here are not continuum extrapolated a small lattice spacing dependence of the vacuum parameters is observed.

Once the values of $\alpha_S$, $\sigma$ and $c$ are set, a fit of the single temperature dependent parameter $m_D$ allows one to reproduce the values of ${\rm Re}[V]$ equally well (solid lines). The errorbands include both the uncertainty of the $T=0$ and $T>0$ fits. interestingly \cref{eq:DL_ReVc,eq:DL_ReVs} describe the functional form of the real-part not only at asymptotically large distances but also in the intermediate regime, where a remnant of the confining behavior persists even above $T_C$. An inspection by eye previously revealed a smooth transition of ${\rm Re}[V]$ from Cornell to Debye screened behavior in \cref{fig:PotLoopAsqtad}. Consistently a finite value of $m_D$ is observed even in the hadronic phase. The best fit values are plotted in the right panel of \cref{fig:GaussLawVet}. Note that on the lattices considered here the critical temperature lies at around $T_C\approx 174$MeV.

The $m_D$ parameter introduced in the Gauss-law parametrization is of phenomenological origin and not equally rigorously related to the concept of screening, as e.g. the screening mass $m_E$ defined in \cref{sec:screeningQCD}. It simply attempts to summarize the in-medium modification as observed in the numerically determined lattice potential but should reduce to the perturbatively defined Debye mass at high temperatures (where the Gauss-law parametrization goes over into the HTL potential). Bearing these differences, as well as the fact that the results are not continuum extrapolated in mind, we may compare $m_D$ to the NLO prediction from perturbation theory as well as the values extracted from both singlet free energies and the electric correlator $C_{E-}$. What we find is that approaching $T_C$ from above, the Gauss-law parameter $m_D/T$ takes on smaller and smaller values, consistent with the fact that ${\rm Re}[V]$ eventually approaches the Cornell form within the hadronic phase. A similar downward trend is not observed in neither $m_S$ nor $m_E$ so far. At temperatures well above $T_C$ the ratio here takes on slightly smaller values than those arising in $m_S$ or $m_E$ but still is larger than the NLO perturbative prediction.

\begin{figure}
\centering
\includegraphics[scale=0.3]{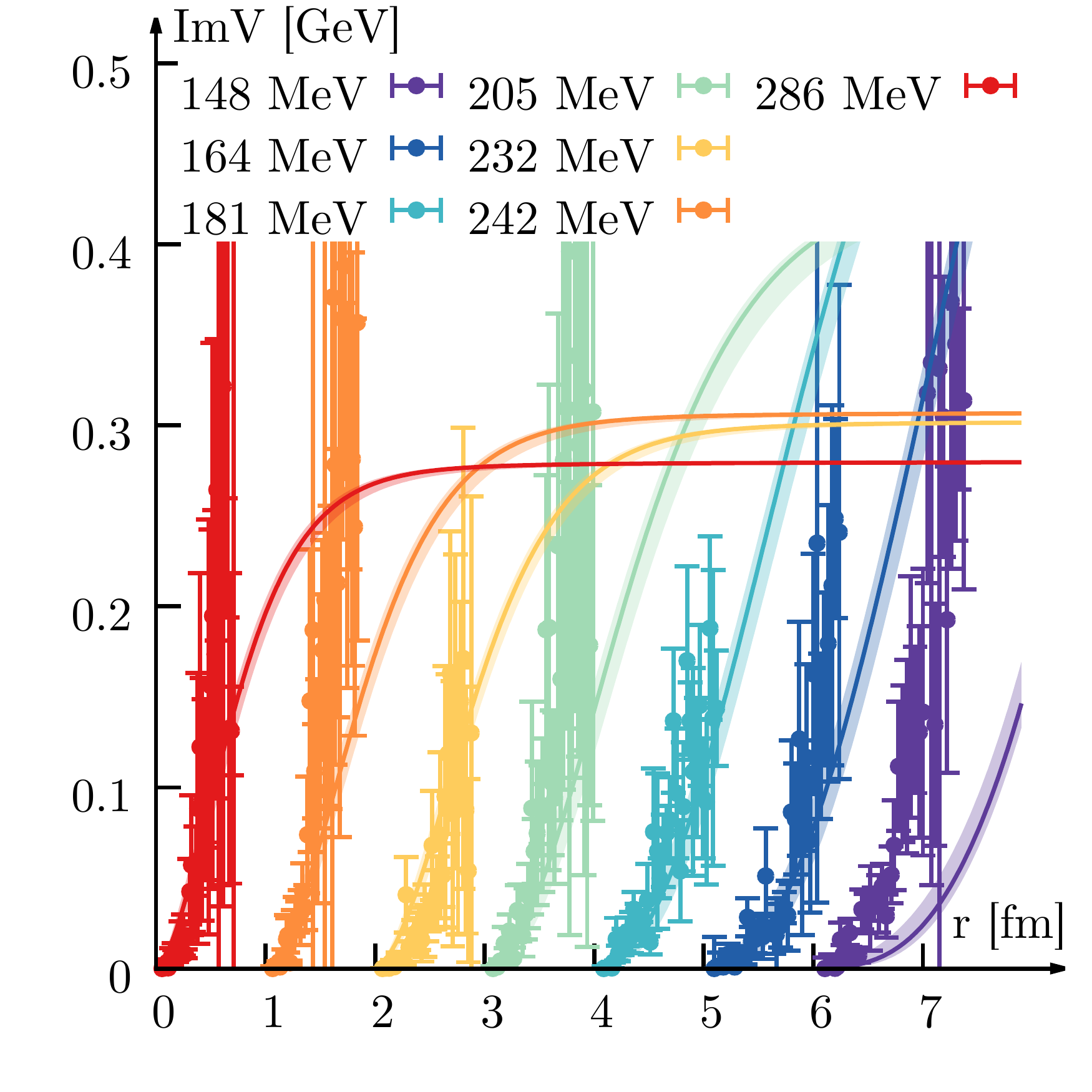}
\caption{The Gauss-law prediction for ${\rm Im}[V]$ (solid lines) from fits of the $m_D$ parameter to ${\rm Re}[V]$. The tentative values of the imaginary part from lattice QCD simulations with the asqtad action are given as colored symbols. Figure adapted from Ref.~\cite{Lafferty:2019jpr}}\label{fig:GaussLawVetIm}
\end{figure}

Since both ${\rm Re}[V]$ as well as ${\rm Im}[V]$ are controlled by $m_D$, the Gauss-law parametrization enables an additional sanity check. Once the values of the real-part have been fitted by tuning $m_D$ we can compute a prediction for the imaginary part. As shown in \cref{fig:GaussLawVetIm}, a (surprisingly) good agreement between the prediction of \cref{eq:DL_ImVc,eq:ImVs_reg} (solid lines) and the tentative values of the imaginary part from the lattice (colored symbols) down to temperatures well within the hadronic phase is indeed found. 

After all it appears that the Gauss-law parametrization provides a efficient prescription to summarize the in-medium behavior of the non-perturbative in-medium heavy quark potential based on three vacuum parameters as well as the temperature dependent Debye mass $m_D$.

\subsubsection*{Non effective field theory definitions of the potential}
\label{sec:NonEFTPot}

So far we have focused on the heavy quark potential defined in the context of non-relativistic effective field theories of QCD. As discussed in \cref{sec:EFTs} the appeal of using such EFTs lies in their systematic power counting, which allows well controlled matching to QCD using expansions in the heavy quark mass (HQET) or heavy quark velocity (NRQCD). The static potential $V^{(0)}_S$ represents the lowest order Wilson coefficient in this sense for pNRQCD. 

On the other hand there are two other types of potentials found in the literature, which we will briefly discuss in the following. The first utilizes a relativistic notion of a potential based on the concept of the Bethe-Salpeter (BS) equation and the corresponding BS wavefunction. The second class defines the potential from the propagation of the time evolution of an analog of heavy quarks using the AdS/CFT conjecture.

The Bethe-Salpeter equation has been developed to treat the scattering, as well as bound state formation of two particles in a manifest relativistic fashion (for a review see e.g. \cite{nakanishi:1969}). The starting point is the time ordered four-point function $D=\langle \Omega | \phi_1 \phi_2\phi_3\phi_4 | \Omega \rangle$ of the involved fields in vacuum $|\Omega\rangle$. It represents the transition amplitude of going from two asymptotic incoming particles with relative four-momentum $p$ and total four momentum $P$ to two asymptotic outgoing ones with relative momentum $q$. Rearranging its perturbative expansion yields an implicit relation
\begin{align}
D(p,q;P)=G(\frac{1}{2}P+p)G(\frac{1}{2}P-p)\delta^{(4)}(p-q) + \int d^4k V(p,k;P)D(k,q;P)G(\frac{1}{2}P+p)G(\frac{1}{2}P-p),\label{eq:BetheSalpeter}
\end{align}
where the sum of all irreducible two-particle diagrams has already been suggestively assigned the letter V. $G(k,P)$ denotes the propagator of the involved fields. In order to describe the formation of a bound state $\psi$ from the particles of $\phi_1$ and $\phi_2$ the BS wavefunction $\Phi=\langle \Omega| \phi_1 \phi_2 |\psi\rangle$ is considered instead, for which a similar recursive definition as in \cref{eq:BetheSalpeter} can be derived. It then acts as the relativistic generalization of the wavefunction of the Schr\"odinger equation.

Since it is not possible to include all diagrams in $V$ in practice, a truncation needs to be employed. The ladder approximation is the most common one. There the $G(k,P)$ are taken to be the free field Feynman propagators of the underlying fields and only single particle interactions are included in $V$. In general the notion of a potential emerges if the diagrams resummed in the interaction Kernel $V$ lead to a time independent and local quantity in position space. Note that for heavy quarks such a potential would contain not only the contributions from static interactions of the $m\to\infty$ limit but also correction due to finite velocity, spin etc. that are absent in $V_S^{(0)}$. 

The BS wavefunction has been used to extract potential-like interaction kernels from non-perturbative $T=0$ lattice QCD simulations, a procedure that involves two steps of reasoning. First, if the time evolution of the BS wavefunction in an interacting theory proceeds according to a Schr\"odinger equation with interaction potential then this potential can be reverse engineered from the computed values of $\Phi$. Secondly, at $T=0$ the evolution in Minkowski time and in Euclidean time are governed by the same potential, i.e. $V$ may be read off from an appropriate imaginary time correlation function. Both in the study of nucleon-nucleon interactions \cite{Aoki:2012tk} (HAL-QCD collaboration) and for charmonium in vacuum this strategy has been deployed \cite{Nochi:2016wqg,Kawanai:2013aca,Kawanai:2011xb,Kawanai:2011jt,Ikeda:2011bs}.

At finite temperature the so called T-matrix approach has been developed to describe both quarkonium and open heavy flavor particles (for a recent overview see \cite{Liu:2017qah}). It starts from the fact that the Bethe-Salpeter transition amplitude is related to the scattering S-matrix, whose nontrivial part is christened the T-matrix. The ladder approximation is deployed for both light and heavy degrees of freedom in the system. The interaction diagrams for the T-matrix are expressed by using the free finite temperature quark and gluon propagators $G$. In this approach the interaction kernel for both light and heavy quarks is modeled by a real-valued two-body potential that implements screening with different screening masses for the Coulombic $m_D$ and string like part $m_D^\prime$, as well as a third parameter $c_s$ to take into account string breaking effects
\begin{align}
V^{\rm RL}=-\alpha_S\frac{e^{-m_D r}}{r}-\frac{\sigma}{m_D^\prime}e^{-m_D^\prime r - (c_s m_D^\prime r)^2}
\end{align}
Note that starting with a real-valued microscopic potential here does not preclude the self energies computed via the T-matrix from becoming complex. In turn scattering effects are included in the evolution of meson-meson correlators and one may still find an imaginary part in the potential governing the evolution of the medium averaged quantities.

To determine the values of $V^{\rm RL}$ the authors of Ref.~\cite{Liu:2017qah} computed several quantities, such as the free energy differences  in the presence of heavy quark pairs in the T-matrix approach and compared those to first-principle simulations on the lattice. In addition they compared their computations to open-heavy flavor \cite{Liu:2018syc} and bottomonium observables \cite{Du:2019tjf}. While the output of the iterative procedure to obtain $V^{\rm RL}$ depended on the initialization  according to a weak- or strong-coupling potential ansatz, it shows consistently larger values than those found for $V^{(0)}_S$ directly on the lattice. Since $V^{\rm RL}$ in the T-matrix approach governs not only the interactions of heavy quarks but also those of the light d.o.f. it is actually surprising that not larger differences compared to the static EFT potential are observed.

There have been proposals put forward in Refs.~\cite{Iida:2011za,Allton:2015ora} to utilize the Euclidean time BS wavefunction at finite temperature to reverse engineer a potential for in-medium quarkonium. While straight forward in implementation, this strategy suffers from the absence of an equally rigorous theoretical connection between the imaginary time and real-time BS wavefunction as at $T=0$.

For completeness let us also mention another more exploratory proposal to determine the static interquark potential non-perturbatively. In a two step process first the real-time gluon propagator is determined from the lattice in Landau gauge and subsequently Fourier transformed to obtain the in-medium potential. A first implementation at $T=0$ has been presented in Ref.~\cite{Serenone:2015qra} and a finite temperature generalization could make use of gluon spectral function results presented e.g. in \cite{Ilgenfritz:2017kkp}.

Instead of staying within QCD and being limited to the Euclidean domain for first principles computations, one may go over to QCD-like systems, where strong-coupling computations can be carried out directly in real-time. One such strategy to approximate the physics of QCD at finite temperature and density is based on the AdS/CFT conjecture (for a review see \cite{CasalderreySolana:2011us}). Roughly speaking a pair of static test charges propagating in four-dimensional space-time is interacting via a string, which spans between the constituents and which extends into a fifth dimensions, called the bulk. Finite temperature fluctuations are implemented via the presence of a Hawking-radiating black-hole in that extra dimension. At small t'Hooft coupling quarkonium in AdS/CFT is expected to be hydrogenlike as discussed in Ref.~\cite{Hong:2003jm}, so the potential at small distances will be dominated by a Coulombic behavior. At large coupling on the other hand the behavior is more involved. It is found in Ref.~\cite{Karch:2002xe} that even though the Wilson loop does not exhibit an area law, i.e. confinement is absent in the traditional sense, the asymptotic states of the theory are color neutral with the color charge being screened. The latter fact is understood from a process in analogy with string breaking. Separating two quarks analogues far enough apart breaks the string between them and leads to the generation of a light quark analogue pair, which then compensates for the color charge of the heavy color sources.

The propagation of the static pair of test charges can be on the one hand described by the real-time Wilson loop and on the other hand may be related to the Nambu-Goto action for the connecting string (the first works in this regard being Refs.~\cite{Rey:1998bq,Brandhuber:1998bs}). The further the charges are spatially separated the deeper the string extends into the bulk, approaching the horizon of the black hole. Following e.g. Refs.~\cite{Noronha:2009da,Hayata:2012rw,Ewerz:2016zsx}, one finds that the finite temperature potential remains purely real until at a threshold distance, an imaginary part abruptly sets in arises. At this point the string has not yet touched the black hole horizon. Once it does, the computations as of yet cannot be meaningfully continued to further separation distances. For details on the involved difficulties related to identifying the appropriate string configurations contributing to the evolution of the Wilson loop see \cite{Ewerz:2016zsx} and references therein.

The decisive advantage of the AdS/CFT computations is that they are carried out directly in a real-time setting in the four-dimensional gauge theory. This gives access to the values of the AdS/CFT counterpart of the QCD static potential defined from the late-time behavior of the real-time Wilson loop. Comapring the definition of the proper real-time potential $V^{(0)}_S$ and the free energies $F_1$ it has to be stated that what is computed in e.g. Ref.~\cite{Ewerz:2016zsx} and previous studies is not the free energies but the real-time potential itself. This fact is not correctly reflected in the title of these studies. An important contribution to the field was made in Ref.~\cite{Ewerz:2016zsx} as the correct renormalization of the potential has been addressed. It leads to a ${\rm Re}[V]$, which, as physically expected, reduces to the $T=0$ result at small separation distances at any temperature of the thermal medium. Prior to this study a temperature dependent renormalization scheme had been deployed. One current challenge lies in extending the results for the potential in the original AdS/CFT framework to geometries, in which the conformal symmetry is broken, in order to accommodate more realistically the behavior of QCD.

\begin{summary}
The proper real-time potential between static quarks at finite temperature in general takes on complex values and its ${\rm Re}[V]$ shows Debye screening at high temperature. The potential can be extracted non-perturbatively from imaginary time simulations by inspecting the spectral functions $\rho_\square$ of the Wilson loop $W_\square$. If a well defined lowest lying spectral feature is present and takes on the shape of a skewed Breit-Wigner, its position and width encode the real- and imaginary part of $V^{(0)}_S$ respectively. Using improved Bayesian spectral reconstruction methods (BR method) as well as the Pade approximation on Wilson line correlators in Coulomb gauge, first estimates of ${\rm Re}[V]$ and ${\rm Im}[V]$ in quenched and dynamical QCD with $N_f=2+1$ light flavors have been obtained. In both cases well defined spectral peaks have been observed at all temperatures investigated. A transition from a Cornell to a Debye screened behavior is found in ${\rm Re}[V]$ and as expected it proceeds relatively abruptly in the quenched and smoothly in the dynamical case. All studies so far report indications of a finite imaginary part for $T>T_C$. The in-medium behavior of the potential can be well reproduced by means of simple potential models. One successful example is the Gauss-law parametrization. Combining self consistently the Cornell ansatz at $T=0$ with the HTL medium permittivity it reproduces the $T$ and $r$ dependence of the lattice data using a single temperature dependent fit parameter $m_D$, identified with the Debye mass in the high temperature limit. $m_D/T$ on the lattice is found to approach zero shortly below $T_C$ and at $T>T_C$ takes on larger values than predicted by NLO perturbation theory. Alternative non-EFT potentials based on the Bethe-Salpeter equation, the T-matrix approach and the AdS/CFT correspondence have been studied in detail in the literature.
\end{summary}

\subsection{Quarkonium in-medium properties}
\label{sec:QQbarequilprop}

Up to this point we have investigated quark binding and screening of color fields in the context of static color sources. With these preparations at hand, we take the next step in this section, and shed light on the in-medium properties of realistic quarkonium with finite mass. To this end we will investigate quarkonium spectral functions, comparing those at $T=0$ with those at $T>0$ so that changes in their peak structures reveal the influence that the medium exerts. This section is divided into four subsections presenting computations of heavy quarkonium spectral functions according to the involved compromise between accuracy and precision.

We will start out with quarkonium formulated fully relativistically in lattice QCD. This approach does not entail any truncations and in principle is able to reproduce in-medium quarkonium properties highly accurately. We will see however that the numerical effort required to extract spectral functions is extremely high and has so far limited the quantitative insight into in-medium spectral properties of individual states. To obtain an understanding of the overall in-medium modification among quarkonium states with the same quantum numbers, the Euclidean time correlation functions will also be studied. 

The second approach we consider is the effective field theory NRQCD discretized on the lattice. While still challenging, the extraction of spectral functions from its meson current correlation functions is less demanding than in full QCD and first quantitatively robust determinations of in-medium ground state properties indeed have been achieved. The price to pay is that the physics of the heavy quarks is only captured up to the order in which the NRQCD expansion is implemented numerically.

The third approach uses the static in-medium potential defined from the effective field theory pNRQCD evaluated non-perturbatively on the lattice. Combining the static potential with a kinetic term with a finite mass, the Schr\"odinger equation governing the real-time point split forward meson correlator is computed. Taking the imaginary part of its Fourier transform after removing the spatial splitting yields an approximation of the meson spectral function. This approach provides the most precise determination of in-medium spectral functions but misses so far any corrections of the in-medium potential according to finite velocity and spin.

In the fourth subsection we will briefly survey results on quarkonium in-medium spectral functions that are not based on lattice QCD or effective field theory. These include QCD sum rules, the T-matrix approach and the AdS/CFT correspondence.

\subsubsection*{In-medium quarkonium from relativistic lattice QCD}

Relativistic meson current correlators in Euclidean time \footnote{Some works also consider spatial correlation functions along a single axis. These are related to the spectral function via a Fourier transform 
\begin{align}
D^{\rm s}_M(x_3,\mathbf{p}_\perp,\omega_n)=\sum_{x_1,x_2,\tau}{\rm exp}[-i\mathbf{p}_\perp (x_1,x_2)-i\omega_n \tau] D_E(\mathbf{x},\tau)
\end{align}
and their exponential decay at large spatial separation distances is connected to the concept of spatial {\it hadronic screening mass}, which is helpful in investigations of symmetry properties of finite temperature QCD (see e.g. Refs.~\cite{Ding:2012pt,Brandt:2015sxa}). We will not consider these correlators further here.} that can be simulated in lattice QCD at finite temperature in principle contain a wealth of vital information on in-medium heavy quarks. On the one hand they encode the in-medium quarkonium spectral functions, whose peak structures inform us about the changes in mass and lifetime of each individual state. On the other hand, whenever a correlator is related to a conserved charge, in addition, the spectral function may contain a so called transport peak, describing the propagation of that conserved charge over large distance scales. This is the case for the vector, scalar and axial vector channel. While the vector channel $\Gamma=\gamma_i$ is related to the approximately conserved heavy flavor quark number, the scalar channel with $\Gamma=1$ is related to the fermionic contribution to the trace anomaly. 

The challenge for lattice QCD lies in the fact that the discretization artifacts for heavy quarks go in powers of $a_s m_Q$, requiring extremely finely spaced grids for the study of bottomonium. So far, most dynamical QCD  studies have thus focussed on charmonoium, while bottomonium is treated mainly in the quenched approximation, where large enough lattices can be efficiently simulated. A wealth of studies has scrutinized correlation functions and spectra over the past two decades. The first spectral reconstructions in this context have been pioneered in Ref.~\cite{Asakawa:2003re}. On the one hand there are those works that remain in the quenched approximation (see e.g. Refs.~\cite{Umeda:2002ur,Datta:2003ww,Jakovac:2006sf,Ding:2012sp,Ikeda:2016czj}), which allows them to deploy large grids and recently even perform continuum extrapolations (see Ref.~\cite{Burnier:2017bod}). On the other hand heavy quarkonium correlators and spectra have been investigated using isotropic and anisotropic dynamical QCD scenarios in Refs.~\cite{Aarts:2007pk,Borsanyi:2014vka,Kelly:2018hsi}. All studies at $T>0$ so far deploy a form of improved Wilson fermions to discretize the heavy d.o.f. The extraction of spectral functions from relativistic correlators is particularly challenging, since the input data are symmetric around $\tau=1/2T$ and thus only half the points simulated on the lattice provide independent information. The presence of a transport contributions leads to an additional challenge. As was first pointed out by Ref.~\cite{Umeda:2007hy}, the physics of the transport peak and the in-medium modification of the bound state spectra become intertwined in the correlator and particular care is needed to disentangle them. 

To gain insight into the in-medium properties without having to deal with the additional uncertatinty from a spectral reconstruction one starts with an inspection of the imaginary time correlators themselves. Naively one might expect that the ratio of the in-medium correlation function and the vacuum correlation function is a good quantity in this regard. In the relativistic formulation however the current correlator according to \cref{eq:Euclrep} contains two sources of temperature dependence. One arises from the kernel and is not informative, the other is from changes in the spectral function which is our main focus. To eliminate the effect of the temperature dependence of the kernel in ratios of correlators, according to Ref.\cite{Datta:2003ww,Ding:2012sp} one may construct the so called {\it reconstructed correlator}
\begin{align}
D_{\rm rec}(\tau, T, T^\prime)= \sum_{\tau^\prime/a=\tau/a, \Delta \tau^\prime/a = N_\tau}^{N_\tau^\prime-N_\tau+\tau/a}  D_E(\tau^\prime, T^\prime).
\end{align}
Here $T^\prime$, $\tau^\prime$ and $N_\tau^\prime$ denote the temperature, imaginary time and Euclidean extent for the simulated correlator $D_E$. The quantity $D_{\rm rec}$ now represents the Euclidean correlator where the same spectral function underlying $D_E$ is encoded with a kernel at $T$ and correspondingly $N_\tau$. A ratio of unity between the in-medium correlator $D_E(\tau,T>T_C)$ and $D_{\rm rec}(\tau,T,T^\prime <T_C)$ indicates that no medium modification is present. It has to be kept in mind however that changes in the underlying spectral function in different frequency regimes may cancel out, once the convolution with the in-medium kernel is taken. This is a manifestation of the inherent exponential information loss induced by the convolution.

Another quantity that is considered in the literature is the midpoint subtracted correlators $D_{\rm sub}(\tau)=D_E(\tau)-D(\tau=1/2T)$. If the transport contribution is of the free theory form of \cref{eq:freespecfunccont}, i.e. $\rho_{\rm transp}(\omega)=\omega\delta(\omega)$ only, then subtracting the midpoint $D(\tau=1/2T)$ from the Euclidean corelator (or taking its derivative) will remove it completely. On the other hand in an interacting theory the transport peak is expected to be of Breit-Wigner type \cite{Petreczky:2005nh} and subtractions will only remove part of its contribution.

\begin{figure}
\centering
\includegraphics[scale=0.4]{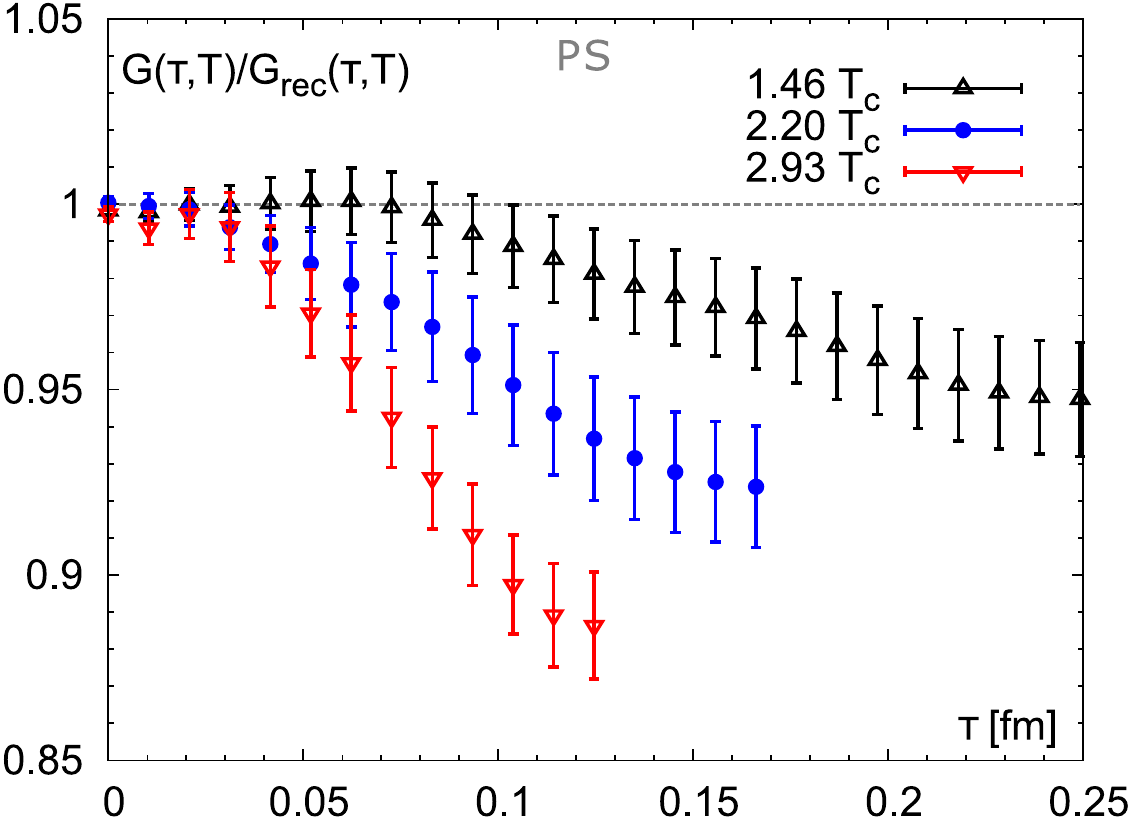}
\includegraphics[scale=0.2, clip=true, trim=0 1.5cm 0 2cm]{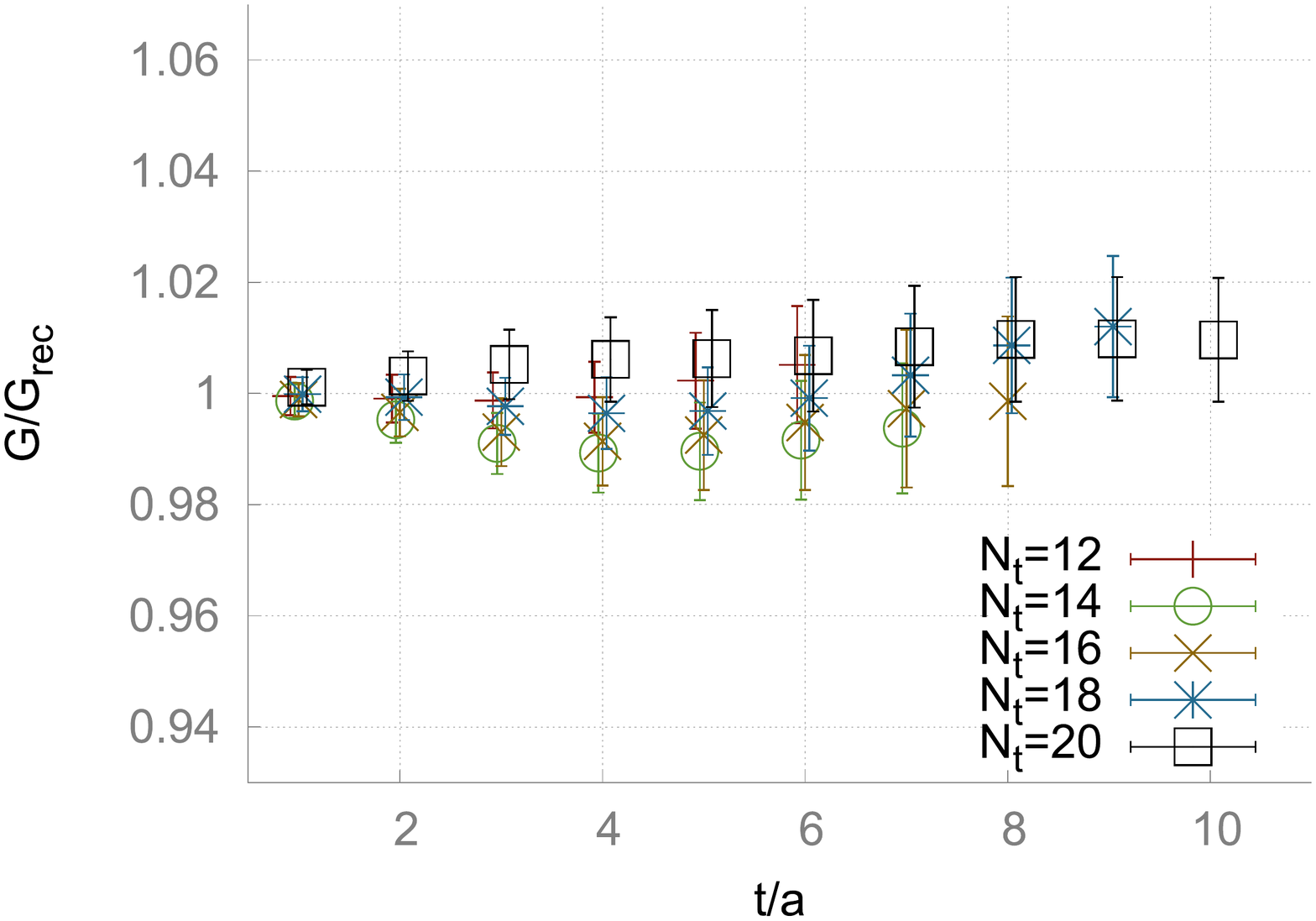}\\
\includegraphics[scale=0.4]{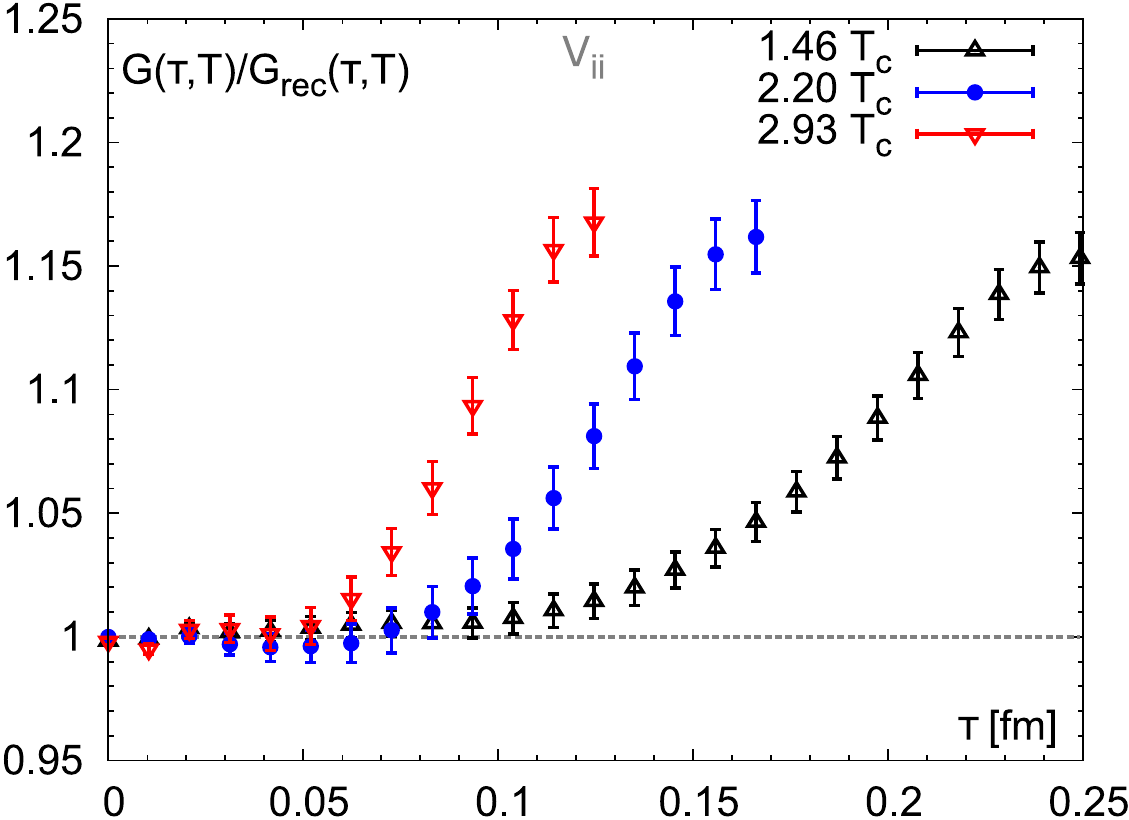}
\includegraphics[scale=0.2, clip=true, trim=0 1.5cm 0 2cm]{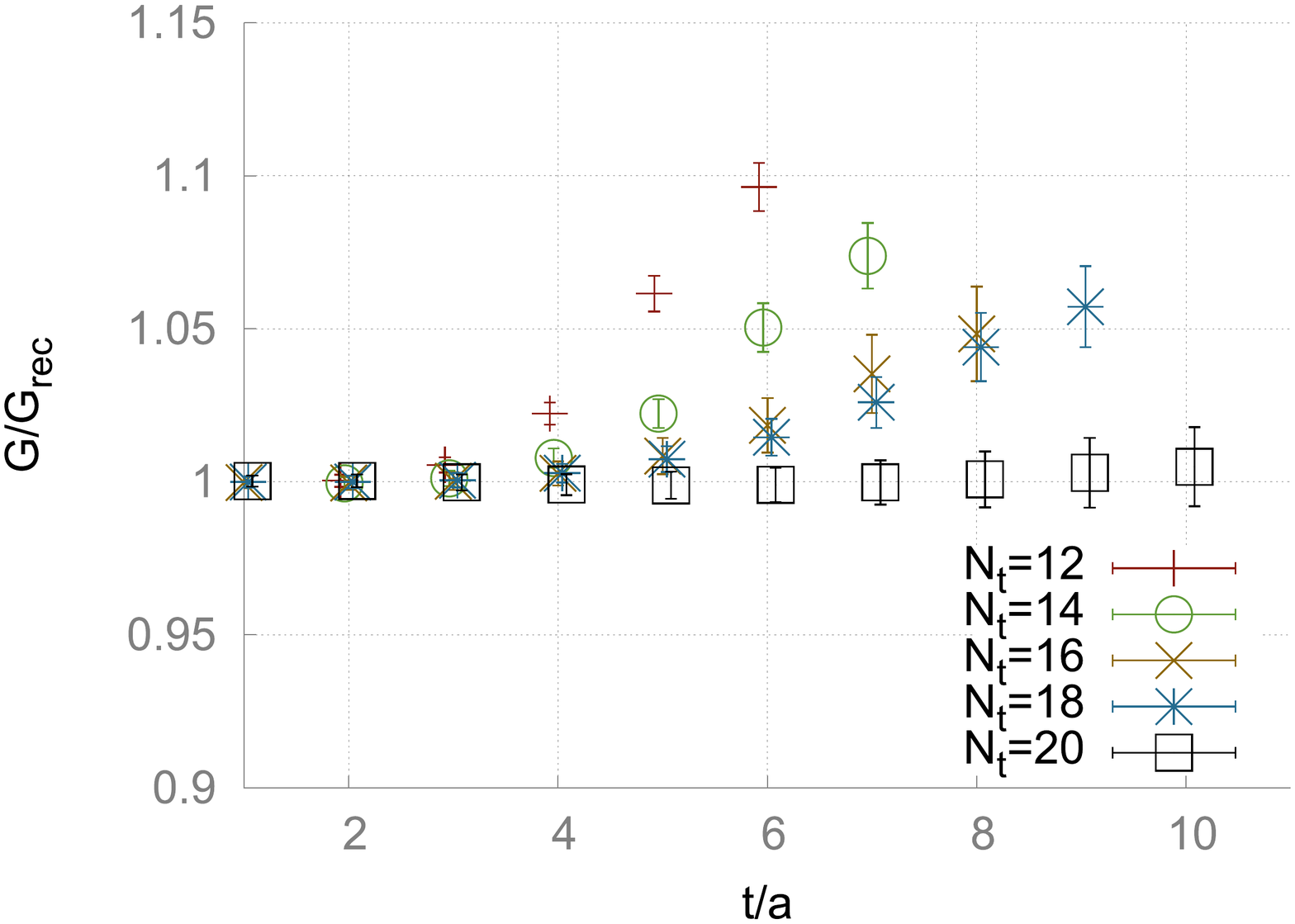}
\caption{The in-medium correlator divided by the low temperature reconstrcuted correlator for the pseudoscalar channel (top row) and the vector channel (bottom row). (left column) results from a quenched QCD simulation on large isotropic $N_s=128$ lattices with $a=0.01$fm. The comparison temperature for $G_{\rm rec}$ is $T^\prime=0.73T_C$. (right column) results from isotropic lattices with $N_f=2+1$ dynamical wilson fermions on $N_s=64$ lattices. The pion mass is relatively heavy $m_\pi=545$MeV and $T^\prime\approx 0.6T_c$. Figures adapted from Refs.~\cite{Ding:2012sp,Borsanyi:2014vka}}\label{fig:GrecRatios}
\end{figure}

In \cref{fig:GrecRatios} the results for the ratio of the in-medium and a low temperature reconstructed correlator from two representative studies of charmonium on isotropic lattices are shown. In both cases the particles are considered at rest, i.e. the correlators are summed over all spatial positions. The left column shows data from Ref.~\cite{Ding:2012sp} on large quenched QCD lattices $N_s=128$ with a fine spacing of $a=0.01$fm, where the low reference temperature is $T^\prime=0.73T_C$. On the right hand side the plots were obtained in Ref.~\cite{Borsanyi:2014vka} in simulations with $N_f=2+1$ dynamical Wilson fermions with still relatively large pion masses of $m_\pi=545$MeV on $N_s=64$ lattices. The reference temperature is $T^\prime=0.6T_C$. The top row shows the pseudoscalar channel, containing the two particles $\eta_c(1S)$ and $\eta_c(2S)$ below the $D\bar D$ threshold. The bottom row on the other hand corresponds to the vector channel, which houses $J/\psi$ and $\psi^\prime$ below threshold. We have to keep in mind that even though the $\eta_c$ and $J/\psi$ have significantly different decay widths \cite{Tanabashi:2018oca}, the latter being of keV magnitude, while in the former being on the MeV level, the spectral function underlying the correlators shown here only relate to electromagnetic decays \cite{Bodwin:1994jh}. I.e. dilepton decay $J/\psi \to \ell^-\ell^+$ and diphoton decay $\eta_s \to\gamma\gamma$, which are of the same order of keV the latter smaller than the former. Since the mass difference between the $J/\psi$ and $\eta_c$ can be understood as hyperfine splitting, it is only a few tens of MeV with $\eta_c$ being a bit more strongly bound. The excited states $\eta_c^\prime$ and $\psi^\prime$ both lie less than $100$MeV away from the $D\bar D$ threshold, i.e. their vacuum binding energies are similarly small and should be similarly affected by medium effects.

We find that the two studies show very similar behavior, which due to the large pion masses may not be surprising. For the pseudoscalars the ratios at small $\tau$ is consistent with unity and then shows a downward trend, which becomes stronger and sets in at earlier imaginary time as temperature increases. The vector channel shows the opposite behavior, i.e. while starting out at unity for small $\tau$ it bends upward. The full QCD result indicates that the bending up actually becomes stronger as as temperature rises. The deviations from unity in both cases are between $10-20$\% at the highest temperature with those for $\eta_c$ being smaller than those for $J/\psi$. Note that both computations use a fixed scale approach, which means that the UV behavior of the underlying spectral functions is the same at all temperatures.

\begin{figure}
\centering
\includegraphics[scale=0.4]{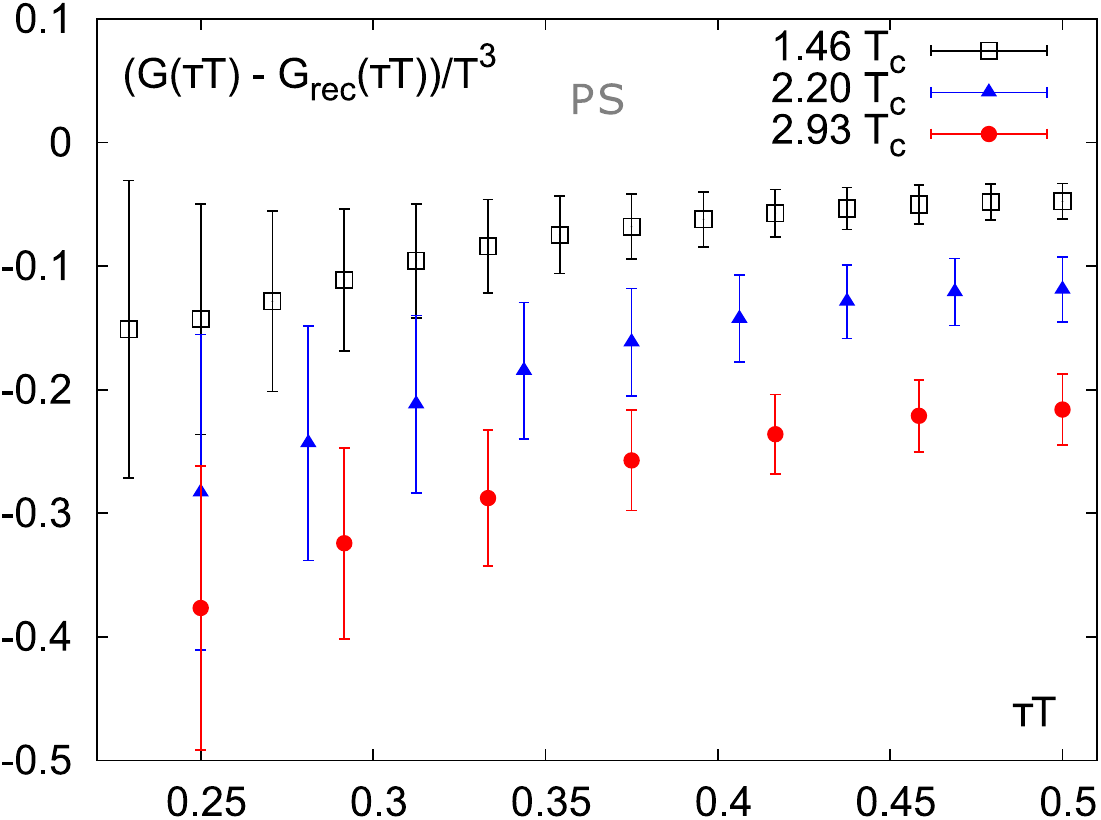}
\includegraphics[scale=0.2, clip=true, trim=0 1.5cm 0 2cm]{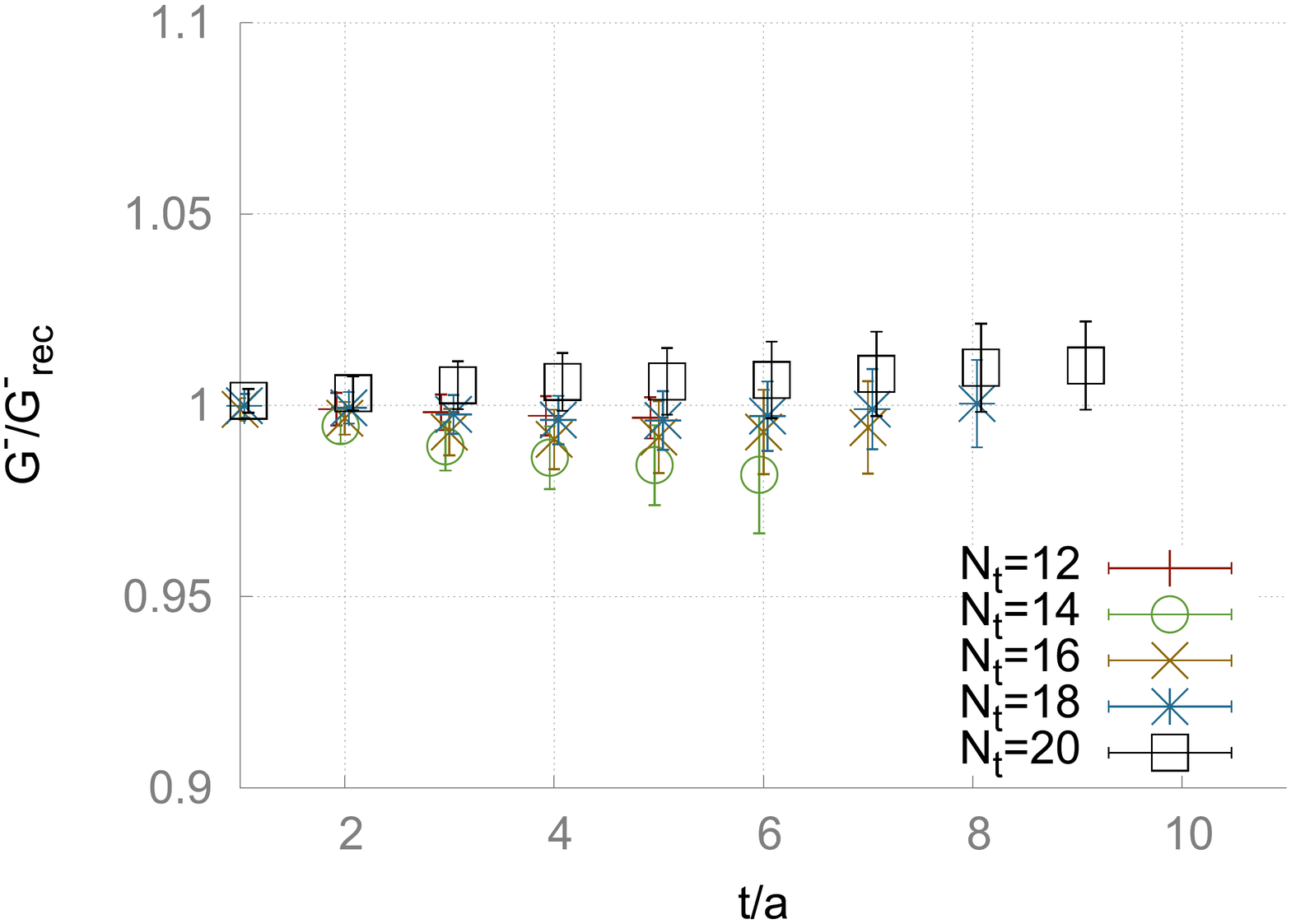}\\
\includegraphics[scale=0.4]{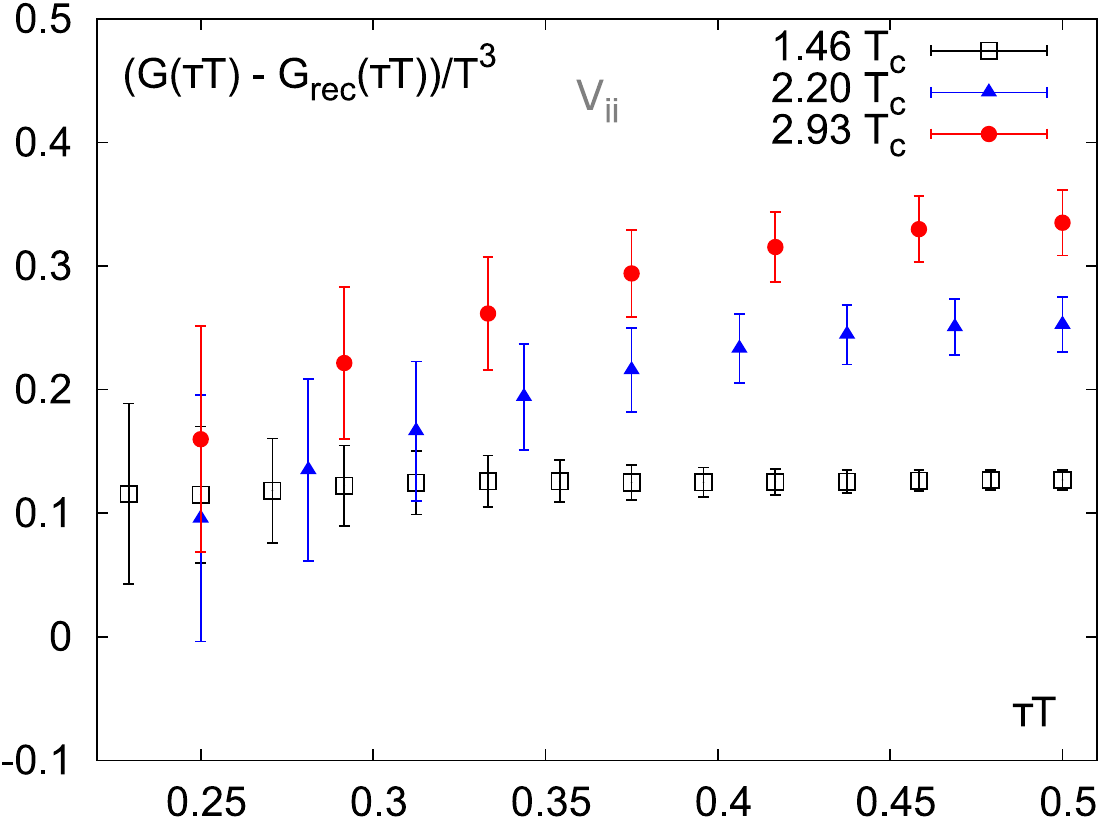}
\includegraphics[scale=0.2, clip=true, trim=0 1.5cm 0 2cm]{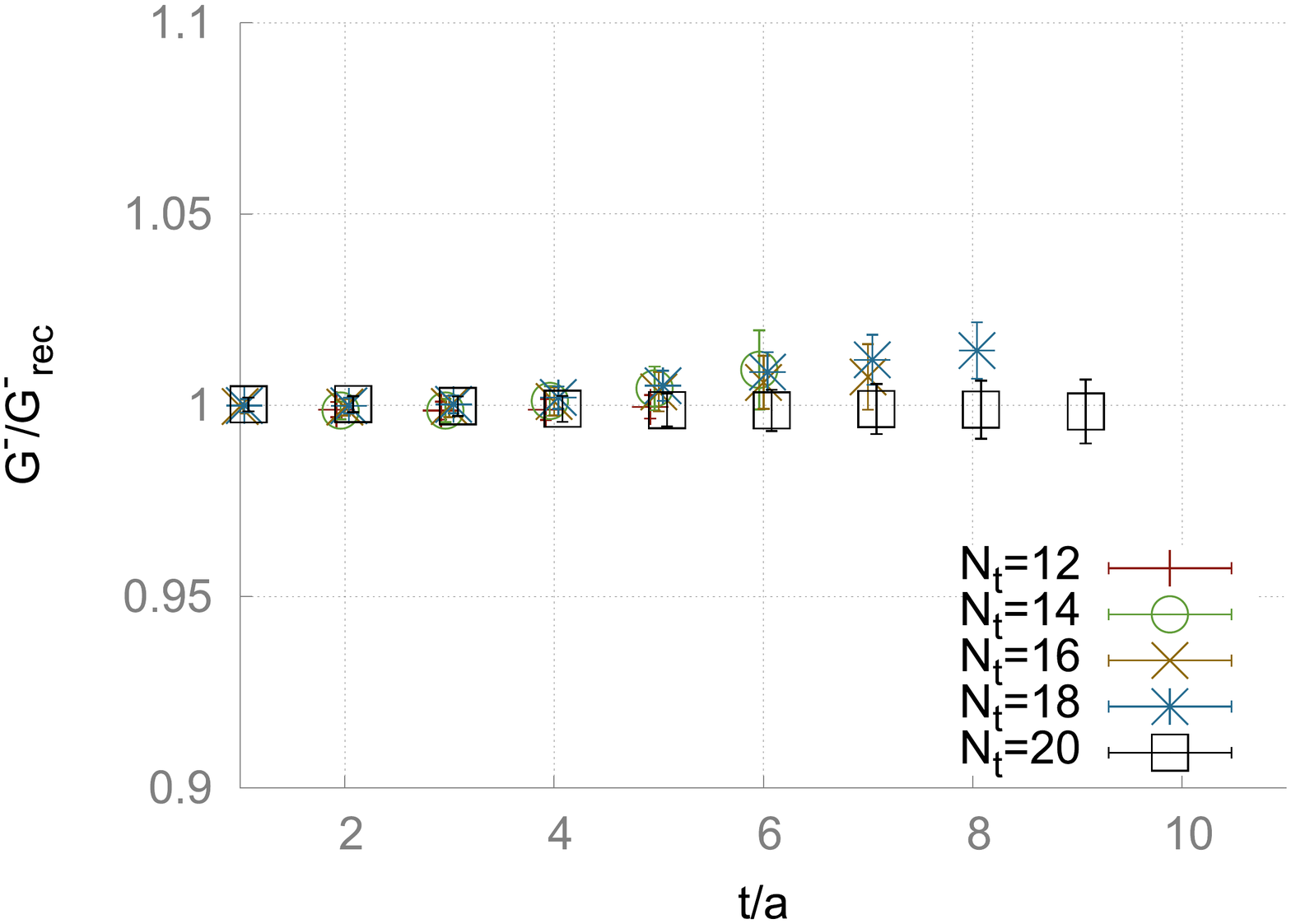}
\caption{Additional comparison of correlators to disentangle the transport contribution and in-medium modificaiton of bound states. (left) The difference between in-medium correlator and the reconstructed correlator divided by $T^3$ for the pseudoscalar (top) and vector channel (bottom) in quenched QCD. (right) The ratio between the midpoint-subtracted in-medium correlator and the midpoint subtracted reconstructed correlator for the pseudoscalar (top) and vector channel (bottom) in $N_f=2+1$ dynamical QCD. Figures adapted from Refs.~\cite{Ding:2012sp,Borsanyi:2014vka}}\label{fig:GrecSubtr}
\end{figure}

From the point of view of the vacuum bound state content alone, such a different behavior is not expected. On the other hand only the pseudoscalar channel is expected to be free from the transport contribution. This has led to attempts to remove the related transport peak from the correlator in order to showcase only the modification of the bound states. In \cref{fig:GrecSubtr} we thus show on the left the values of the in-medium correlator with the reconstructed correlator at a lower reference temperature subtracted in quenched QCD from Ref.~\cite{Ding:2012sp}. On the right the ratio of the midpoint subtracted correlator with the correspondingly subtracted reconstructed correlator is shown in $N_f=2+1$ dynamical QCD from Ref.~\cite{Borsanyi:2014vka}. 

Roughly speaking: the increasing magnitude, combined with being negative, of the difference $D_E-D_{\rm rec}$ in the pseudoscalar channel (top left) is interpreted as indicating that the amplitudes in the underlying spectral function of $D_E$ decreases. This is in agreement with expectations for a weakening bound state content. On the other hand in the vector channel the upward movement is indicative of parts of the spectrum amplitude increasing, which is interpreted as arising from an increasing transport peak. This interpretation is supported by the ratio of the midpoint subtracted correlators on the right, which shows that the difference in the $\eta_c$ channel do not change significantly compared to the unsubtracted ratio, while in the vector channel the deviation from unity is markedly reduced. 

While these above arguments are consistent, it turns out that the upward behavior observed here in the vector channel, which is attributed mainly to the transport peak will be present also in ratios of NRQCD correlators, which are free from transport contributions. In addition in a recent computation \cite{Kelly:2018hsi,Quinn:2019uwq} of charmonium correlators on anisotropic lattices with $N_f=2+1$ dynamical Wilson fermions with a lighter pion mass $m_\pi=380$MeV a much more similar behavior between pseudoscalar and vector channel ratios has been observed. It will be highly interesting to compute these ratios in the future on lattices much closer to the physical pion mass to further elucidate the question of the transport contribution.

Another way of understanding the physics encoded in a correlation function is to compare to those from a model of the underlying spectral function. For the pseudoscalar correlator of both charmonium and bottomonium this has been achieved using continuum resummed perturbation theory for the high energy region and a pNRQCD computation at the threshold. These analytic results are then compared to continuum extrapolated correlators in the quenched approximation. As discussed in detail in Ref.~\cite{Burnier:2017bod} the behavior of the pseudoscalar charmonium channel already at $T=1.1T_C$ can be reproduced using a (slightly rescaled) spectrum that does not show any peak structure besides an enhanced onset of the threshold. At the same time for bottomonium the correlator at $T=1.1T_C$ is compatible with the presence of a well distinguishable in-medium remnant peak. At $T=2.25T_C$ neither for $\eta_b$ nor $\eta_c$ a peak is required to reproduce the correlator within errors. A similar analysis of the vector channel is currently work in progress.

\begin{figure}
\centering
\includegraphics[scale=0.4]{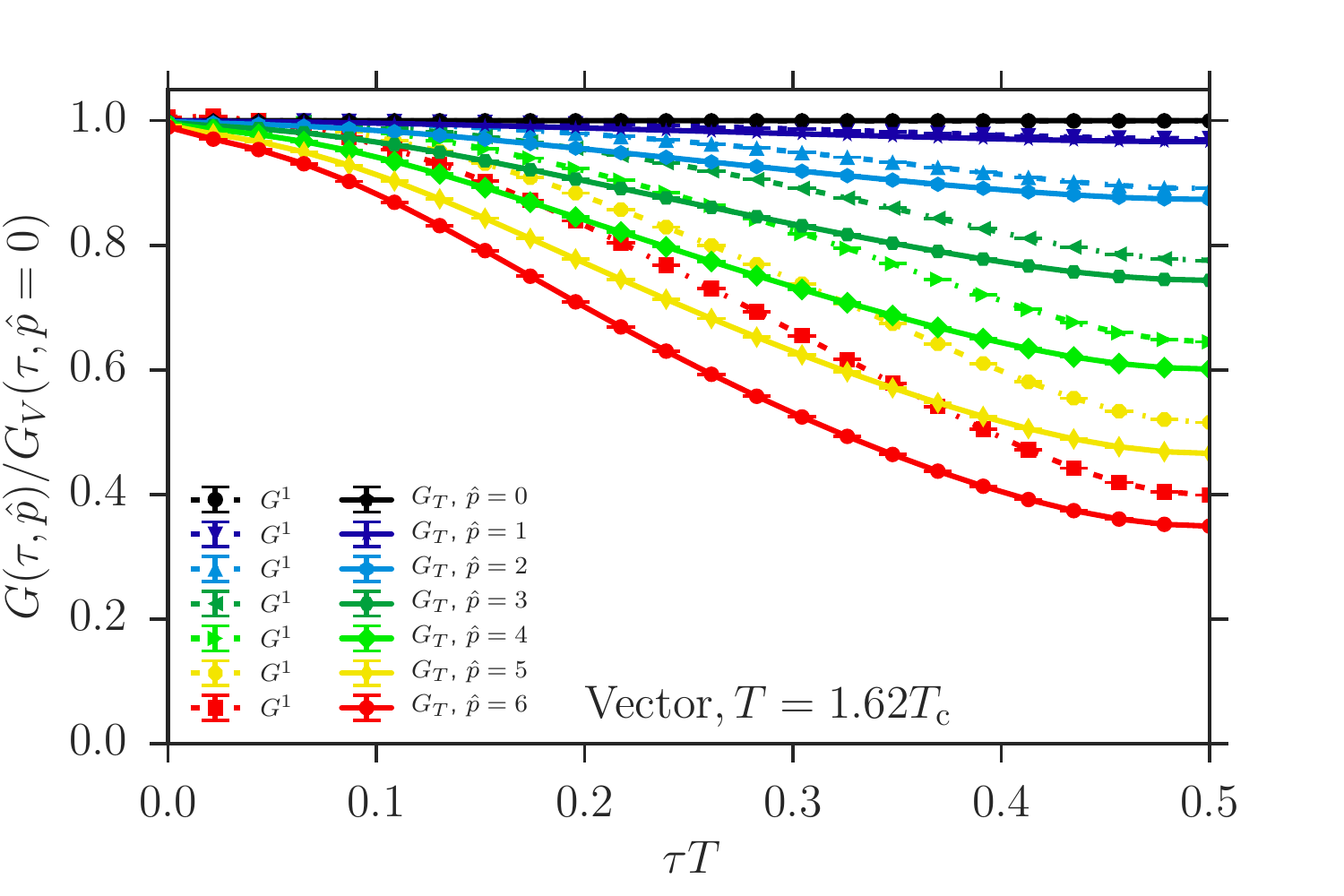}
\caption{Vector channel charmonium correlator in quenched QCD at finite momentum divided by the $p=0$ correlator. Figure adapted from Ref.~\cite{Ikeda:2016czj}}\label{fig:FiniteMomVecQuenched}
\end{figure}

Let us also consider the scalar and axial-vector channel encoding the P-wave states. When computed in the quenched approximation and even more so in the dynamical theory the ratios show an upward bend whose deviation from unity is much larger than in the S-wave channels. It easily reaches 50\% and more, indicating that sizable in-medium changes already occur for these channels just above $T_C$.

Up to this point we have only discussed quarkonium at zero momentum but also the finite momentum situation has been considered, e.g. in the quenched approximation in Ref.~\cite{Ikeda:2016czj} on anisotropic $N_s=64$ lattices with $a_s/a_t=4$ and $a_\tau=0.00975$fm (see also Ref.~\cite{Ding:2012pt}). If a quarkonium bound state peak exists, it will follow a dispersion relation, in which the peak position will move to higher frequencies for larger values of p. The stronger decay in the corresponding Euclidean correlator is expected to lead to a lower than unity ratio to the $p=0$ case. As shown in \cref{fig:FiniteMomVecQuenched} this is exactly what is observed in practice. Note that two sets of curves are shown representing the longitudinal and transversal components defined in \cref{eq:translongdecomp}.

\begin{figure}
\centering
\includegraphics[scale=0.06]{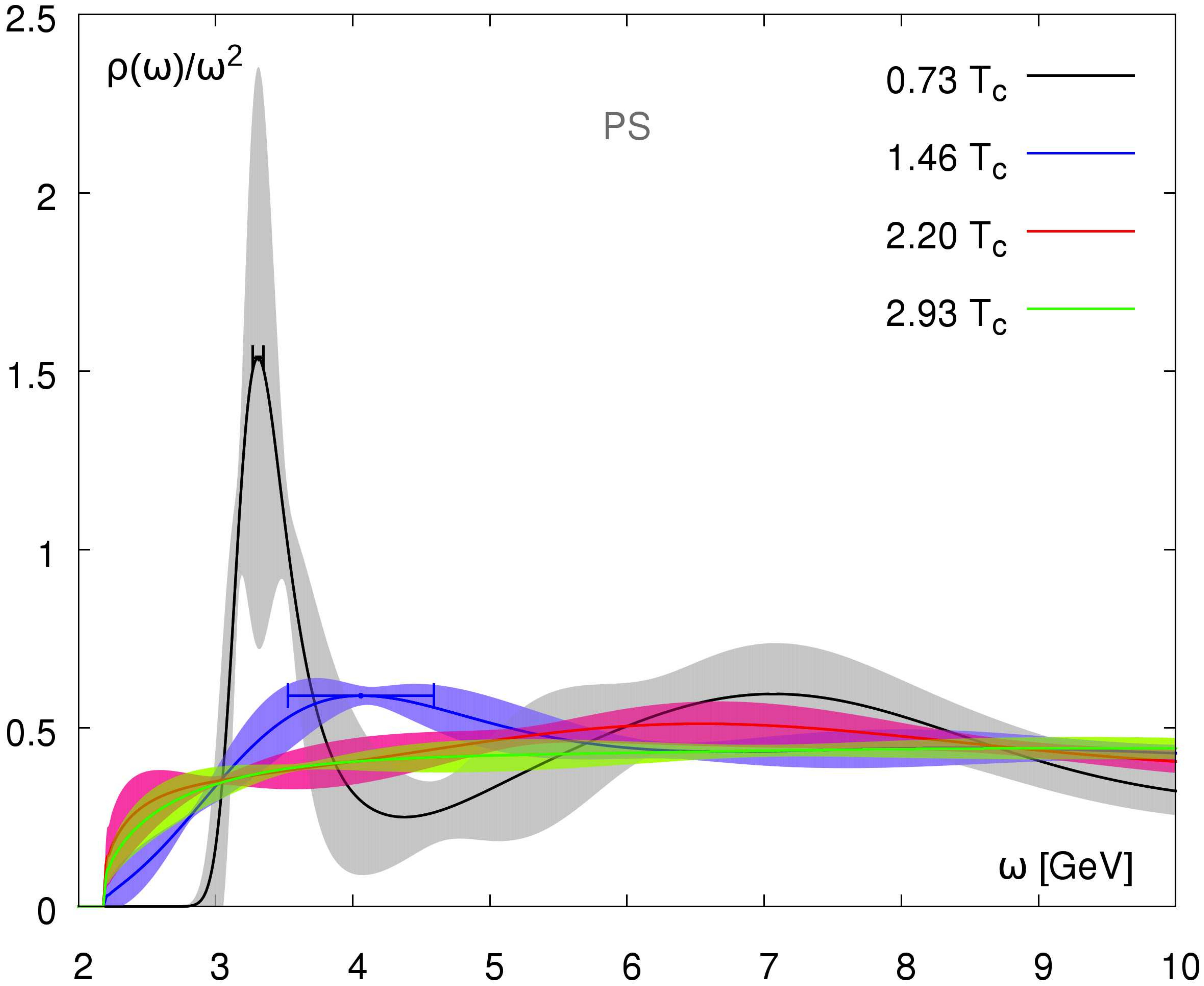}
\includegraphics[scale=0.375]{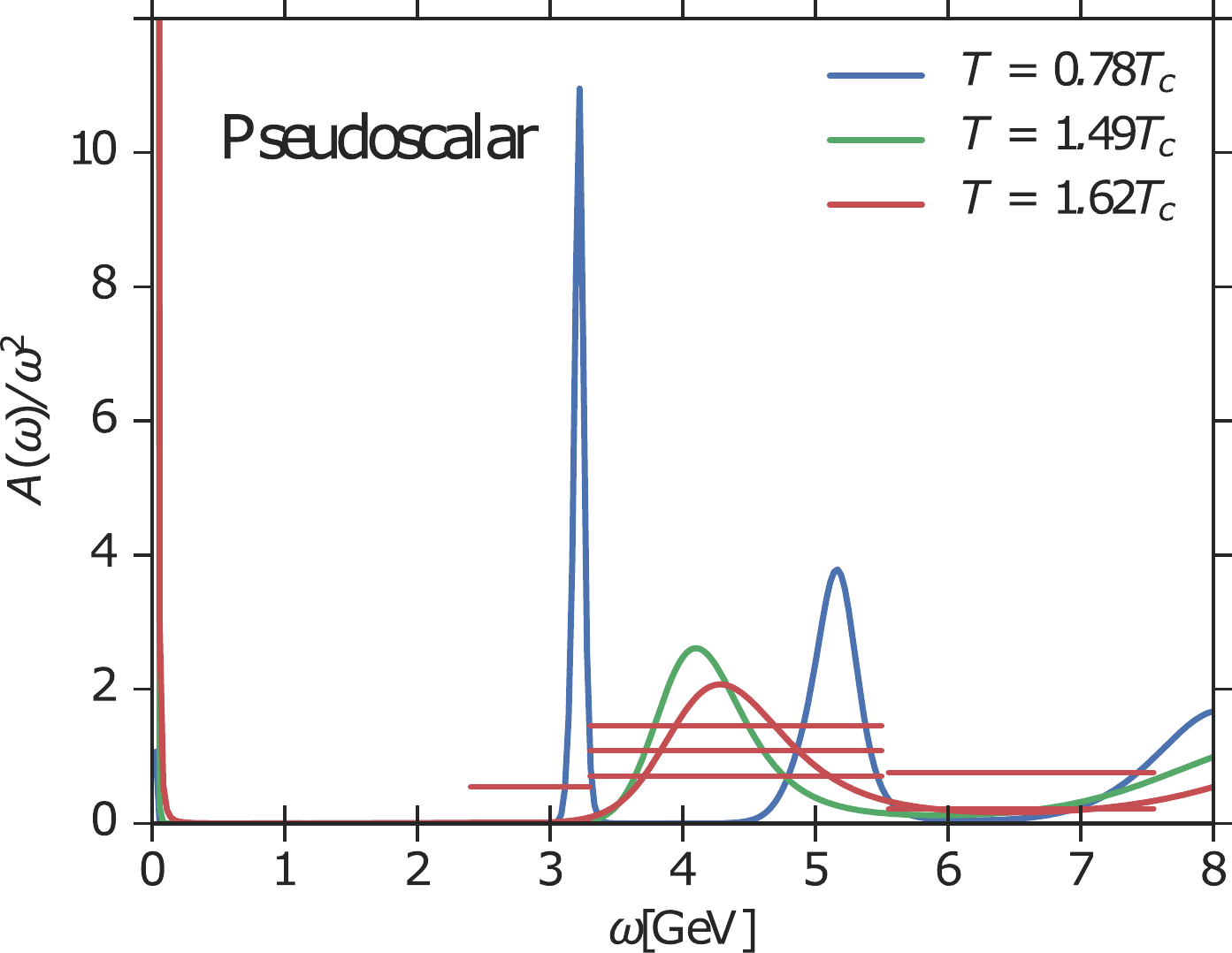}
\caption{Pseudoscalar spectral function from quenched QCD resconstructed via the Maximum Entropy Method. (left) the results from isotropic $N_s=128$ lattices with $a=0.01$fm  and (right) results from anisotropic $N_s=64$ lattices with $a_s=0.039$. Figures adapted from Refs.~\cite{Ding:2012sp,Ikeda:2016czj}}\label{fig:CharmoniumDynamicSpecRelPS}
\end{figure}

Let us next consider the spectral functions themselves. All the studies quoted above have used the in-medium correlation functions to extract the underlying spectral functions deploying mostly the MEM, one also used the BR method. These computations are particularly challenging since only $N_\tau/2$ individual datapoints exist and in addition the physical Euclidean range reduces as temperature increases. In the MEM, as discussed in \cref{sec:BayesRec} the smoothing induced through the restricted SVD subspace becomes stronger, as the number of datapoint is reduced. At the same time the amount of information on the ground state decreases as the imaginary time extend decreases. In the BR method there is no additional smoothing present which in case of a small number of datapoint may instead lead to numerical ringing. I.e. besides the actual in-medium effects manifesting themselves in the spectral function, the reconstruction efficiency of the deployed methods changes at different temperatures. Those effects need to be disentangled. The clearest comparison is obtained when taking the low temperature reference correlator and use it to compute a reconstructed correlator at high temperature. Subsequently the spectral reconstruction is performed both on that $D_{\rm rec}$, as well as the actual high temperature correlator and compared, as suggested and for the first time deployed in Ref.~\cite{Kelly:2018hsi}.

\begin{figure}
\centering
\includegraphics[scale=0.3]{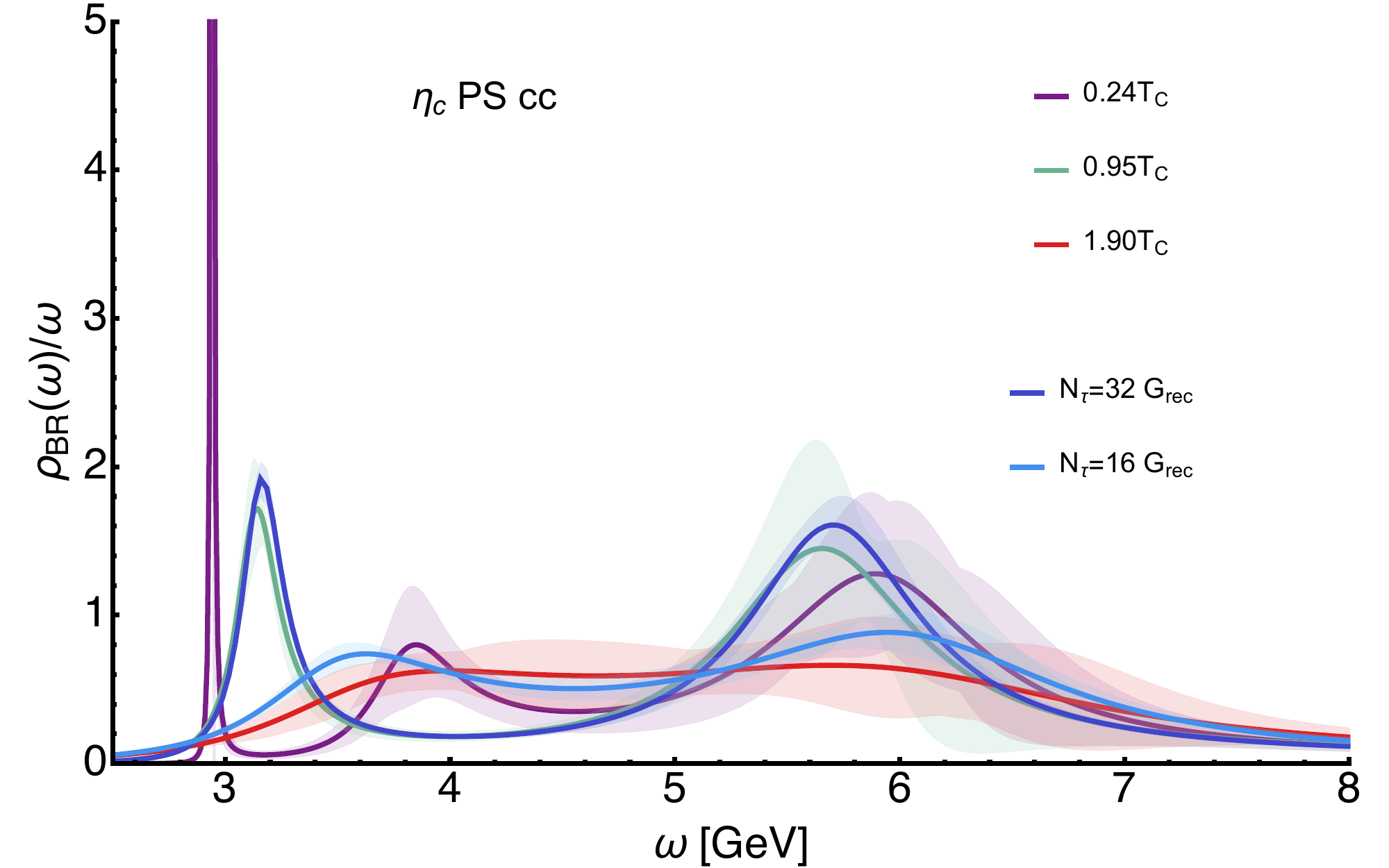}
\includegraphics[scale=0.3]{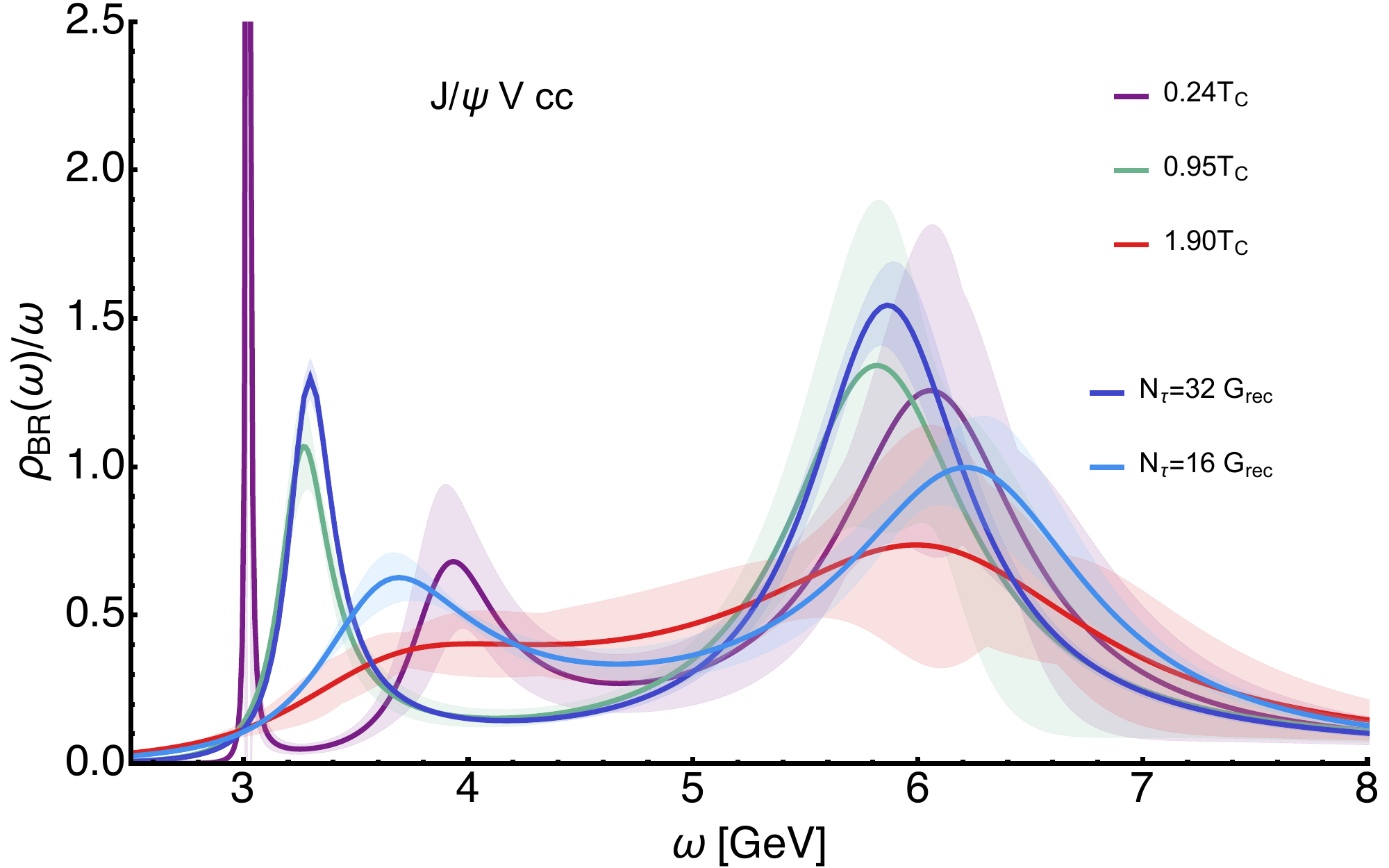}
\caption{(left) pseudoscalar and (right) vector channel charmonium spectral function extracted from $N_f=2+1$ dynamical QCD on anisotropic lattices with the BR method. Comparison between the reconstructed high temperature spectrum and the low temperature spectrum encoded via the reconstructed correlator at the same Eucldiean extend reveals little signs of in-medium modification. Errorbands denote both Jackknife statistical uncertatinty, as well as systematic one due to default model dependence. Figures adapted from Ref.~\cite{Kelly:2018hsi}}\label{fig:CharmoniumDynamicSpecRel}
\end{figure}

It is currently still difficult to arrive at a quantitative interpretation of the reconstructed spectra due to the involved reconstruction uncertainties. As presented in \cref{fig:CharmoniumDynamicSpecRelPS}, on the one hand Ref.~\cite{Ikeda:2016czj} shows clear peaks for both the vector and pseudovector channel charmonium at $T=1.62T_C$, while Ref.~\cite{Ding:2012sp} already at $T=1.46T_C$ shows a very washed out lowest lying feature. In both studies the in-medium spectral feature appears to move to higher frequencies above $T_C$. Note that both studies use a similar MEM reconstruction and that the authors of Ref.~\cite{Ikeda:2016czj} also carried out the reconstruction based on the reconstructed correlator in Ref.~\cite{Kitazawa:2018xbl} indicating that the shift in the peak position may not simply be a methods artifact. The difference between the two results lies in that the former is carried out on anisotropic lattices with roughly twice the spatial volume compared to the latter. On the other hand the UV cutoff is located at a lower energy in the former. Especially in light of the results from modelling the pseudoscalar correlator in the continuum limit, it is paramount to clarify the situation further. Let us note that in Ref.~\cite{Ikeda:2016czj} no significant difference between the longitudinal and transversal reconstructed spectral functions was observed even though at finite temperature these two components need not agree. In the case of dynamical QCD on anisotropic lattices, as shown e.g. in \cref{fig:CharmoniumDynamicSpecRel} from Ref.~\cite{Kelly:2018hsi} (a continuation of the FASTSUM results presented in Ref.~\cite{Aarts:2007pk}), the comparison between the actual high temperature reconstructed spectrum and the one reconstructed from the low temperature reconstrcuted correlator show only very small differences around $T_C$. At $T=1.9T_C$ indications for in-medium modification are visible. While there are hints for the in-medium ground state peak moving to higher frequencies (similar to what was found in the quenched approximation in Ref.~\cite{Ikeda:2016czj}), it will require more robust reconstructions in dynamical QCD to draw a final conclusion on this point. 

\begin{figure}
\centering
\includegraphics[scale=0.3]{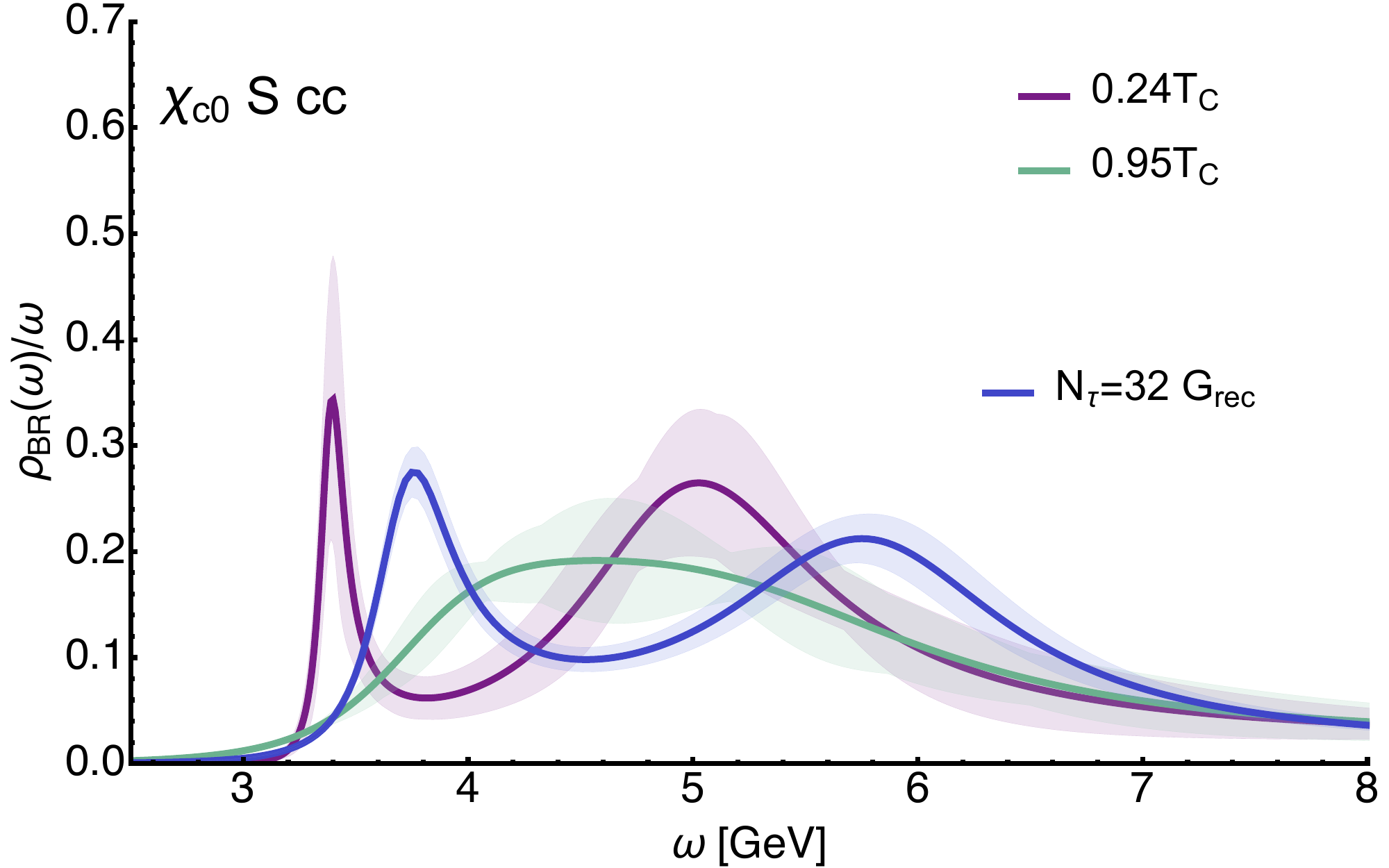}
\includegraphics[scale=0.3]{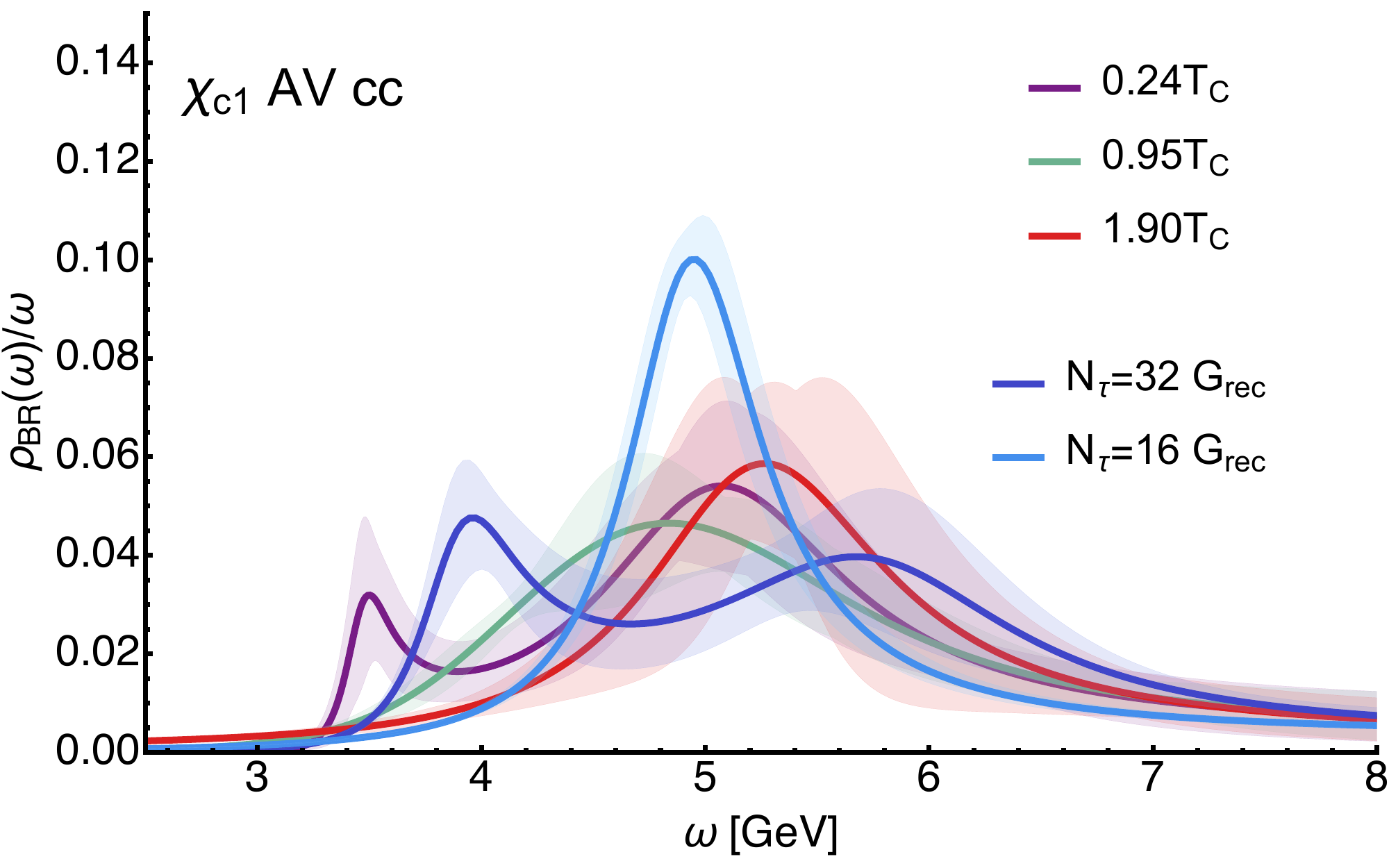}
\caption{(left) scalar and (right) axial vector channel charmonium spectral function extracted from $N_f=2+1$ dynamical QCD on anisotropic lattices with the BR method. Comparison between the reconstructed high temperature spectrum and the low temperature spectrum encoded via the reconstructed correlator at the same Eucldiean extend reveals signs of large in-medium modification already close to $T_C$. Errorbands denote both Jackknife statistical uncertatinty, as well as systematic one due to default model dependence. Figures adapted from Ref.~\cite{Kelly:2018hsi}}\label{fig:CharmoniumDynamicSpecRelSAV}
\end{figure}

The P-wave states in the scalar and axial vector channels, shown in \cref{fig:CharmoniumDynamicSpecRelSAV} due to the smaller signal to noise ratio exhibit larger uncertainties. Again we plot both the reconstruction of the actual in-medium spectrum, as well as the reconstructions from the low-temperature reference reconstructed correlator. The sizable in-medium modification of the correlator translates into significant changes of the spectra already around $T_C$, with no discernible peak structure remaining at $T=0.95T_C$. This is in stark contrast to the S-wave states, which are much less affected. It will be important to improve the signal to noise ratio in the P-wave channels to more robustly ascertain the full extend of the in-medium effects.

The study of the effect of finite momentum in the spectral functions suffers from less systematics, since for a fixed temperature the Euclidean extend stays the same among different momenta. Ref.~\cite{Ikeda:2016czj} confirmed that the less than unit ratio between the $p>0$ and $p=0$ correlator indeed translates into an in-medium peak, which moves to higher frequencies. Extracting the dispersion relation of this peak structure, it was observed that it differed only minutely from the vacuum form, both indicating that this peak actually encodes a particle d.o.f. and that its in-medium modification is apparently weak. 

\begin{summary} A concerted effort is undertaken to elucidate the in-medium properties of quarkonium based on the relativistic formulation in lattice QCD. It is now understood that an investigation of the pseudoscalar sector allows to disentangle transport from bound state physics. Modeling its physics content using perturbation theory and pNRQCD indicates a fast disappearance of any peak structures for charmonium around $T_C$, while the lattice data indicates the presence of a bottomonium remnant peak up to around $T\approx 2T_C$. On the other hand it remains highly challenging to quantitatively extract the in-medium effects on individual states via reconstructed spectral functions. There are clear signs of an in-medium modification in the P-wave states, whose peak structures seem to disappear quickly around $T_C$, while for the S-wave states the final word has not yet been spoken, up to which temperature their features persist. Indications are found that above $T_C$ the in-medium peak structures tend towards higher frequencies. Further progress will depend on improving the spectral reconstructions significantly, which is currently work in progress both based on continuum extrapolated correlators (Bielefeld-CCNU collaboration), as well as improved anisotropic lattices closer to the continuum in full QCD (FASTSUM collaboration).
\end{summary}
 
\subsubsection*{In-medium quarkonium from lattice NRQCD}

In order to progress with currently available lattice simulations one may  decide to leave the relativistic formulation and instead deploy the lattice regularized version of the effective field theory NRQCD. As a theory of non-relativistic Pauli spinors, valid at energies of the order $m_Q v $ and below it offers several advantages over the direct relativistic formulation as discussed in \cref{sec:latNRQCD}. On the other hand it has to be kept in mind that the EFT is only an approximation to QCD and that effects, such as accurately capturing the hyperfine splitting of the S-wave states requires high order expansions in the quark velocity and beyond leading order radiative corrections. As the uncertainties in the spectral decomposition outweigh those systematics, most studies so far deploy the ${\cal O}(v^4)$ NRQCD Hamiltonian and leading order Wilson coefficients with tadpole improvements. There have been exploratory studies of in-medium quarkonium in NRQCD presented early on in Ref.~\cite{Fingberg:1997qd}. The first comprehensive study on in-medium Bottomonium in lattice NRQCD has been undertaken in a series of papers by the FASTSUM collaboration in Refs.~\cite{Aarts:2014cda,Aarts:2013kaa,Aarts:2012ka,Aarts:2011sm,Aarts:2010ek}. A complementary lattice effort has been started with the papers of Refs.~\cite{Kim:2014iga,Kim:2018yhk}, the latter of which extended the use of NRQCD to a study of charmonium properties.

Lattice setups and spectral reconstruction methods differ between the two recent sets of papers. The groups have focused on achieving good accuracy either in the heavy quark or the medium sector. Since in both cases NRQCD matching has not been performed beyond leading order, the result of the different studies do not necessarily have to agree within their statistical errors. 

The latest FASTSUM studies (see e.g.\cite{Aarts:2014cda}) are carried out in a fixed box approach on $N_s=24$ anisotropic lattices of $a_s/a_\tau=3.5$ with $N_f=2+1$ light flavors of clover improved Wilson fermions at $a=0.1227$fm. Using temporal grids $N_\tau=40\ldots16$ a temperature range between $T=141\ldots 352 = 0.76T_C\ldots 1.90T_C$ is covered. For calibration a $N_\tau=128$ ensemble has been obtained. Between 500 and 1000 realizations of the current correlators on the individual lattices are computed. The values of the heavy quark mass parameter are fixed using the dispersion relation of quarkonium S-wave states, maintaining a spin averaged 1S kinetic mass close to the PDG value \cite{Tanabashi:2018oca}.

The benefit of this approach is that the anisotropy allows the NRQCD expansion to be very robust, i.e. a Lepage parameter of $n=1$ has been deployed in the time evolution throughout. The drawback of this approach on the other hand is that the pion mass is rather heavy with $M_\pi\approx 400$MeV so that e.g. the crossover temperature $T_C^{\rm lat}=185$MeV lies above the physical value. 

The second set of works uses lattice ensembles from the HotQCD collaboration originally intended for the study of the crossover transition in Refs.~\cite{Bazavov:2017dsy,Bazavov:2014pvz}, where $N_f=2+1$ light quarks are discretized with the HISQ action on $48^3\times 12$ isotropic grids in a fixed box approach. For each of the different lattice spacings that implement a temperature range between $T=140\ldots 407$MeV a $T\approx0$ ensemble with $N_\tau=32\ldots64$ is present for scale setting and calibration purposes. Using a naively set heavy quark mass of $m_b=4.65$GeV and $m_c=1.275$GeV at $T\approx0$ 400 realizations of the current correlators have been computed, while at high temperatures up to 4000 have been obtained.     

\begin{figure}
\centering
\includegraphics[scale=0.25]{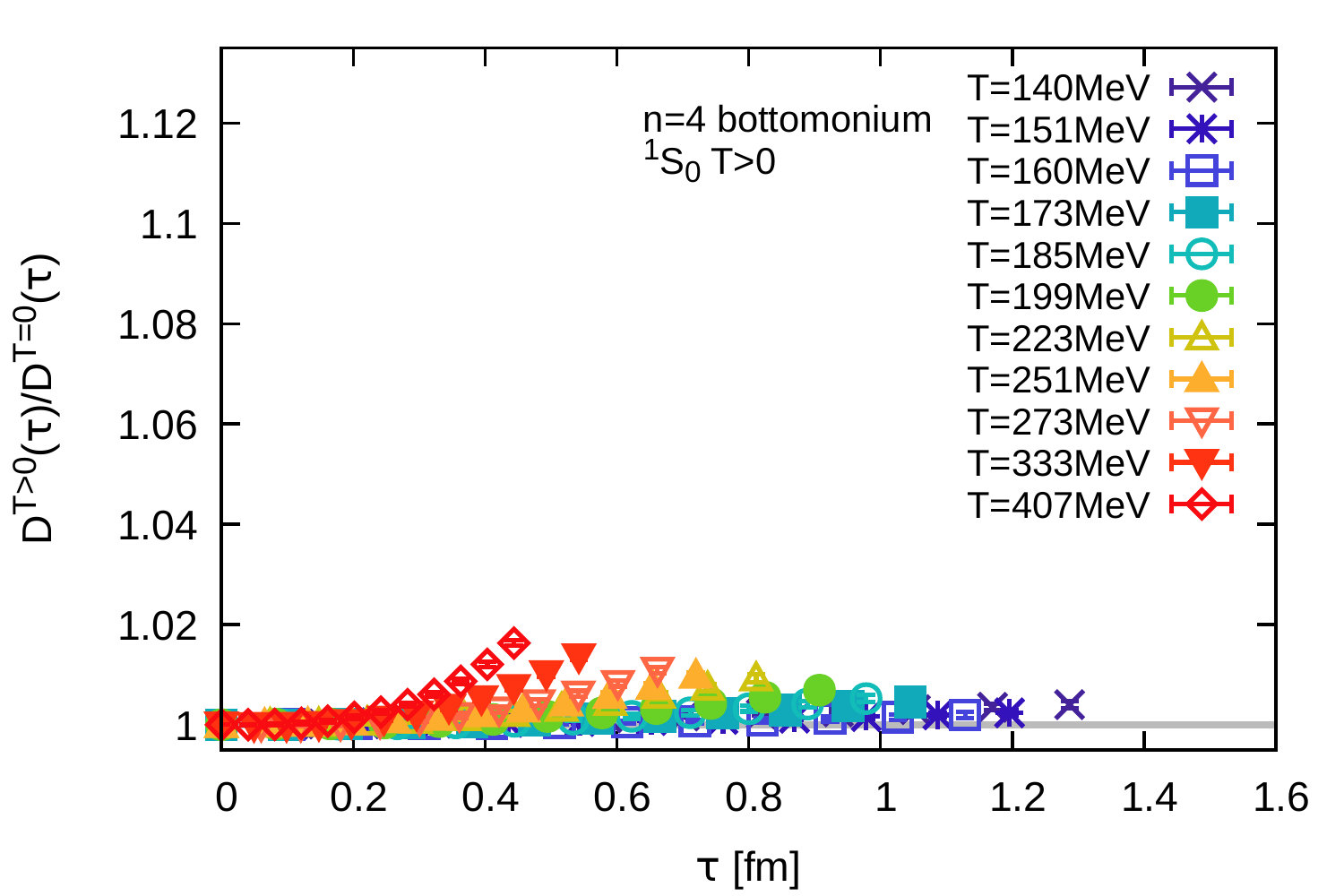}
\includegraphics[scale=0.25]{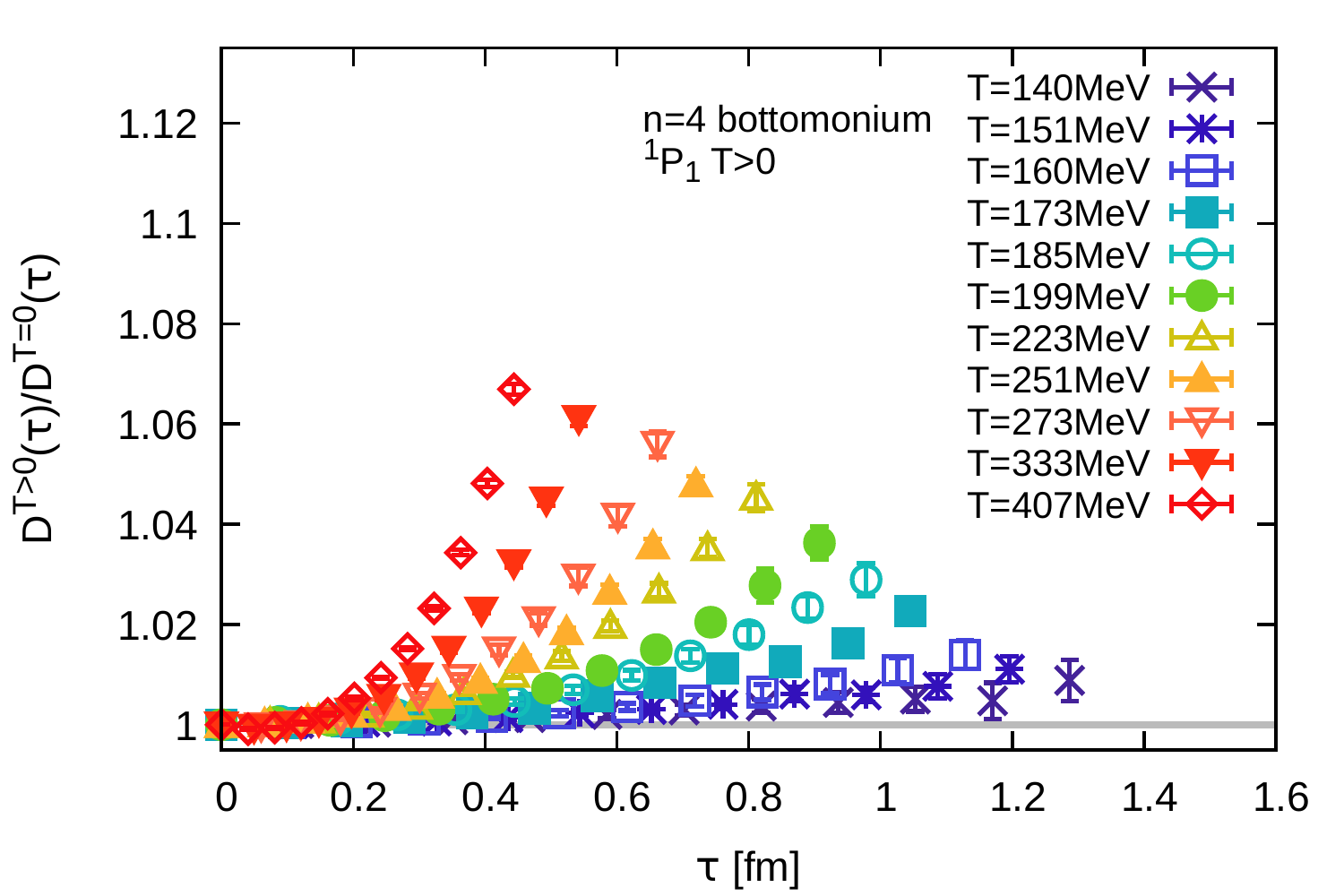}
\hspace{0.5cm} \includegraphics[scale=0.25]{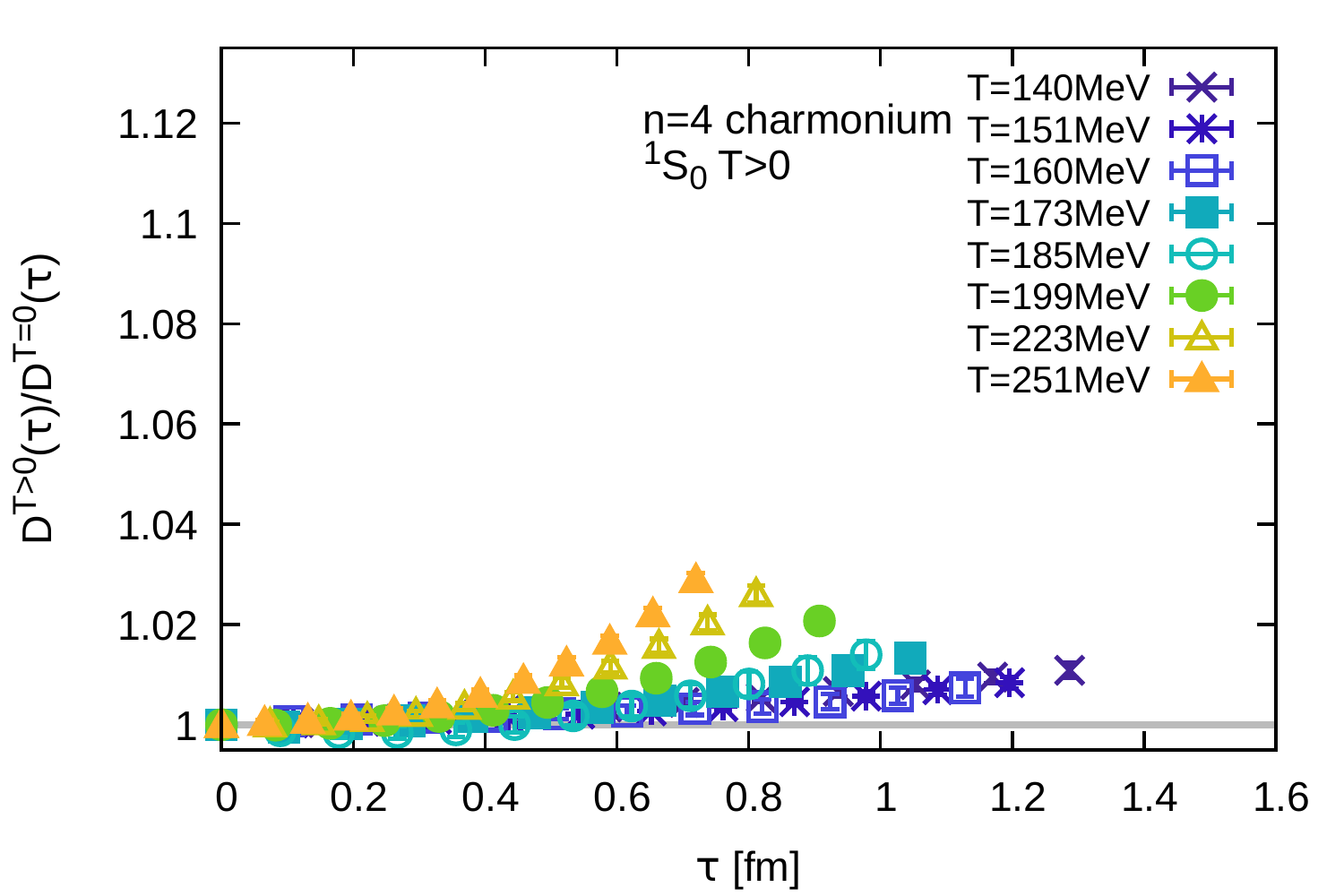}
\includegraphics[scale=0.25]{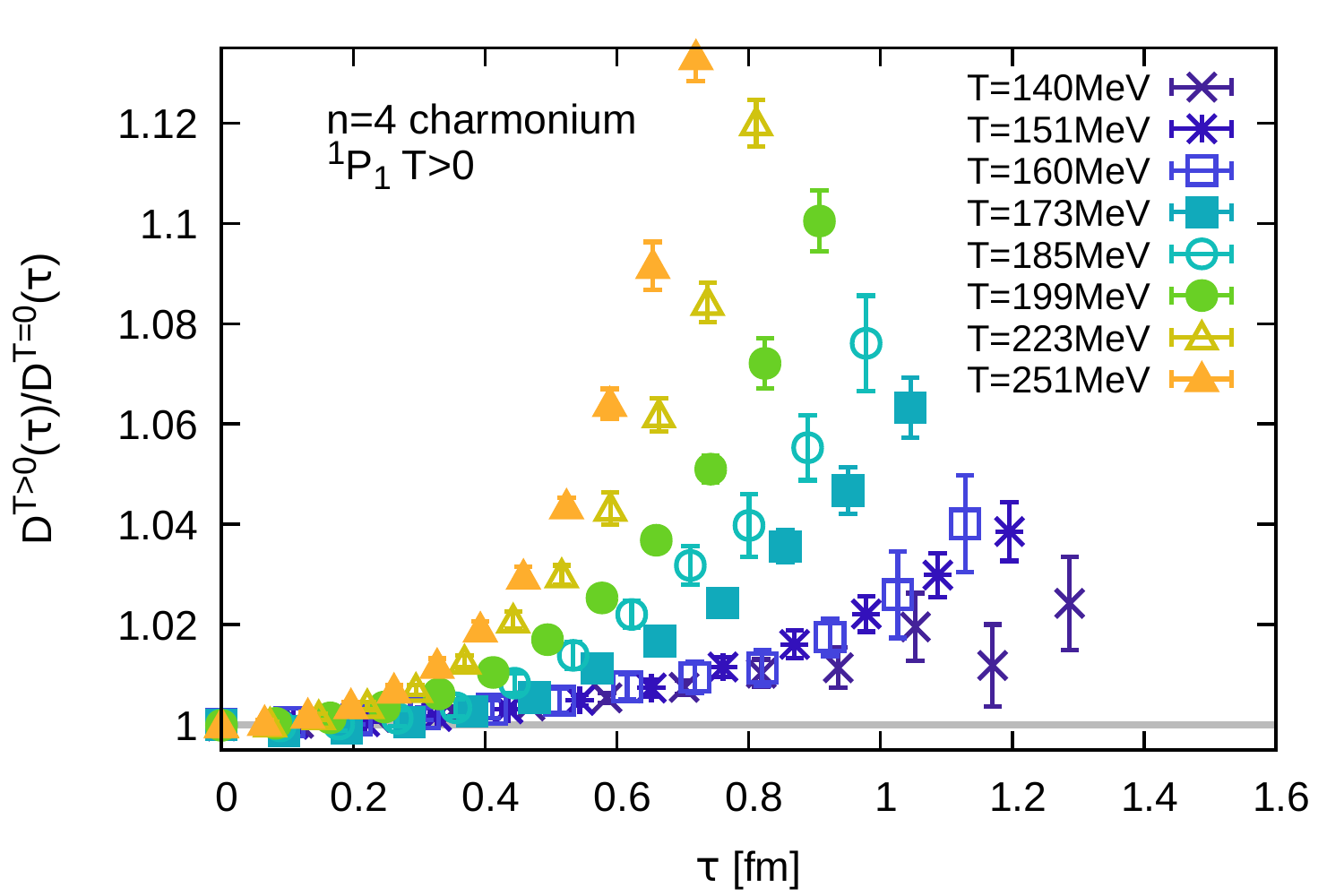}\\
\includegraphics[scale=0.25]{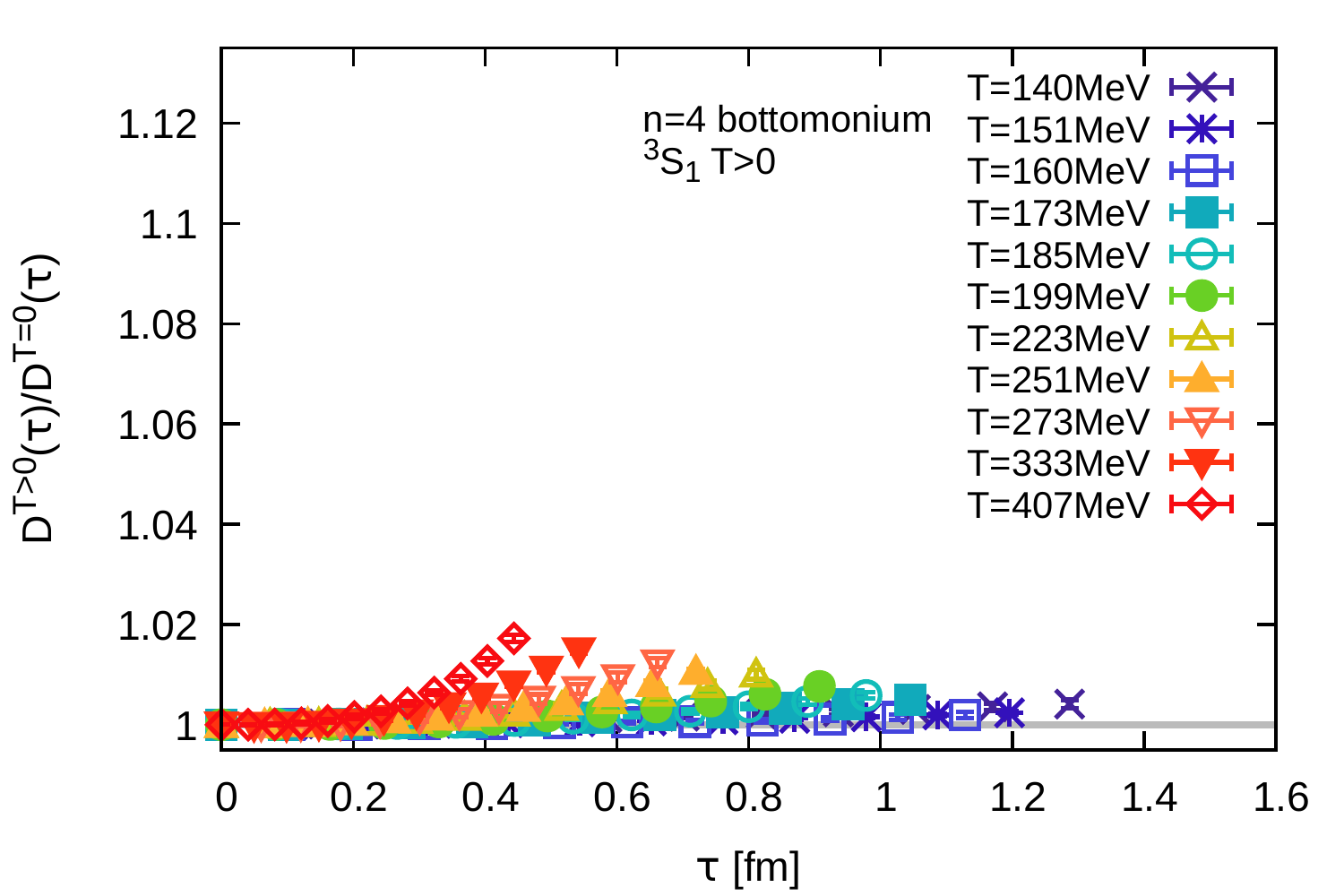}
\includegraphics[scale=0.25]{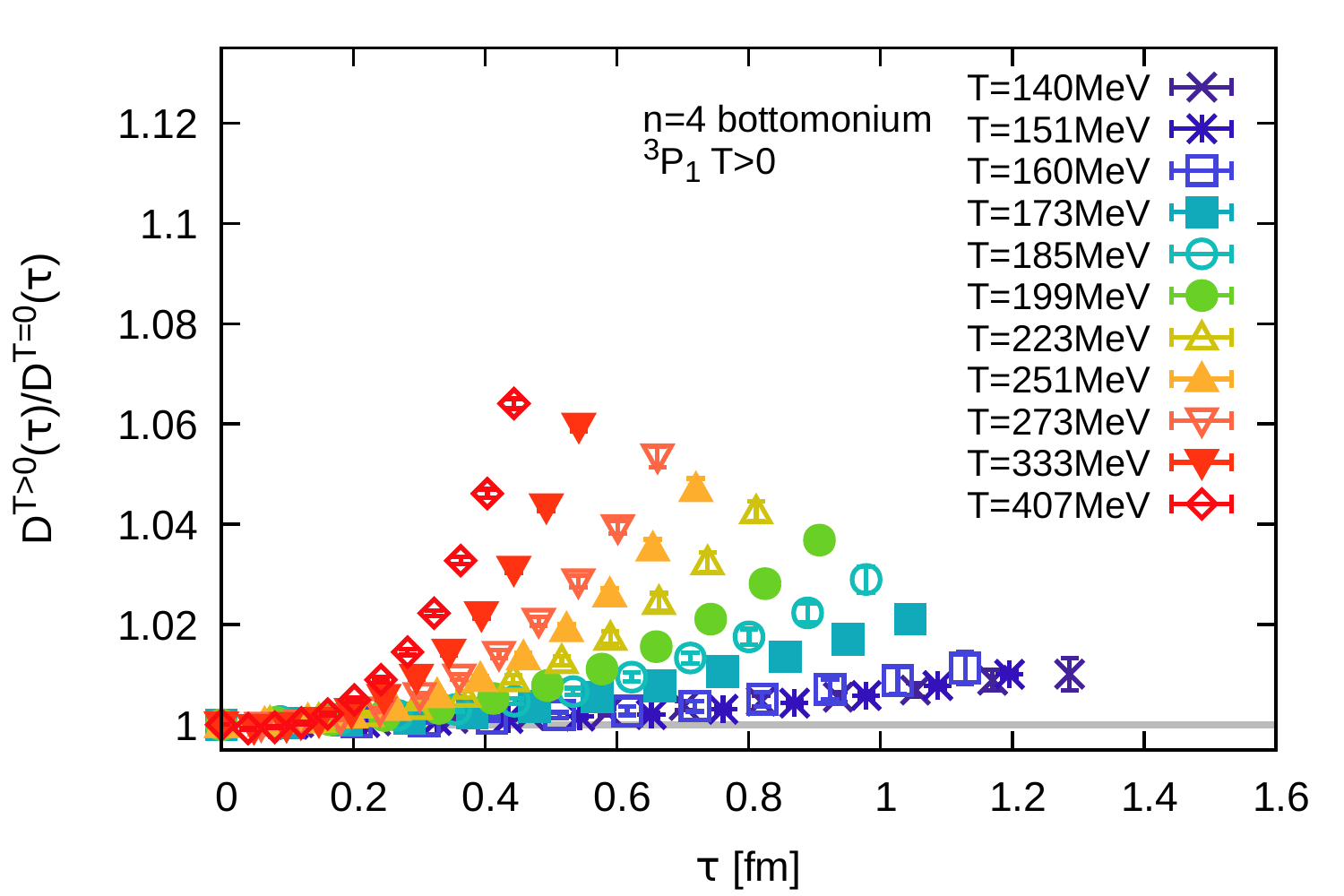}
\hspace{0.5cm} \includegraphics[scale=0.25]{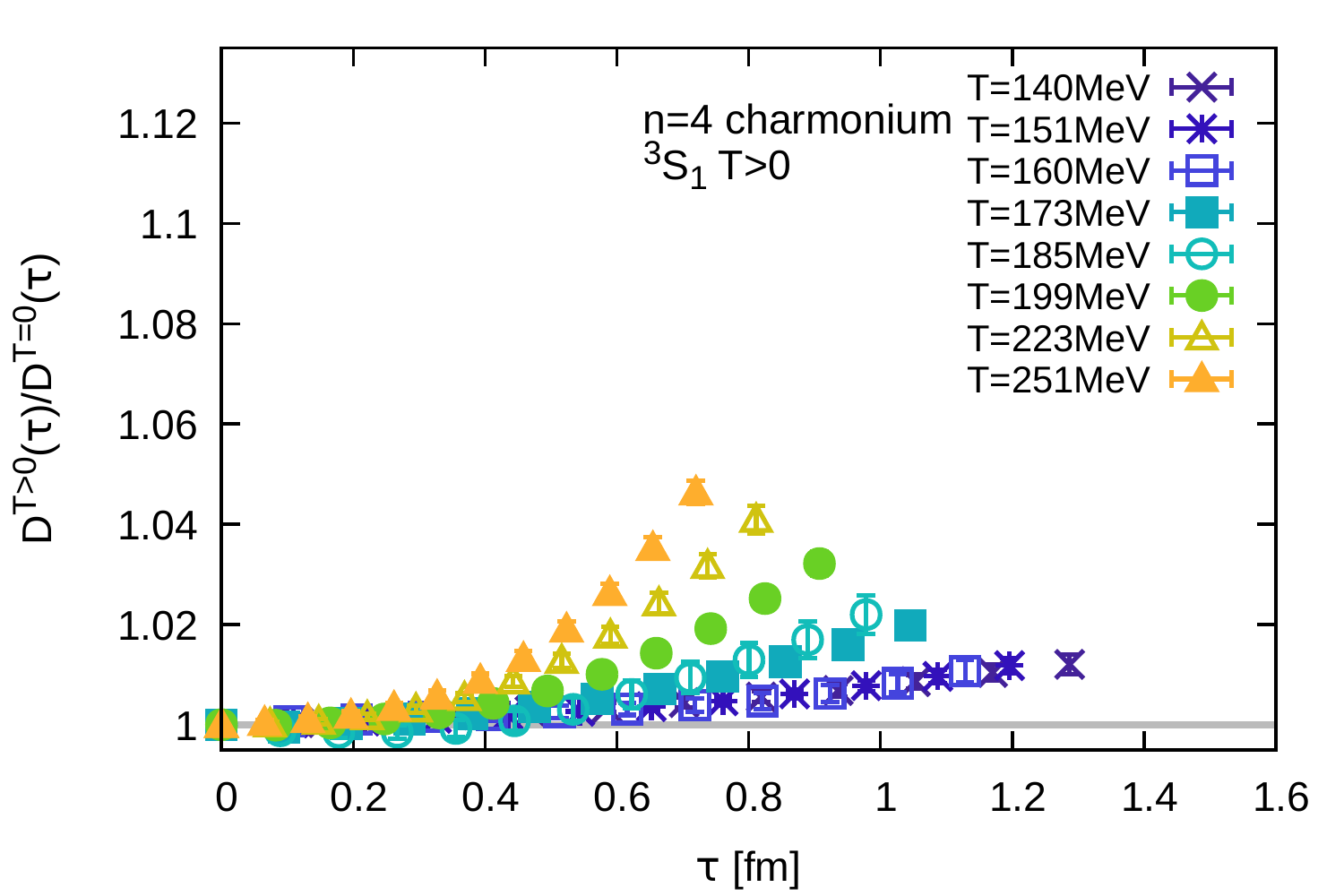}
\includegraphics[scale=0.25]{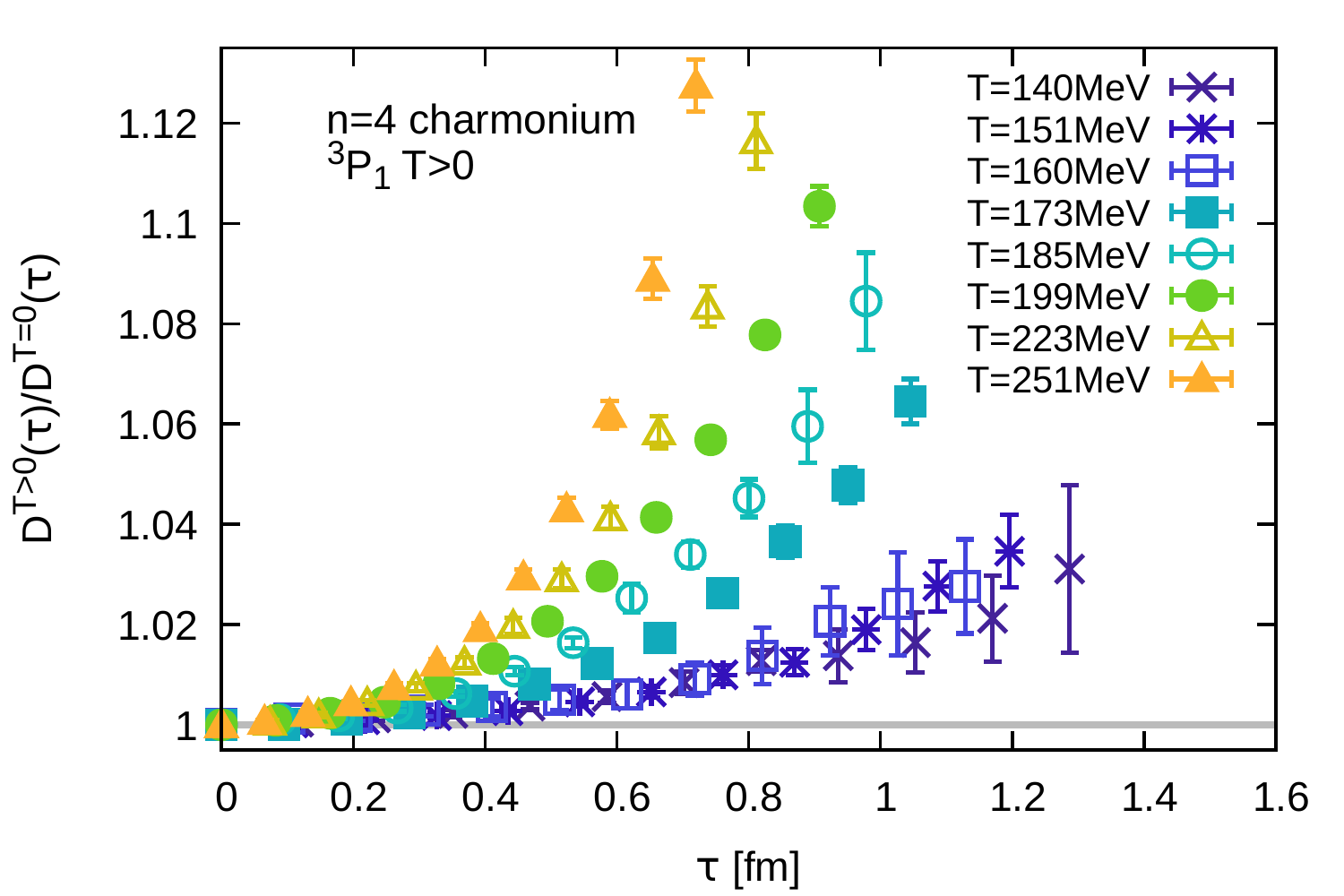}
\caption{Ratios of the in-medium NRQCD quarkonium correlator to its $T=0$ counterpart. In the left block of four panels, we show the bottomonium ratios for temperatures between $T=140\ldots 407$MeV. Counterclockwise from the top left the $^1S_0$ $(\eta_b)$, $^3S_1$ $(\Upsilon)$, $^3P_1$ $(\chi_{b1})$ and $^1P_0$ $(h_b)$ channel is shown. In the right block of four panels, we show the charmonium ratios for $T=140\ldots 251$MeV . Counterclockwise from the top left the $^1S_0$ $(\eta_c)$, $^3S_1$ $(J/\psi)$, $^3P_1$ $(\chi_{c1})$ and $^1P_0$ $(h_c)$channel is plotted. Note that no transport contribution is expected to contribute to any of these NRQCD correlators. Figures partially reproduced from Ref.~\cite{Kim:2018yhk}}\label{fig:NRQCDcorrelatorratios}
\end{figure}
\begin{figure}
\centering
\includegraphics[scale=0.35]{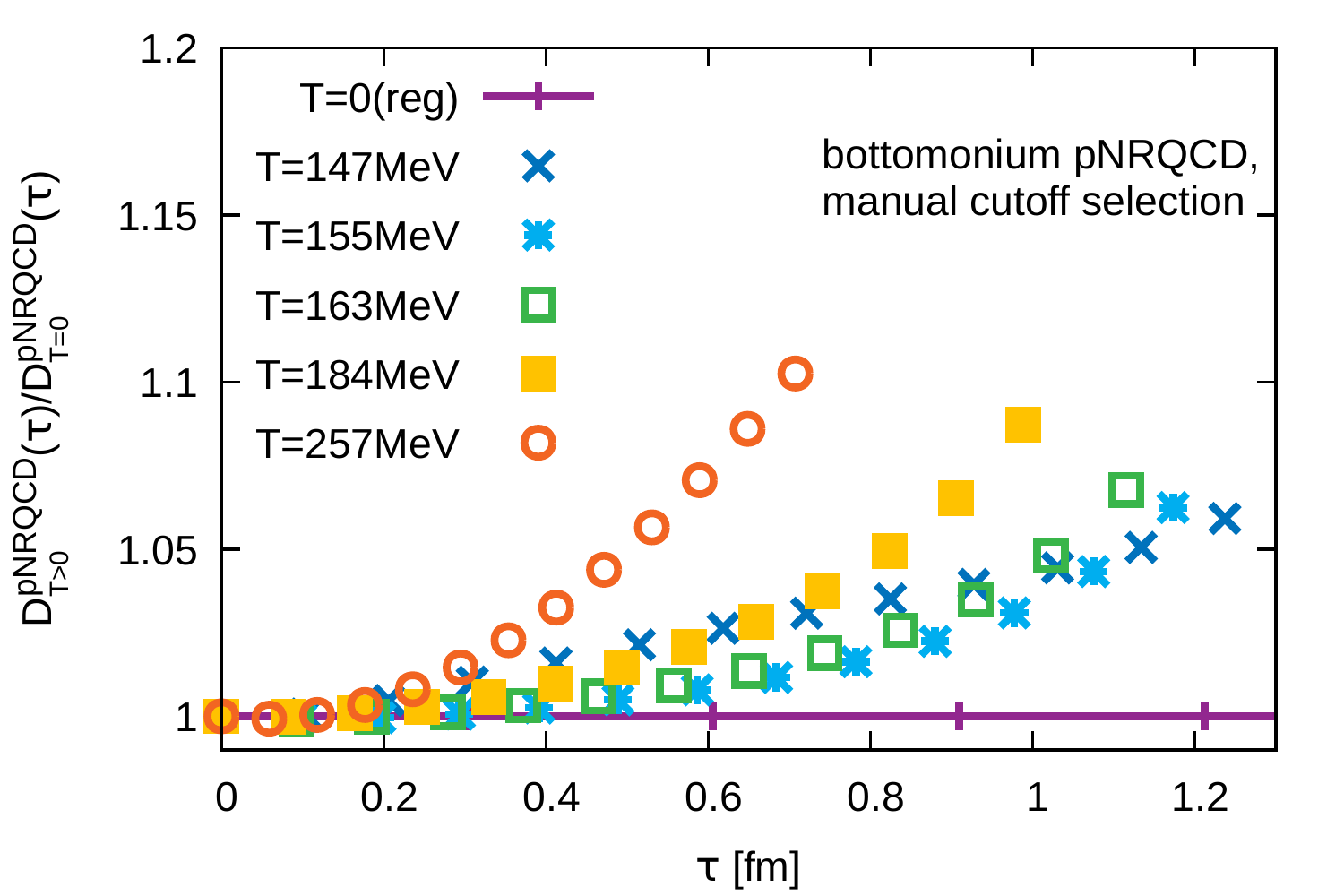}
\includegraphics[scale=0.35]{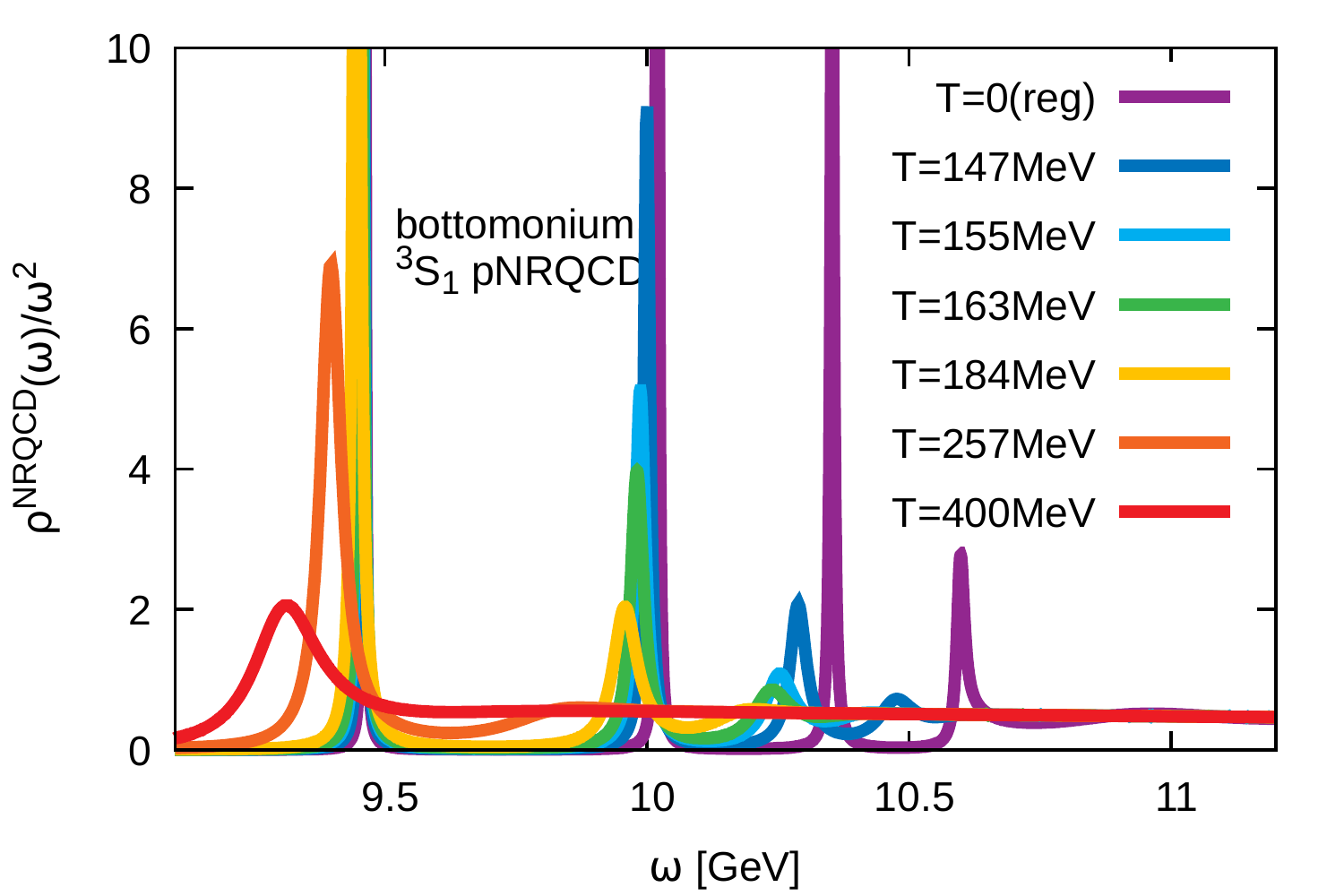}
\caption{(left) the ratio of the Euclidean in-medium $T>0$ correlator and the $T=0$ correlator obtained from a pNRQCD model computation of the bottomonium spectral function (right). A shift of the ground state peak to lower frequencies induces the upward bend in the correlator. Figures reproduced from Ref.~\cite{Kim:2018yhk}}\label{fig:NRQCDCorrRevEng}
\end{figure}

Here the benefit of the approach lies in realistically capturing the physics of the medium d.o.f. with a pion mass of $m_\pi=161$MeV and a corresponding crossover temperature of $T_C^{\rm lat}=159$MeV. On the other hand its drawback lies in the fact that these lattices were not designed originally with NRQCD in mind. Thus the naive expansion parameter $1/m_Qa$ takes on relatively large values making the NRQCD expansion less robust. Hence the study limits its use of charm quarks to temperatures to and below $T=251$MeV and needs to apply Lepage parameters of $n=4$ for bottom and $n=8$ for charm.

Previously there have been tensions reported between Ref.~\cite{Aarts:2014cda} (using MEM) and Ref.~\cite{Kim:2014iga} (using BR) in the in-medium modification of P-wave states that went beyond what was expected to arise simply from the different lattice setups. From a much improved understanding of the uncertainties in the spectral reconstruction, especially the role of ringing and smoothing, these differences have recently been sorted out. The P-wave disappearance is found to occur closer to the earlier results of Ref.~\cite{Aarts:2014cda}. In addition improved quantitative results on the in-medium masses, for the first time consistent with the behavior of the  in-medium correlator ratios, have been obtained in Ref.~\cite{Kim:2014iga}.

Since the results for the correlation functions are quite similar among the different studies in the Bottomoniums sector, we showcase here the recent data from Ref.~\cite{Kim:2018yhk} where also the charm d.o.f. were computed. In NRQCD the integral Kernel of the spectral representation is temperature independent and therefore the in-medium changes in the spectral function may be investigated from the naive ratio of the in-medium correlator to the $T=0$ correlator, i.e. without having to construct a reference reconstructed correlator. For completeness the ratios for all available channels are shown in \cref{fig:NRQCDcorrelatorratios}. There are four panels on the left corresponding to bottomonium for temperatures between $T=140\ldots 407$MeV. The channels starting from the top left panel and going counter clockwise are the $^1S_0$ $(\eta_b)$, $^3S_1$ $(\Upsilon)$, $^3P_1$ $(\chi_{b1})$ and $^1P_0$ $(h_b)$ respectively. The charmonium results in the restricted temperature range of $T=140\ldots 251$MeV are arranged also counterclockwise from the top left as the $^1S_0$ $(\eta_c)$, $^3S_1$ $(J/\psi)$, $^3P_1$ $(\chi_{c1})$ and $^1P_0$ $(h_c)$ channels.

As first qualitative difference to the relativistic case we find that all channels, in particular also the pseudoscalar channel of the $\eta$ particles show the same characteristic upward bending as temperature increases. The difference in the underlying spectral function is two-fold:  no transport peak is present at small frequencies and the UV continuum behavior differs due to different powers dominating at large $\omega$. A full understanding of these differences remains to be established. 

For the S-wave bottomonium states there are virtually no medium effects visible at around $T_C$, while both bottomonium P-waves and charmonium S-waves show deviations from unity at $T=140$MeV beyond statistical error bars. This is a first hint that the overall strength of the in-medium modification is connected with the vacuum binding energies of the encoded states. The higher temperature results confirm this impression. We compare the maximum deviation at $T=407$MeV between the $\Upsilon$ channel (bottom left panel) and the $\chi_{b1}$ channel (bottom right panel). The ground state of the former is very strongly bound with $E_{\rm bind}^{\Upsilon}(T=0)\approx 1.1$GeV the one of the latter with $E_{\rm bind}^{\chi_{b1}}(T=0)\approx 0.64$GeV. And indeed while the $^3S_1$ channel shows below $2$\% deviations in the ratio, the $^3P_1$ channel already shows $6.5$\%. We continue the comparison with charmonium, where the S-wave channel (bottom left panel) with ground state $J/\Psi$ $E_{\rm bind}^{J/\psi}(T=0)\approx 0.64$GeV features a very similar vacuum binding as $\chi_{b1}$. To compare apples to apples, let us take the highest temperature where both bottom and charm are available, i.e. at $T=251$MeV. And indeed both the $\chi_{b1}$ channel, as well as the $J/\Psi$ channel show a very similar $5$\% deviation there. On the other hand the $\chi_{c1}$ channel (bottom right) with a much lower ground state binding energy of $E_{\rm bind}^{\chi_{b1}}(T=0)\approx 0.2$GeV already exhibits a deviation of $12.5$\% at that temperature. These results clearly establish an ordering of the overall in-medium modification with the vacuum binding energy. This is in agreement with our intuition, as a more strongly bound state also features a smaller spatial extent, which in turn makes it more difficult for the medium d.o.f. to interfere with the binding.

Since in NRQCD the change in the correlator is expected to be dominated by bound state modification, we may attempt to reverse engineer the underlying behavior of the quarkonium spectral function. To this end one can construct non-relativistic model spectral functions based on e.g. the lattice static interquark potential (see next subsection) and compute from these the corresponding Euclidean correlator. Having done so in Ref.~\cite{Kim:2018yhk} with the results for bottomonium shown in \cref{fig:NRQCDCorrRevEng}, the authors concluded that the upward behavior can be understood as the vacuum peaks starting to broaden and moving to lower frequencies. Due to the difficulty of modeling e.g. the lattice cutoff in such an approach this result should be understood as qualitative, i.e. the magnitude of the deviation from unity is not captured accurately as of yet.

\begin{figure}
\centering
\includegraphics[scale=0.6]{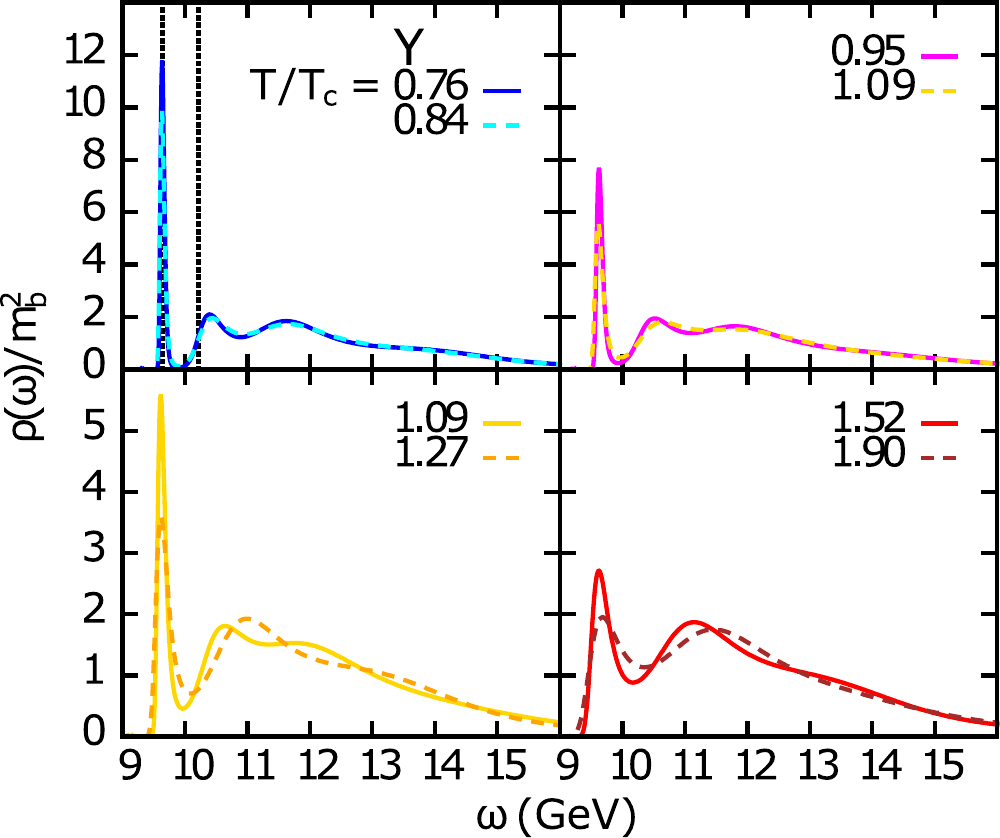}
\includegraphics[scale=0.62]{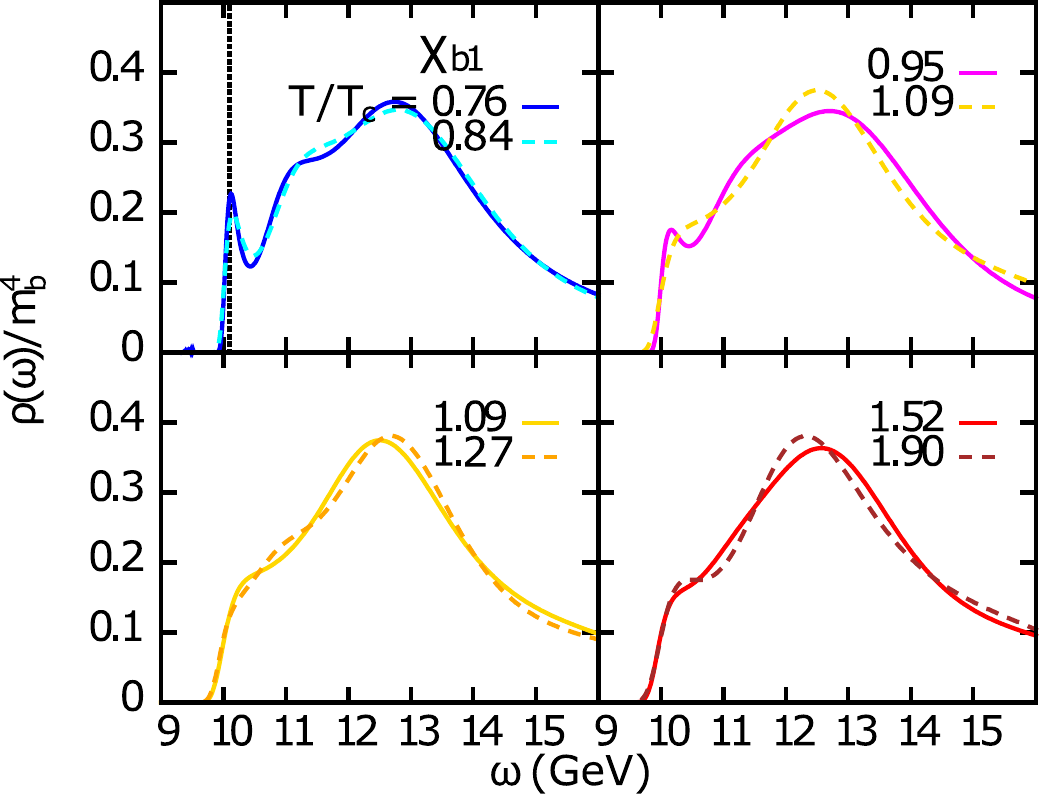}
\caption{(left) S-wave and (right) P-wave in-medium spectral function for bottomonium from anisotropic $N_f=2+1$ dynamical QCD simualtions by the FASTSUM collaboration obtained by the MEM. Figures adapted from Ref.~\cite{Aarts:2014cda}}\label{fig:BottomoniumNRQCDFASTSUM}
\end{figure}

Let us continue with considering the spectral functions as extracted by the MEM in Ref.~\cite{Aarts:2014cda} on anisotropic $N_f=2+1$ lattices and shown in \cref{fig:BottomoniumNRQCDFASTSUM}. The left panel contains the S-wave results, the right panel those for the P-wave. One can clearly see that at low temperatures the ground state peak agrees with the T=0 result indicated by gray vertical lines. The second bump structure summarizes both the first and possible higher excited states, which is why it is shifted above the $T=0$ excited state position. With increasing temperature the ground state amplitude monotonously reduces and only one washed out continuum feature remains at high frequencies. A well defined ground state peak structure at $T=1.9T_C$ is still observed.

The P-wave results, as expected, show a much weaker ground state signal at small temperatures. There are several reason responsible for this difference to the S-wave. On the one hand the signal to noise ratio in the P-wave is lower, due to the larger mass of the ground state. Secondly while the amplitude of the S-wave ground state peak at $T=0$ is related to the radial S-wave wavefunction at the origin squared, the strength of the P-wave state is related to the first derivative of the wavefunction and suppressed by the square of the heavy quark mass. The third issue is related to the different scaling of the continuum which in the S-wave goes as $\omega^{1/2}$, while it is much more dominant in the P-wave with $\omega^{3/2}$. Nevertheless around $T_C$ a ground state feature is visible, which however vanishes into the continuum above $T=1.09T_C$. At the highest temperatures only a shoulder-like feature persists.

On the one hand several crosschecks of the systematic uncertainties, such as dependence on the number of used input datapoints and the shape of the default model, have been carried out in Ref.~\cite{Aarts:2014cda}. These indicate that within the MEM the results are robust. On the other hand these checks did not include a comparison to a different Bayesian reconstruction approach. I.e. while the default model has been varied, the influence of the functional form of the prior probability and the limitation of the search space was not assessed. 

Ref.~\cite{Kim:2014iga} contrasted the MEM to the standard BR method when reconstructing the in-medium spectral functions. A very similar presence of the S-wave ground state peak feature deep in the QGP phase has been observed, while for the P-wave a stronger ground state signal beyond $T=1.09T_C$ has been obtained. During this study it became clear that while the BR method is able to reproduce sharp peak with higher accuracy than the MEM, it may introduce numerical ringing when reconstructing extended structures from a small number of datapoints. Systematic crosschecks were carried out to identify ringing but questions on the strength of the P-wave signal remained.

In the follow up Ref.~\cite{Kim:2018yhk} an improved understanding of the role of ringing and smoothing has been achieved. To this end the newly developed smooth BR method was deployed. In it the strength of smoothing is implemented in the prior probability with an explicit hyperparameter and thus decoupled from the number of datapoints as is the case in the MEM. The tuning of the smoothing parameter as discussed in \cref{sec:BayesRec}, is carried out using the analytically known free NRQCD spectral functions. It has been checked that the choice of $\kappa=1$ both removes ringing artifacts and at the same time still allows to accurately identify the ground state features present at $T=0$. In practice it is then deployed in tandem with the standard BR method. After ascertaining with the smooth method, whether a genuine in-medium peak has been found, the standard BR method is used to extract its peak position.

\begin{figure}
\centering
\includegraphics[scale=0.5]{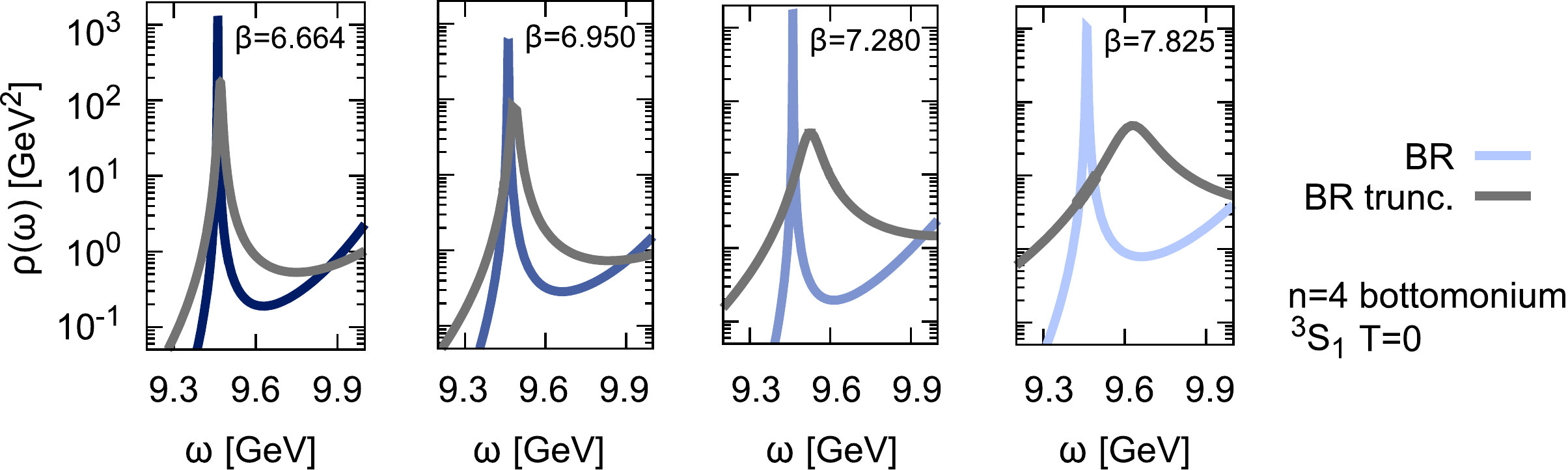}
\caption{The $T=0$ reconstruction of the Bottomoniunm Upsilon channel spectral function using the Euclidean correlator over the whole imaginary time extend available (colored solid) compared to the reconstruction based on truncated input data with the same extent as is available at finite temperature. Note the artificial shift and broadening. Figure adapted from Ref.~\cite{Kim:2018yhk}}\label{fig:BottomoniumTruncTest}
\end{figure}

To interpret the in-medium results particular care was taken to understand the effect of the finite Euclidean extent available on the lattice. In \cref{fig:BottomoniumTruncTest} we show comparisons between the $T=0$ reconstructions (colored solid) both on the full Euclidean correlator, i.e. with input data extending over the full imaginary time extend, as well as using a truncated input dataset with the same Euclidean extent as is available at finite temperature. Obviously the underlying spectral function is the same. On the lattices with coarser lattice spacing e.g. $\beta=6.664$ the differences are not significant. On the other hand for the finely spaced lattices $\beta=7.925$ a clear artificial shift and broadening is observed. These methods artifacts need to be kept in mind when interpreting the in-medium reconstructions.

\begin{figure}
\centering
\includegraphics[scale=0.35]{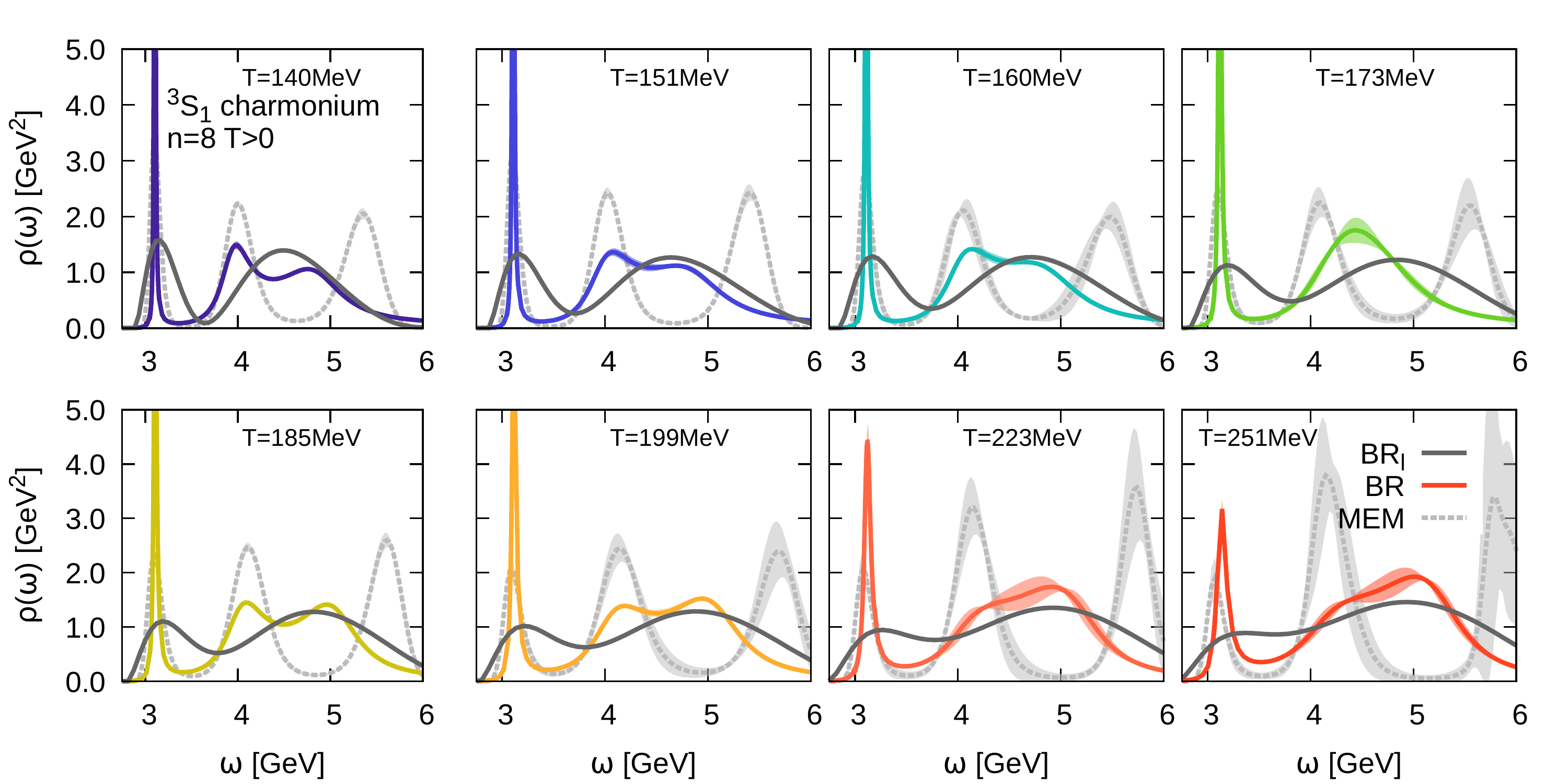}
\includegraphics[scale=0.35]{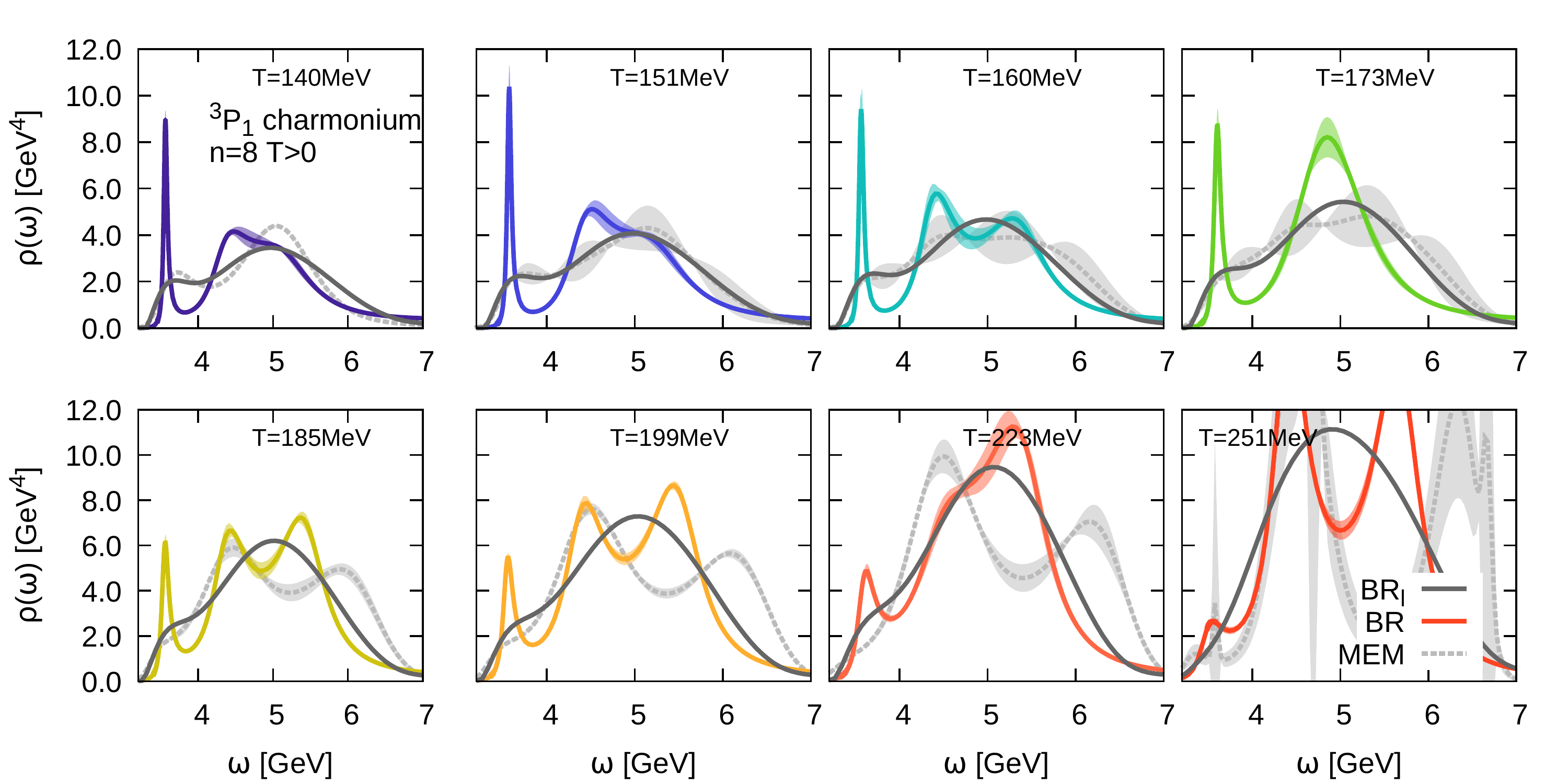}
\caption{Reconstructions of in-medium charmonium spectral functions in the (top two rows) $J/\psi$ channel and the (bottom two rows) $\chi_c$ channel. Figure adapted from Ref.~\cite{Kim:2018yhk}}\label{fig:CharmoniumNRQCD}
\end{figure}

Let us discuss the in-medium results from Ref.~\cite{Kim:2018yhk}. When pitting the standard BR method, the MEM and the smooth BR method against each other it is found that for bottomonium the results of the smooth BR and the MEM agree, in that the Upsilon ground state peak survives up to the highest $T=407$MeV and that $\chi_{b1}$ disappears at around $T=185-210$MeV. These temperatures are similar to those found in Ref.~\cite{Aarts:2014cda}, keeping in mind that the underlying medium description still differs between the two studies. Looking at the charmonium spectral functions shown in \cref{fig:CharmoniumNRQCD}, we find that in the S-wave channel the MEM and the standard BR method both seem to be affected by ringing, while the smooth BR method recovers a smooth continuum regime. The $J/\psi$ peak becomes insignificant between $T=200-210$MeV. For the $\eta_c$ the MEM and the smooth BR show rather similar behavior for the ground state, while only the smooth BR manages to avoid ringing in the continuum regime. Here no peak structures are discernible at around $T=185$MeV.

\begin{figure}
\centering
\includegraphics[scale=0.35]{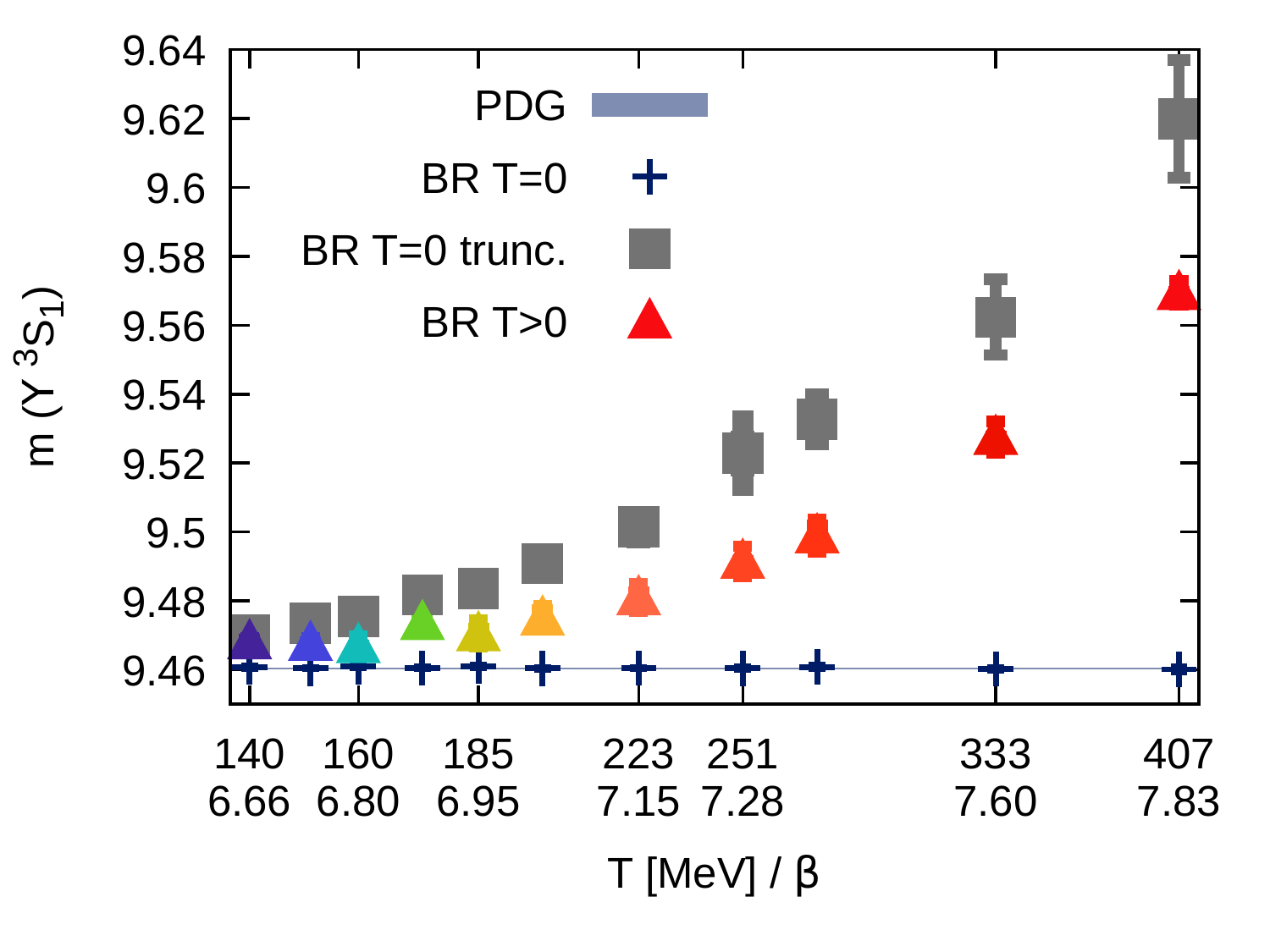}
\includegraphics[scale=0.35]{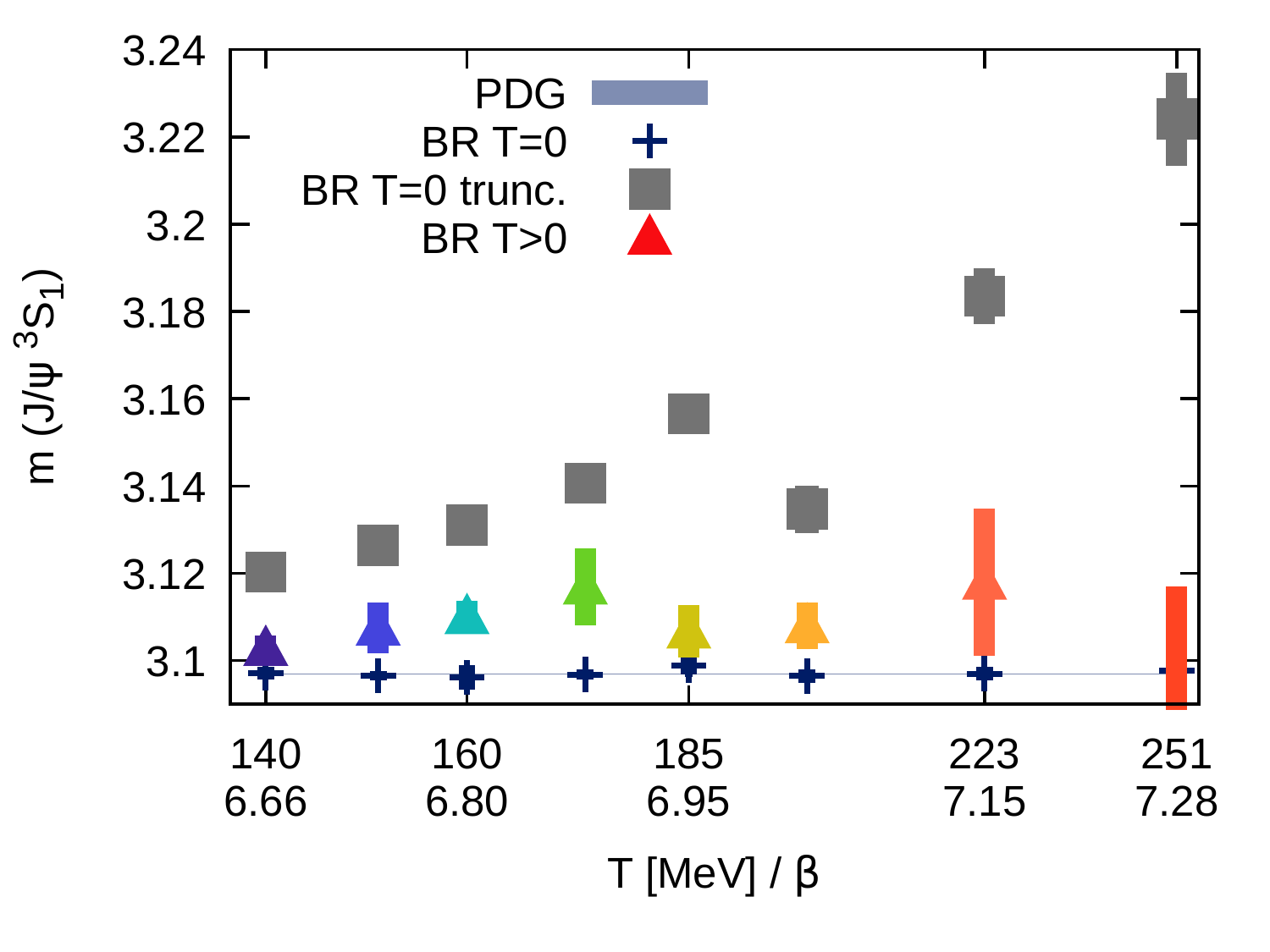}\\
\includegraphics[scale=0.35]{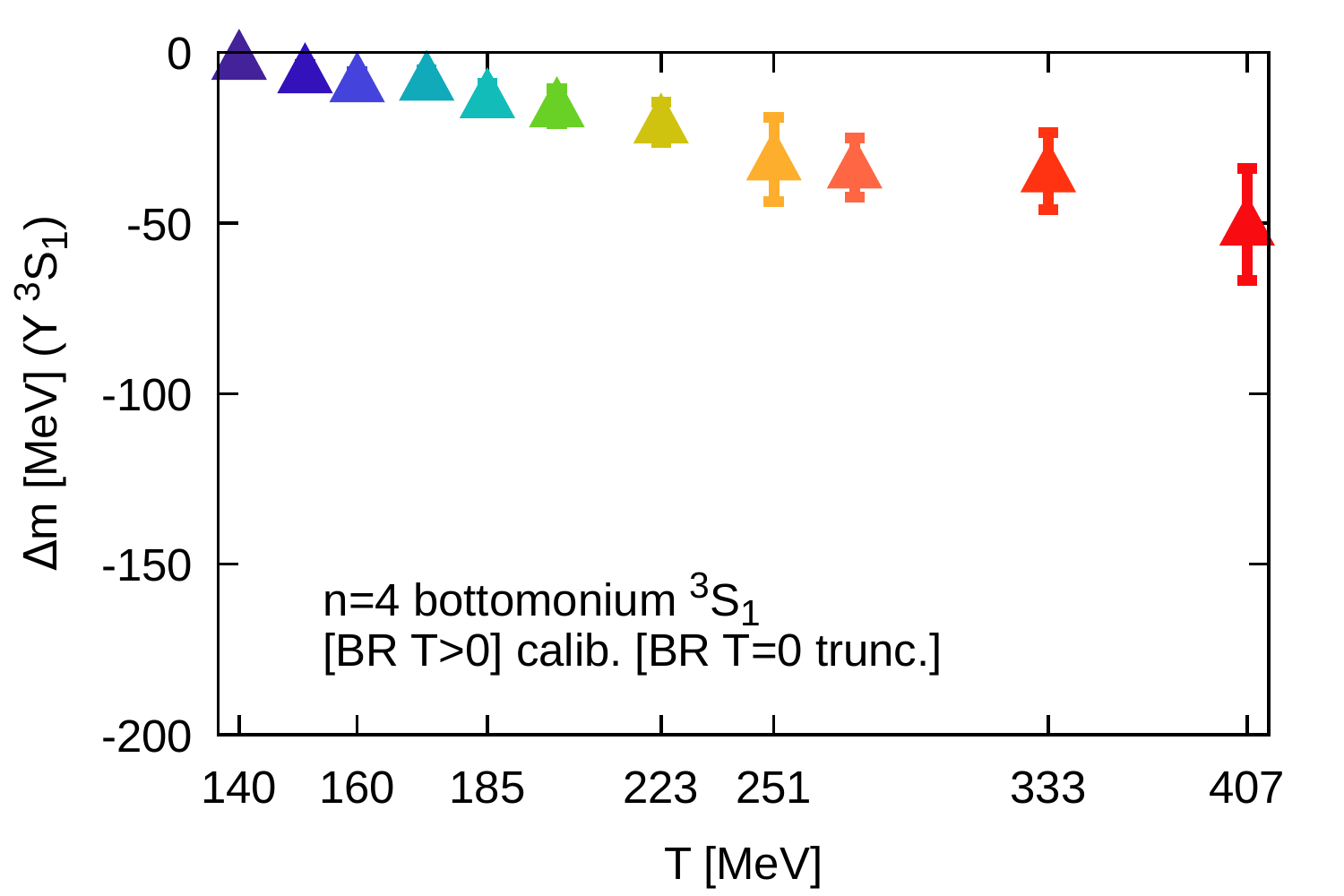}
\includegraphics[scale=0.35]{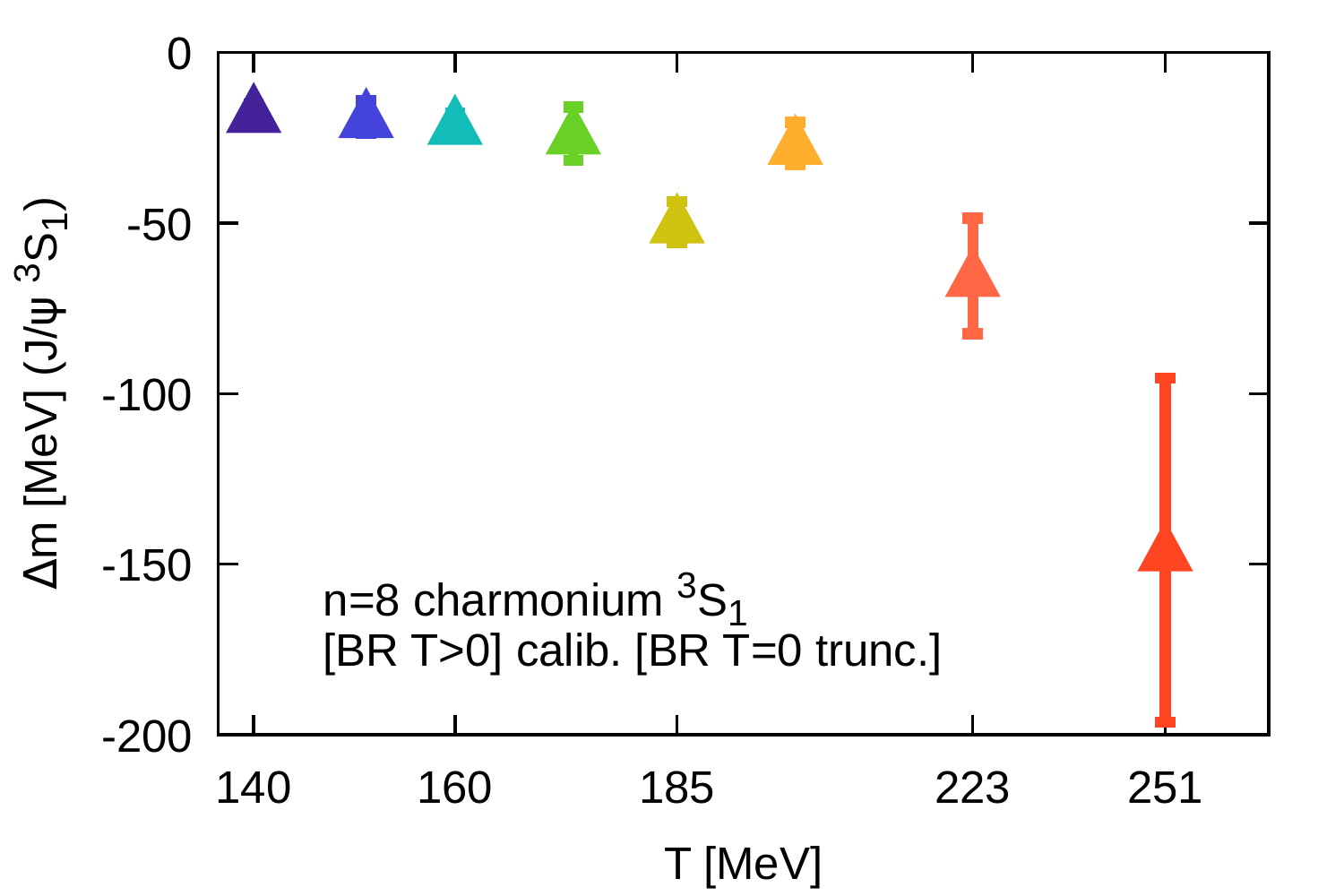}
\caption{(top row) The masses of the $\Upsilon$ (left) and $J/\psi$ ground state particles from the full $T=0$ correlator (blue crosses), from the truncated $T=0$ dataset (gray squares) and from the finite temperature correlator (colored symbols). (bottom row) The corresponding negative in-medium mass shifts. Figures adapted from Ref.~\cite{Kim:2018yhk}}\label{fig:NRQCDInmediumMass}
\end{figure}

The main quantitative finding of Ref.~\cite{Kim:2018yhk} is an improved determination of the in-medium mass shifts of the ground state quarkonium particles. In the top row of \cref{fig:NRQCDInmediumMass} the raw masses  entering the analysis are shown for bottomonium (left) and charmonium (right) S-wave channels. Errorbars include both statistical and systematic uncertainties from the Bayesian spectral reconstruction and the ground state peak fit. The blue crosses denote the ground state mass obtained from reconstructing the full $T=0$ datasets, while the gray boxes are obtained from the truncated $T=0$ datasets, which feature the same imaginary time extent as the data at $T>0$. The actual $T>0$ mass estimates are given by the colored symbols. A first naive comparison by eye of the blue crosses and colored symbols would lead to the (premature) conclusion that the in-medium masses lies above the vacuum ones. Since at $T>0$ not only the spectral function changes but also the Euclidean time extend is significantly reduced the authors of Ref.~\cite{Kim:2018yhk} argue that instead the gray boxes should be taken as correct reference point. This changes the conclusion significantly leading to negative mass shifts, which are plotted in the bottom row. As we will see in the next section the values obtained here are compatible with the results obtained from non-relativistic spectral functions computed using model potentials. Note that for bottomonium at $T=140$MeV  virtually no difference between the colored triangle and the gray box is found, consistent with no in-medium modification seen in the correlator ratio. For charmonium on the other hand one finds a difference beyond the uncertainties, which again corresponds qualitatively with the changes present in the correlator ratios. The fact that the mass shift is negative agrees with the upward bend observed in the correlator ratios.

With several consistent arguments presented here for the mass shifts in NRQCD to be negative it has to be understood how this stack up to the results in the relativistic formulation. Crosschecks with the reconstructed correlator performed in Ref.~\cite{Kitazawa:2018xbl} e.g. seem to indicate that a positive mass shift is obtained instead. Whether this difference is related to a deficiency of the lattice NRQCD implementation (e.g. due to radiative corrections) or whether it is a reconstruction artifact of the MEM used in the relativistic study, remains to be seen.

\begin{figure}
\centering
\includegraphics[scale=0.35]{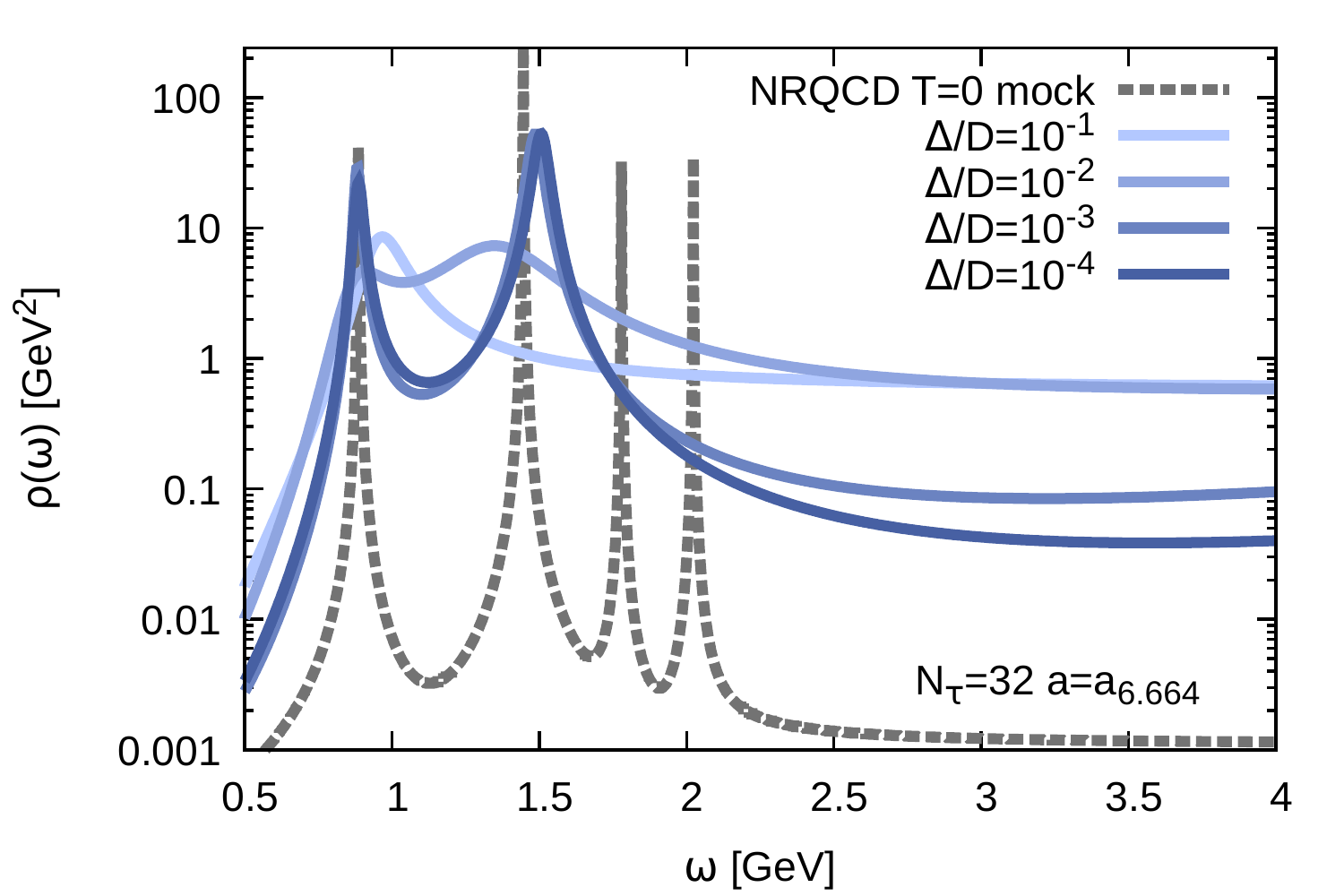}
\includegraphics[scale=0.35]{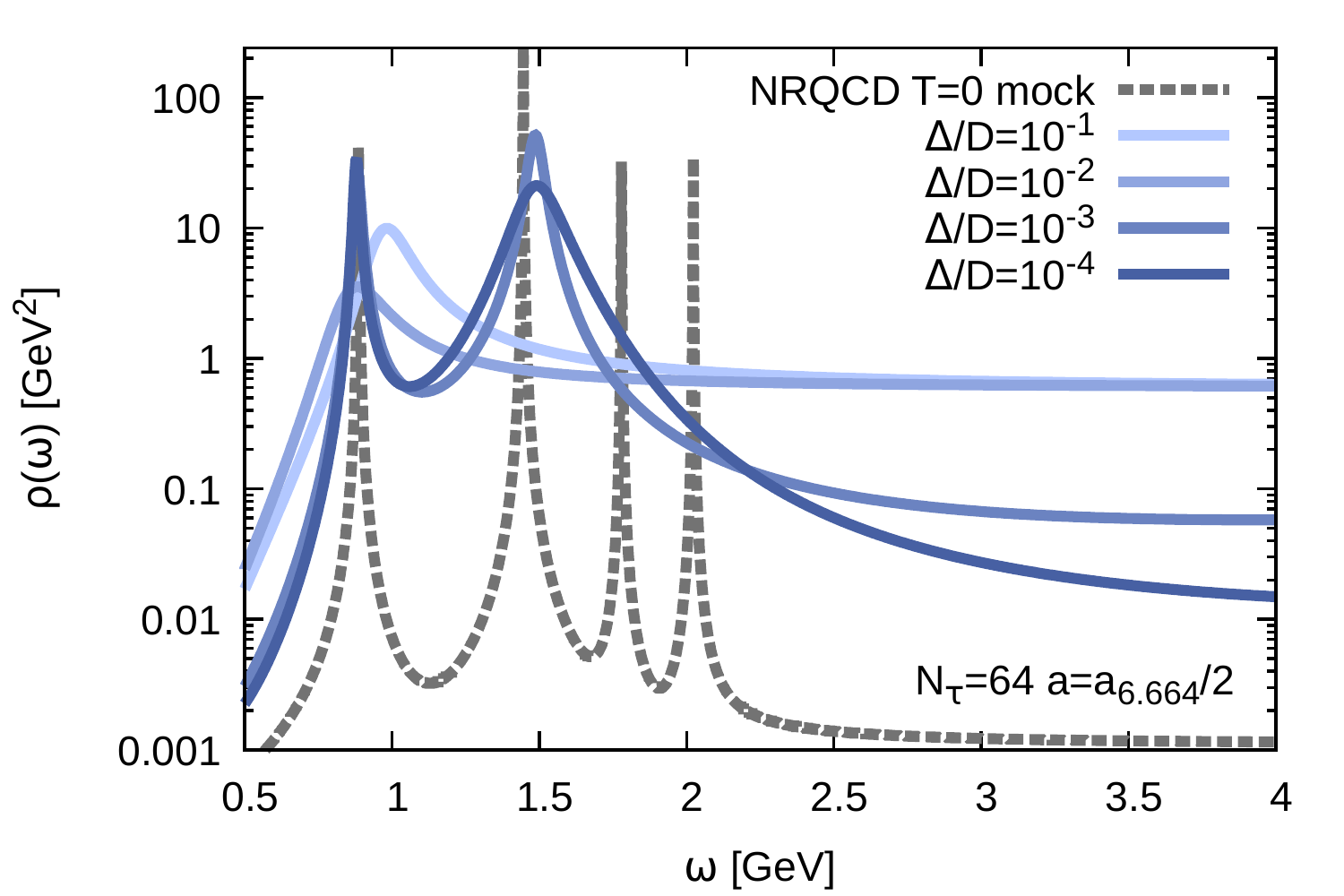}
\caption{(left) mock data test of reconstructing a NRQCD bottomonium S-wave like spectrum without a continuum contribution using $N_\tau=32$ points at a realistic coarse lattice spacing of $a=0.117$fm. (right) Reconstruction of the same mock spectrum using double the number of Eulcidean input datapoints and half the lattice spacing. Figures adapted from Ref.~\cite{Kim:2018yhk}}\label{fig:MockNRQCDtst}
\end{figure}

Up to this point only ground state properties have been discussed in finite temperature NRQCD, since the excited state contributions already at $T=0$ are difficult to pin down in spectral reconstructions. To quantify the challenge ahead one can take a look at the mock data tests shown in \cref{fig:MockNRQCDtst}. The left panel shows the BR method reconstructions (colored solid) of a $T=0$ bottomonium spectrum, as expected to be present in an NRQCD setting. To focus on the bound state reconstruction, no continuum has been added in the mock input spectrum. Discretized with $N_\tau=32$ and a realistic $a=0.117$fm one finds that sub-percent precision in the input data is required to get hold of the second peak. In the right panel we show the same mock spectrum now reconstructed with double the number of input points at half the lattice spacing. The bad news is that no significant improvement in the bound state reconstruction is obtained. On the other hand if a continuum structure is added to the mock spectrum it is seen that a smaller lattice spacing does improve the reconstruction of the UV part of the spectral function. One might think that by improving the quality of the reconstruction algorithm the problem might be solved. That this also is not the magic bullet has been shown by tests of the information content of the correlators performed in Ref.~\cite{Kim:2018yhk}. After subtracting from the $T=0$ correlator the ground state peak contribution, as well as the second exponential falloff, usually associated with the first excited state, only a very small number of convex points remain from which all the intricate structure of the higher lying states, as well as the continuum need to be determined. Even with a perfect reconstruction algorithm this will remain challenging. One possibility is to consequently deploy anisotropic lattices (as is done by the FASTSUM collaboration and in planning at the HotQCD collaboration), to have the UV part of the spectrum better resolved. In the long run it appears that improvements in simulation algorithms e.g. an extension of the multilevel algorithm to dynamical QCD would be required for substantial progress in the reconstruction of the bound state features.

\begin{summary}
Lattice NRQCD provides an alternative discretization of heavy quarks on the lattice, which has been applied to the study of both bottomonium and charmonium at finite temperature. Due to the absence of transport contributions and full $N_\tau$ individual correlator datapoints available, a robust picture of ground state in-medium modification has emerged. Correlator ratios show a hierarchical ordering of the overall in-medium modification with the vacuum binding energy of the states in that channel. Their temperature dependence also hints at the in-medium particles to become lighter as temperature increases. Spectral reconstructions have been carried out using a range of different Bayesian methods, which has significantly improved the understanding of the involved methods uncertainties. In addition the effects of diminishing access to Euclidean time have been elucidated. In turn very similar patterns for the disappearance of bound state features in the bottomonium sector are found among different groups. In medium mass shifts extracted from the reconstructed charmonium and bottomonium spectral functions show negative values, consistent with the findings of the correlator ratios, as well as with spectral functions computed with model potentials in the next section. Improvements of the spectral reconstruction results obtained so far, especially for excited states will require substantial efforts by the whole lattice community.
\end{summary}

\subsubsection*{In-medium quarkonium from a lattice vetted potential model}
\label{sec:specfrompNRQCD}

While the use of the effective field theory NRQCD on the lattice has already led to significantly improved understanding of the ground state in-medium properties in both bottomonium and charmonium, we have seen that so far excited states, as well as the continuum are not well captured. To progress in this direction we can turn to the effective field theory pNRQCD, which allows to derive the proper real-time in-medium potential systematically from QCD.

The study of quarkonium in-medium properties based on potentials has a long history. Until recently however only model potential were utilized, starting with works that used purely real potential models (for an overview see e.g. Ref.~\cite{Satz:2005hx}), such as the color singlet free energies (see e.g. Ref.~\cite{Mocsy:2007jz}) or internal energies. With the realization that the proper real-time potential is complex, first computations of the in-medium spectral functions using the perturbatively evaluated real-time potential were carried out in Ref.~\cite{Burnier:2007qm}, followed by models combining a lattice inspired real part with the perturbative imaginary part \cite{Mocsy:2013syh}. Shortly after the first non-perturbative lattice determination of the proper real-time potential had been achieved, it became a vital ingredient in spectral function computations in Refs.~\cite{Burnier:2015tda,Burnier:2016kqm}.

There are several generic features of in-medium spectral functions, which can directly be related to the form of the potential. At $T=0$ the energy at which string breaking sets in in ${\rm Re}[V]$ indicates the position of the open-heavy flavor threshold. Generically finite temperature effects lead to a weakening of the real part. For very small Debye masses, string breaking dominates the asymptotic flattening off, while at some point the inverse of the Debye mass becomes smaller than the string breaking radius and screening takes over. Note that contrary to a purely Coulombic term that always asymptotes to zero, the asymptotic value of ${\rm Re}[V]$ reduces monotonously from the $T=0$ value, indicating that the continuum threshold moves to lower and lower energies. As the binding energy of a state, defined via spectral functions, is computed from the energy distance between the continuum threshold and the position of the bound-state peak a weakening of the in-medium binding is thus expected to occur. While for a purely real potential genuine bound state peaks remain at finite temperature, the presence of an imaginary part leads to a finite width (and essentially additional no shift) in the spectral function, indicating the dynamical nature of the $Q\bar{Q}$ pair in the medium.

The computation of spectral functions proceeds via solving the Schr\"odinger equation for the unequal time and point split meson correlator \(D^>(t;\mathbf{r},\mathbf{r'})\). In the vector channel it reads 
\begin{equation}
\label{eq:spectral_schro}
i\partial_t D^{>}(t;\mathbf{r},\mathbf{r'})=\left[\hat{H}\mp i\lvert\mathrm{Im}V(r)\rvert\right]D^{>}(t;\mathbf{r},\mathbf{r'}); \quad t\gtrless 0,
\end{equation}
with the Hamiltonian defined as
\begin{equation}
\label{eq:spectral_ham}
\hat{H}=2m_Q-\frac{\nabla_{\mathbf{r}}^2}{m_Q}+\frac{l(l+1)}{m_Qr^2}+\mathrm{Re}V(r).
\end{equation}
Note that the in-medium potential which we will use to implement this time evolution was defined in the static limit in \cref{eq:realtimepotedef}, i.e. it denotes the leading order contribution in a systematic expansion in the finite heavy quark velocity. This entails that at this point no finite velocity, e.g. spin dependent, corrections have been included. In the following we will use the static potential to approximately describe the evolution of quarkonium with a finite mass. After evolving to late enough Minkowski time the Fourier transform can be reliably computed
\begin{equation}
\label{eq:cor_Four}
\tilde{D}(\omega,\mathbf{r},\mathbf{r'})=\int_{-\infty}^{\infty}\mathrm{d}t\; e^{i\omega t}D^{>}(t;\mathbf{r},\mathbf{r'}).
\end{equation}
The vector channel spectral function follows by considering its imaginary part and taking the limit of vanishing quark-antiquark separation, which recovers the appropriate frequency space current correlator
\begin{equation}
\label{eq:spectral_limit}
\rho^{V}(\omega)=\lim_{|\mathbf{r}-\mathbf{r'}|\to 0}\frac{1}{2}\tilde{D}(\omega;\mathbf{r},\mathbf{r'}).
\end{equation}

In the Appendix A of Ref.~\cite{Burnier:2007qm} an efficient implementation of the above prescription has been worked out for both vector channel (S-wave) and scalar channel (P-wave) spectral functions. It is directly formulated in frequency space and carefully considers the involved limiting procedure. The reference  provides a readily implementable prescription for practical use. Since to low order in the heavy quark velocity expansion it can be argued that 
\begin{equation}
\label{eq:rho_all}
\rho^P\simeq -\frac{1}{3}\rho^{V};\quad\rho^{A^0}\simeq -\frac{1}{3}\rho^V;\quad\rho^{\mathbf{A}}\simeq 2\rho^S,
\end{equation}
all relevant spectra can be obtained. In case of a purely Coulombic potential, due to its additional symmetries, it may happen that the vector and scalar channel mix with a numerically small contribution.

To achieve an accurate description of the in-medium properties of quarkonium all input parameters of the above Schr\"odinger equation need to be evaluated in a realistic setting. For the potential this means that we require continuum extrapolated lattice data, as well as an appropriately renormalized quark mass. While there are ongoing efforts to extract the static interquark potential on ensembles close to the continuum (see e.g. Ref.~\cite{Petreczky:2018xuh}) no genuine continuum extrapolation has been computed so far. This necessitates manual continuum corrections and we discuss here the strategy introduced in Ref.~\cite{Burnier:2015tda}. 

As it was found that the Gauss-law parametrization successfully reproduces the real part of the lattice potential and shows good agreement with the tentatively extracted values of the imaginary part, it has been used in the literature to implement the in-medium effects. In a first step its vacuum parameters are tuned in a phenomenological fashion. I.e. they are varied until the solution of the corresponding S-wave and P-wave Schr\"odinger equation reproduces the PDG values of the different ground state masses. The finite temperature physics is then introduced by using appropriately rescaled values of the Debye mass found on the lattice.

For the quark mass parameter we consider first the bottomonium system, as here pNRQCD is expected to work reliably. In addition the matching to QCD can be implemented perturbatively, since the bottom mass is much larger than $\Lambda_{\rm QCD}$. As has been worked out in Ref.~\cite{Pineda:2001zq} the appropriate mass to use in the context of the Schr\"odinger equation is the renormalon subtracted mass. Within this renormalization scheme, the ambiguities in the perturbative definition of the pole mass are reshuffled into the definition of the constant part of the potential, which contains the same ambiguity, effectively canceling the two and in turn leading to a well defined Schr\"odinger equation. For bottom quarks the renormalon subtracted mass takes on the value
\begin{equation}
\label{eq:m_B}
m_b^{RS'}=4.882\pm 0.041~\mathrm{GeV}.
\end{equation}
Note that this value is different from both the often deployed bottom pole mass $m_b^{\rm pole}=4.93$GeV, as well as the conventionally perturbatively defined $\overline{MS}$ mass $m_b^{\overline{MS}}(m_b^{\overline{MS}})=4.18$GeV (for a relation between the latter two see e.g. Refs.~\cite{Melnikov:2000qh,Marquard:2007uj}). Indeed we find that $m_b^{RS'}$ among the three different masses provides the best $\chi^2$ for the vacuum parameter fit.

This mass now enters the tuning procedure for the vacuum parameters of the Gauss-law parametrization that allows one to successfully reproduce the masses of the four low lying S-wave states \(\Upsilon(1S)-\Upsilon(4S)\). By manually introducing a flattening of the string-like part of the Cornell potential at a characteristic string-breaking scale $R_{\rm sb}$ also the location of the B meson threshold can be reproduced. The consistency of the tuning procedure has been checked by subsequent comparison to the spin averaged mass of the P-wave states \(\chi_{b0}(1P)-\chi_{b0}(3P)\) not included in the original fit. The best fit values obtained in the most recent study in Ref.~\cite{Lafferty:2019jpr} are given by
\begin{equation}
\alpha_S =0.513\pm 0.0024~\mathrm{GeV}, \quad \sqrt{\sigma}=0.412\pm 0.0041~\mathrm{GeV},\quad c=-0.161\pm 0.0025~\mathrm{GeV}, \quad r_{\rm SB}=1.25\pm0.05{\rm fm} \label{eq:phenopotbestfit}
\end{equation}
and the resulting vacuum properties of bottomonium states are listed in \cref{tab:b_fam}.

Once the values for the static vacuum potential are fixed, only the charm mass remains to be set. Since its physical value lies much closer to $\Lambda_{\rm QCD}$ than the bottom mass a similar robust perturbative renormalon subtracted mass is not available. Instead one may use the fact that the static potential in pNRQCD is universal among different flavors and thus the parameters of \cref{eq:phenopotbestfit} may also be used to compute the charmonium vacuum states. This allows to find the optimal charm quark mass by fitting to the PDG masses of the S-wave states (\(J/\psi, \psi^\prime)\). As expected finite velocity corrections do already play a role for charmonium and thus the masses are less accurately reproduced than in the bottomonium case using just the static potential. The corresponding best fit value is 
\begin{equation}
\label{eq:charm_mass_fit}
m_c^{\mathrm{fit}}=1.4692~\mathrm{GeV}.
\end{equation} 
and the vacuum properties of the resulting states are given in \cref{tab:charm_fam}.

\begin{table}
\begin{center}
\begin{tabular}[t!]{| c | c | c | c | c || c | c | c | }
\hline
 & \(\Upsilon(1S)\) & \(\Upsilon(2S)\) & \(\Upsilon(3S)\) & \(\Upsilon(4S)\) & \(\chi_{b}(1P)\) & \(\chi_{b}(2P)\) & \(\chi_{b}(3P)\) \\ \hline\hline
\(m~\mathrm{[GeV]}\) & 9.4603 & 10.023 & 10.355 & 10.569 & 9.931 & 10.273 & 10.534 \\ \hline
\(m^{PDG}~\mathrm{[GeV]}\) & 9.4603 & 10.023 & 10.355 & 10.579 & 9.888 & 10.252 & 10.534 \\ \hline
\(\langle r\rangle~\mathrm{[fm]}\) & 0.2918 & 0.5878 & 0.8697 & 1.0999 & 0.48 & 0.786 & 1.017 \\ \hline
\(\bar{m}^{PDG}_{B\bar{B}}-m~\mathrm{[GeV]}\) & 1.1 & 0.535 & 0.203 & -0.011 & 0.627 & 0.286 & 0.024 \\ \hline
\end{tabular}
\end{center}
\caption{Vacuum properties of relevant bottomonium particles based on the best fit parameters of \cref{eq:phenopotbestfit} and the renormalon subtracted mass of \cref{eq:m_B}.}
\label{tab:b_fam}
\end{table}

\begin{table}
\begin{center}
\begin{tabular}[t!]{| c | c | c ||  c | c | }
\hline
 & \(J/\psi\) & \(\psi^\prime\) & \(\chi_{c}(1P)\) & \(\chi_{c}(2S)\)\\ \hline\hline
\(m~\mathrm{[GeV]}\) & 3.0969 & 3.6632 & 3.5079 & 3.775 \\ \hline
\(m^{PDG}~\mathrm{[GeV]}\) & 3.0969 & 3.6861 & 3.4939 & 3.9228 \\ \hline
\(\langle r \rangle ~\mathrm{[fm]}\) & 0.565 & 1.249 & 0.672 & 1.109 \\ \hline
\(\bar{m}^{PDG}_{D\bar{D}}-m~\mathrm{[GeV]}\) & 0.642 & 0.076 & 0.231 & -0.036 \\  \hline
\end{tabular}
\end{center}
\caption{Vacuum properties of relevant charmonium particles based on the best fit parameters of \cref{eq:phenopotbestfit} and the best estimate of the charm mass of \cref{eq:charm_mass_fit}.}
\label{tab:charm_fam}
\end{table}

\begin{figure}
\centering
\includegraphics[scale=0.25]{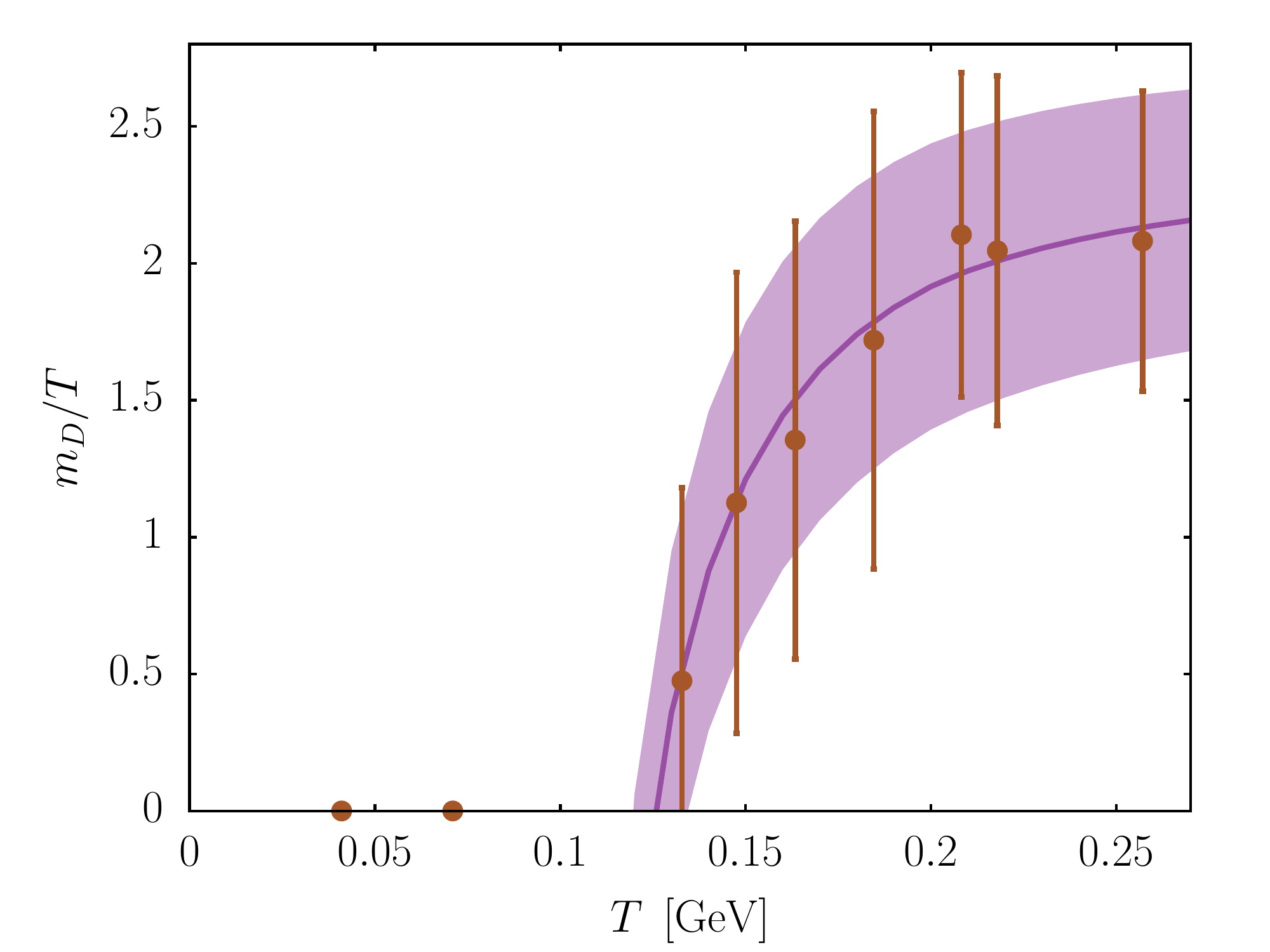}
\caption{The continuum corrected Debye mass parameter for the Gauss-law parametrization of the in-medium static interquark potential (points). The band denotes the best fit from a HTL inspired interpolation formula for the Debye mass. Figure adapted from Ref.~\cite{Lafferty:2019jpr}}\label{fig:DebyeMassContCorr}
\end{figure}

With the vacuum sector set up, the focus turns to the question of how to translate the Debye masses extracted at finite lattice spacing to values representative of continuum physics. The discretization leads to two main effects: first, since the masses of light quarks do not yet take their physcial values the crossover temperature also lies above its physical value. Secondly, the vacuum parameters of the Cornell potential contain a small lattice spacing dependence and also are not yet at their physical value. In Refs.~\cite{Burnier:2015tda,Lafferty:2019jpr} these artifacts are counteracted by considering the dimensionless ratio of the lattice Debye mass and the square root of the lattice string tension. It is multiplied by the square root of the physical string tension and then evaluated at a rescaled temperature where $T_C$ lies at $155$MeV.
\begin{align}
\label{eq:mD_cont}
m_D^{\mathrm{phys}}(t=T/T_c^{\mathrm{lat}})=\frac{m_D^{\rm lat}(t)}{\sqrt{\sigma^{\rm lat}(t)}}\sqrt{\sigma^{\mathrm{phys}}},
\end{align}
The values of the continuum corrected Debye mass with errorbars including both uncertainties from the lattice extraction and those from the correction procedure itself are shown in \cref{fig:DebyeMassContCorr}. The characteristic bending down of the ratio around the (now physical) crossover temperature is visible. In order to clearly expose the strength of the in-medium modification in the spectral functions, a temperature scan needs to be carried out for which an interpolation of the Debye mass will be a prerequisite. Amending the NLO expression for the Debye mass in \cref{eq:DebMassNLO} by a second non-perturbative correction term one ends up with the following expression
\begin{align}
\label{eq:mD_fit}
m_D^\mathrm{phys}=T g(\Lambda)\sqrt{\frac{N_c}{3}+\frac{N_f}{6}}&+\frac{N_cTg(\Lambda)^2}{4\pi}\mathrm{log}\Big[\frac{1}{g(\Lambda)\sqrt{\frac{N_c}{3}+\frac{N_f}{6}}}\Big]+ \kappa_1 Tg(\Lambda)^2+\kappa_2 Tg(\Lambda)^3,
\end{align}
taking here \(\Lambda=2\pi T\) as renormalization scale. The four loop results from Ref.~\cite{Vermaseren:1997fq} have been used to implement the running of the strong coupling $g$. For evaluating $g$, an appropriate value of \(\Lambda_{QCD}=0.2145~\mathrm{GeV}\) is deployed, initializing the renormalization group flow downwards from energies, where $N_f=5$ flavors are active. The two non-perturbative parameters $\kappa_1$ and $\kappa_2$ may now be fitted to the continuum corrected values of $m_D$. Ref.~\cite{Lafferty:2019jpr} reports best fit values of
\begin{equation}
\label{eq:mD_fit_res}
\kappa_1=0.686\pm 0.221,\qquad \kappa_2=-0.317\pm 0.052 .
\end{equation}
which implement the deviation from the perturbative result, needed to describe the downward trend of $m_D/T$ around $T_C$.

\begin{figure}
\centering
\includegraphics[scale=0.35]{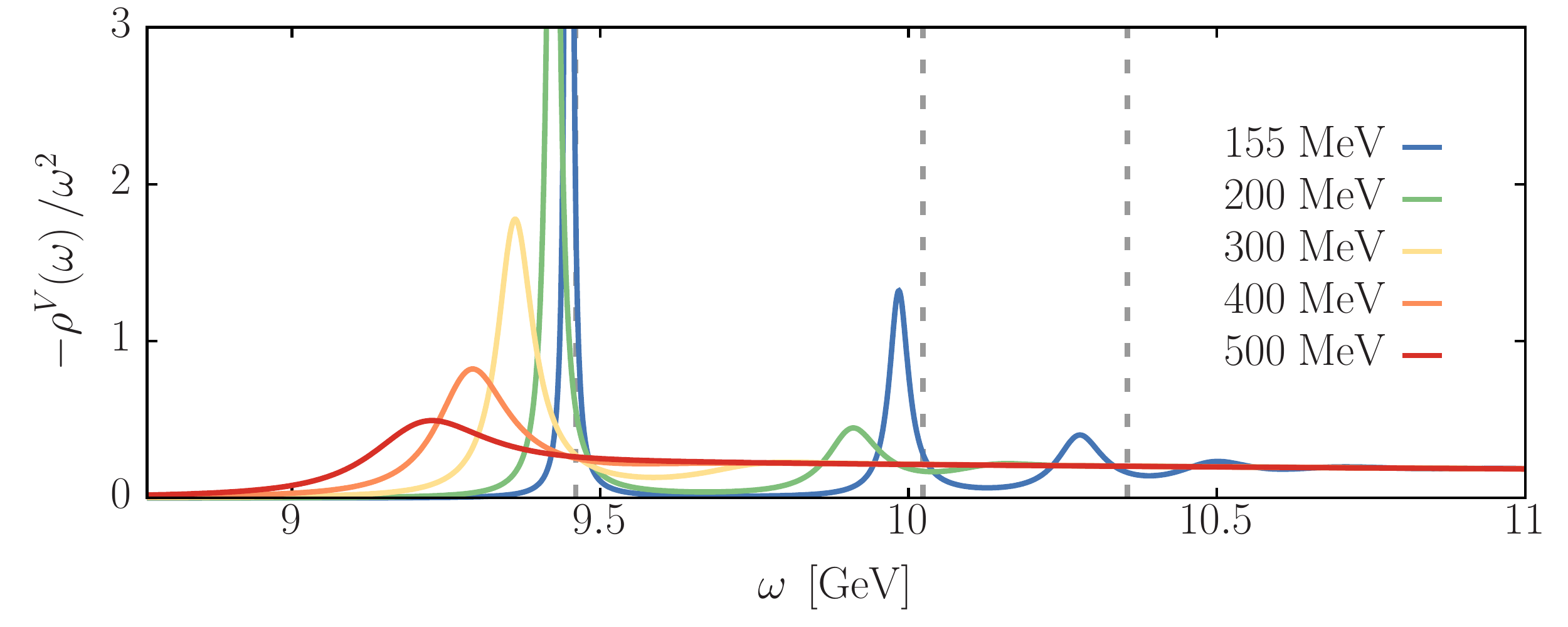}
\includegraphics[scale=0.35]{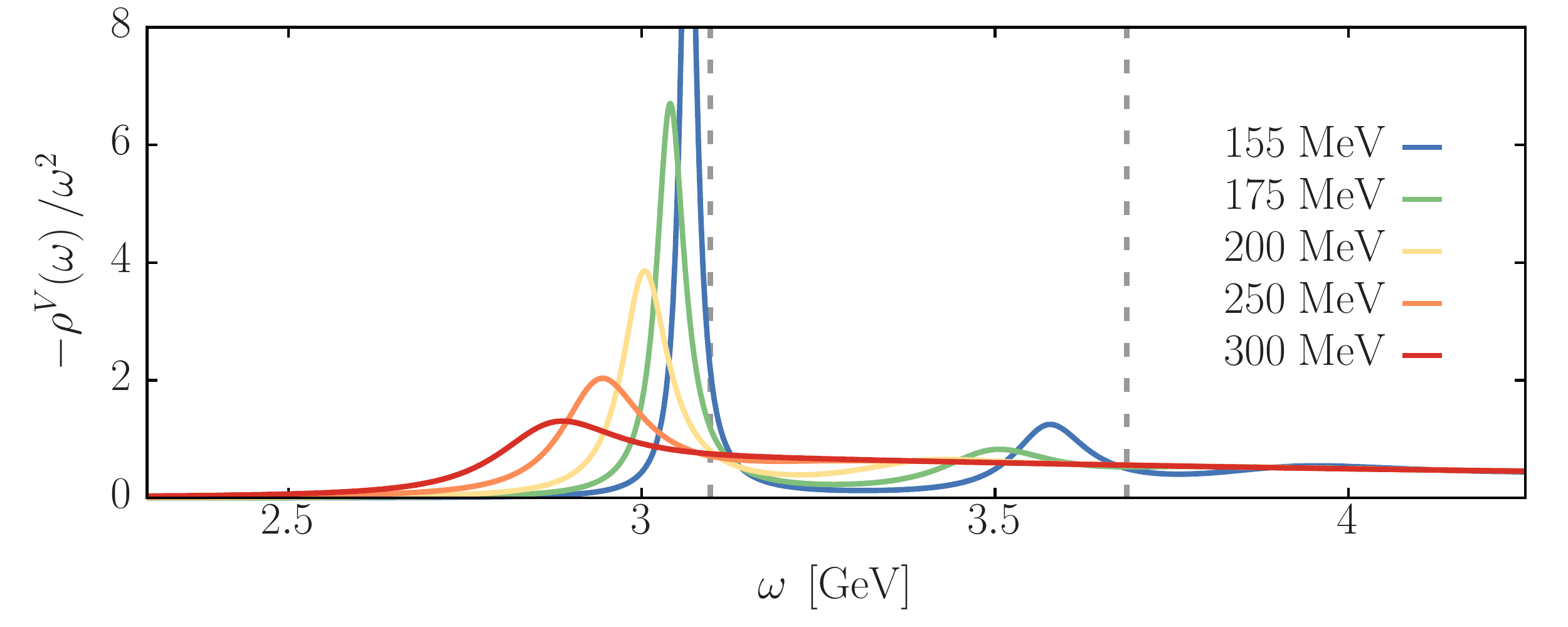}
\caption{The in-medium heavy quarkonium spectral functions in the S-wave channel for (top panel) bottomonium and (bottom panel) charmonium evaluated at different temperatures. The computation is based on the lattice vetted Gauss-law parametrization for the static in-medium potential. Figures adapted from Ref.~\cite{Lafferty:2019jpr}}\label{fig:pNRQCDSwaveSpectralFunc}
\end{figure}

All ingredients have thus been collected for the evaluation of the Schr\"odinger equation of \cref{eq:spectral_schro}, using a lattice vetted parametrization of the static in-medium potential via the Gauss-law parametrization. The latest results for the S-wave spectral functions are shown in \cref{fig:pNRQCDSwaveSpectralFunc}. Characteristic in-medium effects are clearly visible. The bound state peaks at $T=0$ are delta-peaks, as here only the strong interaction contribution to the physics is included. The peaks start to broaden at finite temperature and move to lower frequencies, indicating that the particles become lighter in-medium. Since at the same time the continuum also moves to lower energies, the in-medium binding energy actually reduces. Eventually a bound state remnant will merge with the continuum leading to a threshold enhancement, which reflects remnant correlations between the quark antiquark pair even if no genuine bound state structure is discernible. The in-medium modifications are found to be ordered hierarchically with the vacuum binding energy of each individual state.

\begin{figure}
\centering
\includegraphics[scale=0.25]{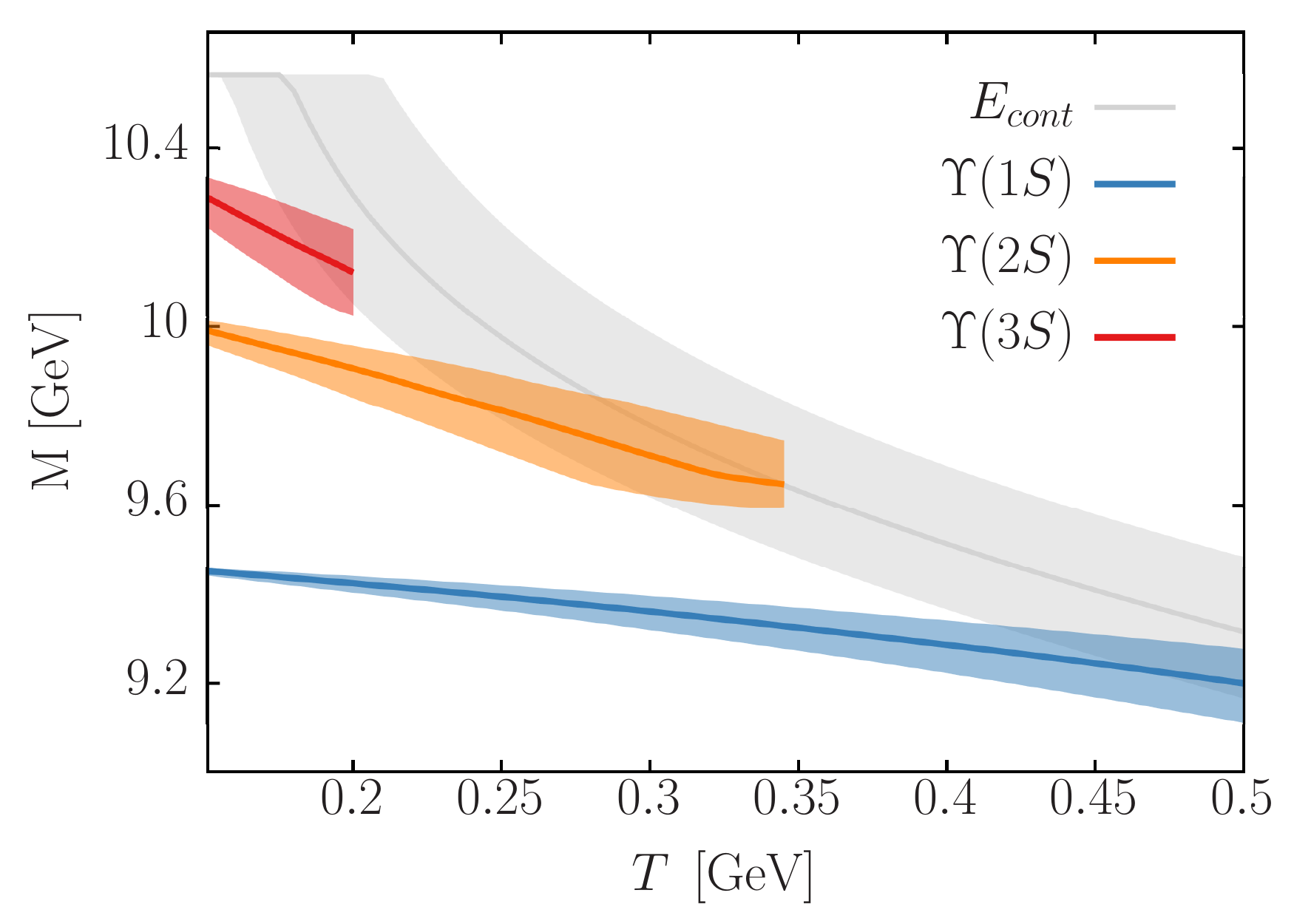}
\includegraphics[scale=0.25]{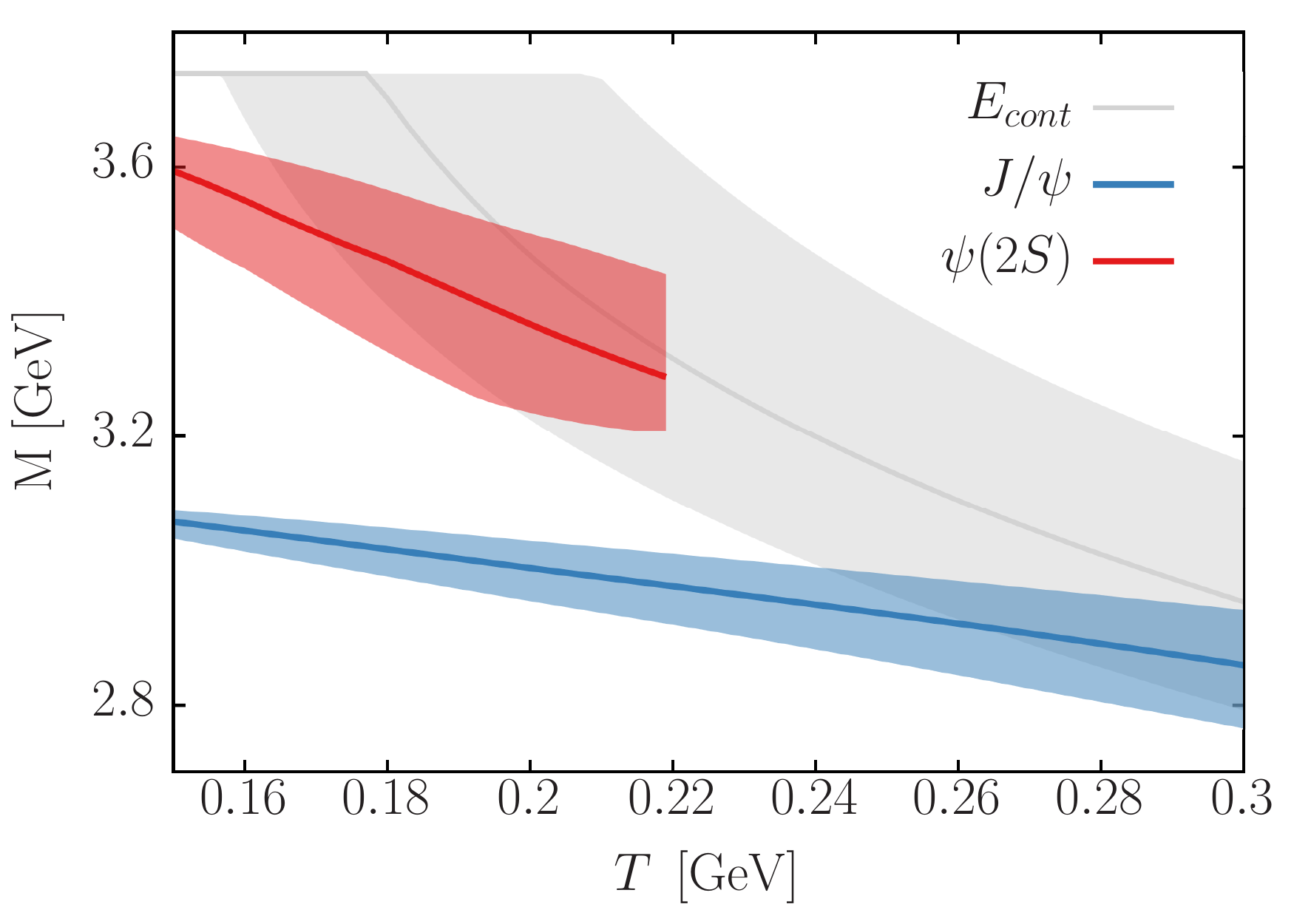}\\
\includegraphics[scale=0.25]{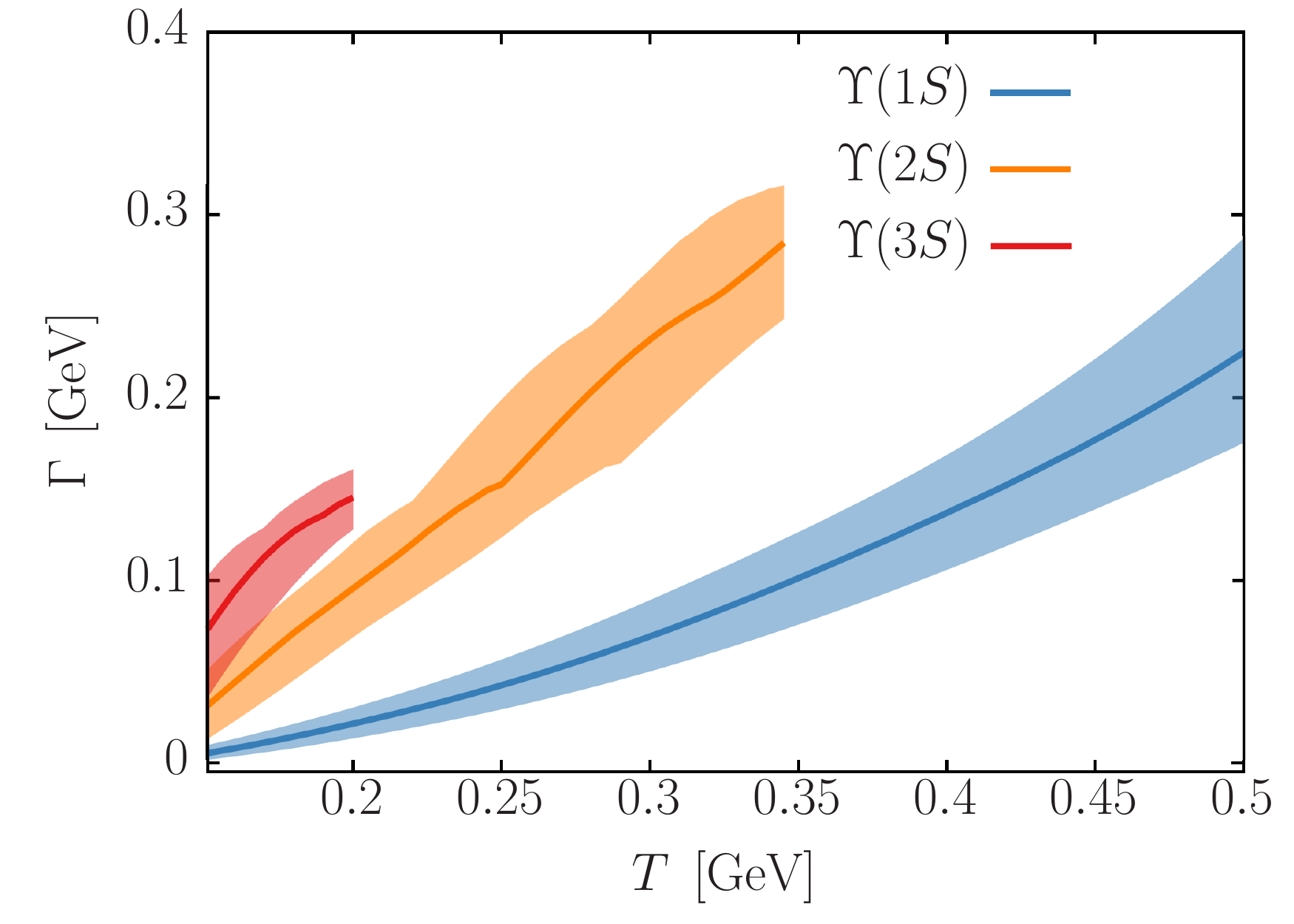}
\includegraphics[scale=0.25]{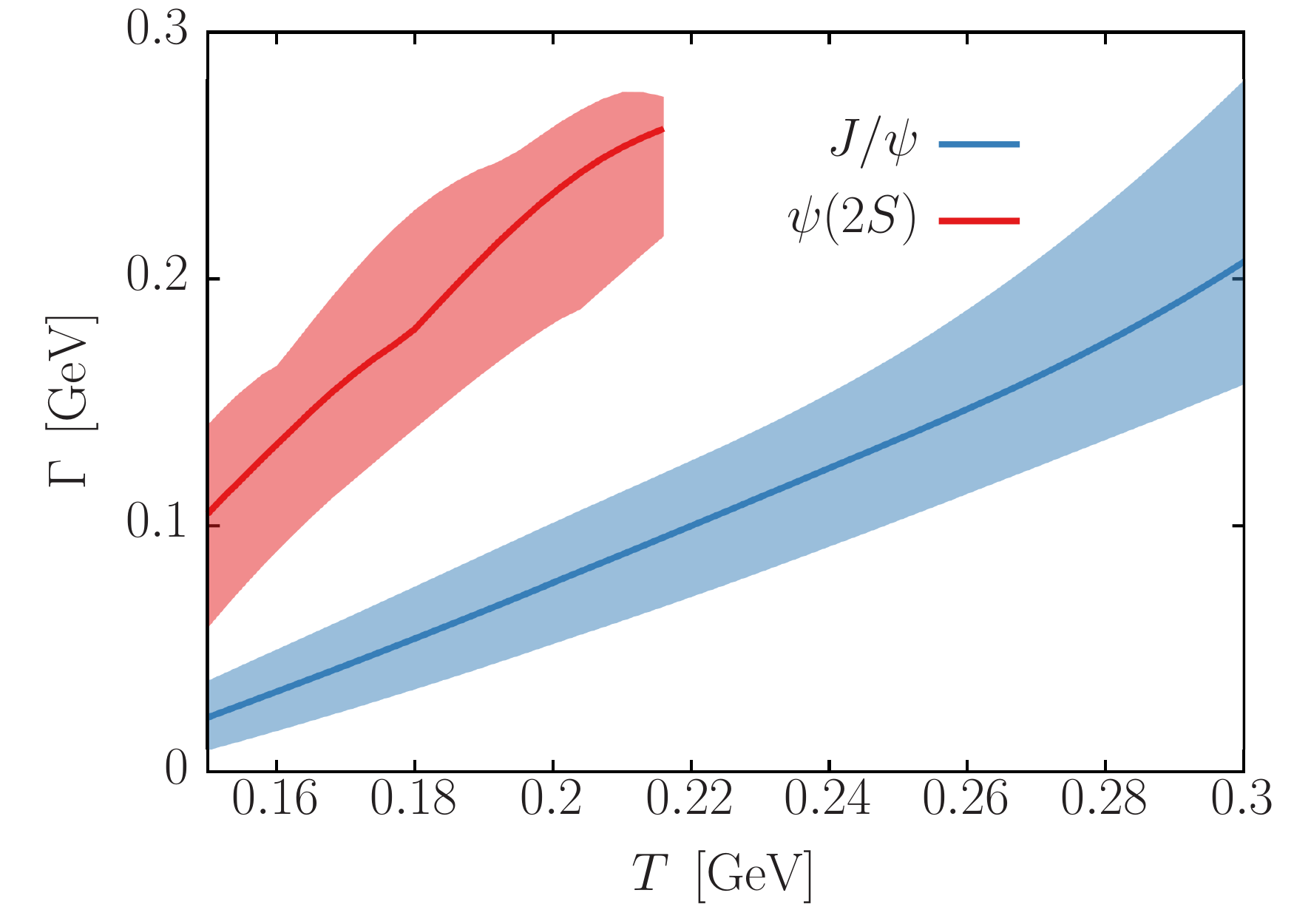}
\caption{ The in-medium properties of (left column) bottomonium and (right column) charmonium extracted via Breit-Wigner fits from spectral functions, computed from the lattice vetted Gauss-law parametrization of the complex in-medium static potential. The top row shows the in-medium masses, the bottom row the thermal widths. Figures adapted from Ref.~\cite{Lafferty:2019jpr}}\label{fig:pNRQCDInMediumProp}
\end{figure}

For a more quantitative exploration of the in-medium properties, the peak structures can be fitted to extract the corresponding particle mass and thermal width. Scattering theory suggests \cite{Taylor:1972} to use a skewed Breit Wigner of the following form
\begin{equation}
\label{eq:skewed_BW}
\rho(\omega\approx E)=C\frac{\left(\Gamma/2\right)^2}{\left(\Gamma/2\right)^2+\left(\omega-E\right)^2}+2\delta\frac{\left(\omega-E\right)\Gamma/2}{\left(\Gamma/2\right)^2+\left(\omega-E\right)^2}+C_1+C_2\left(\omega-E\right)+\mathcal{O}(\delta^2),
\end{equation}
with a skewing fit parameter $\delta$, as well as possible background terms $C_i$. The values for the in-medium mass and width according to the spectral functions of \cref{fig:pNRQCDSwaveSpectralFunc} are shown in \cref{fig:pNRQCDInMediumProp}. The left column corresponds to bottomonium, the right column to charmonium. In the top row the in-medium masses for the different states that are bound in vacuum are shown as colored solid lines, with the error bands being dominated by the uncertainty in the Debye mass determination. The lines end at the temperature where the in-medium peak remnant becomes too washed out for a fit to succeed. The gray line denotes the onset of the continuum.

The reduction in the in-medium quarkonium mass at first may appear counter intuitive, as perturbative pNRQCD computations (see e.f. Refs.~\cite{Brambilla:2010vq,Burnier:2007qm}) based on a Coulombic potential, predict the opposite behavior. Non-perturbatively the reduction emerges from the subtle interplay of the medium modification of the string part and the Coulombic part of the vacuum Cornell potential. In addition one may ask how the lowering of the in-medium quakronium mass compares to the mass gain $\delta m_Q$ for an individual parton, which takes place at finite temperature. That effect enters as a first correction to the static limit, i.e. it is supressed with the heavy quark mass. As argued in \cite{Burnier:2015tda} using perturbation theory, the expected values at $T=200$MeV would be $\delta m_c\approx7$MeV and $\delta m_b\approx2$MeV, which are insignificant compared to the shifts of tens to hundreds of MeV observed in \cref{fig:pNRQCDInMediumProp}. (Recently a prescription to compute the in-medium mass shift non-perturbatively from a Euclidean correlators has been put forward in Ref.~\cite{Eller:2019spw}.)

Note that the mass shifts observed here for the ground state particles are in agreement within the relatively large errorbands, with those extracted from direct lattice QCD determinations of the spectral function using the NRQCD discretization of the heavy quarks in \cref{fig:NRQCDInmediumMass}. This is reassuring, as it signals consistency between the two different non-relativistic approaches.

\begin{figure}
\centering
\includegraphics[scale=0.25]{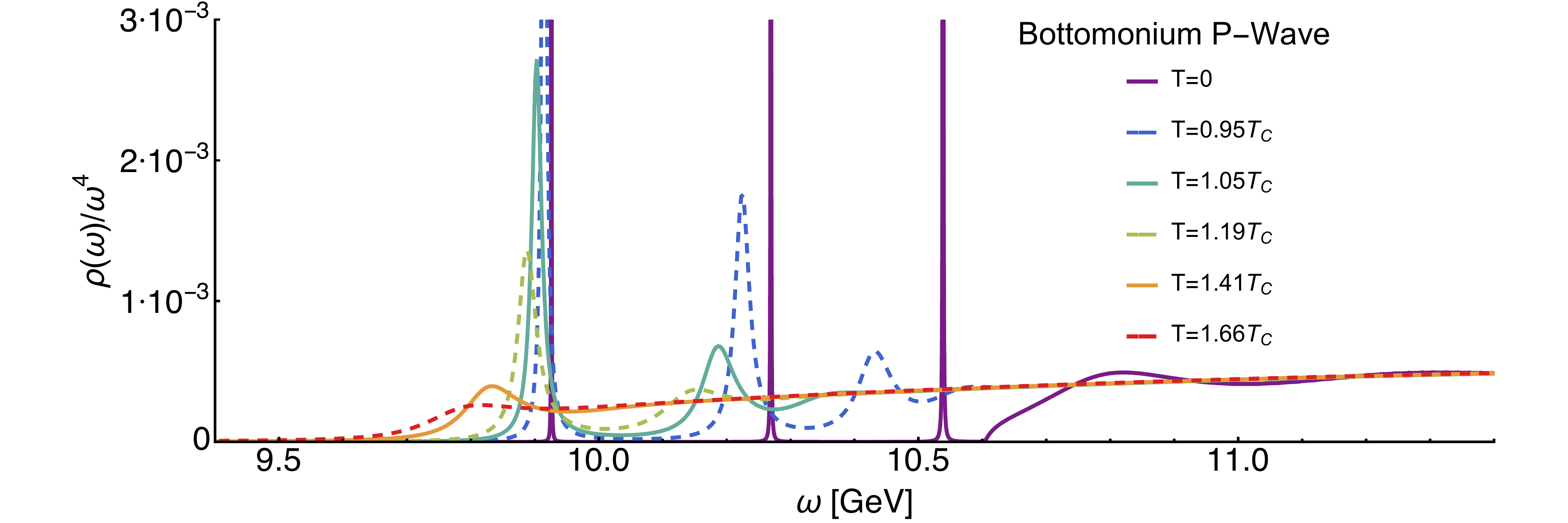}
\includegraphics[scale=0.25]{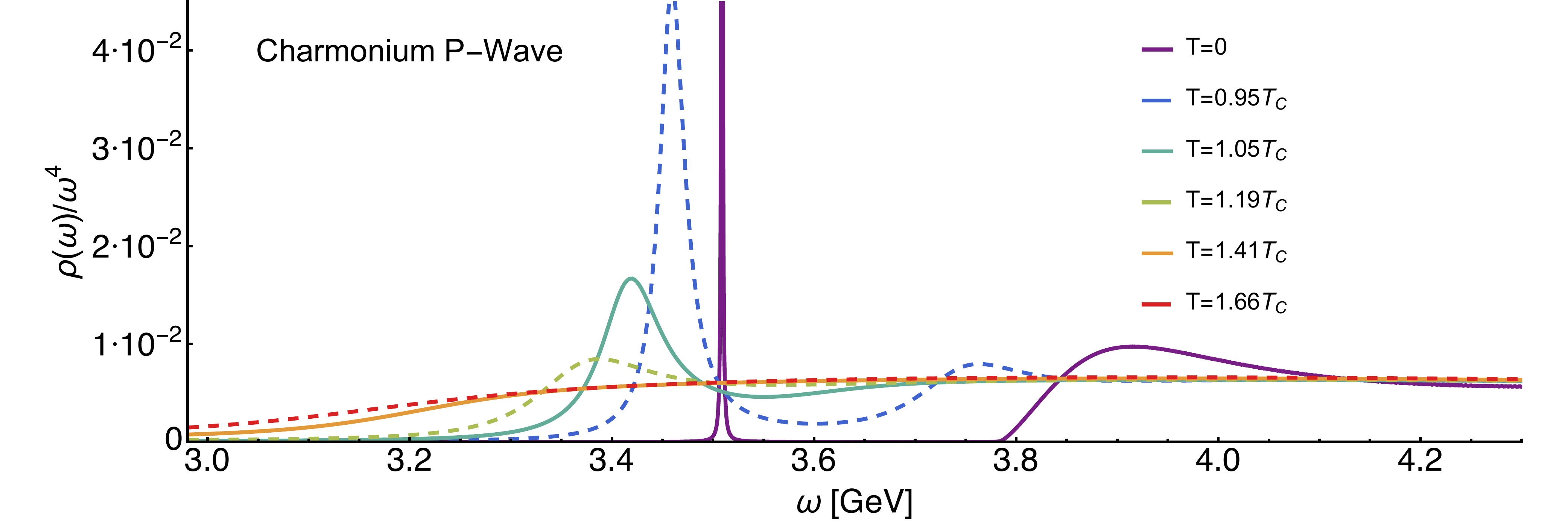}
\caption{The in-medium heavy quarkonium spectral functions in the P-wave channel for (top panel) bottomonium and (bottom panel) charmonium evaluated at different temperatures. The computation is based on a legacy implementation of the Gauss-law parametrization for the static in-medium potential. Figures adapted from Ref.~\cite{Burnier:2016kqm}}\label{fig:pNRQCDPwave}
\end{figure}

For the P-wave states, spectral functions have been computed using a legacy version of the Gauss-law parametrization in Ref.~\cite{Burnier:2016kqm} and the results are shown in Fig.\ref{fig:pNRQCDPwave}. The main difference to the S-wave case is the presence of the centrifugal term in the underlying Schr\"odinger equation. It was found that the centrifugal barrier, which it induces, leads to a slightly different pattern of weakening of the in-medium states. In the S-wave case the continuum approaches the in-medium peaks with increasing temperature and eventually swallows them, leaving only a threshold enhancement. For a P-waves its bound state remnant feature persists after being engulfed by the continuum, before it also washes out eventually.

\begin{summary}
The computation of in-medium quarkonium spectral functions from a Schr\"odinger equation with the static potential at finite temperature provides precise information on not only ground state properties but also those of in-medium excited states and the continuum. The price to pay is reduced accuracy in that so far no finite velocity corrections have been included.
The values of the potential available in lattice QCD can be incorporated via the Gauss-law parametrization vetted on appropriately continuum corrected numerical data. The in-medium effects manifest themselves in a characteristic manner: the delta-like vacuum state peaks broaden and shift to lower energies, before being swallowed by the continuum, which even more quickly moves downward in energy as temperature rises. This behavior is opposite to the predictions from perturbative pNRQCD and arises from an interplay of the medium modification of the string and Coulombic part of the Cornell potential. The values for the resulting negative in-medium mass shift are compatible with those obtained recently in direct lattice NRQCD studies, signalling consistency between different non-relativistic approaches.
\end{summary}

\subsubsection*{Further approaches to in-medium quarkonium}

Having focused on direct lattice QCD and effective field theory based strategies in the previous subsections, let us briefly touch on other related approaches which have contributed to improving our understanding of in-medium quarkonium.

The first approach to mention is based on finite temperature QCD sum rules, developed in their modern form in Refs.~\cite{Hatsuda:1991ez,Hatsuda:1992bv,Lee:1997zta}, which generalize the well established sum rules in vacuum (see e.g. Ref.~\cite{Shifman:1978bx}). By combining the operator product expansion of the in-medium meson current correlation function with a Borel transformation, this approach allows to probe differently weighted convolutions of the in-medium spectral functions. Non-perturbative information enters e.g. via the (quenched) lattice QCD determination of the matrix elements constituting the OPE. In particular the scalar and twist-2 gluon condensates, derived from the Lorentz decomposition of the field strength tensor play a role, which in practice are extracted via thermodynamic relations.

In the past decade, progress has been achieved by further combining the sum rule approach with Bayesian reconstruction methods, such as the MEM, in Ref.~\cite{Gubler:2011ua}. This in turn allows to determine charmonium in-medium spectral function and provides insight into the ground state medium modification. In addition sum rules have been used to derive constraints on the in-medium spectral function in Refs.~\cite{Gubler:2016hnf,Gubler:2017qbs}, which may be used as prior information in studies of spectral functions directly in lattice QCD. The sum rule approach has been used to study the real-part of the in-medium heavy quark potential in Ref.~\cite{Lee:2013dca} indicating preference for a value close to the color singlet free energies in agreement with findings in direct lattice QCD. A sum rule approach valid in the presence of strong magnetic fields has been introduced in Ref.~\cite{Cho:2014exa} with a special focus on the mixing of the S-wave hyperfine doublet $\eta_c$ and $J/\psi$ as discussed in Ref.~\cite{Cho:2014loa}. Extending the parameter of the Borel transformation, the Borel mass, into the complex plane, the sum rule approach has been generalized in Ref.~\cite{Araki:2014qya}, making possible to also extract spectral information on the excited charmonium states at finite temperature \cite{Araki:2017ebb}. The sum rule approach at $T=0$ reproduces the ground and excited state vacuum masses within $50-150$MeV. In addition, consistent with the non-relativistic approaches discussed before, it is found that the vacuum peaks move to lower energies as temperature is increased, the magnitude of the shift being a bit weaker than that found from the lattice QCD potential.

In the second approach the T-matrix formalism is used to compute bottomonium in-medium spectra from a real-valued interaction potential that enters a local interaction kernel of a Bethe-Salpeter equation. As the potential is provided as external input and it is not uniquely determined by comparisons with lattice data of e.g. the system free energies, two scenarios, the weak-binding and the strong-binding one are considered. In earlier studies (see e.g. Ref.\cite{Riek:2010fk}) a temperature dependent renormalization of the potential had been deployed which leads to significant in-medium mass shift to higher energies, incompatible with the results from the non-perturbative EFT potential. In the most recent state-of-the-art realization of the T-matrix approach reviewed in Ref.~\cite{Liu:2017qah} the potential is renormalized such that it agrees in the short distance regime at all temperatures. In turn the resulting charmonium spectra for both the weak and strong binding scenario are in qualitative agreement with the EFT potential computations showing negative mass shifts at finite temperature. The authors conclude that the strong coupling scenario, in which the potential governing the interactions among both light and heavy degrees of freedom is significantly larger than the color singlet free energies is preferred by their results. Since this potential is not directly related to the EFT potential derived for static quarks, there is apriori no tension between these results.

The third approach we mention here is based on the AdS/CFT correspondence. While we had briefly mentioned that the real-time potential from the Wilson loop may be computed in this approach it is also possible to directly compute the in-medium spectral functions. The challenge here lies in the fact that as discussed in Ref.~\cite{Hong:2003jm} by adding heavy quark analogues to ${\cal N}=4$ super Yang-Mills theory one obtains heavy-light meson states, that would be bound so deeply that it is instead favorable to produce light quark pairs to shield the charge of the heavy constituent. In addition the spatial extent of quarkonium analogues appears to be independent (at least in order of magnitude) of their radial excitation. It is difficult to identify these bound states with actual QCD quarkonium particles. In order to overcome these limitations of the top-down approach, instead AdS/CFT models, such as the soft-wall model (see e.g. \cite{Kim:2007rt,Fujita:2009wc}) have been constructed from the bottom up (see also \cite{Grigoryan:2010pj}) to incorporate e.g. additional scales that allow to produce a more realistic quarkonium behavior. 

The computation of spectral functions may be implemented via the identification of a vector field in the bulk with the vector meson current on the gauge theory side. Some model computations in the holographic approach, such as in Ref.\cite{Fujita:2009ca} produce in-medium modifications of quarkonium spectral functions that are qualitatively compatible with the QCD results, in that the in-medium peaks move to lower energies as temperature is increased. On the other hand there are also computations found in the literature, which produce opposite behavior, such as in Ref.~\cite{Braga:2017bml}. It appears that there is not yet a final consensus reached on the in-medium modification of quarkonium within the holography community.

\begin{summary}
The development of QCD sum rules in thermal equilibrium has opened a complementary route to investigating quarkonium in-medium properties. The approach has matured over the past two decades and its results on in-medium modification are consistent with those obtained directly from lattice QCD and effective field theory. The improved renormalization of the interaction potential in the T-matrix approach has allowed to obtain in-medium quarkonium spectral functions that are also in qualitative agreement with the EFT potential based results. The apparent discrepancy between the larger values of the T-matrix potential and the EFT potential is partially related to the fact that the two quantities are not the same, the former also governing the physics of the light degrees of freedom. In the bottom-up approach to the holographic description of QCD, quarkonium in-medium spectral functions have been computed. Different implementations of the bulk physics produce opposite in-medium modifications of the bound state features and it appears that a consensus in the community is still outstanding.
\end{summary}

\subsection{Quarkonium melting}
\label{sec:QQbarmelting}

In the preceding sections we have compiled a wealth of insight into the in-medium behavior of quarkonium immersed in a heat bath at finite temperature by studying the in-medium spectral functions and associated Euclidean correlation functions. This preparation in turn allows us to approach the question of quarkonium melting in this section.

Historically the theoretically well posed question of the stability of quarkonium states at finite temperature has been intimately connected to the experimental measurements of quarkonium yields in relativistic heavy-ion collisions. Starting with the seminal work by Matsui and Satz in Ref.\cite{Matsui:1986dk} on quarkonium suppression by QGP formation (see also Ref.\cite{Hashimoto:1986nn}), as well as through the proposal of the sequential suppression scenario in Ref.~\cite{Karsch:2005nk} this has lead to an unfortunate entangling of the question of quarkonium in-medium binding in equilibrium with the intricacies related to quarkonium production in a highly non-equilibrium environment, such as present in a heavy-ion collisions. 

Without diminishing the important role played by quarkonium in the study of heavy-ion collisions, we shall try to disentangle the two, by clearly distinguishing between {\it quarkonium melting} and {\it quarkonium suppression}. The former we will discuss in this section the latter we will return to in the context of of heavy-ion collisions in \cref{sec:qqbarprodhic}.

For more than 20 years, intuition surrounding the question of in-medium quarkonium stability has been acquired in terms of real-valued potential models (for an overview see e.g. \cite{Satz:2005hx} and references therein). In these models the thermal environment leads to a monotonous weakening of the potential, which was used to compute the eigenfunctions and energies of a non-relativistic Hamiltonian. Instead of the bound stationary states of the vacuum Hamiltonian, at $T>0$ one considered stationary states of the in-medium Hamiltonian. At low temperatures these states remain bound but, at well defined thresholds $T_{\rm melt}$, one by one go over into unbound states, as can be cleanly identified e.g. using the {\it complex scaling method} reviewed in Ref.~\cite{Moiseyev:1998}. This entirely static setup has given rise to the intuitively appealing idea of {\it sequential melting}, i.e. vacuum states that are more deeply bound will dissociate more easily at finite temperatures than those that are more weakly bound at $T=0$. In addition a strong focus was placed on determining the stability parameters of quarkonium states in this model approach, the melting temperatures $T_{\rm melt}$.

In the following we argue that the concept of sequential melting, interpreted as a dynamical statement, remains a valid guiding principle in a modern understanding of in-medium quarkonium. On the other hand the concept of melting temperature turns out to be less informative than originally anticipated.   

Our understanding of in-medium quarkonium changed fundamentally with the realization that the in-medium interquark potential is complex valued. It forced the research community to acknowledge that the $Q\bar{Q}$ system is far from stationary and instead should be understood as inherently dynamical. It also reinforced the fact that the potential computed in the effective field theory pNRQCD does not govern the time evolution of the $Q\bar{Q}$ wavefunction but instead that of the unequal time point-split meson correlator. As we saw in \cref{sec:specfrompNRQCD} the complex static interquark potential hence allows us to compute an approximation of the $T>0$ spectral function, from which in-medium properties of individual states can be read off. It is this information that QCD theory brings to the table and from which vital insight into the stability and eventual melting of heavy quarkonium in thermal equilibrium is deduced.

An important feature at finite temperature is the presence of thermal broadening, induced by ${\rm Im}[V]\neq0$. The width $\Gamma$ of an in-medium spectral peak represents the finite probability for that state to transition into some other state over a characteristic time scale $\sim1/\Gamma$. While the lattice simulation may provide insight into the values of $\Gamma$, it does not tell us about the processes involved in generating it. On the other hand perturbative studies have revealed that depending on the hierarchy of energy scales e.g. Landau damping and gluon absorption may contribute to the thermal broadening. Unfortunately, from the spectral function alone one cannot determine into which state the quarkonium transitions, i.e. whether it stays in the singlet channel as a more highly excited (or deexcited) state or whether it turns into an unbound color octet. This information is hidden in higher order correlation functions, similar to the density matrix of the quarkonium system, which only recently has been studied starting from first principles. In \cref{sec:qqbarrealtime} we will discuss how the changes in occupancies of individual can states be described in real-time and will find that the phenomenon of decoherence is an essential ingredient in understanding the survival of heavy quarkonium in medium. It is clear that the question of quarkonium stability cannot be answered in a static picture and quarkonium melting is an inherently time dependent process. 

The stability of in-medium quarkonium of course is closely related with its in-medium binding energy, defined from the distance between the in-medium peak and the continuum threshold. Qualitatively similar to what had been observed in the historic potential models, we found in \cref{sec:specfrompNRQCD} that screening of ${\rm Re}[V]$ leads to a decrease of the in-medium binding energy from an interplay of a reduction in the bound-state mass and the onset of the continuum threshold. As the formerly long lived quarkonium state is heated up its binding is weakened. The associated increase in spatial extent leads to a higher chance of scattering with medium partons that in turn may kick it into a different state. I.e. the reduction in $E_{\rm bind}$ goes hand in hand with an increase of $\Gamma$. 

What the computations based on the lattice vetted complex potential show, supported by the consistent correlator ratios in NRQCD, is that the weakening of the in-medium quarkonium states indeed proceeds hierarchically ordered with the vacuum binding energy. The more weakly bound states start to broaden first, while the more deeply bound vacuum states remain as narrow structures up to high temperatures. In this dynamical sense quarkonium melting is a sequential process. 

The fact that the in-medium peaks are broadened and smoothly go over into the continuum structure at high temperatures also tells us that the concept of melting temperature in this dynamical picture is not uniquely defined. For historic reasons one may still wish to single out where melting happens, selecting a single temperature at which individual quarkonium states are already highly susceptible to transitions due to kicks from the medium. To this end it has been suggested in e.g. Ref.~\cite{Laine:2006ns} to choose a $T_{\rm melt}$, where the in-medium binding energy is equal to the thermal width. This represents a situation where the correlations between a meson state identified at time $t_0$ have diminished by a factor $1/e$ after having evolved by a time $\sim 1/E_{\rm bind}^{\rm med}$. For the latest estimates based on this criterion and the spectral functions computed in Ref.~\cite{Lafferty:2019jpr} see \cref{tab:MeltTbbbar} and \cref{tab:MeltTccbar}. It is important to note that as of yet only studies based on an in-medium potential are able to provide melting temperatures defined in this quantitative manner, as it requires knowledge of the onset of the continuum structure. In other words, melting temperatures quoted in studies based on lattice NRQCD and relativistic lattice QCD may take on different values, as they are based solely on a visual inspection of the disappearance of discernible peak structures.

\begin{table}
\begin{tabular}[c]{|c|c||c|c|c|c|c|}\hline
& $J/\Psi$ & $\psi^\prime$ & $\Upsilon(1S) $ & $\Upsilon(2S) $ & $\Upsilon(3S) $& $\Upsilon(4S) $ \\ \hline
$T_{\rm melt}$[GeV] & $0.267^{+0.033}_{-0.036}$ & $<0.147$ & $0.440^{+0.080}_{-0.055}$ & $0.250^{+0.050}_{-0.035}$ & $0.200^{+0.045}_{-0.053}$ & $<0.147^{+0.068}$ \\ \hline
\end{tabular}
\caption{Melting temperatures for bottomonium states estimated from the spectral functions of \cref{fig:pNRQCDSwaveSpectralFunc} and Ref.~\cite{Lafferty:2019jpr}.}\label{tab:MeltTbbbar}
\end{table}
\begin{table}
\begin{tabular}[c]{|c|c|c||c|c|c|}\hline
& $\chi_{c0}(1P)$ & $\chi_{c0}(2P)$ & $\chi_{b0}(1P) $ & $\chi_{b0}(2P) $ & $\chi_{b0}(3P) $ \\ \hline
$T_{\rm melt}$[GeV] & $0.183^{+0.009}_{-0.09}$ & $<0.147$ & $0.264^{+0.030}_{-0.078}$ & $0.210^{+0.033}_{-0.024}$ & $0.191^{+0.011}_{-0.021}$\\ \hline
\end{tabular}
\caption{Melting temperatures for charmonium states estimated from the spectral functions of \cref{fig:pNRQCDSwaveSpectralFunc} and Ref.~\cite{Lafferty:2019jpr}.}\label{tab:MeltTccbar}
\end{table}

One of the main reasons underlying the interest in melting temperatures was related to phenomenological modeling. In e.g. rate equation based approaches to heavy quarkonium in heavy ion collisions, melting temperatures provided a convenient way to incorporate non-perturbative information from QCD. In an era, where reliable access to the full spectral function is available and where genuine real-time descriptions of quarkonium are developed that directly interface with EFTs and lattice QCD, as discussed in \cref{sec:qqbarrealtime}, the concept of melting temperature thus has become less and less relevant. 

The interplay of scattering and screening also complicates stability estimates, which had been common in the era of real-valued potential models. The inverse of the Debye mass of the QCD medium still represents the characteristic length scale beyond which color sources do not communicate with each other. I.e. if the spatial extent of the quark antiquark pair becomes similar to this distance, it can be considered fully decorrelated. At the same time the kicks of the medium partons already at closer separation distances may have induced a transition to an unbound configuration. Stability arguments based on the Debye mass alone are thus understood to provide only upper limits. In \cref{sec:qqbarrealtime} we will show in more detail how the effect of in-medium kicks can be understood in terms of decoherence of the quarkonium system.

The appearance of the thermal width in the quarkonium spectral function is a clear indication that we are dealing with a truly dynamical problem even in thermal equilibrium. While access to spectral functions has helped to shed light on the overall stability of kinetically equilibrated quarkonium, it does not yet allow us to answer even the simple straight forward questions of what happens to a vacuum quarkonium state as it is thrown into a thermal QCD bath. A comprehensive understanding of quarkonium stability and quarkonium melting thus requires a genuine real-time description, which will be the focus of the next section.
  
\section{Quarkonium in-medium real-time evolution}
\label{sec:qqbarrealtime}  
  
The study of meson spectral functions and the in-medium heavy quark potential has revealed the dynamical nature of in-medium quarkonium, even in an idealized setting such as in thermal equilibrium. The first task at hand thus is to develop a description that is capable of shedding light on how individual quarkonium states evolve in real-time in the presence of a QCD medium. This will allow us to form a more detailed picture of dynamical quarkonium melting. The second step will be to use such a framework to support the interpretation of heavy quarkonium measurements in a heavy-ion collision.
  
Phenomenological modelling of heavy quarkonium real-time dynamics has a long history. For charmonium the method of choice is to consider Boltzmann type transport equations from which, under additional time scale separation assumptions, coupled rate equations for the different quarkonium states are obtained (see Refs.~\cite{Grandchamp:2002wp,Grandchamp:2003uw,Rapp:2008tf,Zhao:2010nk,Du:2018wsj} and Refs.~\cite{Zhou:2014kka,Chen:2016dke,Zhou:2016wbo}, for application to bottmonium see e.g. Ref.~\cite{Du:2017qkv}). In this approach the dissociation of quarkonium and recombination of individual quarks is both included. Arguments based on detailed balance, potential model computations for quarkonium binding and perturbative insight on dissociation provide the input values for the strength of each term. A step towards a more systematic treatment of the individual contributions was taken in Refs.~\cite{Yao:2017fuc,Yao:2018sgn} and Ref.~\cite{Hong:2018vgp}, where screening, dissociation and recombination contributions have been elucidated based on the effective field theory pNRQCD. 

For Bottomonium where an in-medium potential picture is expected to provide an accurate description, the most common approach to date is to solve some form of Schr\"odinger equation. Either one considers a deterministic linear equation to which a complex model potential is supplied (see e.g. Refs.~\cite{Strickland:2011aa,Krouppa:2015yoa,Krouppa:2016jcl,Krouppa:2017jlg}) or one implements the dynamics via a non-linear stochastic Schr\"odinger equation as in Ref.~\cite{Katz:2015qja}.

For both charmonium and bottomonium the above strategies have been successfully applied to reproduce the experimental measurements of quarkonium states in heavy-ion collisions. 

Developing a genuine first principles based understanding of quarkonium real-time evolution thus serves multiple roles. On the one hand it will lead to an intrinsically better understanding of quarkonium stability in idealized situations, such as in thermal equilibrium, but it will also reduce the amount of modeling input necessary when describing quarkonium in complex scenarios such as a heavy-ion collision. 
  
A genuine real-time description of quarkonium forces us to amend the theoretical guiding principle based on separation of {\it energy scales} by taking into account the different {\it time scales} present in the system under consideration. As discussed in \cref{sec:OQS} the open quantum systems approach offers a promising road towards a systematic treatment of heavy quarkonium in this respect. In thermal equilibrium a non-perturbative understanding of various aspects of heavy quarkonium has been achieved, by connecting first principles lattice QCD and effective field theories. However an equally non-perturbative implementation of the real-time dynamics is not available yet. Since the sign problem prevents direct real-time simulations on the lattice, one central focus of theory today lies in establishing how to non-perturbatively match effective field theories, such as pNRQCD to QCD. This in turn may then be used as a starting point to derive quarkonium real-time evolution equations formulated in the language of open-quantum-systems. More concretely, it needs to be established whether and, if so, how a master equation for the $Q\bar{Q}$ density matrix can be derived which systematically incorporates the information on the proper real-time heavy-quark potential, as well as the other non-local Wilson coefficients of pNRQCD. 

\begin{figure}
\centering
\includegraphics[scale=0.45, clip=true, trim= 1cm 2cm 10cm 2cm ]{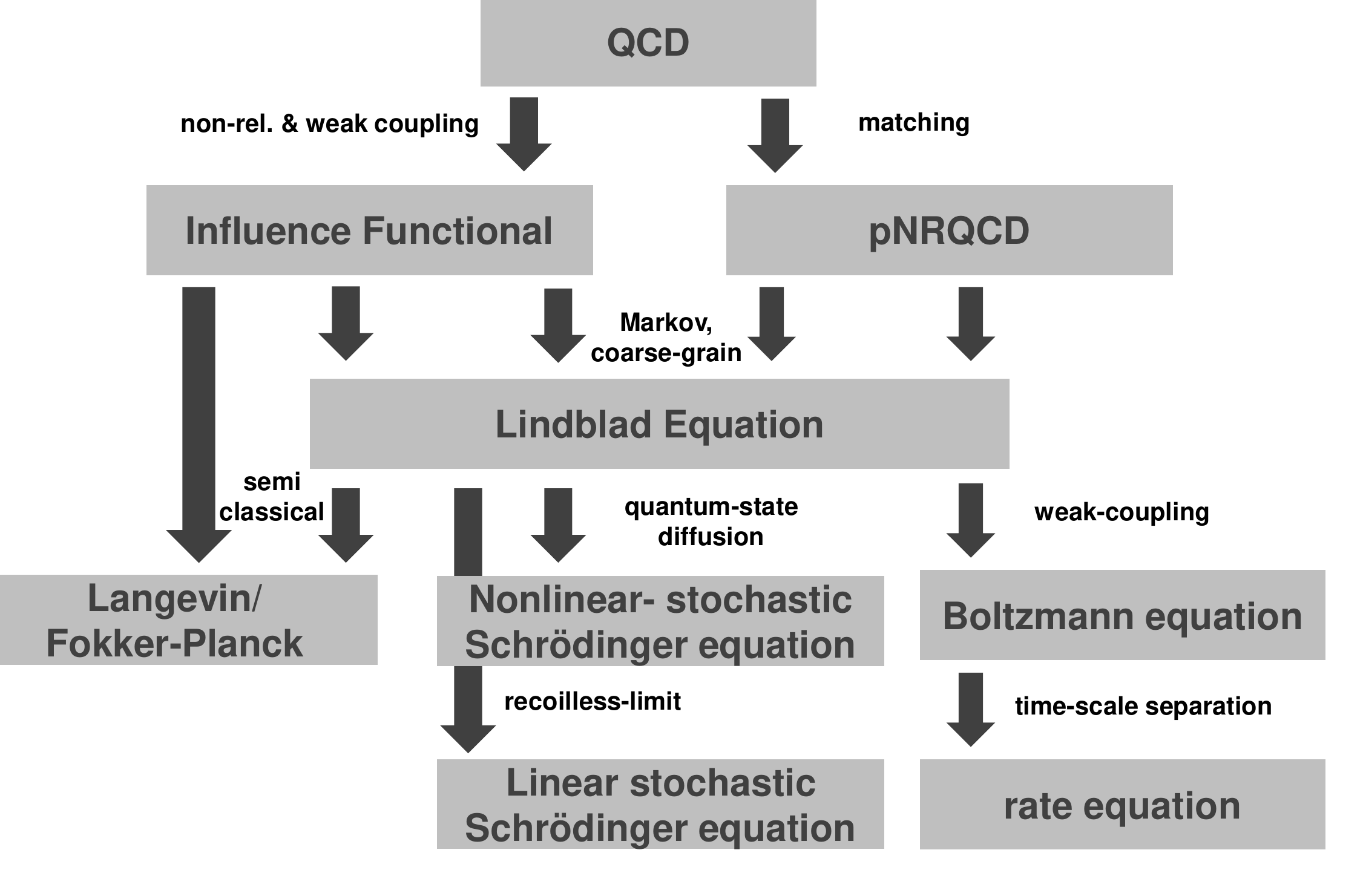}
\caption{Relationship between QCD and the different real-time approaches to heavy quarkonium currently deployed or in development. Flowchart adapted and expanded from Ref.~\cite{DeBoni:2017ocl}.}\label{fig:RealTimeApproach}
\end{figure}  
  
The past five years have already seen major progress in the context of an open-quantum-systems description of heavy quarkonium. Indeed it is now possible to derive from QCD, using a well controlled set of approximations, a non-linear stochastic Schr\"odinger equation describing the quantum dissipative evolution of the quark antiquark pair in a hot thermal medium (see \cref{fig:RealTimeApproach}). This for the first time provides a field theoretical basis to phenomenological models of this type. It is further understood how computations based on deterministic Schr\"odinger equations can be systematically improved to become more accurate. At the same time first proposals have been put forward to derive the Boltzmann equation in a weakly coupled medium directly from QCD via the open-quantum systems approach, not only providing insight on the range of validity of transport approaches for quarkonium but opening up new venues for eventually deriving a fully non-perturbative equation of motion for quarkonium distribution functions.

In the following we briefly survey the current activities in the field from which we will highlight two sets of studies that both make close contact to the effective field theory concepts encountered previously in equilibrium and offer concrete paths to the improvement of current phenomenological real-time approaches to in-medium quarkonium. 
  
\subsection*{Open Quantum System based approaches}
\label{sec:REalTimeOQS}

The first steps in the direction of an open-quantum systems based understanding of in-medium  quarkonium were inspired by the Caldeira-Leggett model in Ref.~\cite{Young:2010jq}, but did not yet attempt to derive a real-time description. Subsequently Refs.~\cite{Borghini:2011yq,Borghini:2011ms} explored the role played by transitions induced between different quarkonium states by the medium and suggested that an open-quantum systems approach (called a quantum dissipative system in their paper) would be most appropriate to capture the relevant physics. Independently Ref.~\cite{Akamatsu:2011se} proposed the first open-quantum systems description to real-time heavy quarkonium that made direct connection with the complex real-time potential. It suggested an interpretation of the imaginary part of the in-medium potential in terms of decoherence induced by the medium kicks on the quarkonium particle. The paper proposed that the values of ${\rm Im}[V]$ manifest themselves in the local correlations of a real valued noise term added to an otherwise real-valued potential, in turn leading to a unitary stochastic time evolution. This early stochastic potential approach has been subsequently used to take a first look at bottomonium evolution as open quantum system in Ref.~\cite{Rothkopf:2013kya}. 

The next decisive step was achieved by Akamatsu in Ref.~\cite{Akamatsu:2012vt} in deriving a master equation for quarkonium at high temperatures directly from the QCD path integral via the Feynman-Vernon influence functional. In turn he showed that based on single gluon exchange the dynamics in the recoilless limit can be reduced to a stochastic potential, corresponding to the first order in a systematic gradient expansion of the equation of motion. It also clarified how the non-local correlations of the stochastic noise are related to the imaginary part of the potential. Note that this colored stochastic potential does not simply reduce to the color singlet version introduced in Ref.~\cite{Akamatsu:2011se}, since singlet to singlet transitions are only possible via multi-gluon exchanges.

A refined analysis, with a particular focus on the implementation of time coarse graining presented in Ref.~\cite{Akamatsu:2014qsa} then delivered for the first time a genuine Lindblad master equation for quarkonium at high temperatures. Having established that the in-medium real-time evolution of quarkonium can be captured in the well established open-quantum systems framework using a clearly defined set of approximations, the community took notice and concerted efforts arose to explore the dynamical aspects of in-medium quarkonium.

Recently several research groups are actively exploring the open quantum systems approach to heavy quarkonium. Two of these (see Refs.~\cite{Brambilla:2016wgg,Brambilla:2017zei,Brambilla:2019tpt} and Ref.~\cite{Yao:2018nmy}) take as starting point directly the effective field theory pNRQCD. The benefit of this approach is that the effective field theory is systematically derived from QCD based on matching with the intermediate effective theory NRQCD. On the other hand a drawback for phenomenology is that pNRQCD is derived for a certain scale hierarchy. Often this hierarchy is taken to apply to the S-wave ground state but excited states and P-wave states may not adhere to the same hierarchy, thus complicating a concurrent treatment of all the states present in a heavy-ion collision. A central difference between the two sets of studies is that in the former the medium degrees are treated as strongly coupled $m_D\sim T$ and the information on the in-medium dynamics is formulated in terms of non-perturbative transport-coefficients, possibly accessible on the lattice \cite{Brambilla:2019tpt}. In the latter, pNRQCD is matched to NRQCD perturbatively, which in turn is argued to allow a simplification of the equations of motions up to the point where the perturbative Boltzmann equation is obtained. Such a systematic derivation of the Boltzmann equation from QCD is an important contribution to putting phenomenological modeling on a solid footing by exposing the underlying chain of required approximations.

Other groups approach the derivation of the quarkonium real-time evolution from a weakly coupled perspective without direct reference to pNRQCD. A central aspect of the works of Refs.~\cite{Blaizot:2015hya,Blaizot:2017ypk} is to explore the effects of dynamical dissociation and recombination in a semi-classical language reducing the master equation further to an effective Langevin form. The first paper considers the effects common to Abelian and non-Abelian plasmas, while the second focuses on the additional effects played by the color degrees of freedom present in QCD. Ref.~\cite{Akamatsu:2015kaa} also investigated the classicalization of the heavy quark dynamics. While the dynamics of quarkonium can become classical if decoherence acts efficiently, it needs to be ascertained in each case whether remnant quantum superpositions among quarkonium state can be neglected on the timescales under consideration. The authors of Ref.~\cite{Blaizot:2018oev} have taken first steps beyond the Markovian approximation in this context, which may e.g. be relevant for the description of quarkonium at intermediate temperatures. Ref.~\cite{DeBoni:2017ocl} attempted an alternative derivation of the e.o.m. at high temperature, but since no explicit Lindblad operators could be constructed, doubts remain whether this study has produced a genuine Lindblad master equation. A thorough reinvestigation of the stochastic potential approximation to the full dynamics has been presented in Ref.~\cite{Kajimoto:2017rel} focusing on the role of decoherence played on the stability of quarkonium. \Cref{fig:RealTimeApproach} attempts to summarize the currently available open-quantum systems based approached to quarkonium.

\subsection{The Feynman-Vernon influence functional at high temperature}
\label{sec:FVIFOQS}

In \cref{sec:OQS} we considered two iconic models of open quantum systems, the quantum optical master equation, as well as quantum Brownian motion at high temperature. While the latter model appears at first to be well suited to the description of heavy quarkonium it needs to be kept in mind that its evolution equation was derived for a single particle, i.e. for a system whose extend is smaller than the correlation length of the medium. This is in general not the case for heavy quarkonium, where depending on the interplay of temperature and binding energy the extent of the in-mediums state may easily be of the same order or larger than the medium correlation length. 

In Ref.~\cite{Akamatsu:2014qsa} a Lindblad master equation for in-medium quarkonium has been derived based on the Feynman-Vernon influence functional. It rests on three sets of approximations, which have been justified apriori using the scales present in Coulombically bound quarkonium states. The first refers to the non-relativistic limit, requiring that both $\sqrt{T/m_Q}\ll1$ as well as $\alpha_S\ll 1$. The second requires the medium to be weakly coupled so that $g\ll1$ and the third is related to how the coarse graining in time is implemented, assuming $1/gT\ll 1/M\alpha^2$. This last relation is nothing but the condition $\tau_E\ll\tau_S$ also used in the quantum Brownian motion master equation. If fulfilled it tells us that the acceleration of the heavy quarks, i.e. the change in the relative velocity $\dot v$ induced by the individual interactions with the medium is small.

Starting from the Schwinger Keldysh path integral for the system of gauge fields $A$, light fermions $q$ and heavy quarks $Q$, summarized as $\phi=(A,q,Q)$ with $\phi_1$ on the forward branch and $\phi_2$ on the backward branch, one may write down the generating functional with external source terms $\eta_1$ and $\eta_2$ as
\begin{align}
Z[\eta_1,\eta_2]=&\int d\phi_1(0) d\phi_2(0) \langle \phi_1(0)| \rho_{\rm tot} | \phi_2(0)\rangle\\
&\times \int_{\phi_1(0)}^{\phi_2(0)} {\cal D}[\phi_1,\phi_2] {\rm exp}\Big[i\int d^4x \big( {\cal L}_{\rm tot}[\phi_1] -\phi_1\eta_1\big)\Big]{\rm exp}\Big[i\int d^4x \big( {\cal L}_{\rm tot}[\phi_2] -\phi_2\eta_2\big)\Big].
\end{align} 
Assuming the factorization of the initial density matrix, defining the heavy quark color color current $j^{a\mu}=\bar{Q}T^a\gamma^\mu Q$ and setting the external sources to zero one arrives at the partition function
\begin{align}
Z=\int dQ_1(0) dQ_2(0) \langle Q_1(0)| \rho_{Q\bar{Q}} | Q_2(0)\rangle \int_{Q_1(0)}^{Q_2(0)} {\cal D}[Q_1,Q_2] {\rm exp}\big[ iS_{\rm kin}[Q_1] -iS_{\rm kin}[Q_2] + iS_{FV}[j_1,j_2]\big] 
\end{align} 
where the all terms that couple the heavy fields with the gauge fields are contained within
\begin{align}
&e^{iS_{FV}[j_1,j_2]}=\int d[A_1(0),A_2(0),q_1(0),q_2(0)] \langle A_1(0),q_1(0)|\rho_{\rm med}^{\rm eq}| A_2(0),q_2(0) \rangle \\
&\times \int_{A_1(0),q_1(0)}^{A_2(0),q_2(0)} {\cal D}[A_1,A_2,q_1,q_2]
{\rm exp}\Big[i\int d^4x \big( {\cal L}_{med}[A_1,q_1] -gj_1^{a\mu}A_{1,\mu}^a\big)\Big]{\rm exp}\Big[i\int d^4x \big( {\cal L}_{med}[A_2,q_2] -gj_2^{a\mu}A_{2,\mu}^a\big)\Big]. \label{eq:FVinflfull}
\end{align}
Note that the medium is assumed to be in thermal equilibrium at this point and that the currents $j_1$ and $j_2$ are not external currents but represent the heavy color sources to which the gauge field couples dynamically. 

By performing the non-relativistic approximation first, the coupling between the heavy quarks and the gauge fields is simplified to only include the heavy quark color density, i.e. an interaction term of the form $gj^{a0}A_0^a$. The subsequent perturbative expansion of \cref{eq:FVinflfull} to second order in the strong coupling allows to relate the interactions between quarkonium and the medium in terms of the different gluon propagators $G$ defined on the Schwinger-Keldysh contour
\begin{align}
iS_{FV}[j_1,j_2]=-\frac{g}{2}\int d^4x d^4y (j_1^{0a}(x),j_2^{0a}(x))\left[ \begin{array}{cc} G_{ab,00} (x-y) & -G^<_{ab,00}(x-y) \\ -G^>_{ab,00} (x-y) & \tilde G_{ab,00} (x-y)\end{array} \right] \left( \begin{array}{c} j_1^{0b}(y)\\j_2^{0b}(y)\end{array}\right)+{\cal O}(g^3).
\end{align}

One of the important contributions of Ref.~\cite{Akamatsu:2014qsa} is to identify that with a particular choice of time variables, i.e. $t={\rm max}(x^0,y^0)$ in addition to the relative time $s=|x^0-y^0|$, coarse graining of the influence functional in time leads to interaction terms that only contain the retarded correlator $G^R$ and the related gluon spectral function denoted here by $\rho_G$. In particular this means that all in-medium information is contained in three quantities
\begin{align}
&V(\mathbf{r})\delta_{ab}=-g^2{\rm Re} G^R_{ab,00}(\omega=0,\mathbf{r}),\\
&D(\mathbf{r})\delta_{ab}=-g^2T\frac{\partial}{\partial\omega}\rho_{G,ab,00}(\omega=0,\mathbf{r}), \quad A(\mathbf{r})\delta_{ab} = -\frac{g^2}{6T}\frac{\partial}{\partial\omega}\rho_{G,ab,00}(\omega=0,\mathbf{r}).
\end{align}
The first one, related to the Fourier transform of the retarded gluon correlator at vanishing frequencies, is nothing but the real-part of the proper real-time potential evaluated perturbatively. The second and third one is related to the gluon spectral function, which in the perturbative language is given by the appropriately shifted imaginary part of the real-time potential, i.e. $D(\mathbf{r})={\rm Im}[V](r)- {\rm Im}[V](r=\infty)$. Combining the weak coupling expansion and time coarse graining together with the Markovian approximation the influence functional may be rewritten as
\begin{align}
S_{FV}= S_{\rm fluct}[D] + S_{\rm diss}[D] + S_{\rm L}[A]
\end{align}
In order to make the coupling between gluons and heavy quarks local one must neglect possible overlap of interactions, similar to taking the ladder approximation in the Bethe-Salpeter equation. The individual terms have been suggestively named from an analogy with the quantum Brownian motion case. $S_{\rm fluct}$ encodes the effects of fluctuations, where e.g. kicks of the medium particles transfer energy into the quarkonium system. $S_{\rm diss}[D]$ on the other hand describes the physics of dissipation, by which the quarkonium state may release energy back to the medium. The combination of the two makes it possible for the quarkonium system to eventually thermalize with its environment. Only thanks to the additional term $S_{\rm L}[A]$, a genuine Lindblad master equation emerges, that preserves all relevant properties of the density matrix under time evolution. Note that here in contrast to the Caldeira-Leggett model this last term is not added by hand but arises due to a careful choice of the coarse graining procedure.

At this point the influence functional is formulated in terms of actions and heavy quark fields, while we are ultimately interested in an evolution equation for the density matrix operator for the two-body system in the position basis. To this end the stratgey put forward in Ref.~\cite{Akamatsu:2014qsa} is based on using the properties of coherent states as generating functionals for heavy quarkonium particles. Using the same notation for the quark and antiquark field operators as in NRQCD, i.e. $\hat \psi$ and $\hat \chi$, the matrix elements of the reduced density matrix in the coherent state basis can be written as
\begin{align}
&\sigma_{Q\bar{Q}}\big[t,J_\psi^1,J_\chi^1,J_\psi^2,J_\chi^2\big]=\langle J_\psi^1,J_\chi^1| \sigma_{Q\bar{Q}}(t)|J_\psi^2,J_\chi^2\rangle\\
&\langle J_\psi^1,J_\chi^1|=\langle \Omega | {\rm exp}\Big[ -\int d^3x \Big( \hat \psi(\mathbf{x}) J_\psi^1(\mathbf{x}) + \hat \chi(\mathbf{x}) J_\chi^1(\mathbf{x})\Big)\Big],\\
&| J_\psi^2,J_\chi^2\rangle= {\rm exp}\Big[ -\int d^3x \Big( J_\psi^2(\mathbf{x}) \hat \psi^\dagger(\mathbf{x})  + J_\chi^2(\mathbf{x}) \hat \chi^\dagger(\mathbf{x}) \Big)\Big] |\Omega \rangle
\end{align}
where the sources $J^i$ are just complex numbers. Using the path integral representation of $\sigma_{Q\bar{Q}}\big[t,J_\psi^1,J_\chi^1,J_\psi^2,J_\chi^2\big]$ the position space basis density matrix can be obtained via functional differentiation
\begin{align}
\sigma_{Q\bar{Q}}^{abcd}(t,\mathbf{x}_1,\mathbf{x}_2,\mathbf{y}_1,\mathbf{y}_2&= \langle \mathbf{x}_1,a;\mathbf{x}_2,b| \hat \rho_{Q\bar{Q}}(t)|\mathbf{y}_1,c;\mathbf{y}_2,d\rangle \propto \langle \Omega| \hat Q^a(\mathbf{x}_1)\hat{\bar{Q}}^b(\mathbf{x}_2)\hat\rho_{Q\bar{Q}}(t) \hat{\bar{Q}}^{d\dagger}(\mathbf{y}_2)\hat Q^{c\dagger}(\mathbf{y}_1)|\Omega\rangle \\
&=\frac{\delta}{\delta J_\psi^a(\mathbf{x}_1)}\frac{\delta}{\delta J_\chi^b(\mathbf{x}_2)}\frac{\delta}{\delta J_\chi^d(\mathbf{y}_2)}\frac{\delta}{\delta J_\psi^c(\mathbf{y}_1)}\left. \sigma_{Q\bar{Q}}\big[t,J_\psi^1,J_\chi^1,J_\psi^2,J_\chi^2\big] \right|_{J=0}.
\end{align}

Collecting all the terms obtained in the evolution equation for the position basis density matrix, these can be summarized as a Lindblad equation with two distinct Lindblad operators. Using the constituent heavy quark position $\mathbf{\hat x}$ and momentum operator $\mathbf{\hat p}$, as well as the antiquark operators $\mathbf{\hat y}$ and $\mathbf{ \hat q}$, the Lindblad operators and their coefficients take the form
\begin{align}
&\hat L^1_{\mathbf{k},a}=e^{i\mathbf{k}\mathbf{\hat x}/2}\Big(1-\frac{\mathbf{k}\mathbf{\hat p}}{4MT}\Big)e^{i\mathbf{k}\mathbf{\hat x}/2}(T^a\otimes 1) - e^{i\mathbf{k}\mathbf{\hat y}/2}\Big(1-\frac{\mathbf{k}\mathbf{\hat q}}{4MT}\Big)e^{i\mathbf{k}\mathbf{\hat y}/2}(1\otimes T^{a*}), \quad  \gamma^1_{\mathbf{k},a}=-\frac{D(\mathbf{k})}{L^3}>0,\\
&\hat L^2_{\mathbf{k},a}=e^{i\mathbf{k}\mathbf{\hat x}/2}\Big(\frac{\mathbf{k}\mathbf{\hat p}}{4MT}\Big)e^{i\mathbf{k}\mathbf{\hat x}/2}(T^a\otimes 1) - e^{i\mathbf{k}\mathbf{\hat y}/2}\Big(\frac{\mathbf{k}\mathbf{\hat q}}{4MT}\Big)e^{i\mathbf{k}\mathbf{\hat y}/2}(1\otimes T^{a*}), \quad  \gamma^2_{\mathbf{k},a}=-\frac{1}{L^3}\Big( 8T^2  A(\mathbf{k}) -  D(\mathbf{k})\Big)>0, \label{eq:LindbladQQbar}
\end{align}
where $V=L^3$ refers to a large but finite volume of the combined system of heavy quarks and medium. In addition to the Lindblad operators,  a non-trivial Hamiltonian for the reduced quarkonium system is obtained
\begin{align}
\hat H_{Q\bar{Q}} = \frac{ \mathbf{\hat p}^2 + \mathbf{\hat q}^2 }{2m_Q} -V(\mathbf{x}-\mathbf{y})(T^a\otimes T^{a*})+\frac{1}{8MT}\Big\{ (\mathbf{\hat p} - \mathbf{\hat q}), {\bf \nabla} D( \mathbf{x}-\mathbf{y} ) \Big\} (T^a\otimes T^{a*}).\label{eq:HamiltonianQQbar}
\end{align}

Several remarks are in order. Similar to assigning physical roles to the individual terms in $S_{FV}$ we can identify the corresponding contributions in the Lindblad operators. The physics of fluctuations is encoded in the terms independent of the heavy quark momentum, while the effects of dissipation reside in those that carry an explicit $\mathbf{p}$ or $\mathbf{q}$ dependence. Interestingly this tells us that the center of mass motion and relative motion both influence dissipation. It turns out that in case that dissipative effects are non-negligible the physics of relative and finite center of mass momenta cannot be fully separated. Note that the coefficients of the Lindblad operators depend on the gluon spectral function and thus on the imaginary part of the potential. They are both positive, as required for the Lindblad formalism to be valid. In case of $\gamma^2$ this is only possible due to the contributions of the $A$ term arising from $S_L$ in the expansion of the influence functional. In practice it is often assumed that the term $8T^2 A-D$ is small so that only $L^1$ remains active. It is important to recognize that the Hamiltonian features both the familiar term containing the screened potential $V$ but also harbors another contribution which arises from the fluctuation part and which may not be neglected apriori. It is interesting to contemplate how and under which circumstances such an additional term can be related to the concept of entropic force (discussed in detail in Refs.~\cite{Satz:2015jsa,Kharzeev:2014pha}) which too is understood as an emergent phenomenon induced by medium fluctuations.

With the derivation of the Lindblad operators and the Hamiltonian for a system at high temperature, it is now possible to state, how the real- and imaginary part of the in-medium potential govern the microscopic evolution of the quarkonium state. The latter encodes the physics of fluctuations and dissipation and together with the real part determines the stability of the quarkonium state. Establishing these roles in a genuine non-perturbative fashion however remains a central open research question.

Note that the Lindblad equation explicitly takes into account the physics of both constituent quark and antiquark. This in turn means that the corresponding density matrix not only contains information about possible color singlet bound states but also captures the physics of (possibly decorrelated) individual color charged quarks. In other words it possesses the necessary knowledge to not only answer questions about dynamical melting of quarkonium but also to eventually answer how probable the recombination of individual quarks and antiquarks is. (As has been explicitly pointed out in the related works of Refs.~\cite{Blaizot:2017ypk,Blaizot:2018oev} the recombination probability actually becomes zero in an infinite volume, but since the fireball of a heavy-ion collision is finite, also the probabilities are.)

Solving the equation of motion defined by \cref{eq:LindbladQQbar,eq:HamiltonianQQbar} head on is still too costly. There are however efforts underway to derive a reduced equation of motion \cite{Akamatsu:2019} focussing on the relative coordinates (see also Ref.~\cite{DeBoni:2017ocl}). Preliminary findings suggest that as anticipated above, relative and center of mass momentum cannot be fully decoupled in the dissipative terms, the latter however appears simply as an external parameter. Even such a reduced density matrix carries a dependence on six spatial dimensions, i.e. the direct simulation of the corresponding partial differential equation is very demanding. On the other hand the quantum state diffusion approach discussed in \cref{sec:OQS} has already been successfully implemented for the description of the Lindblad equation of single heavy quarks in Ref.~\cite{Akamatsu:2018xim} and promises a viable path towards unravelling the evolution of the quarkonium density matrix in terms of an ensemble of wavefunctions. Including both fluctuation and dissipation effects such a simulation in principle will be able to implement genuine thermalization of a heavy quarkonium state with its surrounding. Conceptually the QSD approach for the first time provides a QCD derived non-linear stochastic Schr\"odinger equation for in-medium quarkonium, which so far has only been introduced based on phenomenological modeling.

Since the thermalization of quarkonium states can be captured with the Lindblad equation based on \cref{eq:LindbladQQbar,eq:HamiltonianQQbar}, the resulting late time behavior will provide a direct connection to the equilibrium results discussed in \cref{sec:QQbarequilprop}. These may then e.g. function as a benchmark of the real-time evolution. 

\subsection{Decoherence and Dynamical quarkonium melting}
\label{sec:DecDynMelt}

After discussing that quarkonium melting is an inherently dynamical process in \cref{sec:QQbarmelting}, we are now able to shed light on some of its details by using the open-quantum systems approach. And even though simulations of the full dissipative Lindblad dynamics are still work in progress, we can already consider what has been learned about the in-medium evolution from truncations of \cref{eq:LindbladQQbar,eq:HamiltonianQQbar}. In particular this has lead to insight into the role played by the phenomenon of decoherence on the stability of in-medium quarkonium and its melting.

As it allows us to identify already many relevant aspects of quarkonium stability we start out by considering the recoilless limit, which corresponds to neglecting all dissipative contributions. In turn this amounts to setting $S_{FV}=S_{\rm fluct}$ or simply taking the momentum independent terms of \cref{eq:LindbladQQbar}. At this first order in the gradient expansion of the influence functional one can completely integrate out the center of mass coordinates. This leads to a stochastic linear Schr\"odinger equation coupling the color singlet and octet sector. For the purpose of this section one may consider an even simpler scenario, where only transitions from the color singlet to singlet sector are implemented. This is nothing but the stochastic potential description originally proposed in \cite{Akamatsu:2011se}. Note that while the absence of dissipation will prevent the thermalization of the quarkonium system at late times, this approximation is expected to work well, even quantitatively, for tightly localized states at early times. 

Let us follow the exposition given in Ref.~\cite{Kajimoto:2017rel}, where one considers a quark located at $\mathbf{x}=\mathbf{R}+\frac{\mathbf{r}}{2}$ and an antiquark at $\mathbf{y}=\mathbf{R}-\frac{\mathbf{r}}{2}$. At this level of the approximation, the dynamics can be written in terms of a real-valued in-medium potential $V(\mathbf{r})$ and two noise terms $\eta$, whose spatial correlations are intimately connected to the imaginary part of the potential
\begin{align}
H(\mathbf{r},t)=-\frac{\nabla^2_{\mathbf{r}}}{m_Q} + V(\mathbf{r}) + \underbracket{\eta(\mathbf{R}+\frac{\mathbf{r}}{2},t)-\eta(\mathbf{R}-\frac{\mathbf{r}}{2},t)}_{\Theta(\mathbf{R},\mathbf{r},t)}.
\end{align}   
The origin of the two noise terms lies in the separate fluctuation contributions to the quark and antiquark, similar to the two fluctuation terms present in the high temperature Lindblad equation. These also carry opposite color, which leads to a relative minus sign in the singlet sector. Since the Markovian approximation has been used, the noise is Gaussian but carries non-trivial spatial structure given by
\begin{align}
\langle \eta(\mathbf{x},t)\rangle =0, \quad \langle \eta(\mathbf{x},t)\eta(\mathbf{x}^\prime,t^\prime)\rangle =D(\mathbf{x}-\mathbf{x}^\prime)\delta(t-t^\prime), \label{eq:StochPotNoise}
\end{align}
keeping in mind that at high temperature $D(r)={\rm Im}[V](r)-{\rm Im}[V](r=\infty)$. Consistent with Ito calculus, one finds that the noise terms scale with $\eta\sim dt^{-1/2}$. This requires one to take into account one additional term, absent for deterministic variables, when deriving the wavefunction evolution equation from the time evolution operator. The following stochastic differential equation ensues
\begin{align}
\psi_{Q\bar{Q}}(\mathbf{r},t+\Delta t)= 1 - i\Delta t \Big( - \frac{\nabla^2_{\mathbf{r}}}{m_Q} + V(\mathbf{r}) + \Theta(\mathbf{R},\mathbf{r},t) - \frac{i\Delta t}{2}\Theta(\mathbf{R},\mathbf{r},t)^2 \Big)\psi_{Q\bar{Q}}(\mathbf{r},t).
\end{align}
Since the time evolution operator $U={\rm exp}[-i\Delta t H]$ is unitary, the above equation of motion also preserves the norm of the evolving microscopic wavefunction. Nevertheless considering the ensemble average of the wavefunction, one obtains the following Schr\"odinger equation
\begin{align}
i\frac{d}{dt}\langle \psi_{Q\bar{Q}}(\mathbf{r},t)\rangle = \Big( - \frac{\nabla^2_{\mathbf{r}}}{m_Q} + V(\mathbf{r}) - i\underbracket{\{ D({\bf 0})- D(\mathbf{r})\}}_{=|{\rm Im}[V]|} \Big)\langle \psi_{Q\bar{Q}}(\mathbf{r},t)\rangle. \label{eq:schroedavg}
\end{align}
This simple example shows that the presence of an imaginary part in the potential governing the time evolution of a thermally averaged correlation function (corresponding to $\langle \psi_{Q\bar{Q}}(\mathbf{r},t)\psi_{Q\bar{Q}}^*(\mathbf{r},0)\rangle$) does not automatically reflect a dampening of the underlying microscopic wavefunction. ( Note that the full dissipative dynamics are on the other hand irreversible and can indeed lead to a dampening of the microscopic wavefunction. )

The effect of ${\rm Im}[V]$ here instead is a manifestation of the phenomenon of {\it wavefunction decoherence}, where the fluctuations of the medium successively perturb the individual realizations of the wave function so that after a characteristic time $\tau_D$ there is only faint resemblance among each other and the average washes out. While decoherence does not necessarily imply a decay of occupations, in the case of in-medium quarkonium, destabilization of formerly stable bound states is one of its central effects. Interestingly there is an additional scale hiding in plain sight in the imaginary part of the potential that determines the strength of decoherence and its role in heavy quarkonium stability. 

\begin{figure}
\centering
\includegraphics[scale=0.1]{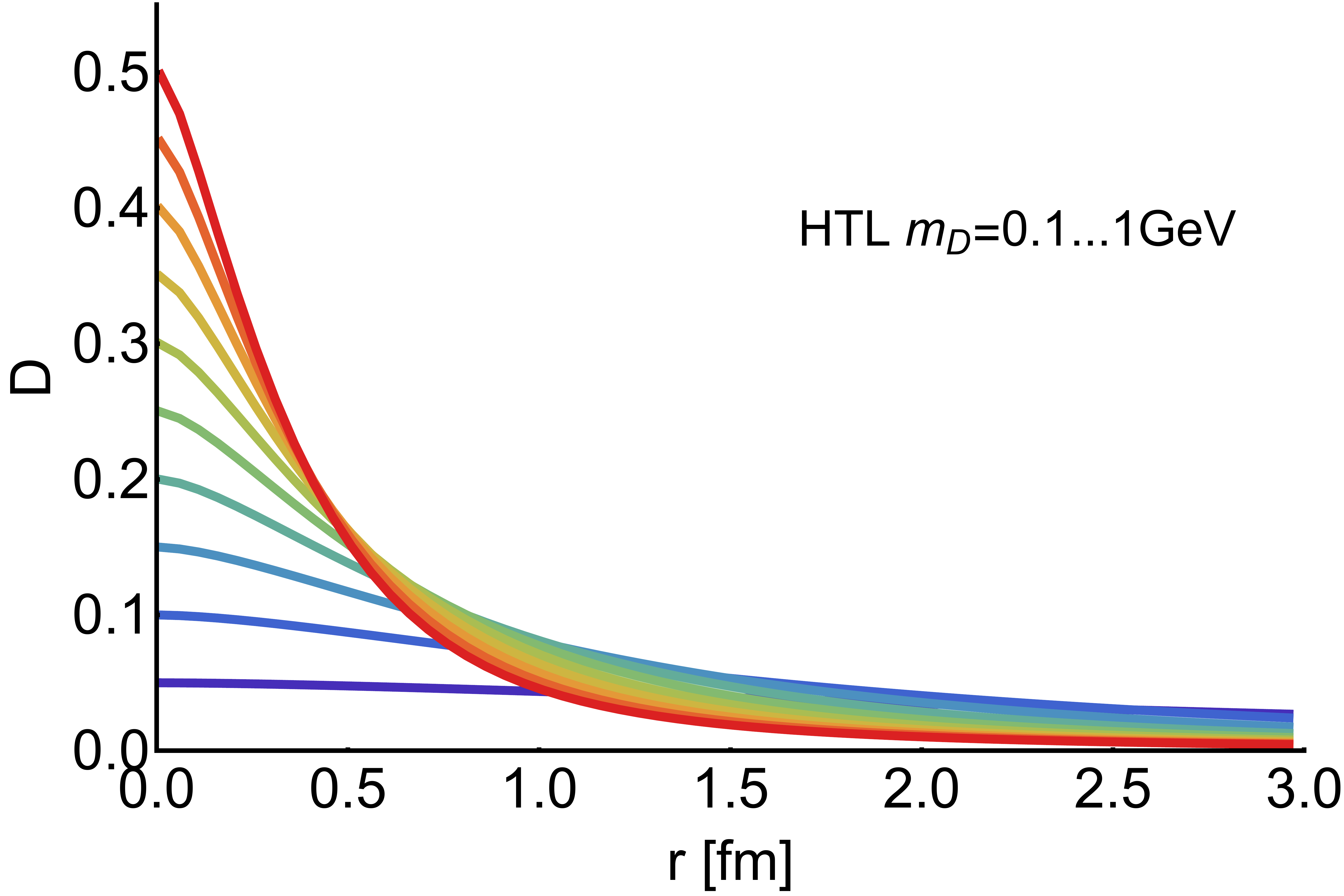}
\caption{The function $D(r)={\rm Im}[V](r)-{\rm Im}[V](r=\infty)$ evaluated in thermal equilibrium in the hard thermal loop approximation. Note its peaked shape around the origin, which invites the definition of the medium correlation length $\ell_{\rm med}$.}\label{fig:DFuncHTL}
\end{figure}

As an example consider the function $D$ evaluated in HTL perturbation theory, as shown in \cref{fig:DFuncHTL}. Its spatial dependence exhibits a well defined peak around the origin, which allows to define a characteristic scale $\ell_{\rm med}$ governing the spatial extend of the noise correlations in \cref{eq:StochPotNoise}. While in perturbation theory at high temperatures the correlation length and temperature are closely related $\ell_{\rm med}^{\rm PT}\sim 1/gT$, in a non-perturbative setting e.g. close to the crossover transition in QCD the relation may be quite different. In other words $\ell_{\rm med}$ represents an independent additional scale in the system.

Depending on the size of the quarkonium system under consideration (denoted in the following by $\ell_{\Psi_0}$) and the size of the correlation length $\ell_{\rm med}$ the effects on the bound state are quite different, as has been discussed in detail in Refs.~\cite{Kajimoto:2017rel,DeBoni:2017ocl}. Indeed based on how well the medium is able to resolve the internal structure of the quarkonium system three distinct regimes can be identified.

If the correlation length $\ell_{\rm med} \gg \ell_{Q\bar{Q}} $ is large compared to the intrinsic length scale of the $Q\bar{Q}$ system it corresponds to very soft momentum kicks $\Delta p_{\rm med}\sim h/\ell_{\rm med}$ on the small subsystem. In the limit of negligible momentum transfer one simply ends up with reversible dynamics, described by the von-Neumann like part of the master equation. Decoherence in this case is inefficient in destroying the quantum superposition present in the system.

Once the correlation length $\ell_{\rm med} \lesssim \ell_{Q\bar{Q}} $ becomes of the same order of the size of the quarkonium system or smaller the full dissipative dynamics of quantum Brownian motion sets in. This entails that decoherence becomes efficient, which in turn means that superpositions of quantum states in the system are gradually transferred into a probabilistic mixture of classical states. After a characteristic decoherence time scale $\tau_D$ the system may then be described by semi-classical means. 

Up to now we have only considered one characteristic scale for the quarkonium system $\ell_{Q\bar{Q}}$. However different states, due to their separated values of the binding energy, exhibit quite different spatial extends and therefore are affected differently by the medium kicks underlying decoherence. An appropriate analogy suggested in Ref.~\cite{DeBoni:2017ocl} is that a medium with $\ell_{\rm med}$ acts as a quarkonium sieve: easily destabilizing those states with $\ell_{\rm med} \ll \ell_{Q\bar{Q}} $ and only weakly affecting those with  $\ell_{\rm med} \gg \ell_{Q\bar{Q}} $.

At very high temperatures the correlation length will become smaller than all of the quarkonia and thus the medium is able to resolve even the most deeply bound states. Taken to the asymptotic limit, all states will thus be destabilized equally efficiently.

\begin{figure}
\centering
\includegraphics[scale=0.45]{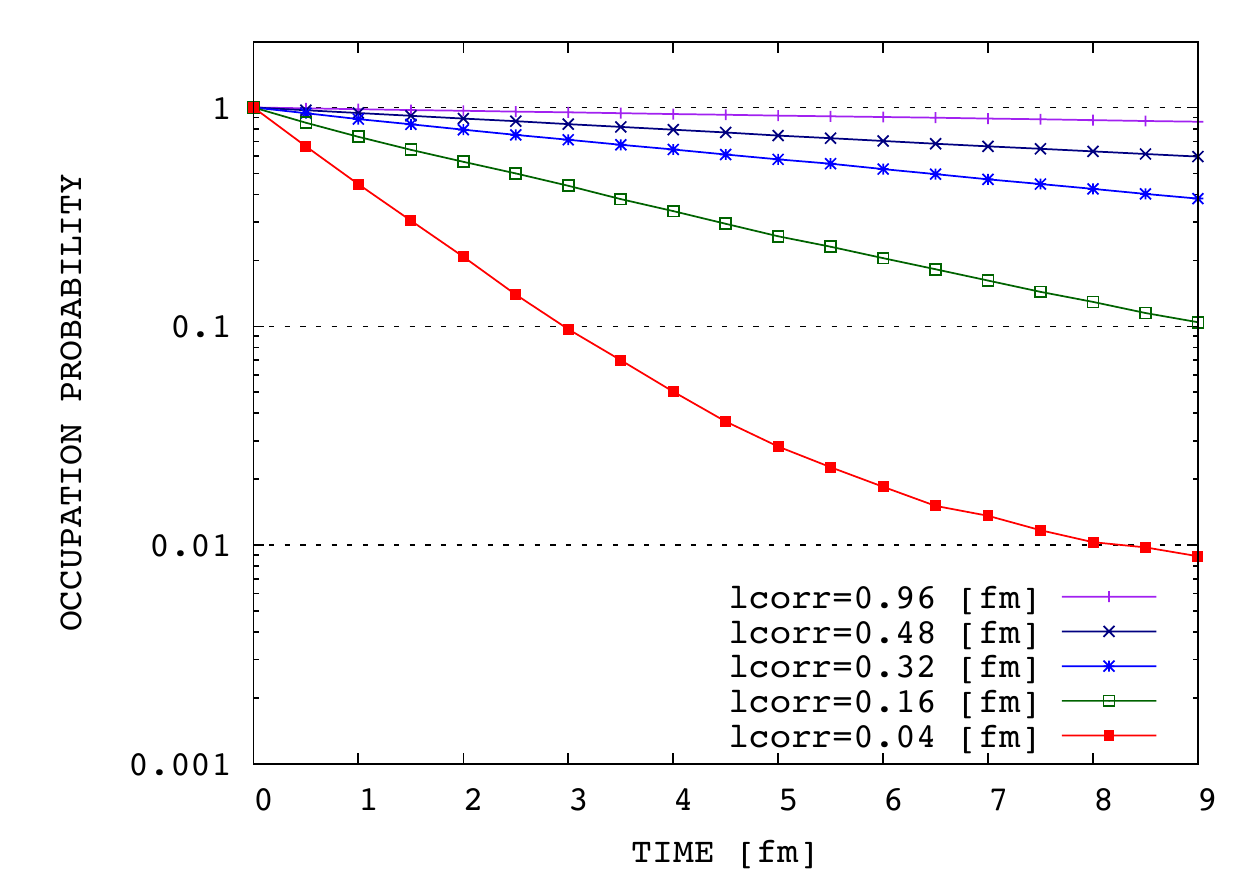}
\includegraphics[scale=0.45]{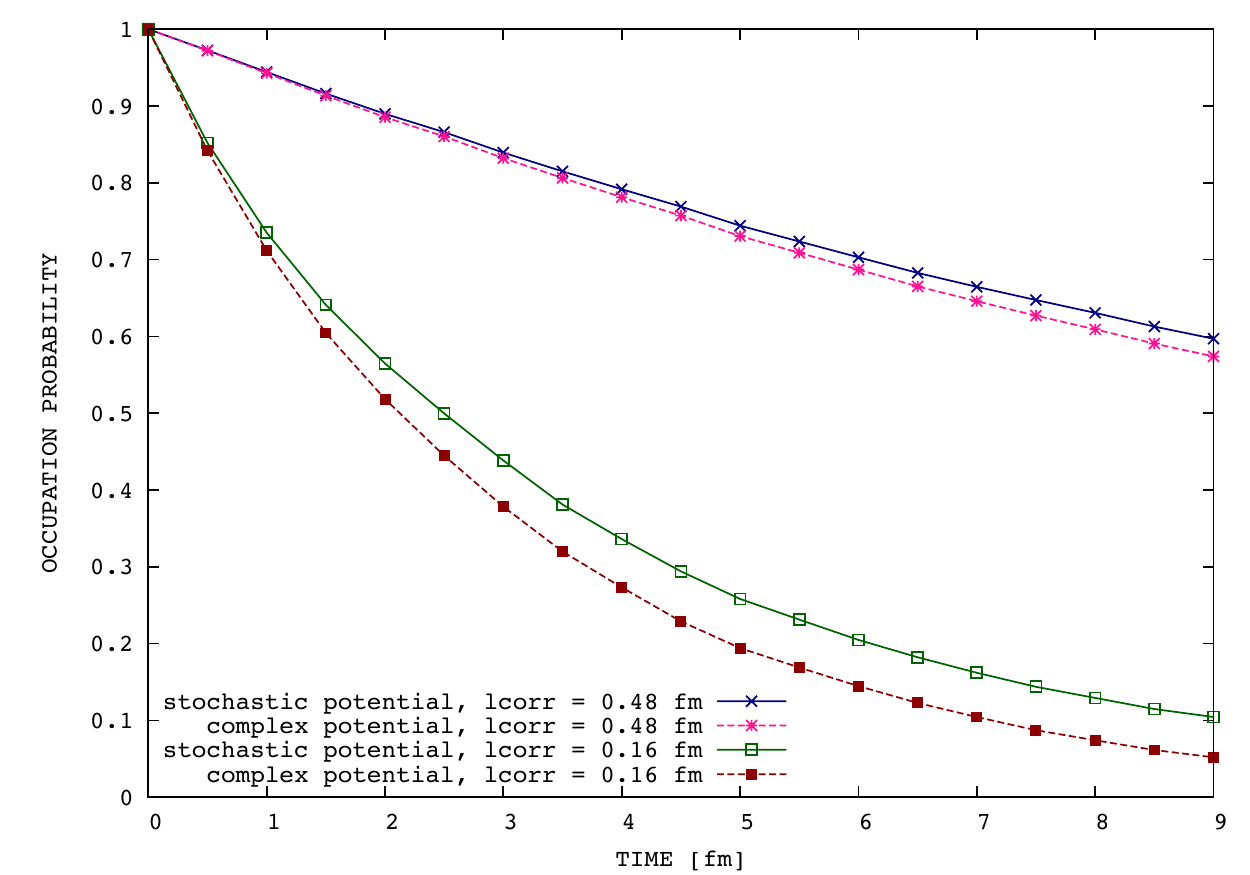}
\caption{(left) the survival probability for the lowest stationary state of $\tilde H_S$, evolved via the stochastic potential prescrtiption for different values of the medium correlation length. Note the transition from inefficient to efficient decoherence as $\ell_{\rm med} \approx \ell_{Q\bar{Q}} $. (right) Comparison of the ground state survival probability for the stochastic potential and the adiabatic approximation. The latter systematically underestimates the in-medium survival. Figures adapted from Ref.~\cite{Kajimoto:2017rel}}\label{fig:StochPotFig}
\end{figure}

Numerical simulations support the conclusions drawn above. In the left panel of \cref{fig:StochPotFig} a simple 1+1 dimensional computation has been performed, where the in-medium stationary ground state of a Hamiltonian with a Debye screened potential is evolved using the stochastic potential prescription. The noise correlations are given by a Gaussian function $D$  with characteristic scale $\ell_{\rm med}$. In such a scenario in the absence of medium induced fluctuations the survival probability $\langle 1S|\sigma_{Q\bar{Q}}(t)|1S\rangle$ of a subsystem eigenstate would remain unity (for details see Ref.~\cite{Kajimoto:2017rel}). Keeping the temperature and thus the strength of the screening fixed and only changing the correlation length of the medium fluctuations leads to the different curves shown. As the radius of the initial state is around $\ell_{Q\bar{Q}}=0.2$fm we expect and also find that a correlation length of $\ell_{\rm med}=0.96$fm is unable to efficiently destabilize the bound state. A much more rapid decay (note the log scale) sets in once the correlation length drops below $\ell_{\rm med}=0.32$fm, becoming equal to the bound state size. While not plotted here, the excited states in the system will be populated more and more strongly, the shorter the correlation length is. 

Note that survival probabilities with values much smaller than unity ensue, not only for the ground but also for the excited states, signaling that a large admixture of the system is actually made up of unbound states in the continuum. An interesting aspect of these dynamics is that continuously excitations and deexcitations are taking place between different states. Such a reshuffling of states will lead to the emergence of a thermalized fixed point of the (full dissipative) dynamics at late times. The availability of transitions between bound and scattering states also entails that if one were to start out from an unbound initial state, efficient decoherence can also lead to a re-population of bound states over time. The probability for this inverse process however is volume dependent, since only in a finite volume the number of unbound states remains finite and thus a non-vanishing probability exists to return into one of the few bound states. By treating explicitly the dynamics between all accessible states the open-quantum systems approach naturally incorporates the physics of recombination. It is important to note however that this form of recombination only considers the two quarks of the initially prepared quarkonium bound state. In order to understand what happens in the presence of multiple $Q\bar{Q}$ pairs a higher order density matrix, such as $\sigma_{Q\bar{Q}Q\bar{Q}}$ needs to be considered, which, while in principle possible, will be a challenge for future studies.

At this point it is informative to consider how the approximate stochastic evolution differs from the adiabatic approximation often employed in phenomenological modeling. In that case the Schr\"odinger equation \cref{eq:schroedavg} is used to solved for the microscopic wavefunction. In the right hand panel of \cref{fig:StochPotFig} the two different computations are shown for a 1+1 dimenisonal setup at two distinct values for $\ell_{\rm med}$. What is found is that the adiabatic approximation works better for cases where the influence of noise is weak. In addition the adiabatic result systematically underestimates the stochastic result, a fact that needs to be kept in mind when such simulations are compared to actual experimental data. Currently first steps are taken to incorporate the insight gained from the open-quantum-systems approach into phenomenological simulations of quarkonium, going beyond the adiabatic approximation \cite{Boyd:2019arx}. 

Access to the (approximate) real-time dynamics of heavy-quarkonium has thus already provided vital insight into their dynamical melting process, in which wavefunction decoherence plays an important role. The historic intuition of sequential melting is elevated to a genuine real-time picture, where depending on the size of individual states, medium fluctuations can efficiently resolve the bound state and induce transitions into higher excited bound or scattering states. The phenomenon of quarkonium melting thus can only be formulated as an initial value problem, inquiring how many states of an initial collection of quarkonia will survive in a medium after a given time. 

In the study of single heavy quarks in the open-quantum systems approach it has been pointed out e.g. in Ref.~\cite{Akamatsu:2015kaa} that the dynamics in phase space, i.e. governed by the momentum operator, in general cannot be disentangled from those in color space and that their interplay under decoherence is essential to understanding of the classicalization of the quarkonium evolution. Thus for a more comprehensive understanding of dynamical quarkonium melting it is paramount that master equations are derived and simulated, which incorporate the dynamics of color degrees of freedom explicitly. This challenge is currently taken on by several groups. For a selection of published master equations including color degrees of freedom see Refs.~\cite{Akamatsu:2014qsa,Brambilla:2016wgg,Blaizot:2017ypk,Yao:2018nmy}.

\subsection{The Boltzmann transport equation from pNRQCD}
\label{sec:BoltzmannEqOQS}

In \cref{sec:FVIFOQS} we discussed the derivation of a Lindblad master equation from QCD at high temperature. It both provided insight into the range of validity of simple Schr\"odinger equation based descriptions of in-medium quarkonium and offered possible routes for their improvement. In this section we will review another open-quantum systems based approach to quarkonium originally presented in Ref.~\cite{Yao:2018nmy}. It starts from the perturbatively matched effective field theory of pNRQCD and translates the operators, states and Wilson coefficients of that EFT into the language of a Lindblad equation. Introducing a Wigner transform of the density matrix, the quarkonium distribution functions are defined, for which transport equations of Boltzmann type are derived under certain time scale separation assumptions.

Starting point of the derivation is pNRQCD considered in thermal equilibrium under the scale hierarchy
\begin{align}
m_Q\gg m_Q v \gg m_Q v^2 \gtrsim T \gtrsim m_D,
\end{align} 
which corresponds to a scenario where the medium still allows for a well defined bound state to exist (c.f. also Ref.~\cite{Brambilla:2011sg}). Taking the S-wave ground state charmonium binding energy as ultrasoft scale $Mv^2\sim E_{\rm bind}(1S)=600$MeV it is indeed larger than the temperatures currently reached in heavy-ion experiments. As made explicit in \cref{eq:pNRQCDcont} the interactions between heavy quarkonium states and the ultrasoft gluons in pNRQCD is to lowest order mediated via a dipole interaction. In a perturbative matching to NRQCD it can be shown (see Ref.~\cite{Pineda:2000gza}) that the corresponding Wilson coefficient $V_A(r)$ does not run in the renormalization group sense and thus at the matching scale can be set to unity $V_A(r; \mu=Mv)=1$ to leading order.

In order to connect to the open-quantum-systems picture formulated in terms of states in a Hilbert space, we need to consider what degrees of freedom are present in such a perturbative pNRQCD setup. Since the color singlet potential is attractive there may be either bound or unbound singlet states present, while due to the repulsive octet potential only unbound colored states exist. Their time evolution in the absence of coupling to ultrasoft gluons can be cast into the form of a Schr\"odinger equation governed by the Hamiltonians
\begin{align}
H_{s,o} = \frac{\mathbf P^2}{4M} + \frac{\mathbf p^2}{M} + V_{S,O}^{(0)} + \frac{V_{S,O}^{(1)}}{M} + \frac{V_{S,O}^{(2)}}{M^2} +  {\cal O}(v^3),
\end{align}
arising from the terms in the first and second line of the pNRQCD Lagrangian in \cref{eq:pNRQCDcont}. Note that here $V_{S,O}$ are purely real and can be perturbatively computed from QCD. Under the present scale separation the kinetic terms $\mathbf{P}^2/4M$ related to the center of mass motion are subdominant so that the following eigenstates for the relative motion ensue
\begin{align}
| \psi_{nl} \rangle, \quad | \psi_{{\mathbf p}} \rangle, \quad | \Psi_{{\mathbf p}} \rangle.
\end{align}

\noindent Ref.~\cite{Yao:2018nmy} now considers creation $(\dagger)$ and annihilation operators for all three distinct composite particle entities 
\begin{align}
a^{(\dagger)}_{nl}(\mathbf{P}): {\rm bound \, singlet}, \quad  b^{(\dagger)}_{{\mathbf{p}}}(\mathbf{P}): {\rm unbound\,singlet}, \quad c^{a(\dagger)}_{{\mathbf{p}}}(\mathbf{P}): {\rm unbound\, octet}
\end{align}
that fulfill the canonical commutation relations. The corresponding single particle states are 
\begin{align}
|\mathbf  K, nl, 1\rangle = a^{\dagger}_{nl}(\mathbf K) | 0 \rangle , \quad 
 |{\mathbf P}, {\mathbf p} ,1\rangle  = b^{\dagger}_{{\mathbf p}}({\mathbf P}) | 0 \rangle, \quad
 |{\mathbf P}, {\mathbf p}, a\rangle  = c^{a\dagger}_{{\mathbf p}}({\mathbf P})  | 0 \rangle\,.
\end{align}
The set of wavefunctions and particle operators allows the authors to write an explicit representation of the singlet and octet wavefunctions as
\begin{align}
|S(\mathbf{R}, t) \rangle =& \int\frac{d^3 P}{(2\pi)^3}  e^{-i(Et-\mathbf{p} \cdot \mathbf{R})} \bigg( \sum_{nl} a_{nl}(\mathbf{P}) \otimes | \psi_{nl} \rangle   + \int\frac{d^3 p}{(2\pi)^3} b_{\mathbf{p}}(\mathbf{P}) \otimes | \psi_{\mathbf{p}} \rangle \bigg), \\
|O^a(\mathbf{R}, t) \rangle =&  \int\frac{d^3 P}{(2\pi)^3} e^{-i(Et-\mathbf{p}\cdot \mathbf{R})}  \int\frac{d^3 p}{(2\pi)^3} c^a_{\mathbf{p}}(\mathbf{P}) 
\otimes | \Psi_{\mathbf{p}} \rangle \,.
\end{align}
Defining the matrix elements in the relative position space for the singlet and octet sector as
\begin{align}
& \langle {\mathbf r} | S({\mathbf R}, t) \rangle \equiv S({\mathbf R}, {\mathbf r}, t), \quad \langle {\mathbf r} | O^a({\mathbf R}, t) \rangle  \equiv O^a({\mathbf R}, {\mathbf r}, t),\\
& \langle S({\mathbf R}, t) | f({\mathbf r}) | O^a({\mathbf R}, t) \rangle \equiv \int d^3r S^\dagger({\mathbf R}, {\mathbf r}, t) f({\mathbf r}) O^a({\mathbf R}, {\mathbf r}, t)\,,
\end{align}
the interaction terms of the pNRQCD Lagrangian can be reexpressed (see Ref.~\cite{Fleming:2005pd}) in an alternative form, which directly invites the connection to the open-quantum systems formulation. Take for example the singlet to octet term
\begin{align}
{\cal L}_{{\rm int},so} =  \sqrt{\frac{T_F}{N_C}}\Big( \langle O^a(\mathbf R, t) | \mathbf{ r} \cdot g{\mathbf E}^a(\mathbf R, t) | S(\mathbf R, t)\rangle + {\rm h.c.} \Big).
\end{align}
It can be reinterpreted according to a $H_{\rm int}=\sum_m \Sigma_m \otimes \Xi_m$ as defining 
\begin{align}
\Sigma_m\equiv \langle S(\mathbf R, t) | r_i  | O^a(\mathbf R, t) \rangle +  \langle O^a(\mathbf R, t) | r_i  | S(\mathbf R, t)\rangle , \quad \Xi_m \equiv \sqrt{\frac{T_F}{N_C}} g E_i^a(\mathbf R, t),
\end{align}
where the subscript $m$ refers to the center of mass coordinate $\mathbf R$, the vector components $i$ and the color degrees of freedom $a$. 

It is at this point that the Markovian approximation is invoked. In the scale hierarchy considered here, it is justified by comparing the characteristic medium time scale $1/T$ with the inverse of the dissociation rate, in turn allowing one to neglect all memory integrals in the master equation. (For technical details see the appendices of Ref.~\cite{Yao:2018nmy})

As we saw e.g. in the derivation of the quantum optical master equation in \cref{eq:DefGamma}, the two-point correlation functions of the $\Xi$ operators can be split into two contributions. One $S_{mn}$ will lead to a modification of the subsystem Hamiltonian and the other $\gamma_{mn}$ provides the prefactors of the Lindblad operators.

\begin{figure}
\centering
\includegraphics[scale=0.45]{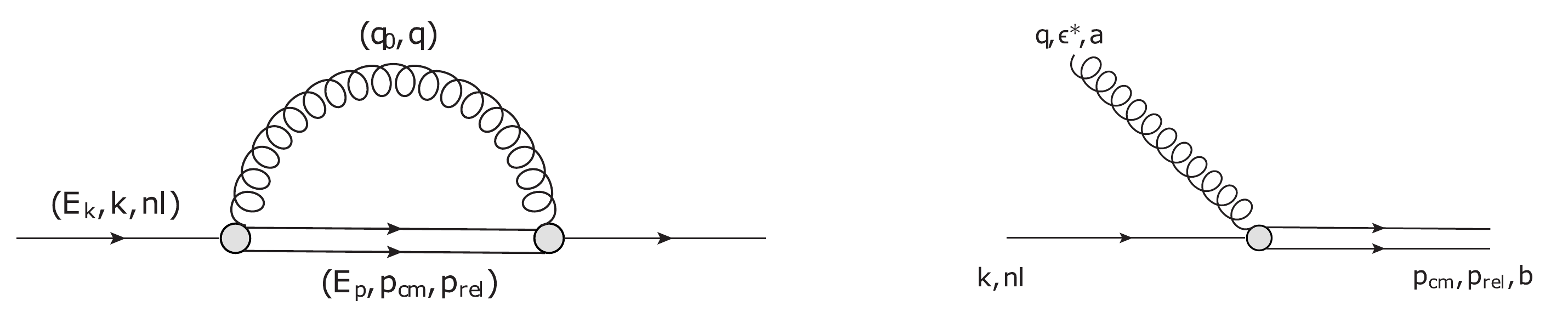}
\caption{(left) The pNRQCD self energy diagram responsible for the correction of the real-part of the in-medium potential, as well as for (left) inducing quarkonium dissociation and recombination via mediating of transitions between the singlet and octet sector.}\label{fig:pNRQCDDiags}
\end{figure}

It turns out that the former can be straight forwardly evaluated in pNRQCD as it arises from the self energy Feynman diagram where a singlet state via interaction with an ultrasoft gluon turns into an intermediate octet state and eventually returns to the singlet by a second interaction (see \cref{fig:pNRQCDDiags}). The real part of this contribution enters as a loop correction to the otherwise real-valued potential and can be used to define the effective subsystem Hamiltonian $\tilde H_S$.  I.e. we are considering a system where part of the genuine field theoretical interactions between the quarkonium and the ultrasoft gluons can be summarized in a correction to a time independent potential.

Before continuing to a more detailed discussion of the contributions from fluctuations onto the quarkonium dynamics one can connect to the language of transport equations. To this end a distribution function for the heavy quarkonium states needs to be defined. Since the main interest lies in the evolution of bound states this may be achieved by taking the Wigner transform of the corresponding matrix elements of the reduced system density matrix in the singlet basis
\begin{align}
f_{nl}({\mathbf X}, {\mathbf K}, t) \equiv \int\frac{d^3K'}{(2\pi)^3} e^{i {\mathbf K}'\cdot {\mathbf X} } \langle  {\mathbf K}+\frac{{\mathbf K}'}{2}, nl,1   | \hat \sigma_S(t)  |   {\mathbf K}-\frac{{\mathbf K}'}{2} , nl, 1\rangle \,.
\end{align}
The strategy here is to derive explicit expressions for $f_{nl}({\mathbf X}, {\mathbf K}, t)$ at finite time $t$ given by terms that posses a linear dependence on time, so that taking the time derivative results in the evolution equation of interest. Essentially one sandwiches many independent small time evolution steps after each other. This trick is applicable due to the Markovian approximation, where e.g. the explicit t dependence of possible integral terms has been removed by setting the upper boundaries to infinity. 

The first step is to consider the effect of the von-Neumann like term in the Lindblad equation on the time evolution of the distribution function.  The commutator with $\tilde H_S$ leads to
\begin{align}
f_{nl}({\mathbf X}, {\mathbf K}, t) 
&= -it\int\frac{d^3K'}{(2\pi)^3} e^{i {\mathbf K}'\cdot \mathbf X} \langle \mathbf K+\frac{{\mathbf K}'}{2}, nl, 1| (\hat{\tilde{H_S}} \hat \sigma_S -\hat \sigma_S \hat{\tilde{ H_S}}) | \mathbf K-\frac{{\mathbf K}'}{2}, nl, 1 \rangle + \cdots\\
&=f_{nl}({\mathbf X}, {\mathbf K}, 0) - it\int\frac{d^3K'}{(2\pi)^3} e^{i {\mathbf K}'\cdot \mathbf X} (E_{\mathbf K +\frac{{\mathbf K}'}{2}} - E_{\mathbf K -\frac{{\mathbf K}'}{2}}) \langle \mathbf K+\frac{{\mathbf K}'}{2}, nl, 1| \hat \sigma_S(0) | \mathbf K-\frac{{\mathbf K}'}{2}, nl, 1 \rangle + \cdots\,.
\end{align}
where the additional terms from the Lindblad operators are not shown explicitly. In conjunction with the Wigner transform this expression can be used as the starting point of a gradient expansion leading eventually to 
\begin{align}
f_{nl}({\mathbf X}, {\mathbf K}, t)
= f_{nl}({\mathbf X}, {\mathbf K}, 0)  - t {\mathbf v} \cdot\nabla_{\mathbf X} f_{nl}(\mathbf X, \mathbf K, 0) + \cdots \,,
\end{align}
where for notational purposes the center of mass velocity is defined as $\mathbf{v}=\mathbf{K}/2m_Q$.

The terms involving Lindblad operators can be divided into two sets. The one including $L^\dagger L \sigma$ and $\sigma L^\dagger L$ is directly related to the perturbative dissociation rate, as it is defined from the dipole transition between the singlet to octet.  I.e. here the imaginary part of the pNRQCD self energy contribution implements the process depicted on the right hand side of \cref{fig:pNRQCDDiags}. This finding is consistent with previous results of Ref.~\cite{Brambilla:2011sg}, where the same diagram was reinterpreted to give an imaginary part to the in-medium heavy quark potential. Note that to this order in the approximation, the potential in the pNRQCD Lagrangian is completely real and that it is the loop corrections from the interaction with the ultrasoft gluons in the scale hierarchy present that induce an imaginary part. 

The last missing term is the one involving $L\sigma L^\dagger$, which is shown to be related to the regeneration of quarkonium states in this scale hierarchy. This can be intuitively understood from the fact that it contains the distribution function of color octet quarkonium states, as well as the dipole transition probability between the singlet and octet sector.

Combined into one expression a Boltzmann equation emerges
\begin{align}
\frac{\partial}{\partial t} f_{nl}({\mathbf X}, {\mathbf K}, t) + {\mathbf v}\cdot \nabla_{\mathbf X}f_{nl}({\mathbf X}, {\mathbf K}, t) = {\cal C}_{nl}^{(+)}({\mathbf X}, {\mathbf K}, t) - {\cal C}_{nl}^{(-)}({\mathbf X}, {\mathbf K}, t)\,,
\end{align}
where ${\cal C}_{nl}^{(+)}({\mathbf X}, {\mathbf K}, t)$ encodes the recombination and ${\cal C}_{nl}^{(-)}({\mathbf X}, {\mathbf K}, t)$ the dissociation processes arising from the Lindblad operators. In the computation of the recombination term often additional approximations are employed. One of these is the so called molecular chaos approximation, where the distribution of the octet quarkonium states is written as a simple product of distribution functions for the individual quark and antiquark degrees of freedom. One may argue from a comparison of the  decorrelation rate of the heavy quarks with the relaxation rate of the quarkonium system that such a simplification is justified. 

A detailed comparison of the involved time scales, finally allows Ref.~\cite{Yao:2018nmy}  to recover in addition to the dissociation term also a regeneration term of the same form as used in previous implementation of the Boltzmann equation in e.g. Ref.~\cite{Yao:2017fuc}. This completes the systematic derivation of the transport approach to heavy quarkonium from the underlying microscopic theory of QCD for a medium, in which the temperature is low enough for well defined bound states to exist.

We find that the open-quantum-systems approach based on perturbative pNRQCD incorporates in a consistent manner the relevant processes for the S-wave ground state dynamics: (1) screening of a real-valued microscopic potential, (2) the effects of scatterings which both include the dissociation of quarkonium due to excitation into octet states (3) recombination from the inverse process where a colored quark antiquark pair can emit a gluon and coalesce back into a bound state. 

Note that Ref.~\cite{Yao:2018nmy} also considered the annihilation of heavy quark antiquark pairs via the four-fermion interactions of the NRQCD Lagrangian, adding their contributions as additional Lindblad operators. It is found that for phenomenologically relevant temperatures and accessible time scales this effect is indeed negligible. This finding reinforces the point that the dileptons measured in a heavy-ion collision are not those emitted from in-medium quarkonium states but actually from vacuum states decaying long after the QGP has ceased to exist.

In many applications of transport theory to heavy quarkonium, instead of the Boltzmann equation, a simpler rate equation is deployed \cite{Rapp:2017chc}. In light of the systematic derivation of the Boltzmann equation from QCD it is an interesting question to ask under which conditions such a rate equation can be subsequently derived as well. Since usually the regeneration term in such an approach is determined from arguments based on detailed balance, the system needs to have had enough time to approach thermal equilibrium for that setup to be valid.

One central open question, similar to the case of the Feynman-Vernon influence functional, is how the connection to microscopic QCD, derived here in a perturbative setting, can be established also in a strongly coupled scenario. In the context of pNRQCD this pertains both to the question of what are the relevant degrees of freedom, as well as how to compute the ensuing Wilson matching coefficients from e.g. lattice QCD simulations. A recent proposal in this direction has been put forward in Refs.~\cite{Brambilla:2017zei,Brambilla:2019tpt}.

\begin{summary}
Significant progress in our understanding of the real-time evolution of heavy quarkonium has been achieved over the past years. With quarkonium in a medium naturally inviting the distinction between a small system and the environment, the open-quantum systems approach has been a vital tool in this regard. It provides a versatile theoretical framework to account for the different timescales present in the system and has lead to an improved understanding of quarkonium melting as a genuine dynamical process. Decoherence induced by the medium fluctuations is found to play an important role in addition to static screening, allowing the medium to act as a bound state sieve. An improved theory understanding of quarkonium real-time evolution has also allowed to put current approaches to phenomenological modeling of quarkonium dynamics on a more solid footing by exposing their underlying assumption. Not only has the stochastic Schr\"odinger equation been derived from a Lindblad equation via the quantum state diffusion method but also the Boltzmann transport equations have been systematically obtained from a pNRQCD based master equation. The quest for a genuine non-perturbative implementation of the open-quantum systems master equations is ongoing.
\end{summary}
  
\section{Heavy quarkonium in relativistic heavy-ion collisions}
\label{sec:qqbarhic}

In the previous sections we have developed an improved theoretical understanding of the properties of kinetically equilibrated heavy quarkonium, as well as its real-time evolution in the background of a thermal medium. It is now time to ask how this insight can help us in shedding light on quarkonium production in heavy-ion collisions.

While the focus of this review is the theoretical understanding of in-medium quarkonium it pertains to only but one part of the puzzle in understanding quarkonium production in heavy-ion collisions (for a recent review including also an extended survey of experimental measurements see Ref.~\cite{Andronic:2015wma}). Indeed such collisions encompass many complex aspects, from the production of the heavy quark antiquark pairs, the evolution of the bulk medium consisting of light partons, the physics of hadronization, as well as the evolution within the hadronic phase. Even in the context of heavy quarkonium alone many open questions remain, some of the most pressing are listed below:
\begin{itemize}
\item How do the energetic partons inside the projectile nuclei interact to form heavy quark antiquark pairs? 
\item Do heavy quarkonium states form in the early stages of the collision, which are dominated by the strong coherent fields of the glasma and if so what are the time scales involved? 
\item If a quarkonium state has formed early on, how does it react to the emergent quark gluon plasma it is immersed in. 
\item How does the interaction with a locally thermalized but steadily cooling medium affect the stability of existing bound states? How efficient is recombination of color octet pairs into singlet bound states in such a setting?
\item Depending on the flavor of the constituent quark and the lifetime of the fireball, to what degree does quarkonium and in turn the heavy quarks equilibrate? 
\item What happens at the crossover transition, where individual colored partons need to come together to form color neutral hadrons?
\item How do the bound states formed at the crossover transition propagate in the hadronic phase, how are their stability properties modified?
\end{itemize}
A true theory understanding of quarkonium in HICs requires a QCD based reply to all of the above, which at the same time should provide an efficient prescription to compute quantitative postdictions of the measured yields in current experiments. Vital insight has been already achieved and further progress is on the horizon based on the steady development of an effective field theory based real-time understanding of heavy-icon collisions. Combining a description of the initial stages of the collisions based on the color glass condensate, of thermalization and evolution of the light degrees of freedom via classical statistical simulations, kinetic theory and relativistic hydrodynamics, of the physics of heavy quarks via NRQCD and that of heavy quarkonium in the open quantum systems approach, all orchestrated in conjunction  with the non-perturbative predictive power of lattice QCD bodes well for such an ambitious task to succeed in the next decade.  

\subsection{Preliminaries: quarkonium in $p+p$ collisions}
\label{sec:qqbarpp}

As a first step one can consider the process of quarkonium production in the absence of a hot medium (for details see e.g. Chap. 2 of Ref.~\cite{Andronic:2015wma}). This area of study has a long history and is still actively pursued both experimentally and theoretically in proton-proton and proton-nucleus collisions at current collider facilities. Quarkonium in $p+p$ collisions constitutes a fascinating subject by itself which allows us to learn e.g. about the complex inner structure of nucleons as well as the intricate dynamics of hadronization. Its understanding also forms the basis for developing insight into the production process in more complex scenarios such as proton-nucleus and nucleus-nucleus collisions.

\begin{figure}
\centering
\includegraphics[scale=0.35, clip=true, trim=0 8cm 22cm 0]{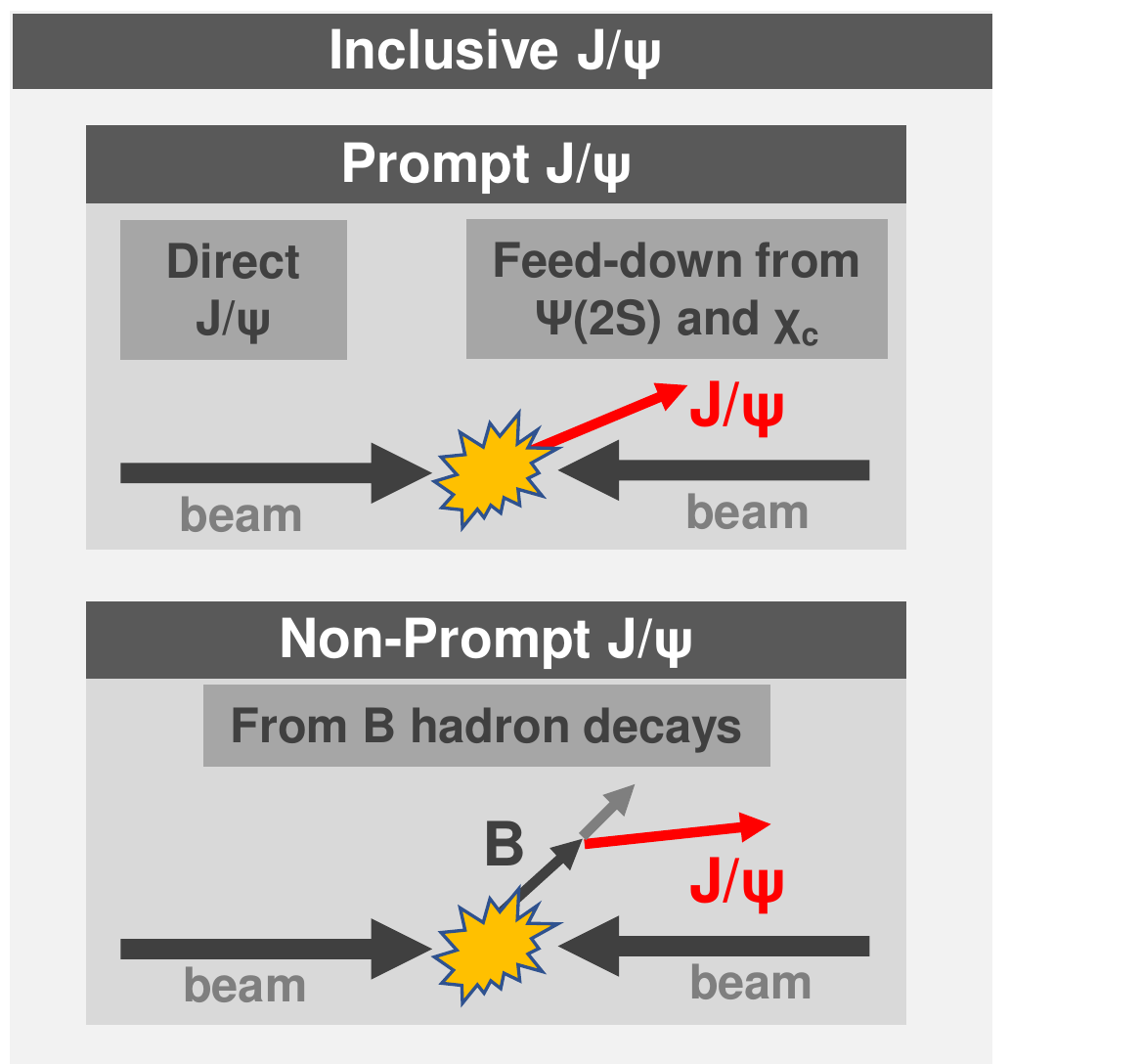}
\includegraphics[scale=0.25, clip=true, trim=0 4cm 12cm 0]{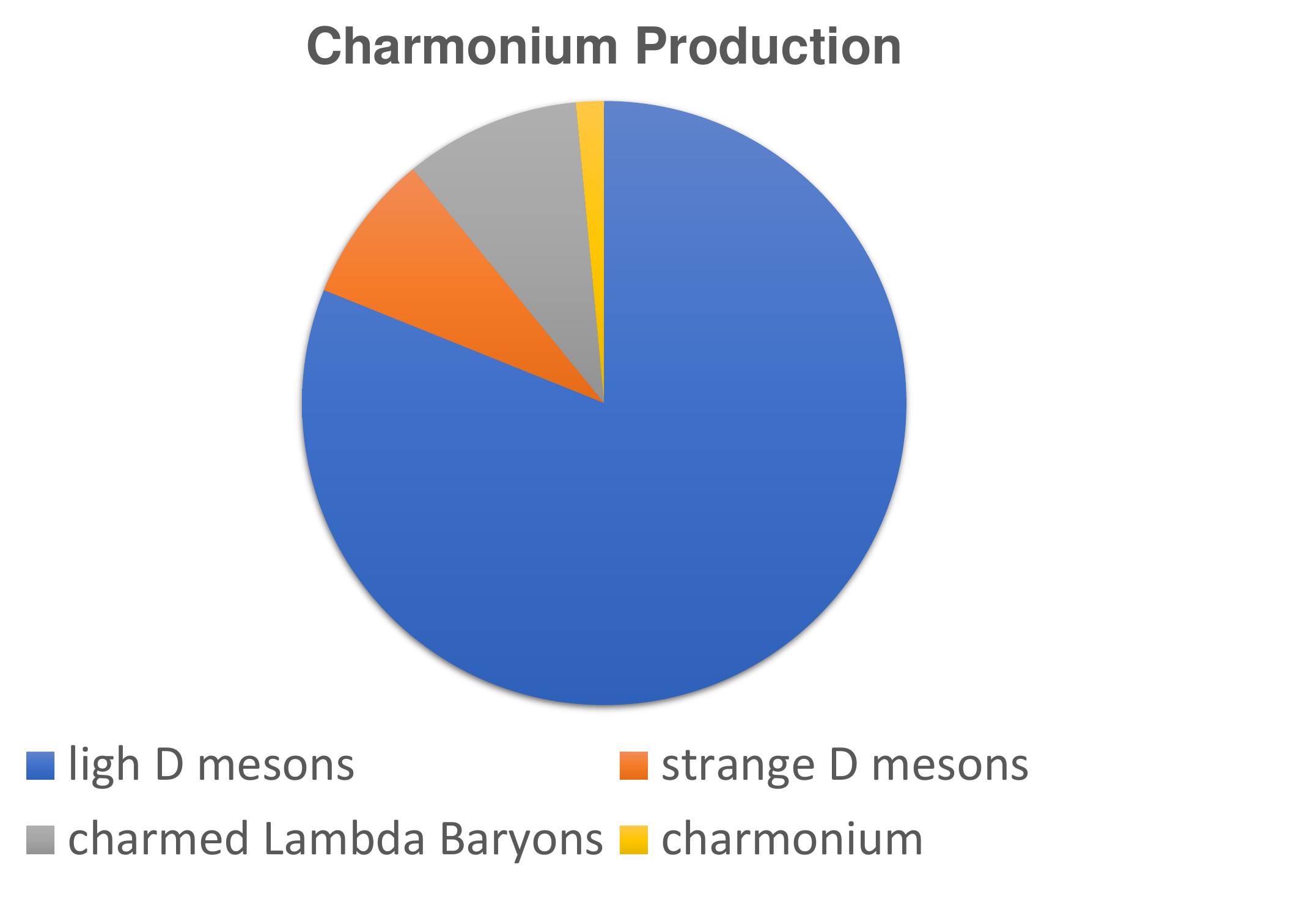}
\includegraphics[scale=0.2, clip=true, trim=0 4cm 5cm 0]{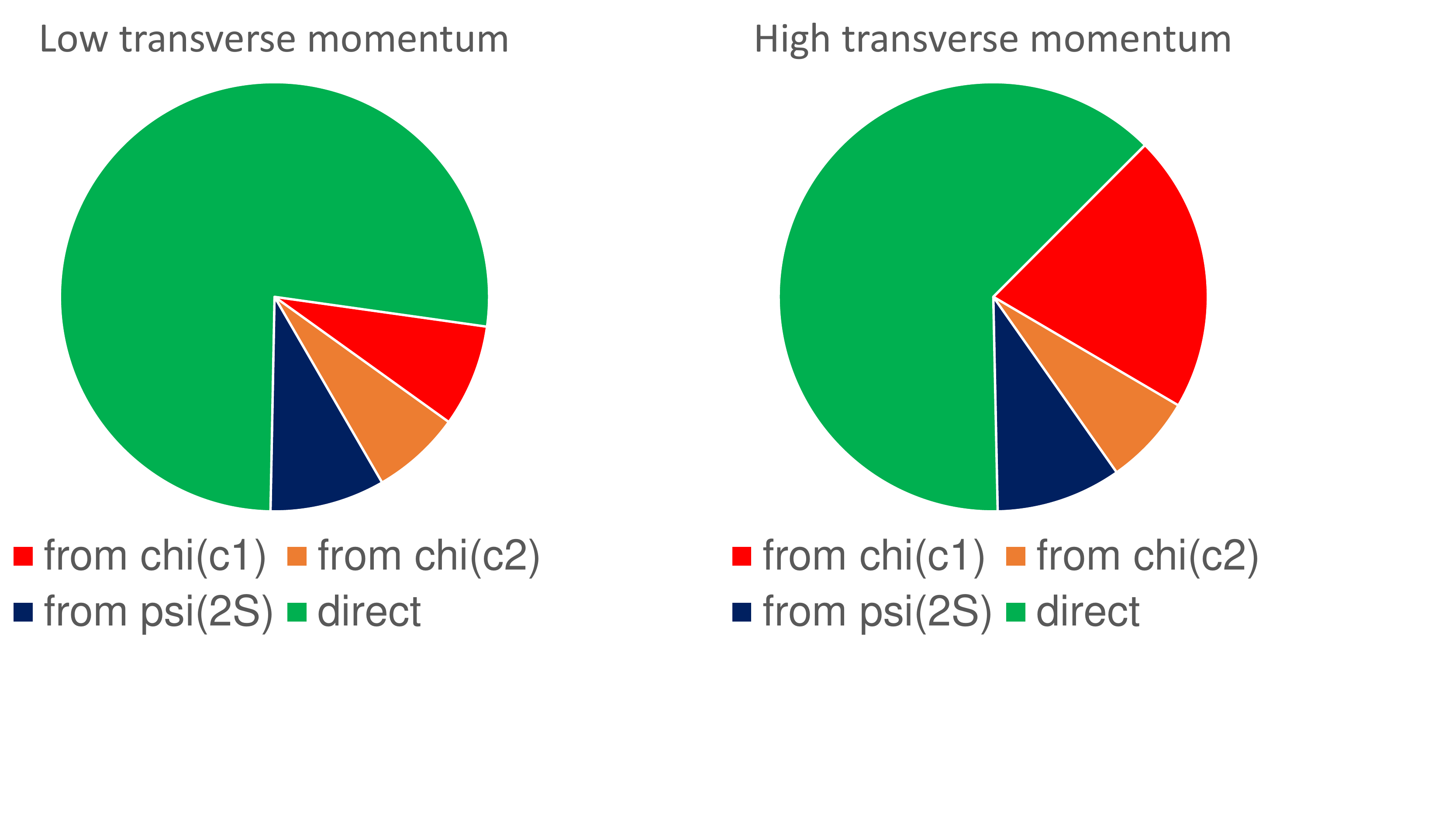}

\caption{(left) Example of inclusive $J/\Psi$ production consisting of both prompt and non-prompt contributions. The former arises from $J/\Psi$ particles formed in the collision region, either directly or as end products of feeddown processes from $\psi(2S)$ and $\chi_c$ states. The latter contribution originates in the decay of B mesons outside of the initial scattering event. (second from left) Typical production probabilities for chamed hadrons in a proton-proton collision. Note the minute share $\sim 2\%$ of charmonium on the overall production yield.(right) Typical values of prompt $J/\Psi$ production in $p+p$ collisions at high and low transverse momentum, sorted according to their origin from direct production or feed-down. (Figure adapted from Ref.~\cite{Andronic:2015wma}) }\label{fig:ppProduction1}
\end{figure}

\begin{figure}
\centering
\includegraphics[scale=0.45, clip=true, trim=0 11cm 12cm 0]{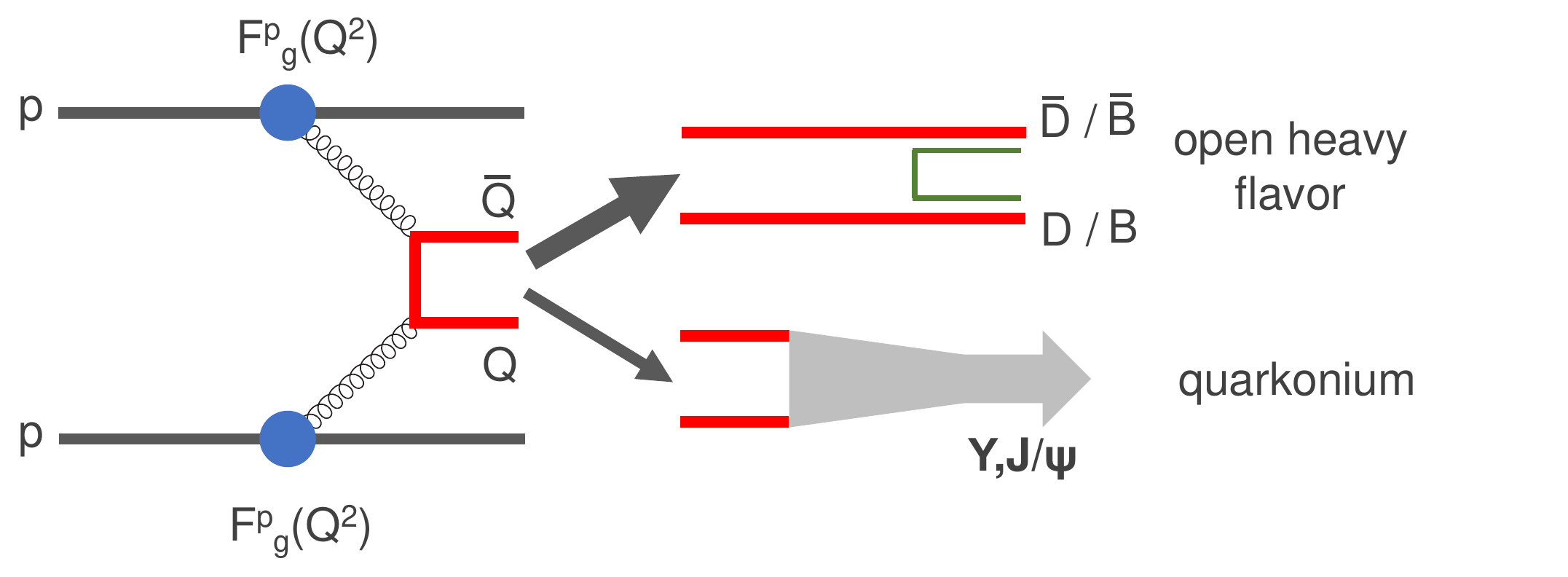}
\caption{ The two stages of the production process of prompt quarkonium: production of the color bearing heavy quark antiquark pair from hard partons and the subsequent formation of either open heavy flavor or quarkonium states. As indicated by the thickness of the arrows, open heavy flavor constitutes the overwhelming majority of produced particles.}
\label{fig:ppProduction2}
\end{figure}

Let us clarify what one actually refers to when talking about quarkonium production. The so called inclusive production contains two independent contributions, prompt and non-prompt. As sketched on the leftmost panel of of \cref{fig:ppProduction1},
for the example of the charmonium ground state $J/\Psi$. The prompt contribution consists of both directly produced $J/\Psi$ and those that follow from feed-down from excited states $\psi^\prime$ as well as P-wave states $\chi_c$. On the other hand the non-prompt contribution refers to $J/\Psi$ that results from the decay of B mesons that initially formed in the collision. 

In general the production of quarkonium only constitutes a small fraction of all mesons that arise from the production of a heavy quark antiquark pair. As indicated in the second panel from the left in \cref{fig:ppProduction1} mostly open heavy flavor mesons emerge from the collision. Even the formation of Baryons with one heavy quark component constitutes a larger share than that of actual quarkonium. If one wishes to understand the yields of the quarkonium ground states, one further needs to understand the feed-down contributions from excited states, which, depending on the transverse momentum involved, actually differ significantly as illustrated in the two right panels of \cref{fig:ppProduction1}. The corresponding charts for Bottomonium including in particular the recent measurements by the LHCb collaboration can be found in Ref.~\cite{Andronic:2015wma}.

The theory understanding of the physics of the production process benefits from the presence of a separation of scales between the hard scale of the heavy quark rest mass and the correspondingly large momenta of the partons in the projectiles, as well as lower scales, such as $\Lambda_{\rm QCD}$ and the binding energies of quarkonium states. In turn the overall production process can be considered as {\it factorized} between individual subprocesses, i.e. the production of a heavy quark and the subsequent formation of color neutral hadrons as indicated in \cref{fig:ppProduction2}. Theory has seen significant progress and success in studying quarkonium production over the past decades, the question of quarkonium polarization however remains a challenging topic.

\begin{figure}
\centering
\includegraphics[scale=0.45, clip=true, trim=0 11cm 1cm 0]{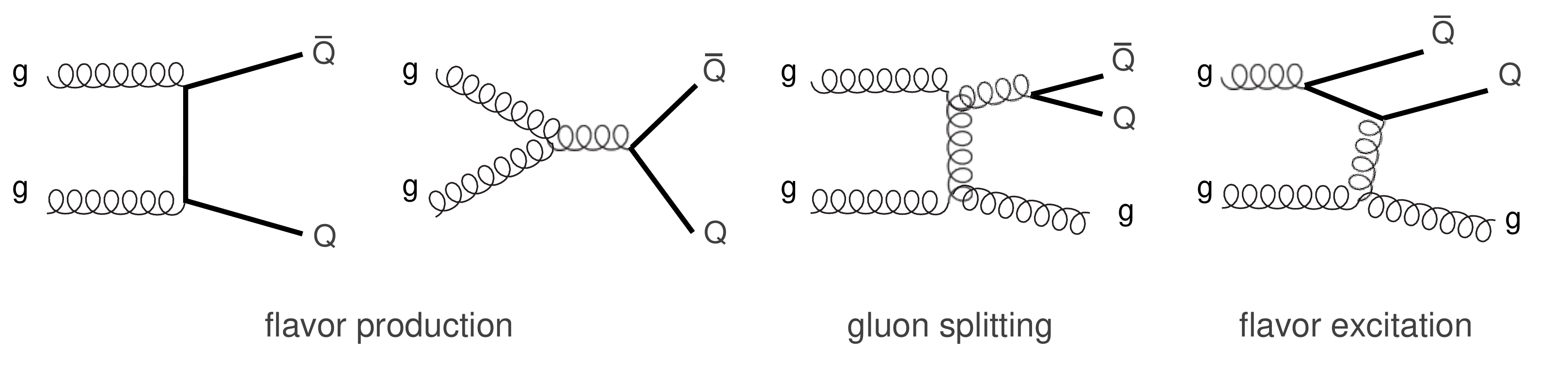}
\caption{Relevant scattering processes in the production of heavy quark anti-quark pairs at current proton-proton colliders.  }\label{fig:ppProduction3}
\end{figure}

The first step consists of understanding how the heavy quarks come into being. The main ingredient is the presence of highly energetic partons in the proton projectiles, which may scatter off of each other to produce a heavy quark antiquark pair. The four most relevant processes in this regard are shown in \cref{fig:ppProduction3}, consisting of s-channel and t-channel flavor production, as well as gluon splitting and flavor excitation. At the end of the interaction a $Q\bar{Q}[n]$ pair  in any of $n$ possible states emerges, which may posses large relative, as well as center of mass momentum.

Due to the heavy quark mass being much larger than $\Lambda_{\rm QCD}$ it might appear that the production of the heavy quark pair is a purely perturbative issue. However we must not forget that the gluons were actually part of a proton, whose highly non-perturbative parton distribution functions $F^p_i(q)$ tell us how probable it is that a gluon with a certain momentum takes part in the scattering process. The cross section  $\sigma^{pp\to Q\bar{Q}[n]+X}$ to produce the $Q\bar{Q}$ from a $p+p$ collision, can actually be approximated to a good degree as a product over the genuinely perturbative cross-section for each of the gluon scattering processes and the gluon distribution functions $F^p_g(q)$ 
\begin{align}
\sigma^{pp\to Q\bar{Q}[n]+X}(q_1,q_2) \overset{q\gg\Lambda_{\rm QCD}}{\approx}  \sum_{X^\prime} F^{p_1}_g(q_1^2)F^{p_2}_g(q_2^2)\cdot \sigma^{gg\to Q\bar{Q}[n]+X^\prime}(q_1,q_2).
\end{align}
The information about the distribution of momenta among the partons within a proton has been studied thoroughly via deep inelastic scattering. Its computation from first principles lattice QCD on the other hand is an active field of research.

The second step is to consider how the in general color charged $Q\bar{Q}$ pair actually forms a color neutral quarkonium meson denoted here by $H$ \cite{Bodwin:1994jh}. Again one benefits from an approximate factorization between the production and the formation stage
\begin{align}
\sigma^{pp\to H +X} \approx  \sum_n\int d\Pi_{Q\bar{Q}} \sigma^{pp\to Q\bar{Q}[n]+X} (\Pi) \cdot M(Q\bar{Q}[n] \to H),
\end{align} 
where $\Pi$ denotes an appropriate integration over the phase space of the quark antiquark pair and $M$ refers to the matrix element of producing the hadron 
\begin{align}
M(Q\bar{Q}[n] \to H)=\sum_X \langle 0 |  \chi^\dagger {\cal K_n} \psi |H+X\rangle\langle H+X| \psi^\dagger {\cal K}_n^\prime \chi\rangle.
\end{align}
Here $X$ denotes any other hadron and ${\cal K}$ appropriate combinations of color and spin matrices, depending on the state $Q\bar{Q}[n]$.

Modeling the formation process has a long history starting from the color singlet model (CSM) and the more recent color evaporation model (CEM). In the former one assumes that a color neutral quarkonium can only arise from an originally color singlet quark antiquark pair. Further, one neglects the relative momentum between the heavy quarks essentially considering the $Q\bar{Q}$ being born at rest on top of each other. Then the radial wavefunction of the quarkonium at the origin $R_H(0)$ can be used to estimate the production cross section
\begin{align}
\sigma^{pp\to H +X}_{\rm CSM} \propto  \sigma^{pp\to Q\bar{Q}[n]+X} (v=0) \otimes |R_H(0)|^2.
\end{align}
In this form the CSM corresponds to the lowest order approximation of NRQCD, i.e. for a relative velocity $v=0$.

In the CEV on the other hand each $Q\bar{Q}$ pair is assigned the same probability to form a quarkonium, as long as the invariant mass of the quark pair is less than that of the meson H. The probability for each meson is contained in a model parameter $F_H$
\begin{align}
\sigma^{pp\to H +X}_{\rm CEM} =F_H \int _{4m_Q}^{4m_H^2}dm_{Q\bar{Q}}^2 \frac{d}{dm_{Q\bar{Q}}} \sigma^{pp\to Q\bar{Q}[n]+X} \cdot  1.
\end{align}
The CEM is able to provide a reasonably good reproduction of e.g. $J/\psi$ production at LHC and it is actively further developed to improve quantitative agreement with data, also in the case of bottomonium. For recent developments see e.g. Refs.~\cite{Cheung:2017osx,Cheung:2018tvq,Cheung:2018upe}, where both the agreement with the measured $p_T$ dependence of $J/\psi$ as well as an improved postdiction of the $\psi^\prime$ yields have been achieved. In turn the first computation of the explicit $p_T$ dependence of the $\psi^\prime$ to $J/\psi$ ratio in the context of the CEM was presented.

The NRQCD approach generalizes the CSM result with systematic corrections of higher power of $v$, which include the contributions in particular from color octet states. At large enough transverse momenta, where the NRQCD scale separation is expected to be valid, one can write the cross section as a sum
\begin{align}
\sigma^{pp\to H +X}_{\rm NRQCD} = \sum_n \sigma^{pp\to Q\bar{Q}[n]+X} \cdot \langle {\cal O}_H^n\rangle,
\end{align}
where the operators ${\cal O}_H^n$ are obtained from the four fermion interactions of NRQCD
\begin{align}
{\cal O}_H^n=\chi^\dagger {\cal K}_n \psi (a_H^\dagger a_H)\psi^\dagger {\cal K}^\prime_n\chi
\end{align}
and contain the creation and annihilation operators of the hadron, as well as the NRQCD heavy quark $\psi$ and antiquark fields $\chi$. For a more detailed discussion and the explicit form of the operators in NRQCD the reader is referred to Ref.~\cite{Bodwin:1994jh}. 

One recent interesting development in the context of quarkonium production is the extension of the NRQCD approach to transverse momenta lower than $5$GeV. Ref.~\cite{Ma:2014mri} proposes to combine heavy quark effective field theory matrix elements with a color glass condensate model to approximate the gluon distributions in the proton projectiles, leading to good agreement with data also for the challenging subject of quarkonium polarization \cite{Ma:2018qvc}. One finds a smooth overlap region of these new results with the standard NLO NRQCD at high $p_T$.  

\subsection{Cold nuclear matter effects}
\label{sec:qqbarcnm}

After gaining a first taste of quarkonium production in $p+p$ collisions let us touch on the next more complex scenario of proton nucleus collisions. Here one wishes to learn about so called {\it cold nuclear matter} effects, which may affect quarkonium production also in a heavy-ion collision. A comprehensive overview of this topic may be found in Refs.~\cite{Albacete:2013ei,Albacete:2016veq,Albacete:2017qng} and chapter 3 of Ref.~\cite{Andronic:2015wma}, a collection of recent results in Ref.~\cite{Ferreiro:2018umi}. 

Experimentally one expresses the deviations in measured yields between a $p+p$ collision and any other type of collision involving a nucleus, be it $p+A$ or $A+A$, with the nuclear modification factor
\begin{align}
R_{xA}(p_T)=\frac{1}{\langle N_{\rm coll}\rangle} \frac{dN_{xA}/dp_T}{dN_{pp}/dp_T},
\end{align}
where the prefactor including the number of binary collisions simply acts as a normalization. If the physics in a collision involving a nucleus were simply that of a large number individual $p+p$ collisions this ratio would stay at unity.

A natural distinction lies between effects which affect the initial-state and the final state. The former are related to differences in the parton composition of nuclei as compared to protons. The latter on the other hand arise from the fact that quarkonium states produced in $p+A$ and $A+A$ collisions may find themselves within a nuclear environment or surrounded by collision fragments, which destabilize the quark antiquark bound state.

\begin{figure}
\centering
\includegraphics[scale=0.25]{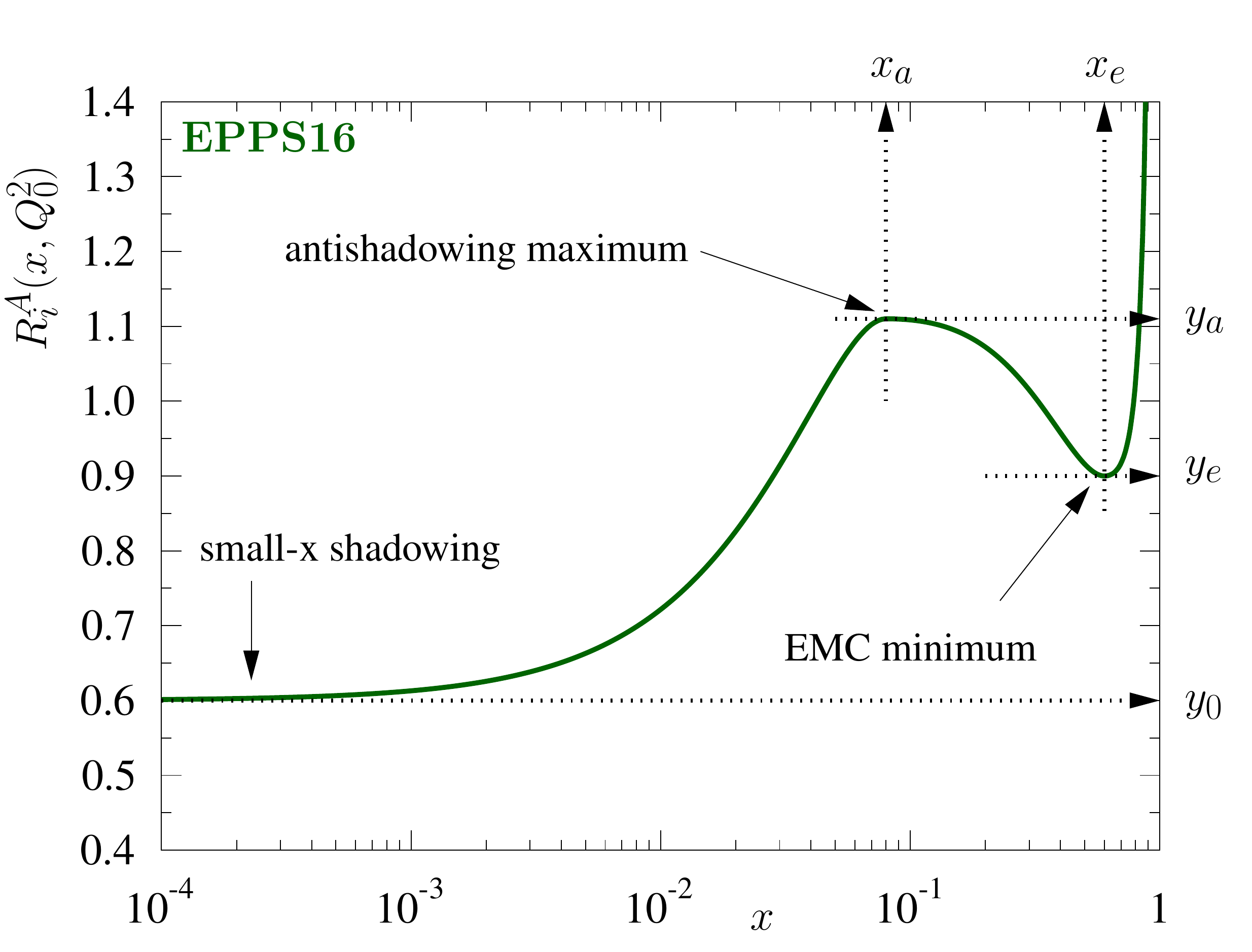}
\caption{The three generic features of nuclear parton distribution functions showcased with the example of EPPS16. The clear suppression below unity at small $x$ is followed by an intermediate anti-shadowing hump, before encountering the EMC minimum at $x\approx0.6$. Figure reproduced from Ref.~\cite{Eskola:2016oht}. }
\label{fig:nPDF}
\end{figure}

The study of parton distribution functions in nuclei is an active research area (for a recent brief overview see e.g. Ref.~\cite{Paakkinen:2018zbs}), which to date relies on an interplay between theory and experiment to extract from measured yields constraints on the momentum fraction carried by quarks and gluons in nucleons embedded in a nucleus. As shown in \cref{fig:nPDF} three generic features emerge: at small values of the parton momentum fraction $x<10^{-2}$ the ratio of nuclear to free parton distribution function $R^A_i=F^{\rm p\in A}_{i}(x,Q^2)/F^{\rm p}_{i}(x,Q^2)$ takes on values smaller than unity, a feature referred to as small-x {\it shadowing}. At intermediate $x\lesssim 0.1$ on the other hand the ratio shows a hump-like structure, which is simply called {\it anti-shadowing}, followed by a dip at $x\approx0.6$ which was first identified by the EMC collaboration and thus has been christened the {\it EMC effect}. There are many different nuclear PDFs on the market that differ in the theory input, as well as the fitting procedure to experimental data, among them EPS09, nCTEQ15 or EPPS16. 

We may now ask how the modification of the nuclear PDF's affects the production of the individual quarkonium states. I.e. we ask how the nuclear modification factor $R_{\rm pA}$ differs from unity. When evaluating $J/\Psi$ and $\Upsilon$ production in $p-Pb$ collisions at $\sqrt{s}=5$TeV the modified PDF's predict \cite{Albacete:2013ei} a {\it suppression} of around $10\%$ for $b\bar{b}$ and $20\%$ for $c\bar{c}$. It is here that we encounter for the first time the notion of {\it quarkonium suppression}. At this point this suppression is not related to the presence of a hot medium and simply reflects changes in the initial projectile. Obviously these numbers need to be kept in mind when investigating production patterns observed in nucleus-nucleus collisions, in order to correctly disentangle the possible effects of a hot QCD medium from those arising from cold-nuclear matter.

On the theory side, the color glass condensate approach \cite{Gelis:2010nm} is used to explore how the phenomenon of {\it parton saturation} affects the dynamics of partons within nuclei \cite{Fujii:2013yja,Fujii:2013gxa, Kang:2013hta,Ma:2015sia}. Already in an individual proton, the parton distribution function at large $Q^2$ but small x is expected not to grow indefinitely but to eventually saturate at the eponymous saturation scale $Q_s$, due to nonlinear effects of many self interacting gluons. As the saturation effect depends on the number $A$ of nucleons present, it is expected that its physics needs to be accounted for in detail if quarkonium production is to be understood in a heavy-ion collision. Eventually the goal for theory has to be to compute nuclear parton distribution functions from first principles, which at the moment however is still computationally unfeasible. 

There are many different phenomenological approaches proposed to describe the final state interactions. At low enough energies, where quarkonium formation may actually take place within the projectile nucleus volume, the bound state may interact with remnants of the nuclear environment leading to its destabilization. Such a scenario, which is considered for SPS beam energies but due to the Loretz contraction of the nuclei at LHC is unlikely to be of relevance there, is referred to as {\it nuclear absorption} or nuclear break-up. On the other hand for LHC energies the effects of coherent energy loss have been discussed originally in Ref.~\cite{Arleo:2010rb} which lead to a suppression of yields compatible with the data and which do not require any modification on the nuclear PDFs. It remains an interesting question of how to reconcile these final state results with the present understanding of nuclear PDFs.

Interestingly the above mentioned initial state effects, as well as coherent energy loss are expected to affect the ground state and final state in an equal fashion. It has however been observed that the excited states of bottomonium and charmonium in $p+A$ collisions suffer from a stronger suppression than the ground states. While at low energies this might be explained by nuclear absorption effects, at LHC energies the comover interaction model \cite{Ferreiro:2012rq,Ferreiro:2014bia} has been proposed and successfully describes this phenomenon. The idea is that the quarkonium state scatters with particles that possess a similar rapidity travelling in its vincinity. Their effects is modelled by a rate equation, which is similar to the treatment in terms of a transport model in \cite{Du:2018wsj} and the final-state interactions in recent computations based on the color-glass condensate and improved color evaporation model in Ref.~\cite{Ma:2017rsu}.  

It is important to note that while theory progress has been made in regards to understanding initial state effects in proton nucleus collisions the question of how cold nuclear matter affects quarkonium production in a nucleus nucleus collision remains a central open question. Vital insight will be gained on the experimental side with the arrival of electron-ion colliders, which will be able to shed light on the parton distribution functions within nuclei with unprecedented precision. On the theory side the question for saturation based computations is how elevate the computations currently carried out for a dilute - dense scenario into the dense - dense sector, relevant for heavy-ion collisions. At the same time it needs to be understood whether and if so how the factorization of subprocesses, a vital tool to make quarkonium production tractable in $p+p$ and $p+A$ collisions, survives in a nucleus nucleus collision.

Besides the modification of how partons in nuclei interact to form a $Q\bar{Q}$ pair, the evolution of such a colored state in the background of highly occupied soft gauge fields in the initial glasma phase of a heavy-ion collisions needs to be elucidated. Classical statistical simulations of the gauge field sector in conjunction with kinetic theory have proven insightful to understand the emergence of hydrodynamical behavior of the bulk matter (see e.g. Refs.~\cite{Berges:2013fga,Kurkela:2018wud,Kurkela:2018vqr}). Combining such a glasma inspired real-time framework with the effective field theory NRQCD is currently considered, to gain new insight how quarkonium formation in the early stages of a nucleus-nucleus collision proceeds.

The goal of this and the preceding subsection is to serve as a reminder that a comprehensive understanding of quarkonium in a heavy-ion collisions relies on an interplay of many different physical mechanisms, many of which are not directly related to quarkonium in a hot medium, but still offering ample opportunity for first principle theory contribution. At the same time some of the effects discussed above may directly affect e.g. the initial conditions from which a real-time simulation of heavy-quarkonium in a hot environment commences. Hence a lively exchange between the communities working on cold and hot nuclear matter effects will be essential to progress towards a genuine microscopic understanding of the physics involved in producing quarkonium states in a heavy-ion collision.

\subsection{Quarkonium production in $A+A$ collisions}
\label{sec:qqbarprodhic}

In this section we arrive at the ultimate challenge for current studies of quarkonium in extreme conditions, i.e. gaining a truly microscopic understanding of its production in relativistic heavy-collisions. As we have discussed in the previous sections, this task consists of many elements besides describing quarkonium dynamics in a hot medium. Both the real-time evolution of quarkonium states discussed in \cref{sec:qqbarrealtime} and their properties in equilibrium discussed in \cref{sec:qqbarequil} constitute central pieces of the puzzle without which no comprehensive picture of their production can arise. At the same time a more detailed knowledge about the partonic make-up of the incoming projectile nuclei and a dynamical picture of hadronization are required to fully account for the observed phenomenology.

\subsubsection{Charmonium in $A+A$ collisions}
\label{sec:ccbarprodhic}

Let us start with a discussion of charmonium. Its study in heavy-ion collisions presents an instructive tale about how only a comprehensive understanding of all stages of the collision can provide us with a full account of the physics of quarkonium production. The first steps towards elucidating charmonium in a heavy-ion collision were made in the context of quarkonium melting. I.e. the classic work of Matsui and Satz \cite{Matsui:1986dk} considered the stability properties of charmonium in a thermal medium, concluding in a first step that the presence of a deconfined QGP will prevent the formation of in-medium bound states. Their arguments at that time were based on a static picture of quarkonium melting, its modern dynamical viewpoint has been discussed in \cref{sec:QQbarmelting}. In general the higher the energy density of the medium, the more efficient the melting will be. At the same time the more weakly the charmonium state is bound in vacuum the more easily the medium will be able to dissolve it. We presented support for this concept of hierarchical ordering of in-medium effects from first principles computations in \cref{sec:QQbarequilprop}. Ref.~\cite{Matsui:1986dk} then goes on to argue in a second step that such a failure of binding will translate into a reduction of charmonium yields, i.e. into quarkonium suppression that follows a similar pattern as that of the in-medium melting. This senario is known as {\it sequential suppression} and has been laid out in detail in \cite{Karsch:2005nk}. From it follows that in central collisions and at small rapidities, i.e. where the density of medium scatterers is highest, the suppression should be strongest. 

The first conclusion on the hierarchical nature of quarkonium melting in its dynamical form remains valid also today, however it is now understood that suppression, especially at high beam energies does not follow automatically. The reason, as we understand it today, lies in both what happens before and what happens at the end of quark-gluon-plasma phase originally considered by Matsui and Satz. Indeed due to the cross section for charm anticharm pairs rising steeply with collision energy a hotter plasma in practice is accompanied also with a higher abundance in $c\bar{c}$ pairs. Primordial bound states formed from these $c\bar{c}$ pairs due to the destabilizing effects of the plasma can melt. At the same time either already during the QGP phase or at hadronization different $c\bar{c}$ pairs can recombine again into charmonium pairs, thus replenishing the yields. This effect had been discussed by Matsui early on (see e.g. Ref.~\cite{Matsui:1987im}) but at that time was discarded due to the small charm cross section accessible experimentally. Several years later with the advent of high energy machines, such as RHIC, on the horizon, regeneration was considered in more detail and it turns out to provide an important piece of the puzzle to understand charmonium production. Is is important to note that for collision systems where the effects of regeneration are less important, e.g. at moderate to large $p_T$, the $R_{AA}$ of different quarkonium species do follow a suppression pattern that is ordered with their in-medium binding energy. We will come back to this issue in the context of bottomonium later on.  

\begin{figure}
\centering
\includegraphics[scale=0.28]{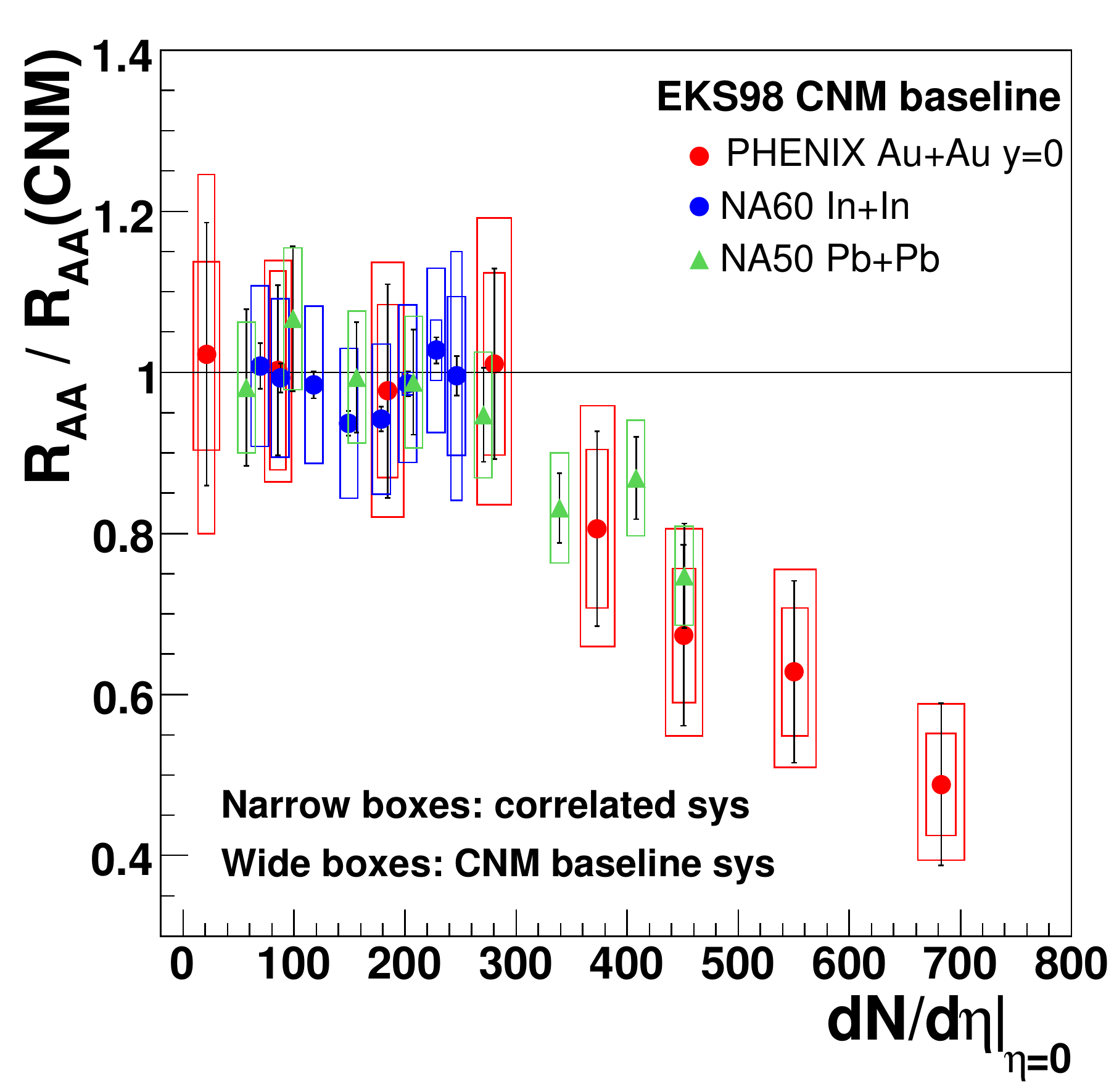}
\includegraphics[scale=0.28]{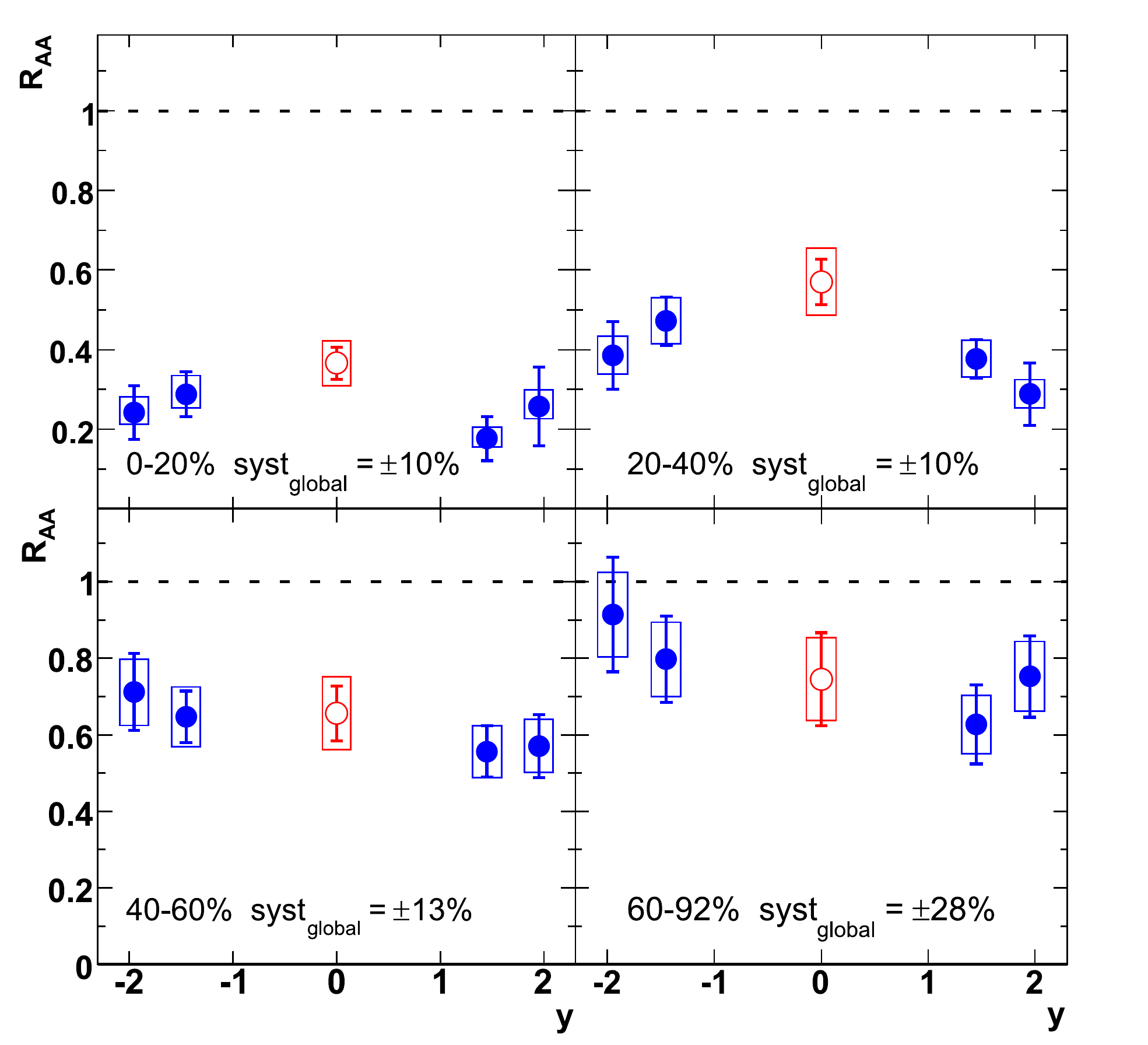}
\includegraphics[scale=0.28]{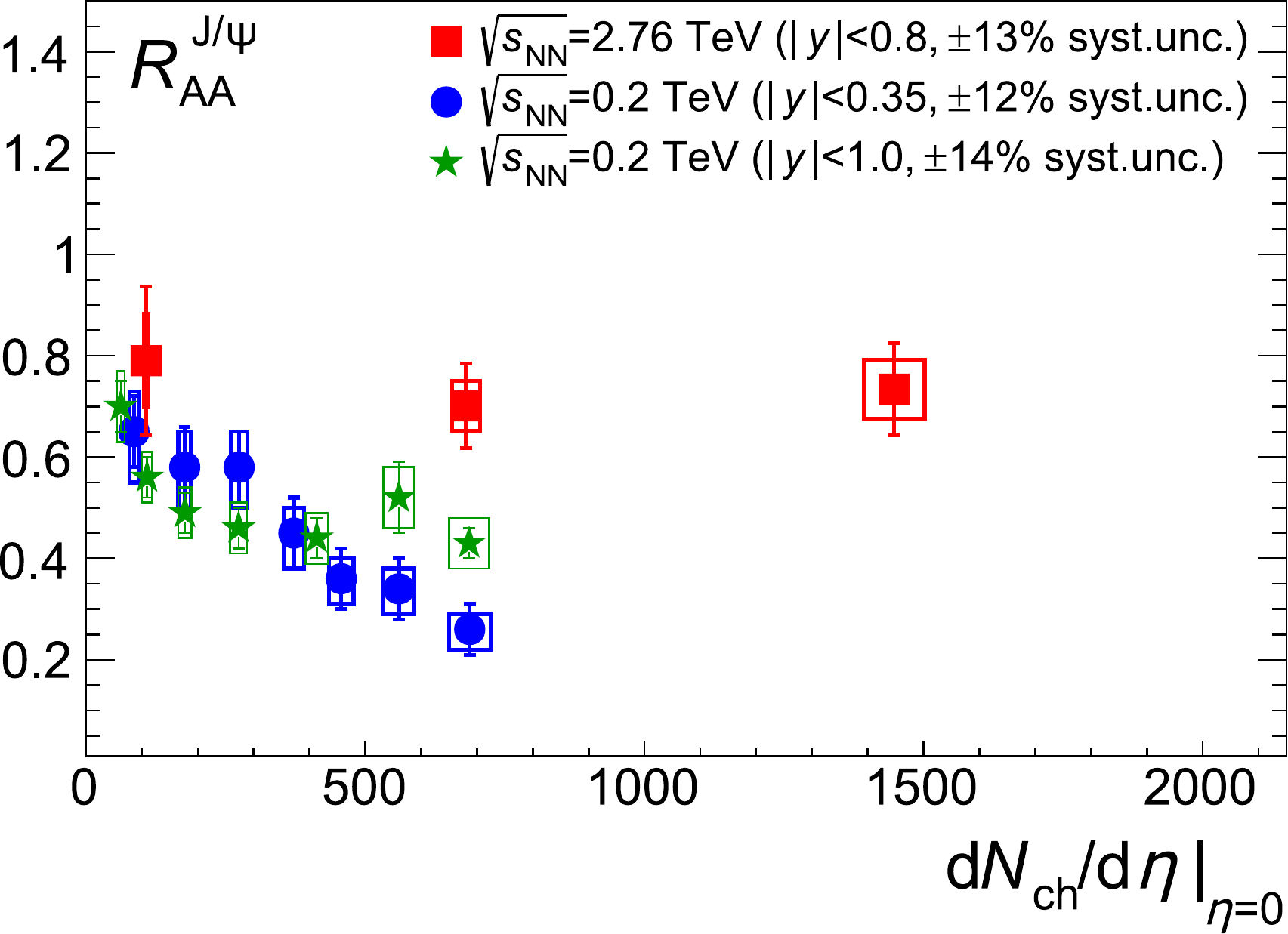}
\caption{(left) comparison of measurements of the nuclear modification factor for $J/\psi$ in terms of charged particle multiplicities from the NA50 and NA60 experiment at SPS, as well as PHENIX at RHIC. Figure reproduced from Ref.~\cite{Brambilla:2010cs}. (center) The measurements of the $J/\psi$ $R_{AA}$ in terms of rapidity measured by PHENIX at RHIC. Figure reproduced from Ref.~\cite{Adare:2006ns}. (right) Comparison of the $J/\psi$ $R_{AA}$ versus multiplicity as measrued from the STAR and PHENIX collaboration at RHIC and the ALICE collaboration at LHC. Figure adapted from Ref.~\cite{Andronic:2017pug}.}\label{fig:qqbarHICmeas1}
\end{figure} 

In \cref{fig:qqbarHICmeas1} we showcase a selection of experimental measurements of charmonium from heavy ion collision history, which have been instrumental in pushing forward our understanding of quarkonium production, as well as to spur advances in the development of phenomenological approaches. The quantity we show is the nuclear modification factor $R_{AA}$ for the $J/\psi$ particle. In the leftmost panel $R_{AA}$ from the NA60 and NA49 experiments at SPS and from the PHENIX experiment at RHIC are compared. Note that x-axis shows the charged particle multiplicity, which is used instead of the historic choice of number of participants to indicate the activity in the collision center ( allowing us to make sense of e.g. signals for collectivity even in $pp$ collisions at high multiplicity ). The values plotted are divided by the predictions for cold nuclear matter effects. While in $In+In$ collisions at SPS with $\sqrt{s_{NN}}=158$GeV only relatively small multiplicities are reached that do not show any suppression beyond cold nuclear matter effects, with the advent of $Pb+Pb$ at SPS a clear additional suppression has been registered. Extending the energy range to $\sqrt{s_{NN}}=200$GeV at RHIC with $Au+Au$ collisions even higher multiplicities are reached and the suppression trend continues. Historically the expectation was that the suppression at RHIC should be stronger than that at SPS, due to the higher beam energies and thus temperatures involved. The question is, can we find indications why the suppression might not increase further?

To this end let us take a look at the center panel of \cref{fig:qqbarHICmeas1}, where the rapidity dependence of $R_{AA}$ (without division by the CNM baseline) is plotted as measured by the PHENIX collaboration at RHIC. In non-central collisions (bottom row) we find a pattern consistent with both arguments from comovers and medium induced melting. I.e. at $y=0$ where the density of potential scatterers is highest the $R_{AA}$ is lowest. On the other hand in more central collisions (top row) the pattern inverts and $R_{AA}$ takes on the largest value at $y=0$. As we will discuss in more quantitative detail below, this effect is what one would expect when regeneration plays a role in the production dynamics. I.e. at $y=0$ the largest number of $c\bar{c}$ pairs is created and will thus lead to the largest recombination probability, in effect overcoming the effects from melting of primordial charmonium pairs. 

That regeneration plays an important role to describe the measured yields at increasing beam energies is further supported by the measurements of $R_{AA}$ by the ALICE collaboration at LHC, shown in the right most panel of \cref{fig:qqbarHICmeas1}. Compared along multiplicities, we see that at $\sqrt{s_{NN}}=2.76$TeV the values of $R_{AA}$ take on significantly larger values than those at measured by STAR and PHENIX at RHIC. A consideration based only on quarkonium melting would suggest that much less quarkonium states should be measured at LHC, while the abundance of $c\bar{c}$ pairs provides a natural explanation of why the values of $R_{AA}$ may even rise compared to RHIC. Indeed the trend continues if one considers the $R_{AA}$ at the highest LHC energies $\sqrt{s_{NN}}=5.02$TeV, where, as shown in Ref.~\cite{Adam:2016rdg}, it takes on slightly larger value than at $\sqrt{s_{NN}}=2.76$TeV.

\begin{figure}
\centering
\includegraphics[scale=0.25]{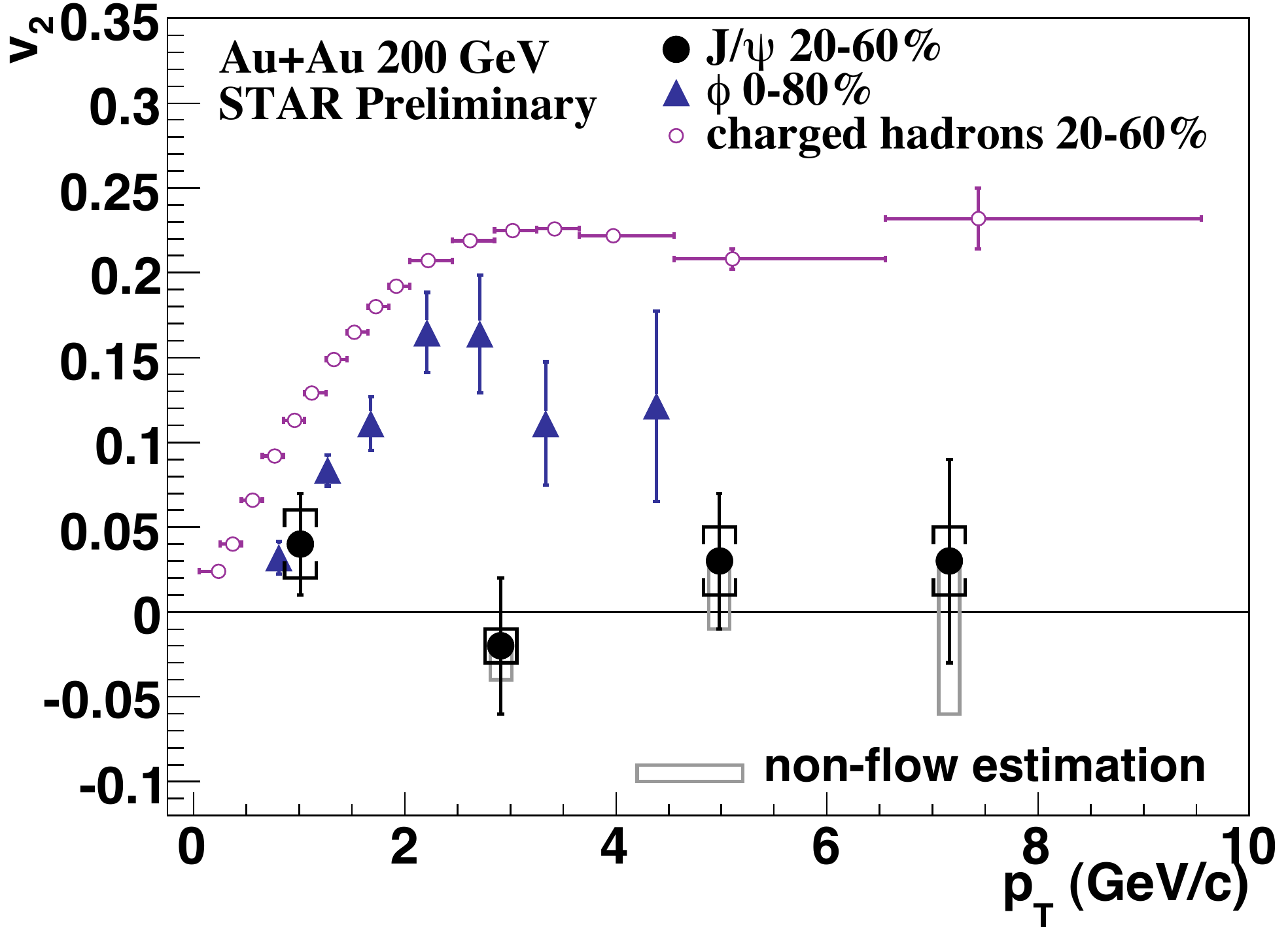}
\includegraphics[scale=0.6]{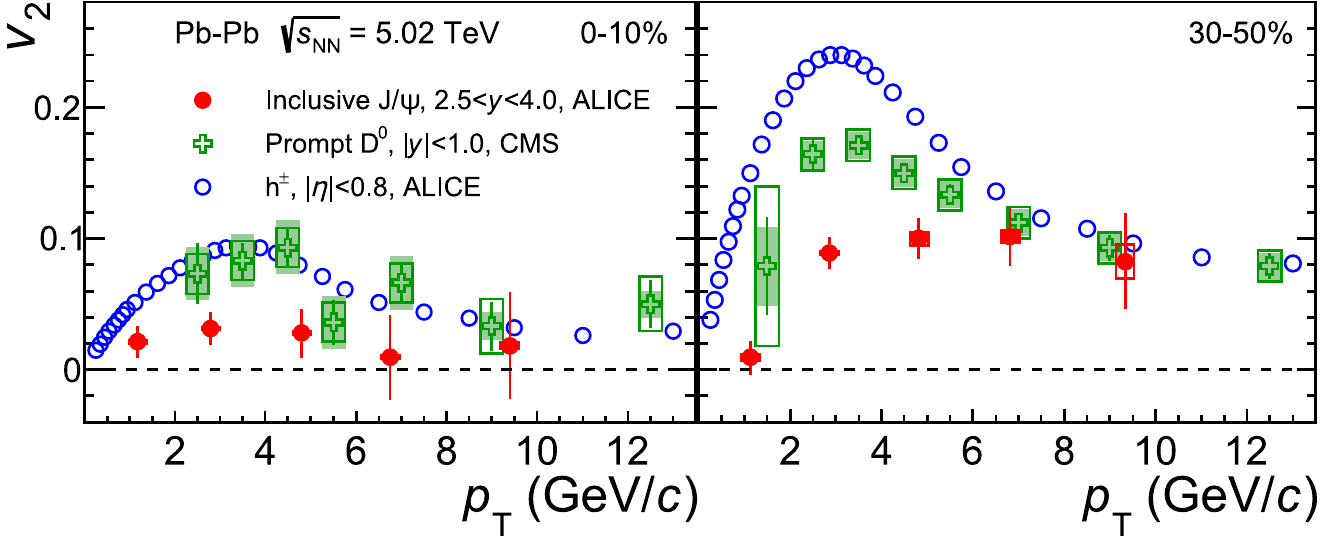}
\caption{Measurements of elliptic flow of $J/\psi$ at RHIC from Ref.~\cite{Powell:2011np} in the left panel and at LHC from Ref.~\cite{Acharya:2018pjd} on the right. While the STAR collaboration at $\sqrt{s_{NN}}=0.2$GeV did not observe a significant $v_2$ more recent measurements by the ALICE collaboration at $\sqrt{s_{NN}}=5.02$TeV show clear signs for a finite elliptic flow of around $50\%$ of that of D mesons.}\label{fig:qqbarHICmeas2}
\end{figure}

As a lot of intuition on quarkonium production originally arose from considering idealized settings in kinetic thermal equilibrium. The question of how efficiently the charm quarks exchange energy and momentum with their surrounding is e.g. of central interest. One possibility to infer this is to look for signs of the produced charmonium to participate in the collective motion of the bulk matter, expressed e.g. in a finite value of {\it elliptic flow} $v_2$ for $J/\psi$. Finite values of $v_2$ would indicate that the charm quarks are in at least partial kinetic equilibrium with their surroundings, which in turn entails a loss of memory of the initial conditions their evolution. In \cref{fig:qqbarHICmeas2} we show measurements of said $v_2$ by the STAR collaboration at RHIC $\sqrt{s_{NN}}=0.2$GeV from Ref.~\cite{Powell:2011np} on the left and from the ALICE collaboration at LHC $\sqrt{s_{NN}}=5.02$TeV from Ref.~\cite{Acharya:2018pjd} on the right. While at the lower beam energies no significant deviation of $v_2$ from zero was observed by START, the situation significantly changes at LHC, where an elliptic flow of around $50\%$ of that of D-mesons has been reported by the ALICE collaboration. In turn we expect that arguments based on thermal equilibrium are a reasonable starting point to develop an understanding of charmonium at LHC.

After this very rough survey of charmonium measurements and their qualitative discussion let us turn to more quantitative means of describing charmonium production. While the ultimate goal for theory is to provide a microscopic description from the initial $c\bar{c}$ pair production over their real-time evolution to dynamical hadronization, such an ambitious framework has yet to be developed. In the meantime it is very instructive to consider what already has been learned about charmonium from describing its production using phenomenological models.

As starting point we can return to the question of why we distinguish heavy quarkonium production from the physics of other lighter flavors. The reason lies in the separation of scales present. In the collider experiments conducted so far, starting at SPS, followed by RHIC and currently at LHC, the temperatures reached in the collision center, as deduced by modeling of the hydrodynamic expansion of the bulk medium, are at maximum $T_{\rm max}\approx 600$MeV and thus much smaller than the charm quark mass. In turn it is consensus that heavy quark pairs are created only in the early stages of the collision and their number remains approximately constant over the short time scales in which a hot medium may form in the collision center. I.e. in a contemporary heavy-ion collision heavy quarks are not chemically equilibrated and their production yields must be understood with this fact in mind.

Experimental data backs up this reasoning when scrutinized through the lens of hadron resonance gas based models, most prominently the {\it statistical hadronization model} first introduced in Refs.~\cite{BraunMunzinger:2000px,BraunMunzinger:2000ep} and reviewed in Ref.~\cite{BraunMunzinger:2003zd}. Let us take a closer look what this model can tell us about equilibration of light and heavy degrees of freedom in heavy-ion collisions (see e.g. Refs.~\cite{Andronic:2005yp,Andronic:2017pug}). The model is based on the grand canonical partition function expressed in terms of an ideal, i.e. non-interacting, gas of hadronic degrees of freedom. Remember that for an ideal quantum gas in the occupation number representation the partition function factorizes
\begin{align}
Z[T,V,\mu]&=\sum_{\{n_{\bf p}\}} {\rm exp}\Big[-\beta \sum_{\bf p} n_{\rm p} \big(\epsilon({\bf p}) - \sum_j \mu_j Q_j\big)\Big]=\sum_{n_{{\bf p}_1}}  \sum_{n_{{\bf p}_2}}\ldots \prod_{i}e^{-\beta n_{{\bf p}_i} (\epsilon({{\bf p}_i}) -\sum_i\mu_iQ_i)}\\
&=\sum_{n_{{\bf p}_1}}  e^{-\beta n_{{\bf p}_1} (\epsilon({{\bf p}_1}) -\sum_i\mu_iQ_i)}  \sum_{n_{{\bf p}_2}} e^{-\beta n_{{\bf p}_2} (\epsilon({{\bf p}_2}) -\sum_i\mu_iQ_i)} \ldots=\prod_{\bf p} \sum_{n_{{\bf p}}}  e^{-\beta n_{{\bf p}} (\epsilon({{\bf p}}) -\sum_i\mu_iQ_i)}\\
&=\prod_{\bf p}\Big( 1 \mp e^{-\beta(\epsilon({\bf p})-\mu_iQ_i)}\Big)^{\mp 1},
\end{align}
with the upper sign for bosons and the lower ones for fermions. Here $\epsilon({\bf p})$ denotes the energy of the eigenmodes of the Hamiltonian $H$ with momentum $\bf p$ and $n_{\bf p}$ refers to the corresponding occupation number of that mode. The chemical potentials $\mu_i$ refer to possible conserved charges $q_i$ that the modes carry. In our case what we call modes of the Hamiltonian will be the different hadrons and their excitations that are present in the low temperature phase of QCD with $\epsilon({\bf p})=\sqrt{{\bf p}^2+m_i^2}$. Here $m_i$ is taken to be the vacuum mass of these states, as temperature effects are implemented via the trace over the thermal density matrix. The conserved charges to which a chemical potential is assigned are strangeness $(S,\mu_s)$, baryon number $(B,\mu_B)$ and the third component of the isospin charge $(I_3,\mu_{I_3})$. In a heavy ion collision with projectiles consisting of protons and neutrons, the overall system has to follow the constraints
\begin{align}
V\sum_i n_iB_i=N_B,\quad V\sum_in_iI_{3i}=I_3^{\rm tot}, \quad V\sum_in_iS_i=0.
\end{align}

The power of hadron resonance gas models lies in the fact that all the non-perturbative information on the strong interactions is included in the values of the masses of the individual contributing hadrons, which one usually takes from measurements cataloged by the PDG, amended by lattice QCD spectroscopy results. Below the crossover transition temperature it has been shown that the equation of state, i.e. the dependence of pressure on temperature arising from the hadron resonance gas model is in good agreement with first principles computations from lattice QCD, lending support to its construction (see e.g. Ref.~\cite{Bazavov:2014pvz}).

The logarithm of the full partition function becomes a sum over all modes ${\rm log}[Z]=\sum_i {\rm log}[Z_i]$, each individual one, after accounting for spin degeneracies with a factor $g_i$, contributes with 
\begin{align}
{\rm log}\big[ Z_i\big]=\frac{V g_i}{2\pi^2}\int_0^\infty dp (\pm p^2){\rm log}\Big[1\pm \lambda_i {\rm exp}\big[-\beta\epsilon_i\big]\Big], \quad \lambda_i={\rm exp}\big[ \frac{B_i\mu_B+S_i\mu_S+q_i \mu_q}{T}\big],\label{eq:hadcontrib}
\end{align}
where the $\lambda_i$ takes on the role of a fugacity. This expression can be further rewritten by expanding the logarithm via its Taylor series and carrying out the momentum integral
\begin{align}
{\rm log}\big[ Z_i\big]= \frac{VTg_i}{2\pi^2} \sum_{k=1}^\infty \frac{(\pm 1)^{k+1}}{k^2} \lambda_i^k m_i^2K_2\big(\frac{k m_i}{T}\big), \quad n_i(T,V{\bf \mu})=\langle N \rangle/V=\frac{Tg_i}{2\pi^2}\sum_{k=1}^\infty \frac{(\pm 1)^{k+1}}{k} \lambda_i^k m_i^2K_2\big(\frac{k m_i}{T}\big).
\end{align}
The function $K_2$ refers to the modified Bessel function. 

To arrive at the particle number densities at constant temperature and volume one exploits the equivalent definition of $\langle N\rangle$ from a derivative with respect to chemical potential $\langle N \rangle = \frac{1}{\beta}\frac{\partial}{\partial \mu} {\rm log}\big[ Z_i\big]$ or  fugacity $\langle N \rangle = \lambda \frac{\partial}{\partial \lambda} {\rm log}\big[ Z_i\big]$. Taking into account the physical constraints, the model thus predicts the particle number densities given three adjustable parameters, the temperature of the system, its volume and Baryo-chemical potential.

This basic formulation of the model can be amended to use another piece of non-perturbative information, which is the decay width of resonance hadronic states, that also contribute to the partition function. Their physics may be incorporated by translating the spectral width into a convolution over different masses in \cref{eq:hadcontrib} 
\begin{align}
n^R_i=\frac{g_i}{2\pi^2}\frac{1}{{\cal N}_{BW}} \int_{M_R}^\infty dm \int_0^\infty dp \frac{\Gamma_i^2}{(m-m_i)^2+\Gamma_i^2/4} \frac{p^2}{{\rm exp}\big[-\beta(\sqrt{{\bf p}^2+m^2}-\mu_iQ_i)\big]\pm 1},
\end{align}
where ${\cal N}_{BW}$ normalizes the integral over the Breit Wigner to unity. The total number of particles measured in experiment furthermore needs to take into account decays of resonances so that 
\begin{align}
\langle N_i^{\rm tot} \rangle[T,V,\mu] = \langle N^{\rm therm}_i \rangle[T,V,\mu] + \sum_j \, \Gamma_{j\to i} \langle N^{R,\rm therm}_j \rangle[T,V,\mu].
\end{align}
Note that once the Baryo-chemical potential becomes large with respect to the temperature, some form of repulsive interactions are often included via an excluded volume prescription.

Apriori it is not clear whether such a model will be able to describe the yields of hadrons composed of the light quarks $u,d$ and $s$. It however has been shown in detailed comparisons with measurements ranging from AGS energies around $\sqrt{s_{NN}}\approx 5$MeV to current LHC energies $\sqrt{s_{NN}}=2.76$TeV that in central heavy-ion collisions a remarkable agreement with a wide variety of particle and antiparticle yields can be achieved, ranging from pions up to Helium and anti-Helium (for a more detailed discussion in particular also of the aspect of strangeness production see Refs.~\cite{Andronic:2005yp,Andronic:2017pug}). At lower beam energies some particle species seem to deviate from the model predictions among them the proton and some kaons. At LHC energies only the protons show slightly less yields than predicted. Interestingly the parameter values with which the yields are reproduced are found to start at values around $T\approx 130$MeV and $\mu_B\approx 540$MeV at AGS energies and quickly saturate from $\sqrt{s_{NN}}\approx 15$MeV onward at  $T^{\rm max}\approx 158$MeV. The Baryo-chemical potential parameter at RHIC energies $\sqrt{s_{NN}}\approx 200$MeV has reduced significantly to $\mu_B\approx 30$MeV but becomes indistinguishable from zero only at LHC energies $\sqrt{s_{NN}}\approx 2.76$TeV. 

Irrespective of the history of the matter in the collision center, at some point it must be possible to describe the end products of the collision in terms of a collection of hadrons, as these are the final measured degrees of freedom. The abundances of these hadrons are fixed at the so called chemical freezeout (see e.g. \cite{Teaney:2002aj} and references therein) after which they may still change their momentum and energy via number preserving reactions such as e.g. $\pi\pi\to\rho\to\pi\pi$. In comparison, number changing processes in a hadron gas are expected to occur on much longer timescales (up to $100$s of fm). Therefore the results by the statistical model are interesting as they tell us that the hadrons at freezeout already exhibit the abundances expected from a chemically equilibrated ensemble, even though the time passed since the collision would not allow for such equilibration to take place in a simple hadron gas. Therefore it has been suggested that the mechanism of hadron production itself can generate such abundances. The fact that hadron yields in $p\bar p$ and $e^+ e^-$ collisions can also be captured by essentially the same statistical model supports such a statistical picture of hadron formation at least for low energy heavy-ion collisions. (Recently ideas have been put forward that relate the thermal distribution of produced particles in elementary collisions to the entanglement between observable and non-observable patches of the light cone in Ref.~\cite{Berges:2017hne}.)

The freezeout temperature as determined from the model fits rises with the beam energy until it saturates around a temperature that is suggestively close to the crossover transition temperature computed from first principles lattice QCD. Hence also from the point of view of the statistical model in current heavy-ion collisions at RHIC and LHC it is very likely that the matter in the collision center has actually persisted in a state of even higher temperatures shortly after the collision. The presence of deconfined quarks and gluons at that point would allow the bulk to more efficiently (locally) chemically and kinetically thermalize, so that at hadronization the grand canonical abundances are reached. 

What about heavy quarks? If we simply extend the hadron resonance gas by including charmed hadrons, i.e. the D mesons, charmed baryons and charmonium states, would we also be able to reproduce the experimental abundances? The answer regarding the absolute yields is a clear no, as first discussed in Ref.~\cite{BraunMunzinger:2000px}. However inspection of the ratio of excited state charmonium to ground state charmonium revealed that for central collisions its value became close to the prediction of the statistical model. I.e. while the overall number of $c\bar c$ pairs produced in the earliest stages of the collision does not correspond to the value expected in chemical equilibrium at the freezeout temperature it appears that relative abundances produced at freezeout are again in agreement with chemical equilibration. How can this be accommodated in the language of the hadron gas? What is needed is to enhance the yields of charmed hadrons relative to their chemical equilibrium levels by introducing an additional charm conservation constraint and the enhancement factor $g_c$
\begin{align}
N_{c\bar{c}}^{\rm direct} &= \frac{1}{2}g_c V \Big[ \sum_i \big( n_{D_i}^{\rm therm} + n_{\Lambda_i}^{\rm therm} \big)\Big]\frac{ I_1 \Big[ g_cV \sum_i \big( n_{D_i}^{\rm therm} + n_{\Lambda_i}^{\rm therm} \big)\Big] } { I_0 \Big[ g_cV \sum_i \big( n_{D_i}^{\rm therm} + n_{\Lambda_i}^{\rm therm} \big)\Big]} + g_c^2 V \sum_i n_{\psi_i}^{\rm therm},\\
  & \overset{  g_cV \sum_i \big( n_{D_i}^{\rm therm} + n_{\Lambda_i}^{\rm therm} \big) \gg 1}{=}  \frac{1}{2}g_c V \Big[ \sum_i \big( n_{D_i}^{\rm therm} + n_{\Lambda_i}^{\rm therm} \big)\Big] + g_c^2 V \sum_i n_{\psi_i}^{\rm therm} .
\end{align}
Here the only relevant terms are those for open heavy flavor mesons, singly charmed Baryons and the charm anticharm mesons. For a small number of singly charmed hadrons the exact conservation of the charm quantum number needs to be correctly treated in the canonical formalism, which leads to additional correction factor involving the Bessel functions $I_1$ and $I_2$, the ratio of which  reduces to unity for large overall number of produced $c\bar{c}$. The total number of charm anticharm pairs for a particular collision centrality is computed by using the total charm cross section taken from $pp$ collisions and scaled with the nuclear overlap factor $T_{AA}$, i.e. $ N_{c\bar{c}}^{\rm direct} = \sigma_{c\bar{c}}^{pp} T_{AA}$. The charm cross section grows rapidly with $\sqrt{s_{NN}}$ so that as shown in Ref.~\cite{Andronic:2006ky} at SPS $g_c\lesssim 5$ while at RHIC it is already  $g_c\approx 10$ and for LHC it takes on even larger values $17 < g_c <30$. Note that $g_c$ in this model is a function of  rapidity and the number of participants are closely related to the total charm cross section, which due to the challenging nature of the underlying measurements unfortunately still to this day carries relatively large uncertainties.

\begin{figure}
\centering
\includegraphics[scale=0.28]{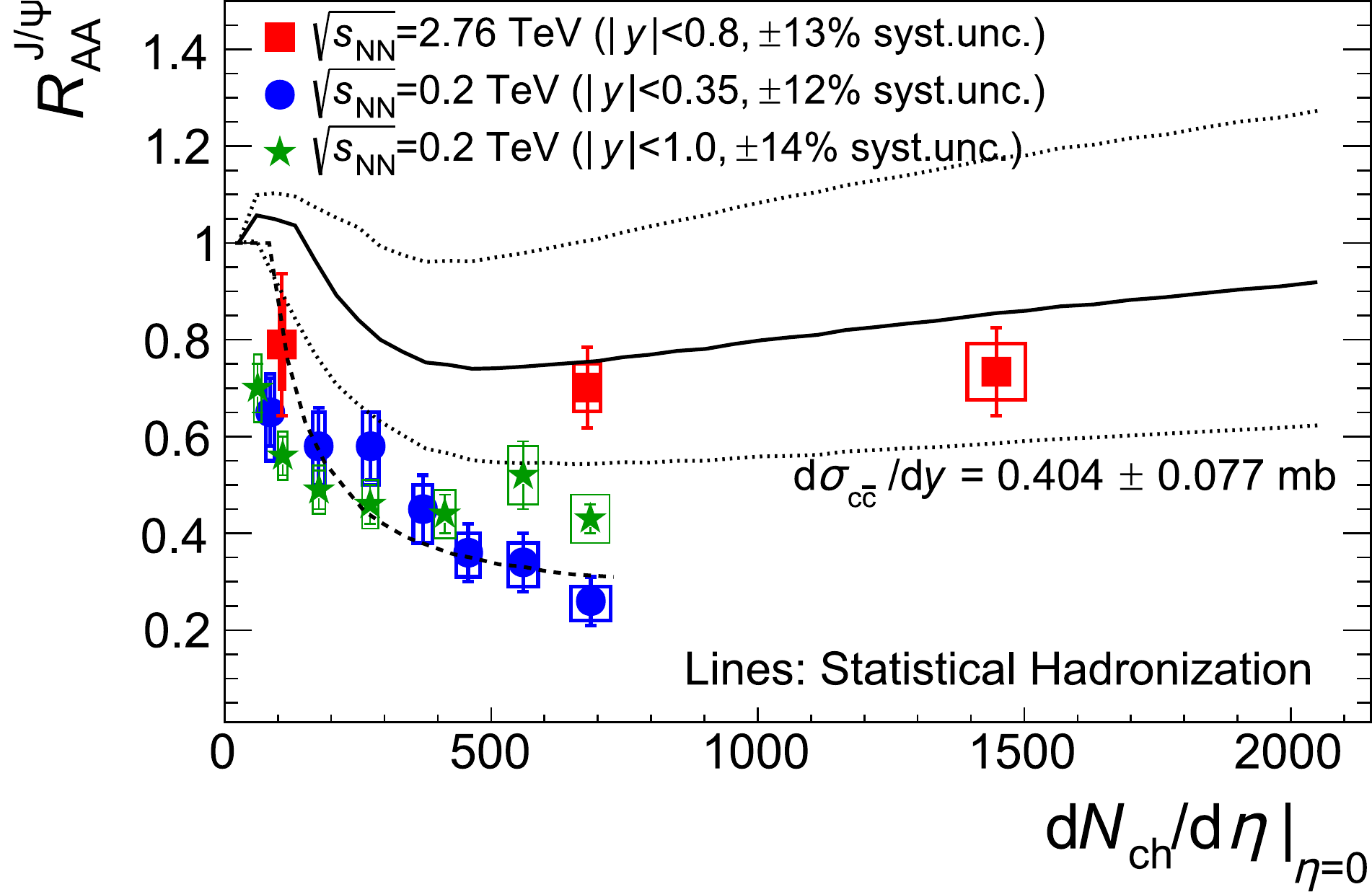}
\includegraphics[scale=0.25]{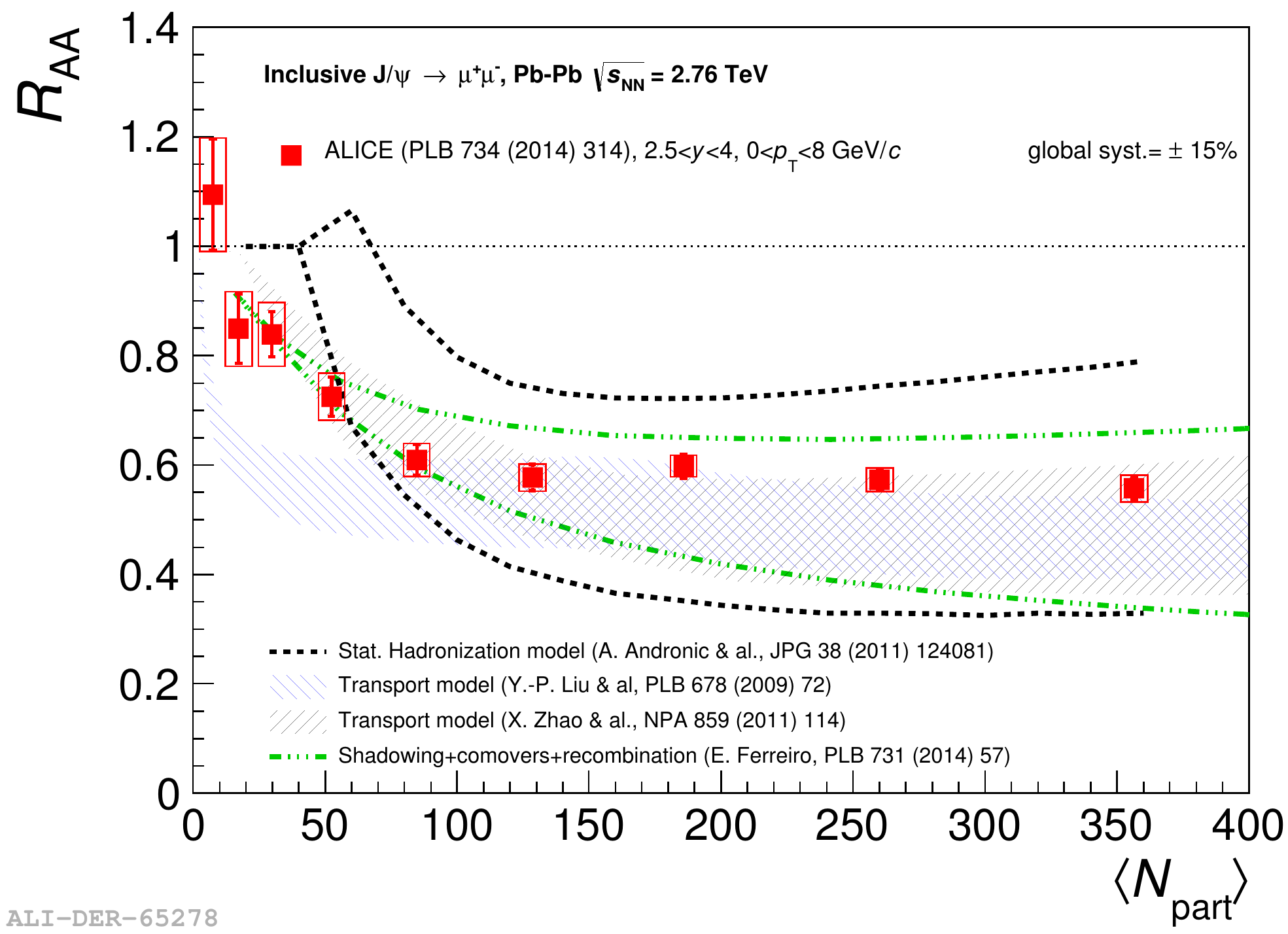}
\includegraphics[scale=0.25]{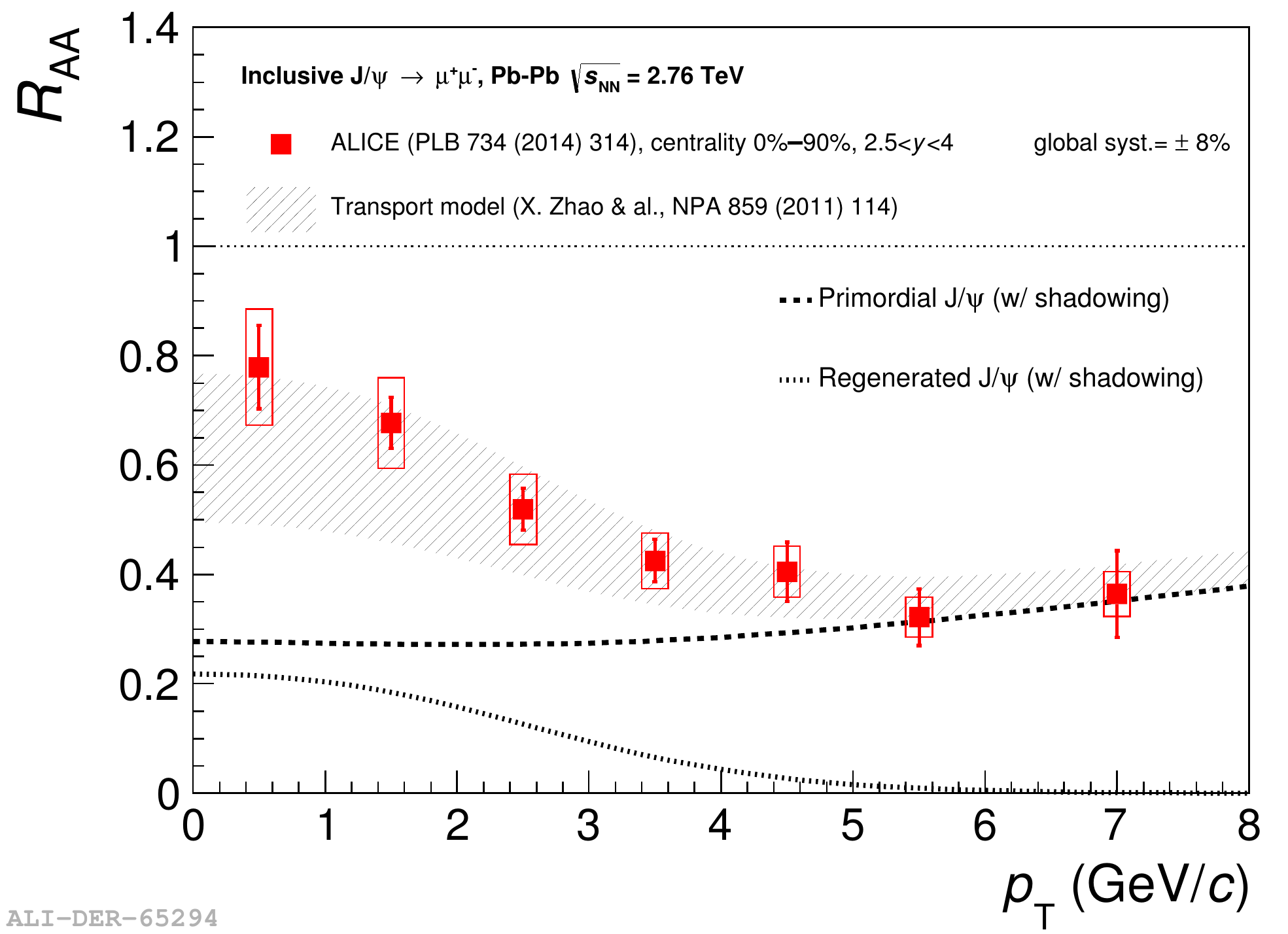}
\caption{A selection of comparisons of measurements of the $J/\psi$ nuclear modification factor and phenomenological models. (left) Measurements from RHIC and LHC (colored points) together with the post- ($\sqrt{s_{NN}}=0.2$TeV) and predictions ($\sqrt{s_{NN}}=2.76$TeV) from the statistical model of hadronization. Figure reproduced from Ref.~\cite{Andronic:2017pug} (center) Example of various models with different underlying physics mechanism being able to reproduce the $J/\psi$ $R_{AA}$ at LHC. Figure reproduced from Ref.~\cite{Andronic:2015wma}. (right) Decomposition of the transport model computation from Ref.~\cite{Zhao:2011cv} into its regeneration and primordial survival component. Figure reproduced from Ref.~\cite{Andronic:2015wma}.}\label{fig:qqbarHICmeas3}
\end{figure}

Combining the fact that heavy quark pair production takes place in the early stages of a collision with the thermal hadron resonance gas model at freezeout enables an efficient description of the measured charmonium data. The interplay of the production yields of singly charmed mesons and baryons together with the overall number of $c\bar{c}$ pairs set by the charm cross section allows to reproduce the decrease of $R_{AA}$ of $J/\psi$ with increasing number of activity in the collision center at RHIC. At the same time it provides a mechanism to explain why the values of $R_{AA}$ at LHC take on significantly larger values, as shown in the left panel of \cref{fig:qqbarHICmeas3}. The rapidity dependence of the total charm cross section, being largest at mid rapidity, furthermore provides an explanation to the measurements in the middle panel of \cref{fig:qqbarHICmeas1} showing that the suppression at RHIC is weakest for small values of $y$. 

The success of the statistical model combined with the fact that charmonium at LHC shows clear signs of partial kinetic equilibration may be taken as indication that the idealization of full kinetic equilibrium employed in first principles lattice QCD based computations is indeed a good starting point. One possible route then to connect QCD results to the measured yields was explored in Refs.~\cite{Burnier:2015tda,Lafferty:2019jpr} where the $\psi^\prime$ to $J/\psi$ ratio has been estimated using thermal spectral functions, computed using a lattice QCD vetted potential. The reason to focus on the ratio of excited to ground state is that it is independent of the enhancement factor $g_c$ related to the actual number of $c\bar{c}$ pairs produced. Remember that the area under the peak structures in the in-medium spectral function encodes the dilepton emission rate from that quarkonium state in a fully equilibrated setting. This however is not what is measured in experiment, which is instead the decay of vacuum quarkonium states long after the QGP has ceased to exist. Thus the question is how to translate the spectral features around the crossover transition into abundances of vacuum states produced at hadronization. Ref.~\cite{Burnier:2015tda} proposed to use an instantaneous freezeout scenario, where the number of produced vacuum particles are estimated from the spectral functions in units of dilepton emission. I.e. one takes the weighted area under in-medium peak for e.g. $J/\psi$ and divides it by the area of the peak in the vacuum spectral function, which is directly related to the radial wavefunction of the $T=0$ $J/\psi$ at the origin. This value is then used as estimate for the number of vacuum states produced. Carrying out the same procedure for $\psi^\prime$ and dividing the two results has then been taken as estimate for the $\psi^\prime$ to $J/\psi$ ratio. It is found that the value obtained in this fashion lies quite close to the statistical model and using a conservative error estimate is currently compatible within the combined theory uncertainties. Such an agreement bodes well for future studies trying to explore quarkonium production from genuine first principle thermal QCD computations.

The ability of the statistical model to reproduce the production yields of light hadrons and charmonium is intriguing and has been instrumental in revealing the role played by regeneration in the production of charmonium at RHIC and LHC. At the same time, being formulated in terms of a hadron gas, this model can only make statements about the physics at and after freezeout. Therefore the next step towards a fully dynamical understanding of charmonium production is to consider phenomenological models that allow us to describe the physics within the QGP realm and to see how they compare to the statistical model. One question in particular which is of current interest is to what degree charmonium states produced early on in a heavy-ion collision, so called primordial charmonium, survive until the end of the QGP phase or whether they are efficiently melted on the way.

Two groups, one based at Texas A\& M and the other at Tsinghua university have developed transport models based on the Boltzmann equation, respectively on a corresponding averaged rate equation. Both approaches have in common that they implement the possibility for dissociation of primordial charmonium particles, as well as the dynamical recombination of such states throughout the QGP evolution. Feed-down from excites state is included at the end of the medium evolution. Non-perturbative information on quarkonium dissociation is implemented in the former approach by computing their stability properties via the T-matrix approach or via melting temperatures in the latter. As we discussed in \cref{sec:NonEFTPot}, the definition of the potential used in a Bethe-Salpeter based approach and its relation to the EFT based potential remain an active area of research and constitute one significant source of uncertatinty. On the other hand we have seen that the definition of a melting temperature in the presence of a thermal width is not uniquely defined and especially difficult when using direct lattice QCD results, which in turn contributes to the overall systmatic uncertatinty. The recombination probability in these models is constructed using arguments of detailed balance, which are expected to work well close to equilibrium but may require corrections at early times far from equilibrium. More details on the construction of loss and gain terms can be found in Ref.~\cite{Liu:2009nb} and Ref.~\cite{Zhao:2011cv}.
The medium evolution is treated somewhat differently among the two models. In the former the concept of entropy conservation in conjunction with the measured particle multiplicities is used to model a temperature profile within an isotropically expanding fireball.  The latter model on the other hand computes a temperature profile directly based on $2+1$ dimensional Bjorken expansion in the QGP phase. In each case a first order transition like regime is used to connect the QGP with a hadron resonance gas phase at lower temperatures.

Both models are able to describe the $R_{AA}$ in terms of centrality and transverse momentum at RHIC and LHC in an equally good fashion as shown for the example of its centrality dependence in the center panel of \cref{fig:qqbarHICmeas3}, see also Ref.~\cite{Andronic:2015wma} (in non-central collisions the model of Ref.~\cite{Liu:2009nb} seems to somewhat underestimate the actual $R_{AA}$). For the $p_t$ dependence we select here for better readability in the right panel of \cref{fig:qqbarHICmeas3} a single results from Ref.~\cite{Zhao:2011cv} compared to the measurements by the ALICE collaboration at $\sqrt{s_{AA}}=2.76$TeV. As indicated by the two different black lines referring to the primordial and the regeneration component of the total $R_{AA}$ there are two different regimes present, which are smoothly connected. At large $p_t$ primordial charmonium appears to contribute a majority of the yield, while at small $p_t$ a significant fraction of produced $J/\psi$ arises from recombination effects. Similar behavior for the $R_{AA}$ dependence on centrality is observed: in non-central collisions primordial $J/\psi$ dominates but for central collisions regeneration plays an almost equally important role. The effects of regeneration at LHC are pronounced but also at RHIC the transport model computations indicate that dissociation of primordial charmonium alone cannot account for the observed patterns in $R_{AA}$.

From considerations of the nuclear modification factor of the charmonium ground state $J/\psi$ we have so far learned that dissociation and regeneration are leading to an intricate pattern of charmonium suppression requiring insight into all stages of the collision. While in the transport models a partial equilibration of the charmonium states occurs, equilibration is assumed to be complete in the statistical model. The question thus remains: is there a way how to distinguish between these two scenarios even though both reproduce the $R_{AA}$ well. This question becomes even more pressing when realizing that also models with a very different physics content are able to reproduce the $R_{AA}$ at LHC, as seen in the green dashed lines in the center panel of \cref{fig:qqbarHICmeas3} which corresponds to a model based solely on shadowing, comovers and the possibility for $c\bar c$ pairs to recombine \cite{Ferreiro:2012rq}. 

\begin{figure}
	\centering
		\includegraphics[scale=0.3]{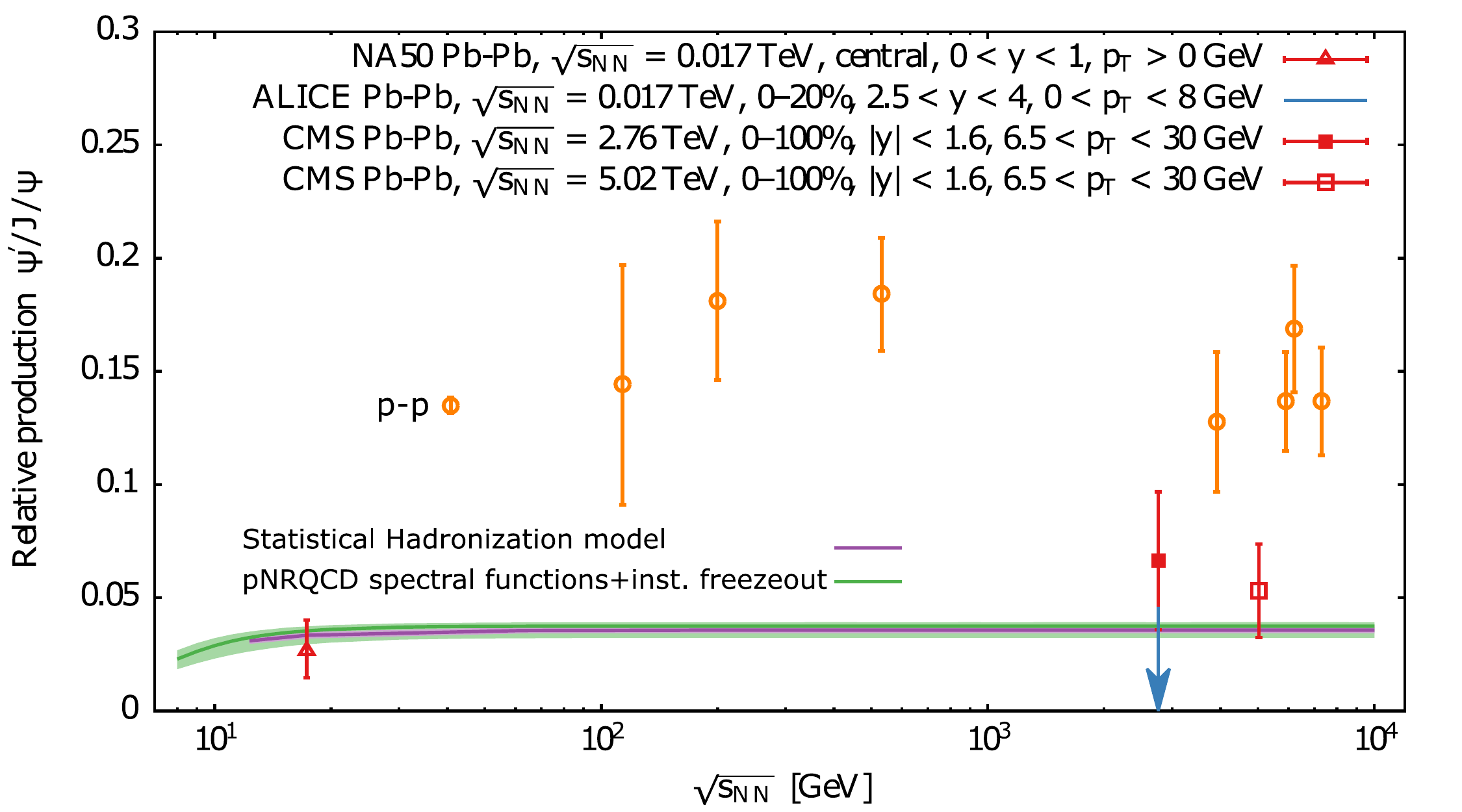}
		\caption{Measurements of the \(\psi^\prime\) to \(J/\psi\) ratio from the NA50 \cite{Alessandro2007}, ALICE \cite{Adam:2015isa} and CMS \cite{Khachatryan:2014bva,Sirunyan:2016znt} collaborations in $Pb+Pb$ collisions. The $pp$ baseline is given as orange datapoints \cite{Andronic:2017pug,DREES2017417}. The green solid line denotes the prediction of the statistical model of hadronization \cite{Andronic:2017pug}, while the purple line corresponds to a computation based on pNRQCD spectral functions combined with an instantaneous freezeout scenario from \cite{Lafferty:2019jpr}.} \label{fig:PsiToJPsiRatio}
\end{figure}

Two paths forward are possible. On the one hand one can look for other experimental observables, which are more discriminatory among the different models. One such quantity of current interest is the ratio of $\psi^\prime$ to $J/\psi$. It is a particularly challenging observable due to the small signal to noise ratio in the measurements of the excited state yields. At the LHC there exist currently measurements by the CMS collaboration (see Refs.~\cite{Sirunyan:2016znt,Khachatryan:2014bva}), however published only as double ratios with respect to $pp$ collisions and upper limits by the ALICE collaboration \cite{Adam:2015isa}. These values together with the $pp$ baseline and earlier measurements at lower beam energies by the NA50 experiment are listed in \cref{fig:PsiToJPsiRatio} together with the predictions from the statistical hadronization model from Ref.~\cite{Andronic:2017pug} as well as the computations of Ref.~\cite{Lafferty:2019jpr} modeled based on pNRQCD spectral functions combined with an instantaneous freezeout scenario.

On the other hand one needs to develop further first principles based real-time descriptions of heavy quarkonium in order to reduce the need for model assumptions. As we discussed in \cref{sec:REalTimeOQS} the open quantum systems approach appears to be a viable candidate  allowing one e.g. to derive evolution equations via pNRQCD from underlying QCD. In the weak-coupling scenario even the Boltzmann equation that forms the basis for the transport model computations has been derived in that fashion. In general in such a setup the dynamical evolution of the quarkonium states is governed by low energy matching coefficients, be it the complex heavy quark potential or other non-perturbative transport coefficients. However all the approaches developed so far make extensive use of a separation of scales between the heavy quark rest mass and other relevant scales, which means that their application to charmonium in current heavy-ion collisions does not rest on as solid foundations as one would like. E.g. the complex in-medium potential currently available corresponds only to the lowest order contribution of the full in-medium potential without finite velocity corrections. For charmonium in heavy-ion collisions such corrections are expected to be relevant, as already at $T=0$ the static potential alone is not able to reproduce the vacuum charmonium bound states in an equally accurate fashion as is possible for bottomonium. The theory community thus need to develop further the effective field theory based approaches to heavy quarkonium, on the one hand to establish their validity in the non-perturbative regime (one recent example being Refs.~\cite{Brambilla:2019tpt,Eller:2019spw}) and furthermore for masses in which the separation of scales is not as pronounced, i.e. to higher order in their respective expansion schemes.

\subsubsection{Bottomonium in $A+A$ collisions}
\label{sec:bbbarprodhic}

Let us now turn to the heavier flavor, bottomonium, which due to its larger mass and the correspondingly smaller $b\bar{b}$ cross section requires higher luminosities or more sensitive detectors at the same beam energies than charmonium. The absence of significant non-prompt contributions to their production on the other hand simplifies the analysis. The most comprehensive results to date, including individual measurements of the $R_{AA}$ for the bottomonium ground state, as well as the first and second excited state, stem from the CMS collaboration at LHC (for the most recent instalment see e.g. Ref.~\cite{Sirunyan:2018nsz}, for a review see Ref.~\cite{Hu:2017pat}). More recently the ALICE collaboration has also published first measurements of the ground state $R_{AA}$ in Ref.~\cite{Acharya:2018mni}. At RHIC, the STAR collaboration has presented recent measurements \cite{STARupsilon} of the ground state $R_{AA}$ as well as the combined $R_{AA}$ of ground and excited states. 

\begin{figure}
\centering
\includegraphics[scale=0.28]{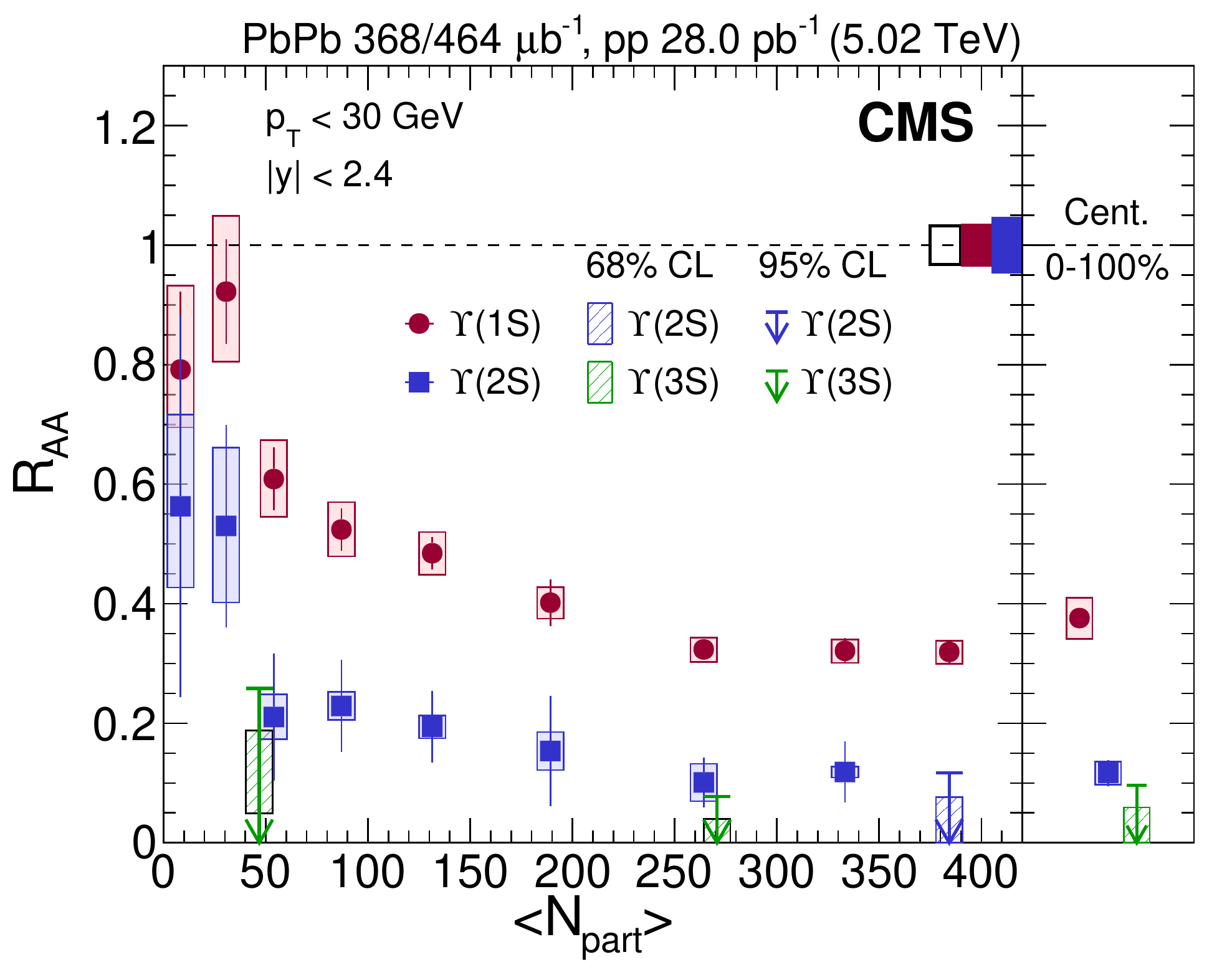}
\includegraphics[scale=0.25]{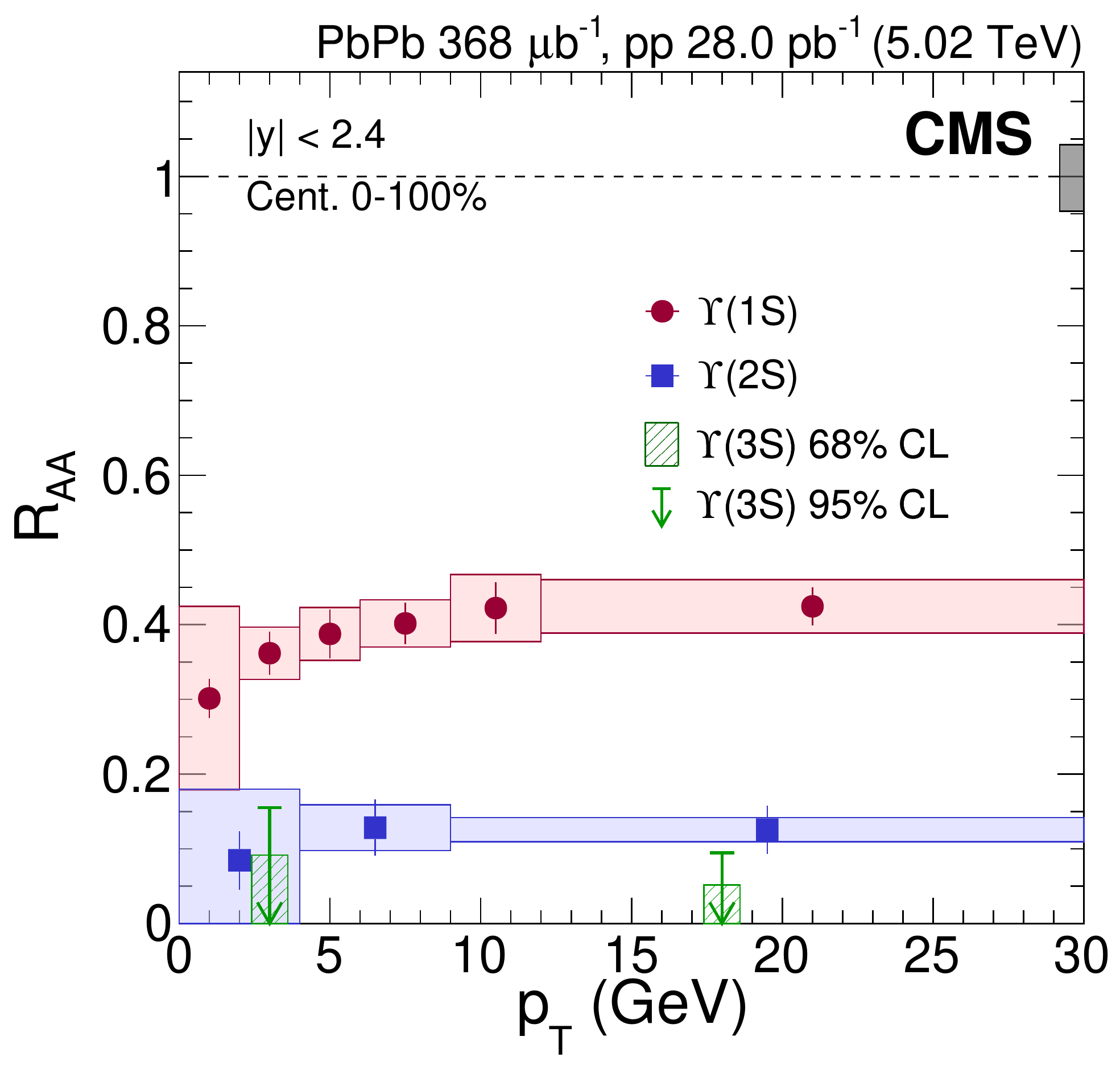}
\includegraphics[scale=0.23]{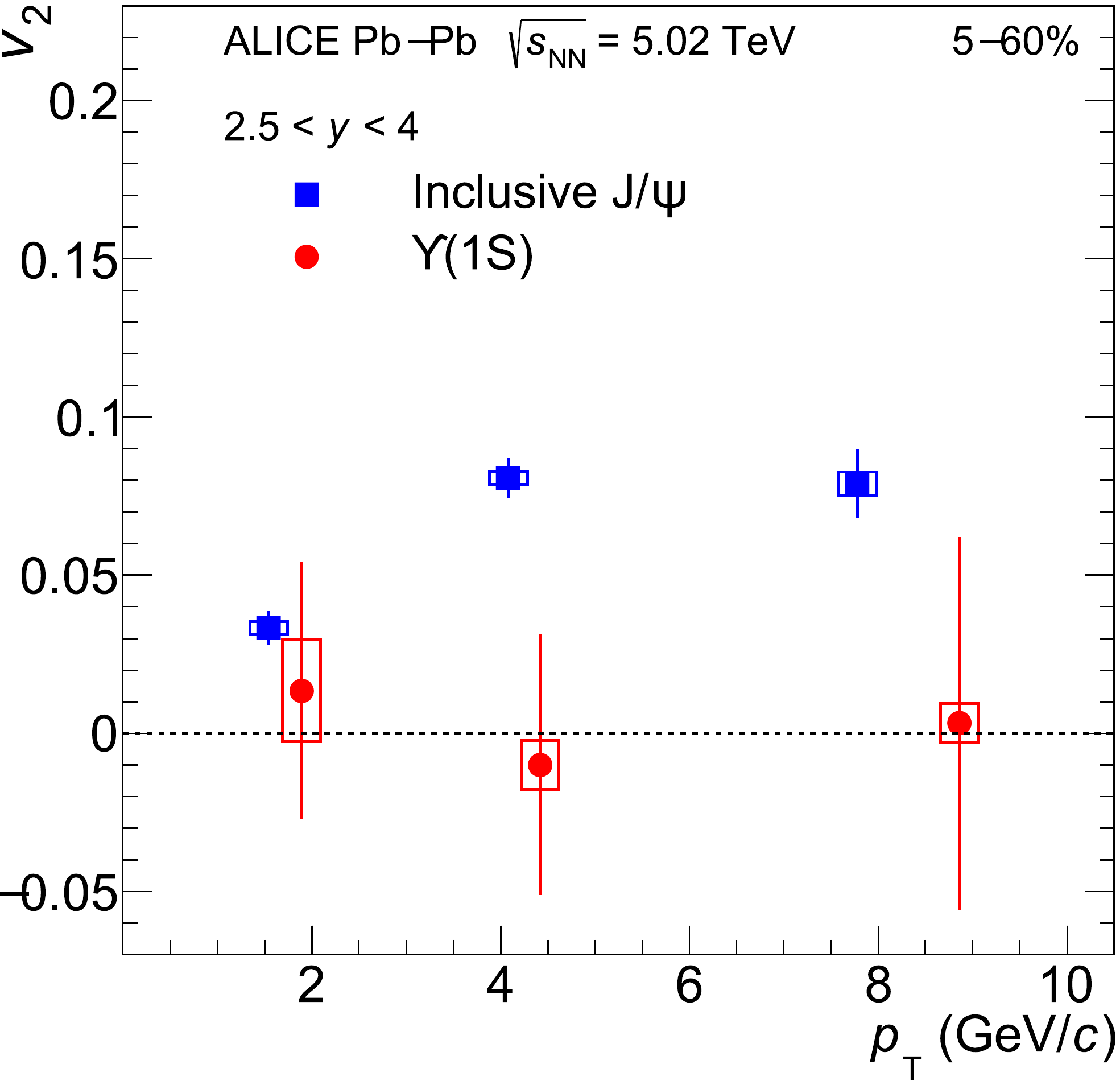}
\caption{(left) The nuclear modification factor for the ground, first and second excited state of vector channel bottomonium versus collisions centrality measured by the CMS collaboration at LHC at $\sqrt{s_{NN}}=5.02$TeV. Note the ordering of the strength of the suppression with the binding energy of the individual states. (center) The same $R_{AA}$ now plotted as a function of transverse momentum. Figures reproduced from Ref.~\cite{Sirunyan:2018nsz}. (left) Recent measurements of the ALICE collaboration of the elliptic flow of $\Upsilon(1S)$ (red) versus transverse momentum at $\sqrt{s_{NN}}=5.02$TeV. The values for $J\/psi$ at the same beam energy are given as blue symbols. Figure adapted from Ref.~\cite{Acharya:2019hlv}.}\label{fig:qqbarHICmeas4}
\end{figure}

In \cref{fig:qqbarHICmeas4} we showcase a selection of recent and characteristic measurements from LHC run2 at $\sqrt{s_{NN}}=5.02$TeV. The left and center panel contain measurements of the $R_{AA}$ of the ground and excited states from the CMS collaboration \cite{Sirunyan:2018nsz} plotted versus the centrality of the collision and versus transverse momentum respectively. The values show clear and consistent suppression patterns. The more central the collision becomes the stronger the suppression. At the same time, the excited states are more strongly suppressed than the more strongly bound ground state. While the uncertainties are still significant at small $p_t$ there appears to be a trend at least for the ground state that the suppression is stronger close to $p_T=0$ than for those at the largest momenta. A similar trend for the first excited state is not visible at the moment. This behavior is decidedly different from that found for charmonium (see e.g. \cref{fig:qqbarHICmeas3}) and at first sight is reminiscent of what a scenario based primarily on quarkonium melting would suggest. The hot medium in the collision center destabilizes the more weakly bound states more thoroughly and this suppression is most pronounced where the quarkonium spends most time in the hot environment. This conclusion is supported by the fact that by going from $\sqrt{s_{AA}}=2.76$TeV to $\sqrt{s_{AA}}=5.02$TeV the suppression in bottomonium is found to become slightly stronger, while it becomes weaker in charmonium. 

While equilibriation with the environment has been an important part of the dynamics of charmonium no significant signs of participation in the collective motion of the bulk have so far been observed for bottomonium, as shown in recent measurements of $v_2$ by the ALICE collaboration in the right panel of \cref{fig:qqbarHICmeas4}. One should however keep in mind that the errorbars on these first measurements are quite large and are compatible with small values of $v_2$ predicted by some models (see e.g. Ref.~\cite{2018arXiv180906235P}).

The naive visual inspection of the experimental results suggest that bottomonium, in contrast to charmonium at LHC, behaves as a genuine non-equilibrium probe of the collision center, for which regeneration effects do not play a major role. Since the $b\bar{b}$ cross section at RHIC is even smaller and the lifetime of the medium produced shorter than at LHC, this conclusion is expected to remain equally valid at $\sqrt{s_{NN}}=0.2$TeV.

On the theory side Bottomonium, due to its larger mass, is a promising candidate for a direct application of effective field theory based approaches. In particular descriptions based on the static in-medium potential are expected to provide a reasonable description of the relevant physics. While the large bottom mass $m_b\gg \Lambda_{\rm QCD}$ allows the partonic processes in $b\bar{b}$ production to be described in a fully perturbative manner, the in-medium evolution may still require non-perturbative insight, as only the binding energy of the ground state Upsilon lies above the characteristic QCD scale $E_{\rm bind}^{\Upsilon(1S)} \gg \Lambda_{\rm QCD}$.

From the point of view of the dynamical description of heavy quarkonium discussed in \cref{sec:qqbarrealtime}, bottomonium modeling has recently entered an phase of rapid evolution. The open quantum systems approach has led to new avenues to implement the real-time evolution of this quarkonium species leading to Lindblad-type master equations derived systematically from QCD. While many of these approaches still use weak-coupling arguments in intermediate steps first fully non-perturbative formulations are under development. At high temperature we discussed in \cref{sec:FVIFOQS} that when one remains within the language of wavefunctions, a non-linear stochastic Schr\"odinger equation needs to be solved to account for the full dissipative dynamics of the quarkonium system in a QCD medium. In terms of distribution functions for the quarkonium states, we have seen in \cref{sec:BoltzmannEqOQS} that a Boltzmann equation can be derived based on a set of clearly specified assumption. Note that all the derivations of real-time dynamics in  \cref{sec:qqbarrealtime} were based on a fully thermal background. When considering an evolving medium, as is present in the case of a heavy-ion collision a new timescale, describing the cooling process enters. If that scale is large enough compared to the other relevant timescales we may regard the temperature of the medium as external parameter, governing the values of e.g. the potential of the stochastic Schr\"odinger equation, an approximation that is used in most models today.

Historically the most common approach to bottomonium modeling is to resort to solving a deterministic Schr\"odinger equation which is governed by an in-medium potential. From our discussion in \cref{sec:DecDynMelt} we saw in \cref{eq:schroedavg} that such an ansatz amounts to a truncation of the dynamics, neglecting dissipative effects, and in addition to an adiabatic approximation that averages over the fluctuations. From the numerical tests within the stochastic potential model, we learned that in such an adiabatic truncation the survival of quarkonium states may be underestimated if the effects of decoherence are relevant. I.e. the $R_{AA}$ computed in that way may underestimate the correct physical value.

the Kent State University group has explored the suppression of bottomonium in heavy-ion collisions in detail, combining the deterministic Schr\"odinger equation with an anisotropic hydrodynamic description of the medium background evolution. They use a a Glauber model based initial distribution of primordial states and take into account the decays from feeddown after hadronization. A full description of the framework is found in Ref.~\cite{Strickland:2011aa}, where in particular the implementation of the potential in an anisotropic background is discussed. Note that no recombination contribution enters in this framework. (A similar model based on isotropic hydrodynamics has been used in Refs.~\cite{Nendzig:2012cu,Nendzig:2014qka})

The potential used in the Schr\"odinger equation itself is an input of the model and so far two ansaetze have been deployed. Both of these models feature a Debye screened Coulombic in-medium part. For the in-medium string part either \cref{eq:Vs_ad_hoc} or the similar legacy Gauss law parametrization is used. The main difference between them lies the form of the imaginary part, which is taken to be the purely Coulombic HTL one in the former \cite{Strickland:2011aa,Krouppa:2015yoa,Krouppa:2016jcl,Krouppa:2017lsw}, while in the latter it contains both contributions from the Coulombic and string in-medium potential \cite{Krouppa:2017jlg}.

A prescription of how to initialize the Schr\"odinger equation for a mixed ensemble of bottomonium states in agreement with the scale separation of NRQCD has been put forward in Ref.~\cite{CasalderreySolana:2012av}. Its proposal for efficient evolution of such an ensemble however remains restricted to unitary time evolution, i.e. when dissipation effects are small.

\begin{figure}
\centering
\includegraphics[scale=0.12]{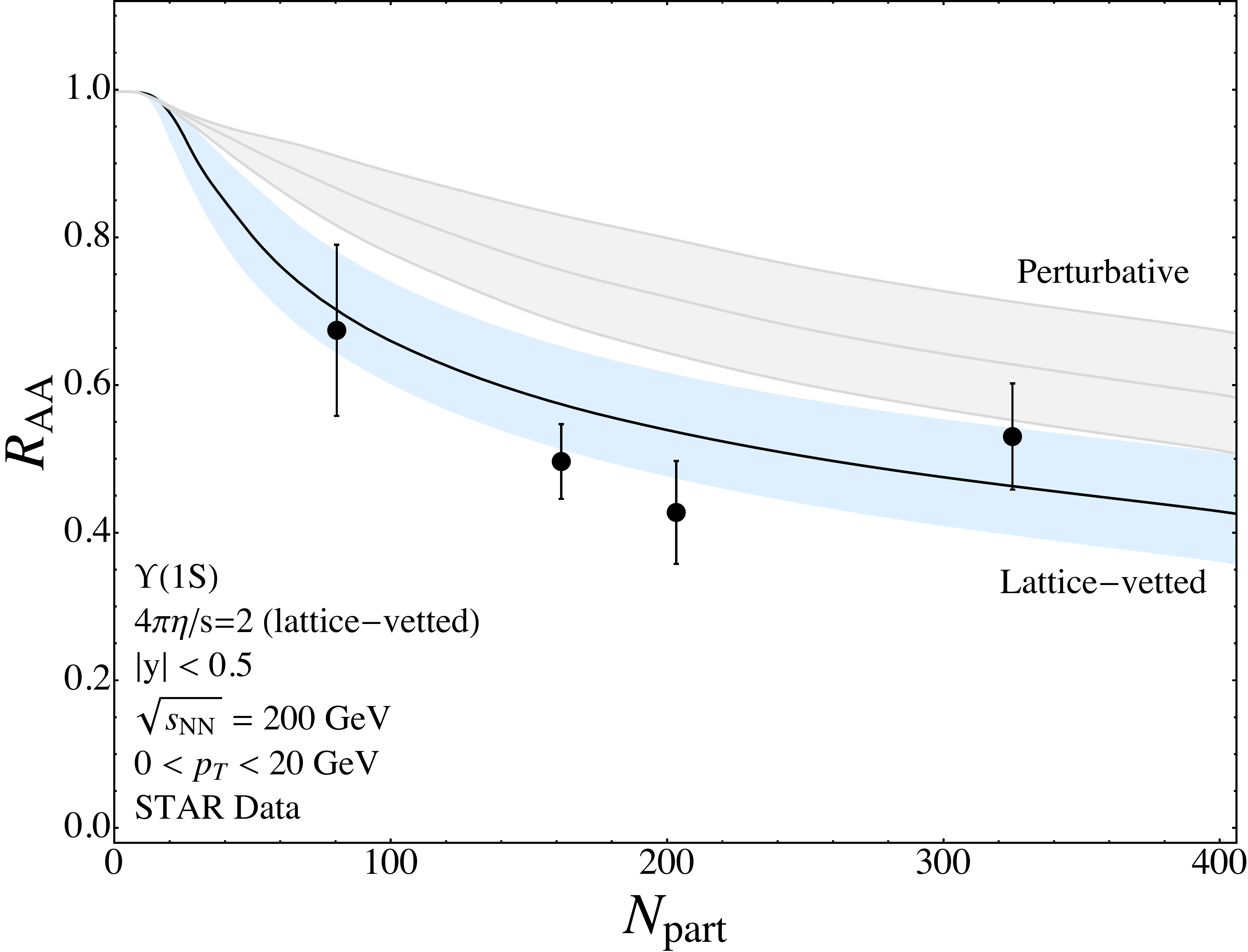}
\includegraphics[scale=0.35]{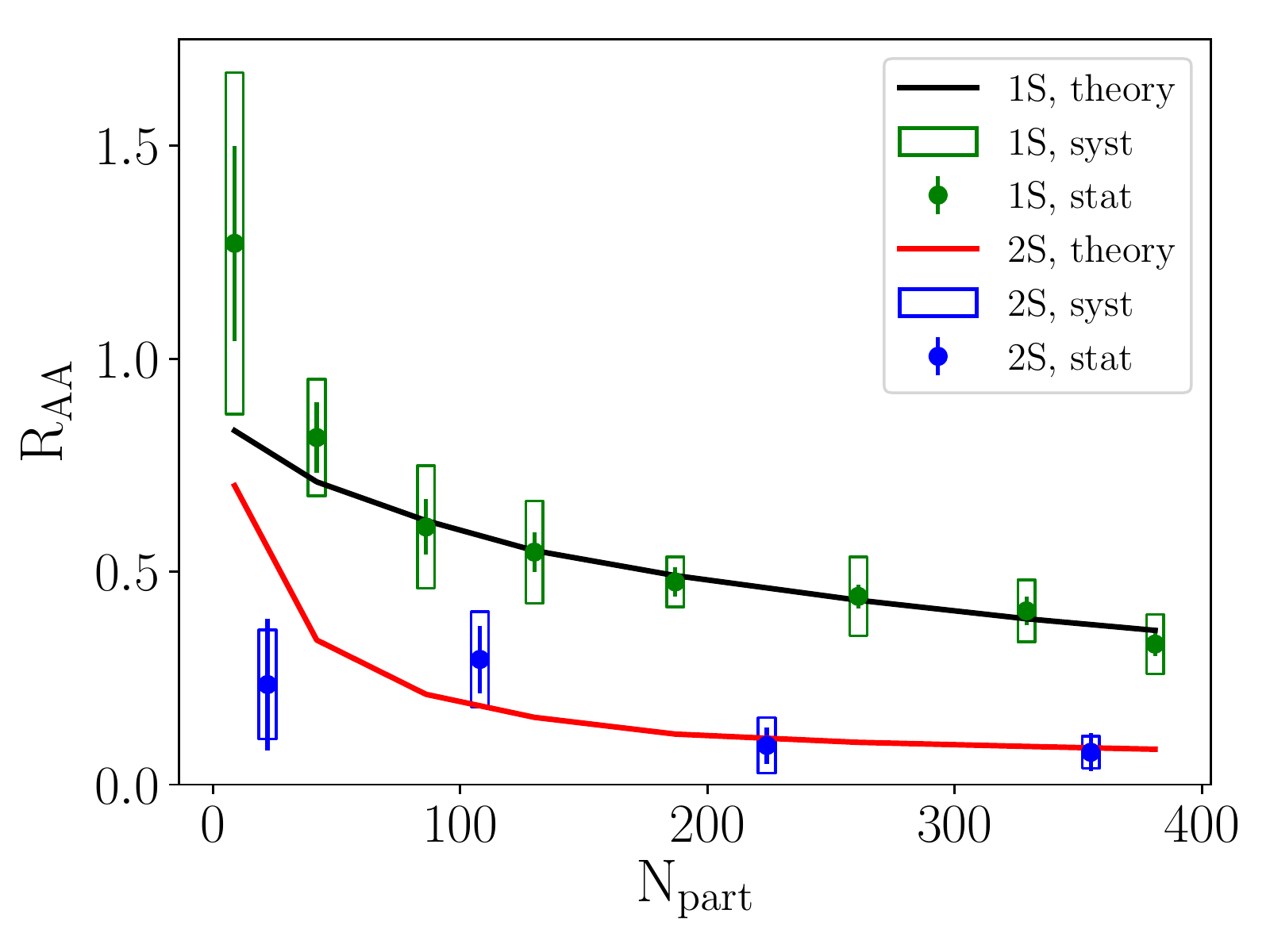}
\includegraphics[scale=0.25]{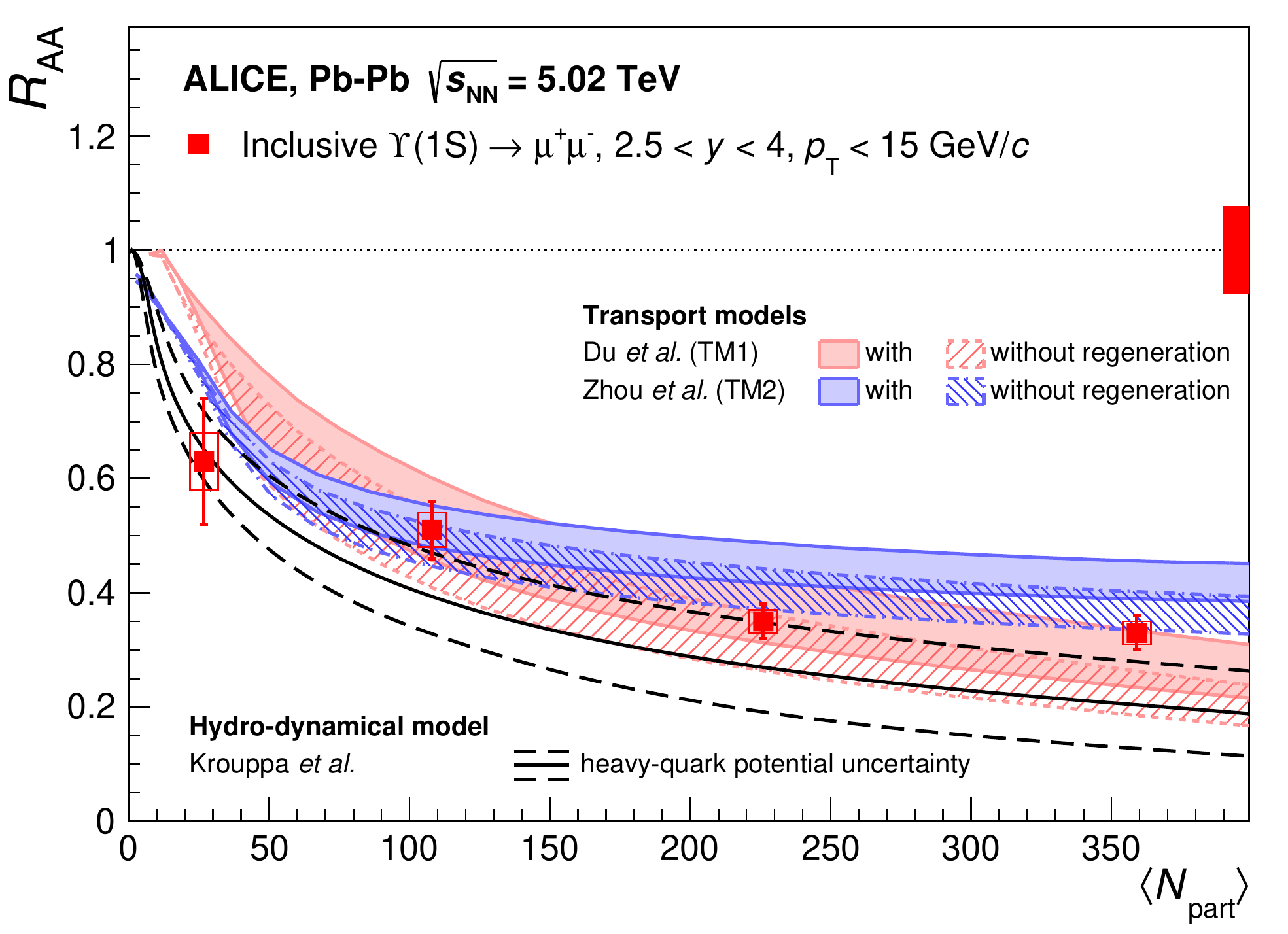}
\caption{A selection of comparisons of the $R_{AA}$ for bottomonium with recent phenomenological model studies. (left) The $\Upsilon(1S)$ nuclear modification factor as measured by STAR compared to a model based on a deterministic Schr\"odinger equation and anisotropic hydrodynamics background. The grey curve denotes the outcome using a perturbative ${\rm Im}[V]$ while the blue curve uses a Gauss-law based complex potential. Figure reproduced from Ref.~\cite{Krouppa:2017jlg}. (center) Comaprison of the CMS measurements for the bottomonium $R_{AA}$'s at $\sqrt{s_{NN}}=5.02$TeV with the combined Boltzmann-Langevin approach of Ref.~\cite{Yao:2018zrg}, from where this figure is reproduced. (right) The $\Upsilon(1S)$ $R_{AA}$ as measrued by the ALICE collaboration compared to transport models  either including or excluding a regeneration component. Figure reproduced from Ref.~\cite{Acharya:2018mni}.}\label{fig:qqbarHICmeas5}
\end{figure}

Let us start with Bottomonium at STAR, since there the separation of scales between the medium temperature created in the collision center and the bottom mass is most pronounced. In addition  bottomonium is expected to act as a true non-equilibrium probe, without reaching any significant degree of kinetic thermalization. In such a scenario, one may speculate that decoherence is not yet very effective and thus the adiabatic approximation may prove already satisfactory. In the left panel of \cref{fig:qqbarHICmeas5} we plot a comparison of the Bottomonium ground state $R_{AA}$ from recent measurements of the STAR collaboration together with the computed values from  Refs.~\cite{Strickland:2011aa,Krouppa:2017jlg}. The gray curve denotes the computation based on the potential using only the Coulombic imaginary part, while the blue shaded curve denotes the outcome from using a lattice vetted Gauss-law potential. One finds that the $R_{AA}$ is well reproduced from the Gauss-law potential whose stronger ${\rm Im}[V]$ leads to more efficient bottomonium destabilization and brings the computed values into agreement with experiment. One important difference between the two results is that the computation based on the Gauss-law potential is much less sensitive to differences in the values of the shear viscosity governing the hydrodynamics back ground evolution. I.e. while the gray band denotes the difference between changing the values for the shear viscosity over entropy ratio $\eta/S$ in the background hydrodynamics from $1/4\pi - 3/4\pi$, such a change has virtually no effect on the blue curve. There the error band stems instead from the uncertainty in determining the Debye mass parameter from lattice QCD simulations. 

In contrast to fully perturbative description of the in-medium evolution, where the contributions from e.g. Landau damping and gluo-dissociation can be explicitly disentangled, here the values of the imaginary part represent the aggregate of all these effects. I.e. the bottomonium state may be excited into another singlet state, bound or undbound or may go over to an unbound octet state. The suppression of e.g. the ground state contains two elements. On the one hand the ground state itself may be destabilized by the medium. On the other hand as argued in e.g. Ref.~\cite{Nendzig:2012cu} the melting of excited state quarkonium leads to a diminished contribution from feed-down, already reducing the $R_{AA}$ in the presence of an otherwise stable ground state.

What happens then at LHC, where bottomonium spends a significantly longer time in a much hotter medium? Comparing the predictions for LHC at $\sqrt{s_{NN}}=5.02$TeV based on the potential with Coulombic ${\rm Im}[V]$ from Refs.~\cite{Krouppa:2016jcl,Krouppa:2017lsw} with the postdictions using the Gauss law potential in Ref.~\cite{Krouppa:2017jlg} an interesting trend emerges. The potential that led to overestimation of $R_{AA}$ at RHIC now manages to describe the yields with very good accuracy at LHC. On the other hand the Gauss-Law model already at $\sqrt{s_{NN}}=2.76$TeV starts to show signs of underestimating $R_{AA}$ which it clearly does at the highest LHC energies. There are several possible mechanism at play here. On the one hand the imaginary part of the potential in the Gauss-law parametrization has only been vetted to tentative values of the lattice ${\rm Im}[V]$ at relatively low temperatures $T<300$MeV. Whether the Gauss law potential describes the imaginary part accurately at higher temperatures still needs to be ascertained, which is work in progress. On the other hand if the Gauss-law potential is indeed the appropriate potential to use then two sources for the underestimation of the $R_{AA}$ come to mind. The first one is related to the effects of fluctuations and dissipation that have been discarded in the adiabatic approximation. Indeed the stochastic potential computations suggest that when these effects are included the $R_{AA}$ should increase. Ongoing efforts towards developing and implementing fully dissipative dynamics in the open-quantum systems language promise to provide the necessary tools to do so in the near future. At the same time one has to investigate whether regeneration effects have already set in at the accessible LHC energies. 

The last question has been investigated in the context transport models to Bottomonium evolution. Among the used models are variants of the two approaches \cite{Zhou:2014hwa,Du:2017qkv,Du:2017hss} we have already met in the study of charmonium, as well as a new approach, which is a combined transport model based on the Boltzmann equation derived from pNRQCD \cite{Yao:2018sgn} and a Langevin evolution \cite{Ke:2018tsh} for bottom quarks in order to allow for dynamical regeneration. Due to the explicit treatment of the heavy quarks it does not rely on detailed balance to implement recombination while still being able to reach equilibrium distributions in the late time limit.
 
All transport models manage to describe the measured ground and first excites state $R_{AA}$'s for bottomonium at LHC in a satisfactory manner, one example from Ref.~\cite{Yao:2018zrg} is shown in the center panel of \cref{fig:qqbarHICmeas5}. In general a small but non-vanishing regeneration component is found to be required when considering the measurements from CMS. For the latest results from the ALICE collaboration on the $\Upsilon(1S)$ suppression, the situation is not as clear, as shown in the center panel of \cref{fig:qqbarHICmeas5}. There the experimental data is better reproduced with regeneration in one and without regeneration in another model. It would be interesting in this context to see the results of Ref.~\cite{Yao:2018zrg} divided into their dissociation and regeneration contributions.  

The path forward for bottomonium studies is clear. On the one hand the potential based studies need to incorporate the effects of dissociation via a genuine stochastic evolution prescription for the wavefunction of the quarkonium system. This will remove a significant source of systematic uncertainty currently inherent in the adiabatic approximation. The explicit coupling of color singlet and color octet sectors, which is work in progress, will eventually allow to self-consistently treat the effects of recombination in the wavefunction language. It will also allow to disentangle the different effects currently summarized in the imaginary part of the singlet potential. In addition the determination of the potential, especially of its imaginary part in non-perturbative lattice QCD needs to be improved, to reduce extrapolation artifacts at high temperatures relevant in the beginning of the QGP evolution. All of these aspects are in principle feasible and are on the agenda of the theory community.

On the side of transport models the derivation of the Boltzmann equation from pNRQCD via the open quantum systems approach has been a major step forward. The next step has to be to compute the matching coefficients required in this context from lattice QCD with high precision. The Langevin based approach to heavy quark evolution will eventually require precision knowledge of e.g. the heavy quark diffusion constant whose extraction from the lattice is also actively pursued.

In both cases efforts need to be intensified to put the derivation of the underlying evolution equations on a non-perturbative footing. The dissipative dynamics of the wavefunction approach currently relies on a Feynman Vernon influence functional, derived at weak coupling, while the Boltzmann equation was obtained with weakly coupled pNRQCD in mind. These are exciting challenges that have the potential elevate phenomenological modeling even closer to first principles theory as has already been possible over the past decade.

\begin{summary}
Heavy quarkonium in heavy-ion collisions constitutes a rich and challenging field of study. A microscopic understanding of quarkonium production requires insight not only into the evolution of quark antiquark pairs in a hot environment but also into the non-perturbative composition of the incoming projectiles and the dynamics of hadronization. Such an all-encompassing view of quarkonium production has already lead to a significantly improved understanding of the phenomenon of quarkonium suppression. It arises from an intricate interplay of quarkonium destabilization  due to the environment present in the collision center, as well as from recombination of heavy quark antiquark pairs, when produced in ample numbers in the earliest moments of the collision. One current focus of the community is to hone in on the quantitative details of the production mechanism for charmonium, shedding light e.g. on the role of primordial versus regenerated yields. This task requires to more strongly discriminate between the various models currently able to reproduce the experimental $J/\psi$, i.e. ground state $R_{AA}$. Both the experimental efforts to measure excited states such as $\psi^\prime$ and the overall charm cross section, as well as theory efforts to reduce the modeling input by bringing transport models closer to first principles QCD, will be essential in this regard. For bottomonium direct effective field theory based approaches either in the language of wavefunctions or distribution functions have already been quite successful in reproducing the hierarchical suppression patterns observed at LHC. With a much more pronounced separation of scales at hand for bottomonium, theory is looking ahead to using the open-quantum systems framework to systematically derive non-perturbative evolution equations treating both the color singlet and octet sector explicitly. With the dynamics of these degrees of freedom governed by non-perturbative matching coefficients, such as transport coefficients and potentials, efficient extraction procedures from QCD are being investigated. Bottomonium thus appears to provide a unique opportunity to develop a genuine non-perturbative and first principles based real-time formalism for a system of strongly interacting matter, eventually capable of connecting microscopic QCD and experimental measurements of heavy quarkonium in extreme conditions.
\end{summary}

\section*{Conclusion}

It is an exciting time to work on heavy quarkonium in extreme conditions (for a recent community perspective see e.g. Ref.~\cite{Aarts:2016hap}). Experiment has amassed a wealth of high precision data on quarkonium production in relativistic heavy-ion collisions at RHIC and LHC in different kinematical regimes that provides a challenging testing ground for theory and phenomenology. These include the nuclear modification factors of the charmonium ground state, as well as those of the bottomonium ground and excited states. Measurements of both $J/\psi$ and $\Upsilon$ elliptic flow are by now also available. Experiments are currently taking aim at even more challenging measurements, such as a detailed study of the excited state of charmonium $\psi^\prime$. The determination of the overall charm and bottom  cross section is equally high up on the agenda. Hopefully luminosities and detector performance in the future will allow to also capture the P-wave states in heavy-ion collisions, which so far have only been studied in proton-proton collisions.

We have seen that charmonium at RHIC and LHC behave quite differently with regeneration taking on a more and more important role at higher energies. In order to learn in more quantitative detail how this change in production mechanism proceeds, and also to prepare for bottomonium at future higher energy colliders, where it is expected to behave very similarly to charmonium at LHC, a series of lower energy collisions at LHC would be very instructive. The planned electron-ion colliders on the other hand will in the near future provide unprecedented insight into the internal make-up of the nuclear projectiles allowing to much better constrain the initial conditions from which heavy-quarkonium production proceeds in a heavy-ion collision.

This report discussed several aspects in which the theory understanding of quarkonium in thermal equilibrium has improved over the last decade. Better access to ground state spectral properties from combined lattice QCD and effective field theory studies, as well as the development of a QCD derived and non-perturbatively evaluated in-medium potential are two examples. While progress has been made, a lot of challenging work remains. Better access to the excited state spectral properties in direct lattice QCD studies will require both advances in simulation algorithms and data analysis strategies. At the same time extending the in-medium potential beyond the lowest order in pNRQCD asks for both conceptual and numerical work on the effective field theory and lattice QCD side.

When it comes to quarkonium real-time descriptions the community is in a state of heightened activity. Several groups concurrently explore complementary ways to derive and implement an open-quantum systems treatment of heavy quarkonium. Some of these efforts in the context of weak-coupling approaches have been highlighted in this report. The central task at hand for the near future is to gain an understanding how the real-time description can be put on a genuine non-perturbative footing. This will entail connecting the transport coefficients and potentials governing the evolution with first principles lattice QCD simulations. Similarly the transport properties of individual heavy quarks, such as the heavy quark diffusion coefficient need to be determined with much higher precision than is possible today. These challenges will require a close collaboration between practitioners in the fields of open-quantum systems, effective field theory and lattice QCD.

Theory is also making inroads into a first principles understanding of the initial conditions of proton-proton collisions. Recent developments in defining and extracting parton distribution functions of nucleons from lattice QCD simulations (the current status is discussed in Ref.~\cite{Lin:2017snn}) promise to provide novel complementary constraints to the momentum and spin distributions of partons in the proton projectiles. While only on the horizon, the final goal would be to eventually understand at least partially the changes induced in the parton distribution functions once more than one nucleon is present in the projectile. With the rapid pace of advances in lattice QCD simulations over the past decade such insight might not be out of reach in the next decade. 

There are many aspects of heavy quarkonium in extreme conditions that did not find their way into this report. One is the physics at finite baryon chemical potential, whose study from first principles is still hindered by lattice QCD simulations suffering from the sign problem. Future collider facilities such as NICA and FAIR, as well as the beam energy scan at RHIC set out to explore this region of the QCD phase diagram in more detail, offering additional motivation for theory to develop novel computational approaches for this regime.

Another topic is the physics of quarkonium at large external magnetic fields (for a review on magnetic fields in heavy-ion collision see e.g. Refs.~\cite{Hattori:2016emy,Andersen:2014xxa}), which recently has attracted attention in the community (see e.g. Refs.~\cite{Bonati:2015dka,Hasan:2017fmf,Fukushima:2015wck}). With first principles computations possible, the question to answer is whether or how such large fields can exert an influence on quarkonium formation in the early stages of a collisions.

This bring us to the third topic which is the formation dynamics of quarkonium at early time. Currently only simple estimates based on the uncertainty principle are used to argue why some quarkonium stated may form early on in the pre-thermal phase. To gain a true first principles understanding of quarkonium production we will however need to develop genuine real-time descriptions for heavy quark pairs in the presence of strong coherent glasma fields. First steps in that direction are being taken at the moment by combining classical statistical simulations of gauge fields with a real-time implementation of the effective field theory of NRQCD \cite{LehmannLattice}.

The study of heavy quarkonium in extreme conditions hence remains an active field of research with many challenging facets for both experiment and theory. The start-up of new colliders in the near future and the continuation of successful programs at current facilities, all with quarkonium on the agenda, promises continued research funding opportunities in the field for the next decade. This support will help the theory community to sustain its activities in the quest for a truly microscopic QCD based understanding of quarkonium in extreme conditions.

\subsubsection*{Acknowledgments}

The author thanks Y. Akamatsu and P. Petreczky for insightful discussions and the nuclear theory group at the University of Tokyo for their hospitality, where part of this manuscript has been composed. Funding from the Research Council of Norway under the FRIPRO Young Research Talent grant 286883 and grant 295310 are gladly acknowledged.

\printcredits

\bibliographystyle{elsarticle-num}

\bibliography{InMediumQuarkonium}


%\bio{figs/Lichtbild}
%Alexander Rothkopf is a tenured associate professor for physics at the University of Stavanger in Norway. After obtaining his PhD from the University of Tokyo he worked as postodoctoral research associate at the Universities of Bielefeld, Bern and Heidelberg, before taking on the role as a principal investigator and scientific manager of the collaborative research center ISOQUANT in Heidelberg. His research focuses on the intersection of in-medium heavy quarkonium, real-time dynamics of strongly correlated quantum systems and lattice QCD. Among his main contributions are the first non-perturbative computation of the complex in-medium heavy quark potential, the design of a novel approach to spectral function reconstruction from lattice QCD simulations and the development of a novel open-quantum systems description of quarkonium real-time dynamics.

\end{document}